\documentclass[10pt,a4paper, openright]{book}
\usepackage{standalone}
\usepackage[a4paper, top=2.0cm, bottom=1.7cm, left=2.0cm, right=1.7cm]{geometry}
\usepackage{amsmath}
\usepackage{amssymb}
\usepackage{graphicx}
\usepackage{array}
\usepackage{multirow}
\usepackage[title]{appendix}
\usepackage{placeins}
\usepackage{floatrow}
\usepackage{caption}
\usepackage{physics}
\usepackage{gensymb}
\usepackage{empheq}
\usepackage{bm}
\usepackage{braket}
\usepackage{booktabs}
\usepackage{tikz}
\usepackage{subfig}
\usepackage{pgfplots}
\usepackage{pgfplotstable}
\usetikzlibrary{plotmarks}
\usepackage{pdfpages}
\usepackage[english]{babel}
\usepackage[utf8]{inputenc}
\usepackage[numbers,sort&compress]{natbib}
\usepackage{url}
\usepackage{doi}
\usepackage{hyperref}

\usepackage[labelfont=bf]{caption}
\usepackage[]{frontespizio}

\usepackage{nomencl}
\makenomenclature

\usepackage{fancyhdr}
\usepackage{scrextend}
\usepackage{mathrsfs}
\usepackage{setspace}
\usepackage{verbatim}
\usepackage{color}
\usepackage{slashed}
\usepackage{xfrac}
\usepackage{enumerate}
\usepackage{physics}
\usepackage{lipsum}  
\usepackage{caption}

\definecolor{torgreen}{rgb}{0.0, 0.5,0.2}  

\newcommand{\ft}[2]{{\textstyle\frac{#1}{#2}}}
\newcommand{\nn}{\nonumber}

\def\be{\begin{equation}}
\def\ee{\end{equation}}
\def\bea{\begin{align}}
\def\eea{\end{align}}
\def\beaq{\begin{eqnarray}}
\def\eeaq{\end{eqnarray}}

\renewcommand{\quote}[1]{``#1''}

\def\a{\alpha} \def\b{\beta} \def\g{\gamma} 
   \def\q{\theta}
 \def\k{\kappa}  \def\m{\mu}
\def\n{\nu}    
\def\s{\sigma} \def\t{\tau}

   \def\L{\Lambda}

\makeindex

\begin{document}

\thispagestyle{empty}

\begin{titlepage}
	\begin{center}
		\includegraphics[width=0.65\textwidth]{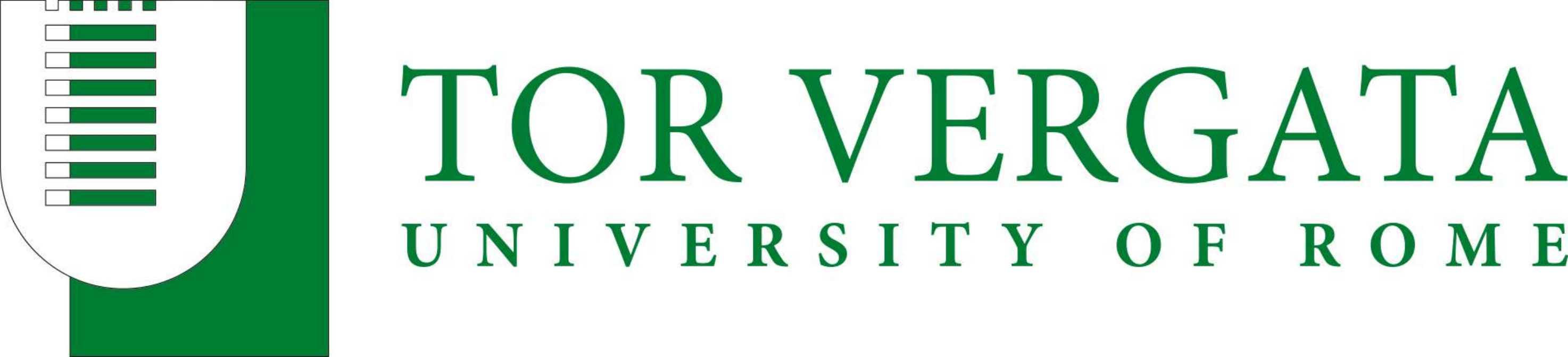}
		
		\vspace*{2.5cm}
		{\color{torgreen} Ph.D. in Physics \\ 
			Ph.D. Cycle: XXXVIII \\}			
		\vspace*{2.5cm}		
		\textbf{{\color{torgreen}\Large{Aspects of $\mathcal{N}=2$ SUSY gauge theories and applications to Quantum Field Theory and Gravity}}} \\		 
		\vspace{2.5cm}				
		\textbf{\large Andrea Cipriani} \\
		Academic Year 2025/2026 \\
		\vspace{3.5cm}

						Advisors: \\
						Prof. Francesco Fucito \\
						Prof. Josè Francisco Morales \\
						Prof. Raffaele Savelli \\
						\vspace{1cm}
						Coordinator of the Ph.D. Course: \\
						Prof. Massimo Bianchi \\
						\vspace{1cm}
						Thesis defended on the $11^{\text{th}}$ February 2026.\\
						Committee composed by: \\
						Prof. Massimo Bianchi \\
						Prof. Alba Grassi \\
						Prof. Julio Parra-Martinez \\

		\vfill
	\end{center}
\end{titlepage}
\clearpage

\newpage
\thispagestyle{empty}
\null\vfill
\begin{flushright}
	\textit{If you will forgive me for being personal,\\
		I do not like your face, Mr. Ratchett.}\\ \bigskip\medskip
	Agatha Christie, \textit{Murder on the Orient Express}
\end{flushright}
\vfill\vfill

\clearpage

\chapter*{Acknowledgements}
\thispagestyle{empty}
When writing the acknowledgements for the work I have completed, it is customary to mention one's supervisors among the first. Dearest \textbf{Francesco} and \textbf{Francisco}, however, the reason why I begin by thanking you is not a matter of convention, but because I am immensely grateful to you for everything you have done for me, and for the scientific and human example you have given me. You are like two adjacent pieces of a puzzle: so different from one another, yet so complementary in contributing to the beauty of the picture they create together. I hope to always carry with me the legacy I have received from you, with the aim of becoming another piece of the puzzle that fits together with yours. Thank you also, \textbf{Raffaele}, for supporting me during the first half of my PhD and encouraging me to understand which path was the right one for me. I would also like to thank all the other Professors and Researchers in our group, \textbf{Massimo Bianchi}, \textbf{Giuseppe Dibitetto}, \textbf{Alejandro Ruipérez}, and \textbf{Mohammad Akhond}, for the fruitful discussions I had the opportunity to share with you over these three years, from which I always learned something, whether great or small. As for the PhD students, I will return to them shortly, together with everyone else.\\
I would also like to thank \textbf{Alba Grassi} and \textbf{Cristoforo Iossa} for giving me the opportunity to interact with them for some time at CERN, and \textbf{Carlo Heissenberg} and \textbf{Rodolfo Russo} for welcoming me into the collaboration we started with Francesco and Francisco. My visit to IPhT was also extremely fruitful, and therefore another reason for which I am grateful to you. Finally, I would like to thank you, \textbf{Alessandro Nagar}, not only for welcoming me into the scientific community in your field and allowing me to interact with people who have been invaluable from a scientific perspective, but also for giving me the opportunity to work with you and learn many things about your area of research.\\

Turning to my dear fellow PhD students, there are many of you, and therefore I will list, in alphabetical order, those who have left some mark on my person and soul over these years, while asking forgiveness from anyone I may forget.\\
\textbf{Alessandro}, getting to know you first and then working with you have been not only sources of great personal and scientific enrichment, but also of joy and personal pride. Your tenacity, precision, and sense of humour are qualities that made me realise that becoming your friend was the right thing to do. I hope to draw increasingly on those qualities throughout my own journey, and so I thank you for always being there. \textit{A tutte l'ore, alligatore! Al primo squillo, coccodrillo!}\\
\textbf{Beatrice}, your origami, memes, and jokes accompanied me during the early stages of this journey, and I treasure memories of them filled with so much joy and fun. Interacting with you was, for me, a healthy exercise in understanding sarcasm better, and so, since you left, I have become a little rusty. In any case, thank you for all the laughs you have always given me.\\
\textbf{Francesca}, the words I am about to use to explain why I am grateful to you do not do justice to what I felt while spending my days with you. You are so kind and selfless, a true member of Hufflepuff; you are funny, reliable, generous (perhaps even too generous), and intelligent. Perhaps the expression “the jewel in the crown of our PhD programme” captures, quite well, though only partly, what I mean. So thank you for the person you are and for all the care and attention you have shown me over these years.\\
\textbf{Giorgio}, you have been an older brother to me in Physics. If it had not been for you, for the affection you showed me and that I felt from you, and for welcoming me into the first projects I worked on, I would not have made it this far. For this reason, you would perhaps deserve to be mentioned at the beginning, but I wanted to place you here because the scientific interest you showed in me was not driven by any institutional obligation attached to your role. Therefore, I not only thank you for all your support and for your patience in working with me, but I also renew my very best wishes that you may achieve everything you hope for.\\
\textbf{Giuseppe}, from the very beginning of our PhD, I have considered you my partner in crime in Physics. Ever since the first Journal Club we did together, I felt as though I had known you for a lifetime, and I was always surprised whenever someone else asked me whether you and I had known each other forever. To me, this was confirmation that the understanding we had developed was truly deep. As you always say, despite being completely different and having literally opposite interests, we have always understood each other with just a glance during conferences, Journal Clubs, and so on. So, in addition to thanking you once again for the first project you allowed me to work on with you, thank you also for all the times you opened up to me, regarding me as someone you could rely on.\\
\textbf{Lorenzo G.}, you have been a revelation to me. Although we have known each other since our Bachelor's degree, I have seen a side of you that really struck me. As I always say, despite being so rigorous, articulate, and intelligent, you always have a joke in \textit{romanaccio} ready at exactly the right moment, provoking uncontrollable laughter in me that still echoes through the walls of our PhD students' room, just like your jokes themselves. So thank you for all the hearty laughs you have always given me, and also for choosing to interact with me during your Master's thesis work.\\
\textbf{Luca}, although you arrived almost halfway through our first year (even if they only started paying you in November), you immediately earned an important place, both in the PhD students' room and in making my days lighter with your sense of humour. You were my first accomplice in decorating our office with Christmas decorations and, despite the brief deviation into Halloween decorations that you put on my desk, you proved that the Christmas Rehabilitation Office works wonderfully. Thank you, truly, for the carefree spirit with which you brightened my days.\\
Finally, I would like to thank \textbf{Alfredo}, \textbf{Antonio}, \textbf{Carlo}, \textbf{Chiara (Camilla)}, \textbf{Claudio}, \textbf{Daniele M.}, \textbf{Daniele T.}, \textbf{Elena}, \textbf{Filippo}, \textbf{Federico}, \textbf{Francesco}, \textbf{Lorenzo B.}, and \textbf{Valentina} for the wonderful moments you chose to share with me, allowing me to enjoy so many peaceful and fun days.\\

As for my friendships outside the working world, I would like to warmly thank \textbf{Matteo} and \textbf{Stefano}, because after we completed our university studies, our bond remained strong, even though we now work in different fields. I have held you in very high esteem ever since our first laboratory class together, and I hope to remain in touch with you for all the years to come. Thank you also, \textbf{Fabio}, for remaining one of the few long-standing friends who has continued to share with me the joys you have experienced over these years, both your own and those of your family. \textbf{Alessia}, seeing how our bond after high school has not only endured but has grown stronger over these years fills me with immense joy. Having a pathologist like you as a friend gives me an indescribable sense of pride, and I wish you all the very best. Finally, my most heartfelt thanks go to you, \textbf{Chiara}, not only because you allowed me to become part of your life in every respect after we rediscovered each other at university, having known each other since primary school, but also because, with your carefree spirit, immeasurable sense of humour, and all the times you opened up to me about your own struggles, you helped me lighten many difficult situations I went through. For this reason, you have been very important to me on a personal level. Thank you, truly, from the bottom of my heart.\\

Turning to the spiritual side of my life, I would first like to thank \textbf{Don Francesco} for the guidance and closeness you have given me, and which I hope you will continue to give me in the years to come. You made me experience things that I would never have experienced on my own, and for that I am grateful as well. \textbf{Enrico}, your closeness to me as my godfather has always touched me, and it has been precious to me over these years; for this, I thank you immensely. I hope that you, too, can feel my closeness to you. Thank you also, \textbf{Roberto}, for all the esteem you have for me, which is entirely undeserved, and perhaps even somewhat exaggerated. Finally, I would like to thank the entire \textbf{Choir}: those who have been there since the beginning, those who were there for a while, and those who have joined more recently. Your presence, through our rehearsals and our singing, has been an important point of reference for me over these years. I will always carry you with me.\\

As the saying goes, \textit{dulcis in fundo}, and so I cannot end without thanking my family, in particular my parents, \textbf{Cinzia} and \textbf{Raffaele}, and my brothers, \textbf{Pietro} and \textbf{Davide}. The words I would need to express my gratitude to you would not be enough to fill this thesis, so I will have to make do with summarising them in the space that follows. If it had not been for your love, your carefree spirit, the fun you brought into my life, your support, comfort, encouragement, and, above all, your patience with my character and all my outbursts, I would never have made it this far. I do not believe I could ever repay everything you have done for me, and so I ask you to accept whatever I will be able to give you, with all my limitations and shortcomings. I love you, and I always will, with all my heart!\\

Goodbye everyone,\\
Andrea

\thispagestyle{empty}

\newenvironment{Abstract}%
{\cleardoublepage\thispagestyle{empty}\null\vfill\begin{center}%
		\bfseries Abstract\end{center}}%
{\vfill\null}
\begin{Abstract}
This thesis lies within the framework of applying techniques developed in supersymmetric gauge theories to the computation of physical quantities in both Quantum Field Theory and Gravity. The main tool employed is the Seiberg–Witten (SW) curve, whose geometric structure enables the description of the low-energy dynamics of $\mathcal{N}=2$ supersymmetric theories. For this reason, the thesis begins with an overview of the essential concepts of Supersymmetry and of the derivation of the SW curve.\\
On the Quantum Field Theory side, the SW curve is used to explore how Argyres–Douglas theories emerge in the moduli space of certain supersymmetric models. The main physical observable studied is the extremal correlator. To understand its instanton corrections, we first analyze a conformal model where the computations for determining this physical quantity can be explicitly performed at the perturbative level. We show that instanton's contribution agrees with expectations from the Effective Field Theory approach. We then discuss implications of our results for analogous observables in Argyres-Douglas theories.\\
On the Gravity side, we employ the quantum version of the SW curve, which is a differential equation that can be mapped to other equations that govern various physical systems. Within this framework, we show the stability of the Topological Star—an object arising in the fuzzball proposal—under massive charged scalar perturbations at the linearized level. We also study charge superradiance when the object presents a horizon, and we derive a novel hypergeometric representation of the confluent Heun function, relevant to the propagation of gravitational waves in the Schwarzschild background. This leads to a new resummed expression for the tail contributions to the waveform, which shows excellent agreement with numerical results.\\
Finally, beyond the SW formalism, we develop a Machine Learning algorithm based on Hamiltonian Neural Networks to reproduce the geodesics of a neutral particle moving in a fuzzball geometry. In particular, for critical trajectories, the proposed method outperforms standard numerical integrators in both efficiency and accuracy.
\end{Abstract}

\frontmatter

{\pagestyle{plain}\tableofcontents}
\mainmatter

\chapter*{Introduction}
\addcontentsline{toc}{chapter}{Introduction}
\markboth{Introduction}{}
Why is Supersymmetry important? How could it play a significant role in Particle Physics? To address these questions, we should consider the contribution of symmetries to the understanding of elementary processes.\\
As we know very well, a symmetry is a transformation on a system that leaves the physical quantities unchanged and it has always played a crucial role in the study of phenomena for several reasons. In the framework of Quantum Field Theory (QFT), symmetries allow us to identify and classify particles via Noether's theorem, using their conserved quantum numbers. Moreover, they determine the structure of interactions through gauge invariance. The basic example is when we want to couple a free scalar field $\phi$ to a gauge field $A_\mu$ with the promotion of a global symmetry to a local one. Another important aspect is the phenomenon of spontaneous symmetry breaking.\\

The theoretical framework that describes the known fundamental particles and interactions in four-dimensional spacetime is the Standard Model (SM), whose validity is supported by an extraordinary number of experimental successes. Nevertheless, it is not a fundamental theory of nature, but rather an effective theory valid up to 13 TeV. Indeed there are many fundamental issues this theory cannot solve:
\begin{itemize}
\item \textbf{Quantum Gravity}: the SM accounts for three of the four fundamental interactions at the quantum level; gravity is treated classically, following Einstein’s theory. Quantum effects of gravity come into play only at energy scales higher than the Planck scale $M_{\text{Pl}} \simeq 10^{19} \text{GeV}$, much higher than the upper bound of validity of the SM and the non-renormalizability of Einstein theory makes it not possible to provide predictions on observables beyond this scale;
\item \textbf{The strong CP problem}: it is notoriously known that, adding to QCD the so called \textit{theta term}, the CP symmetry is broken. Such result has relevant phenomenological physical implications, as the electric dipole moment for the neutron. Experimental measurements can put a constraint on the value of $\theta$. The fact is that the electric dipole of the neutron is not observed and this implies that $\theta < 10^{-10}$, raising the puzzle of why this parameter is so small. Actually the solution was found in 1977 by Peccei and Quinn (\cite{PhysRevLett.38.1440,PhysRevD.16.1791}), who added a new symmetry to the SM, the PQ symmetry, which is spontaneously broken. Later on, it was understood that the introduction of such a symmetry gives rise to a new particle, the \textit{axion}, and this allows the solution of the problem because QCD non-perturbative effects introduce a potential for this particle which has a minimum in correspondence of $\theta = 0$;
\item \textbf{Dark Matter}: several astrophysical and cosmological observations—galactic rotation curves, cluster dynamics, gravitational lensing and structure formation—provide strong evidence for Dark Matter. Many possible candidates have been proposed, but they do not fit in the SM;
\item \textbf{The hierarchy problem}: the weak part of the SM is spontaneously broken at the scale $M_{\text{EW}} \simeq 10^2 \text{GeV}$ due to the Higgs effect. This implies that in the SM there are two relevant energy scales, $M_{\text{EW}}$ and $M_{\text{Pl}}$, that are completely different, indeed $\frac{M_{\text{EW}}}{M_{\text{Pl}}} \simeq 10^{-17}$. This vast difference is un-natural and it requires a severe fine tuning of many orders of magnitude.
\end{itemize}   
These difficulties indicate that an extension of the SM is needed. One promising strategy is to enlarge its symmetry structure, as exemplified by the PQ symmetry for the strong CP problem. We can consider more general \textit{spacetime symmetries} and one of them is Supersymmetry (SUSY), which manifests itself in the exchange of bosons and fermions.\\ 
SUSY solves many of the problems we listed above: it provides candidates for Dark Matter (such as the neutralino), it allows to obtain non-perturbative results in strongly coupled theories; finally it offers a solution to the hierarchy problem through cancellations between bosonic and fermionic contributions. Nonetheless, nowadays there are not observations in Nature of the presence of SUSY (as no superpartners have been observed so far). This is a signal that SUSY should be broken and there are many studies and works aimed at describing the consequences of it.\\

Beyond these physical motivations, supersymmetric gauge theories possess deep mathematical structures that translate into powerful computational tools applicable to both QFT and Gravity. In other words, SUSY provides mathematical tools that allow one to access observables in strongly coupled regimes of QFT, as well as physical quantities relevant in Gravity. The agreement of SUSY-based predictions with results obtained through independent methods confirms the mathematical robustness of this framework.\\
In this Thesis one of these instruments is deeply analyzed and employed, that is the \textit{Seiberg-Witten (SW) curve}, introduced by the physicists Nathan Seiberg and Edward Witten in \cite{Seiberg:1994rs,Seiberg:1994aj} in the context of the $\mathcal{N}=2$ supersymmetric extension of the Yang Mills theory. Due to its huge impact in determining the IR effective action of such model, many other works employ the same tool also for other models. It is an elliptic curve fibered over the moduli space of the theory under examination. The geometric properties of such curve encode the physical nature of the theory at each point of the moduli space. Here we focus only on rank-1 models, meaning that the gauge group is SU(2), and this implies that the curve is a torus. Whenever one or both the homological non-trivial 1-cycles of the torus shrink in correspondence of a given point, the latter is a singularity of the moduli space, attributed to the fact that some massless degrees of freedom have been integrated out in the effective description. In order also to understand how this geometric tool allows the determination of the effective action of the model, we briefly describe here the main steps. The key ingredient is a 1-form, called the \textit{Seiberg-Witten differential}, and, integrating it along the two 1-cycles of the torus, gives rise to the two periods $a(u)$ and $a_D(u)$, both expressed as a function of the holomorphic coordinate $u$ of the moduli space. Inverting one of the two functions, it is possible to write $a_D$ as a function of $a$. Integrating with respect to $a$, we obtain a relevant function, the \textit{prepotential}, which characterizes the Lagrangian of the effective description of the model.\\
The interesting and important aspect of this tool is that the quantities that characterize it provide access to observables in the strongly coupled sector of QFTs and have applications in gravity. For example, it turns out that a suitable quantization of the dual period $a_D$ yields the spectrum of Quasi-Normal Modes (QNMs) of a given gravitational object; the period $a$ enters the solutions of the equations describing the perturbations of Black Holes (BHs).\\ 

The main goal of this Thesis is exploring these correspondences, both conceptually and computationally, providing new analytical tools and physical insights, following the original results of the \cite{Cipriani:2026iws, Cipriani:2024ygw, Cipriani:2025ini, Cipriani:2025ikx, Cipriani:2026xmx}, carried out by the author of the present Thesis in collaboration with senior and young researchers.\\
Given the variety of topics addressed, the remainder of this Introduction outlines the structure of the Thesis and summarizes the content of each Chapter.\\

\textbf{Chapter \ref{chap_1}} reviews the basic notions of SUSY (its algebra, irreducible representations and superfields) in order to explain how $\mathcal{N}=1$ and $\mathcal{N}=2$ theories are constructed. This paves the way to understand the arguments proposed by Seiberg and Witten in \cite{Seiberg:1994rs,Seiberg:1994aj} to build the effective action of the $\mathcal{N}=2$ pure SU(2) Super-Yang Mills theory, whose moduli space is deeply investigated. In the end, the quantum version of this tool is investigated, since quantum corrections allow to smooth out the singularities that arise from the classical description; furthermore the quantum curve plays a vital role in the computation of physical observables in the gravity context.\\

In \textbf{Chapter \ref{chap_2}} we are in the QFT side, with the goal of studying a particular class of correlator in strongly coupled superconformal field theories. To this purpose, we consider the 4-dimensional $\mathcal{N}=2$ supersymmetric extension of SU(2) QCD with a number $N_f$ of hypermultiplets transforming in the fundamental representation of the gauge group equal to 1,2,3,4. Only the case $N_f=4$ is superconformal, while the theories with $N_f=1,2,3$ are not; however, their moduli spaces contain special points associated with Argyres–Douglas (AD) theories. These are isolated superconformal field theories, intrinsically strongly coupled and without a Lagrangian description. In the language of the SW curve introduced earlier, these points correspond to loci where both non-trivial 1-cycles of the torus shrink, giving rise to mutually non-local massless states. For this reason, after briefly reviewing conformal symmetry and the notion of isolated fixed points, we show how these AD theories arise in the moduli space of $\mathcal{N}=2$ SU(2) SQCD with $N_f=1,2,3$ through the SW curve. The resulting models are respectively labelled $\mathcal{H}_0$, $\mathcal{H}_1$, and $\mathcal{H}_2$. After this detour, we then introduce the observable of interest—the extremal correlators—involving Coulomb branch operators. For theories with a Lagrangian description, as for the case with $N_f=4$, these correlators can be computed via localization on the 4-sphere. In order to investigate their behavior in perturbation theory, we consider the genus expansion of the prepotential, which becomes a genuine perturbative expansion only in the so called large R-charge limit. A remarkable feature of this limit is that it also allows the use of Effective Field Theory (EFT) techniques to determine the behaviour of extremal correlators. The main result of this Chapter is showing the agreement with the previous analysis once instanton contributions to the correlators are properly included.\\
Regarding the AD theories discussed above, the literature contains computations of the same type of correlators using the genus expansion supplemented by the large-charge limit. In this context, we report a discussion that leads to formulate an educated guess for a function interpolating between two asymptotic regimes of the partition function of $\mathcal{H}_0,\mathcal{H}_1,\mathcal{H}_2$ models. Furthermore, we also show a numerical computation, under a suitable limit, of some non-universal coefficients characterizing the expansion of the extremal correlator under the large R-charge limit of above. This Chapter is based on the results of \cite{Cipriani:2026iws}.\\

\textbf{Chapter \ref{chap_3}} is the first dealing with gravitational aspects. In particular, the idea unifying this Chapter and the following two is considering a central gravitational object and we want to investigate how we can probe its dynamical properties through perturbations of the geometry, then we desire to determine particle trajectories without emission of gravitational waves (GWs) and finally we consider also GW emission. The present Chapter analyses the first situation, in particular the central object is a Topological Star (TS). It is a solution of Einstein-Maxwell theory in 5 dimensions, whose 4-dimensional reduction is a BH solution of Einstein-Maxwell-Dilaton theory. It enters the so called fuzzball program, aimed at resolving the typical puzzles that arise when Quantum Mechanics is combined with General Relativity (most notably the information paradox). The TS is characterized by two parameters such that, depending on their relative values, it behaves either as a black object (i.e. with an event horizon) or as a horizonless one. After introducing these conceptual motivations and describing how fuzzballs address the associated puzzles, we investigate the properties of the TS and then consider massive, charged scalar perturbations of this background with the goal of determining its stability at the linearized level. Using different methods (numerical integration, the Leaver method, the SW-based approach, the asymptotic matching condition and WKB) we compute the spectrum of QNMs. We find good agreement among all methods and, importantly, that the TS is stable. We furthermore examine whether the TS, when behaving as a black object, exhibits charge superradiance, and we show that this is indeed the case. All these results were obtained for the first time in \cite{Cipriani:2024ygw}.\\

In \textbf{Chapter \ref{chap_4}} we consider as central object a particular kind of fuzzball and we determine the trajectories followed by a neutral particle moving in this spacetime. Here we set aside the SW formalism and instead compute the geodesics using an algorithm from Machine Learning, which is the Hamiltonian Neural Network (HNN). It is a specific type of Physics Informed Neural Network, which is a neural network that, during the training phase, learns to solve the differential equations governing the physical laws of a system. In the case of a HNN, the differential equations are the Hamiltonian ones. All the details of how the process is performed are reported in this Chapter. We then analyze the main features of the geodesics in the geometry under consideration, defining the situation in which the motion occurs along critical geodesics, i.e. stable or unstable bound orbits separating the asymptotically flat region from the inner one of the fuzzball. When the motion is planar, we can thus distinguish three regimes: the overcritical, the critical and the subcritical ones. In the case of a non-planar motion, we restrict to the critical regime. We compute the geodesics in all relevant cases using the HNN and then we compare the results with those obtained from standard numerical integrators (in particular Euler and Runge-Kutta). The main results are that the performances of the HNN are quite the same throughout all the different cases and, more importantly, in the critical regimes the HNN is the only method capable of reproducing the correct behavior. This observation is one of the reasons that favor these Machine Learning-based algorithms over standard integrators, even when their performances appear comparable. Detailed discussions on this aspect are reported in \cite{Cipriani:2025ini}.\\

\textbf{Chapter \ref{chap_5}} treats the final scenario, in which a particle orbits a central Schwarzschild black hole and emits gravitational waves. In particular, the final quantity we aim to compute is the luminosity, i.e. the amount of energy per unit of time emitted by the system towards infinity. The propagation of GWs is governed by the Teukolsky equation. To solve it, we first consider its homogeneous version, which in the Schwarzschild case is a Confluent Heun Equation (CHE). The first part of the Chapter, following the new analysis of \cite{Cipriani:2025ikx}, provides a new representation for the Confluent Heun function, that we call the hypergeometric representation. Indeed, thanks to the method developed in Chapter \ref{chap_1}, the CHE can be understood as the quantum version of the SW curve of a specific gauge theory and, inspired by this correspondence, we express the solution of the homogeneous Teukolsky equation in terms of a hypergeometric function and its first derivative, whose arguments depend on the SW period $a$. From the homogeneous solution, we construct the inhomogeneous one and finally we compute the luminosity. Evaluating this quantity for different values of the particle's tangential velocity, we find excellent agreement with numerical data, confirming the reliability of the method.\\ 
For this reason we then move on writing the waveform we obtain starting from the previous representation in a factorized form as typical in literature, allowing for a direct comparison between the new resummed tail factors (written in terms of exponentials and gamma functions) and those proposed in the literature.The final results show the best agreement with numerical data among the existing approaches, demonstrating the importance of the new hypergeometric representation of the Confluent Heun function. In the end, we also discuss a possible extension of our paradigm to a binary systems with comparable masses. The original results are the ones derived in \cite{Cipriani:2026xmx}.\\

At the end we give our conclusions and summarize the main results obtained throughout the Thesis; we also discuss some possible direction for further investigations. Many Appendices are present in order to give supplementary material to several subjects dealt with in the previous Chapters.\\

Another work the author of this Thesis has also contributed, which is not included but it is strictly related with the latter, is \cite{Cipriani:2026myb}.

\chapter{Seiberg-Witten Curve: from the Classical to the Quantum Version} \label{chap_1}
This Chapter is aimed at introducing the theory of Seiberg-Witten (SW) and the tool that they introduced in \cite{Seiberg:1994rs,Seiberg:1994aj} in order to find the action of this theory in the low-energy regime. The importance of such instrument will be clear in the following Chapters, where its applications are considered in the gravity context. The application to gravity is viable only if one uses the quantum version of the SW curve, which will be discussed at the end of the present Chapter.\\
In order to understand the results treated by Seiberg and Witten, we have to summarize some notions of supersymmetric gauge theories in four dimensions, in particular, quickly starting from the Supersymmetry (SUSY) algebra with its representations on physical states, we can move on to the formalism of superspace and superfields (i.e. representations of SUSY algebra on fields), since it has a significant importance for the study of supersymmetric theories, which lie at the core of the treatment of Seiberg and Witten.\\
The structure of the Chapter is the following: in Section \ref{sec:intro_SUSY} we recall the basic notions of SUSY, in terms of its algebra and irreducible representations on physical states; in Section \ref{sec:chap1_2} we describe the superspace formalism in $\mathcal{N}=1$ SUSY, with all the superfields that can be present; in Section \ref{sec:chap1_3} some $\mathcal{N}=1$ supersymmetric theories are constructed, with the aim of introducing notions useful for the $\mathcal{N}=2$ scenario. Section \ref{sec:N2_theories} deals with $\mathcal{N}=2$ theories, which are the central objects of the thesis. In Section \ref{sec:chap1_5} we investigate the arguments proposed by Seiberg and Witten for the determination of the IR effective action of $\mathcal{N}=2$ pure SU(2) Super-Yang Mills theory; Section \ref{sec:chap1_6} is devoted to the SW curve, the tool thanks to which the previous model can be solved. Finally, Section \ref{sec:qSW_curve} passes to examine how the previous tool can be quantized.

\section{Introduction to Supersymmetry} \label{sec:intro_SUSY}
The starting point is the theorem due to Coleman and Mandula in \cite{PhysRev.159.1251}, according to which under certain conditions, the only possible symmetries of a consistent quantum field theory are (besides C,P,T) the Poincarè symmetries, whose generators are $M_{\mu\nu}$ and $P_\mu$ (respectively the generators of Lorentz transformations, composed of boosts and rotations, and of spacetime translations), and internal symmetries, whose generators commute with the Poincarè ones; furthermore, they are codified only through commutation relations between these generators. Haag, Lopuszanski and Sohnius proved in \cite{Haag:1974qh} that it is possible to relax only one condition of the Coleman-Mandula theorem in order to get a larger set of symmetries that is still consistent with a relativistic quantum field theory: this is the condition that the symmetry algebra admits as product between the generators of the symmetries only the commutator. This means that the (Lie) algebra of such symmetries is promoted to a \textit{superalgebra} (or \textit{graded Lie algebra}), which is characterized by commutation and anticommutation relations between the generators. In particular, the schematic form assumed by these rules is
\begin{equation}
\label{structure}
\{Q,Q'\} = X \qquad [X,X'] = X'' \qquad [Q,X'] = Q''
\end{equation}
where $Q$,$Q'$ and $Q''$ form the anticommuting part of the algebra, while $X$,$X'$ and $X''$ constitute the commuting part, hence they are determined by the Coleman-Mandula theorem as discussed before. For what concerns the anticommuting generators, the three scientists demonstrated (and a good proof is presented in \cite{Wess:1992cp}) that the unique possibility for them in order to have a superalgebra of symmetries consistent with a relativistic quantum field theory is that they are spin-$\frac{1}{2}$ generators, $Q^I_\alpha$ and $\overline{Q}^I_{\dot{\alpha}}$, transforming in the representation $\bigl(\frac{1}{2},0\bigr)$ and $\bigl(0,\frac{1}{2}\bigr)$ of the Lorentz group. They are called \textit{supercharges} and in general $I = 1,_{\cdots}, \mathcal{N}$, while the spinorial indices are $\alpha=1,2, \dot{\alpha} = \dot{1},\dot{2}$.\\
In the following parts of this Section, we briefly present the SUSY algebra and the irreducible representations on physical states. Details cannot be suitably treated in this thesis, but they can be consulted at \cite{Wess:1992cp, Bertolini:2025jhz, Quevedo:2010ui, Terning:2006bq}.\\

\subsection{SUSY algebra}
For what just said, we realize that the SUSY algebra has a part which is exactly equal to the one of the Poincarè algebra, for what concerns the commuting generators, hence it is useful to report only the supersymmetric part, which is
\begin{subequations}
\label{tot_rel}
\begin{align}
[P_\mu, Q^I_\alpha] &= 0 \label{PQ} \\
[P_\mu, \overline{Q}^I_{\dot{\alpha}}] &= 0 \label{PQbar} \\
[M_{\mu\nu}, Q^I_\alpha] &= i(\sigma_{\mu\nu})_\alpha^{\, \, \beta} Q^I_\beta \label{MQ}\\ 
[M_{\mu\nu}, \overline{Q}^{I\dot{\alpha}}] &= i({\overline{\sigma}}_{\mu\nu})^{\dot{\alpha}}_{\, \, \dot{\beta}} \overline{Q}^{I{\dot{\beta}}} \label{MQbar}\\
\{Q^I_\alpha, \overline{Q}^J_{\dot{\beta}} \} &= 2\sigma^\mu_{\alpha \dot{\beta}} P_\mu \delta^{IJ} \label{QQbar}\\
\{Q^I_\alpha, Q^J_\beta \} &= 2 \epsilon_{\alpha \beta} Z^{IJ} \quad \text{with} \quad Z^{IJ} = -Z^{JI} \quad \text{(since $\epsilon_{\alpha,\beta}$ is antisymmetric)}  \label{QQ}\\
\{\overline{Q}^I_{\dot{\alpha}}, \overline{Q}^J_{\dot{\beta}} \} &= 2 \epsilon_{\dot{\alpha} \dot{\beta}} (Z^{IJ})^* \label{QbarQbar}
\end{align}
\end{subequations}
where  $\sigma_{\mu\nu}$ and $\overline{\sigma}_{\mu\nu}$ are defined in the following way
$$(\sigma^{\mu\nu})_\alpha^{\, \, \beta} = \frac{1}{4}(\sigma^\mu_{\alpha \dot{\gamma}} (\overline{\sigma}^\nu)^{\dot{\gamma} \beta} - \sigma^\nu_{\alpha \dot{\gamma}} (\overline{\sigma}^\mu)^{\dot{\gamma} \beta}) \qquad ({\overline{\sigma}}^{\mu\nu})^{\dot{\alpha}}_{\, \, \dot{\beta}} = \frac{1}{4}(({\overline{\sigma}}^\mu)^{\dot{\alpha} \gamma} \sigma^\nu_{\gamma \dot{\beta}} - ({\overline{\sigma}}^\nu)^{\dot{\alpha} \gamma} \sigma^\mu_{\gamma \dot{\beta}})$$  
with $\sigma^\mu = (\sigma^0, \sigma^i) = (\sigma_0, -\sigma_i)$ and ${\overline{\sigma}}^\mu = (\sigma^0, -\sigma^i) = (\sigma_0, \sigma_i)$ where
\begin{equation} \label{eq:Pauli_matrices}
\sigma_1 = 
\begin{pmatrix}
0 & 1 \\
1 & 0
\end{pmatrix}
\quad , \quad \sigma_2 = 
\begin{pmatrix}
0 & -i \\
i & 0
\end{pmatrix}
\quad , \quad \sigma_3 =
\begin{pmatrix}
1 & 0 \\
0 & -1
\end{pmatrix}
\end{equation}
The antisymmetric matrix $Z^{I J}$ that appears in \eqref{QQ} and \eqref{QbarQbar} denotes the \textit{central charges} of the algebra, that is the set of operators that commute with the entire algebra and within themselves. We come back to them later, but here we simply notice that this matrix is $\mathcal{N} \times \mathcal{N}$ and it is non trivial only for $\mathcal{N} > 1$ (\textit{extended supersymmetry}). This can be shown in details, but a quite simple way is that for $\mathcal{N}=1$ (\textit{minimal supersymmetry}) the only $1 \times 1$ anti-symmetric matrix is 0.\\
Within the commutation rules exposed above, it is interesting to focus on \eqref{MQ} and \eqref{MQbar}, since they allow us to understand why it is typically said that SUSY exchanges bosons with fermions. Calling $M_{12} = J_3$ as the $z$-component of the spin, we get
$$[J_3, Q_1^I] = i {\sigma_{12}}_1^{\, \beta} Q_\beta^I = \frac{i}{2}
\begin{pmatrix}
-i & 0 \\
0 & i
\end{pmatrix}
_1^{\, \beta} Q_\beta^I = \frac{Q^I_1}{2} \qquad [J_3, Q_2^I] = -\frac{Q_2^I}{2}
$$
and taking the hermitian conjugate of these relations
$$[J_3, {\overline{Q}}^I_{\dot{1}}] = -\frac{{\overline{Q}}^I_{\dot{1}}}{2} \qquad [J_3, {\overline{Q}}^I_{\dot{2}}] = \frac{{\overline{Q}}^I_{\dot{2}}}{2}$$
This means that $Q^I_1$ and ${\overline{Q}^I_{\dot{2}}}$ increase the $z$-component of the spin by half unit, while $Q^I_2$ and ${\overline{Q}^I_{\dot{1}}}$ decrease it by half unit. The implication is that if a supercharge acts on a bosonic state, we get a fermionic state and viceversa.\\
An important property of supersymmetric theories is the existence of a global (internal) symmetry which rotates the supercharges. This symmetry is named \textit{R-symmetry} and the group associated to it and that can act non trivially on the supersymmetry operators is U$(\mathcal{N})$.

\subsection{Irreducible representations on physical states: supermultiplets} \label{sec:supermultiplets}
Now it is time to describe the irreducible representations of the Supersymmetry algebra on the physical states.\\
From basic notions of QFT, particles are identified by irreducible representations (IRREPS) of the Poincarè algebra and these representations are labelled by the eigenvalues of the Casimir operators. For the Poincarè group, these are 
$$P^2 = P_\mu P^\mu \qquad W^2 = W_\mu W^\mu$$
where $P^\mu$ is the usual 4-momentum and $W^\mu = \frac{1}{2} \epsilon^{\mu\nu\rho\sigma}P_\nu M_{\rho \sigma}$ is the Pauli-Lubanski vector. The eigenvalue of the former is related to the mass, the eigenvalue of the latter to the spin of the particle (if it is massive, otherwise it is the helicity). Consequently, fixing these two eigenvalues, it is possible to find all the states of a given IRREP in which the associated particle can be and these states share the same values of mass and spin (or helicity) and differ in the 3-momentum, for example. For this reason an IRREP associated to a particle is also called a \textit{multiplet}.\\

Passing to the SUSY algebra, as well as a particle is an IRREP of the Poincarè algebra, a \textit{superparticle} is an IRREP of the SUSY algebra.  Since the Poincarè algebra is a subalgebra of the supersymmetric one, an IRREP of the SUSY algebra is also a representation of the Poincarè algebra, but in general it is reducible and hence it contains different particles and not only one. In other words, a superparticle is a collection of particles, related by the action of the supercharges and, for what seen before, they have spins that differ by units of half. For this property, a superparticle is often called \textit{supermultiplet}.\\
In order to better understand this, we notice that also the SUSY algebra has two Casimir operators, one is always $P^2$ (since the supersymmetry generators commute with $P_\mu$), but the other one is not $W^2$ as in the non-supersymmetric case (since $M_{\rho \sigma}$ does not commute with the supercharges), indeed it is $C^2 := C_{\mu \nu} C^{\mu \nu}$, where $C_{\mu \nu} := B_\mu P_\nu - B_\nu P_\mu$ with $B_\mu := W_\mu - \frac{1}{4}\overline{Q}^I_{\dot{\alpha}}({\overline{\sigma}}_\mu)^{\dot{\alpha} \beta} Q^J_\beta \delta_{IJ}$ and this corresponds to the \textit{superspin}. This means that particles inside the same supermultiplet have the same mass and superspin, but different spins.\\

An interesting property is that the numbers of bosons and of fermions in a given supermultiplet are equal. This can be demonstrated by introducing the fermion number operator, which is equal to -1 on a fermionic state and +1 on a bosonic one, and then computing its trace on all the states of a given supermultiplet. It turns out that this quantity is 0 \cite{Bertolini:2025jhz}. This explains also the fact that SUSY is considered as the symmetry that associates to a boson a fermion and viceversa.
Now we briefly discuss the massless and massive IRREPS on physical states.

\subsubsection{Massless supermultiplets}\label{sec:massless_supermultiplets}
Let us start to discuss the massless supermultiplets. We go to the frame where $p_\mu = (E,0,0,E)$ and hence, in this way, from the anticommutation rule of the supercharges (we will show with massive representations that the central charges are vanishing in the massless case)
$$\{Q^I_\alpha, \overline{Q}^J_{\dot{\beta}}\} = 2 (\sigma^\mu P_\mu) \delta^{IJ} = 2E (\sigma^0+\sigma^3) \delta^{IJ} = 4E
\begin{pmatrix}
0 & 0 \\
0 & 1
\end{pmatrix}_{\alpha \dot{\beta}}
\delta^{IJ} \Rightarrow \{Q^I_1, \overline{Q}^J_{\dot{1}}\} =0$$
and this implies that both $Q^I_1$ and $\overline{Q}^I_{\dot{1}}$ are null, indeed
$$0 = \bra{\phi}\{Q^I_1, \overline{Q}^J_{\dot{1}}\} \ket{\phi} = ||Q^I_1\ket{\phi}||^2 + ||\overline{Q}^I_{\dot{1}}\ket{\phi}||^2$$
and the only solution (thanks to the positiveness of the norm) is that $Q^I_1 = \overline{Q}^I_{\dot{1}} = 0$. This has the relevant consequence that in the massless case we are left with only half of the supersymmetry generators.\\
At this point, it is possible to define, from the non trivial supercharges, the operators $a^I$ and $a^{I\dagger}$ ($I=1,\dots,\mathcal{N}$) that satisfy the usual anticommutation relations of the fermionic harmonic oscillators. For this reason it is possible to prove that $a^I$ lower the helicity of the state of half unit, while $a^{I\dagger}$ raise it of the same amount. This has the implication that there exists a state which is annihilated by all the operators $a^I$ and it is called the \textit{Clifford vacuum}. Such state carries some helicity $\lambda_0$ (for this reason it is indicated as $\ket{\lambda_0}$) and, at this point, the full supermultiplet is obtained by acting on the Clifford vacuum with the creation operators, noticing that, due to their anticommutating nature, two equal creation operators acting on the Clifford vacuum give rise to 0 ($a^{I\dagger} a^{I\dagger} \ket{\lambda_0} = 0$). For this reason a massless supermultiplet has a finite number of states and, keeping in account all the multiplicities, it turns out to be
\begin{equation}
 \# \, \, \text{states in a massless supermultiplet} = 2^{\mathcal{N}}
\end{equation}
and half of this number describes bosonic states ad the remaining half fermionic ones. The helicity ranges from $\lambda_0$ to $\lambda_0 + \frac{\mathcal{N}}{2}$.\\

Now we want to list the most relevant massless supermultiplets that will be important in the following, with their field content. Even if we are interested in $\mathcal{N}=2$ SUSY, it is important to consider also the $\mathcal{N}=1$ supermultiplets, since the latter can be used to describe $\mathcal{N}=2$ supermultiplets.\\
For what concerns \textbf{$\mathcal{N}=1$ supermultiplets} (shortly called in the following multiplets), the most common ones are
\begin{itemize}
\item the \textbf{matter} or \textbf{chiral multiplet}, characterized by $\lambda_0 = 0$. It can be shown that the degrees of freedom of this representation are those of \textbf{one (massless) Weyl fermion} and of \textbf{one (massless) complex scalar}. This is the representation where the matter is present in $\mathcal{N}=1$ theories;\\
\item the \textbf{vector} or \textbf{gauge multiplet}, characterized by $\lambda_0 = \frac{1}{2}$ . In this multiplet we have \textbf{one Weyl fermion} and \textbf{one (massless) spin 1 particle}, so \textbf{one vector}. This is the representation that is aimed at describing gauge field in supersymmetric theories.
\end{itemize}
For what concerns \textbf{$\mathcal{N}=2$ multiplets}, the ones that will play a crucial role in our treatment are
\begin{itemize}
\item the \textbf{matter multiplet} or \textbf{hypermultiplet}, which is characterized by $\lambda_0 = -\frac{1}{2}$. This supermultiplet contains \textbf{two Weyl spinors} and \textbf{two complex scalars}. This is the representation where matter is present in $\mathcal{N}=2$ theories. In the language of $\mathcal{N}=1$ theories, this supermultiplet has two chiral multiplets;\\
\item the \textbf{gauge} or \textbf{vector multiplet}, which is characterized by $\lambda_0 = 0$. This multiplet contains \textbf{two Weyl fermions}, \textbf{a complex scalar} and \textbf{a vector,} so it has one $\mathcal{N}=1$ chiral multiplet and one $\mathcal{N}=1$ vector multiplet. We should stress that, since the supercharges commute with gauge transformations, all states in a vector multiplet transform in the adjoint representation of the gauge group.
\end{itemize}

\subsubsection{Massive supermultiplets}\label{massive supermultiplet}
For massive supermultiplets, the logic of the construction of the IRREP is similar to that one for the massless case. Nonetheless, in this case the number of supercharges is not diminished, indeed in the rest frame $p_\mu = (M,0,0,0)$ and we have
$$\{Q^I_\alpha, \overline{Q}^J_{\dot{\beta}}\} = 2M 
\begin{pmatrix}
1 & 0 \\
0 & 1
\end{pmatrix}_{\alpha \dot{\beta}} 
\delta^{IJ}$$
so no supersymmetric generators are trivially realized. Another difference is that both Casimir operators are different from 0, in particular $P^2 = M^2$ and for $C^2$ we have $C^2 = 2m^4 Y^i Y_i$, where
$$Y_i = J_i - \frac{1}{4M} \overline{Q}^I \overline{\sigma}_i Q^J \delta_{IJ} \qquad \text{satisfying} \qquad [Y_i, Y_j] = i \epsilon_{ijk} Y_k$$
is the superspin. As a consequence, massive supermultiplets are identified by the values of mass and superspin. A given Clifford vacuum, as also any other state of the same supermultiplet, will be defined by mass $M$ and superspin number $y$. It is possible to show that for this particular state the spin number $s_0$ and the superspin number $y$ are the same.\\

For what concerns the \textbf{$\mathcal{N}=1$ supermultiplet}, there are no more things to say with respect to the procedure depicted in the massless case, also because in this situation the central charges continue to be trivial. For this reason we limit ourselves to list the relevant massive $\mathcal{N}=1$ multiplets:
\begin{itemize}
\item the \textbf{matter multiplet}, characterized by the value of the spin of the Clifford vacuum equal to 0: $s_0 = 0$. The multiplet is made of \textbf{one (massive) Majorana fermion} and \textbf{one (massive) complex scalar};\\
\item the \textbf{gauge} or \textbf{vector multiplet}, characterized by $s_0 = \frac{1}{2}$. In this supermultiplet we have \textbf{a (massive) vector}, \textbf{one (massive) Dirac fermion} and \textbf{one (massive) real scalar}. The representation is longer than that of a massless vector multiplet, as expected. 
\end{itemize} 

Moving on to the \textbf{$\mathcal{N}>1$ supermultiplets}, here the central charges can be different from 0, hence we have to distinguish between two principal cases. The first one is when $Z^{IJ} = 0$ and hence we can define $2\mathcal{N}$ creation and annihilation operators from the supercharges $Q^I_\alpha$ and $\overline{Q}^I_{\dot{\alpha}}$. This leads to a supermultiplets with $2^{2 \mathcal{N}}$ states. We notice that these multiplets are exactly large as the double of massless ones, due to the fact that, in that case, half of the supersymmetry generators vanishes.\\
The case with $Z^{IJ} \ne 0$ is much more interesting and in order to arrive to a very important result, we define the annihilation and creation operators. Since the matrix $Z^{IJ}$ is antisymmetric, with a U$(\mathcal{N})$ rotation one can put it in the standard block-diagonal form
$$Z^{IJ} =
\begin{pmatrix}
0 & Z_1 & 0 & 0 & \cdots & \cdots & 0 & 0 \\
-Z_1 & 0 & 0 & 0 & \cdots & \cdots & 0 & 0 \\
0 & 0 & 0 & Z_2 & \cdots & \cdots & 0 & 0 \\
0 & 0 & -Z_2 & 0 & \cdots & \cdots & 0 & 0 \\
\vdots & \vdots & \vdots & \vdots & \ddots &  & 0 & 0\\
0 & 0 & 0 & 0 & \cdots & \cdots & 0 & Z_{\frac{\mathcal{N}}{2}} \\
0 & 0 & 0 & 0 & \cdots & \cdots & -Z_{\frac{\mathcal{N}}{2}} & 0 \\
\end{pmatrix}
$$
(we have assumed for simplicity $\mathcal{N}$ even; if $\mathcal{N}$ is odd we simply have to add a row and a column of zeros to the $(\mathcal{N}-1)\times(\mathcal{N}-1)$ matrix), where $Z_i$ are taken real for every $i = 1, \dots, \frac{\mathcal{N}}{2}$. Now we define the operators
$$a^1_\alpha = \frac{1}{2}[Q^1_\alpha + \epsilon_{\alpha \beta}(Q^2_\beta)^\dagger] \qquad b^1_\alpha = \frac{1}{2}[Q^1_\alpha - \epsilon_{\alpha \beta}(Q^2_\beta)^\dagger]$$
$$a^2_\alpha = \frac{1}{2}[Q^3_\alpha + \epsilon_{\alpha \beta}(Q^4_\beta)^\dagger] \qquad b^2_\alpha = \frac{1}{2}[Q^3_\alpha - \epsilon_{\alpha \beta}(Q^4_\beta)^\dagger]$$\\
$$\cdots=\cdots$$
$$a^{\frac{\mathcal{N}}{2}}_\alpha = \frac{1}{2}[Q^{\mathcal{N}-1}_\alpha + \epsilon_{\alpha \beta}(Q^{\mathcal{N}}_\beta)^\dagger] \qquad b^{\frac{\mathcal{N}}{2}}_\alpha = \frac{1}{2}[Q^{\mathcal{N}-1}_\alpha - \epsilon_{\alpha \beta}(Q^{\mathcal{N}}_\beta)^\dagger]$$
and they satisfy the algebra of creation and annihilation operators
$$\{a^r_\alpha, (a^s_\beta)^{\dagger} \} = (M+Z_r) \delta^{rs} \delta_{\alpha\beta} \quad \{ b^r_\alpha, (b^s_\beta)^{\dagger} \} = (M-Z_r)\delta^{rs} \delta_{\alpha\beta} \quad \{a^r_\alpha, (b^s_\beta)^{\dagger} \} =\{a^r_\alpha,a^s_\beta\} = _{\cdots} = 0$$
where $r,s=1,_{\cdots},\mathcal{N}/2$. We have $2\mathcal{N}$ creation operators $(a^r_\alpha)^\dagger, (b^r_\alpha)^\dagger$ with $r=1,_{\cdots},\mathcal{N}/2$ and $\alpha = 1,2$ and also $2\mathcal{N}$ annihilation operators. We can use the former to construct massive representations starting from the Clifford vacuum. From these definitions, creation operators with spinorial index $\alpha = 1$ lower the spin by half unit, while those with spinorial index $\alpha = 2$ raise it.\\
Now we obtain a relevant result. Consider a state $\ket{M,Z_r}$ of unit norm and for the positiveness of the Hilbert space, from the anticommutation rule of the $b$'s operators we get
$$\bra{M,Z_r}b^r_\alpha(b^r_\alpha)^\dagger\ket{M,Z_r} + \bra{M,Z_r}b^r_\alpha(b^r_\alpha)^\dagger\ket{M,Z_r} =(M-Z_r) \geq 0 \Rightarrow$$
\begin{equation}
M \geq  Z_r \quad \forall r= 1,_{\cdots},\frac{\mathcal{N}}{2}
\end{equation}
By repeating the same procedure with $\{ a_\alpha^r, (a_\alpha^r)^\dagger \}$, we arrive at getting $M \geq -Z_r$, hence we can write in general
\begin{equation}
\label{central}
M \geq  |Z_r| \quad \forall r= 1,_{\cdots},\frac{\mathcal{N}}{2}
\end{equation}
This means that the mass of a given IRREP is always larger or equal than the modulus of any central charge eigenvalue. The first important consequence is that for massless representations the central charges are always 0 and this explains the fact that in the massless case we did not use the central charge matrix.\\
Another important consequence is that upon those states whose mass saturates the r-th bound \eqref{central}, either $\{ a_\alpha^r, (a_\alpha^r)^\dagger \}$ or $\{ b_\alpha^r, (b_\alpha^r)^\dagger \}$ act trivially. In particular, if $M = Z_1 = \dots = Z_k$ with $0 \leq k \leq \mathcal{N}/2$, the dimension of the representation reduces to $2^{2 \mathcal{N}-2 k}$. When:
\begin{itemize}
\item $k=0$ we have a \textit{long multiplet}, with $2^{2 \mathcal{N}}$ states;
\item $0 < k < \mathcal{N}/2$ we have a \textit{short multiplet}, with $2^{2 \mathcal{N}-2 k}$ states;
\item $ k = \mathcal{N}/2$ we have a \textit{ultra-short multiplet}, with $2^{\mathcal{N}}$ states, which is exactly the same as that of the massless multiplet
\end{itemize}
The ultra-short multiplet is also called \textit{BPS state} (due to Bogomol'ny-Prasad-Sommerfeld). Notice that, being the mass bound \eqref{central} an algebraic result, it is left untouched by quantum corrections, meaning that mass and central charges are renormalized in such a way to preserve this relation. In other words, the size of a multiplet cannot be changed by quantum corrections, since it cannot acquire further degrees of freedom from them.\\
Restricting to the case of $\mathcal{N}=2$, we only have one central charge and, since $k$ can be equal to $0,1$, then we have either long or ultra-short multiplets. The condition \eqref{central} for these two cases becomes $M \geq |Z|$ and it is the BPS bound that can be derived also in Electromagnetism (look at Appendix \ref{app:EM_duality} for details, which are going to be recalled later on). Thanks to this fact, the BPS states that we will encounter in the weak coupling regime of the theory studied by Seiberg and Witten correspond to monopoles and dyons.

\section{Superspace and Superfields in $\mathcal{N}=1$ SUSY}  \label{sec:chap1_2}
We want to arrive at seeing how we can build supersymmetric theories with matter and gauge multiplets, starting with $\mathcal{N}=1$. The case with $\mathcal{N}=2$ will be analyzed later because it uses the stategy that is going to be analyzed here. In order to do so we need a useful description of SUSY, which is the \textit{superfield} formalism. This is defined in a space where SUSY is manifest and for this reason this cannot be the simple Minkowsky spacetime, but the so called \textit{superspace}.\\
It is the promotion of Minkowsky spacetime to a new space parametrized by the following set of coordinates: $(x^\mu, \theta_\alpha, \theta^{\dot{\alpha}})$, where $x^\mu$ indicates the usual spacetime coordinates and so they are associated to the $P_\mu$ generator, while $(\theta_\alpha, \overline{\theta}^{\dot{\alpha}})$ are anti-commuting Grassmann variables (which are introduced in order to express the whole supersymmetry algebra \eqref{tot_rel} in terms of only commutators) and they are associated to the supercharges $Q_\alpha, \overline{Q}_{\dot{\alpha}}$.\\

At this point, as a field is a function of Minkowsky spacetime coordinates, a \textit{superfield} is a function of the superspace coordinates, namely $S(x^\mu,\theta,\overline{\theta})$. We can get its general expression exploiting the anticommuting nature of $\theta_\alpha$ and $\overline{\theta}^{\dot{\alpha}}$, indeed due to this fact we can consider at most expressions with a second power in $\theta$ and in $\overline{\theta}$, in the sense that $\theta_\alpha \theta_\beta \theta_\gamma = 0$ since at least two indices are the same and $\theta^2_\alpha = -\theta^2_\alpha \rightarrow \theta^2_\alpha = 0$ (which is different by $\theta \theta = \theta^\alpha \theta_\alpha$). In particular, by Taylor expanding $S(x^\mu,\theta,\overline{\theta})$ in $\theta$ and in $\overline{\theta}$ around their null values we obtain
\begin{equation}
\label{superfield}
S(x^\mu,\theta,\overline{\theta}) = \phi(x) + \theta \psi(x) + \overline{\theta} \, \overline{\chi}(x) + \theta \theta m(x) + \bar{\theta} \bar{\theta} n(x) + (\theta \sigma^\mu \overline{\theta}) v_\mu(x) + \theta \theta \bar{\theta} \bar{\lambda}(x) + \bar{\theta} \bar{\theta} \theta \rho(x) + \theta \theta \bar{\theta} \bar{\theta} D(x)
\end{equation}
where each function of $x$ is a field (some of them having non-trivial tensor structure). From this we understand that a superfield is a collection of ordinary fields.\\
Now, thanks to the state-field correspondence, a superfield has to represent a given supermultiplet, but the general superfield \eqref{superfield} does not represent any supermultiplet, due to the fact it has too many degrees of freedom (i.e. fields). Consequently, we have to impose apposite constraints on $S$ that have to respect the supersymmetry nature of the superfield itself. This means that these constraints have to be supersymmetric invariant. In this way they restrict the form of $S$ in \eqref{superfield} in such a way that the reduced set of fields that we obtain is still collected in a superfield. In order to do so, we need to understand how a superfield transforms under SUSY transformations (i.e. under the action of supercharges) and this can be achieved if we give a realization for $Q_\alpha$ and $\overline{Q}_{\dot{\alpha}}$.
Recall what happens with the translation operator $P_\mu$: we consider a field $\varphi(x)$ on which we act with the translation generated by the abstract operator $P_\mu$ with infinitesimal parameter $a^\mu$. In this way we get
\begin{equation}
\label{traslazione1}
\varphi(x+a) = e^{-i a^\mu P_\mu} \varphi(x) e^{i a^\mu P_\mu} = (1 -i a^\mu P_\mu) \varphi(x) (1 + i a^\mu P_\mu) = \varphi(x) + i a^\mu [\varphi(x), P_\mu] + \mathcal{O}(a^2)
\end{equation}
On the other hand, we can Taylor expand $\varphi(x+a)$ for $a$ close to 0
\begin{equation}
\label{traslazione2}
\varphi(x+a) = \varphi(x) + a^\mu \partial_\mu \varphi(x) + \mathcal{O}(a^2)
\end{equation}
and so, equating the RHSs of \eqref{traslazione1} and \eqref{traslazione2} we gain
\begin{equation}
\label{P_representation}
[\varphi(x), P_\mu] = -i \partial_\mu \varphi(x) \equiv \mathcal{P}_\mu \varphi(x)
\end{equation}
where $\mathcal{P}_\mu \equiv -i \partial_\mu$ is a realization of the operator $P_\mu$. Therefore, a translation of a field by a parameter $a^\mu$ induces a change on the field itself given by
\begin{equation}
\label{variation_translation}
\delta_a \varphi = \varphi(x+a) - \varphi(x) = i a^\mu \mathcal{P}_\mu \varphi
\end{equation}
The same procedure can be applied to the superfield \eqref{superfield}. First of all the parameters of an infinitesimal transformation generated by the supercharges have to posses a spinorial nature and hence we indicate them as $(\epsilon_\alpha, \overline{\epsilon}_{\dot{\alpha}})$; furthermore, as we showed in \eqref{tot_rel}, the anticommutator of the two kinds of supercharges is a spacetime translation and this means that the action of a supersymmetry transformation on a superfield produces a change not only on the coordinates $\theta$ and $\overline{\theta}$, but also of $x$. It can be shown (details can be found in \cite{Bertolini:2025jhz}), following the same procedure just discussed for the operator $P^\mu$, the variation of a superfield under a supersymmetry transformation with parameters $(\epsilon, \overline{\epsilon})$, is
\begin{equation}
\label{importante}
\delta_{\epsilon, \overline{\epsilon}}S \equiv S(x+\delta x, \theta + \delta\theta, \overline{\theta} + \delta\overline{\theta}) - S(x,\theta,\overline{\theta}) = i(\epsilon \mathcal{Q} + \overline{\epsilon} \, \overline{\mathcal{Q}}) S
\end{equation}
where $\epsilon \mathcal{Q} \equiv \epsilon^\alpha \mathcal{Q}_\alpha$, $\overline{\epsilon} \, \overline{\mathcal{Q}} \equiv \overline{\epsilon}_{\dot{\alpha}} \, \overline{\mathcal{Q}}^{\dot{\alpha}}$ (see Appendix \ref{app:convention} for the convention adopted for the contractions between spinorial objects and the importance of the position of the indices) and $\mathcal{Q},  \overline{\mathcal{Q}} $ are the differential operators that represent the supercharges
\begin{align}
\mathcal{Q}_\alpha = -i\partial_\alpha - \sigma^\mu_{\alpha \dot{\beta}} \overline{\theta}^{\dot{\beta}} \partial_\mu \label{supercharge_Q_diff_operators}\\
\overline{\mathcal{Q}}_{\dot{\alpha}} = i \overline{\partial}_{\dot{\alpha}} + \theta^\beta \sigma^\mu_{\beta \dot{\alpha}} \partial_\mu \label{supercharge_Qb_diff_operators}
\end{align}
($\partial_\alpha \equiv \frac{\partial}{\partial \theta^\alpha}, \partial_{\dot{\alpha}} \equiv \frac{\partial}{\partial \overline{\theta}^{\dot{\alpha}}} $). \eqref{supercharge_Q_diff_operators} and \eqref{supercharge_Qb_diff_operators} can be checked because $\mathcal{Q}^\dagger_\alpha = \overline{\mathcal{Q}}_{\dot{\alpha}}$ and they satisfy for example: $\{\mathcal{Q}_\alpha, \overline{\mathcal{Q}}_{\dot{\beta}} \} = - 2 i \sigma^\mu_{\alpha \dot{\beta}} \partial_\mu = 2 \sigma^\mu \mathcal{P}^\mu$. Comparing with \eqref{variation_translation} we see that the supersymmetry generators implement a translation in the superspace.\\
An important remark that will be useful for later is the following. While $\partial_\mu S$ is a superfield, it can be demonstrated that $\partial_\alpha S$ is not, in the sense that, under a supersymemtry transformation, it does not transform according to the right hand side (RHS) of \eqref{importante}. In order to make this object a superfield, we need to define the following \textit{covariant derivatives}
$$\mathcal{D}_\alpha := \partial_\alpha + i \sigma^\mu_{\alpha \dot{\beta}} \overline{\theta}^{\dot{\beta}} \partial_\mu \qquad \overline{\mathcal{D}}_{\dot{\alpha}} = \overline{\partial}_{\dot{\alpha}} + i \theta^\beta \sigma^\mu_{\beta \dot{\alpha}} \partial_\mu$$
and they anticommute with the supersymmetric generators. Furthermore
$$\{\mathcal{D}_\alpha, \overline{\mathcal{D}}_{\dot{\beta}}\} = 2i\sigma^\mu_{\alpha \dot{\beta}} \partial_\mu = -2 \sigma^\mu_{\alpha \dot{\beta}} P_\mu \qquad \{\mathcal{D}_\alpha, \mathcal{D}_\beta\} = \{\overline{\mathcal{D}}_{\dot{\alpha}}, \overline{\mathcal{D}}_{\dot{\beta}}\} = 0$$
So $\mathcal{D}_\alpha S$ is a superfield and, analogously, also $\overline{\mathcal{D}}_{\dot{\alpha}}S$ is a superfield.\\
We are going to see two kinds of supersymmetric invariant constrains to be imposed on \eqref{superfield}, studying the corresponding superfields.

\subsection{Chiral and Anti-chiral superfields} \label{chiral_superfields}
Since $\mathcal{D}_\alpha S$ and $\overline{\mathcal{D}}_{\dot{\alpha}} S$ are superfields, then $\mathcal{D}_\alpha S = 0$ and $\overline{\mathcal{D}}_{\dot{\alpha}} S = 0$ are supersymmetric invariant conditions that can be imposed on the generic superfield \eqref{superfield} in order to reduce the number of its components.\\
A \textit{chiral superfield} is a superfield $\mathbf{\Phi}$ that satisfies the following constraint
\begin{equation}
\label{chiral_field}
\overline{\mathcal{D}}_{\dot{\alpha}} \mathbf{\Phi} = 0 
\end{equation} 
while an \textit{anti-chiral superfield} is a superfield $\mathbf{\Psi}$ satisfying the other costraint
\begin{equation}
\label{anti_chiral_field}
\mathcal{D}_\alpha \mathbf{\Psi} = 0 
\end{equation} 
Now, if $\mathbf{\Phi}$ is chiral, $\overline{\mathbf{\Phi}}$ is anti-chiral, as it can be obtained by hermitian conjugation of \eqref{chiral_field}.\\
As well as we have determined the general expression for the general superfield $S$ in \eqref{superfield}, we want to do the same for the chiral superfield $\mathbf{\Phi}$. In order to do so, it is convenient to apply the following change of coordinates for $x^\mu$
$$y^\mu := x^\mu + i \theta \sigma^\mu \overline{\theta}$$
indeed, expressing $\mathbf{\Phi}$ as a function of $y^\mu,\theta,\overline{\theta}$ and keeping in count that $y^\mu$ depends on $\overline{\theta}$, we get
\begin{equation*}
\begin{split}
\overline{\mathcal{D}}_{\dot{\alpha}} \mathbf{\Phi} & = \overline{\partial}_{\dot{\alpha}} \mathbf{\Phi} + \frac{\partial \mathbf{\Phi}}{\partial y^\mu} \frac{\partial y^\mu}{\partial \overline{\theta}^{\dot{\alpha}}} + i\theta^\beta (\sigma^\mu)_{\beta \dot{\alpha}} \frac{\partial \mathbf{\Phi}}{\partial y^\mu} = \\
& = \overline{\partial}_{\dot{\alpha}} \mathbf{\Phi} + \frac{\partial \mathbf{\Phi}}{\partial y^\mu} (-i \theta^\beta \sigma^\mu_{\beta \dot{\alpha}}) + i\theta^\beta (\sigma^\mu)_{\beta \dot{\alpha}} \frac{\partial \mathbf{\Phi}}{\partial y^\mu} = \overline{\partial}_{\dot{\alpha}} \mathbf{\Phi} = 0
\end{split}
\end{equation*}
which means that $\mathbf{\Phi}(y^\mu,\theta^\alpha,\overline{\theta}_{\dot{\alpha}})$ depends explicitly only on $(y^\mu,\theta_\alpha)$, but not on $\overline{\theta}_{\dot{\alpha}}$, whose dependece enters implicitly via $y^\mu$. Consequently the general expression for the chiral superfield in these coordinates is
$$\mathbf{\Phi}(y,\theta) = \phi(y) + \sqrt{2} \theta \psi(y) + \theta \theta F(y)$$
where the coefficient $\sqrt{2}$ for the $\psi$ field is used in order to guarantee the right behavior of the superfield under the supersymmetry transformation. Before passing to the old coordinates, we see why this superfield is called \textit{chiral}: it contains the same degrees of freedom of a chiral multiplet, indeed $\phi$ is a scalar field, $\psi$ is a fermion of spin $\frac{1}{2}$ and $F$ is an auxiliary boson field. Its role is the one of making SUSY manifest off-shell, while on-shell, since SUSY is already present, this field can be expressed in terms of the other fields, not giving new degrees of freedom (so it is not dynamical).\\
Taylor-expanding the above expression around $y = x$ we get the expression of the chiral superfield in the $(x^\mu, \theta_\alpha, \overline{\theta}^{\dot{\alpha}})$ coordinates
\begin{equation}
\label{general_chiral}
\mathbf{\Phi}(x,\theta,\overline{\theta}) = \phi(x) + \sqrt{2} \theta \psi(x) + \theta \theta F(x) + i \theta \sigma^\mu \overline{\theta} \partial_\mu \phi(x) - \frac{i}{\sqrt{2}} (\theta \theta) \partial_\mu \psi(x) \sigma^\mu \overline{\theta} - \frac{1}{4} (\theta \theta)(\bar{\theta}\bar{\theta})\Box \phi(x)
\end{equation}
where $\Box = \partial_\mu \partial^\mu$. By hermitian conjugation we get the corresponding expression of the general anti-chiral superfield $\bar{\mathbf{\Phi}}(x,\theta,\overline{\theta})$.

\subsection{Vector superfield}
Now we need a constraint to be imposed on the general superfield $S$ in \eqref{superfield} in such a way the resulting superfield contains the same degrees of freedom of a vector multiplet. In particular it has to contain the vector field $v_\mu(x)$, which is real. This last condition is known as \textit{reality condition} and the superfield that satisfies it
\begin{equation}
\label{vector_field}
V(x,\theta,\overline{\theta}) = \overline{V}(x,\theta,\overline{\theta})
\end{equation}
is called \textit{vector superfield}. Starting from the general expression \eqref{superfield}, the expansion for $V$ is
\begin{equation}
\label{general_vector}
\begin{split}
V(x,\theta,\overline{\theta}) & = C(x) + i \theta \chi(x) - i \bar{\theta}\bar{\chi}(x) - \theta \sigma^\mu \bar{\theta}v_\mu(x) + \frac{i}{2} \theta \theta (M(x) + iN(x))-\frac{i}{2} \bar{\theta}\bar{\theta}(M(x)-iN(x)) \\
& + i \theta \theta \bar{\theta} \Bigl(\bar{\lambda}(x) + \frac{i}{2} \bar{\sigma}^\mu \partial_\mu \chi(x) \Bigr) - i \bar{\theta} \bar{\theta} \theta \Bigl(\lambda(x) + \frac{i}{2} \sigma^\mu \partial_\mu \bar{\chi}(x) \Bigr) + \frac{1}{2}\theta \theta \bar{\theta}\bar{\theta}\Bigl(D(x)+\frac{1}{2}\Box C(x) \Bigr)
\end{split}
\end{equation}
and, in order \eqref{vector_field} is satisfied, the scalar fields $C(x)$,$M(x)$,$N(x)$ and $D(x)$ have to be real. Consequently there are 8 bosonic degrees of freedom and 8 fermionic ones ($\chi_\alpha$, $\overline{\chi}_{\dot{\alpha}}$, $\lambda_\alpha$ and $\overline{\lambda}_{\dot{\alpha}}$).\\
The next step is the introduction of the supersymmetric generalization of gauge transformations acting on vector superfield. First of all, we notice that, given $\mathbf{\Phi}$ a chiral superfield, the superfield $\mathbf{\Phi} + \overline{\mathbf{\Phi}}$ is a vector superfield. It can be seen that the coefficient of its $\theta \sigma^\mu \bar{\theta}$ component is a total derivative term and this motivates us to define the supersymmetric generalization of a gauge transformation on the vector superfield $V$ in the following way
\begin{equation}
\label{SUSY_gauge_transformation}
V \rightarrow V + \mathbf{\Phi} + \overline{\mathbf{\Phi}}
\end{equation}
Under this transformation, the component field in \eqref{general_vector} transform as follows
\begin{subequations}
\begin{align}
C & \rightarrow C + \phi + \overline{\phi} \\
\chi & \rightarrow \chi - i \sqrt{2}\psi \\
M + iN & \rightarrow M + iN - 2iF \\
v_\mu & \rightarrow v_\mu -i \partial_\mu(\phi-\overline{\phi}) \label{gauge1} \\
\lambda & \rightarrow \lambda \label{gauge2} \\
D & \rightarrow D \label{gauge3}
\end{align}
\end{subequations}
Notice that \eqref{gauge1} is precisely how an ordinary (abelian) gauge transformation acts on a vector field; for this reason \eqref{SUSY_gauge_transformation} is a natural definition for the supersymmetric version of a gauge transforamtion.\\
By properly choosing the chiral superfield $\mathbf{\Phi}$, that is
$$\text{Re} \phi = -\frac{C}{2} \quad \psi = -\frac{i}{\sqrt{2}} \chi \quad \text{Re} F = \frac{N}{2} \quad \text{Im} F = -\frac{M}{2}$$
we see that we are setting $C = M = N = \chi = 0$. This corresponds to a specific gauge, called \textit{Wess-Zumino gauge} and in it the vector superfield can be written as
\begin{equation}
\label{vector_wz}
V_{\text{WZ}} = -\theta \sigma^\mu \bar{\theta} v_\mu(x) + i \theta \theta \bar{\theta} \bar{\lambda}(x) - i \bar{\theta}\bar{\theta}\theta \lambda(x) + \frac{1}{2} \theta \theta \bar{\theta}\bar{\theta} D(x)
\end{equation}
From all of this we see that the physical components of a vector superfield are: a vector field $v_\mu$ representing a gauge boson, then the fermionic superpartner of the gauge boson $\lambda$ and finally an auxiliary scalar field $D(x)$. We conclude this Subsection by noticing that in \eqref{vector_wz} each term contains at least one $\theta$, therefore 
$$V^2_{\text{WZ}} = -\frac{1}{2}\theta \theta \bar{\theta}\bar{\theta} v_\mu v^\mu \qquad  V^n_{\text{WZ}} = 0 \quad \text{for}\quad n \geq 3$$
This simplifies things a lot when we are going to construct supersymmetric gauge actions

\section{$\mathcal{N}=1$ SUSY theories} \label{sec:chap1_3}
At this point we are ready to construct $\mathcal{N}=1$ supersymmetric theories. The case with $\mathcal{N}=2$ will be analyzed later, but it uses notions that are typically introduced in the minimal SUSY setup. \\
First of all we need to know how to derive a supersymmetric invariant action. For this it is enough that the Lagrangian transforms as a scalar under Poincarè transformations (as it happens for non supersymmetric QFT) and its variation under a SUSY transformation is a total spacetime derivative. Here we can understand the power of the superfield notation: the integral in superspace of any superfield 
\begin{equation}\label{eq:SUSY_action}
\int \mathrm{d}^4 x \mathrm{d}^2 \theta \mathrm{d}^2 \overline{\theta} S(x,\theta,\overline{\theta}) \equiv \int \mathrm{d}^4 x \int \mathrm{d}^2\theta \mathrm{d}^2 \overline{\theta} S(x,\theta,\overline{\theta})
\end{equation}
is a supersymmetric invariant quantity. Let us see why it is true. We have defined
$$\mathrm{d}^2 \theta := \frac{1}{2} \mathrm{d}\theta^1 \mathrm{d}\theta^2 \qquad \mathrm{d}^2\overline{\theta} := \frac{1}{2} \mathrm{d}\overline{\theta}^{\dot{2}} \mathrm{d}\overline{\theta}^{\dot{1}}$$
and we do not report the usual operations with Grassman variables, for which the reader is referred to \cite{Bertolini:2025jhz}, \cite{Quevedo:2010ui} and \cite{Cabibbo:Gauge}. Now, thanks to the invariance under translation of the integration measure
$$\int \mathrm{d}\theta \theta = \int \mathrm{d}(\theta + \epsilon) (\theta + \epsilon) = 1$$
we have
\begin{align*}
\delta_{\epsilon,\overline{\epsilon}} \int \mathrm{d}^4x \, \mathrm{d}^2\theta \, \mathrm{d}^2\overline{\theta} S(x,\theta,\overline{\theta}) & = \int \mathrm{d}^4x \, \mathrm{d}^2\theta \, \mathrm{d}^2\overline{\theta} \delta_{\epsilon,\overline{\epsilon}} S(x,\theta,\overline{\theta}) \underset{\text{\eqref{importante},\eqref{supercharge_Q_diff_operators},\eqref{supercharge_Qb_diff_operators}}}{=} \\
& = \int \mathrm{d}^4x \, \mathrm{d}^2\theta \, \mathrm{d}^2\overline{\theta}\left(\epsilon^\alpha \partial_\alpha S + \overline{\epsilon}_{\dot{\alpha}} \overline{\partial}^{\dot{\alpha}} S + \partial_\mu[-i(\epsilon \sigma \overline{\theta} - \theta \sigma \overline{\epsilon})S]\right)
\end{align*}
Now, since the first two terms have at most terms with only one $\theta$ or $\overline{\theta}$, their integration in $\mathrm{d}^2\theta$ and in $\mathrm{d}^2\overline{\theta}$ is 0. The last term survives but it is a total spacetime derivative, therefore
$$\delta_{\epsilon,\overline{\epsilon}} \int \mathrm{d}^4x \, \mathrm{d}^2\theta \, \mathrm{d}^2\overline{\theta} S(x,\theta,\overline{\theta}) = 0$$
The integrand in the original quantity \eqref{eq:SUSY_action}, including also the integration in $\mathrm{d}^2\theta$ and in $\mathrm{d}^2\overline{\theta}$, transforms, under SUSY, in a total spacetime derivative. For this reason we define the Lagrangian of the theory as the superspace integral (not considering the spacetime $x^\mu$ coordinates) of a suitable superfield
\begin{equation}
\label{super_L}
\mathcal{L} = \int \mathrm{d}^2\theta \, \mathrm{d}^2\overline{\theta} S(x,\theta,\overline{\theta})
\end{equation} 
where by suitable we mean that it is such that $\mathcal{L}$ is real, it has dimension four and it transforms as a scalar under Poincarè transformations. Automatically, the supersymmetric invariant action of the theory is
\begin{equation}
\label{super_S}
\mathcal{S} = \int \mathrm{d}^4x \mathcal{L}
\end{equation}
Now, let us briefly focus on three classes of supersymmetric $\mathcal{N}=1$ theories: the first considers only the presence of matter (i.e of chiral superfields), the second only the vector superfield and the third examines the interaction between the previous two ones. 

\subsection{Matter Lagrangians} \label{matter actions}
From what we have discussed in the previous Section, in order to have a matter action the superfield in the integral of \eqref{super_L} is, in general, a function of $\mathbf{\Phi}$ and $\mathbf{\overline{\Phi}}$ (we are now considering the presence of one chiral superfield with its antichiral counterpart)
\begin{equation}
\label{kinetic_L}
\mathcal{L}_1 = \int \mathrm{d}^2\theta \, \mathrm{d}^2\overline{\theta} K(\mathbf{\Phi},\overline{\mathbf{\Phi}}) 
\end{equation}
As we are going to see, this is not the most general case. In order to make \eqref{kinetic_L} a good Lagrangian, it has to satisfy the properties we listed before ($\mathcal{L}_1$ has to be a scalar, real and has to have mass dimension equal to 4), which give some constrains on the function $K$. One of these is that it has to be a superfield, in order to for the Lagrangian to be supersymmetry invariant. Without entering in details (which can be consulted in \cite{Bertolini:2025jhz}), we can say that the most general expression for this function is
\begin{equation} \label{eq:general_Kahler}
K(\mathbf{\Phi},\overline{\mathbf{\Phi}}) = \sum_{m,n = 1}^\infty a_{nm}\overline{\mathbf{\Phi}}^n \mathbf{\Phi}^m \quad \text{with} \quad a^*_{mn} = a_{nm} \qquad \text{(from the reality condition)}
\end{equation}
and $a_{nm}=0$ for $n \ne m$. This function $K$ is called \textit{K\"ahler potential} and we will see later the reason of this name.\\
Now, for \textbf{renormalizable} theories, it can be demonstrated that \eqref{eq:general_Kahler} reduces to $K(\mathbf{\Phi},\overline{\mathbf{\Phi}}) = \overline{\mathbf{\Phi}} \, \mathbf{\Phi}$ and it is called \textit{canonical K\"ahler potential}. Even if we are going to be interested in \textbf{non-renormalizable} formulations, at the moment let us see what \eqref{kinetic_L} describes in the former case. The reason of it is that when we examine its generalization to the non-renormalizable situation, the physical information will be the same. Substituting the canonical K\"ahler potential into \eqref{kinetic_L}, we get
\begin{equation}\label{eq:ren_kin_term}
\mathcal{L}_1 = \int \mathrm{d}^2\theta \, \mathrm{d}^2\overline{\theta} \, \overline{\mathbf{\Phi}} \mathbf{\Phi} = \partial_\mu \overline{\phi} \partial^\mu \phi + \frac{i}{2}(\partial_\mu \psi - \psi \sigma^\mu \partial_\mu \overline{\psi}) + \overline{F}F
\end{equation}
and it is a kinetic term, hence we are describing, in this way, a free chiral superfield: $\mathcal{L}_1 = \mathcal{L}_{kin}$.\\
At this point we want to see how to describe interactions between chiral superfields. In order to understand it, we have to say that when we deal with chiral superfields, there is another possibility to get supersymmetric invariant action with all the desired properties we previously analyzed. Call $\Sigma$ a generic chiral superfield, then
$$\int \mathrm{d}^4x \int \mathrm{d}^2\theta \, \mathrm{d}^2\overline{\theta} \Sigma = 0$$
since the corresponding term that is selected with the first integral is a total derivative (as it can be seen in \eqref{general_chiral}). Nonetheless, if we consider
$$\int \mathrm{d}^4x \int \mathrm{d}^2\theta \Sigma$$
it is not 0 anymore, since the term that is selected from the first integral is the $\theta\theta$ component, which is not a total spacetime derivative. It is the term corresponding to the $F$ field. In literature, the two integrals
$$\int \mathrm{d}^2\theta \, \mathrm{d}^2 \overline{\theta} S \qquad \int \mathrm{d}^2\theta S$$
where $S$ is a general superfield are respectively called \textit{D-terms} and \textit{F-terms}.\\

Coming back to the problem of describing interactions between chiral superfields, we focus on F-terms. From the properties of Section \ref{chiral_superfields} on chiral superfields, it can be seen that any \textit{holomorphic} function of $\mathbf{\Phi}$ is a chiral superfield. We call this function $W(\mathbf{\Phi})$ and it is such that $\partial W / \partial \overline{\mathbf{\Phi}} = 0$ (simply requiring that $\overline{\mathcal{D}}_{\dot{\alpha}} W(\mathbf{\Phi})=0$).\\
Thanks to all of this, the term that describes interactions in a SUSY theory of a chiral superfield is
\begin{equation}
\label{interacting_L}
\mathcal{L}_{int} = \int \mathrm{d}^2\theta W(\mathbf{\Phi}) + \int \mathrm{d}^2\overline{\theta} \, \overline{W}(\overline{\mathbf{\Phi}})
\end{equation}
where we have added the h.c of the first integral in such a way $\mathcal{L}_{int}$ is real. The function $W$ is called \textit{superpotential}.\\
As well as the K\"ahler potential has to satisfy some properties, also $W$ has to do the same. One of these, as already said, is that it has to be a holomorphic function of the chiral superfield. Studying all the other constraints (on mass dimension, R-charge, etc...) we can arrive at the following general expression for the superpotential 
\begin{equation}\label{eq:general_superpotential}
W(\mathbf{\Phi}) = \sum_{n=1}^\infty b_n \mathbf{\Phi}^n
\end{equation}
and, for renormalizable models, it has to be at most cubic.\\
The computation of \eqref{interacting_L} can be done by Taylor-expanding the superpotential around $\mathbf{\Phi} = \phi$, having a power series in $\theta$
\begin{equation*}
\begin{split}
W(\mathbf{\Phi}) & = W(\phi) + (\mathbf{\Phi} - \phi) \frac{\partial W}{\partial \phi} + \frac{1}{2} (\mathbf{\Phi}-\phi)^2 \frac{\partial^2 W}{\partial \phi^2} \\
& = W(\phi) + \sqrt{2}\theta \psi \frac{\partial W}{\partial \phi} + \theta \theta \Bigl(\frac{\partial W}{\partial \phi} F - \frac{1}{2} \frac{\partial^2 W}{\partial \phi \partial \phi} \psi \psi\Bigr)
\end{split}
\end{equation*}
Thanks to this we have, also for non-renormalizable theories
\begin{equation}\label{eq:genral_int_term}
\mathcal{L}_{int} = \frac{\partial W}{\partial \phi} F - \frac{1}{2} \frac{\partial^2 W}{\partial \phi \partial \phi} \psi \psi + h.c
\end{equation}
We can now put toghether the kinetic term \eqref{kinetic_L} with the interaction one \eqref{eq:genral_int_term} in order to get the most generic supersymmetric matter Lagrangian
\begin{equation}
\label{matter_L}
\mathcal{L}_{matter} = \mathcal{L}_{kin} + \mathcal{L}_{int} = \int \mathrm{d}^2 \theta \, \mathrm{d}^2 \overline{\theta} K(\mathbf{\Phi},\overline{\mathbf{\Phi}}) + \int \mathrm{d}^2\theta W(\mathbf{\Phi}) + \int \mathrm{d}^2\overline{\theta} \, \overline{W}(\overline{\mathbf{\Phi}})
\end{equation}
For the particular case of renormalizable theories, we have to consider \eqref{eq:ren_kin_term} for the kinetic term and the interaction term \eqref{eq:genral_int_term} keeping in mind that $W$ is at most cubic.\\

For further purposes, it is quite relevant to quickly see how to find the scalar potential from this model (we consider the renormalizable situation for simplicity). As we hinted in the Section \ref{chiral_superfields}, the auxiliary fields $F$ and $\overline{F}$ are relevant only off-shell, hence, when we go on-shell, we should integrate them out, meaning that we have to substitute them with their equations of motion. The part of the Lagrangian depending on these fields takes the form
$$\mathcal{L}_{(F)} = \overline{F}F + \frac{\partial W}{\partial \phi} F + \frac{\partial \overline{W}}{\partial \overline{\phi}} \overline{F}$$
and imposing that $\partial \mathcal{L}_{(F)} / \partial F = \partial \mathcal{L}_{(F)} / \partial {\overline{F}} = 0$, we obtain the following expressions for the auxiliary fields
$$\overline{F} = - \frac{\partial W}{\partial \phi} \qquad F = - \frac{\partial \overline{W}}{\partial \overline{\phi}}$$ 
In this way the on-shell renormalizable Lagrangian is
$$\mathcal{L}^{on-shell}_{matter} = \partial_\mu \overline{\phi} \partial^\mu \phi + \frac{i}{2}(\partial_\mu \psi \sigma^\mu \overline{\psi} - \psi\sigma^\mu\partial_\mu\overline{\psi}) - \Bigl|\frac{\partial W}{\partial \phi}\Bigr|^2 - \frac{1}{2}\frac{\partial^2 W}{\partial \phi \partial \phi}\psi \psi - \frac{1}{2}\frac{\partial^2 \overline{W}}{\partial \overline{\phi} \, \partial \overline{\phi}} \overline{\psi} \, \overline{\psi}$$
From this we read the scalar potential, which is
$$V(\phi,\overline{\phi}) = \Bigl|\frac{\partial W}{\partial \phi}\Bigr|^2 = \overline{F}F$$
All this treatment can be extended to a set of chiral superfields $\mathbf{\Phi}^i$, with $i=1,_{\cdots},n$, having as Lagrangian
\begin{equation}\label{eq:general_matter_multiple}
\mathcal{L}_{matter} = \int \mathrm{d}^2\theta \mathrm{d}^2\overline{\theta} K(\mathbf{\Phi}^i, \overline{\mathbf{\Phi}}_i) + \int \mathrm{d}^2\theta W(\mathbf{\Phi}^i) + \int \mathrm{d}^2\overline{\theta} \, \overline{W}(\overline{\mathbf{\Phi}}_i)
\end{equation}
Now a last comment is in order, regarding the K\"ahler potential $K(\mathbf{\Phi}^i, \overline{\mathbf{\Phi}}_i)$. We said that we will be interested in non renormalizable theories since we are going to consider effective models, more precisely supersymmetric non-linear sigma models. In order to understand what a sigma model is, notice that a general field $\eta^i$ is a map from the spacetime $\mathcal{M}$ to the \textit{target space} $T$, which is the space of the configurations assumed by the field $\eta^i$ itself
\begin{equation} \label{field_map}
\eta^i : \, \mathcal{M} \, \to \, T
\end{equation}
We can equip these two spaces with metric tensors $h_{\mu \nu}$ and $g_{ij}$, respectively. The Lagrangian density of a sigma model can be written using a metric of the spacetime $\mathcal{M}$ which is obtained as the pullback of the metric on $T$ through the map $\eta^i$. In other words
\begin{equation}
 \mathcal{L} = \frac{1}{2} g_{ij} (\eta) \, h^{\mu\nu} \, \partial_\mu \eta^i \, \partial_\nu \eta^j
\end{equation}
where $\partial_\mu$ are the spacetime derivatives and $g_{i j} \partial_\mu \eta^i \partial_\nu \eta^j$ is another metric on $\mathcal{M}$ and it is exactly the pullback of the metric $g$ through the map $\eta^i$. For a non-linear sigma model, the fields that we consider as the maps \eqref{field_map} are scalar fields and, for supersymmetric non-linear sigma model, they are the scalar components of the superfields in the SUSY extension. So, for our analysis, these are the $\phi^i$ fields.  At this point, it can be shown that the metric of the space of scalar fields configurations for these theories is defined as the second derivative of the K\"ahler potential with respect to the scalar components of the chiral superfields $\phi$ and $\overline{\phi}$ (which represent, for what just explained, the coordinates on the target space). Indeed, by Taylor expanding $K(\mathbf{\Phi}^i, \overline{\mathbf{\Phi}}_j)$ around $\mathbf{\Phi}^i = \phi^i$ and $\overline{\mathbf{\Phi}}_j = \overline{\phi}_j$ and using for $\mathbf{\Phi} - \phi(x)$ what derives from \eqref{general_chiral}, we get
\begin{equation}\label{eq:kahler1}
K(\mathbf{\Phi}^i, \overline{\mathbf{\Phi}}_j) = K(\phi^i, \overline{\phi}_j) + K_i \Delta^i + K^j \overline{\Delta_j} + \frac{1}{2} K_{i j} \Delta^i \Delta^j + \frac{1}{2} K^{i j} \overline{\Delta}_i \overline{\Delta}_j + K^j_i \Delta^i \overline{\Delta}_j + \dots 
\end{equation}
where
\begin{align}\label{eq:kahler2}
& \Delta^i \equiv \mathbf{\Phi}^i - \phi^i =  \sqrt{2} \theta \psi^i(x) + \theta \theta F^i(x) + i \theta \sigma^\mu \overline{\theta} \partial_\mu \phi^i(x) - \frac{i}{\sqrt{2}} (\theta \theta) \partial_\mu \psi^i(x) \sigma^\mu \overline{\theta} - \frac{1}{4} (\theta \theta)(\bar{\theta}\bar{\theta})\Box \phi^i(x) \notag \\
& \overline{\Delta}_j \equiv \overline{\mathbf{\Phi}}_j - \overline{\phi}_j= \sqrt{2} \bar{\theta} \bar{\psi}_j(x) + \bar{\theta} \bar{\theta} \, \overline{F}_j(x) - i \theta \sigma^\mu \overline{\theta} \partial_\mu \bar{\phi}_j(x) + \frac{i}{\sqrt{2}} (\bar{\theta} \bar{\theta}) \theta \sigma^\mu \partial_\mu \overline{\psi}_j(x) - \frac{1}{4} (\theta \theta)(\bar{\theta}\bar{\theta})\Box \overline{\phi}_j(x) \notag \\
& K_i \equiv \frac{\partial K}{\partial \phi^i}, \qquad K^j \equiv \frac{\partial K}{\partial \overline{\phi}_j}, \qquad K^j_i = \frac{\partial^2 K}{\partial \phi^i \, \partial \overline{\phi}_j}
\end{align}
Selecting the $\theta \theta \bar{\theta} \bar{\theta}$ component in the term $K^j_i \Delta^i \overline{\Delta}_j$ as required in \eqref{eq:general_matter_multiple}, we obtain
$$K^j_i \Delta^i \overline{\Delta}_j = K^j_i \left(F^i \overline{F}_j + \frac{1}{2} \partial_\mu \phi^i \partial^\mu \overline{\phi}_j - \frac{i}{2} \psi^i \sigma^\mu \partial_\mu \overline{\psi}_j + \frac{i}{2} \partial_\mu \psi^i \sigma^\mu \overline{\psi}_j \right)$$ 
and it can be seen that the mixed derivatives $K^j_i$ are coefficients of the kinetic terms; furthermore it can be proven that $K^j_i$ are hermitian, positive definite and non-singular (see Appendix C of \cite{Wess:1992cp}), which are all necessary properties for the components of a metric of a manifold, which, in the present case, is subjected to certain conditions that make it a K\"ahler manifold. This explains the origin of the name K\"ahler potential for $K(\mathbf{\Phi},\mathbf{\overline{\Phi}})$.

\subsection{$\mathcal{N}=1$ SuperYang-Mills} \label{pure SYM}
Now it is the turn of the vector superfield, that is we want to determine the supersymmetric invariant action describing the kinetic term of a vector superfield, having in this way the $\mathcal{N}=1$ supersymmetric generalization of Yang-Mills theory. Since we will consider only the non-abelian case, we report here the analysis for this situation and at the end we will explain how to find the corresponding quantities in abelian models.\\
We denote the gauge group as $G$ and, as already said in Section \ref{sec:massless_supermultiplets}, the vector multiplet (i.e. superfield $V$, in this case) transforms in the adjoint representation of $G$, meaning that
$$V = V_a T^a \qquad {\text{with}} \quad a=1,_{\cdots},n = \text{dim}(G)$$
where $T^a$ are hermitian generators of $G$. At this point, one has to work out the right gauge transformation that generalizes \eqref{SUSY_gauge_transformation}, valid in the abelian scenario. It turns out that such transformation is
\begin{equation}
\label{general_gauge_transformation}
e^V \rightarrow e^{i\overline{\Lambda}} e^V e^{-i\Lambda}
\end{equation}
where $\Lambda = \Lambda_a T^a$ is a matrix of chiral superfields $\Lambda_a$ and this transformation satisfies the requested property because, identifying $\mathbf{\Phi} = -i\Lambda$ (obviously $\mathbf{\Phi} = \mathbf{\Phi}_a T^a$), we have, at leading order in $V$
$$1 + V \rightarrow (1+\overline{\mathbf{\Phi}})(1+V) (1+\mathbf{\Phi}) = 1 + V + \mathbf{\Phi} + \overline{\mathbf{\Phi}} + \text{commutators between superfields}$$
finding \eqref{SUSY_gauge_transformation} when all the commutators are 0 (i.e. in the abelian case). Now we have to find the supersymmetric generalization of the field strength, since it enters in the Lagrangian (and so in the action). We consider the following chiral superfield, with its anti-chiral counterpart
\begin{equation} \label{eq:super_field_strength}
\mathcal{W}_\alpha = -\frac{1}{4}\overline{\mathcal{D}} \, \overline{\mathcal{D}}\bigl(e^{-V}\mathcal{D}_\alpha e^V\bigr) \qquad \overline{\mathcal{W}}_{\dot{\alpha}} = -\frac{1}{4} \mathcal{D} \mathcal{D} \bigl(e^V \overline{\mathcal{D}}_{\dot{\alpha}} e^{-V}\bigr)
\end{equation}
It can be shown that, at leading order in $V$, the first object becomes (the other one behaves in a similar way, but with complex conjugation)
$$\mathcal{W}_\alpha = -\frac{1}{4} \overline{\mathcal{D}} \, \overline{\mathcal{D}}\mathcal{D}_\alpha V + \text{commutators between the superfields}$$
from which we can easily read the expression of this superfield in the abelian case. Furthermore, it can be shown that under the gauge transformation \eqref{general_gauge_transformation} these two superfields transform in a covariant way, meaning that
$$\mathcal{W}_\alpha \to e^{i \Lambda} \mathcal{W}_\alpha e^{- i \Lambda} \qquad \overline{\mathcal{W}}_{\dot{\alpha}} \rightarrow e^{i\overline{\Lambda}} \overline{\mathcal{W}}_{\dot{\alpha}} e^{-i\overline{\Lambda}}$$
from which it is clear that, in the abelian case, these two superfields are invariant under \eqref{SUSY_gauge_transformation}. In order to see why $\mathcal{W}_\alpha$ is the right supersymmetric generalization of the field strength, let us find its components. With this goal it is very convenient to work in the Wess-Zumino gauge with the coordinates $(y,\theta,\overline{\theta})$, for which the vector superfield is
$$V_{\text{WZ}} = -\theta \sigma^\mu \bar{\theta} v_\mu(y) + i\theta \theta \bar{\theta} \bar{\lambda}(y) - i\bar{\theta}\bar{\theta}\theta \lambda(y) + \frac{1}{2}\theta \theta \bar{\theta}\bar{\theta}[D(y)- i \partial_\mu v^\mu (y)]$$
As we saw, $V_{\text{WZ}}^n = 0$ for $n \geq 3$, thus we automatically get that 
$$e^{V_{\text{WZ}}} = 1 + V_{\text{WZ}} + \frac{V^2_{\text{WZ}}}{2}$$
In this way it can be proven that
\begin{equation}
\mathcal{W}_\alpha = -\frac{1}{4} \overline{\mathcal{D}} \, \overline{\mathcal{D}}\mathcal{D}_\alpha V_{\text{WZ}} + \frac{1}{8} \overline{\mathcal{D}} \, \overline{\mathcal{D}}[V_{\text{WZ}},\mathcal{D}_\alpha V_{\text{WZ}}]
\end{equation}
where
$$\frac{1}{8} \overline{\mathcal{D}} \, \overline{\mathcal{D}}[V_{\text{WZ}},\mathcal{D}_\alpha V_{\text{WZ}}] = \frac{1}{2} (\sigma^{\mu\nu}\theta)_\alpha[v_\mu, v_\nu]-\frac{i}{2}\theta \theta \sigma^\mu_{\alpha \dot{\beta}}[v_\mu,\overline{\lambda}^{\dot{\beta}}]$$
Putting everything together, it turns out that
\begin{equation}
\label{W_components_nonab}
\mathcal{W}_\alpha = -i\lambda_\alpha (y) +\theta_\alpha D(y) + i(\sigma^{\mu\nu}\theta)_\alpha F_{\mu\nu}+\theta\theta(\sigma^\mu D_\mu \overline{\lambda}(y))_\alpha
\end{equation}
where
$$F_{\mu\nu} = \partial_\mu v_\nu -\partial_\nu v_\mu - ig[v_\mu,v_\nu] \qquad D_\mu = \partial_\mu - ig[v_\mu,]$$
$g$ is the gauge coupling. All of this confirms the fact that $\mathcal{W}_\alpha$ is the supersymmetric field strength of our theory. Its expression in the abelian case is obtained by turning off all the commutators.\\

Now we are able to see how to construct the Lagrangian that describes the kinetic term of our vector superfield. Since $\mathcal{W}_\alpha$ is a chiral superfield, then also $\mathcal{W}^\alpha \, \mathcal{W}_\alpha$ is a chiral superfield and so we can consider the integral of this quantity with measure $\mathrm{d}^2 \theta$. Actually it is the case for the abelian scenario, due to the fact that here $\mathcal{W}^\alpha \, \mathcal{W}_\alpha$ is a Lorentz scalar, but now we are in the non-abelian scenario and we have understood the index of the gauge group. Keeping in mind that: $\lambda_\alpha = \lambda_{\alpha a}T^a$, $v_\mu = v_{\mu a}T^a$ and $D = D_a T^a$, we can write, more explicitly (we have understood the dependence from $y$)
$$\mathcal{W}_{\alpha a} = -i\lambda_{\alpha a} +\theta_\alpha D_a + i(\sigma^{\mu\nu}\theta)_\alpha F_{\mu\nu a}+\theta\theta(\sigma^\mu D_\mu \overline{\lambda}_a)_\alpha$$
Recall that
$$[T^a,T^b] = i f^{ab}_{\,\,\,\,c}T^c$$
with the generators normalized such that $\text{Tr}(T^a T^b) = \delta^{ab}$ and so
$$F_{\mu\nu a} = \partial_\mu v_{\nu a} - \partial_\nu v_{\mu a} + gv_{\mu b} v_{\nu c}f^{bc}_{\,\,\,\, a} \qquad D_\mu \overline{\lambda}_a = \partial_\mu \overline{\lambda}_a + g v_{\mu b}\overline{\lambda}_c f^{bc}_{\,\,\,\, a}$$
 For this reason we have to consider, as quantity inside the integral, the trace over the gauge group of $\mathcal{W}^\alpha \, \mathcal{W}_\alpha$ (which is also gauge invariant). The gauge invariant Lagrangian which is supersymmetric and that describes the kinetic term of vector superfields is
\begin{equation}
\label{SYM_L}
\begin{split}
\mathcal{L}_{gauge}^{ren} & = \text{Im} \left[ \frac{\tau}{16 \pi i} \int \mathrm{d}^2\theta \text{Tr}(\mathcal{W}^\alpha \mathcal{W}_\alpha)\right] = \\
& = \text{Tr}\Bigl[-\frac{1}{4g^2} F_{\mu\nu}F^{\mu\nu}-\frac{i}{g^2} \lambda^\alpha \sigma^\mu_{\alpha\dot{\beta}} D_\mu \overline{\lambda}^{\dot{\beta}} + \frac{1}{2g^2} D^2 \Bigr] + \frac{\theta}{32\pi^2} \text{Tr}(F_{\mu\nu}\widetilde{F}^{\mu\nu})
\end{split} 
\end{equation}   
where $\widetilde{F}^{\mu\nu} = \frac{1}{2}\epsilon^{\mu\nu\rho\sigma}F_{\rho\sigma}$ and
$$\tau = \frac{\theta}{2\pi} + \frac{4\pi i}{g^2}$$
is the \textit{complexified gauge coupling}. $\theta$ is the $\theta$ term that concerns the strong CP problem, not the superspace coordinate. Thanks to the previous normalization of the generators, we have that $\text{Tr}(F_{\mu\nu}F^{\mu\nu}) = F_{\mu\nu, a}F^{\mu\nu,a}$ and so on for all the other objects. The gauge coupling $g$ appears in an holomorphic way in $\tau$, so it is called \textit{holomorphic gauge coupling}, which is not the physical one, because it is obtained when the kinetic term of the gauge field is canonically normalized. This can be done with the following rescaling of the fields
\begin{equation} \label{eq:rescalings}
v_\mu \to g \, v_\mu \qquad \lambda \to g \, \lambda \qquad D \to g \, D
\end{equation}
Up to now we have considered a renormalizable Lagrangian. As done in the previous Subsection \ref{matter actions} for chiral superfields, we can also consider a kinetic term of gauge multiplets in a non-renormalizable model. It can be written as
\begin{equation} \label{gauge_L}
\mathcal{L}_{gauge} = \text{Im} \int \mathrm{d}^2\theta \mathcal{F}_{a b} \mathcal{W}^{\alpha a} \mathcal{W}^b_\alpha
\end{equation}
where $\mathcal{F}_{a b}$ is the generalization of the complexified gauge coupling, holomorphic in the chiral superfields, and it will play a crucial role when passing to $\mathcal{N}=2$ SUSY.

\subsection{$\mathcal{N}=1$ Gauge matter actions} \label{gauge matter actions}
Now it is time to couple matter superfields with vector ones, having a more complete theory. For this purpose, we have first to see how the matter sector behaves in this framework.\\
Let us consider one chiral superfield $\mathbf{\Phi}$ transforming in some representation $R$ of the gauge group $G$. Under the gauge transformation \eqref{general_gauge_transformation} we get that
$$\mathbf{\Phi} \rightarrow \mathbf{\Phi}' = e^{i\Lambda}\mathbf{\Phi}$$
where $\Lambda = \Lambda_a T^a_R$ ($T^a_R$ are the generators of the gauge group in the representation $R$). The problem of this is that the kinetic term of the matter sector we introduced above, that is $\overline{\mathbf{\Phi}} \mathbf{\Phi}$, is not gauge invariant anymore, since
$$\overline{\mathbf{\Phi}} \mathbf{\Phi} \rightarrow \overline{\mathbf{\Phi}} e^{-i\overline{\Lambda}} e^{i\Lambda} \mathbf{\Phi} \ne \overline{\mathbf{\Phi}} \mathbf{\Phi}$$
This shows that the kinetic term has to be changed and that one which is supersymmetric invariant and gauge invariant happens to be
$$\overline{\mathbf{\Phi}} e^V \mathbf{\Phi}$$
With this modification we have that
$$\mathcal{L}_{matter} = \int \mathrm{d}^2\theta \mathrm{d}^2\overline{\theta} \, \overline{\mathbf{\Phi}}e^V \mathbf{\Phi} + \int \mathrm{d}^2\theta W(\mathbf{\Phi}) + \int \mathrm{d}^2\overline{\theta} \, \overline{W}(\overline{\mathbf{\Phi}})$$
Obviously the superpotential has to be gauge invariant. The computation of the D-term of this Lagrangian, after having done the rescaling $V \rightarrow 2gV$ (analogous to \eqref{eq:rescalings}), brings to the following result
$$\int \mathrm{d}^2\theta \mathrm{d}^2\overline{\theta} \, \overline{\mathbf{\Phi}} e^{2gV} \mathbf{\Phi} = (\overline{D_\mu \phi})D^\mu \phi - i\psi\sigma^\mu D_\mu \overline{\psi}+ \overline{F}F + i\sqrt{2} g \overline{\phi} \lambda \psi - i\sqrt{2} g \overline{\psi} \, \overline{\lambda} \phi + g\overline{\phi}D\phi$$
where the covariant derivative $D_\mu = \partial_\mu - igv^a_\mu T^a_R$ is present to guarantee the gauge covariance of this kinetic term. We see how this D-term not only provides good matter kinetic terms, but also interaction terms between matter fields $\phi,\psi$ and the fermionic partner of the vector field, that is $\lambda$ (called \textit{gaugino}). Making the indices explicit, we have
$$\overline{\phi} \lambda \psi = \overline{\phi}_i(T^a_R)^i_{\,j}\lambda_a \psi^j$$
At this point we can assemble all the ingredients \footnote{In supersymmetric theories a new ingredient can be added to the Lagrangian \eqref{eq:total_lag} when the gauge group $G$ contains U$(1)$ factors. It is called \textit{Fayet-Iliopulos term}, but we do not consider it since it is not relevant for our discussion.} to write down the most general $\mathcal{N}=1$ renormalizable supersymmetric Lagrangian
\begin{align} \label{eq:total_lag}
\mathcal{L} & = \mathcal{L}_{matter} + \mathcal{L}_{gauge} =  \int \mathrm{d}^2\theta \mathrm{d}^2\overline{\theta} \, \,\overline{\mathbf{\Phi}}e^{2gV} \mathbf{\Phi} + \int \mathrm{d}^2\theta W(\mathbf{\Phi}) + \int \mathrm{d}^2\overline{\theta} \, \, \overline{W}(\overline{\mathbf{\Phi}}) + \text{Im} \left[ \frac{\tau}{16 \pi i} \int \mathrm{d}^2\theta \text{Tr}(\mathcal{W}^\alpha \mathcal{W}_\alpha)\right] = \notag \\
& = \text{Tr}\Bigr[-\frac{1}{4}F_{\mu\nu}F^{\mu\nu} - i\lambda\sigma^\mu D_\mu \overline{\lambda} +\frac{1}{2}D^2 \Bigr] + \frac{\theta}{32\pi^2}g^2 \text{Tr} \left(F_{\mu\nu}\widetilde{F}^{\mu\nu}\right) + (\overline{D_\mu \phi}) D^\mu \phi -i \psi \sigma^\mu D_\mu \overline{\psi} + \overline{F}F + i \sqrt{2} g \overline{\phi} \lambda \psi + \notag \\
& -i\sqrt{2} g \overline{\psi} \, \, \overline{\lambda} \phi + g \overline{\phi}D\phi + \frac{\partial W}{\partial \phi^i}F^i + \frac{\partial {W}}{\partial \overline{\phi}^i}\overline{F}_i -\frac{1}{2}\frac{\partial^2 W}{\partial \phi^i \partial \phi^j} \psi^i \psi^j - \frac{1}{2}\frac{\partial^2 \overline{W}}{\partial \bar{\phi}_i \partial\overline{\phi}_j} \overline{\psi}_i \overline{\psi}_j
\end{align}
The auxiliary fields $F^i$ and $D^a$ can be integrated out, substituting their expression obtained by their equations of motion
$$\frac{\partial \mathcal{L}}{\partial F_i} = 0 \Rightarrow \overline{F}_i = - \frac{\partial W}{\partial \phi^i} \qquad \frac{\partial \mathcal{L}}{\partial D^a} = 0 \Rightarrow D^a = -g\overline{\phi}T^a\phi$$
Plugging back these relations in the Lagrangian, one obtains the on-shell supersymmetric Lagrangian and the scalar potential that is obtained is
\begin{equation}
\label{scalar_potential}
V(\phi,\bar{\phi}) = \frac{\partial W}{\partial \phi^i} \frac{\partial \overline{W}}{\partial \overline{\phi}_i} + \frac{g^2}{2} \sum_a |\overline{\phi}_i (T^a)^i_{\, j} \phi^j|^2 = \overline{F}F + \frac{D^2}{2}
\end{equation}
The importance of the scalar potential is the possibility to define a \textit{moduli space}, which is the space of scalar field vacuum expectation values (VEVs) in correspondence of which the scalar potential itself vanishes 

$$F_i (\braket{\phi}) = 0 \quad D^a(\braket{\phi}, \braket{\overline{\phi}}) = 0$$
In presence of a superpotential only the second set of equations has to be satisfied to describe this kind of space. Since we are dealing with gauge theories, the VEVs that can be related one with the other via a gauge transformation describe exactly the same physics, hence the moduli space is more properly defined as
\begin{equation}
\label{moduli space}
\mathcal{M}_{cl} = \{\braket{\phi_j} | D^a(\braket{\phi_j},\braket{\overline{\phi}_j}) = 0, F_i(\braket{\phi_j}) = 0\}/\text{gauge transformation}
\end{equation} 
Notice that we are dealing here with a \textbf{classical} moduli space because we are describing the vacua at the classical or tree level, without adding quantum corrections, which are only non-perturbative in supersymmetric theories with 8 supercharges.

\section{$\mathcal{N}=2$ SUSY theories} \label{sec:N2_theories}
Now we want to examine the structure of the theories with $\mathcal{N}=2$ SUSY, which are the central objects of interest of this thesis. As done in the previous Section, we desire to see how to construct the most general $\mathcal{N}=2$ supersymmetric action in four dimensions and, at the moment, we are going to use the $\mathcal{N}=1$ superspace formalism.\\
We have to consider $\mathcal{N}=2$ superfields (vector superfield and hypermultiplets) and what we discussed in terms of supermultiplets in Section \ref{sec:supermultiplets} can be used also for superfields. This implies that a $\mathcal{N}=2$ vector superfield can be seen as the direct sum of a $\mathcal{N}=1$ chiral superfield $\mathbf{\Phi}$ and a $\mathcal{N}=1$ vector superfield $V$; similarly, an hypermultiplet can be built from two $\mathcal{N}=1$ chiral superfields, $\mathbf{H}_1$ and $\mathbf{H}_2$. Schematically we have
\begin{align}
& \mathcal{N}=2 \quad \text{vector superfield}: \quad \mathbf{\Phi} = (\phi,\psi_\alpha, F) \oplus V=(\lambda_\alpha, v_\mu, D) \notag \\
& \mathcal{N}=2 \quad \text{hypermultiplet}: \quad \mathbf{H}_1 = (h_1,\psi_{1,\alpha}, F_1) \oplus \overline{\mathbf{H}}_2 =(\overline{h}_2, \overline{\psi}_{2 \dot{\alpha}}, \overline{F}_2) 
\end{align}
(notice that $\mathbf{H}_1$ and $\overline{\mathbf{H}}_2$ transform in the same representation).\\

Let us start considering \textbf{pure SYM}, which is the object examined in the following Section using the analysis of Seiberg-Witten. In order to construct the Lagrangian, we have two requirements to impose. The first is that, since $\mathbf{\Phi}$ has to transform in the same representation of $V$ and they are elements of a vector superfield, then $\mathbf{\Phi}$ has to transform in the adjoint representation of the gauge group (hence $\phi = \phi_a T^a, \psi_\alpha = \psi_{\alpha a} T^a$ and $F = F_a T^a$). Furthermore, the Lagrangian has to be invariant under the exchange of the two fermions $\lambda_\alpha \leftrightarrow \psi_\alpha$. This because the $R$-symmetry of this kind of theories is U$(2) \simeq \text{SU}(2)_R \times \text{U}(1)_R$ and its compact component has to be a symmetry of the Lagrangian itself. Now, all the boson fields $\phi, F, v_\mu, D$ are singlets under SU$(2)_R$, while $\lambda_\alpha$ and $\psi_\alpha$ transform in the fundamental representation of this group, hence $(\lambda_\alpha, \psi_\alpha)$ is a doublet. The reason for this transformation property is that these fields are obtained by acting once with the supercharges on the Clifford vacuum and these operators transform in the fundamental representation.\\
Thanks to all of this, the Lagrangian reads
\begin{align} \label{eq:SYM_lag_N2}
\mathcal{L}_{SYM} & = \int \mathrm{d}^2\theta \mathrm{d}^2\overline{\theta} \, \,\text{Tr}\left(\overline{\mathbf{\Phi}}e^{2gV} \mathbf{\Phi}\right) + \text{Im} \left[ \frac{\tau}{16 \pi i} \int \mathrm{d}^2\theta \text{Tr} \, (\mathcal{W}^\alpha \mathcal{W}_\alpha)\right] = \notag \\
& = \text{Tr}\left\{-\frac{1}{4}F_{\mu\nu}F^{\mu\nu} - i\lambda\sigma^\mu D_\mu \overline{\lambda} -i \psi \sigma^\mu D_\mu \overline{\psi}  +\frac{1}{2}D^2 + \frac{\theta}{32\pi^2}g^2 F_{\mu\nu}\widetilde{F}^{\mu\nu} + (\overline{D_\mu \phi}) D^\mu \phi + \overline{F}F + \notag \right. \\
& \left. + i \sqrt{2} g \overline{\phi} \{\lambda ,\psi\} - i\sqrt{2} g \{\overline{\psi}, \overline{\lambda} \} \phi + g D [\phi, \overline{\phi}] \right\}
\end{align}
We can immediately see some differences with respect to the $\mathcal{N}=1$ Lagrangian \eqref{eq:total_lag} (which, even if describes a different theory with respect pure SYM, it has a form similar to \eqref{eq:SYM_lag_N2}). The first is the presence of commutators and anticommutators. This is due to the fact that all the fields transform in the adjoint representation of the gauge group. By definition, it means that $(T^a_{adj})^{bc} = - i f^{abc}$ and so the term $\overline{\phi} \lambda \psi$ becomes
$$\overline{\phi}_b \, \lambda_a \, (T^a_{adj})^{b c} \, \psi_c = - i \overline{\phi}_b \, \lambda_a \, f^{abc} \, \psi_c = i \overline{\phi}_b \, \lambda_a \, f^{bac} \, \psi_c = \overline{\phi}_b \, \lambda_a \, \psi_c \, T^b [T^a,T^c] = \text{Tr} \left(\overline{\phi} \{\lambda, \psi\}\right)$$
and so on for all the other terms. The other difference is the absence of a superpotential. This is due to the symmetry of the Langrangian under the exchange of $\lambda_\alpha$ with $\psi_\alpha$: a superpotential would give interactions and/or mass terms only for the $\psi$ field and this would not be compatible with the SU$(2)_R$ symmetry.\\

As seen for the $\mathcal{N}=1$ SUSY, the scalar potential is obtained by the auxiliary fields' contribution, which is
$$V=\text{Tr} \left\{\frac{1}{2} D^2 + \overline{F} F + g D \, [\phi, \overline{\phi}] \right\}$$
with minima given by
\begin{equation}
F^a = 0 \qquad D^a = -g [\phi, \overline{\phi}]^a
\end{equation}
As we can see, if the superpotential is 0, the auxiliary fields $F^a$ are trivial. This explains what we said at the end of the previous Section concerning the reason why the moduli space is described only by the set of equations for the $D$ field when the superpotential is not present. The scalar potential hence reads
\begin{equation}
V(\phi,\overline{\phi}) = -\frac{1}{2} D^a D_a = -\frac{1}{2} g^2 \text{Tr} \left([\phi,\overline{\phi}]^2\right)
\end{equation}
and the moduli space is described by the equations
\begin{equation}
[\phi, \overline{\phi}] = -2 i [\phi_1, \phi_2] = 0 
\end{equation}
where we set $\phi = \phi_1 + i \phi_2$. The solutions of these equations are, by definition, elements of the Cartan sub-algebra of the gauge group $G$ (remember that, since our fields in consideration transform in the adjoint representation of $G$, they are elements of its gauge algebra $\mathcal{G}$ and the parametrization of the moduli space is done, at this stage, through the elements of the Cartan sub-algebra). Due to the fact that $\phi_1$ and $\phi_2$ are hermitian (being linear combinations of the Cartan generators over the real field), then they can be simultaneously diagonalized and this can be done in a suitable choice of the basis of the Cartan sub-algebra itself. This corresponds to fix the gauge, which is necessary to correctly parametrize the space of inequivalent vacua, as explained at the end of Section \ref{gauge matter actions}.\\

Let us see an example of what just said. Consider as gauge group $G=\text{SU}(N)$, whose rank (the number of independent Cartan generators) is equal to $\text{rk}(G) = N-1$ and dimension is $\text{dim}(G) = N^2-1$. It is known that its generators can be represented as traceless hermitian $N \times N$ matrices. In this scenario, a representative of a vacuum configuration would be a $N \times N$ diagonal matrix of the form $\phi = (\phi_{1,j} + i \phi_{2,j}) h^j$ (where the $h$'s denote the $N-1$ generators of the Cartan sub-algebra) with $N$ complex eigenvalues. Anyway, the number of independent parameters which identify the vacuum configuration is $N-1 = \text{rk}(G)$, thanks to the constraint $\text{Tr}\phi = 0$. From all of this we learn that the moduli space is a $\text{rk}(G)$-dimensional complex manifold whose points represent inequivalent vacuum configurations and that everywhere, except the origin of this space, the gauge group is spontaneously broken to 
$$\text{SU}(N) \to \text{U}(1)^{\text{rk}(G)}$$
For this reason, thanks to Goldstone's theorem, to each of the $\text{rk}(G)$ of broken generators of the symmetry group corresponds a massless Goldstone boson which eventually gives mass to one of the vector bosons of the theory thanks to the Higgs mechanism. This means that the low energy effective dynamics of a theory of this kind is that of $\text{rk}(G)$ massless vector multiplets and $\text{dim}(G) - \text{rk}(G)$ massive vector multiplets whose masses depend on the scalar fields VEVs. We will come back to this aspect later on.\\
At this point we have to say that the eigenvalues of the vacuum configuration are not a good parametrization of the moduli space. This is due to the fact that the action of the gauge group $G$ on the Cartan sub-algebra is codified by the Weyl group (hence assumed here to be a subgroup of $G$), which in this way relates several points of the vacuum space we have described up to now and this is not what we want. In order to identify these points and hence getting a good description of the moduli space, we have to work out combinations of the eigenvalues that behave as singlets under the action of the Weyl group. Let us see the case for SU$(2)$, which is the simplest one and also the case of our interest for the rest of the thesis.\\
The Cartan sub-algebra of this group has only one generator, which is the Pauli matrix $\sigma_3$ in \eqref{eq:Pauli_matrices}, and hence the classical vacuum (a point of the complex 1-dimensional moduli space) is represented by
\begin{equation}
\phi = \frac{1}{2} \, a \sigma_3
\end{equation}
with $a \in \mathbb{C}$. The Weyl group of SU$(2)$ is the symmetric group $S_2 = \mathbb{Z}_2$, whose action is the one that exchanges the basis (real) vectors of the Cartan sub-algebra. Since we are dealing with a matrix group, its action with an element $U$ on an element $\phi$ of its algebra is $U \phi \, U^\dagger$ and choosing as element of the present Weyl group
$$U = \begin{pmatrix}
0 & -1 \\
1 & 0
\end{pmatrix}$$
we get that
\begin{equation}
U \phi \, U^\dagger = \frac{1}{2} a  
\begin{pmatrix}
0 & 1 \\
-1 & 0
\end{pmatrix} = - \phi
\end{equation}
In other words, the Weyl group sends $a$ into $-a$ and so the two vacua corresponding to these two values of the complex parameter are related. They can be identified if we choose as parametrization of the moduli space the Casimir
\begin{equation} \label{eq:classical_SU2}
u = \text{Tr} \left(\phi^2\right) = \frac{1}{2} a^2
\end{equation}
(the number of Casimir's is equal to the rank of the gauge group and in this case we only have the quadratic one), which is clearly invariant under the Weyl group.\\
Concluding, the \textbf{classical} moduli space for SU$(2)$ SYM is a copy of $\mathbb{C}$ and the identity $u = \frac{1}{2} a^2$ is valid for every value of $u$. In the \textbf{quantum} case, the fields are promoted to operators and for this reason the Weyl invariant quantities (i.e. the Casimir's) are expressed as scalar field VEVs, hence for SU$(2)$ we have
\begin{equation}\label{eq:quantum_SU2}
u = \braket{\text{Tr} \left(\phi^2\right)}
\end{equation}
This time the classical relation \eqref{eq:classical_SU2} holds only in the weak coupling limit (where quantum corrections are less important and hence the VEVs of quantum operators approach the classical results) of the IR effective theory. Such limit corresponds to the region around infinity of the $u$-plane, indeed the theory is asymptotically free and hence at high energies (i.e. when $a/\Lambda \to \infty$, where $a = \braket{\phi}$ denotes the scalar field VEV and it sets the scale of SU$(2)$ spontaneous breaking to U$(1)$, while $\Lambda$ is the dynamically generated energy scale) the theory is weakly coupled and this limit is reached at infinity of the $u$-plane. The crucial consequence of this, that will be analyzed, is that $a^2$ is a good coordinate only to describe the region at infinity of the quantum moduli space. We are going to show how Seiberg and Witten were able to completely uncover the structure of this space through the determination of the IR effective action of SU$(2)$ SYM. Introducing the $\mathcal{N}=2$ superspace formalism, we are going to see that the geometry of the space of vacua and the dynamics of vacuum configurations are completely determined by a holomorphic function of the $\mathcal{N}=2$ vector multiplet, dubbed \textbf{prepotential} and that will play a crucial role in our study.\\

Before passing to this formalism and even if we are not going to examine in many details what happens when we add hypermultiplets (since we are going to simply use known results in literature obtained following the analysis of SW of the following Sections for other applications), it is instructive to see the $\mathcal{N}=2$ matter Lagrangian. The analysis on the supermultiplets leads to the conclusion that the scalar fields $h_1$ and $h_2$ form a SU$(2)_R$ doublet. Furthermore, in renormalizable theories, hypermultiplets cannot interact between themselves since no cubic SU$(2)$ invariant is possible and for this reason also in this case a superpotential is not allowed, in such a way that the interactions turn out to be all gauge ones. Supposing at this point that matter transforms under some representation $R$ of the gauge group, the corresponding $\mathcal{N}=2$ hypermultiplet Lagrangian is
\begin{equation} \label{eq:N2_matter_action}
\mathcal{L}^{\mathcal{N}=2}_{matter} = \int \mathrm{d}^2\theta \mathrm{d}^2\overline{\theta} \, \,\left(\overline{\mathbf{H}}_1 \, e^{2gV_R} \mathbf{H}_1 + \overline{\mathbf{H}}_2 \, e^{-2gV_R} \mathbf{H}_2 \right) + \int \mathrm{d}^2\theta g \mathbf{H}_1 \mathbf{\Phi} \mathbf{H}_2 + h.c. 
\end{equation} 
The coupling of the hypermultiplets with the chiral multiplet $\mathbf{\Phi}$ (which belongs to the $\mathcal{N}=2$ vector multiplet) is somehow the supersymmetric partner of the kinetic terms that couple hypermultiplets to the $\mathcal{N}=1$ vector superfield. For this reason a cubic interaction arises, but it has a gauge nature, in the sense that as $g \to 0$ this vanishes. Notice how in this situation a mass term is possible and it has the form
\begin{equation} \label{eq:mass_term}
m h_1 h_2
\end{equation}
but this one is present only for BPS hypermultiplets.

\subsection{$\mathcal{N}=2$ Superspace and the Prepotential} \label{sec:N2_Superspace}
The generalization of the superspace formalism to $\mathcal{N}=2$ SUSY is done through the introduction of two sets of spinor coordinates, $\left(\theta, \overline{\theta}\right)$ associated to $\left(Q^1_\alpha, \overline{Q}^1_{\dot{\alpha}}\right)$ and $\left(\tilde{\theta}, \overline{\tilde{\theta}}\right)$ associated to the supercharges $\left(Q^2_\alpha, \overline{Q}^2_{\dot{\alpha}}\right)$. In this way a superfield is a function of $\left(x,\theta,\overline{\theta},\tilde{\theta}, \overline{\tilde{\theta}}\right)$. It can be shown that a $\mathcal{N}=2$ vector multiplet is represented, in this new superspace, by a superfield $\mathbf{\Psi}$, which is chiral with respect to both the $\theta$'s and the $\overline{\theta}$'s (meaning that $\overline{\mathcal{D}}_{\dot{\alpha}} \mathbf{\Psi} = 0$ and $\overline{\widetilde{\mathcal{D}}}_{\dot{\alpha}} \mathbf{\Psi} = 0$; notice that the covariant derivatives in the new spinor coordinates have the same form as in the $\mathcal{N}=1$ case) and can be expanded in the $\tilde{\theta}$'s yielding
\begin{equation}
\mathbf{\Psi} = \mathbf{\Phi}(y,\theta) + \sqrt{2} \, \tilde{\theta}^\alpha \, \mathcal{W}_\alpha (y,\theta) + \tilde{\theta}^\alpha \tilde{\theta}_\alpha \Sigma(y,\theta)
\end{equation}
where the first component is a $\mathcal{N}=1$ chiral superfield, the second is the $\mathcal{N}=1$ supersymmetric field strength and the third is
$$\Sigma(y,\theta) = -\frac{1}{2} \int \mathrm{d}^2 \overline{\theta} \, \overline{\mathbf{\Phi}}(y-2 i \theta \sigma \overline{\theta}, \overline{\theta}) e^{2 V(y,\theta, \overline{\theta})}$$
where in this scenario $y^\mu = x^\mu + i \theta \sigma^\mu \, \overline{\theta} + i \tilde{\theta} \sigma^\mu \, \overline{\tilde{\theta}}$. It can be shown that the renormalizable SYM Lagrangian \eqref{eq:SYM_lag_N2} can be written now as
\begin{equation}\label{eq:ren_SYM_Lag}
\mathcal{L}_{SYM} = \frac{1}{16 \pi} \text{Im} \left[ \text{Tr} \left(\int \mathrm{d}^2 \theta \, \mathrm{d}^2 \tilde{\theta} \frac{1}{2} \tau \mathbf{\Psi}^2 \right) \right]
\end{equation}
As done for $\mathcal{N}=1$ cases and since we will be interested to an effective description of the theory, we can generalize this Lagrangian to non-renormalizable formulations. In particular we can write
\begin{equation} \label{eq:general_nonren_Lag_N2}
\mathcal{L} = \frac{1}{16 \pi} \text{Im} \left[ \text{Tr} \left(\int \mathrm{d}^2 \theta \, \mathrm{d}^2 \tilde{\theta} \mathcal{F}(\mathbf{\Psi}) \right) \right] = \frac{1}{32 \pi} \text{Im} \left(\int \mathrm{d}^2 \theta \mathcal{F}_{ab}(\mathbf{\Phi}) \mathcal{W}^{\alpha a} \mathcal{W}^b_\alpha + 2 \int \mathrm{d}^2 \theta \mathrm{d}^2 \overline{\theta} (\overline{\mathbf{\Phi}} e^{2 V})^a \mathcal{F}_a \right)
\end{equation}
where $\mathcal{F}(\mathbf{\Psi})$ is the aforementioned prepotential and it has been expanded around the chiral superfield components in the RHS of this equation, hence
\begin{equation}
\mathcal{F}_a \equiv \frac{\partial \mathcal{F} (\mathbf{\Phi})}{\partial\mathbf{\Phi}^a}, \qquad \mathcal{F}_{a b} \equiv \frac{\partial^2 \mathcal{F}(\mathbf{\Phi})}{\partial \mathbf{\Phi}^a \partial \mathbf{\Phi}^b}
\end{equation}
\eqref{eq:general_nonren_Lag_N2} is a $\mathcal{N}=2$ supersymmetric form of a non-linear sigma model, whose target space is parametrized by the lowest components of the chiral superfields. In particular, comparing the two terms in the RHS with, respectively, the kinetic term for the $\mathcal{N}=1$ vector multiplet \eqref{gauge_L} and the kinetic term for the $\mathcal{N}=1$ chiral multiplet \eqref{eq:general_matter_multiple} (with $W=0$, obviously), we see that the following identifications can be made 
\begin{equation} \label{eq:important_identifications}
\mathcal{F}_{ab} = \tau \quad , \quad \text{Im} \left[(\mathbf{\Phi}^\dagger e^{2V})^a \mathcal{F}_a \right] = \widetilde{K}(\mathbf{\Phi}, \overline{\mathbf{\Phi}})
\end{equation}
that is $\mathcal{F}_{ab} (\mathbf{\Phi})$ can be seen as a field dependent complexified coupling $\tau$, while $\text{Im} \left[(\mathbf{\Phi}^\dagger e^{2V})^a \mathcal{F}_a \right]$ as a particular K\"ahler potential $\widetilde{K}(\mathbf{\Phi}, \overline{\mathbf{\Phi}}$). Remembering that the second derivative of the K\"ahler potential with respect to $\overline{\mathbf{\Phi}}$ and $\mathbf{\Phi}$ evaluated on the scalar components gives the metric on the target space \eqref{eq:kahler1}-\eqref{eq:kahler2}, then, in the present situation, such metric is
\begin{equation}
\widetilde{K}_{a b}(\phi, \overline{\phi}) = \text{Im}(\mathcal{F}_{ab}) = \text{Im} \tau
\end{equation}
and this will constitute an important constraint in the study of the moduli space, as we are going to see. It is interesting to say that the fact that the kinetic terms of $\mathcal{N}=1$ chiral and vector multiplets are specified in terms of the same function $\mathcal{F}$ is a direct consequence of $\mathcal{N}=2$ SUSY. In particular, for SU$(2)$, the constraints coming from $\mathcal{N}=2$ SUSY make it possible to find explicit expression for the prepotential which is exact to all orders in perturbation theory. This is the analysis performed by Seiberg and Witten and that now we are going to review.

\section{Seiberg-Witten theory} \label{sec:chap1_5}
Now we are ready to investigate the arguments proposed in \cite{Seiberg:1994rs,Seiberg:1994aj} for the determination of the IR effective action of $\mathcal{N}=2$ pure SU$(2)$ SYM theory, as already hinted in Section \ref{sec:N2_theories}. For a deeper understanding many reviews are present, as \cite{Alvarez-Gaume:1996ohl,Bilal:1995hc}.\\
We recall that (as already expressed in Section \ref{sec:N2_theories}) we want to write an effective action that describes the dynamics of our theory under the scale of the SU$(2)$ spontaneous symmetry breaking to U$(1)$, given by the scalar field VEV $a = \braket{\phi}$. This action is intrinsically bounded by a cutoff scale, whose expression can in principle be obtained by integrating out all the degrees of freedom that become massive at the cutoff scale. This process involves a renormalization of the bare couplings of the UV action, as we will see. In the present case, the IR action should describe a U$(1)$ effective theory and under the scale of spontaneous symmetry breaking we have just one massless gauge boson. Moreover, when we explore low energy configurations we need that the massless scalar field is constrained on the moduli space, having a supersymmetric non-linear sigma model. All of this leads us to start to consider the Lagrangian \eqref{eq:general_nonren_Lag_N2} with gauge group U$(1)$
\begin{equation}
\mathcal{L}_{\text{U}(1)} = \frac{1}{32 \pi} \text{Im} \left(\int \mathrm{d}^2 \theta \frac{\partial^2 \mathcal{F}}{\partial \mathbf{\Phi}^2}(\mathbf{\Phi}) \mathcal{W}^{\alpha} \mathcal{W}_\alpha + 2 \int \mathrm{d}^2 \theta \mathrm{d}^2 \overline{\theta} \, \overline{\mathbf{\Phi}} \frac{\partial \mathcal{F}}{\partial \mathbf{\Phi}} \right)
\end{equation}
where $e^{2 V}$ is trivially realized on the chiral superfields when they transform in the adjoint representation of an abelian gauge group, indeed $e^{2 V} \mathbf{\Phi} = (1 + 2V) \mathbf{\Phi} = \mathbf{\Phi}$.\\
We have previously discussed how the geometry of the moduli space of this theory (which is a K\"ahler manifold) is encoded in one single function which is the prepotential, in particular the identifications \eqref{eq:important_identifications} now reduce to
\begin{equation}
\widetilde{K} \equiv K = \text{Im} \left(\mathbf{\Phi}^\dagger \frac{\partial \mathcal{F}}{\partial \mathbf{\Phi}} \right) \qquad \tau = \frac{\partial^2 \mathcal{F}}{\partial \mathbf{\Phi}^2}
\end{equation}
We have also discussed how $a$ is a good coordinate only for the region around infinity of the quantum moduli space. Locally, the K\"ahler metric can be expressed as
\begin{equation} \label{eq:metric_a}
\mathrm{d}s^2 = K_{a \overline{a}} \, \mathrm{d} a \mathrm{d} \overline{a} = \text{Im} \left( \frac{\partial^2 \mathcal{F}}{\partial a^2} \right)(a) \, \mathrm{d} a \mathrm{d} \overline{a} = \text{Im} (\tau(a)) \, \mathrm{d} a \mathrm{d} \overline{a}
\end{equation}
The present goal is determining the function $\mathcal{F}(a)$. This is obtained by adding perturbative quantum corrections to its classical bare counterpart (which is $\frac{1}{2} \tau_0 \mathbf{\Phi}^2$, as it can be read in \eqref{eq:ren_SYM_Lag}). In particular, setting
\begin{equation} \label{eq:tau_F_a}
\tau(a) = \frac{4 \pi i}{g(a)^2} + \frac{\theta(a)}{2 \pi}
\end{equation}
the complexified coupling is renormalized at the scale $a$ according to the beta function of the running of the gauge coupling $g$. Notice that this beta function is one-loop exact thanks to $\mathcal{N}=2$ SUSY. Once that the form of $\tau(a)$ is known, we can calculate the perturbative contribution to the prepotential simply by integrating the equation
\begin{equation} \label{eq:final_F_a}
\tau(a) = \frac{\partial^2 \mathcal{F}}{\partial a^2}
\end{equation}
Actually, also non-perturbative corrections should be taken into account and we will discuss also them.

\subsection{R-symmetry discussions} \label{sec:U(1)toZ8toZ2}
Even if we did not focus so much on R-symmetry of the theories we examined, some discussions are now useful in order to understand some properties that we are going to use later. As we have already said for the construction of $\mathcal{N}=2$ SUSY theories, the R-symmetry is $\text{U}(2) \simeq \text{SU}(2)_R \times \text{U}(1)_R$, where the $\text{U}(1)_R$ component is the same as the $\mathcal{N}=1$ SUSY, meaning that it acts on each of the supercharges as a phase shift
$$Q^I_\alpha \to e^{i \alpha} Q^I_\alpha \qquad \overline{Q}^I_{\dot{\alpha}} \to e^{-i \alpha} \overline{Q}^I_{\dot{\alpha}} \quad \text{for} \, \, I=1,2$$
while the SU$(2)_R$ component acts on the supersymmetry index $I$ of the supercharges. We have also said that the two fermions of a $\mathcal{N}=2$ vector multiplet transform in the fundamental representation of SU$(2)_R$, implying that this part of the R-symmetry is kept manifest in not mixing these two fermions. This is due to the U$(1)_J$ component of SU$(2)_R$, under which $(\lambda, \psi) \to (e^{i \alpha} \lambda, e^{-i \alpha} \psi)$. For what concerns the U$(1)_R$ symmetry, at the classical level it acts as
\begin{align*}
& \phi \to e^{2 i \alpha} \phi \\
& (\lambda, \psi) \to e^{i \alpha} (\lambda, \psi) \\
& v_\mu \to v_\mu
\end{align*}
from which we read that the R-charge of the scalar field $\phi$ is equal to 2 and $R[\psi] = R[\lambda] = 1$ and the vector field is neutral. In general, the R-charge of a chiral superfield is equal to the one of its scalar component and so $R[\mathbf{\Phi}]=2$. If we now consider the Dirac fermion $(\lambda, \overline{\psi})$, it can be shown that U$(1)_J$ sends it into $e^{i \alpha} (\lambda, \overline{\psi})$, while the U$(1)_R$ part into $e^{i \alpha \gamma_5} (\lambda, \overline{\psi})$. This shows that U$(1)_R$ is a chiral global symmetry and this kind of symmetry is anomalous at the quantum level. It means that the measure of the functional integral, relevant when we consider the quantum corrections, is not invariant under this transformation. An anomalous global symmetry does not introduce inconsistencies in the quantum theory, but it gives contributions to the running law of the gauge coupling through the one-loop diagrams. It translates in a non trivial Jacobian for the measure in the path integral. We cannot give too many details, for which the interested reader can consult \cite{Peskin:QFT, Fujikawa:1979ay, Kapustin:2014lwa, Yamaguchi:2018xse}.\\
By exponentiating the non trivial Jacobian we can extract the contribution to the Lagrangian of the theory from this anomaly. It turns out that the variation of the effective Lagrangian under U$(1)_R$ is
\begin{equation} \label{eq:variation_Lag}
\delta \mathcal{L}_{\text{U}(1)} = - \frac{\alpha N_c}{8 \pi^2} F_{\mu \nu} \widetilde{F}^{\mu \nu} = - 4 N_c \alpha \left(\frac{1}{32 \pi^2} F_{\mu \nu} \widetilde{F}^{\mu \nu} \right) = - A \alpha k \qquad k \in \mathbb{Z}
\end{equation}
where $N_c$ is the number of colors (2 for SU$(2)$), $A = 4 N_c = 8$ is the anomaly coefficient and $k$ corresponds to the instanton number associated to the background field configuration (details of instantons can be read in Appendix \ref{app:instantons} and we are going to come back on them later). Keeping in mind that \eqref{eq:variation_Lag} is present at the exponent, we see that U$(1)_R$ is (explicitly) broken, at the quantum level, to $\mathbb{Z}_8$, indeed $e^{i \delta \mathcal{L}_{\text{U}(1)}} = e^{- 8 i \alpha k}$ and the expression assumed by $\alpha$ in order to have a symmetry is $\alpha = 2 \pi n/8$, with $n=0,\dots,7$ and it is a parameter of $\mathbb{Z}_8$. Notice that the U$(1)_R$ R-charge of $u \simeq \phi^2$ is 4, hence the independent values of $n$ for this quantity under the same transformation are only $0,1$. In other words, $u$ is charged under $\mathbb{Z}_8$ only with respect to $\mathbb{Z}_8 / \mathbb{Z}_4 = \mathbb{Z}_2$. At the end of this story, the U$(1)_R$ part of the R-symmetry is non trivially realized on the moduli space of the theory we are studying only as the $\mathbb{Z}_2$ subgroup of $\mathbb{Z}_8$. This will have an important consequence on the structure of singularities of this space.\\
It can be shown that U$(1)_R$ symmetry produces a shift in the theta term of the classical effective Lagrangian, then a suitable chiral rotation on fermion fields can set to 0 this contribution, in such a way that the complexified coupling turns out to simply be
\begin{equation} \label{eq:coupling_simplified}
\tau (a) = \frac{4 \pi i}{g(a)^2}
\end{equation}
Once we have the knowledge on $g(a)$, we can simply compute the coupling through this formula and then passing to the perturbative corrections to the prepotential as explained before.

\subsection{Computation of perturbative corrections to the prepotential}
In order to compute this kind of contributions, we have to determine the renormalized gauge coupling $g(a)$ and this can be done by integrating the one-loop beta function, which is \cite{Seiberg:1988ur}
\begin{equation}
\beta(\mu) = \mu \frac{\partial g(\mu)}{\partial \mu} = - \frac{b}{16 \pi^2} g^3(\mu)
\end{equation}
where $\mu$ is the running energy scale and $b= A/2 = 4$. As previously done, we call $\Lambda$ the dynamically generated scale, which, by definition, is the value of the energy where the coupling diverges. For this reason we can integrate the previous equation from $\Lambda$ to the scale $a$, getting
\begin{equation} \label{eq:gauge_coupling_weak_coupling}
\frac{1}{g^2(a)} = \frac{1}{4 \pi^2} \log \left(\frac{a^2}{\Lambda^2}\right)
\end{equation}
Thanks to this we can compute the complexified gauge coupling thanks to \eqref{eq:coupling_simplified}
\begin{equation} \label{eq:coupling_weak_coupling}
\tau(a) = \frac{i}{\pi} \log \left(\frac{a^2}{\Lambda^2}\right)
\end{equation}
Integrating twice this expression according to \eqref{eq:final_F_a} we gain the asymptotic behavior of the prepotential in the weak coupling limit
\begin{equation} \label{eq:prep_weak_coupling}
\mathcal{F}(a) \underset{a \to \infty}{\simeq} \frac{i}{2 \pi} a^2 \log \left(\frac{a^2}{\Lambda^2}\right) \Leftrightarrow \mathcal{F}(\mathbf{\Phi}) \underset{a \to \infty}{\simeq} \frac{i}{2 \pi} \mathbf{\Phi}^2 \log \left(\frac{\mathbf{\Phi}^2}{\Lambda^2}\right)
\end{equation}
Notice that we have neglected linear terms in $\mathbf{\Phi}$ because they would produce the following change in the K\"ahler potential
$$\mathcal{F}' = \mathcal{F} + \alpha \mathbf{\Phi} + \beta \Rightarrow K' = \text{Im} \left(\mathbf{\Phi}^\dagger \frac{\partial \mathcal{F}}{\partial \mathbf{\Phi}}\right) + \frac{i}{2} (\overline{\alpha} \mathbf{\Phi} - \alpha \overline{\mathbf{\Phi}})$$
Setting $\Omega = \frac{i}{2} \overline{\alpha} \mathbf{\Phi}$ we find the so called K\"ahler transformation
\begin{equation}
K' = K + \Omega + \overline{\Omega}
\end{equation}
and it is easy to prove that any combination of $\mathcal{N}=1$ scalar superfields $\Omega + \overline{\Omega}$ does not contribute to the Lagrangian density, since its $\theta^2 \bar{\theta}^2$ component is a total derivative.\\
The same result \eqref{eq:prep_weak_coupling} can be obtained also if we use the variation of the Lagrangian under the U$(1)_R$ \eqref{eq:variation_Lag}.

\subsection{Computation of non-perturbative corrections to the prepotential} \label{sec:non_pert_corrections}
Apart from the 1-loop correction we have just analyzed, there exists a non-perturbative contribution to the prepotential which is due to instantons. Since a central role will be played by them in the next Chapter, Appendix \ref{app:instantons} is devoted to introducing them and determining the main properties we simply use here.\\
In gauge theories instantons are solitonic solutions, i.e. field configurations with finite energy that solve the equations of motion, and they have codimension 1 with respect to the ambient space (other solitons are, for example, magnetic monopoles, which have codimension equal to 3, or vortices, with codimension 2). In particular they are defined as solutions to the Yang-Mills theory formulated in the Euclidean sector and, to be precise, a self dual field strength represents an instanton, while an anti-self dual one is said to be an anti-instanton. For these field configurations the action is finite
\begin{equation} \label{eq:inst_action}
S_{\text{inst}} = \frac{8 \pi^2 k}{g^2}
\end{equation}
where $k$ is an integer number denoting the instanton number, being positive for instantons and negative for anti-instantons. It is the same $k$ that appears in \eqref{eq:variation_Lag} for what shown in Appendix \ref{app:instantons}. For this reason the instantons' finite action contributions to the functional integral has the form
\begin{equation}
\sum_{k=1}^\infty a_k \, e^{- \frac{8 \pi^2 k}{g^2}}
\end{equation}
exponentially decaying for $g \to 0$. The $a_k$ constants can be computed with path integral calculation. Using \eqref{eq:gauge_coupling_weak_coupling}, we see that
$$e^{-\frac{8 \pi^2 k}{g^2}} = \left(\frac{\Lambda}{a}\right)^{4 k}$$
and hence we can write the instantons' contribution to the prepotential as
\begin{equation} \label{eq:inst_prep}
\mathcal{F}_{\text{inst}}(a) = \sum_{k=1}^\infty \mathcal{F}_k (a) \left(\frac{a}{\Lambda}\right)^{-4 k} \Leftrightarrow \mathcal{F}_{\text{inst}}(\mathbf{\Phi}) = \sum_{k=1}^\infty \mathcal{F}_k (\mathbf{\Phi}) \left(\frac{\mathbf{\Phi}}{\Lambda}\right)^{-4 k}
\end{equation}
For what concerns the dependence of the coefficients $\mathcal{F}_k$ from $\mathbf{\Phi}$, a R-symmetry argument comes to our help. Indeed, from the perturbative result \eqref{eq:prep_weak_coupling}, we see that $\Lambda$ can be treated as the scalar component of a $\mathcal{N}=2$ vector multiplet. Since the R-charge of the prepotential has to be equal to 4 (in order for R-symmetry to be exact), then we have to impose that the R-charge of $\Lambda$ is equal to 2. This implies that the coefficients $\mathcal{F}_k(\mathbf{\Phi})$ can only be proportional to $\mathbf{\Phi}^2$. For this reason we can set $\mathcal{F}_k (\mathbf{\Phi}) = \mathbf{\Phi}^2 \mathcal{F}_k$ (with $\mathcal{F}_k$ numbers) and the full prepotential is hence equal to
\begin{equation} \label{eq:final_prep}
\mathcal{F} (\mathbf{\Phi}) = \frac{i}{2 \pi} \mathbf{\Phi}^2 \log \left(\frac{\mathbf{\Phi}^2}{\Lambda^2}\right) + \mathbf{\Phi}^2 \sum_{k=1}^\infty \mathcal{F}_k \left(\frac{\mathbf{\Phi}}{\Lambda}\right)^{-4 k}
\end{equation}

\subsection{Positivity constraint} \label{sec:positivity_imt}
In \eqref{eq:metric_a} we wrote a local expression for the K\"ahler metric of the moduli space. The reason why that expression in terms of the coordinate $a$ cannot be global is the following. The holomorphicity of $\tau$ implies that $\text{Im}\tau$ is an harmonic function, so, if it were globally defined, it would have no minimum. The implication of this is the existence of regions of the moduli space where $\text{Im} \tau(a)$ would be negative, which is forbidden because a negative kinetic term would violate the unitarity of the theory (in addition to the fact that $\text{Im} \tau \simeq 1/g^2$ and $g$ is a real quantity). From this reasoning we conclude that regions where $\text{Im}\tau(a)$ breaks the positivity condition should be described by a different coordinate system. We want to show that in this new coordinate the metric assumes the same form \eqref{eq:metric_a}.\\
Let us introduce the Legendre transforms of $a$, $a_D$, and of the prepotential, $\mathcal{F}_D$, which are, respectively
\begin{equation} \label{eq:def_Leg_transf}
a_D = \frac{\partial \mathcal{F}}{\partial a} (a) \qquad a = -\frac{\partial \mathcal{F}_D}{\partial a_D}
\end{equation}
From \eqref{eq:final_F_a} we gain that 
\begin{equation} \label{eq:tau_aD_a}
\tau = \frac{\partial a_D}{\partial a}
\end{equation}
and so the K\"ahler metric can be written as
\begin{equation} \label{eq:metric_a_aD}
\mathrm{d}s^2 = \text{Im} (\mathrm{d} a_D \, \mathrm{d} \overline{a}) = -\frac{i}{2} (\mathrm{d} a_D \, \mathrm{d} \overline{a} -\mathrm{d} \overline{a}_D \, \mathrm{d} a)
\end{equation} 
This form is manifestly invariant under SL$(2,\mathbb{R})$ transformations plus a translation by a constant vector $\vec{v} = (v_1, v_2)^T$. For physical reasons that are going to be clear in the following, we impose $\vec{v}=\vec{0}$ and then we restrict from SL$(2,\mathbb{R})$ to SL$(2,\mathbb{Z})$, whose generators $S$ and $T$ are analyzed in Appendix \ref{app:EM_duality} for the extension of electric-magnetic duality for gauge theories with magnetic monopoles and dyons. In particular there we analyze the action of these generators on the field strength $F$, the coupling $\tau$ and the charge vector $(n,m)^T$. Here we report their effect on the coordinate vector $(a_D, a)^T$, knowing that
\begin{equation}
S = \begin{pmatrix}
0 & 1 \\
-1 & 0
\end{pmatrix}  \qquad T= \begin{pmatrix}
1 & 1 \\
0 & 1
\end{pmatrix}
\end{equation}
Similarly on the charge vector $(n,m)^T$, we obtain that
\begin{equation} \label{eq:duality_transformations_a_aD}
\begin{pmatrix}
a_D \\
a
\end{pmatrix} \underset{S}{\to}  \begin{pmatrix}
a'_D \\
a'
\end{pmatrix} = \begin{pmatrix}
-a \\
a_D
\end{pmatrix} \qquad \begin{pmatrix}
a_D \\
a
\end{pmatrix} \underset{T}{\to} \begin{pmatrix}
a'_D \\
a'
\end{pmatrix} = \begin{pmatrix}
a_D + a \\
a
\end{pmatrix}
\end{equation}
Plugging the transformed variables $(a'_D,a')$ inside the metric \eqref{eq:metric_a_aD}, it is possible to show that this metric is invariant in form under $S^{-1}$ and $T$. Furthermore, if we substitute $a = - \frac{\partial \mathcal{F}_D}{\partial a_D}$ inside the same metric \eqref{eq:metric_a_aD}, we obtain
\begin{equation}
\begin{cases}
\mathrm{d} s^2 = -\frac{i}{2} \left(-\frac{\partial^2 \overline{\mathcal{F}}_D}{\partial \overline{a}_D^2} \mathrm{d} \overline{a}_D \mathrm{d} a_D + \frac{\partial^2 \mathcal{F}_D}{\partial a_D^2} \mathrm{d} \overline{a}_D \mathrm{d} a_D\right) = \text{Im} \tau_D \, \mathrm{d} \overline{a}_D \mathrm{d} a_D \\
\tau_D = \frac{\partial^2 \mathcal{F}_D}{\partial a_D^2} = -\frac{\partial a}{\partial a_D} = -\frac{1}{\tau}
\end{cases}
\end{equation}
From this we conclude that the coordinate change $a \to a_D$ allows to write the metric in a local form similar to \eqref{eq:metric_a_aD}, but with a transformed metric coefficient $\tau_D = -1/\tau$. The relation between the original complexified coupling and the new one suggests that $a_D$ provides a good description of the moduli space in the region corresponding to the strong coupling regime of the effective theory. $a$ well describes the region around infinity of the vacuum space, associated to the weak coupling regime, instead $a_D$ provides the good tool to parametrize the region corresponding to the strong coupling. This is the effect of electric-magnetic duality discussed in Appendix \ref{app:EM_duality}. In particular, in \eqref{eq:S_duality}, we show that the duality transformation $S$ has the effect of turning magnetic charges into electric ones and viceversa. This means that magnetic monopoles, which appear as non local gauge field configurations in the weak coupling limit of the original theory, are described as fundamental degrees of freedom in the dual theory.\\

In order to explain the reason why we restricted SL$(2,\mathbb{R})$ to SL$(2,\mathbb{Z})$, let us consider the SL$(2,\mathbb{R})$ transformation
\begin{equation}
T_c = \begin{pmatrix}
1 & c \\
0 & 1
\end{pmatrix}
\end{equation}
with $c \in \mathbb{R}$. This sends, as it can be seen in \eqref{eq:T_duality}
\begin{equation}
\tau \to \tau + c = \frac{4 \pi i}{g^2} + \frac{\theta}{2 \pi} + c
\end{equation}
and it produces a shift of $2 \pi c$ in the $\theta$ angle. As reviewed in Appendix \ref{app:instantons}, in order to preserve the physics, the $\theta$ angle can only be shifted by integer multiples of $2 \pi$, forcing $c \in \mathbb{Z}$. Alternatively, the effect of $T_c$ on the electric-magnetic charge vector $(n,m)^T$ is
\begin{equation}
\begin{pmatrix}
n \\
m
\end{pmatrix} \to 
\begin{pmatrix}
n' \\
m'
\end{pmatrix} = \begin{pmatrix}
n+cm \\
m
\end{pmatrix}
\end{equation}
and a non integral $c$ would produce a violation of the integrality of the electric charge.\\
For what concerns the translation by the constant vector $\vec{v}$, we have first to highlight the connection of the BPS bound obtained in the supersymmetric case in Section \ref{massive supermultiplet} and in the Appendix \ref{app:EM_duality} \eqref{eq:BPS_final_result} for gauge theories with monopoles and dyons. Considering in this scenario the energy functional $\mathcal{E}$ as the mass of the supermultiplet, we can make the following identification for the central charge (that will play an important role later)
\begin{equation} \label{eq:central_charge_coupling}
Z = a g (n + \tau m)
\end{equation}
As we said in Section \ref{massive supermultiplet}, the mass and central charge are renormalized in such a way to preserve the BPS relation also when we add quantum corrections. The point is that if we apply the translation using the vector $\vec{v} = (v_1, v_2)^T$ to the coordinate vector $(a_D,a)$, the central charge \eqref{eq:central_charge_coupling} is modified in a different way with respect to the mass (in absence of hypermultiplets, which is what we are considering) and so the BPS mass formula is not valid anymore. This forces us to set the vector $\vec{v}$ to zero.

\subsection{BPS states and the mass formula} \label{sec:BPS_mass_formula}
At the end of Appendix \ref{app:EM_duality}, we have shown that monopoles and dyons in the particular model analyzed there are BPS states; then, recalling the analogous result obtained in the supersymmetric case, we have identified the central charge in the latter scenario with the final outcome of Appendix \ref{app:EM_duality}, leading to the identification \eqref{eq:central_charge_coupling}. Obviously, all of this is valid if the central charge arises as an effect of the soliton solutions of the present theory, i.e. $\mathcal{N}=2$, SU(2) SYM. It is possible to prove that the identification \eqref{eq:central_charge_coupling} is well approximated in the weak coupling limit of the effective theory. This can be done by a direct computation of the anti-commutator $\{Q_{1 \alpha}, Q_{2 \beta} \}$. We directly give the final result and details of the explicit computation can be consulted in \cite{Alvarez-Gaume:1996ohl}.\\
It turns out that
\begin{equation}
\{Q_{1 \alpha}, Q_{2 \beta} \} = -2 i \epsilon_{\alpha \beta} \, a \left(n + \frac{4 \pi i m}{g^2}\right)
\end{equation}
and recalling also the result from the supersymmetry algebra \eqref{QQ}, we get
\begin{equation}
Z = - i a \left(n + \frac{4 \pi i m}{g^2}\right)
\end{equation}
Neglecting the factor $-i$, which is a phase of $Z$ and it is convention dependent, and shifting the electric charge as expected from the Witten effect for the presence of the $\theta$ term (look at Appendix \ref{app:EM_duality} for details), we obtain the identity \eqref{eq:central_charge_coupling}. Notice that the gauge coupling $g$ does not appear with the convention adopted in the Lagrangian of SUSY scenario, since it is absorbed in the definition of the vector multiplet.\\

Let us see how this expression is modified when we add matter. Suppose that we want to couple an electric charge $n$ to the SYM effective action. This should be represented by a hypermultiplet, whose structure in terms of $\mathcal{N}=1$ superfields has already been analyzed in Section \ref{sec:N2_theories}. We consider the case in which $\mathbf{H}_1$ (and $\overline{\mathbf{H}}_2$) transforms in the fundamental of SU$(2)$, in such a way that $\overline{\mathbf{H}}_1$ (and $\mathbf{H}_2$) transforms in the antifundamental. The coupling between these superfields and the chiral one of the $\mathcal{N}=2$ vector multiplet is implemented by the superpotential written as second integrand in \eqref{eq:N2_matter_action}, in particular as
\begin{equation} \label{eq:coupling_nSYM}
n \mathbf{\Phi} \mathbf{H}_2 \mathbf{H}_1
\end{equation}
All of this in the abelian case (since the present effective theory has as gauge group U$(1)$). In the non-abelian case this term becomes
$$\mathbf{H}_{2 i} \mathbf{\Phi}^a (T^a)^i_j \mathbf{H}_1^j$$
with $a$ the index of the adjoint of the gauge group, lower $i$ runs in the antifundamental and upper $j$ in the fundamental. Expanding around a non trivial vacuum (i.e. $\braket{\phi} = a \ne 0$, with $\phi$ the scalar component of $\mathbf{\Phi}$), we can see that \eqref{eq:coupling_nSYM} gives a mass term for the hypermultiplet of the form
\begin{equation}
n a \mathbf{H}_2 \mathbf{H}_1
\end{equation}
that is $M = na$. Recalling that the massive ultra-short multiplets saturate the BPS bound \eqref{central}, we deduce that $Z=na$. Since the physics does not change under the application of the $S$ duality, which sends the electric charge into a magnetic monopole, $n \to m$ (see \eqref{eq:S_duality}) and $a \to a_D$ (see \eqref{eq:duality_transformations_a_aD}), then we have $Z = m a_D$. Since the central charge is additive, for a dyon we have
\begin{equation} \label{eq:central_charge_formula}
Z = na + m a_D
\end{equation}
This central charge formula plays a crucial role in the characterization of the singularities of the moduli space, as we are going to see. Let us make two interesting observations.\\
The first is that \eqref{eq:central_charge_formula} can be expressed as the scalar product $(m,n) (a_D, a)^T$, which is explicitly SL$(2,\mathbb{Z})$ invariant. The second is that \eqref{eq:central_charge_formula} gives the correct result in the weak coupling limit. Indeed, when $a \to \infty$, we got in \eqref{eq:prep_weak_coupling}
$$\mathcal{F}(a) \underset{a \to \infty}{\simeq} \frac{i}{2 \pi} a^2 \log \left(\frac{a^2}{\Lambda^2}\right)$$
and hence
\begin{align*}
& a_D = \frac{\partial \mathcal{F}}{\partial a}(a) \underset{a \to \infty}{\simeq} \frac{i}{\pi} a \left[ \log \left(\frac{a^2}{\Lambda^2}\right) +1 \right] \\
& \tau(a) = \frac{\partial a_D}{\partial a} (a) \underset{a \to \infty}{\simeq} \frac{i}{\pi} \left[ \log \left(\frac{a^2}{\Lambda^2}\right) +3 \right] \simeq \frac{i}{\pi} \log \left(\frac{a^2}{\Lambda^2}\right) \simeq \frac{a_D}{a}
\end{align*}
In this way \eqref{eq:central_charge_formula} gives
$$Z = a \left(n + \frac{a_D}{a} m\right) \underset{a \to \infty}{\simeq} a (n + \tau(a) m)$$
This expression is identical to \eqref{eq:central_charge_coupling} (with the usual caveat on the coupling), except for the fact that here $\tau$ is renormalized at the scale $a$.

\subsection{Structure of the moduli space} \label{sec:monodromies}
Thanks to all the discussions conducted up to this point, we are now able to investigate in details the structure of the moduli space of the theory under examination. We are going to show that this space contains a specific number of singularities, whose physical nature will be investigated.\\
We have seen that the variable $u$ (which is $\braket{\text{Tr} (\phi^2)}$), is a local holomorphic coordinate on the moduli space. By expressing the quantities $a, \mathcal{F}, a_D$ and $\tau$ as functions of $u$, in the weak coupling regime, where $u \simeq a^2/2$, we have
\begin{equation} \label{eq:behaviors_infinity}
\begin{cases}
a \simeq \sqrt{2 u} \\
\mathcal{F} \simeq \frac{i}{\pi} u \log \left(\frac{2 u}{\Lambda^2} \right) \\
a_D \simeq \frac{i}{\pi} \sqrt{2 u} \left[ \log \left(\frac{2 u}{\Lambda^2} \right) +1 \right] \\
\tau \simeq \frac{i}{\pi} \log \left(\frac{2 u}{\Lambda^2} \right)
\end{cases}
\end{equation}
These functions are all multivalued, enjoying a non trivial monodromy around the point at infinity of the $u$-plane. Indeed, after a turn along a closed loop around $u \simeq \infty$, i.e. $ u \to e^{2 \pi i} u$, we obtain
\begin{equation} \label{eq:shifts_monodromy}
\begin{cases}
a \to -a\\
a_D \to -a_D + 2 a \\
\tau \to \tau - 2
\end{cases}
\end{equation}
In this way we can identify the following monodromy matrix
\begin{equation} \label{eq:monodromy_inf}
M_\infty = P T^{-2} =\begin{pmatrix}
-1 & 2 \\
0 & -1
\end{pmatrix}
\end{equation}
where $P$ is the matrix $- \mathbb{I}$. In this way, the action of such matrix on the coordinate vector $(a_D,a)^T$ gives the shifts listed in \eqref{eq:shifts_monodromy} \footnote{It is interesting to see that the action of this monodromy on $\tau$ can also be computed in the following way. Remember that $M_\infty \in \text{SL}(2,\mathbb{Z})$ and $\tau$ transforms as \eqref{eq:tau_lin_frac} under this type of transformation, so setting $a=-1, b=2, c=0, d=-1$ we get $\tau \to \tau-2$.}. $P$ corresponds to the non trivial element of the $\mathbb{Z}_2$ subgroup of the unbroken $\mathbb{Z}_8$ symmetry, as already discussed in Section \ref{sec:U(1)toZ8toZ2}, and it reverses the sign of the $a$ coordinate. As shown also in Section \ref{sec:N2_theories}, this symmetry has the same effect of the Weyl transformation on the $a$ variable. It can be shown that, if we work on the $a$-plane (which is the double cover of the $u$-plane through the relation $u = a^2/2$), keeping the gauge redundancy, it is possible to compensate the action of the $P$ symmetry with a Weyl transformation; by the way, the effect of $P$ becomes manifest when we quotient the Weyl redundancy, ending on the $u$-plane. So, the contribution of this matrix to the monodromy relies on the structure of the space, while $T^{-2}$ arises from the logarithm introduced by the analyzed corrections. It induces a shift in the electric charge of the BPS states and this is due to the Witten effect. \\
In order to explicitly see this, we have to understand how the matrix $M_\infty$ acts on the charge vector $(m,n)$. The first observation of the previous Subsection outlined the fact that the central charge can be written as the scalar product $(m,n) (a_D,a)^T$ and, since the coordinate vector transforms according to $M_\infty$ (as said above), it is necessary that $(m,n)$ transforms with the inverse of the monodromy matrix, $M^{-1}_\infty$. In this way, the central charge formula \eqref{eq:central_charge_formula} is invariant, which is what we want since $Z$, which is the mass for BPS states, cannot be affected by monodromies. So, the way in which $(m,n)$ is modified is
\begin{equation}
(m \, \, \, \, n) \to (m \, \, \, \, n) \begin{pmatrix}
-1 & -2 \\
0 & -1
\end{pmatrix} = (m \, \, \, \, -n-2m)
\end{equation}
seeing that the magnetic charge is not modified, but the electric charge gets changed due to the Witten effect. All of this forces the monodromy matrices to have integer values. In the end it turns out that the \textit{monodromy group} of our moduli space $\mathcal{M}$ (which is the fundamental group of this space, $\pi_1 (\mathcal{M})$) can be at most SL$(2,\mathbb{Z})$.\\

We have shown that $u = \infty$ corresponds to a branch point and now we claim that, under this situation, there should be at least another singularity elsewhere on the $u$-plane. This is due to the presence of the $\mathbb{Z}_2$ symmetry discussed so far (the quotient group $\mathbb{Z}_8 / \mathbb{Z}_4$ seen in Section \ref{sec:U(1)toZ8toZ2}), which acts non trivially on the $u$-plane as $u \to -u$, So, for each singularity $u \ne 0$, also $-u$ is another singular point. Now, if we had just one singularity in addition to the infinity, this would be the origin. But, in this way (that is in presence of just two singularities), the monodromy of the moduli space would be trivial. Indeed, the loop that determines the monodromy at the origin could be continuously deformed in such a way to encircle the point at infinity. This implies that the two monodromies coincide. As a consequence, $a^2$ would be a global coordinate, which is forbidden due to the positivity condition on $\text{Im} \tau$ discussed in Section \ref{sec:positivity_imt}. Actually, this constraint forces to exclude the possibility of an abelian monodromy group, since this would imply the existence of a frame of globally defined coordinates. Two singularities automatically lead to an abelian monodromy group. We will see that the picture of two singularities together with the point at infinity is the only option yielding a consistent solution.\\

Let us focus now on the physical interpretation of such singularities. They typically arise when massless degrees of freedom have been integrated out. Indeed, we know that an effective description of the low energy theory is obtained by integrating out all the massive states from the action. This obviously requires a renormalization procedure that requires the reabsorption in the couplings of all the higher energy effects (that involve massive fields). Such a procedure is performed by introducing an upper energy cutoff in the integral for the action. In the present case this bound is, as already said different times, the scale $a = \braket{\phi}$ at which spontaneous symmetry breaking takes place, in such a way gluons acquire mass through the Higgs mechanism. Now, since the massive states are functions of $a$ (as hinted in the previous Subsection), which is itself a local function of $u$, it may happen that there are values of $u$ for which some of the massive states become massless and which cannot be removed in our effective description. For this reason at these points the effective theory breaks down, failing in describing all the massless degrees of freedom. Hence, the presence of a singularity in the metric of a non-linear sigma model reflects the fact that the structure of the moduli space has not been completely captured by the effective description, having removed particles from our description that are instead massless.\\
We have to understand what is the nature of the states that are massless at the singularities in $\mathcal{M}$, which correspond to $u = \braket{\text{Tr}(\phi^2)} \ne 0$. For sure these cannot be gluons, because a non zero expectation value of the observable $\text{Tr}(\phi^2)$ would introduce a scale and thus it would violate the conformal invariance which is instead expected in the IR limit of theories with massless non-abelian vector bosons. Using some general results derived from superconformal invariance, we can even exclude the possibility of having massless gauge bosons at $u = 0$. The only states which can possibly become massless hence have spin $\leq 1/2$, but since the theory does not display fundamental states of this kind, these massless states should be monopoles and dyons, which are extended configurations in the description we are using and they are bosons. In a $\mathcal{N}=2$ SUSY theory massive states with spin $\leq 1/2$ are described by short multiplets, which saturate the BPS bound. Therefore, monopoles and dyons must be in hypermultiplets $(\mathbf{H}_1, \mathbf{H}_2)$. As shown in Appendix \ref{app:EM_duality}, monopoles and dyons constitute soliton solutions of the equations of motion and they cannot be coupled to the fundamental gauge field in a local way, but we also know that the theory admits a dual description where these appear as fundamental states, which naturally couple to a dual gauge field through their electric charge. It is interesting to know that the dual electric charge can only be unitary. In order to show this, we consider the charge vector $(m,n)$ and, in order to go to the dual frame, we have to apply a SL$(2,\mathbb{Z})$ transformation. Once here, we impose that the dual charge vector is $(0,k)$, with $k \in \mathbb{Z}$
$$\begin{pmatrix}
m \\
n
\end{pmatrix} \to 
\begin{pmatrix}
a & b \\
c & d
\end{pmatrix}\begin{pmatrix}
m \\
n
\end{pmatrix} = \begin{pmatrix}
0\\
k
\end{pmatrix} \Rightarrow \begin{cases}
am + bn = 0 \\
cm + dn = k
\end{cases}
$$
Introducing $l \in \mathbb{Z}$, from the first equation we can set $a = ln$ and $b = -lm$. In this way
$$\begin{cases}
a = ln, \, \, b=-lm \\
-\frac{c b}{l} + \frac{da}{l} = k \Rightarrow ad-bc = kl
\end{cases}
$$
Now, since $a,b,c,d \in \mathbb{Z}$ and $ad-bc =1$, we obtain $k l = 1$, meaning that $l = k = \pm 1$. Since the coupling is implemented by the term $\mathbf{\Phi}' \mathbf{H}_1 \mathbf{H}_2$ (where $\mathbf{\Phi}'$ is the chiral multiplet in the new duality frame), the mass of the hypermultiplet in the effective theory is the absolute value of $\braket{\phi'}(u) = a'(u)$, as already hinted in Section \ref{sec:BPS_mass_formula}.\\

Assuming that two singularities are located at $u_0, -u_0$ (chosen within the strong coupling region), we want to determine the relative monodromies in the dual frame. Here we can use perturbation theory, since the theory is weakly coupled in the IR in this dual frame. When a hypermultiplet (where the massless monopole/dyon is put) of unitary charge is coupled to the dual gauge multiplet near the point $u_0$ such that $a'(u_0)=0$, the effective theory reduces to $\mathcal{N}=2$ SQED (a theory with a U$(1)$ vector multiplet coupled to a hypermultiplet) after having integrated out all the other massive degrees of freedom. In this case the beta function is known to be
$$\beta(g') = \frac{g'^3}{8 \pi^2}$$ 
This theory is not asymptotically free in the UV, due to the fact that $\beta(g') >0$, as expected (the UV in this theory corresponds to the IR in the original version, which is strongly coupled). Integrating this equation, we find
$$\frac{1}{g'^2(a')} = -\frac{1}{4 \pi^2} \log \left(\frac{a'}{\Lambda_{\text{UV}}}\right)$$
where we have neglected the term $1/g'^2(\Lambda_{\text{UV}})$. Thanks to the relation $\tau(a') = 4 \pi i/g'^2(a')$, we can determine the complexified coupling in the dual frame, for $u \to u_0$
\begin{equation}
\tau(a'(u)) \underset{u \to u_0}{\simeq} - \frac{i}{\pi} \log \left(\frac{a'}{\Lambda_{\text{UV}}}\right)
\end{equation}
Integrating this equation with respect to $a'$, we can obtain $a'_D(a')$
\begin{equation}
a'_D(a') \underset{u \to u_0}{\simeq} a'_D(0) - \frac{i}{\pi} a' \log \left(\frac{a'}{\Lambda_{\text{UV}}}\right) + \frac{i}{\pi} a'
\end{equation}
Notice that the constant $a'_D(0)$ cannot be equal to 0, otherwise, when $a'=0$, also $a'_D$ would vanish, implying that at $u_0$ we would have both magnetically and electrically charged states simultaneously massless, which are mutually non local. This would not allow a Lagrangian description, which is not what we want to investigate in this Chapter, postponing this analysis to the next one.\\
Now, since $a'$ tends to 0 when $u \to u_0$, it is a good holomorphic coordinate in the neighborhood of $u_0$. Consequently, we can write $a' \underset{u \to u_0}{\simeq} c_0 (u-u_0)$. At this point we can evaluate the monodromy around $u_0$, hence when $u-u_0 \to (u-u_0) e^{2 \pi i}$, we obtain
\begin{equation}
\begin{cases}
a' \to a'\\
a'_D \to a'_D + 2 a' \\
\tau' \to \tau' + 2
\end{cases}
\end{equation}
from which we can obtain the monodromy matrix around $u_0$ in the dual frame
$$M'_{u_0} = \begin{pmatrix}
1 & 2 \\
0 & 1
\end{pmatrix}$$
We desire to see how this monodromy acts on the original variables $(a_D,a)$. Let $(m,n)$ be the electric-magnetic charge vector of the dyon which becomes massless at $u_0$. We choose the dual frame in such a way this dyon has charge vector $(0,1)$. In order to keep $Z = n a + m a_D$ invariant, a column vector $(m,n)^T$ transforms with the inverse transpose of the matrix acting on $(a_D,a)^T$, for what discussed above. Hence we consider
\begin{align}
& \begin{pmatrix}
a'_D \\
a'
\end{pmatrix} = 
\begin{pmatrix}
p & q \\
r & s
\end{pmatrix}\begin{pmatrix}
a_D \\
a
\end{pmatrix} \notag \\
& \begin{pmatrix}
0\\
1
\end{pmatrix}= 
\begin{pmatrix}
s & -r \\
-q & p
\end{pmatrix}\begin{pmatrix}
m \\
n
\end{pmatrix}
\end{align}
From the second system we get $ms - nr = 0 \Rightarrow s=n, r=m$. Therefore
\begin{equation}
\begin{cases}
a'_D = p a_D + q a \\
a' = m a_D + n a
\end{cases}
\qquad
\begin{cases}
a_D = n a'_D - q a' \\
a = -m a'_D + p a'
\end{cases}
\end{equation}
From these two we can read the monodromy on $(a_D,a)$, indeed
\begin{equation}
\begin{cases}
a_D \to n (a'_D + 2 a') - q a' = \overbrace{n a'_D - q a'}^{a_D} + 2 n a' = a_D + 2 n(m a_D + n a) \\
a \to -m (a'_D + 2 a') + p a' = \underbrace{-m a'_D + p a'}_{a} - 2 m a' = a - 2 m (m a_D + na)
\end{cases}
\end{equation}
equivalent to
\begin{equation} \label{eq:monodromy_M1}
\begin{pmatrix}
a_D \\
a
\end{pmatrix} \to 
\begin{pmatrix}
1+ 2n m & 2 n^2 \\
-2 m^2 & 1 - 2n m
\end{pmatrix}\begin{pmatrix}
a_D \\
a
\end{pmatrix} 
\end{equation}
Now, in order to determine $(m,n)$ let us set $u_0=1$. A second dyon becomes massless at $-u_0 = -1$ with charges $(m^*,n^*)$. Since a loop around $u_0$ and $-u_0$ corresponds to a loop around infinity (but with opposite orientations), we have to impose the following consistency condition
\begin{equation} \label{eq:consistency_condition}
M_1 (m,n) M_{-1}(m^*, n^*) = M_\infty
\end{equation}
Plugging the expression of $M_1$ obtained in \eqref{eq:monodromy_M1} and of $M_\infty$ in \eqref{eq:monodromy_inf}, we have
$$\begin{pmatrix}
1+ 2n m & 2 n^2 \\
-2 m^2 & 1 - 2n m
\end{pmatrix} = \begin{pmatrix}
-1 & 2 \\
0 & -1
\end{pmatrix} \begin{pmatrix}
1- 2n^* m^* & - 2 (n^*)^2 \\
2 (m^*)^2 & 1 +  2n^* m^*
\end{pmatrix}$$
which gives rise to the following system
\begin{equation}
\begin{cases}
(m^*)^2 = 1 \\
m^2 = (m^*)^2 = 1 \\
n^2 = (n^* \pm 1)^2
\end{cases}
\end{equation}
Its set of solutions is
\begin{equation} \label{eq:final_solutions}
\begin{cases}
m^* = 1, \quad m = \pm 1, \quad n = \pm (n^*+1) \\
m^* = -1, \quad m = \pm 1, \quad n = \pm (n^*-1)
\end{cases}
\end{equation}
We can see that the monodromy matrices, as $M_1$ in \eqref{eq:monodromy_M1}, are all congruent to the identity matrix modulo 2 \footnote{A matrix $M$ is congruent to the identity matrix modulo 2 if there exists a matrix $P$ such that $$P M P^T = \begin{pmatrix}
\text{odd} & \text{even}\\
\text{even} & \text{odd}
\end{pmatrix}$$}.
In this way they constitute a subgroup of SL$(2,\mathbb{Z})$ and it corresponds to the first homotopy group of $\mathcal{M}$, which is equivalent to the complex plane with three punctures, hence it has three non trivial homotopy classes. Indeed the monodromy at a point $u_0$ corresponds to a cycle that goes around the same point, positively oriented with respect to the region that contains $u_0$ and it is a representative of a certain homotopy class in $\pi_1(\mathcal{M})$. Obviously cycles that encircle different singularities are non-homotopic when the number of singularities is greater than 2, because they cannot be continuously deformed one into the other. In this case the relative monodromies form a non-abelian group, but it is always possible to express the monodromy at infinity as the composition of the monodromies around all the other singularities (as seen in \eqref{eq:consistency_condition}). Notice that the ordering of the monodromies in this equation depends on the choice of the base-point for the curves that encircle the singularities.\\
As an example, let us consider the simplest solution of \eqref{eq:final_solutions}, that is $m^* = m = 1, n=0$ and $n^*=-1$. So we have a massless magnetic monopole $(1,0)$ in $u_0 = 1$ and a massless dyon $(1,-1)$ in $-u_0=-1$. Their monodromies are
\begin{equation} \label{eq:simplest_solution}
M_1 (1,0) = \begin{pmatrix}
1 & 0 \\
-2 & 1
\end{pmatrix}, \quad M_{-1} (1,-1) = \begin{pmatrix}
-1 & 2 \\
-2 & 3
\end{pmatrix}
\end{equation}
From the central charge formula \eqref{eq:central_charge_formula}, in order for the monopole $(1,0)$ and the dyon $(1,-1)$ to become massless we must impose $a_D=0$ at $u_0=1$ and $a_D - a=0$ at $-u_0 = -1$.

\section{Solution of the model: the Seiberg-Witten curve} \label{sec:chap1_6}
At this point we are ready to see how Seiberg and Witten in \cite{Seiberg:1994rs, Seiberg:1994aj} solved the model we have been studying so far. By "solving" we mean determining the effective action of $\mathcal{N}=2, \text{SU}(2)$ SYM and this is equivalent to totally describe the moduli space of such theory. Indeed, the possible paths we have depicted are the following. The first is expressing $u$, the holomorphic local coordinate of $\mathcal{M}$, as function of $a$ (the scalar field VEV), indeed the former is the (unique, being our theory of rank equal to 1) Casimir $\braket{\text{Tr}(\phi^2)}$, but we managed to find only the classical contribution to it, that is
\begin{equation} \label{eq:first_expr_uvsa}
u = a^2 \left(\frac{1}{2} + \dots\right)
\end{equation}
where in $\dots$ there are quantum corrections. The other way is finding $a$ and $a_D$ as functions of $u$, considered as parameter of the theory, and then, in order to come back to the physical scenario, we would invert these relations. In this regards, in Section \ref{sec:monodromies}, we managed to find only $a'$, that is the scalar field VEV in the dual frame, in terms of $u$, but not $a$ and $a_D$. What Seiberg and Witten did is finding the functions $a(u)$ and $a_D(u)$. Then, they managed to do the inversion only in the weak coupling limit, in order to get \eqref{eq:first_expr_uvsa}. The quantum corrections are not captured in their limit, which is indeed denoted in literature as \textit{SW limit}, and we are going to see in the next Section how to go to the quantum version of their treatment.\\
The method they depicted is based on a geometric tool and for this reason it is vital to spend some words on the mathematical side of all the quantities we have introduced up to now. 

\subsection{Mathematical picture and elliptic curves}
\begin{figure}[H] 
\centering
\includegraphics[scale=0.5]{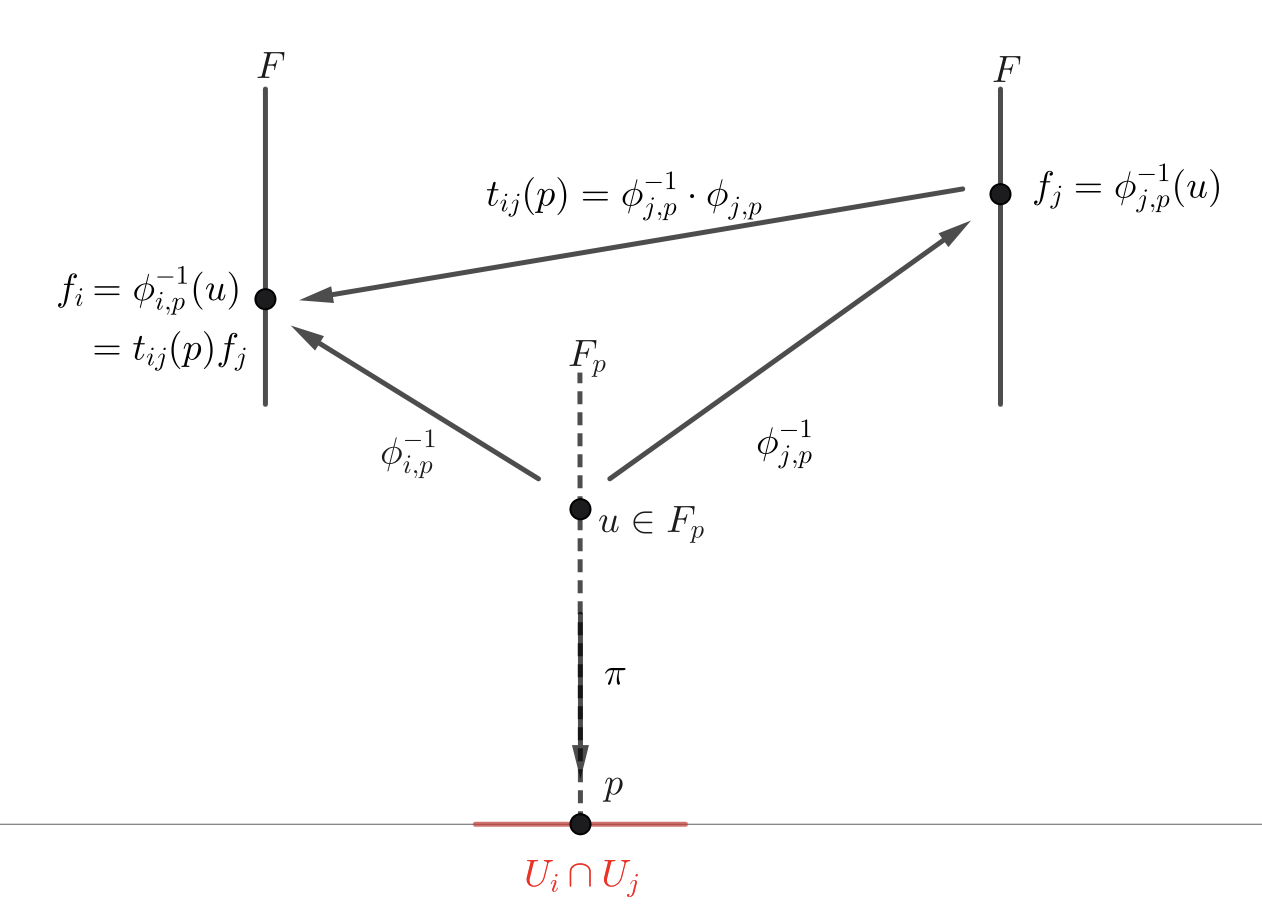}
\caption{Representation of all the elements that characterize a fibre bundle. The grey line at the bottom is the base space $M$.}
\label{im:fibre_bundle}
\end{figure}

We start with the definition of differentiable fiber bundle. We follow the explanation given by \cite{Nakahara:2003nw}.\\
A differentiable fiber bundle $(E,\pi,M,F,G)$ consists of the following elements
\begin{itemize}
\item a differentiable manifold $E$, called \textbf{total space};
\item a differentiable manifold $M$, called \textbf{base space};
\item a differentiable manifold $F$, called \textbf{fibre};
\item a surjiective map $\pi: E \to M$ called \textbf{projection}. The inverse image, $\pi^{-1}(p) = F_p \cong F$ is called \textbf{fibre at $p$}. So $F_p$ is an entire subset of $E$ that corresponds to the point $p$ of the base space through $\pi^{-1}$ and it is isomorphic to $F$;
\item a Lie group $G$, called \textbf{structure group}, which acts on $F$ on the left ($g \in G : F \to F$ such that $f\mapsto g \cdot f$);
\item a set of open covering $\{U_i\}$ of $M$ with a diffeomorphism $\phi_i: U_i \times F \to \pi^{-1}(U_i)$ such that $\pi \circ \phi(p,f) = p$. The map $\phi_i$ is called \textbf{local trivialization}, since $\phi^{-1}_i$ maps $\pi^{-1}(U_i)$ onto the direct product $U_i \times F$;
\item a set of functions $t_{ij}$ called \textbf{transition functions}. If we fix the point $p$ on $M$, the local trivialization becomes the following map $\phi_{i,p} : F \to F_p$ and it is a diffeomorphism. On $U_i \cap U_j \ne \emptyset$, we require that the map $t_{ij}(p) \equiv \phi^{-1}_{i,p} \circ \phi_{j,p} : F \to F$ is an element of $G$. In this way $\phi_i$ and $\phi_j$ are related by the smooth map $t_{ij}: U_i \cap U_j \to G$ as
$$\phi_j(p,f) = \phi_i(p, t_{ij}(p) f)$$
\end{itemize}
In Figure \ref{im:fibre_bundle} we represent all the elements introduced in this definition. Typically we refer to $E$ as the fibre bundle, in place of $(E,\pi,M,F,G)$.

Strictly speaking, the definition of a fibre bundle should be independent of the special covering $\{U_i\}$ of $M$. This aspect is used to define a coordinate bundle $(E,\pi,M,F,G,\{U_i\},\{\phi_i\})$ and hence a fibre bundle is an equivalence class of coordinate bundles. This is a mathematical detail, since in physics a certain definite covering is always employed and no distinction between coordinate and fibre bundle is made.\\
The transition functions have to satisfy the following consistency conditions
\begin{enumerate}
\item $t_{ii}(p)$ is the identity map on $F$, for $p \in U_i$;
\item $t_{ij}(p) = t_{ji}(p)^{-1}$, for $p \in U_i \cap U_j$;
\item $t_{ij}(p) \cdot t_{jk}(p) = t_{ik}(p)$, for $p \in U_i \cap U_j \cap U_k$ 
\end{enumerate}
If these conditions are not satisfied, local pieces of a fibre bundle cannot be glued together consistently. If all the transition functions are the identity for a given fibre bundle, this one is called a \textbf{trivial bundle}, which is the direct product $M \times F$.\\
Now we have to define another ingredient. A \textbf{section} is a smooth map $s: M \to E$ which satisfies $\pi \circ s = \text{id}_M$. Clearly, $s(p) = s|_p$ is an element of $F_p$. If $U \subset M$, we may talk of a local section, which is defined only on $U$. So the difference between $\phi^{-1}$ and $s$ is that the former associates to a single point $p$ of $M$ the entire fibre on $p$, instead the latter associates to the point $p$ a specific element of $F_p$. For this reason a local section is a particular case of local trivialization.\\
A \textbf{vector bundle} $E \overset{\pi}{\to} M$ is a fibre bundle whose fibre is a vector space. Let $F$ be $\mathbb{R}^k$ (or $\mathbb{C}^k$) and, on a chart $U_i$, the piece $\pi^{-1}(U_i)$ is trivial, as already said, $\pi^{-1}(U_i) \equiv U_i \times \mathbb{R}^k$. We can choose $k$ linearly independent sections $\{e_1(p), \dots, e_k(p) \}$ over $U_i$ and these define a \textbf{frame} over $U_i$.\\

At this point we can translate all the elements we have introduced up to this point for our specific scenario. Our base space is the moduli space $\mathcal{M}$, parametrized by $u$, the total space is denoted as $V$ and so the fibre associated to the point $u$ of $\mathcal{M}$ is the space $V_u$. In this way the coordinates $(a_D,a)$ form a section of $V$. Acting as local trivializations, they are
$$(a_D(u), a(u)): V_u \to \mathbb{C}^2$$
due to the fact that $a_D$ and $a$ are in general complex. In other words, for each value of $u$, $a_D$ and $a$ define a map from the fibre $V_u$ to $\mathbb{C}^2$, introducing in this way a frame on $V_u$. The intention is identifying the vector space $V_u$ with the first cohomology group, $H^1(S_u,\mathbb{C})$, of a torus $S_u$ fibered on the $u$-plane. This is equivalent to say that we are considering an \textit{elliptic fibration} over the moduli space. An \textit{elliptic curve} is a torus with a marked point, but this is not important for what we are going to study, so we can simply say that this curve is a torus. In this way the elliptic fibration on the $u$-plane is a family of elliptic curves, each one associated to every point $u$, and they may become singular for some values of $u$. The power of this description is that the moduli space can be interpreted as the moduli space of a genus one Riemann surface, on which the complex structure modulus varies holomorphically. We will not enter many details for complex bundles, homology and cohomology, for which the interested reader can consult \cite{Nakahara:2003nw, Griffiths:Algebraic_Geometry}.\\

In order to describe what said right now in terms of equations, we choose some coordinates $(x,y) \in \mathbb{C}^2$ (notice that this $\mathbb{C}^2$ is different from the fibre on which $(a_D(u),a(u))$ are defined), which allow us to parametrize each torus as $u$ varies. Since the torus is a bidimensional surface, in $\mathbb{C}^2$ it is described by a single complex equation. So an elliptic curve of the fibration (also called \textit{Seiberg-Witten curve} in our scenario) is described by the following algebraic equation
\begin{equation}
y^2 = f(x,u)
\end{equation}
where $f(x,u)$ is a polynomial in $x$ and the coordinate $u$ is a parameter of this curve, so different $u$'s correspond different tori. Let us focus on the theory under our study. We showed that $\mathcal{M}$ is topologically equivalent to the complex plane with three punctures, located at $u=-1,1,\infty$, and we determined also the monodromy matrices around these points. In correspondence of these points something which is particular has to take place with respect all the other regular points. Hence it is reasonable to parametrize the SW curve through $u$ in the following way
\begin{equation} \label{eq:SW_SYM_SU2}
y^2 = (x-1)(x+1)(x-u)
\end{equation}
Let us see why this equation describes a torus, for a fixed $u$ (at the moment different from $1,-1$ or $\infty$; later we are going to see what happens in these points). This equation is quadratic in $y$ and, if we consider $x$ an element of the simple complex plane, $y$ is a multi-valued function, whose branch points are the zeros of the polynomial $f(x,u) = (x-1)(x+1)(x-u)$ and also $\infty$. This can be seen by considering a closed loop around each of the three points $x=-1,1,u$, for which we obtain $y \to -y$, and the same takes place also along another loop around all these three points (equivalent to a closed path around $\infty$), because we have an odd number of singularities. Notice that, even if $x = \infty$ is not an apparent zero of \eqref{eq:SW_SYM_SU2}, if we use the quartic parametrization for this curve, for which $f(x,u)$ is a quartic polynomial in $x$, then the point at infinity is sent into another finite point, which is a zero of this quartic form, and in this way it is a clear branch point. In order to have $y$ single-valued, then the space in which $x$ is defined has to be the double covering of the complex plane $\mathbb{C}$, with the inclusion of the point at infinity. Consider one cut that connects $x=-1$ with $x=1$ and another one that goes from $x=u$ to $x = \infty$. This is represented in the following Figure

\begin{figure}[H] 
\centering
\includegraphics[scale=0.5]{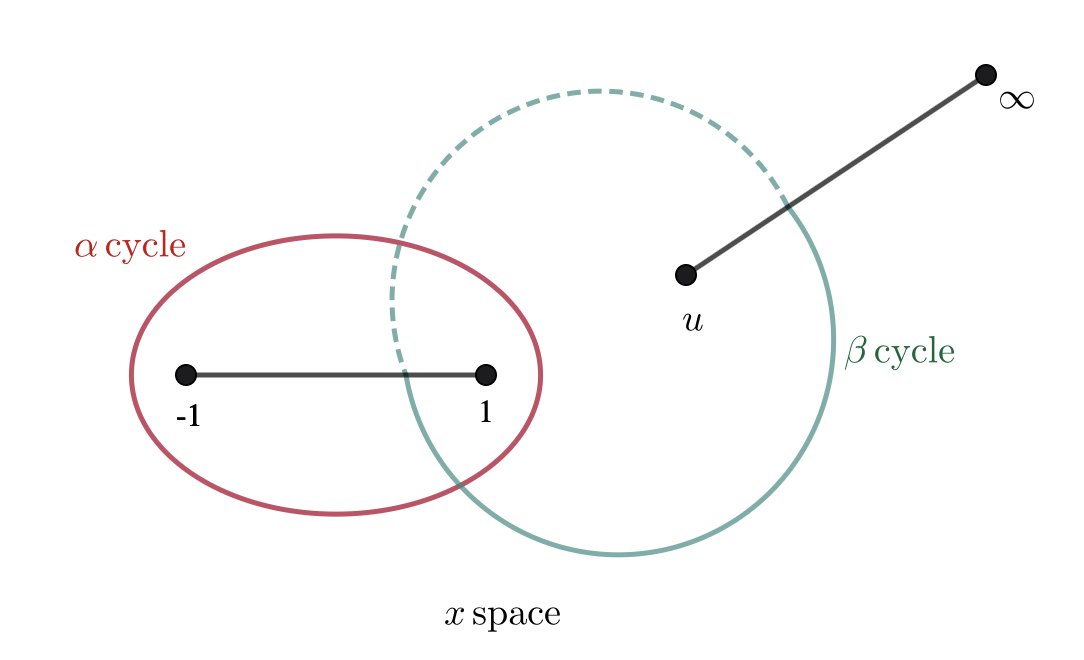}
\caption{$x$ space as double cover of the complex plane $\mathbb{C}$. It can be seen as two copies of $\mathbb{C}$ with two cuts connecting the branch points $x=-1,1,u,\infty$ in pairs and joining them along the cuts. When a line is dashed it means that it lies in the other sheet.}
\label{im:x_space}
\end{figure}

\begin{figure}[H] 
\centering
\includegraphics[scale=0.5]{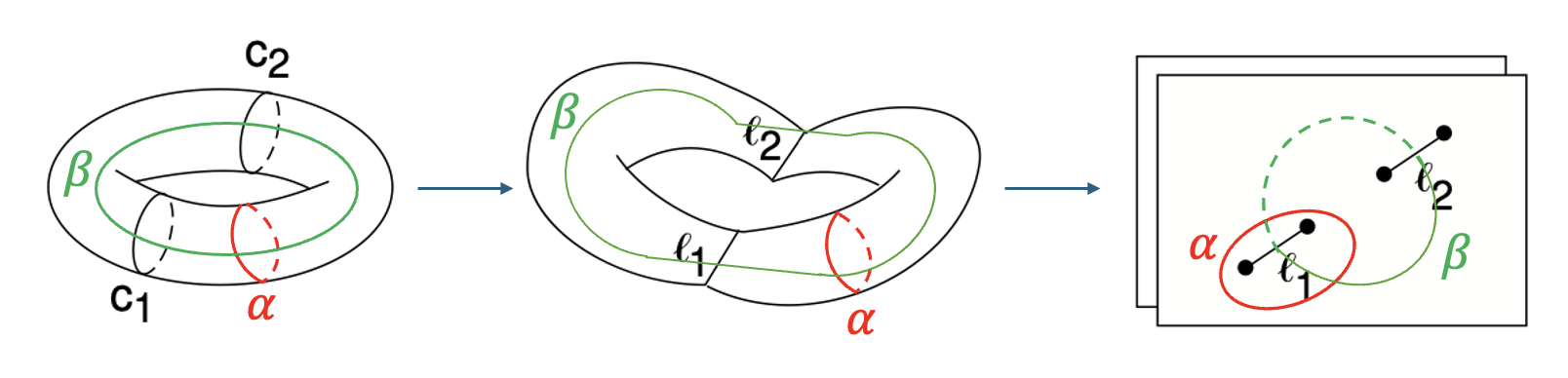}
\caption{Representation that shows why the $x$-space as depicted in Figure \ref{im:x_space} is topologically equivalent to a torus. $\alpha$ and $\beta$ are the two homologically non-trivial 1-cycles of the torus, which are going to play an important role later on. Image taken from \cite{Alvarez-Gaume:1996ohl} and suitably modified.}
\label{im:torus_procedure}
\end{figure}

This curve represented here is exactly the torus associated to the point $u$ of the moduli space. This can be seen in Figure \ref{im:torus_procedure}, where the circle $c_1$, developed along the $\alpha$-cycle, has been translated into $c_2$ in order to describe the torus. At this point, by squashing $c_1$ and $c_2$ into segments $l_1$ and $l_2$ respectively, we divide the torus into two halves, attached along the two segments themselves. By cutting the two halves, we can open them, getting exactly the surface represented in Figure \ref{im:x_space}; $l_1$ and $l_2$ are exactly the two cuts and the point at infinity is mapped to a finite point.

From this explanation we understand that the loop that entirely lies in one of the two sheets, encircling one of the two couples of branch points joined together, corresponds to the $\alpha$-cycle of the torus, highlighted in red. On the other hand, the loop that goes across both cuts (hence it goes around $x=1$ and $x=u$) corresponds to the $\beta$ cycle of the torus, depicted in green.\\
Thanks to what has been explained, it is clear what happens when $u$ becomes equal to $x=-1,1$ or $\infty$, that is when we are in a singularity of the moduli space $\mathcal{M}$: one of the two independent cycles of the torus shrinks to 0 and the curve becomes degenerate. In our case, when $u=\infty$, the segment $l_2$ in Figure \ref{im:torus_procedure} reduces to a point and so the $\alpha$ cycle shrinks to 0; instead, when $u=1$ or $u=-1$ it is the turn of the $\beta$ cycle to shrink. Summarizing, the singularities of the $u$-plane are those points where the corresponding elliptic curve develops a trivial cycle, becoming singular. We will understand at the end of this Section why this is related to the fact that some degrees of freedom become massless and this allows us also to treat singular points on $\mathcal{M}$ whose associated torus has both cycles shrunk to zero. These points are very peculiar and they will be analyzed in the next Chapter.\\

Due to the large importance played by the torus, it is useful to see some of its properties. It is a compact Riemann surface defined as
\begin{equation}
T^2 \cong \frac{\mathbb{C}}{L_\tau} = \{[z] | z \in \mathbb{C}, z \simeq z + n + m \tau \}
\end{equation}
where $L_\tau= \mathbb{Z} + \tau \mathbb{Z}$ is a lattice. The form of such lattice is determined by the complex parameter $\tau$, which satisfies $\text{Im}\tau >0$ and it is called \textit{complex structure modulus}. It is defined as
$$\tau = \frac{\eta_1}{\eta_2}$$
with $\eta_1$ and $\eta_2$ two complex numbers that are independent over the reals. This $\tau$ has not to be confused, at this stage, with the complexified parameter, but we are going to show they are the same quantity. It can be shown that the group of isomorphism of the torus if SL$(2,\mathbb{Z})$. Indeed, given $C = \begin{pmatrix}
a & b \\
c & d
\end{pmatrix}$, acting on the lattice coordinates $(m,n)^T$ as we have already seen many times gives
$$\begin{pmatrix}
m'\\
n'
\end{pmatrix} = \begin{pmatrix}
d & -c \\
-b & a
\end{pmatrix} \begin{pmatrix}
m \\
n
\end{pmatrix}$$
Recalling how $\tau$ transforms under this SL$(2,\mathbb{Z})$ according to \eqref{eq:tau_lin_frac}, we find
\begin{equation}
L_\tau \ni m \tau + n = -bm + an + (dm-cn) \left(\frac{a \tau + b}{c \tau + d}\right) = n' + m' \tau' \in L_{\tau'}
\end{equation}
concluding that $L_\tau = L_{\tau'}$.

\subsection{Homology and cohomology elements}
We know very well that $T^2$ is classified as a Riemann surface of genus one. This topological property in encoded in the homology of the Riemann surface. Indeed it can be shown (look at Theorem 3.4 of \cite{Nakahara:2003nw}) that homology groups are topological invariants, in the sense that two homeomorphic spaces have isomorphic homology groups. We are interested on the first homology group of a Riemann surface $X$ with genus $g$, $H_1(X,\mathbb{Z})$, and it is an abelian group generated by all the classes of closed curves (that are 1-cycles) on $X$ defined modulo boundaries. This means that two 1-cycles $z$ and $z'$ are in the same equivalence class (and so in a given element of $H_1(X,\mathbb{Z})$) if they differ only by a boundary. Now, there is a theorem that states that $H_1(X,\mathbb{Z})$ and the fundamental group $\pi_1(X)$ (which, as already seen, groups the curves that are homotopically equivalent, in the sense that they can be deformed one into another) are related in the following way
$$H_1(X,\mathbb{Z}) \cong \frac{\pi_1(X)}{F}$$
where $F$ is the commutator subgroup of $\pi_1(X)$. Given a group $G$ with elements $\{x_i \}$, the commutator subgroup is a group generated by the elements of the form $x_i x_j x_i^{-1} x_j^{-1}$. For this reason we can say that $H_1(X,\mathbb{Z})$ is isomorphic to the abelianization  of the fundamental group $\pi_1(X)$. This reflects that the group operation of the latter group, which is the usual composition of curves, is promoted to the sum in $H_1(X,\mathbb{Z})$, meaning that given a set of independent generators of $H_1(X,\mathbb{Z})$, $\{[\gamma_k ]\}, k \in (1,\dots, 2g)$, then a general cycle $[\gamma] \in H_1(X,\mathbb{Z})$ can be written as
$$[\gamma] = n_1 [\gamma_1] + n_2 [\gamma_2] + \dots + n_{2 g} [\gamma_{2 g}], \quad n_1,\dots, n_{2 g} \in \mathbb{Z}$$
A topological invariant that allows to find a convenient basis of cycles is the \textit{intersection number} between curves. This quantity is defined for any two cycles $\gamma$ and $\eta$ in $H_1$ (from now on we will drop the square brackets when we consider a representative of a certain class) in the following way
\begin{equation}
\gamma \cdot \eta = \sum_{P \in \gamma \cap \eta} \nu (P)
\end{equation}
Once that the curves have been deformed to intersect in a transverse way at each intersection point $P$, $\nu(P)$ is equal to 1 if the vectors tangent to $\gamma$ and $\eta$ in $P$ form a frame with the same orientation of the Riemann surface $X$, instead it is equal to -1 when the orientation is opposite to that of $X$. From this definition we see that the intersection defines a map $H_1(X,\mathbb{Z}) \times H_1(X,\mathbb{Z}) \to \mathbb{Z}$ and it can be shown it is skew-symmetric, bilinear and non-degenerate and it depends only on the homology class of the curves, from the representative of each of them. This allows to identify a set of cycles $\alpha_1,\dots, \alpha_g, \beta_1, \dots, \beta_g$ such that
\begin{equation} \label{eq:canonical_basis}
\alpha_i \cdot \alpha_j = \beta_i \cdot \beta_j = 0, \quad \alpha_i \cdot \beta_j = \delta_{ij}, \quad i,j=1,\dots,g
\end{equation}
These form a canonical basis, in terms of which a generic cycle $\gamma$, with $\gamma \cdot \alpha_i = s_i$ and $\gamma \cdot \beta_i = k_i$, can be decomposed as
\begin{equation}
\gamma = \sum_{i} k_i \alpha_i - \sum_i s_i \beta_i
\end{equation}
Let us specify now to the case of the torus, that is $X = T^2$. Since $g=1$, there are only two non trivial 1-cycles, that are the ones already depicted in Figures \ref{im:x_space}-\ref{im:torus_procedure} and analyzed in the previous Subsection. In this way the condition \eqref{eq:canonical_basis} for $\alpha$ and $\beta$ can be written as
\begin{equation}
\begin{pmatrix}
\alpha \\
\beta
\end{pmatrix} \cdot \begin{pmatrix}
\alpha & \beta 
\end{pmatrix} = \begin{pmatrix}
0 & 1 \\
-1 & 0
\end{pmatrix} \in \text{SL}(2,\mathbb{Z})
\end{equation}
and this relation is invariant under any change of the homology basis of the form
$$\begin{pmatrix}
\alpha' \\
\beta'
\end{pmatrix} = \begin{pmatrix}
a & b \\
c & d 
\end{pmatrix} \begin{pmatrix}
\alpha \\
\beta
\end{pmatrix}, \quad  \begin{pmatrix}
a & b \\
c & d
\end{pmatrix} \in \text{SL}(2,\mathbb{Z})
$$
as expected.\\

We have said that the couple $(a_D,a)$ can be seen as a local trivialization for each complex vector space $V_u$ (which is a copy of the fibre $\mathbb{C}^2$ at $u$) and we identify it with the first cohomology group of the torus $S_u$ associated to $u$, $H^1(S_u,\mathbb{C})$. This group consists of all closed 1-forms on $S_u$ modulo exact forms, hence two 1-forms are in the same cohomology class if they differ only by an exact form. According to the Hodge decomposition theorem, a generic complex differential form on a $n$-dimensional complex manifold can be labelled with two integers $(p,q)$, where $p$ stands for the holomorphic components $\mathrm{d}z^i$ and $q$ for the anti-holomorphic ones $\mathrm{d}\overline{z}^j$. Thanks to this, $H^1(S_u, \mathbb{C})$ is equivalent to the direct sum
$$H^1(S_u, \mathbb{C}) \cong H^{(1,0)} \oplus H^{(0,1)}$$
where $H^{(1,0)}$ represents the group of holomorphic 1-forms defined on $S_u$ and $H^{(0,1)}$ is related to the anti-holomorphic ones. It is known that for a Riemann surface of genus $g$, the dimension of the first cohomology group is $2 g$. Since
$$\text{dim} (H^{(0,1)}) = \frac{\text{dim} (H^1(S_u, \mathbb{C}))}{2} = 1$$
we conclude that $S_u$ admits only one holomorphic 1-one form (modulo multiplication by a constant, obviously).\\
It must be said that the dimension of the first homology group of the torus is the same as that of the first cohomology one. This is due to the existence of a one-to-one correspondence between the elements of the two groups. This isomorphism is realized through the pairing, defined in the following way
\begin{align} \label{eq:pairing}
\oint:\, & H^1(S_u, \mathbb{C}) \times H_1(S_u, \mathbb{C}) \to \mathbb{C} \notag \\
& (\lambda, c) \mapsto \oint_c \lambda
\end{align}
and this is bilinear and non-degenerate. Using Stokes' theorem and the non-degeneracy, it can be shown that the pairing is invariant as $\lambda$ varies in the same cohomology class and, analogously, as $c$ varies in the same homology class. Furthermore, since
$$(\cdot, \lambda) : H_1(S_u,\mathbb{C}) \to \mathbb{C}$$
defines a functional over $H_1(S_u,\mathbb{C})$, we can identify the cohomology group as the dual of the homology group. Thanks to the symmetric nature of the pairing, we can also say that $H_1(S_u,\mathbb{C})$ is the dual vector space of $H^1(S_u,\mathbb{C})$.\\
At this point, if we choose as canonical basis of homology cycles $\{c_1, c_2 \}$ such that $c_1 \cdot c_2 = 1$, we can define the \textbf{periods} of the holomorphic differential
\begin{equation} \label{eq:holomorphic_form1}
\omega_1 = \frac{\mathrm{d} x}{y}
\end{equation}
that generates $H^{(1,0)}$ the following integrals
\begin{equation}
b_1 = \oint_{c_1} \omega_1 , \qquad b_2 = \oint_{c_2} \omega_1
\end{equation}
The complex structure modulus associated to the torus $S_u$, $\tau_u$, is then given by the ratio of these two periods
\begin{equation}
\tau_u = \frac{b_1}{b_2}
\end{equation}
and it obeys to $\text{Im} \tau_u >0$.\\
In order to generate all $H^1(S_u,\mathbb{C})$, we can follow two possible paths. The first relies on finding a generator for the anti-holomorphic component. The second one consists in representing the elements of $H^1(S_u,\mathbb{C})$ as a meromorphic $(1,0)$-form $\lambda$, with vanishing residue. This last request implies that the integral pairing \eqref{eq:pairing} is invariant under deformations of the integration path even across a pole. In order to build this form, we first need a basis for such forms and we choose $\omega_1$ as defined in \eqref{eq:holomorphic_form1} and
\begin{equation} \label{eq:holomorphic_form2}
\omega_2 = \frac{x \mathrm{d} x}{y}
\end{equation}
which has a double pole at $x =\infty$. This can be seen through a parametrization of the Riemann surface in the neighborhood of $x = \infty$. In particular, we express the $(x,y)$ coordinates around the point at infinity on the Riemann surface in terms of the complex parameter $\xi = 1/\sqrt{x}$. Indeed from
$$x = \frac{1}{\xi^2}, \quad y = \frac{1}{\xi^3} \sqrt{(1+\xi^2)(1-\xi^2)(1-u \xi^2)}, \quad \mathrm{d}x = -\frac{2}{\xi^3} \mathrm{d}\xi$$
we obtain that
$$\omega_2 = \frac{1}{\xi^2} \frac{-2}{\sqrt{(1+\xi^2) (1-\xi^2) (1-u \xi^2)}}$$
and it displays a double pole at $\xi=0 \Leftrightarrow x=\infty$. At this point we can write $\lambda$ in terms of the basis \eqref{eq:holomorphic_form1}+\eqref{eq:holomorphic_form2}
\begin{equation}
\lambda = f_1(u) \omega_1 + f_2(u) \omega_2
\end{equation}
where the couple $(f_1,f_2)$ is a generic section of the bundle $V$. It can be proven that there exists a basis of cycles $\{\gamma_1, \gamma_2 \}$ such that
\begin{equation} \label{eq:sections_periods}
f_1 = \oint_{\gamma_1} \lambda \qquad , \qquad f_2 = \oint_{\gamma_2} \lambda
\end{equation}

\subsection{Seiberg-Witten analysis}
Thanks to \eqref{eq:sections_periods} and the fact that $(a_D, a)$ has to be a local section of $V$, we can identify the periods of $\lambda$ (called in this way the \textbf{Seiberg-Witten differential}) with $a_D(u)$ and $a(u)$. In order to do it, we have to find the suitable constraints for $\lambda$ and for the homology cycles basis in order to reproduce the section $(a_D, a)$ with the monodromies derived from the asymptotic analysis as shown in the previous Section and such as to guarantee the positivity of $\text{Im}\tau(u)$ (where $\tau(u)$ is the complexified coupling).\\
For what just said, we can do the following identification
\begin{equation}
a_D = \oint_{c_1} \lambda \qquad , \qquad a = \oint_{c_2} \lambda
\end{equation}
where $c_1$ and $c_2$ constitute a generic basis of cycles (normalized to have unitary intersection number). Deriving with respect to $u$, we get
\begin{equation}
\frac{\mathrm{d} a_D}{\mathrm{d} u} = \oint_{c_1} \frac{\mathrm{d} \lambda}{\mathrm{d} u} \qquad , \qquad \frac{\mathrm{d} a}{\mathrm{d}u} = \oint_{c_2} \frac{\mathrm{d} \lambda}{\mathrm{d}u} 
\end{equation}
Assuming $\frac{\mathrm{d} \lambda}{\mathrm{d} u} = f(u) \omega_1$, with $f(u)$ a function of the sole $u$ variable, we get that the complexified coupling is
\begin{equation}
\tau(u) = \frac{\mathrm{d} a_D/\mathrm{d} u}{\mathrm{d} a/\mathrm{d} u} = \frac{\mathrm{d} a_D}{\mathrm{d} a} = \frac{f(u) \oint_{c_1} \omega_1}{f(u) \oint_{c_2} \omega_1} = \tau_u
\end{equation}
and since $\text{Im}\tau_u >0$ for what said above, we also get $\text{Im}\tau(u) >0$, as it should. Conversely, if we suppose that $\text{Im}\tau(u)>0 \, \, \forall u$, we conclude that $\tau(u)$ is the complex structure modulus of a family of elliptic curves. This is demonstrated in details by Seiberg and Witten in their works \cite{Seiberg:1994rs, Seiberg:1994aj}. So the complexified coupling for a given $u$ and the structure modulus with the same $u$ are the same quantity.\\
The correct form for $f(u)$ is the constant one, due to the fact that a not constant function would introduce unwanted poles or zeros, and in particular it has been proven that $f(u) = -\frac{\sqrt{2}}{4 \pi}$. Integrating $\frac{\mathrm{d} \lambda}{\mathrm{d} u} = f(u) \omega_1$ with respect to $u$ we obtain the expression of the SW differential $\lambda$ for this theory
\begin{equation}
\lambda = \frac{\sqrt{2}}{2 \pi} (\omega_2 - u \omega_1) = \frac{\sqrt{2}}{2 \pi} \frac{\mathrm{d}x \, y}{x^2 -1} = \frac{\sqrt{2}}{2 \pi} \frac{\sqrt{x-u}}{\sqrt{x^2-1}} \mathrm{d} x
\end{equation}
From this expression it is clear that the residues of $\lambda$ vanish, since $\omega_1$ is holomorphic and $\omega_2$ has only a double pole at infinity.\\
We are left with choosing a basis of homology cycles that reproduces the desired behavior for $a_D$ and $a$ near the singularities, as studied in the previous Section. We consider the $x$-plane with the cuts and the $\alpha$ and $\beta$ cycles as represented in Figure \ref{im:x_space}. With this choice the two periods $a_D$ and $a$ are given by
\begin{align} \label{eq:final_expression_periods_pure_model}
a_D = \oint_\beta \lambda =\frac{\sqrt{2}}{\pi} \int_1^u \frac{\sqrt{x-u}}{\sqrt{x^2-1}} \mathrm{d} x \notag \\
a = \oint_\alpha \lambda =\frac{\sqrt{2}}{\pi} \int_{-1}^1 \frac{\sqrt{x-u}}{\sqrt{x^2-1}} \mathrm{d} x
\end{align}
Inverting these two relations, we obtain $u(a)$ as in \eqref{eq:first_expr_uvsa}. Let us test their asymptotic behavior near $u \simeq \infty$ and $u \simeq 1$, since the trend around $u=-1$ is related to that for $u=1$ through a SL$(2,\mathbb{Z})$ transformation, as already seen. For this goal it is convenient to change variable in $a_D$, in particular we introduce $z$ such that $x = u z$ and hence
$$a_D = \frac{\sqrt{2 u}}{\pi} \int_{1/u}^1 i \, \mathrm{d} z \frac{\sqrt{1-z}}{z \sqrt{1 - \frac{1}{u^2 z^2}}}$$
and this integral enjoys a logarithmic divergence for $z \to 0$. In this way, extracting the divergent terms, we get
\begin{equation}
u \simeq \infty \to \begin{cases}
a_D \simeq \frac{i}{\pi} \sqrt{2 u} \log(2u) \\
a \simeq  \frac{\sqrt{2 u}}{\pi} \int_{-1}^1 \frac{\mathrm{d} x}{\sqrt{1-x^2}} = \sqrt{2 u}
\end{cases}
\end{equation}
in agreement with \eqref{eq:behaviors_infinity}. Passing to $u \simeq 1$, we have
\begin{equation}
u \simeq 1 \to \begin{cases}
a_D \simeq 0\\
a \simeq \frac{\sqrt{2}}{\pi} \int_{-1}^1 \frac{\mathrm{d} x}{\sqrt{x+1}} = \frac{4}{\pi}
\end{cases}
\end{equation}
In order to catch the logarithmically divergent contribution for $a$, we notice it comes from $\mathrm{d}a /\mathrm{d} u$ evaluated near $u \simeq 1$
$$\frac{\mathrm{d} a}{\mathrm{d} u} = -\frac{\sqrt{2}}{2 \pi} \int_{-1}^1 \frac{\mathrm{d} x}{\sqrt{(x^2-1)(x-u)}} \simeq -\frac{1}{2 \pi} \log(u-1)$$
and so the correct result for $a$ is
\begin{equation}
a = \frac{4}{\pi} - \frac{u-1}{2 \pi} \log(u-1) + \dots
\end{equation}
Since $a_D$ starts with 0 in $u=1$, then we can set
\begin{equation}
a_D = c_0(u-1) + \dots
\end{equation}
The constant $c_0$ can be choosen in order to reproduce the correct monodromy. For example, with the choice of the solutions discussed in \eqref{eq:simplest_solution}, we set $c_0 = i/2$, in such a way that
\begin{equation}
a \underset{u-1 \to e^{2i\pi} (u-1)}{\longrightarrow} a-2a_D
\end{equation}
in agreement with the monodromy matrix $M_1(1,0)$ written in \eqref{eq:simplest_solution}.

\subsection{Summary: linking between Physics and Geometry}
We can summarize the main similarities between the physical quantities of the model and the geometric tools introduced by Seiberg and Witten.
\begin{enumerate}
\item Physically, the coordinates $(a_D,a)$ give a local parametrization of the moduli space of inequivalent vacuum configurations. Mathematically, this couple is a section of the two complex $V$ bundle with structure group SL$(2,\mathbb{Z})$. Thanks to the integral pairing \eqref{eq:pairing}, the elements of the fibers of this bundle are in one-to-one correspondence with the elements of homology group $H_1(S_u,\mathbb{C})$. Hence, choosing a canonical basis $(\alpha,\beta)$ of this group, we can write
\begin{equation} \label{eq:final_exp_periods}
a_D = \oint_\beta \lambda \qquad a = \oint_\alpha \lambda
\end{equation}
The linearity of the pairing implies that any change of homology basis implemented by an element of SL$(2,\mathbb{Z})$ gives rise to the same transformation on $a_D$ and $a$. This is consistent with the fact that $(a_D,a)$ is a section of the bundle, as previously said. Furthermore
\begin{equation}
\frac{\mathrm{d}a_D}{\mathrm{d} u} \simeq \oint_\beta \omega_1 \qquad \frac{\mathrm{d}a}{\mathrm{d} u} \simeq \oint_\alpha \omega_1
\end{equation}
with $\omega_1$ the only holomorphic 1-form on the torus $S_u$.
\item SL$(2,\mathbb{Z})$, from a physical point of view, corresponds to electric-magnetic duality (as deeply seen in Appendix \ref{app:EM_duality}), while from a mathematical side it is the group of diffeomorphisms of the holomorphic vector bundle $V$, whose fibers $V_u$ are isomorphic to the first homology groups of a family of complex tori $\{S_u\}$, fibered on the moduli space.
\item The complexified coupling $\tau(u)$, whose imaginary part is forced to be positive to preserve the unitarity of the theory, is identified with the complex structure modulus of the aforementioned family of complex tori $\{S_u \}$.
\item The electric-magnetic charge vector $(m \, \, \, n)$ associated to a dyon state can be seen, from a geometric point of view, as the set of winding numbers of some cycles in $H_1(T^2,\mathbb{Z})$. In order to understand this, consider the cycle
\begin{equation}
\nu = m \beta + n \alpha = \begin{pmatrix}
m & n
\end{pmatrix} \begin{pmatrix}
\beta \\
\alpha 
\end{pmatrix}
\end{equation}
and we know the central charge formula \eqref{eq:central_charge_formula}
\begin{equation}
Z = m a_D + n a = \begin{pmatrix}
m & n
\end{pmatrix} \begin{pmatrix}
a_D \\
a 
\end{pmatrix}
\end{equation}
Thanks to \eqref{eq:final_exp_periods}, we can write $Z$ as the period of the Seiberg-Witten differential over the cycle $\nu$
\begin{equation}
Z = m \oint_\beta \lambda + n \oint_\alpha \lambda = \oint_{m \beta + n \alpha} \lambda \equiv \oint_\nu \lambda
\end{equation}
From this we see that when $\nu \to 0$, the central charge generated by the dyon state $(m \, \, \, n)$ vanishes. This implies that a vanishing cycle (which gives rise to a singularity of the curve) can be associated to a dyon becoming massless (which, as said many times, corresponds to a singularity on the moduli space).\\
We can use this identification of cycles with dyons in order to interpret the Schwinger-Swanziger equation \eqref{eq:schwinger_equation} in terms of intersection number of two cycles. Consider $\nu_1 = n_1 \alpha + m_1 \beta$ and $\nu_2 = n_2 \alpha + m_2 \beta$. Their intersection number is
\begin{equation}
\nu_1 \cdot \nu_2 = n_1 m_2 - m_1 n_2 = k \in \mathbb{Z}
\end{equation}
Putting $n_{1,2} = q_{1,2}/e$ and $m_{1,2} = e g_{1,2}/(2 \pi)$, we find \eqref{eq:schwinger_equation}.
\end{enumerate}

\section{Quantum Seiberg-Witten (qSW) curve} \label{sec:qSW_curve}
It is time to include quantum corrections in the description of the SW curve. The main importances of this aspect are the following. The first is that, as already hinted in the previous Section, in order to compute the function $u(a)$ as in \eqref{eq:first_expr_uvsa} we need to include higher order terms. The second reason is that the presence of singularities in the moduli space, whose origin has been extensively explained in the previous Sections, can be smoothed out by adding these quantum corrections, as usual. Finally, the addition of these quantities will have a crucial role in studying and computing physical observables in the gravity context, as it will be clear starting from Chapter \ref{chap_3}.\\
The way in which quantum corrections are turned on is deforming our spacetime with the so called \textbf{$\Omega$ background}. In particular, we start from our flat 4-dimensional spacetime considered up to now, $\mathbb{R}^{1,3}$, which can be conceived as $\mathbb{R}^4$. Then we deform it in such a way that the topology remains the same, but the metric changes. This deformation is produced by this background, which is parametrized by two parameters, $\epsilon_1$ and $\epsilon_2$, implying that the metric depends on them. A very naive way to realize how to introduce this kind of deformation consists in noticing that the isometry group of the original spacetime is SO$(4)$ which, in terms of algebra, gives rise to
$$so(4) \cong su(2) \oplus su(2) \cong so(3) \oplus so(3)$$
Taking the spacetime as $\mathbb{R}^2 \times \mathbb{R}^2$, we act on each of these two planes with the rotations described by each $so(3)$ algebra and parametrized by $\epsilon_1$ and $\epsilon_2$.\\
Other reasons to introduce the $\Omega$ background in quantum field theories are the following:
\begin{itemize}
\item the computation of partition function from a quantum point of view requires the path integral, which is divergent on $\mathbb{R}^4$. One way to regularize it is introducing the parameters $\epsilon_1$ and $\epsilon_2$;
\item thanks to this background we can apply the \textbf{localization} technique, that will be quickly analyzed in the following Chapter. In few words, this method allows to transform the path integral into an integral over a finite number of degrees. SUSY guarantees that this huge change on the type of integral does not produce any loss of information for the final computation
\end{itemize}
Depending from the values assumed by $\epsilon_1$ and $\epsilon_2$, the $\Omega$ background's slice takes different geometries. For example:
\begin{itemize}
\item when $\epsilon_1 = \epsilon_2=\epsilon$, the slice is a 4-sphere, whose radius is $R = \epsilon^{-1}$;
\item when $\epsilon_2=0$ and $\epsilon_1 \ne 0$ (or viceversa) we have the \textbf{Nekrasov-Shatashvili} (NS) background;
\item when $\epsilon_1 = -\epsilon_2$, we have a self-dual slice
\end{itemize}
We are mainly interested on the NS $\Omega$ background, as we are going to see in this Section, since it is useful for applications on gravity that will be dealt from Chapter \ref{chap_3}. Only in the next Chapter we consider the first case, that will be useful for the computation of some correlators.

\subsection{Hanany-Witten Setup} \label{sec:hanany_witten}
Before proceeding to this quantic scenario, it is interesting and useful to see a way of deriving the SW curve (even at classical level) in terms of objects naturally defined in String Theory, that is branes. Such realization is known as \textbf{Hanany-Witten setup} \cite{Hanany:1996ie}. A complete treatment is not possible and the reason why we are focusing on this description is that it facilitates the understanding of the correspondence between the several types of differential equations that come out when quantizing the SW curve and the number of hypermultiplets that are present in our theory, as it will be clear later. For this reason, even if we have considered up to now only $\mathcal{N}=2, \text{SU}(2)$ SYM theory, it is time now to introduce also matter, in particular fundamental hypermultiplets, in number equal to $N_f$. We start with $\mathcal{N}=2, \text{SU}(2)$ SYM with $N_f=4$ (so it is generically called also SQCD), in such a way that all the cases with $N_f < 4$ can be deduced immediately from the present scenario. Excellent references that clearly expose this setup are \cite{Bianchi:2021mft, Bianchi:2024mlq}.\\

The objects that are involved in this construction are D4-branes and NS5-branes. In general,  D$p$-branes are $p+1$ dimensional stable objects that extend both along the time direction of spacetime and along $p$ spatial dimensions. They carry mass and electric charge and, consequently, are subjected to the action of forces. So we consider the case with $p=4$ and NS5-branes are another type of branes, in particular they are 6-dimensional stable objects that carry a magnetic charge, instead of an electric one.\\
So, the configuration that has to be considered is depicted in the following Figure

\begin{figure}[H] 
\centering
\includegraphics[scale=0.5]{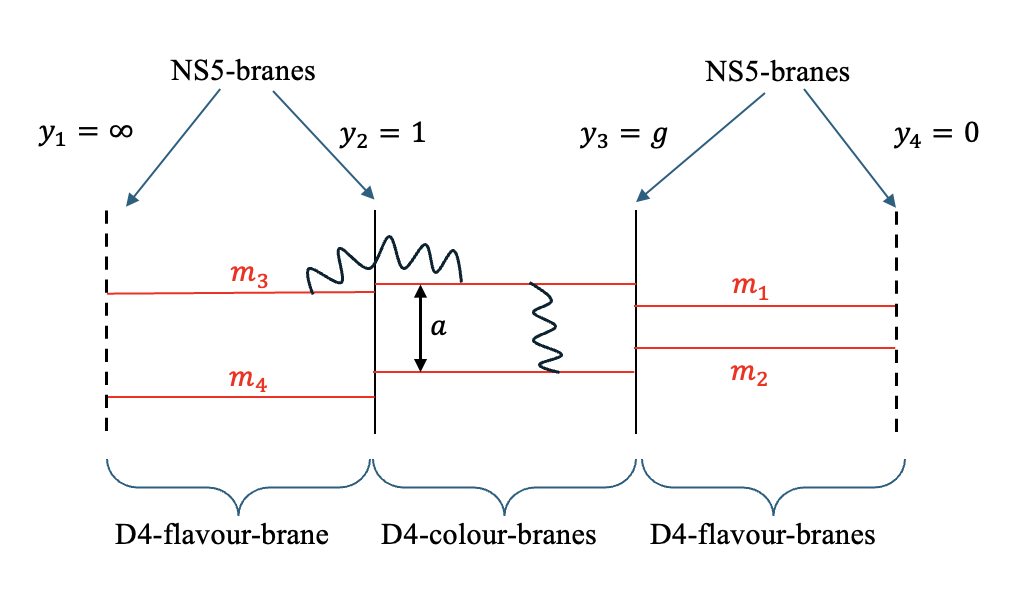}
\caption{Brane realization for $\mathcal{N}=2$ SYM with SU$(2)$ gauge group and $N_f=4$. The QFT lives on a stack of two D4-branes (horizontal red lines) suspended between two non dynamical NS5-branes (vertical black lines). The external D4-branes that extend to infinity (hence here there are some fictitious NS5-branes) provide the matter hypermultiplets of the theory, with masses $m_1,m_2,m_3,m_4$.}
\label{im:hanany_witten}
\end{figure}

D4-branes (horizontal red lines) are suspended between stacks of NS5-branes, distributed along a line, and then we consider open strings that connect all these D4-branes. We denote by $y_i$, with $i=1,2,\dots,N$, the positions of the NS5-branes. Open strings that connect D4-branes between the two stacks of NS5-branes put at a finite distance produce gauge degrees of freedom, meaning vector multiplets, while those connecting D4-branes belonging to two adjacent intervals gives rise to hypermultiplets, with masses $m_1,m_2,m_3,m_4$ (for this reason we have considered $N_f=4$). As notoriously known in String Theory, the tension of strings connecting branes is related to the mass of the states that are produced. For this reason the positions of D4-branes along the vertical direction parametrize the masses (for the hypermultiplets) and the scalar VEVs (for the vector multiplets, $a$ indicated in Figure \ref{im:hanany_witten} and studied so far in the previous Sections). For what concerns the gauge couplings, they are related to the distance between the NS5-branes. In particular, they are specified by the ratio $g_i = y_{i+1}/y_i$. Thanks to this, D4-branes starting at $y_1 = \infty$ or at $y_4 = 0$ and ending at $y_2=1$ or $y_3 = g$ have zero gauge couplings, $g_1=g_4=0$. For this reason they can be viewed as flavour branes, different from the D4-branes suspendend between $y_2$ and $y_3$, which instead have a non trivial gauge coupling and hence called colour branes. This also explains why strings connecting two colour branes give rise to gauge degrees of freedom and those going from a flavour brane to a colour one produce hypermultiplets \footnote{From this description it is quite simple to understand what we can do if we want to reproduce a gauge theory with gauge group SU$(2)^2$. The situation is the same as in Figure \ref{im:hanany_witten}, but we have to add another stack of NS5-branes in between the two ones located in $y_2 = 1$ and $y_3=g$. Denoting by $z$ the position of this new stack, then the two SU$(2)$ groups arise from the colour D4-branes extending along the $[1,z]$ and $[z,g]$ intervals. This could be kept in mind in Chapter \ref{chap_5}.}.\\
This system can be lifted to M-theory \cite{Witten:1997sc} and realized in terms of a single M5-brane wrapping a two-dimensional Riemann surface, which is exactly our SW curve. Its classical form is given by
\begin{equation} \label{eq:general_classical_SW_curve}
q \, y^2 P_L(x) + y P_0(x) + P_R(x) = 0
\end{equation}
with
\begin{equation} \label{eq:polynomials}
P_0(x) = x^2 - u + q p_0(x) \quad , \quad P_R(x) = (x-m_1)(x-m_2) \quad , \quad P_L(x) = (x-m_3)(x-m_4)
\end{equation}
where $q=e^{2 \pi i \tau}$, with $\tau$ the complexified gauge coupling, $u = \frac{1}{2} \braket{\text{Tr} \phi^2}$ as already known and $p_0(x)$ is a quadratic polynomial in $x$ that is determined later, when passing to the quantum scenario. As we can see, the zeros of $P_0(x)$ are associated to the positions of the colour D4-branes, while those of $P_{L,R}(x)$ to the positions of the flavour D4-branes. The degrees of $P_0(x)$ and $P_{L,R}(x)$ are given by the rank of the colour and flavour group plus 1, respectively.\\
Let us see how, in this new viewpoint, we can write the periods $a$ and $a_D$, always at the classical level, following \cite{Bianchi:2021mft} \footnote{Notice that the formalism followed here, more closely related to the literature, involves different norrmalizations with respect to the one used in the previous Sections, for these reasons some differences in terms of constants $i$ and $\pi$ will be present.}. Solving \eqref{eq:general_classical_SW_curve} for $y$ we obtain
\begin{equation} \label{eq:ypm}
y_\pm = \frac{1}{2 q P_L} \left(-P_0 \pm \sqrt{P^2_0 - 4 q P_L P_R}\right)
\end{equation}
As discussed in the previous Section, this elliptic curve can be viewed as the double cover of the complex plane with branch points $e_i$, that are the zeros of the discriminant of \eqref{eq:ypm}, so, using the quartic parametrization, we have
\begin{equation}
P^2_0 - 4 q P_L P_R = \prod_{i=1}^4 (x-e_i)
\end{equation}
The SW differential $\lambda$ is hence given by
\begin{equation} \label{eq:classical_sw_differential_1}
\lambda = \frac{1}{2} (\lambda_+ - \lambda_-) \qquad \text{with} \qquad \lambda_\pm = \frac{1}{2 \pi i} x \partial_x \log y_\pm(x) \, \mathrm{d}x
\end{equation}
in such a way that
\begin{equation}
a = \oint_\alpha \lambda \qquad a_D = \oint_\beta \lambda
\end{equation}

\subsection{The quantum curve}
Now it is time to finally see how to quantize our SW curve. As already said, we consider the gauge theory on a curved NS $\Omega$ background \cite{Nekrasov:2002qd,Flume:2002az,Bruzzo:2002xf,Nekrasov:2009rc,Poghossian:2010pn,Fucito:2011pn} and the net effect of this is making the curve "quantum". In particular, we set $\epsilon_1 = \hbar$ (which keeps track of the quantum side of the description) and $\epsilon_2=0$ and the quantum curve is obtained by the classical one \eqref{eq:general_classical_SW_curve} after promoting $x$ and $y$ to non-commuting operators $\hat{x}$ and $\hat{y}$ such that
\begin{equation} \label{eq:commutation_relation}
[\hat{x}, \hat{y}] = \hat{y} \hbar
\end{equation}
In particular, when passing to the quantum curve, we have to distribute the powers of $\hat{y}$ symmetrically \cite{Zenkevich:2011zx}, getting
\begin{equation} \label{eq:first_step_quantum}
\left[q \hat{y}^{\frac{1}{2}} P_L(\hat{x}) \hat{y}^{\frac{1}{2}} + P_0(\hat{x}) + \hat{y}^{-\frac{1}{2}} P_R(\hat{x}) \hat{y}^{-\frac{1}{2}} \right] U = 0
\end{equation}
with $P_0, P_L,P_R$ given in \eqref{eq:polynomials} and
\begin{equation} \label{eq:p0_polynomial}
p_0(x) = x^2 - \left(x+\frac{\hbar}{2}\right) \sum_i m_i + u + \sum_{i<j} m_i m_j + \frac{\hbar^2}{2}
\end{equation}
This function is determined by requiring that the quantum SW differential, which will be defined later, behaves at large $x$ as \cite{Poghossian:2010pn,Fucito:2011pn}
$$\lambda_+ = \sum_{n=0}^\infty \frac{\braket{\text{Tr} \phi^n}}{x^n} = 2 + \frac{2 u}{x^2} + \dots$$
Using the commutation relation \eqref{eq:commutation_relation} to bring all the dependence of $\hat{y}$ on the left and setting
\begin{equation} \label{eq:first_rule}
\hat{x} = \hbar y \partial_y \qquad \hat{y} = y \mathbb{I}
\end{equation}
with $y$ a variable, the equation \eqref{eq:first_step_quantum} can be viewed as an ordinary differential equation in the $y$-variable
\begin{equation} \label{eq:final_sw_curve_N4}
\left[q y^2 P_L \left(\hat{x} + \frac{\hbar}{2}\right) + y P_0 (\hat{x}) + P_R \left(\hat{x} - \frac{\hbar}{2}\right)\right] U(y) = \left[A(y) \hat{x}^2 + B(y) \hat{x} + C(y) \right] U(y) = 0
\end{equation}
with
\begin{align}
&A(y) = (1+y)(1+qy) \qquad, \qquad B = -m_1-m_2-\hbar+ q y \left[y (\hbar - m_3 -m_4) - \sum_i m_i \right] \notag \\
&C(y) = \left(m_1+\frac{\hbar}{2}\right)\left(m_2+\frac{\hbar}{2}\right) - uy + qy \left[ u + \sum_{i<j} m_i m_j - \frac{\hbar}{2} \sum_i m_i + \frac{\hbar^2}{2} + y \left(m_3 - \frac{\hbar}{2}\right) \left(m_4 - \frac{\hbar}{2}\right)\right]
\end{align}
The solution of this differential equation can be related to the partition function of the SU$(2)$ quiver gauge theory.\\
The reason why we have called the variable with the same name as the one appearing in the classical SW curve is the following: the differential equation \eqref{eq:final_sw_curve_N4} has regular singularities for $y=\infty,1,g,0$, which are exactly the positions of the NS5-branes in the Hanany-Witten setup described before. For this reason this differential equation is an \textbf{Heun equation} and this aspect will be important later and in particular when we will move to analyze the applications in gravity. Indeed, all the differential equations that govern the gravitational phenomena that are going to be investigated from Chapter \ref{chap_3} (as the propagation of gravitational waves) are of Heun type or confluences of it, hence they can be matched with the quantum SW curve we have introduced right now (or with confluences of it, depending from the phenomenon under consideration). This gives rise to the so called \textbf{gauge/gravity dictionaries}, that will be extensively used later on, since they allow to map gravitational quantities into all the quantities that are defined in supersymmetric gauge theories that we have been studying so far and hence, through the methods we are depicting in order to compute the latter, we can determine the former.\\
In order to obtain these dictionaries, it is important to bring the differential equation into the canonical or the Schr\"odinger like form
\begin{equation} \label{eq:canonical_form}
\Psi''(y) + Q(y) \Psi(y) = 0
\end{equation}
In general, given the differential equation $\chi''(z) + N(z) \chi'(z) + M(z) \chi(z) = 0$, this can be recast in the previous normal form if we introduce the new function $\Phi(z)$ related to the previous one through the following relation
\begin{equation}
\chi(z) = e^{-\frac{1}{2} \int^z N(z') \mathrm{d}z'} \Phi(z) 
\end{equation}
By doing so for the equation \eqref{eq:final_sw_curve_N4}, the function $U(y)$ is related to the canonical one $\Psi(y)$ through
$$U(y) = \frac{1}{\sqrt{y}} e^{-\frac{1}{2 \hbar} \int^y \frac{B(y')}{y' A(y')} \mathrm{d}y'} \Psi(y)$$
and the $Q$ function, now denoted as $Q_{22}$ (the subscript 22 refers to the number of D4-branes on the left and on the right in the Hanany-Witten setup), is
$$Q_{22}(y) = \frac{4 C A - B^2 + 2 \hbar y(B A'- A B') + \hbar^2 A^2}{4 \hbar^2 y^2 A^2}$$
For what concerns the computation of the quantum periods $(a_D,a)$, this can be done exactly in $\hbar$ and perturbatively in $q$. The most convenient way is writing the quantum SW curve \eqref{eq:first_step_quantum} in the form of a difference equation. This can be done in the $x$ rather than in the $y$-plane, in other words, in place of \eqref{eq:first_rule}, we set \cite{Poghossian:2010pn,Fucito:2011pn}
\begin{equation}
\hat{x} = x\mathbb{I} \qquad \hat{y} = e^{-\hbar x}
\end{equation}
In this way \eqref{eq:first_step_quantum} becomes
\begin{equation}
\left[q P_L \left(x-\frac{\hbar}{2}\right)\hat{y} + P_0(x) + P_R \left(x+\frac{\hbar}{2} \right) \hat{y}^{-1}\right] \widetilde{U}(x) = 0
\end{equation}
and we introduce the functions
\begin{equation}
W(x) = \frac{1}{P_R\left(x+\frac{\hbar}{2}\right)} \frac{\widetilde{U}(x)}{\widetilde{U}(x+\hbar )} \qquad M(x) = P_L\left(x-\frac{\hbar}{2}\right) P_R\left(x-\frac{\hbar}{2}\right)
\end{equation}
in such a way the previous equation reduces to
\begin{equation} \label{eq:difference_equation}
q M(x) W(x) W(x-\hbar) + P_0(x) W(x) + 1=0
\end{equation}
which is a difference equation in $W(x)$. Now we see two equivalent ways of computing the quantum period $a$.\\
The last equation can be recursively solved order by order in the instanton fugacity $q$. In the small $q$ limit we can write $W(x)$ as a continuous fraction
\begin{equation} \label{eq:first_way_Wx}
W(x) = - \frac{1}{P_0(x) + q M(x) W(x-\hbar)} = - \frac{1}{P_0(x) - \frac{q M(x)}{P_0(x-\hbar) - \frac{q M(x)}{P_0(x-2 \hbar) - \dots}}}
\end{equation}
where, if we want the correct expression at order $q^k$, we have to apply the recursive scheme for $W(x)$ $k$ times. For example, up to order 2 in $q$ we obtain
$$W(x) = - \frac{1}{P_0(x)} \left(1 + \frac{q M(x)}{P_0(x) P_0(x-\hbar)} + \mathcal{O}(q^2)\right)$$
It can be easily seen that, at order $q^k$, the function $W(x)$ has poles at points $x_n^+ = \sqrt{u} + n \hbar$ or $x_n^- = -\sqrt{u} - n \hbar$, with $0 \leq n \leq k$. The quantum period $a(u)$ can therefore be written as a sum over residues of $\lambda_+$
\begin{equation} \label{eq:long_procedure}
a(u) = \oint_\alpha \lambda = 2 \pi i \sum_{n=0}^\infty \text{Res}_{\sqrt{u}+n \hbar} \lambda_+(x)
\end{equation}
where now $\lambda_+$ is the component of the $\hbar$-deformed SW differential
\begin{equation}
\lambda_+(x) = - \frac{x}{2 \pi i} \mathrm{d} (\log W(x))
\end{equation}
that coincides with \eqref{eq:classical_sw_differential_1} at $\hbar=0$. Inverting the previous equation we can find $u(a)$ order by order in $q$. The same thing can be done for $a_D$ along the other cycle, or we can also use its definition in terms of the prepotential, as we are going to see. Indeed, this method is not so advantageous if we want to arrive at quite high order in the instanton fugacity \footnote{I thank Giorgio Di Russo for discussion on this aspect}, which is instead one important aspect for the computations that are performed from Chapter \ref{chap_3}. The other procedure consists in solving the difference equation \eqref{eq:difference_equation} also for $M(x-\hbar)$ and then we equate the expressions obtained in these two alternate ways. In particular, from \eqref{eq:first_way_Wx} we obtain
\begin{align} \label{eq:first_way_Wxmp}
& W(x-\hbar) = - \frac{1}{P_0(x-\hbar) + q M(x-\hbar) W(x-2 \hbar)} \notag \\
& W(x+\hbar) = - \frac{1}{P_0(x+\hbar) + q M(x+\hbar) W(x)} \Rightarrow (W(x))^{-1} = - \frac{q M(x+\hbar)}{P_0(x+\hbar) W(x+\hbar) + \frac{1}{W(x+\hbar)}}
\end{align}
Then, solving \eqref{eq:difference_equation} with respect to $M(x-\hbar)$ we get
\begin{equation}
W(x-\hbar) = - \frac{1}{W(x) qM(x)} - \frac{P_0(x)}{q M(x)}
\end{equation}
and comparing with the first equation of \eqref{eq:first_way_Wxmp}, we obtain
$$P_0(x) = \frac{q M(x)}{P_0(x-\hbar) - \frac{q M(x-\hbar)}{P_0(x-2 \hbar) - \dots}} - \frac{1}{W(x)}$$
and using the second result of \eqref{eq:first_way_Wxmp}, we arrive at the following consistency condition \cite{Poghosyan:2020zzg}
\begin{equation} \label{eq:consistency_condition_a}
P_0(a) = \frac{q M(a)}{P_0(a-\hbar) - \frac{q M(a-\hbar)}{P_0(a-2 \hbar) - \dots}} + \frac{q M(a+\hbar)}{P_0(a+\hbar) - \frac{q M(a+2 \hbar)}{P_0(a+2 \hbar) - \dots}}
\end{equation}
The solution of this equation is exactly the quantum period $a(u)$ (which is a function also of the masses $m_i$ and of $q$), as it can be seen by directly comparing with the result coming from \eqref{eq:long_procedure}. The advantage of this last procedure is that it allows us to reach very high order in $q$ in the expression of the function $a(u)$.\\
Since in this new recipe we have not used the integration of the SW differential, the only way to obtain the other quantum period $a_D(u)$ is through its definition, which uses the prepotential. The NS prepotential $\mathcal{F}_{\text{NS}}(a,q,m_i,\hbar)$, which is defined as
$$\mathcal{F}_{\text{NS}}(a,q,m_i,\hbar) = \underset{(\epsilon_1,\epsilon_2) \to (\hbar,0)}{\text{lim}} \epsilon_1 \epsilon_2 \log Z_{\text{Nekrasov}} (a,\epsilon_1, \epsilon_2, q, m_i)$$
where $Z_{\text{Nekrasov}} (a,\epsilon_1, \epsilon_2, q, m_i)$ is the Nekrasov partition function (that we do not want to discuss it here, but we will come back in the following Chapters), is obtained from the quantum version of the Matone relation \cite{Matone:1995rx, Flume:2004rp}
\begin{equation}
u = -q \frac{\partial \mathcal{F}_{\text{NS}} (a,\hbar,q)}{\partial q}
\end{equation}
after integrating upon $q$. The integration $q$-independent constant is obtained from the one-loop prepotential, whose details can be seen in Appendix A of \cite{Bianchi:2021mft}. In this way we can write
\begin{equation}
\mathcal{F}_{\text{NS}}(a,q,m_i,\hbar) = \mathcal{F}_{\text{tree}}(a,q,m_i,\hbar) + \mathcal{F}_{1-\text{L}}(a,m_i,\hbar) + \mathcal{F}_{\text{inst}}(a,q,m_i,\hbar)
\end{equation}
where the tree level and instantonic parts are computed by solving \eqref{eq:consistency_condition_a} for $a(u,m_i,q)$, in particular we find
\begin{align} \label{eq:prep_pieces_N4}
& \mathcal{F}_{\text{tree}} = -a^2 \log q \\
& \mathcal{F}_{\text{inst}} = q \left[-\frac{4 a^2 + 3 \hbar^2}{8} - \frac{2 m_1 m_2 m_3 m_4}{4 a^2 + \hbar^2} + \frac{\hbar}{2} \sum_i m_i - \frac{1}{2} \sum_{i<j} m_i m_j \right] + \mathcal{O}(q^2) \\
& \frac{\partial \mathcal{F}_{1-\text{L}}}{\partial a} = \hbar \log \left[\frac{\Gamma^2\left(1+\frac{2a}{\hbar}\right)}{\Gamma^2\left(1-\frac{2a}{\hbar}\right)} \prod_{i=1}^4 \frac{\Gamma\left(\frac{1}{2}+\frac{m_i-a}{\hbar}\right)}{\Gamma\left(\frac{1}{2}+\frac{m_i+a}{\hbar}\right)}\right]
\end{align}
At this point the $a_D$ period is given by
\begin{equation}
a_D(u,m_i,q,\hbar) = -\frac{1}{2 \pi i} \frac{\partial \mathcal{F}_{\text{NS}}}{\partial a}
\end{equation}
This is the procedure we are going to follow in the gravitational context in order to compute these periods, relevant for the determination of several quantities. For example, the quantization of the $a_D$ period is related to the quasi-normal modes (QNMs) spectrum of a given gravitational object and we are going to see also all the other connections.

\subsection{Theories with $N_f<4$}
We have analyzed, in this new description for the quantum scenario, the theory $\mathcal{N}=2$ SYM with gauge group SU$(2)$ and $N_f=4$. For what we are going to be interested later on and also in order to match with what analyzed in the previous Sections where $N_f=0$, it is important to inspect how the same procedure can be applied to the same class of theories but with $N_f < 4$. These can be obtained by decoupling some of the hypermultiplets by sending their mass to infinity. From the Hanany-Witten setup in Figure \ref{im:hanany_witten}, this means that the external D4-branes from which the open strings generate the integrated out hypermultiplets are located at infinity, along the vertical direction. In sending the mass of one hypermultiplet $m \to \infty$, we have to send also $q \to 0$, in such a way we can keep finite the product $m q$, which becomes the gauge coupling of the new theory and that will be renamed $q$ for simplicity. It must also be noticed that, every time a flavour in $P_R$ decouples, the $y$ variable in \eqref{eq:general_classical_SW_curve} must be rescaled to $\tilde{y} = - y/m$ in order to keep the SW curve in the same form.\\
For all these reasons, we label the general case by $N_f=(N_L,N_R)$ (in all the cases of interest for this thesis, $N_L,N_R \leq 2$, which leads to second order differential equations, and they respectively refer to the number of flavour D4-branes on the left and on the right, according to Figure \ref{im:hanany_witten}), with $N_f = N_L + N_R \leq 3$. The curve is given as in \eqref{eq:general_classical_SW_curve}, but now
\begin{align}
& P_L(x) = \prod_{i=3}^{2+N_L} (x-m_i) \qquad P_R(x) = \prod_{i=1}^{N_R} (x-m_i) \\
& P_0(x) = x^2 - u + q \delta_{N_f,3} \left(x - \sum_i m_i + \frac{\hbar}{2}\right) + q \delta_{N_f,2}
\end{align}
The correct expression for the NS prepotential can be obtained from \eqref{eq:prep_pieces_N4} performing the double limit $q \to 0, m \to \infty$. For example, for $N_f=(1,2)$ (that will be the case of our interest in Chapter \ref{chap_3} and \ref{chap_5}) one finds
\begin{align}
& \mathcal{F}_{\text{tree}} = - a^2 \log \left(-\frac{q}{\hbar}\right) \quad , \quad \mathcal{F}_{\text{inst}} = q \left[\frac{1}{2} \sum_{i=1}^3 m_i + \frac{2 m_1 m_2 m_3}{4a^2-\hbar^2} - \frac{\hbar}{2}\right] + \mathcal{O}(q^2) \notag \\
& \frac{\partial \mathcal{F}_{1-\text{L}}}{\partial a} = \hbar \log \left[\frac{\Gamma^2 \left(1 + \frac{2 a}{\hbar}\right)}{\Gamma^2 \left(1 - \frac{2 a}{\hbar}\right)} \prod_{i=1}^3 \frac{\Gamma \left(\frac{1}{2} + \frac{m_i-a}{\hbar}\right)}{\Gamma \left(\frac{1}{2} + \frac{m_i+a}{\hbar}\right)}\right]
\end{align}
and the extra contribution in the tree level component is produced by the one-loop term in the decoupling limit.\\
The interesting aspect for these theories is that the differential equations associated to the quantum SW curves are confluences of the Heun equation for the $N_f=(2,2)$ case analyzed before. In particular \cite{Bianchi:2022wku}
\begin{itemize}
\item the Heun Equation (HE) with 4 regular singular points corresponds to $N_f = 4= (2,2)$, as already seen;
\item the Confluent Heun Equation (CHE) with 2 regular and 1 irregular singular points corresponds to $N_f = 3 = (2,1)$ (or $(1,2)$);
\item the Reduced Confluent Heun Equation (RCHE) with 2 regular and 1 irregular singular points is associated to $N_f = 2 = (2,0)$ (or $(0,2)$);
\item the Doubly Confluent Heun Equation (DCHE) with 2 irregular singular points is mapped to $N_f = 2 = (1,1)$ qSW curve;
\item the Reduced Doubly Confluent Heun Equation (RDCHE) with 2 irregular singular points corresponds to $N_f = 1 =(1,0)$ (or $(0,1)$);
\item the Doubly Reduced Doubly Confluent Heun Equation (DRDCHE) with 2 irregular singular points corresponds to $N_f=0$, that is pure SYM which is the theory deeply analyzed in the previous Sections
\end{itemize}
The gravitational phenomena that we are going to study from Chapter \ref{chap_3} are mainly governed by differential equations of the CHE type and for this reason this is the only case that will be discussed later on.\\

In order to end this topic, it must be said that there is another previous tool available in the study of $\mathcal{N}=2$ SYM, that is the AGT correspondence, due to Alday-Gaiotto-Tachikawa \cite{Alday:2009aq}. Thanks to this duality, we can relate quantities in the gravitational and in the gauge theory sides to a two-dimensional conformal field theory (CFT), called Liouville theory. We will recall in the next Chapter what conformal symmetry is and, since this correspondence will be the background of all our treatments but it will not be explicitly used so much (differently from the correspondence between gravity and gauge theories with SW curve), we limit ourselves only to cite what the main quantities we have encountered so far are mapped in the CFT side. In the latter, one defines some chiral operators $V_{\alpha_i}$ and the gauge partition function is related to the 4-point function of these operators
$$Z = e^{\frac{\mathcal{F}}{\epsilon_1 \epsilon_2}} = \braket{V_{\alpha_1}(y_1) V_{\alpha_2}(y_2)V_{\alpha_3}(y_3)V_{\alpha_4}(y_4)}$$
Then the characteristic function $Q_{22}$ encountered before is mapped to a specific ratio of correlators of these operators and the wave function $\Psi(y)$ entering in the differential equation corresponds to a 5-point correlation function, where 4 points correspond to the aforementioned chiral operators and one instead is associated to a degenerate field. Details can be seen, for example, in \cite{Bianchi:2021mft}.
\chapter{Instantons and Partition Function of Strongly Coupled Superconformal Field Theories}\label{chap_2}
This Chapter focuses on some theories belonging to the family which we have extensively analyzed in the previous Chapter, i.e. 4-dimensional $\mathcal{N}=2, \text{\text{SU(2)}}$ SQCD with $1 \leq N_f \leq 4$. In particular, in the case of $N_f=4$ and massless hypermultiplets, the theory is conformal. The role of instantons in the determination of a specific class of correlators, the so called \textit{extremal correlators}, is analyzed, explicitly proving in a new way the results obtained in a completely different scenario in \cite{Hellerman:2018xpi, Hellerman:2020sqj, Hellerman:2021yqz}. In particular, these last references furnish an expression for this kind of correlators under the large R-charge limit, in such a way that tools of Effective Field Theory (EFT) can be used. By doing so, the authors claim where the dependece on the coupling $\tau$ should be present in those quantities. In \cite{Grassi:2019txd} it is shown that the expectation of \cite{Hellerman:2018xpi} is valid using the language of random matrix theory, but no instantons are considered. The main result of \cite{Cipriani:2026iws} is taking into account such field configurations and, through an analysis similar to the one of \cite{Grassi:2019txd}, it is shown that instantons impact on the structure of extremal correlators as expected in \cite{Hellerman:2018xpi, Hellerman:2020sqj, Hellerman:2021yqz}.\\
For what concerns the cases with $N_f = 1,2,3$, these theories are not conformal, but in their moduli spaces some very peculiar points exist, which describe isolated, superconformal field theories that are intrinsically strongly coupled and without a Lagrangian. These are called \textbf{Argyres-Douglas (AD) theories} \cite{Argyres:1995jj, Argyres:1995xn}. We are going to see not only how they arise, but also an educated guess for a function \textit{interpolating} between two asymptotic regimes of the partition function of these theories. Furthermore, we are going to present a numerical determination of the large-VEV contributions to the non-universal coefficients introduced in \cite{Hellerman:2018xpi}, following \cite{Cipriani:2026iws}.\\
The present Chapter is hence structured in the following way: Section \ref{sec:chap2_1} recalls the basic concepts of Conformal Symmetry and Conformal Field Theories (CFTs), since the theories we will be interested enjoy this kind of symmetry; Section \ref{sec:chap2_2} is devoted to introduce the notion of isolated theories, since AD models enter this class. In Section \ref{sec:chap2_3} we present the notion of Coulomb branch of the moduli space of a given theory, paving the way to understand how AD theories, with all the properties we listed above, appear; Section \ref{sec:chap2_4} introduces the quantity we compute in this scenario, that is the extremal correlator, that involves Coulomb branch operators, which are thus defined here. Section \ref{sec:EFT} reports the result of \cite{Hellerman:2018xpi} for the extremal correlator using the EFT language for $\mathcal{N}=2, \text{\text{SU(2)}}$ SQCD with $N_f=4$ and it is the reference expression when comparing the final results. In Section \ref{sec:chap2_6} we present the calculations reported in \cite{Cipriani:2026iws}, up to 2 instantons, for the determination of the extremal correlator in the same theory, showing the agreement with the expectations of \cite{Hellerman:2018xpi}. Section \ref{sec:chap2_7} focuses on the regime of large VEVs: in Subsection \ref{subsec:chap2_7_1}, we perform a computation up to 4 instantons to show how the agreement with \cite{Hellerman:2018xpi} is still present, then in Subsection \ref{subsec:chap2_7_2} we match our results for the non-universal coefficients to the ones in the literature \cite{Hellerman:2020sqj, Hellerman:2021yqz}; Subsection \ref{subsec:chap2_7_3} analyses the large-VEV sector \textit{exactly} in the gauge coupling. Finally, in Section \ref{sec:chap2_8} we offer a speculative discussion about large-charge two-point functions in Argyres-Douglas SCFTs and a numerical computation of the non-universal coefficients for rank-one isolated models.

\section{Conformal Symmetry and Conformal Field Theories} \label{sec:chap2_1}
\begin{figure}[H] 
\centering
\includegraphics[scale=0.8]{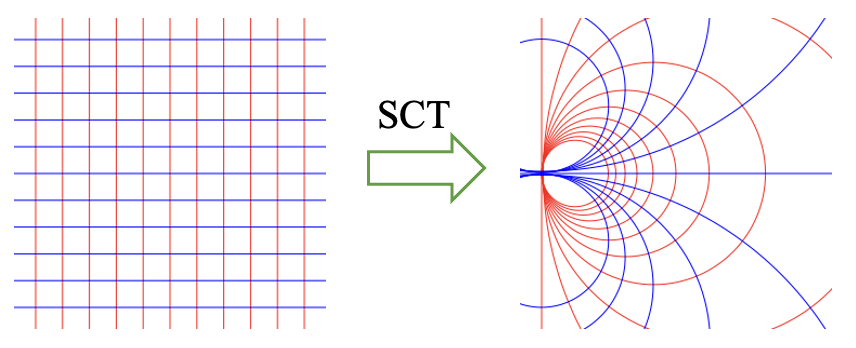}
\caption{Representation of how a special conformal transformation (SCT) acts on the Cartesian plane. Image taken from Wikipedia, \href{https://en.wikipedia.org/wiki/Special_conformal_transformation}{Special conformal transformation}.}
\label{im:sct_transformation}
\end{figure}

Consider a general spacetime of dimension $d$ with metric $g_{\mu\nu}$. A conformal transformation is a transformation that leaves the metric invariant up to an arbitrary positive spacetime dependent parameter, called \textit{scale factor}
\begin{equation} \label{eq:conf_transf}
g_{\mu\nu}(x) \rightarrow \Omega^{-2}(x) g_{\mu\nu}(x)
\end{equation}
It can be shown that this kind of transformation changes the length of an infinitesimal spacetime interval, but the angles are left invariant and also the causal structure of the spacetime is preserved \cite{Birrell:1982ix}; exploiting this fact one builds the so called Penrose diagrams, that are useful to understand the causal properties in an easier way. So we can say that conformal transformations are the most general local transformations that preserve causality and they form a group, the \textbf{conformal group}. We are interested in the case in which the spacetime is flat ($g_{\mu\nu}=\eta_{\mu\nu}$). In this situation we get that the Poincarè algebra is extended to the conformal one.\\
As usual, in order to obtain the generators of the conformal algebra we consider a generic infinitesimal conformal transformation
\begin{equation}
x^\mu \rightarrow x^\mu + \epsilon^\mu (x)
\end{equation}
with the infinitesimal parameter $\epsilon^\mu(x)$ satisfying the Killing equation
\begin{equation}
\label{Killing_equation}
[\eta_{\mu\nu}\partial_\rho \partial^\rho + (d-2) \partial_\mu \partial_\nu] \partial_\sigma \epsilon^\sigma(x) = 0
\end{equation} 
We notice that we have to distinguish between two cases: $d > 2$ and $d = 2$. The difference is huge, indeed in the former case the algebra is finite-dimensional, while in the latter is infinite-dimensional (see \cite{DiFrancesco:1997nk} for details). In this case \eqref{Killing_equation} reduces to the Cauchy-Riemann equations. We are not considering $d=1$ since in this situation we only have time and hence the Poincarè algebra is extended to the conformal one by simply including temporal dilatations (in 1 dimension there are no angles). So we focus only on the first case, specializing then to $d=4$.\\
The equation \eqref{Killing_equation} is solved only if $\epsilon^\mu(x)$ is at most of second order and, following \cite{DiFrancesco:1997nk} we get
\begin{equation}
\epsilon^\mu (x) = a^\mu + \omega^\mu_{\,\,\nu} x^\nu + \lambda x^\mu + b^\mu x^\nu x_\nu - 2(b^\nu x_\nu) x^\mu
\end{equation}
We have to understand the meaning of these parameters. We see that $a^\mu$ and $\omega^\mu_{\,\,\nu}$ are the parameters of translations and Lorentz transformations; the corresponding generators, respectively $P_\mu$ and $M_{\mu\nu}$, are the usual generators of the Poincarè algebra, also encountered in the previous Chapter. They leave the metric unchanged (so $\Omega =1$ in reference to \eqref{eq:conf_transf}). For what concerns the remaining terms, we notice that they cause a rescaling of the metric at each point, in particular the operator corresponding to the parameter $\lambda$, since this one does not depend on the spacetime point, is the \textit{dilatation operator} $D$ (it provokes a rescaling of the metric independently from the spacetime point), while the generators corresponding to the parameters $b^\mu x^\nu x_\nu - 2(b^\nu x_\nu) x^\mu$, that we indicate as $K_\mu$, give rise to a rescaling that depends on the point. In particular these are the generators of the \textit{special conformal transformations} (SCTs) or \textit{conformal boosts}. This kind of transformation can be seen as the composition of an inversion ($x^\mu \to y^\mu = \frac{x^\mu}{x^2}$), a translation ($y^\mu \to z^\mu = y^\mu - b^\mu$) and another inversion ($z^\mu \to x'^\mu = \frac{z^\mu}{z^2}$), so, in conclusion, under a SCT
$$x^\mu \to x'^\mu = \frac{x^\mu - b^\mu x^2}{1 - 2b \cdot x + b^2 x^2}$$  
The action of such transformation on the standard Cartesian plane is represented in Figure \ref{im:sct_transformation}

At this point the entire conformal algebra turns out to be
\begin{subequations}
\begin{align}
& [M_{\mu\nu}, M_{\rho \sigma}] = i (\eta_{\mu \rho} M_{\nu \sigma} + \eta_{\nu \sigma} M_{\mu \rho} - \eta_{\nu \rho} M_{\mu \sigma} - \eta_{\mu \sigma} M_{\nu \rho}) \label{Poincare1}\\
& [M_{\mu\nu}, P_\rho] = i (\eta_{\mu\rho}P_\nu - \eta_{\rho \nu} P_\mu) \\
& [P_\mu , P_\nu] = 0 \label{Poincare2}\\
& [K_\mu, K_\nu] = 0 \\
& [D, M_{\mu\nu}] = 0 \\
& [D, P_\mu] = i P_\mu \label{increasing}\\
& [D, K_\mu] = -i K_\mu \label{decreasing}\\
& [M_{\mu\nu}, K_\rho] = i(\eta_{\mu\rho}K_\nu - \eta_{\nu \rho} K_\mu) \\
& [K_\mu, P_\mu] = -2i(\eta_{\mu\nu}D - M_{\mu\nu})
\end{align}
\end{subequations}
We recognize from \eqref{Poincare1} to \eqref{Poincare2} the relations that characterize the Poincarè algebra, already discussed in the previous Chapter.\\

A conformal invariant theory, usually called \textit{conformal field theory} (CFT), is a theory that does not change under the transformations of the conformal group, in particular it is the same under dilatations. For this reason, the fields in a CFT are classified according to their scaling dimension $\Delta$ (which labels the IRREPS of the conformal algebra, like for example the spin for Lorentz algebra). The scaling dimension codifies the properties of deformation of a field under a dilatation, that is
\begin{equation} \label{eq:dilatation}
D \phi(x) = \lambda^{-\Delta} \phi(x)
\end{equation}
hence in this situation the fields are eingenstates of the dilatation operator.\\
The interesting aspect of a CFT is that in it we can simply consider particular fields, the so called \textit{conformal primaries}. These are fields that commute with the SCT operators. The reason is that via \eqref{increasing} we see that $P_\mu$ increases the scaling dimension by 1, while $K_\mu$ decreases it always by 1, as it can be seen from \eqref{decreasing}. Since in CFTs a lower bound for the scaling dimensions of fields must exist, it follows that any conformal representation must contain operators of lowest dimension, that have to be annihilated by $K_\mu$. These are the primary fields or conformal primaries. All other fields, dubbed as \textit{conformal descendants}, are obtained by acting with $P_\mu$ on the primary ones.

\subsection{Superconformal algebra and SCFTs} \label{sec:SCFT}
In the case of superconformal algebra (meaning that the conformal theory is supersymmetric), the generators can be grouped into the ones of the conformal algebra ($P_\mu$, $M_{\mu\nu}$, $D$ and $K_\mu$), and the ordinary supercharges $Q^I_\alpha$ and $\overline{Q}^{\dot{\alpha}}_I$. However this is not the full set of generators. In fact it turns out that they do not close the entire superconformal algebra. In particular the commutator between the supercharges and the SCT generators force us to introduce a new set of supercharges, called \textit{conformal supercharges} $S^I_\alpha$ and $\overline{S}^{\dot{\alpha}}_I$, in a number which is equal to the number of ordinary supercharges. The entire algebra is made up of about thirty commutators and anticommutators, so we do not report them, but the reader that is interested can consult \cite{Ammon:2015wua}. For this reason, in SCFTs, supercharges are doubled. It is interesting to say that this supersymmetric extension of the conformal algebra can be applied only if $2 \leq d \leq 6$.\\

As in the case of conformal algebra in which there are the primary fields, also here we can consider basic fields, called \textit{superconformal primary fields}, and these are the objects with the lowest scaling dimension within the superconformal multiplets. From the superconformal algebra it turns out that, besides $K_\mu$, also the conformal supercharges $S^I_\alpha$ and $\overline{S}^{\dot{\alpha}}_I$ lower the scaling dimension, while ordinary supercharges, as $P_\mu$, raise it (in particular, $K_\mu$ lowers by 1, the conformal supercharges lower by $\frac{1}{2}$, $P_\mu$ raises by 1 while the ordinary supercharges raise by $\frac{1}{2}$). Hence from these properties we learn that superconformal primary fields  are annihilated by $K_\mu, S_\alpha^I$ and $\overline{S}^{\dot{\alpha}}_I$ and the application of $P_\mu, Q_\alpha^I$ and $\overline{Q}^{\dot{\alpha}}_I$ gives rise the descendants.\\
Now, the superconformal primary fields form a multiplet. In the following we will focus on \textbf{chiral} superconformal primaries, which enjoy the additional property to be annihilated by the ordinary supercharges $\overline{Q}^{\dot{\alpha}}_I$. These operators differ in the spin and we are going to consider the \textbf{scalar} ones, due to the fact that they are the only ones to acquire VEVs and this allows to parametrize the moduli space. This will be important later.

\section{Isolated theories} \label{sec:chap2_2}
\begin{figure}[H] 
\centering
\includegraphics[scale=0.69]{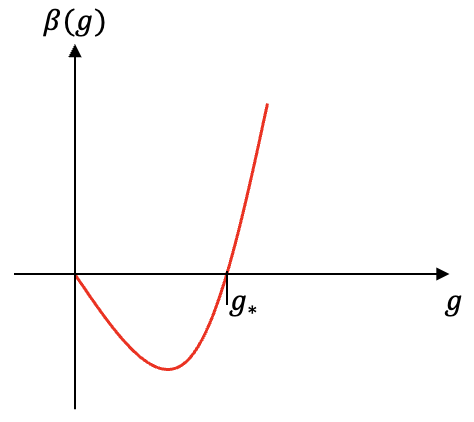}
\caption{$\beta$ function of the QCD theory of the Standard Model}
\label{im:beta_function}
\end{figure}

In order to understand what an isolated theory is, it is useful to give the formulation of conformal symmetry in terms of the $\beta$ function, that we used in the previous Chapter. Here we want to understand its meaning at a conceptual level.\\
Assuming to have, at the moment, a gauge theory, we recall that the definition of this function is
\begin{equation}
\beta(g) = \frac{\mathrm{d} g}{\mathrm{d} \log(\mu)} = \mu \frac{\mathrm{d} g}{\mathrm{d} \mu}
\end{equation}
where $g$ is the gauge coupling and $\mu$ is the renormalization energy scale which is defined in the context of renormalization in order to get rid of divergences that would come out when passing to consider the quantum treatment of the present theory. From this definition, we see that the $\beta$ function encodes how the theory changes as the energy scale changes. For example, if we consider the example of QCD of the Standard Model, we get that the behavior of this function is represented in Figure \ref{im:beta_function}, from which we easily read that the energy of the theory decreases as the coupling $g$ increases. For this reason a gauge theory with this kind of $\beta$ function is strongly coupled in the IR and asymptotically free.

The value $g_*$ of the gauge coupling corresponds to the situation in which $\beta(g_*)=0$. Points with this property are called \textit{fixed points} and they correspond to theories that enjoy conformal invariance at the quantum level. Indeed, for what explained above, if the $\beta$ function vanishes, then the associated theory is the same at the different energy scales, it does not change and so, practically, it does not have a reference scale. All of this implies that the theory is invariant under dilatations, which are part of the conformal algebra. The general argument is that if we have a theory with a given RG flow and if it has a fixed point, then it will flow to a conformal or superconformal theory in that fixed point.\\

Now, there are several ways in which the $\beta$ function vanishes at the quantum level: one could be that it is zero for any value of the gauge coupling, otherwise it could happen for a specific value of it. In the former case the theory enjoys a \textit{conformal manifold}, in the latter is said to be \textbf{isolated}. Let us see an example of the first case. It turns out that $\mathcal{N}=2$ SQCD with gauge group SU$(N)$ and a number of massless hypermultiplets transforming in the fundamental representation $N_f = 2N$ is superconformal for all the values of the gauge coupling and this because $\beta \simeq N_f - 2N$. In the case of $N=2$ we have been considering from the previous Chapter, it is clear that the gauge coupling $\tau$ parametrizes the conformal manifold in such a way that every single point on it describes a different copy of the original theory, where the difference in each of these copies enters in the entity of the interactions codified by the coupling itself. The metric of this manifold is the \textit{Zamolodchikov metric}, related to a specific kind of correlator. Even if we are going to study another type of correlators, this highlights the importance to understand how to compute correlation functions in this kind of theories.\\
If the situation just described does not take place, meaning that our theory does not have one or more free parameters (more properly called \textit{exactly marginal couplings}) that allow to describe a conformal manifold, then such theory is isolated. For this reason isolated models are never gauge theories, since they do not have a free gauge coupling, which instead is fixed to a specific value. This implies that any symmetry enjoyed by this kind of theories is only global. Furthermore, if an isolated theory is interacting, it is intrinsically strongly coupled, since it does not have a coupling that can be taken small enough in order to go to a weak coupling region.

\section{Coulomb branch and Argyres-Douglas theories}\label{sec:chap2_3}
We have said, at the beginning of this Chapter, that Argyres-Douglas (AD) theories are isolated, superconformal field theories that are intrinsically strongly coupled and without Lagrangian description. While the first two properties are now clear, we have to see why this kind of models is interacting and does not admit a Lagrangian description. In order to do it, we have to include the notion of \textit{Coulomb branch}.\\

In the previous Chapter we saw that $\mathcal{N}=2$ theories admit a space of degenerate vacua, the moduli space $\mathcal{M}$. What we implicitly did was focusing only on a region of it, indeed we considered only the part of $\mathcal{M}$ parametrized by the VEVs of the scalar component of the vector multiplets. But we also have hypermultiplets, which posses scalar components, and consequently also their VEVs can describe the moduli space. In general, the set of vacua associated to the VEVs of the scalar components of the vector multiplets $\phi^I$ is called \textbf{Coulomb branch} (CB), while the one described by the VEVs of the scalar components of the hypermultiplets $h^i$ is called \textbf{Higgs branch} (HB). Branches where both $a^I = \braket{\phi^I}$ and $\mathfrak{h}^i = \braket{h^i}$ are non vanishing are called mixed branches. At this point we see that the dimension of the CB is equal to the rank of the gauge group, due to the fact that the number of independent scalar field VEVs $a^I$ is related to the number of the Casimir of the gauge group and it is equal to the rank of this group $I = 1,\dots, \text{rk}(G)$. We stress that in the previous Chapter we focused only on the CB of $\mathcal{M}$, also when we explained that in a point of the moduli space the gauge group $G=\text{SU}(N)$ was spontaneously broken down to $\text{U}(1)^{\text{rk}(G)}$. The reason will become clear shortly, but this will allow us to consider only the CB in all the future discussions of our study.\\

\begin{figure}[H]
    \centering
    \begin{minipage}{0.45\textwidth}
        \centering
        \includegraphics[width=\linewidth]{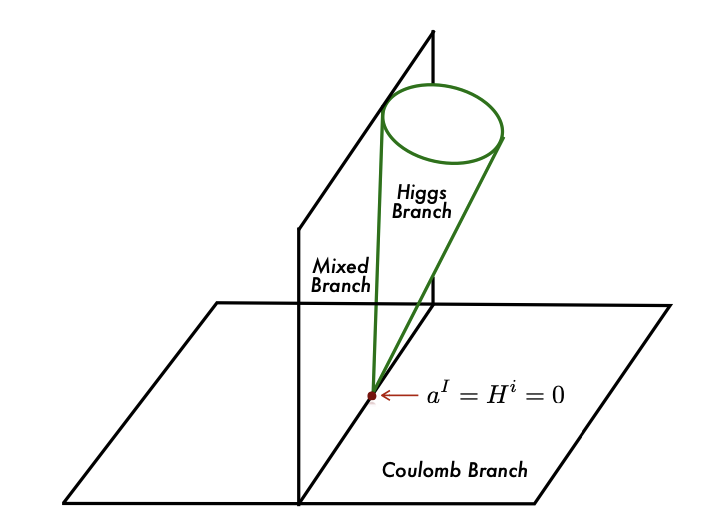}
    \end{minipage}
    \hfill
    \begin{minipage}{0.45\textwidth}
        \centering
        \includegraphics[width=\linewidth]{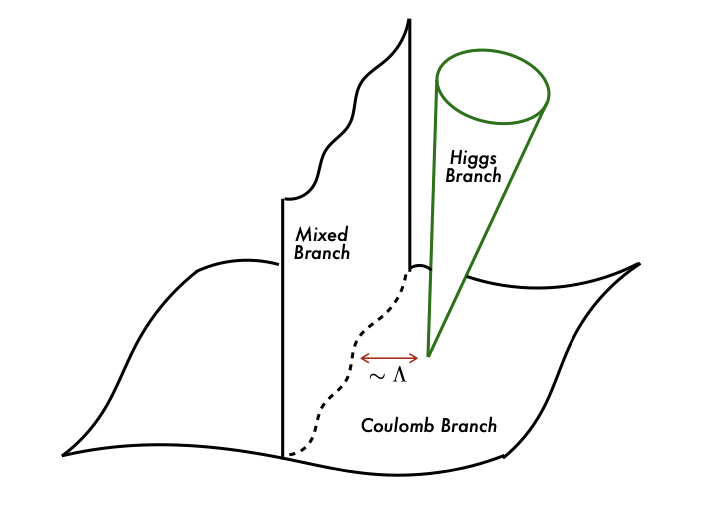}
    \end{minipage}
\caption{Cartoons representing the classical (on the left) and the quantum (on the right) $\mathcal{N}=2$ moduli space. The $H^i$ that appear are the scalar field VEVs $\braket{h^i}$, that we called $\mathfrak{h}^i$ in the text. Images taken from \cite{Bertolini:2025jhz}.}
\label{im:moduli_spaces_classical_quantum}
\end{figure}

From the definitions given above, we understand that, for example, the CB and HB meet at the origin of the moduli space. This gives rise to a singularity of the metric, corresponding, as already said, to massless degrees of freedom that should be included in the low energy effective description. But this is true only at the classical level. When we include quantum corrections the story is quite different. We cannot go into many details of them, for which the reader can consult \cite{Bertolini:2025jhz}.\\
The fact that, even under quantum corrections, CB and HB continue to exist is guaranteed by $\mathcal{N}=2$ SUSY. Then, a very interesting property, is the fact that the HB remains classically exact even after the addition of quantum corrections. In order to understand this, we have to recall that we are considering a supersymmetric non-linear sigma model, so the metric on the CB and on the HB is, respectively, a function of the scalar fields $\phi^I$ and $h^i$. From what we learnt in the previous Chapter, the metric on the CB is related to the imaginary part of the generalized complexified coupling $\tau_{I J}$ (now it is a matrix because we are considering generic rank), but since the latter undergoes renormalization (at one loop and non perturbatively), its quantum corrected expression becomes a certain function of the strong coupling scale $\Lambda$. Since $\Lambda$ appears in $\tau_{I J}$, it appears in the Lagrangian in the same way as a VEV of the scalar belonging to the vector multiplet would. Since instead the metric on the HB does not depend on vector multiplet scalars, it does not depend on $\Lambda$, either. Taking the classical limit $\Lambda \to 0$, one concludes that the metric on the HB is classically exact. Hence, the finale message is that the metric on the CB can receive quantum corrections, while that on the HB is classically exact. This has the consequence that the exact low energy effective action will be described by a quantum corrected CB and a classically exact HB. Solving the quantum theory consists in determining the geometry on the CB and this is exactly what we did in the previous Chapter when we passed to consider the SW curve. This explains why we focused only on the CB. Another important property that makes $\mathcal{N}=2$ theories special is that a moduli space always survives at quantum level. This is clear for the HB, hence it means that the CB always survives at the quantum level. A picture showing the classical and quantum corrected moduli spaces is reported in Figure \ref{im:moduli_spaces_classical_quantum}.

As already discussed, the classical moduli space has singularities and these could also be present at the quantum level. We saw in the previous Chapter that, in the quantum scenario, these singularities do not describe massless gauge fields, but rather massless dyons and monopoles. We are going to see that some of these points can also describe mutually non-local particles that become simultaneously massless. These are known as AD points and a Lagrangian description is not possible, as we are going to clearly declare later.\\
Summarizing, apart from special points where a Lagrangian description is not available (and that we are about to analyze), the structure of the $\mathcal{N}=2$ low energy effective action is (using a slightly different normalizations with respect to the previous Chapter, only to emphasize the final message)
\begin{equation} \label{eq:lagrangian_effective}
\mathcal{L} = K_I^{\, J}(\phi, \overline{\phi}) \partial_\mu \overline{\phi}^I \partial^\mu \phi_J + K_i^{\, j}(h, \overline{h}) \partial_\mu h^i \partial^\mu \overline{h}_j + \frac{1}{2} \text{Im} \left[\tau_{I J} (\phi) \mathcal{W}^I_{\mu \nu} \mathcal{W}^{J \mu \nu}\right] + \text{fermions}
\end{equation}
where $K_I^{\, J}$ is the metric on the CB, $K_i^{\, j}$ is the metric on the HB and the complexified gauge coupling is related to the prepotential as already seen in the previous Chapter and the we report here for simplicity
\begin{equation}
\tau_{I J} = \frac{\partial^2 \mathcal{F}}{\partial \phi^I \partial \phi^J}
\end{equation}
At this point we can easily understand what we meant, in the previous Chapter, by solving the theory: determining the quantum expression of the prepotential $\mathcal{F}$, in such a way that we can find the gauge coupling and then the Lagrangian of our model \eqref{eq:lagrangian_effective}. The SW solution that we analyzed in the previous Chapter does exactly this, indeed we found $a(u)$ and $a_D(u)$ in terms of some integrals. We then can invert $a(u)$ in order to obtain $u(a)$ and, substituting into $a_D(u)$, we get $a_D$ as a function of $a$. Integrating with respect to $a$ yields to the prepotential $\mathcal{F}(a)$, which is what we need to determine the Lagrangian.\\

As already hinted before, we have considered, up to now, a model that admits a Lagrangian description, but we have said that it is not always the case. For this reason it is quite important to define the CB and the HB in a way that is independent on the VEVs of fundamental fields of our theory (which instead could also not exist), also because the existence of such branches is a consequence of $\mathcal{N}=2$ SUSY. The intrinsic property of any theory with SUSY is the R-symmetry. We said that for $\mathcal{N}=2$ it is $\text{U}(2)_R \simeq \text{SU(2)}_R \times \text{U(1)}_R$ and the HB is the set of vacua that spontaneously break $\text{SU(2)}_R$ of the R-symmetry, while the CB is made up of all the vacua that spontaneously break $\text{U(1)}_R$. It is important to stress that, for generic rank of the gauge group, the CB is parametrized by the local coordinates we introduced in the previous Chapter $(a_{D_I}, a_I)$, with $I = 1, \dots, \text{rk}(G)$. In what follows and in all the rest of the thesis, by the way, we are interested only on the CB with $\text{rk}(G) = 1$ (since $G = \text{SU(2)}$), leading back to the scenario we have studied in the previous Chapter.

\subsection{Argyres-Douglas theories} \label{sec:AD_theories}
Consider $\mathcal{N}=2, \text{SU(2)}$ SQCD with one massive hypermultiplet in the fundamental representation ($N_f=1$). We follow in this Section the conventions adopted in \cite{Bertolini:2025jhz}, but we are about to see that the way in which the SW curves can be written is not unique, thanks to changes of rescaling and normalizations. This hence explains why the SW curve for a given model appears in different forms in several references. The SW curve that we consider is
\begin{equation} \label{eq:Nf1_curve_bertolini}
y^2 = x^2 (x-u) - \Lambda_1^6 + 2 m \Lambda_1^3 x
\end{equation}
where $m$ is the mass of the hypermultiplet, $\Lambda_1$ is the dynamically generated scale (the subscript 1 refers to $N_f$).\\ First of all, we want to check that, according to the decoupling procedure we described at the end of the previous Chapter, we recover the curve for the pure $\text{SU(2)}$ theory we analyzed. So we have to send $m \to \infty, \Lambda_1 \to 0$ in such a way that $m \Lambda_1^3 = \Lambda_0^4$ is fixed and it is the new coupling of the final theory. Then \eqref{eq:Nf1_curve_bertolini} reduces to
\begin{equation} \label{eq:Nf0_curve_bertolini}
y^2 = x^2(x-u) + 2 \Lambda^4_0 x
\end{equation}
This is the expected curve for the unflavored theory, even if it is written in a different parametrization and normalization with respect to \eqref{eq:SW_SYM_SU2}. In order to see how these two forms are related, we have to say that in a normalization where charged fields transforming in the adjoint representation of the gauge group have integer charges, those fields transforming in the fundamental representation have charges that are half integers. For this reason, with the goal of comparing a gauge theory with and without matter, it is convenient to change the normalization we used to treat pure SYM theory. In particular, in the central charge formula \eqref{eq:central_charge_formula}, we multiply $n$ by 2, in such a way that the final charge is only integer, and divide $a$ by 2,  ensuring the invariance of the formula itself. This change of conventions implies the following transformation of the vector $(a_D,a)$
\begin{equation}
\begin{pmatrix}
a_D \\
a
\end{pmatrix} \to 
\begin{pmatrix}
1 & 0\\
0 & \frac{1}{2}
\end{pmatrix} \begin{pmatrix}
a_D \\
a
\end{pmatrix}
\end{equation}
and the following transformation on the generic monodromy matrix
\begin{equation}
\begin{pmatrix}
a & b \\
c & d
\end{pmatrix} \to 
\begin{pmatrix}
1 & 0\\
0 & \frac{1}{2}
\end{pmatrix} \begin{pmatrix}
a & b \\
c & d
\end{pmatrix}\begin{pmatrix}
1 & 0 \\
0 & 2
\end{pmatrix} = \begin{pmatrix}
a & 2 b \\
\frac{c}{2} & d
\end{pmatrix}
\end{equation}
In this way the three monodromy matrices \eqref{eq:monodromy_inf}-\eqref{eq:simplest_solution} (remember that we set $u_0 = 1$, but it was only for simplicity; restoring the correct units it is $u_0 = \Lambda_0^2$) become
\begin{equation}
M_\infty = \begin{pmatrix}
-1 & 4 \\
0 & -1
\end{pmatrix}, \qquad 
M_{\Lambda^2_0} = \begin{pmatrix}
1 & 0\\
-1 & 1
\end{pmatrix}, \qquad
M_{-\Lambda^2_0} = \begin{pmatrix}
-1 & 4 \\
-1 & 3
\end{pmatrix}
\end{equation}
The elliptic curve associated to these matrices is
\begin{equation} \label{eq:Nf0_curve_bertolini_alternative}
y^2 = x^2 (x-u) + \frac{1}{4} \Lambda^4_0 x = x \left(x^2 - u x + \frac{1}{4} \Lambda_0^4 \right)
\end{equation}
This equation coincides with \eqref{eq:Nf0_curve_bertolini} after a rescaling of $\Lambda_0$.\\

The form of the equation \eqref{eq:Nf0_curve_bertolini} for the pure case makes less clear which are the singular points on the moduli space, with respect to the form \eqref{eq:SW_SYM_SU2}. But we have to recall what we learnt from the previous Chapter: singularities on this space arise when branch points of the elliptic curve collide, provoking a degeneracy of the curve itself. In order to find these situations, independently from the parametrization that is used, we have to impose the vanishing of the discriminant of the elliptic curve. This is exactly what we mentioned when treating the quantum curve. Indeed the discriminant $\Delta$ of the previous polynomial can be written as
$$\Delta = \prod_{i < j} (\alpha_i - \alpha_j)$$
where $\alpha_i$ are the roots of the polynomial and hence the branch points of the curve and then we have to find the values of $u$ for which some roots coincide, making $\Delta = 0$. Applying this to \eqref{eq:Nf0_curve_bertolini_alternative}, we get
$$\Delta = \frac{1}{16} \Lambda_0^8 \left(u^2 - \Lambda_0^4\right)$$
which vanishes at $u = \pm \Lambda^4_0$, as expected from the analysis of the previous Chapter.\\
More generally, the roots $\alpha_i$ of the $x$-polynomial are functions of the parameters of the theory which, in pure SYM $\text{SU(2)}$, are just $u$ and $\Lambda_0$. The form of these functions depends from the employed parametrization. Indeed, in the form used in \eqref{eq:SW_SYM_SU2} the branch points are $\alpha_1 = \Lambda_0^2, \alpha_2 = - \Lambda_0^2, \alpha_3 = u$, while in the new parametrization \eqref{eq:Nf0_curve_bertolini_alternative} these are
$$\alpha_1 = 0, \qquad \alpha_2 = \frac{u + \sqrt{u^2 - \Lambda_0^4}}{2}, \qquad \alpha_3 = \frac{u - \sqrt{u^2 - \Lambda_0^4}}{2}$$
What cannot change are the \textbf{number} and the \textbf{locations} of the singularities on the $u$ plane (meaning the values of $u$ for which two or more roots coincide) nor the \textbf{nature} of the particle that become massless. In the present case, indeed, when $u = \Lambda_0^4$ we have $\alpha_1 = \alpha_3$ in the first parametrization and $\alpha_2 = \alpha_3$ in the second one, while when $u = - \Lambda_0^4$ we have that $\alpha_2 = \alpha_3$ in the first representation and still $\alpha_2 = \alpha_3$ in the second one. Notice that there does not exist a value of $u$ for which $\alpha_1 = \alpha_2 = \alpha_3$, in none of the parametrizations that we can write. Furthermore, the other singularity $u = \infty$ cannot be explicitly seen in these cubic parametrizations, as already said in the previous Chapter.\\

Now we can go back to the case with $N_f=1$, referring to the curve \eqref{eq:Nf1_curve_bertolini}. The expectation is that we have three singularities at finite distance, not as in the unflavored case where one of them was located at infinity. This can be found both in the regimes of $m \gg \Lambda_1$ and $m = 0$:
\begin{itemize}
\item in the former case, below the scale $|m|$, the theory effectively behaves as pure $\text{SU(2)}$ SYM. For this reason the structure of the moduli space in the region $|u| \ll |m^2|$ should be the same of the pure theory. with two singularities locates at $u = \pm \Lambda_0^4$, as just discussed. Furthermore, since we are in the limit where $m$ lies in the region where $u \simeq a^2$ (the classical limit), we expect that a third singularity arises when $u \simeq m^2$. Indeed, by sending $m \to \infty$, this third singularity is pushed away to infinity, recovering the moduli space of pure $\text{SU(2)}$ SYM;
\item in the latter case, the R-symmetry argument comes to rescue. It can be proved that in the massless case the surviving R-symmetry is $\mathbb{Z}_6$, but remembering from the previous Chapter that each point on the $u$ plane is $\mathbb{Z}_2$ invariant, we deduce that the singularities are related by the broken $\mathbb{Z}_3$. This implies that they are located in the strong coupling region, being at finite distances.
\end{itemize}
These considerations can be checked from the curve \eqref{eq:Nf1_curve_bertolini}. This is done by computing $\Delta$, its zeros and then expanding these ones for large and small $m$. It turns out that
\begin{equation}
\Delta = 4 m^2 u^2 \Lambda_1^6 - 32 m^3 \Lambda_1^9 - 4 u^3 \Lambda_1^6 - 27 \Lambda_1^{12} + 36 u m \Lambda_1^9
\end{equation}
If we evaluate this quantity with $u = m^2, \pm \Lambda_0^4$, we find that, expanding for $\Lambda_1/m \ll 1$, $\Delta = 0 + \mathcal{O}[(\Lambda_1/m)^3]$ in the first case and $\Delta = 0 + \mathcal{O} [(\Lambda_1/m)^{3/2}] \Lambda_0^4$ in the second one, showing what we said in the first point above. If instead we set $m = 0$, the discriminant reduces to $\Delta = -4 u^3 \Lambda_1^6 - 27 \Lambda_1^{12}$ and its zeros, i.e. the position of the singularities, are
$$u = \left(-\frac{27}{4} \Lambda_1^6 \right)^{\frac{1}{3}}$$
confirming that, in this limit, the three singularities are related by $\mathbb{Z}_3$ rotations.\\
For what concerns the nature of the hypermultiplets that become massless at these singularities, this obviously depends on the value of $m$. Indeed the roots of the $x$-polynomial in this theory are functions of $u,\Lambda_1$ and $m$: $\alpha_i = \alpha_i(u,\Lambda_1,m)$. When $m$ is large, for example, the two strong coupling singularities are practically the same as the pure theory, hence the corresponding massless particles are a monopole and a dyon, respectively (as seen in the previous Chapter). At $u \simeq m^2$, the object that turns to be massless is electrically charges. As we decrease $|m|$, the three singularities become closer and closer and are related by the $\mathbb{Z}_3$ symmetry. For this reason there is no more a clear distinction between hypermultiplets coming from solitons (monopoles or dyons) or from elementary (i.e. fundamental) objects.\\

We have to stress, nonetheless, that in the present model there exists a \textit{specific} choice of the parameters which makes one of the singular points of the $u$-plane more special than others and it is a difference with respect to the case of $N_f=0$. Indeed, for generic values of $m$ and $\Lambda_1$ we have three different ways to collide two branch points, $\alpha_1 = \alpha_2$, $\alpha_1 = \alpha_3$ or $\alpha_2 = \alpha_3$, which correspond to the three singularities discussed so far. In correspondence of these, as seen in the previous Chapter, one of the cycles of the torus between $\alpha$ (not to be confused with the roots of the $x$-polynomial), $\beta$ or $\alpha + \beta$ collapses and we are making coincident only two out of the three branch points. It turns out that it is possible to make all the three branch points simultaneously coincident. As we said above before the conclusion of the $N_f=0$ case, this situation is not possible for the pure model.\\
Let us choose the mass of the hypermultiplet to be $m = \frac{3}{2} \Lambda_1$ and in this way the SW curve \eqref{eq:Nf1_curve_bertolini} becomes
\begin{equation} \label{eq:SW_curve_special}
y^2 = x^2 (x-u) + 3 x -1
\end{equation}
where we have set $\Lambda_1 = 1$ for simplicity (at the end of the treatment we will restore it with all the necessary units). The discriminant turns out to be
$$\Delta = 9 u^2 - 108 - 4 u^3 + 54 u - 27$$
In this case only two singularities on the $u$-plane arise. One of them is located at $u = -15/4$, in corresponding of which two roots of the $x$-polynomial coincide, while the other one at $u = u_* = 3$, where \textbf{all} the three branch points coincide. Indeed in these two situations we get that
\begin{align*}
& y^2 = (x+2)^2 \left(x-\frac{1}{4} \right) \qquad \text{for} \quad u=-\frac{15}{4} \\
& y^2 = (x-1)^3 \qquad \text{for} \quad u=3
\end{align*} 
This means that the hypermultiplet becoming massless at $u=-15/4$ has the same nature as the ones already encountered. Instead, since for $u=3$ the three branch points collide at $x=1$, something peculiar happens. In particular, at $u=3$ \textbf{both} $\alpha$ and $\beta$ cycles shrink and the corresponding massless hypermultiplets enjoy an important property. In order to understand this, we have to recall the Schwinger-Swanziger equation \eqref{eq:schwinger_equation}
$$q_1 g_2 - q_2 g_1 = 2 \pi n$$
and, only if the RHS vanishes the corresponding states are local with respect to each other. In general, two fields are \textit{mutually non local} if they do not commute at spacelike separations. This implies that for such fields we cannot write a manifestly local Lagrangian that describes their interaction. So, while two electrically charged states are local with respect to each other, an electrically charged state and a magnetic monopole (or a dyon) are not. For this reason, an effective low energy theory where such mutually non-local objects are both present cannot admit a Lagrangian description. Coming back to the hypermultiplets that become massless at $u=3$, since here both the cycles of the elliptic curve are 0, then these massless degrees of freedom, from the central charge formula, are mutually non local.\\

Summarizing what we have found up to now: on the moduli space of $\mathcal{N}=2, \text{SU(2)}$ SYM with a massive fundamental hypermultiplet, there exists a particular choice of the mass for which three branch points of the $x$-polynomial collide. At this point the theory admits mutually non-local degrees of freedom that become simultaneously massless and hence they cannot be integrated out, making a Lagrangian description for this theory not possible. This kind of model is the already mentioned AD theory. As we are going to say at the end of this Section, it is not a peculiar aspect of $\text{SU(2)}$ SQCD with one flavor.\\

Now it is important to understand what is the nature of such theories. It can be shown that at points where mutually non-local objects become simultaneously massless the theory is an interacting CFT, hence it is an interacting fixed-point along the RG flow of another theory. At first glance this might contradict the Coleman-Gross theorem \cite{Coleman:1973sx}, according to which in four dimensions any theory of scalars, spinors and abelian gauge field is IR-free, not admitting a fixed point. Our low energy effective theory is abelian (recall that we are on the CB and here we have the breaking of the gauge group $G$ down to $\text{U(1)}^{\text{rk}(G)}$) and we have already said that there are not other points on the quantum moduli space where extra gauge bosons become massless. For this reason it does not seem that there could be the possibility for an interacting fixed point. The fact is that on the points we are now studying both the homologically non-trivial 1-cycles of the elliptic curve shrink and this implies, as already said, the impossibility of having a Lagrangian description. This does not permit to prove the Coleman-Gross theorem and, for this reason, no contradiction is present.\\
We do not prove that the AD theories correspond to an interacting fixed-point. The interested reader can see \cite{Bertolini:2025jhz}. The take-home message is that in a CFT any interacting field strength must couple both to electric and magnetic charged objects. This can be obtained by using unitarity and conformal symmetry. For example, if we consider in QED both elementary monopoles and electric charges, then this theory admits a non-trivial fixed point in the IR, while if it does not include elementary monopoles then it cannot possess a non-trivial fixed point (in agreement with the theorem of Coleman-Gross).\\
In all this discussion $\mathcal{N}=2$ SUSY does not play any role, but the extra constraints imposed by it allows to get more informations about this interacting fixed point, for example the exact scaling dimension of some CFT operators. In fact, we have learnt that AD theories are superconformal and so they admit all the classes of operators we discussed in Section \ref{sec:SCFT}. In particular, we are going to focus, in the following Section, on all the scalar chiral superconformal primary operators that parametrize the CB via their VEVs. These are called \textit{Coulomb branch operators} (CB operators) and we are going to see how to get their scaling dimensions for the AD theory we are studying in this Section (which is, from the previous discussion, the IR fixed-point along the RG flow of the theory $\mathcal{N}=2, \text{SU(2)}$ SQCD with $N_f=1$), called $\mathcal{H}_0$. In particular we are going to do it with a simple dimensional analysis of the SW curve in the neighborhood of the AD point. In order to do so, we recall that the special point is in correspondence of $u=u_*=3$ and $m=3/2$, so we shift these variables and $x$ by some quantities and then we expand the SW curve \eqref{eq:SW_curve_special} in terms of the shifted variables. In particular, we consider \footnote{We stress that these equalities have sense only because $\Lambda_1 =1$, but later we are going to reintroduce it in such a way to make the dimensional agreement evident.}
\begin{equation} \label{eq:shifted_variables}
m = \frac{3}{2} + \widetilde{m} \qquad u = 3 + \widetilde{u} + 2 \widetilde{m}  \qquad x = \frac{1}{3} u + \widetilde{x}
\end{equation}
As it is going to be clear in the next Section, the conformal dimension of the CB operator is related to the scaling dimension of $\widetilde{u}$ (sometimes denoted also as $u_{\text{AD}}$, since it plays the same role of $u$ but for the AD theory; similarly $\widetilde{m} \equiv m_{\text{AD}}$ and $\widetilde{x} \equiv x_{\text{AD}}$, \cite{Fucito:2023txg}), so we want to determine it. Expanding the curve in the shifted variables $(\widetilde{m}, \widetilde{u}, \widetilde{x})$, we obtain
\begin{equation} \label{eq:SW_curve_neighborhood}
y^2 = \widetilde{x}^3 - 2 (\widetilde{m}+\widetilde{u})\widetilde{x} - \left(\widetilde{u} + \frac{4}{3} \widetilde{m}^2\right) + \dots
\end{equation}
where the dots refer to higher order terms in $\widetilde{m}$ and $\widetilde{u}$, which are small in the neighborhood of the fixed point. In order to find the cubic singularity at $\widetilde{u} = \widetilde{m} = 0$, we must impose the following constraints on the scalings of the shifted variables
\begin{equation} \label{eq:first_constraints_dimensions}
[\widetilde{m}] = 2 [\widetilde{x}] \qquad [\widetilde{u}] = 3 [\widetilde{x}]
\end{equation}
Thanks to this, we can neglect the terms $\widetilde{x} \, \widetilde{u}$ and $\widetilde{M}^2$ in \eqref{eq:SW_curve_neighborhood}, getting the following expression of the SW curve
\begin{equation} \label{eq:SW_curve_H0}
\widetilde{y}^2 = \widetilde{x}^3 - 2 \widetilde{m} \widetilde{x} - \widetilde{u}
\end{equation}
at the SCFT. At this point we are ready to determine $[\widetilde{u}] \equiv d$: recalling that $[a] = 1$ (as it can be seen from the central charge formula) and that $a \simeq \widetilde{u} \frac{\mathrm{d} \widetilde{x}}{\widetilde{y}}$, we see that
$$1 = d + [\widetilde{x}] - [\widetilde{y}] \underset{\text{from} \, \eqref{eq:first_constraints_dimensions}}{=}  4 [\widetilde{x}] - [\widetilde{y}]$$
Furthermore, from \eqref{eq:SW_curve_H0} we also read that $2 [\widetilde{y}] = 3 [\widetilde{x}]$, consequently, together with the equation just found, we get
\begin{equation}
\frac{5}{2} [\widetilde{x}] = 1 \Rightarrow [\widetilde{x}] = \frac{2}{5}
\end{equation}
ending with
\begin{equation}
[\widetilde{u}] = \frac{6}{5} \qquad [\widetilde{m}] = \frac{4}{5} \qquad [\widetilde{y}] = \frac{3}{5}
\end{equation}
In light of this, restoring $\Lambda_1$, the definition of the shifted variables in \eqref{eq:shifted_variables} is more properly given by the following equalities \footnote{Notice that from \eqref{eq:Nf1_curve_bertolini} it can be seen that $[x] = 2$ and $[y] = 3$.}
\begin{equation}
m = \frac{3}{2} \Lambda_1 + \widetilde{m} \Lambda_1^{\frac{1}{5}} \qquad u = 3 \Lambda_1^2 + \widetilde{u} \Lambda_1^{\frac{4}{5}} + 2 \widetilde{m} \Lambda_1^{\frac{6}{5}} \qquad x = \frac{1}{3} u + \widetilde{x} \Lambda_1^{\frac{8}{5}}
\end{equation}
Since the conformal dimension of the CB operator is given by $[\widetilde{x}]$ (as already hinted), we see that this quantity is larger than 1. This confirms that we are at an interacting fixed point \cite{Terning:2006bq}, as expected.\\
Until this point we have been discussing the moduli space, but let us simply see why the AD theory we are examining is also isolated. Without entering into details (that can be clearly consulted in the Appendix A of \cite{Bissi:2021rei}), it is possible to compute the complexified gauge coupling $\tau$ in correspondence of the $\mathcal{H}_0$ theory and it turns out (using the expressions for $a$ and $a_D$ for SQCD with $N_f = 1$, from which the AD model comes from) that
\begin{equation}
\tau_{\mathcal{H}_0} = e^{i \frac{\pi}{3}}
\end{equation}
and it is fixed, not free. This means that $\mathcal{H}_0$ does not admit a conformal manifold and hence it is isolated.\\

As already remarked, AD fixed points do not exist only for the theory we have analyzed so far. For instance, remaining in the realm of $\mathcal{N}=2, \text{SU(2)}$ SQCD, a similar situation takes place when $N_f = 2,3$: in correspondence of these values it is possible to find other AD theories, respectively indicated as $\mathcal{H}_1$ and $\mathcal{H}_2$, on the moduli space. They share the same features encountered (and that we are going to summarize); the differences lie for example on the specific forms of the SW curve (look at Appendix A of \cite{Fucito:2023txg}), the values of the conformal dimension of the CB operator, which turn out to be
$$d_{\mathcal{H}_1} = \frac{4}{3} \qquad d_{\mathcal{H}_2} = \frac{3}{2}$$
and other aspects. Furthermore it turns out that \cite{Bissi:2021rei}
$$\tau_{\mathcal{H}_1} = i \qquad \tau_{\mathcal{H}_2} = e^{i \frac{\pi}{3}}$$
Notice that this phenomenon does not take place for $N_f=0$, since we need matter coupled to pure SYM (as it can be deduced from the discussion of above and also declared before of \eqref{eq:SW_curve_special}), and neither for $N_f=4$, which admits a conformal manifold.
AD models come out also from other generalizations of the original SW model, namely pure SYM with gauge group $\text{SU}(3)$ (which is actually the first instance where this phenomenon was discovered, as the title of \cite{Argyres:1995jj} stresses).\\
Concluding, we have seen that AD theories are 4-dimensional $\mathcal{N}=2$ supersymmetric field theories which are isolated and admit mutually non-local massless degrees of freedom. This implies that these theories do not admit a Lagrangian description and they are interacting and conformal; due to the fact they are interacting and also isolated, they are intrinsically strongly coupled.

\section{Coulomb branch operators and extremal correlators}\label{sec:chap2_4}
It is time to see what are the physical quantities that we want to focus on. After having introduced and described how to compute them, we are going to see the details of the computation for the case of $\mathcal{N}=2, \text{SU(2)}$ SQCD with $N_f=4$ and all the flavors are massless, since the computations in this theory can be explicitly performed at the perturbative level (in a sense that will be clear). For the AD theories $\mathcal{H}_0, \mathcal{H}_1$ and $\mathcal{H}_2$, due to their intriguing properties, the computations cannot be performed in an explicit way, from what the author knows, and hence only an educated guess for a function interpolating between two asymptotic regimes of the partition function of these theories has been possible, leaving to the future the developments in this direction.

\subsection{Operator Product Expansion (OPE) and extremal correlators} \label{sec:OPE_extremal_correlators}
We have already introduced what the CB operators are: they are scalar chiral superconformal primaries that parametrize the CB of the moduli space of our theory via their VEVs. In order to see what important property they satisfy (and that spontaneously leads us to consider the extremal correlators in the following Subsection), we have to introduce in a brief way what the Operator Product Expansion (OPE) in (S)CFTs is.\\
As we have understood, in order to completely know a theory enjoying particular symmetries we should specify the commutation rules between the generators of those symmetries. In case of a (super)conformal theory, we can simply focus on commutators between the primary operators, but there is a nice property peculiar of these theories that allows us to avoid it, passing through the knowledge of other quantities, called OPE coefficients.\\
Consider a basis of primary operators $\{\mathcal{O}_{\Delta_a}\}$, with $\Delta_a$ the conformal dimension of each element of the basis, and pick two of them, $\mathcal{O}_{\Delta_1}(x)$ and $\mathcal{O}_{\Delta_2}(y)$. We define their product through the so called radial ordering, that is
\begin{equation}
\mathcal{O}_{\Delta_1}(x) \mathcal{O}_{\Delta_2}(y) = \begin{cases}
\mathcal{O}_{\Delta_1}(x) \mathcal{O}_{\Delta_2}(y) \, \, \, \text{if} \, \, \, \, \, |x|<|y| \\
\mathcal{O}_{\Delta_2}(y) \mathcal{O}_{\Delta_1}(x) \, \, \, \text{if} \, \, \, \, \, |y|<|x|
\end{cases}
\end{equation}
It is natural to ask what happens when $|x|=|y|$ and the answer depends from the nature of the two operators. Indeed, if they commute, then there is not any problem when they come across each other and hence nothing which is pathological takes place. Conversely, if they do not commute, then something which is ill-defined happens in the situation when the two operators meet each other, which means that the aforementioned product has a singularity in $x=y$ when the operators do not commute. Hence we can write
\begin{equation}
\mathcal{O}_{\Delta_1}(x) \mathcal{O}_{\Delta_2}(y) \underset{x \to y}{\simeq} [\mathcal{O}_{\Delta_1}(x), \mathcal{O}_{\Delta_2}(y)] + \dots
\end{equation}
where the dots refer to regular terms when $x \to y$. In this way if the operators commute the product is well defined, if instead they do not commute, there is a singularity related to the commutator and its structure is encoded in the so colled OPE (Operator Product Expansion)
\begin{equation}
\label{eq:OPE}
\mathcal{O}_{\Delta_1}(x) \mathcal{O}_{\Delta_2}(y) \underset{x \to y}{\simeq} \sum_{a} \frac{\lambda_{12a} \mathcal{O}_{\Delta_a}(y)}{|x-y|^{h}}+\dots
\end{equation}
and the coefficients $\lambda_{12a}$ are called OPE coefficients. The take-home message is that a (S)CFT is completely determined by its OPE coefficients, which are completely equivalent to commutation relations and this statement can be proven in total generality.\\
Notice that the conformal symmetry allows us also to determine the exponent $h$ at denominator. Indeed, thanks to \eqref{eq:dilatation}, we have that $\mathcal{O}_{\Delta_1}(x) \mathcal{O}_{\Delta_2}(y)$ transforms as $\rho^{-\Delta_1-\Delta_2}$, while $\mathcal{O}_{\Delta_a}$ as $\rho^{-\Delta_a}$ and $|x-y|^h$ as $\rho^h$, hence we get
$$\rho^{-\Delta_1-\Delta_2} = \frac{\rho^{-\Delta_a}}{\rho^{h}} \Rightarrow h = \Delta_1+\Delta_2-\Delta_a$$
All this argument is valid for any couple of primary operators of the original basis.\\

Coming back to our CB operators, it can be demonstrated that their conformal dimension is related to their R-charge under the $\text{U(1)}_R$ part of the R-symmetry as $\Delta = R/2$. For this reason, in the following, we are about to use as index of the CB operator its R-charge. The important property that characterizes this kind of operators (and more in general scalar chiral superconformal primaries) is that their OPE contains only regular terms, meaning that the following ring structure is satisfied
\begin{equation} \label{eq:OPE_CB}
\mathcal{O}_a (x) \mathcal{O}_b (0) \simeq C_{a b}^{\, d} \mathcal{O}_d(0) + \dots
\end{equation}
where the dots refer to terms that vanish as $x \to 0$. Supposing that in the CB of the moduli space of our theory we have a fixed point (as the AD points in the case of $\mathcal{N}=2, \text{SU(2)}$ SQCD with $N_f = 1,2,3$, as said before), on this point the R-symmetry and the conformal one are preserved, hence on the RHS of \eqref{eq:OPE_CB} the summation on the index $d$ involves a certain number of operators with the same dimension $\Delta_d = \Delta_a + \Delta_b$. This number is equal to the rank of our theory, which is equal to the number of generators of the ring we are considering. It can be shown that, if the generators of the CB are free (meaning that they do not satisfy any constraint), then there exists a basis of generators, $\{\mathcal{O}_i'\}$, such that
\begin{equation} \label{eq:OPE_CB_simplified}
\mathcal{O}'_a (x) \mathcal{O}'_b (0) \simeq C_{a b}^{' \, d} \mathcal{O}'_d(0) + \dots
\end{equation}
with $C_{a b}^{' \, d} =1$ for a given $d$ and they are 0 for all the other possible values taken by $d$. In other words, in this basis we only have one operator in the RHS with the suitable dimension. Nowadays $\mathcal{N}=2$ theories with the ring \eqref{eq:OPE_CB} not freely generated are not known and for this reason we always consider the new basis in such a way \eqref{eq:OPE_CB_simplified} is valid.\\
Specializing to rank 1, everything becomes simpler, indeed the only operator that appears in the RHS with dimension $\Delta_a + \Delta_b$ is $\mathcal{O}_{a+b}$, which is obtained (normalizing the dimension of the generator to 1 and hence $a$ and $b$ now are non negative integers) by exponentiating the generator to $a+b$
$$\mathcal{O}_a(x) \mathcal{O}_b(0) \simeq \mathcal{O}_{a+b} (0) + \dots$$
From this we clearly see that, if we compute correlation functions of operators involving just the $\mathcal{O}_a$ at the fixed point, they vanish, because $\braket{\mathcal{O}_{a+b}} = 0$ thanks to the conservation of the R-symmetry \footnote{The same conclusion can be reached by using the selection rule coming from the conservation of the $\text{U(1)}_R$ part of the R-symmetry in the fixed point.}. The one-point functions of CB operators are non zero only in points different from the fixed one. This is depicted in the following Figure

\begin{figure}[H] 
\centering
\includegraphics[scale=0.6]{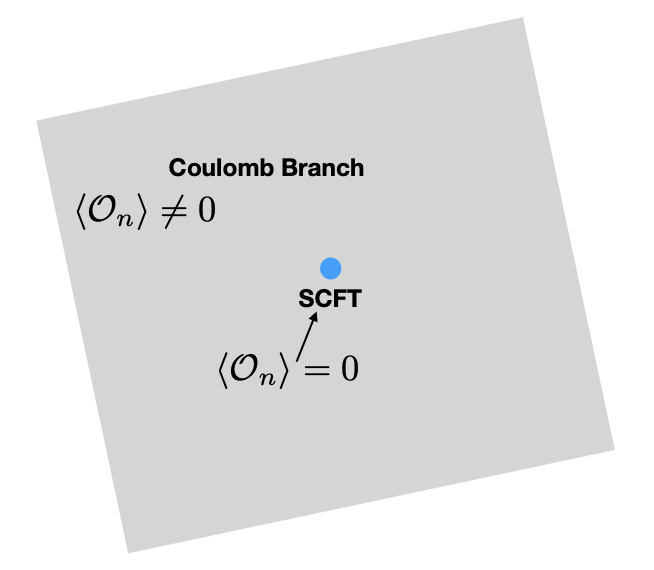}
\caption{Coulomb branch of the moduli space for theories like $\mathcal{N}=2, \text{SU(2)}$ SQCD with $N_f = 1,2,3$. The grey area is parametrized by non-zero VEVs of the CB operators, leading to the spontaneous breaking of conformal and $\text{U(1)}_R$ symmetries. Image taken from \cite{Grassi:2019txd}}
\label{im:CB_moduli_space}
\end{figure}

We want to stress that, for the conformal theory $\mathcal{N}=2, \text{SU(2)}$ SQCD with massless flavors and number $N_f=4$, the already mentioned Zamolodchikov metric is obtained from the two point correlation function between primary operators, not scalar chiral primaries.\\

There are non-trivial correlation functions in the theories we are considering that involve CB operators and anti-CB operators (meaning chiral and anti-chiral ones). In particular, if the number of the latter is equal to one, the final correlation functions are called \textbf{extremal correlators}. We are going to study two-point extremal correlators, meaning that also the number of insertions (in different spacetime point) of CB operator is equal to 1, namely $\braket{\mathcal{O}_i (x) \overline{\mathcal{O}}_j (y)}_{\text{SCFT}}$ (by the subscript SCFT we mean that this correlation function is evaluated at the fixed point of the moduli space as represented in Figure \ref{im:CB_moduli_space} or we are considering a SCFT, as the aforementioned $\mathcal{N}=2, \text{SU(2)}$ SQCD with $N_f=4$). For the theories with rank 1 we are examining, it turns out \cite{Grassi:2019txd} that the matrix $\braket{\mathcal{O}_i (x) \overline{\mathcal{O}}_j (0})$ is diagonal and so the space dependence of the correlation function we are inspecting is \cite{Papadodimas:2009eu}
\begin{equation} \label{eq:2_point_extremal_correlator}
\braket{\mathcal{O}_j (x) \overline{\mathcal{O}}_j (y)}_{\text{SCFT}} = G_{2 j} \frac{1}{(y-x)^{2 j d}}
\end{equation}
where $d \equiv \Delta_{\mathcal{O}}$ is the conformal dimension of the CB operator we already met in the previous Section and the dependence of the function $G_{2 j}$ depends if the theory admits a Lagrangian formulation or not. By the way, in the following we will not be interested in the spacetime dependence and for this reason we will indicate the correlation function simply as $G_{2 j}$.\\
The way in which we are going to compute this kind of object is using \textbf{supersymmetric localization}. Its field of applicability concerns only Lagrangian theories, for what we said in the previous Chapter when introducing the quantum SW curve in Section \ref{sec:qSW_curve}, and so we will quickly say how to extrapolate the ideas from localization also for non-Lagrangian theories (as done in \cite{Bissi:2021rei}).

\subsection{Lagrangian theories} \label{sec:Lag_theories_correlators}
The localization technique was introduced in \cite{Nekrasov:2002qd, Flume:2002az, Bruzzo:2002xf} for gauge theories on $\mathbb{R}^4$ and later, thanks to the work of Pestun \cite{Pestun:2007rz}, to the 4-sphere $S^4$. We are going to focus only on the usual $\mathcal{N}=2, \text{SU(2)}$ SQCD with $N_f=4$, for which the desired function $G_{2 n} (\tau, \overline{\tau})$ is a function of the complexified gauge coupling and its hermitian conjugate, as we are going to see. The shape of the $\Omega$ background is the one of the 4-sphere, hence $\epsilon_1 = \epsilon_2 = R^{-1}$, with $R$ the radius of the 4-sphere.\\
The idea of Pestun is that any SCFT on $\mathbb{R}^4$ can be put on $S^4$ without altering the final results of the original theory, due to the existence of a conformal map that goes from the former space to the latter. In this way, the supersymmetric partition function on $S^4$ can be written as an integral over the imaginary part of $a$ (the same quantity we already saw many times) and then, with the rotation $a \to i a$ (with the final $a$ a real variable), the final integral is convergent (we are going to see why). In the end we can write this partition function for the theory under consideration as
\begin{equation} \label{eq:part_func_4sphere}
Z_{S^4} (\tau, \overline{\tau}) = \int_{\mathbb{R}} \mathrm{d} a \,\, 4 a^2 \, \left| Z_{\text{Nekrasov}}(a, R, q) \right|^2
\end{equation}
where $q = e^{2 \pi i \tau}$ is the usual instanton fugacity and $Z_{\text{Nekrasov}}(a, R, q)$ is the Nekrasov partition function we quickly introduced in the previous Chapter and that we are going to inspect later for our theory. Indeed, at the moment, we want to describe how, starting from the $S^4$ partition function \eqref{eq:part_func_4sphere}, we can arrive at obtaining the extremal correlation functions we introduced above.\\

The first step is that it is possible to prove \cite{Gerchkovitz:2016gxx} that a derivative of the logarithm of $Z_{S^4}$ with respect to $\tau$ brings down an insertion of the chiral ring generator $\mathcal{O}_1$ at the north pole of the sphere and a derivative with respect to $\overline{\tau}$ produces an insertion of $\overline{\mathcal{O}}_1$ at the south pole. Repeating the first operation $i$ and the second $j$ times (hence $i$ and $j$ are non negative integers), we arrive to 
\begin{equation} \label{eq:correlation_function_S4}
\braket{\mathcal{O}_i (N) \overline{\mathcal{O}}_j (S)}_{S^4} = \frac{1}{Z_{S^4}(\tau, \overline{\tau})} \frac{\partial^{i+j} Z_{S^4} (\tau, \overline{\tau})}{\partial \, \tau^i \, \partial \, \overline{\tau}^j}
\end{equation}
where $\mathcal{O}_i = (\mathcal{O}_1)^i$ and similarly for the other object. The next point is that even if $\mathbb{R}^4$ and $S^4$ are conformally equivalent, a direct comparison between this correlation function and the one we need on $\mathbb{R}^4$ is not possible. This is due to the so called \textit{operator mixing}: the CB operator on $S^4$ mixes with a full tower of lower-dimensional operators as follows
$$\mathcal{O}_n \to \mathcal{O}_n + \alpha_1 (\tau, \overline{\tau}) \mathcal{O}_{n-1} R^{-2} + \alpha_2 (\tau, \overline{\tau}) \mathcal{O}_{n-2} R^{-4} + \dots$$
and this mixing violates the $\text{U(1)}_R$ symmetry which is present in $\mathbb{R}^4$, but it preserves the supersymmetry which is defined in $S^4$. Due to this problem, we cannot identify \eqref{eq:correlation_function_S4} with $G_{2n} (\tau,\overline{\tau})$ we are looking for. A way to solve this issue is treating the two-point function on $S^4$ as an infinite dimensional matrix $M_{i j} \equiv \braket{\mathcal{O}_i (N) \overline{\mathcal{O}}_j (S)}_{S^4}$ and then we consider the following ratio of determinants \footnote{The factor 16 which is present is due to the fact that we are following also the convention of \cite{Gerchkovitz:2016gxx}. Since the same procedure can be followed for $\mathcal{N}=4$ SYM, the relation between the normalizations of $\mathcal{N}=2$ and $\mathcal{N}=4$ chiral ring generator is important. It turns out that the latter is four times the former and, repeating this for the chiral and anti-chiral component, we obtain 16.}
\begin{equation} \label{eq:final_correlator}
G_{2 n}(\tau, \overline{\tau}) = 16^n \frac{\det M^{(n)}}{\det M^{(n-1)}}
\end{equation}
where $M^{(n)}$ denotes the upper-left $(n+1) \times (n+1)$ (since $i, j = 0,\dots, n$) submatrix of $M$. In this way we correctly map the operator insertions from $S^4$ to $\mathbb{R}^4$, solving the operator mixing problem, and for this reason we express $G_{2n}$ as the previous ratio. It can be shown that this procedure is equivalent to the Gram-Schmidt orthogonalization \cite{Gerchkovitz:2016gxx}.\\
In \cite{Grassi:2019txd} there are some examples of the application of the previous formula, but we want to focus on the usual $\mathcal{N}=2, \text{SU(2)}$ SQCD with $N_f=4$, specializing the Nekrasov partition function in \eqref{eq:part_func_4sphere}. In \cite{Nekrasov:2002qd} it was shown that this object can be written as a product of a tree level, a 1-loop and an instantonic contributions as
\begin{equation} \label{eq:expansion_Z_coupling}
Z_{\text{Nekrasov}}(a, R, q) = Z_{\text{tree}}(a,R,q) \, Z_{1\text{L}}(a,R) \, Z_{\text{inst}}(a,R,q)
\end{equation}
where, for the theory under consideration
\begin{align}
&  Z_{\text{tree}}(a,R,q) = e^{2 \pi i \tau a^2 R^2} \label{eq:Z_tree} \\
&  |Z_{1\text{L}}(a,R)|^2 = \frac{H(2 i a R) H(-2 i a R)}{|H(i a R)H(-i a R)|^4} \label{eq:Z_1_loop}
\end{align}
where $H(x) \equiv G(1+x) G(1-x)$, with $G(x)$ the Barnes $G$ gamma function. For what concerns the remaining term, it is the instanton partition function on the $\Omega$ background \cite{Nekrasov:2002qd,Moore:1997dj,Nekrasov:2003rj}. Nekrasov managed to write it as a series in $q$ (a.k.a $\tau$ or the coupling $g_{\text{YM}}$ of the theory) which is convergent. Its expansion in $q$ grows quickly in complexity, for example up to 2-instantons
\begin{equation} \label{eq:Z_inst_2_inst}
Z_{\text{inst}}(a,R,q) = 1 + \frac{q}{2} (a^2 R^2 -3) + \frac{q^2}{4} \frac{8 a^8 R^8 + a^6 R^6 - 91 a^4 R^4 - 60 a^2 R^2 +132}{(4 a^2 R^2 + 9)^2} +\dots
\end{equation}
We will come back to this expression later. Due to these complicated expressions of $Z_{1\text{L}}(a,R)$ and $Z_{\text{inst}}(a,R,q)$, the extremal correlators admit a non-trivial perturbative expansion (which agrees with the results coming from standard Feynman perturbation theory) and non-perturbative terms, that we will not consider in the following. It is interesting to stress that, if we ignore at the moment the instantons, the derivatives for computing \eqref{eq:correlation_function_S4} simply lead to insertions of $a^2$. In fact, for the theory we are examining, the CB generator is simply $\mathcal{O} = \frac{1}{2} \text{Tr} \phi^2$, that is it is related to the usual $u \simeq a^2$. From this we also infer that $d=2$ for this theory. This reasoning will be important for the case of non-Lagrangian theories.\\

It is possible to compute the perturbative expansion of the correlator $G_{2n}$ with fixed $n$, but the aim of this Chapter is shedding light on its dependence on $n$, studying also the interplay with perturbation theory (for a reason that will be clear). This is quite difficult to do by expanding in the coupling. For this reason we have to see another useful expansion for the Nekrasov partition function, keeping in mind that \eqref{eq:expansion_Z_coupling} is related to an expansion in $\tau$. The new strategy relies on the so called \textit{large radius expansion}. This consists in taking the radius of the 4-sphere much larger than $a^{-1}$, indeed, due to the fact our theory is conformal, the integrand inside \eqref{eq:part_func_4sphere} depends only on the dimensionless combination $a R$, as it can be seen from the expressions of the ingredients in \eqref{eq:Z_tree},\eqref{eq:Z_1_loop} and \eqref{eq:Z_inst_2_inst}. Considering $a R \gg 1$ is equivalent to have $\epsilon \ll a$ and it is analogous to the \textbf{genus expansion} of topological string theory \cite{Klemm:2002pa} and for this reason we will refer to this new expansion in the latter way.\\
For dimensional reason, we can write the Nekrasov partition function as
\begin{equation} \label{eq:relation_Z_prep}
Z_{\text{Nekrasov}}(a, R, q) = e^{R^2 \mathcal{F}(a,R,q)}
\end{equation}
and the genus expansion consists in writing $\mathcal{F}(a,R,q)$ in the following way
\begin{equation} \label{eq:genus_expansion_prep}
\mathcal{F}(a,R,q) = \sum_{g \geq 0} a^{2-2g} R^{-2g} \mathcal{F}_g(\tau)
\end{equation}
with $\mathcal{F}_g(\tau)$ dimensionless functions of the instanton fugacity. This implies that the 4-sphere partition function \eqref{eq:part_func_4sphere} becomes
\begin{equation} \label{eq:expression_ZS4_genus}
Z_{S^4} (\tau, \overline{\tau}) = \int_{\mathbb{R}} \mathrm{d} a \,\, 4 a^2 \, \exp \left[ \sum_{g \geq 0} R^{2-2g} a^{2-2g} \left(\mathcal{F}_g(q)+\overline{\mathcal{F}_g(q)}\right) \right]
\end{equation}
We are going to come back to the meaning of the genus expansion \eqref{eq:genus_expansion_prep} shortly, but before we want to discuss the relation between the two expansions we have introduced so far and also the component $\mathcal{F}_g$ appearing in \eqref{eq:genus_expansion_prep}.\\

Comparing the expansions \eqref{eq:genus_expansion_prep} and \eqref{eq:expansion_Z_coupling}, we see that each quantity $\mathcal{F}_g$ contains contributions coming from three factors, in particular $Z_{\text{tree}}$ depends only on $\mathcal{F}_0$, as it can be seen from its expression in \eqref{eq:Z_tree}, while $Z_{\text{1L}}$ and $Z_{\text{inst}}$ are relevant for all the $\mathcal{F}_{g \geq 0}$. In general, we have that $\mathcal{F}_0 = \mathcal{F}_0^{\text{tree}} +\mathcal{F}_0^{\text{1-L}} + \mathcal{F}_0^{\text{inst}}$, $\mathcal{F}_1 = \mathcal{F}_1^{\text{1-L}} + \mathcal{F}_1^{\text{inst}}$ and so on with all the other ones. It is important to stress that each $\mathcal{F}_g^{\text{inst}}$ is a resummation of an infinite series of instanton contributions. Some explicit examples of how the different factors in \eqref{eq:expansion_Z_coupling} reflects on the components of the genus expansion are shown in Appendix \ref{app:interplay}. For what concerns the meaning of the $\mathcal{F}_g$, it can be shown that $a^2 \mathcal{F}_0$ is exactly the prepotential that can be obtained from the SW argument (so in flat space) of the previous Chapter for the theory we are examining, while all the other terms $a^{2-2g} \mathcal{F}_g$ account for the gravitational corrections to the prepotential coming from the fact that our spacetime is deformed by the $\Omega$ background, whose curvature affects this quantity.\\
The form of $\mathcal{F}_0$ for this theory can be inferred in a very simple way using the superconformality and the fact that we are in a rank 1 theory. This gives a very simple structure for this quantity with respect to the one we saw in the previous Chapter for the pure model (for which the prepotential can be expressed in terms of hypergeometric functions, since the integrals that define $a$ and $a_D$ in \eqref{eq:final_expression_periods_pure_model} of the previous Chapter can be expressed as hypergeometric functions). In particular, it turns out that
$$\mathcal{F}_0 = 2 \pi i \tau_{\text{R}}$$
where $\tau_{\text{R}}$ is the renormalized coupling. If our theory was not superconformal, other scales would be present and that would modify this expression; if instead our theory had higher rank, we would have, in addition to $a_1$, also $a_2$, modifying the final form of this object. We have to stress that the renormalized coupling that is appearing here is not the bare one $\tau$ that appears in the Lagrangian of $\mathcal{N}=2, \text{SU(2)}$ SQCD with $N_f=4$ and so in all the previous formulae (it is also usually denoted as UV coupling), in fact we have to add to the latter the 1-loop $\tau_{1-L}$ (which does not depend on $g_{\text{YM}}$) and the instantonic corrections $\tau_{\text{inst}}$, which is a complicated function of $\tau$. In particular, following the reasoning exposed in Appendix \ref{app:interplay}, we arrive at getting the following expression
\begin{equation} \label{eq:F0_expression_coupling}
\mathcal{F}_0 = 2 \pi i \tau_{\text{R}} = 2 \pi i \tau - \log 16 + \left( \frac{1}{2} q + \frac{13}{64} q^2 + \frac{23}{192} q^3 + \frac{2701}{32768} q^4 + \mathcal{O}(q^5)  \right)
\end{equation}
which is in agreement with (B.29) of \cite{Alday:2009aq}. From this equation we can easily identify the 1-loop and the instantonic contributions to the coupling. For this reason, even if the form of $\mathcal{F}_0$ is very simple, its expression in terms of the coupling that we are using is far from being trivial. By the way, from this result we can better understand Pestun's prescription to perform the rotation of $a$ into $i a$. Indeed, from \eqref{eq:expression_ZS4_genus}, we see that (we analyze only the contribution of $g=0$ for simplicity, but the same argument holds also for all the other terms)
$$Z_{S^4} \ni \int_{\mathbb{R}} \mathrm{d} a \, 4 a^2 e^{R^2 a^2 \, 2 \pi i (\tau - \overline{\tau})} = \int_{\mathbb{R}} \mathrm{d} a \, 4 a^2 e^{- 4 \pi R^2 a^2 \text{Im} \tau}$$
and since $\text{Im} \tau >0$, the integral is convergent. Without the previous rotation, instead, we would have had $e^{4 \pi R^2 a^2 \text{Im} \tau}$, which is not acceptable. It is interesting to say that the huge result reached by Nekrasov with his partition function, expressed as a convergent infinite series in the instanton fugacity, is that he managed to determine the SW prepotential $\mathcal{F}_0$ through a computation of gauge theory and not from geometrical arguments as done by SW and explained in the previous Chapter.\\
For what concerns instead $\mathcal{F}_1$, since it multiplies $(a R)^0$, dimensional analysis cannot be used in order to constraint its expression. By the way, exploiting the modular invariance of the partition function, one can arrive at the following analytical expression for $\mathcal{F}_1$, expressed in terms of the SW periods and the discriminant $\Delta(u)$ of the SW geometry \cite{Shapere:2008zf, Billo:2013fi, Fucito:2013fba, Nakajima:2003uh, Witten:1995gf, Moore:1997pc, Manschot:2019pog} (specialized here for $\text{SU(2)}$ SQCD with $N_f=4$)
\begin{equation} \label{eq:u_plane_formula_F1}
\mathcal{F}_1(u) = -\frac{1}{2} \log \left(\frac{\partial a}{\partial u} \right) + \frac{1}{12} \log \Delta(u)
\end{equation}
and in terms of $q$ it turns out that
\begin{equation}
\mathcal{F}_1(q) = \frac{1}{6} (3 \log a + 36 \log \gamma_G - 3 - \log 2) - \frac{3}{2} q - \frac{23}{32} q^2 - \frac{15}{32} q^3 - \frac{2839}{8192} q^4 + \mathcal{O}(q^5)
\end{equation}
All the other $\mathcal{F}_{g \geq 2}$ assume an analytical expression that derives from iteratively solving the so called \textit{holomorphic anomaly equation}, as shown in \cite{Fucito:2023txg}.\\

Let us come back to the meaning of the genus expansion \eqref{eq:genus_expansion_prep}. Since our theory is conformal, it does not depend on the value of the radius $R$ of the sphere, since it is insensitive to scale transformations. In other words, on $\mathbb{R}^4$ this model does not admit any scale, but going to the 4-sphere, which possesses a natural scale which is $R$, there must be another scale that compensates this last one and it is $a$. Practically, this means that if we set, in suitable units, $R=1$ or $R=10^6$, the theory does not change. The important implication is that that the series \eqref{eq:genus_expansion_prep} is not perturbative, but asymptotic, in the sense that in principle all the terms contribute at the same level; it is not true that higher-order terms $\mathcal{F}_{g \geq 2}$ are less important than $\mathcal{F}_0$ and $\mathcal{F}_1$.\\
The important point is the following: in defining the genus expansion, we said that $R \gg a^{-1}$, but we did not specify anything on $a$. Indeed, there is a particular limit, that will become crucial for the final results of this Chapter, under which the genus expansion becomes a perturbative one. In order to see this, let us consider the matrix element $M_{nn}$ and we assume that it admits a saddle point for large $a$. The consequence of this can be caught by simply keeping the leading order term in $a$ for $a \to \infty$ of the three ingredients entering in $Z_{\text{Nekrasov}}$ in \eqref{eq:expansion_Z_coupling}. We neglect for simplicity the instantonic part, since, as we are going to see, it does not provoke a modification of the final result and we set $R=1$ for the sake of clearness of the final quantity. It can be shown that, for $a \to \infty$, $|Z_{\text{1L}}(i a)|^2 \simeq e^{-8 a^2 \log 2} a$ and, keeping in mind that the $2 n$ derivatives bring down insertions of $a^2$, we get, at the end of the story
$$M_{nn} \simeq \int_{\mathbb{R}} \mathrm{d} a \, a^2 e^{-4 \pi \text{Im} \tau a^2 - 8 \log 2 \, \,a^2} a \, \, a^{4 n}$$
If we now apply the saddle-point method, we find that the largest part of the contribution to the integral comes from the values of $a$ for which
$$\frac{4 n + 3}{a} - (8 \pi \text{Im} \tau + 16 \log 2) \, a = 0 \Rightarrow a \simeq \sqrt{n}$$
Consequently, if we take the limit of large $n$, the initial assumption of the existence of a saddle point for large $a$ is compatible and this allows us to treat the genus expansion as a perturbative one. Since $n$ is the number of insertions of our chiral/anti-chiral CB operator and it is related to the R-charge of the final operators that are obtained by exponentiating the chiral ring generator, this limit is denoted as \textbf{large R-charge limit} (or simply large charge limit), which will be the core of the final results we are going to discuss. This also because, under this limit, there exists another method for which the extremal correlators can be computed, as we are going to see in Section \ref{sec:EFT}.\\
There is only a subtle point to stress already at this point and that we will reiterate later. As clearly shown in \cite{Grassi:2019txd}, the large charge limit is stronger than the limit of large $a$, in the sense that there are effects that derive from the first limit but not from the second one and these effects are crucial in order to make the agreement between the method we have introduced with the one we are going to recall in Section \ref{sec:EFT}.
\subsection{Non Lagrangian theories} \label{sec:non_lag_theories}
When we go to non Lagrangian theories as the AD ones, we cannot use the first approach described in the previous Subsection in the context of supersymmetric localization to compute the extremal correlators. Indeed we do not have a gauge coupling with respect to which we can derive the 4-sphere partition function. Hence one could try to adopt the genus expansion, eventually accompanied by the large charge in order to see if there is compatibility between the two expansions also for this class of theories. The important step in this direction was done in \cite{Bissi:2021rei}. In particular, the crucial observation was suggested in \cite{Russo:2014nka,Russo:2019ipg}, according to which for specific choices of the masses, the partition function on $S^4$ of $\text{SU}(N)$ gauge theories with fundamental matter is dominated at large radius by saddle points, some of which are precisely the AD points, suggesting that the large-radius physics is described by a superconformal field theory. Thanks to this, the authors of \cite{Bissi:2021rei} extracted the extremal correlators (and also their OPE coefficients) for some non Lagrangian theories, as the AD ones $\mathcal{H}_0,\mathcal{H}_1,\mathcal{H}_2$.\\
Following \cite{Gerchkovitz:2016gxx,Billo:2017glv,Billo:2019job}, the authors arrive at an expression for the extremal correlator which is similar to \eqref{eq:final_correlator}, but in the present case $G_{2n}$ is not anymore a function of $\tau, \overline{\tau}$, which are not present. The difference enters, obviously, the expression of the elements of the matrix $M$, in particular
\begin{equation} \label{eq:M_ij_non_Lag}
M_{ij} = \frac{\int_\mathbb{R} \mathrm{d} a\, O_i(a,q) \, \overline{O}_j (a, \overline{q}) |e^{\sum_{g \geq 0} R^{2-2g} a^{2-2g} \mathcal{F}_g}|^2}{\int_\mathbb{R} \mathrm{d} a\, |e^{\sum_{g \geq 0} R^{2-2g} a^{2-2g} \mathcal{F}_g}|^2}
\end{equation}
where now $q = \Lambda^{2 N - N_f}/4$, the denominator is the 4-sphere partition function, analogous to \eqref{eq:expression_ZS4_genus}, and $O_i$ is the one-point function of the chiral ring generator exponentiated at the $i$-th power (we are going to discuss it soon).\\

It turns out that, in the large radius limit, the integral \eqref{eq:M_ij_non_Lag}, for specific choices of the masses, is dominated by a saddle point corresponding to an AD conformal theory; for this reason $G_{nn}$ will therefore refer to the extremal correlator of that conformal theory. The two main difficulties are the following. The first is that the saddle point lies in the strong coupling region, requiring the non-perturbative formulae for the prepotential that resum the entire tower of instanton contributions. The notion of instantons as we analyzed in the previous Subsection is not well defined in this situation, being the model under consideration isolated, and also for this reason the partition function appearing inside the modulus squared in \eqref{eq:M_ij_non_Lag} cannot be written a la Nekrasov. This explains why we directly used the genus expansion. The second problem, related to this last aspect, is that in principle all the $\mathcal{F}_g$ are important in the computation, but also $\mathcal{F}_0$ and $\mathcal{F}_1$ are known exactly and non recursively. For this reason the authors of \cite{Bissi:2021rei} simply considered only the $\mathcal{F}_0$ and $\mathcal{F}_1$ contributions, allowing to perform the computation in an explicit way. This is, in principle, an unjustified truncation, which acquires sense only under the large charge limit; in fact the authors proved that in this situation their results are in agreement with those derived, under the same limit, from the EFT method, as we are going to recall in the following Section.\\

For what concerns the 1-point functions $O_i$, one could be puzzled for the fact that, as we explained in Section \ref{sec:OPE_extremal_correlators}, this should be 0 at the SCFT, where we are. The fact is that, as already hinted in the previous Subsection, when we go to the 4-sphere, we need something that contrasts the appearance of the scale $R$ and now it is exactly the 1-point function, which is indeed related to $a$. In order to understand this, we take as generator of the chiral ring the following operator
$$\mathcal{O}(x) = \Lambda^{d-2} \left(\frac{1}{2} \text{Tr} \phi^2 - u_*\right) = \widetilde{u}$$
with the shift chosen such that $\braket{O} = 0$ at the SCFT point $u=u_*$ and the normalization chosen such that all the dependence on the scale $\Lambda$ drops out in the extremal correlator. Furthermore, when $d=2$, we find exactly the generator we introduced in the previous Subsection. In that conformal theory, as the derivatives produce insertions of $a^2$ when we do not consider instantons, in the same way we need insertions of $a^d$ and these are obtained with the 1-point function of all the operators in the chiral ring. In fact, these last ones are obtained as powers $\mathcal{O}_i (x) = \mathcal{O}(x)^i$ of the chiral ring generator and in this way
$$O_i(u) = \braket{\mathcal{O}(x)^i} = \widetilde{u}^i = a^{d i}$$ 
In writing this expression we are not considering all the gravitational corrections to the 1-point function coming from the $\Omega$ background that deforms the original spacetime $\mathbb{R}^4$ and this is something left for the future. What we report, in the final Section of this Chapter (obtained in \cite{Cipriani:2026iws}), is a discussion that leads to formulate an educated guess for a function interpolating between two asymptotic regimes ($\epsilon \ll a $ and $\epsilon \gg a$) of the partition function, appearing in \eqref{eq:M_ij_non_Lag},  of these theories. 

\section{Results for the extremal correlator from the Effective Field Theory (EFT)} \label{sec:EFT}
As we have already hinted several times, the importance of the large charge limit is that it allows also to use tools of Effective Field Theory (EFT) in order to compute extremal correlators. More precisely, in the case in which we have a Lagrangian, it can be shown (and also inferred from what said until this point) that the mass which is acquired by the CB scalar field with the Higgs mechanism is $m \simeq \braket{\phi} g_{\text{YM}}$. Since $\braket{\phi}$ is related to $a$ which, at large $n$, scales as $\sqrt{n}$, we obtain that
$$m \simeq \sqrt{n} \, g_{\text{YM}}$$
From this we understand that the EFT description is valid only in the limit of $n \to \infty$ keeping the coupling $g_{\text{YM}}$ fixed. In a double scaling limit $n \to \infty, g_{\text{YM}} \to 0$ in such a way that $n \, g^2_{\text{YM}}$ is fixed the mass is finite and so we cannot use this alternative method.\\

Under this assumption, the authors of \cite{Hellerman:2018xpi} proved that the expression of the logarithm of extremal correlators defined above, using the so called Toda equations \cite{doi:10.1143/JPSJ.57.1901} together with some EFT computations, is
\begin{equation}\label{eq:log_corr_general}
\log(G_{2n}(\tau,\bar{\tau})) \simeq [A(\tau,\bar{\tau})+ 4 \log2] \, \,n+\tilde{B}(\tau,\bar{\tau}) + \log\left[\Gamma\left(d n + \frac{3}{2} d - \frac{1}{2} \right)\right]
\end{equation}
where $A$ and $\tilde{B}$ \footnote{We stick with the notation of ref.~\cite{Hellerman:2021yqz}: $A$ depends on how the chiral-ring generator is normalized and $\tilde{B}$ on the normalization of the unit operator in a way that is independent of the renormalization scheme adopted.} are universal functions, that we will discuss later for the case of $\mathcal{N}=2, \text{SU(2)}$ SQCD with $N_f=4$. The huge importance of this expression is that it is perturbatively exact in powers of $1/n$, meaning that it is exact in $n$ but it does not capture non-perturbative contributions (that scale as $e^{-n}$), and all the dependence on the coupling $\tau$ is present only inside the two functions $A$, $\tilde{B}$ and then also in the non-perturbative sector. In other words, except for the term linear and constant in $n$ (in addition to the non-perturbative terms), all the analytic objects in the expansion around $n = \infty$ are independent on the coupling constant.\\

Specializing on the case of $\mathcal{N}=2, \text{SU(2)}$ SQCD with $N_f = 4$ (which is the only example we are going to deeply discuss), the expansion of \eqref{eq:log_corr_general} for large $n$ produces (remembering that $d=2$)
\begin{align} \label{eq:original_exp}
 \log(G_{2n}(\tau,\bar{\tau}))&  \underset{n \to \infty}{\simeq} 2 n \log n+ 2 \log(n) + n[-2+2 \log 2 + A(\tau,\bar{\tau})+ 4 \log2] + \left[\frac{5}{2} \log 2 + \frac{\log \pi}{2} + \tilde{B}(\tau,\bar{\tau})\right] + \\
 & + \frac{47}{48 n} + \mathcal{O}\left(\frac{1}{n^2}\right) \notag
\end{align}
In \cite{Grassi:2019txd} a different method relying on matrix models \cite{Yang_Chen_1998, BASOR2001214, Basor_2011, 6200861} was employed in order to compute \eqref{eq:final_correlator} and it was proved that the perturbative part of final result (hence neglecting also the linear and constant terms in $n$) is in total agreement with \eqref{eq:original_exp}. Now we can understand what we said at the end of Section \ref{sec:Lag_theories_correlators}. Indeed in getting this agreement the authors of \cite{Grassi:2019txd} firstly performed the computation by keeping only the contribution of large $a$ and, by doing so, they do not reproduce the term going as $n^{-1}$. The final result is obtained by adding also the contribution of the small values of $a$, even if the same limit of large $n$ is considered. This is what we meant by saying that there are effects coming from the limit of large $n$ that are not obtained by simply the limit of large $a$.\\

The only incomplete part of the computation done in this reference is that no instantons were considered (and for this reason the functions $A$ and $\tilde{B}$ have not been analyzed), but they are present in the limit of large $n$ with $g_{\text{YM}}$ fixed \footnote{Instantons are exponentially suppressed under the double scaling limit we hinted before, indeed since $Z_{\text{inst}}$ is a series in $q$, we see that 
\begin{equation*}
q = e^{2 \pi i \tau} = e^{2 \pi i \text{Re}(\tau)} e^{-2 \pi \text{Im}(\tau)} \simeq e^{2 \pi i \text{Re}(\tau)} e^{-2 \pi/g^2_{\text{YM}}}
\end{equation*} and in the double scaling limit $g_{\text{YM}} \to 0$.}. The original result of \cite{Cipriani:2026iws} is verifying that instantons really do not contribute in the perturbative part of \eqref{eq:log_corr_general}, flowing only inside the $A$ and $\tilde{B}$ terms. The following Sections are devoted to show it. We are not going to consider the non-perturbative terms, which instead have been studied for this theory in \cite{Hellerman:2021duh}.\\
A recent quite related study on extremal and integrated correlators for $\mathcal{N}=2$ SQCD and $\mathcal{N}=4$ SYM with gauge group SU$(3)$ is \cite{Grassi:2024bwl} and another reference which focuses on the relations between large charge limit and properties of $\mathcal{N}=4$ SYM is \cite{Brown:2025cbz}. \\

For what concerns AD theories, it is not known if \eqref{eq:log_corr_general} is still valid. Obviously, $A$ and $\tilde{B}$ would not be functions of the coupling anymore and, as checked in \cite{Grassi:2019txd} for the $\mathcal{H}_0$ theory and as it can be repeated also for $\mathcal{H}_1$ and $\mathcal{H}_2$, in the large charge expansion of the correlator the terms that multiply $n \log n$ and $\log n$ are correctly reproduced as deriving from \eqref{eq:log_corr_general}, but what happens further is nowadays not known. At the end of this Chapter (following \cite{Cipriani:2026iws}), we are going to present what we can say for these theories under the limit in consideration.

\section{Analysis up to 2 instantons}\label{sec:chap2_6}
In this Section, we will scrutinize how exactly gauge-theory instantons affect the two-point function $G_{2n}$ by performing a fully general analysis up to 2 instantons. Once that it is clear what happens, we mode to 4 instantons.  After introducing the relevant quantities, we first analyse the polynomial terms in the instanton partition function in Subsection \ref{subsec:chap2_6_1} and then we go over the more intricate case of the non-polynomial remainder in Subsection \ref{subsec:chap2_6_2}. \\
Out of simplicity, in all the subsequent passages we denote by $a$ simply the dimensionless quantity $a R$ (or $a/\epsilon$).\\

We recall the instanton partition function up to 2 instantons, already presented in \eqref{eq:Z_inst_2_inst} and that we rewrite here for convenience
\begin{equation}
    Z_{\text{inst}} (a,\tau) = 1 + \frac{q}{2} (a^2-3) + \frac{q^2}{4} \frac{8 a^8+a^6-91 a^4-60 a^2+132}{(4 a^2+9)^2} + \mathcal{O}(q^3)
\end{equation}
Except the 1 instanton part which is polynomial in $a$, all the other components of $Z_{\text{inst}}$ are rational functions of $a$ and hence they admit a non trivial expansion for $a \gg 1$: for the $k$-th instanton contribution, it is a series in $a^{2(k-p)}$, with $p=0,1,\dots$. From what it has been shown in Appendix \ref{app:interplay}, the $\mathcal{F}_0$ and $\mathcal{F}_1$ components in the genus expansion of the prepotential of the instanton sector correspond to keep only the terms with non negative powers in $a$ (hence $p$ can assume only the values from 0 to $k$). All the terms with $p \geq k+1$ flow to $\mathcal{F}_{g \geq 2}$ components.\\
In view of this, it is convenient to isolate the polynomial part \footnote{By "polynomial part" we simply mean all terms with non-negative powers of $a$ in the Taylor expansion of the instanton partition function around $a=\infty$. At the level of the free energy, as already said, this corresponds to isolating the instanton contributions to $\mathcal{F}_0$ and $\mathcal{F}_1$} in $a$ and write to write the 2-instantons partition function as
\begin{equation}
    Z_{2-\text{inst}}(a,\tau) = \frac{q^2}{4}\Bigl(\frac{a^4}{2}-\frac{35}{16} a^2 + \frac{13}{8}\Bigr) + \frac{q^2}{4} \Bigl(\frac{3}{64 (4 a^2+9)}-\frac{3}{64 (4 a^2+9)^2}\Bigr)
\end{equation}
in such a way that
\begin{equation}
    Z_{\text{inst}}(a,\tau) = \underbrace{1 + \frac{q}{2}(a^2-3) + \frac{q^2}{4}\Bigl(\frac{a^4}{2}-\frac{35}{16} a^2 + \frac{13}{8}\Bigr)}_{Z_{\text{pol}}(a,\tau)} + \underbrace{\frac{q^2}{4} \frac{3(a^2+2)}{16 (4 a^2+9)^2}}_{Z^{(2)}_{\text{rest}}(a,\tau)} + \mathcal{O}(q^3)
\end{equation}
Taking derivatives in order to construct the matrix $M^{(n)}_{ij}$, we get

\begin{equation} \label{derivatives_Zinst}
\begin{aligned}
    &\partial_\tau^i \partial_{\bar{\tau}}^j \left[e^{-4\pi \text{Im}\tau a^2} |Z_{\text{inst}}|^2\right] = (-1)^j(2\pi i)^{i+j} e^{-4\pi \text{Im}\tau a^2}  \Bigl[a^{2i} + \frac{q (a^2+1)^i}{2} \, (a^2-3) + \Bigr. \\ & \Bigl. + \frac{q^2 (a^2+2)^i}{4} \, \Bigl(\frac{a^4}{2}-\frac{35}{16} a^2 + \frac{13}{8}\Bigr) +  (a^2+2)^i Z^{(2)}_{\text{rest}}(a,\tau)\Bigr] (i \to j, q \to \bar{q})
\end{aligned}
\end{equation}
In this way the matrix elements are
\begin{equation}\label{eq:M_ij}
   M^{(n)}_{i j}  = \frac{1}{Z_{S^4}}  \int_\mathbb{R} \, \mathrm{d}a \, 4 a^2 \, \left|Z_{\text{1L}}(i a)\right|^2 \, \, \eqref{derivatives_Zinst}
\end{equation}
Now we use a fundamental identity, called \textit{Andreièf identity}, which relates the determinant of a matrix whose elements are integrals to a multivariate integral over determinants. Let us quickly recall it. Given two sets of $n+1$ functions $\{f_k(y)\}_{k=0}^{n}, \{g_k(y)\}_{k=0}^{n}$ and a certain measure $\mathrm{d} \mu(y)$, then
\begin{equation}
    \int \prod_{j=0}^n \mathrm{d} \mu(y_j) \, \,\underset{a \, b}{\det} (f_a (y_b)) \, \,\underset{c \, d}{\det} (g_c (y_d)) = (n+1)! \, \, \underset{a \, b}{\det} \int \mathrm{d}\mu(y) f_a(y) g_b(y)
\end{equation}
Applying this identity to the determinant of \eqref{eq:M_ij}, we make the following identifications: $a \to y_k$, \\
$\mathrm{d}a \, 4 a^2 \, \left|Z_{\text{1L}}(i a)\right|^2 \, e^{-4 \pi \text{Im}\tau \, a^2} \to \mathrm{d}\mu(y_k)$, the first squared bracket of \eqref{derivatives_Zinst} with $a$ replaced by $y_l$ (denoted in the following as eigenvalues of the random matrices underlying the method we are following) corresponds to the set of functions $\{f_i(y_l)\}_{i=0}^{n}$ and the remaining term corresponds to $\{g_j(y_m)\}_{j=0}^{n}$.
In this way we get
\begin{align}\label{eq:first_expr_det}
    \det M^{(n)} = \frac{(2\pi)^{n(n+1)}}{Z_{S^4}^{n+1} \, (n+1)!} \int_{\mathbb{R}^{n+1}} & \prod_{k=0}^n \mathrm{d}y_k \, 4y^2_k \, e^{-4 \pi \text{Im}\tau \, y^2_k} \, \left|Z_{\text{1L}}(i y_k)\right|^2 \underset{i,l=0,\dots,n}{\det}((\sum_{s=0}^2 \mathcal{Z}^{(s)} + \mathcal{H}^{(2)})_{i,l}) \times \notag \\
  & \times \underset{j,m=0,\dots,n}{\det}((\sum_{s=0}^2 \overline{\mathcal{Z}}^{(s)} + \overline{\mathcal{H}}^{(2)})_{j,m})
\end{align}
where the matrices $\mathcal{Z}^{(s)}$ and $\mathcal{H}^{(2)}$ are defined as follows
\begin{align}\label{A-matrices}
   & \mathcal{Z}^{(0)}_{i,l} = (y^2_l)^i \qquad , \qquad \mathcal{Z}^{(1)}_{i,l} = \frac{q}{2}(y^2_l+1)^i (y^2_l-3) \qquad , \qquad \mathcal{Z}^{(2)}_{i,l} = \frac{q^2 (y^2_l+2)^i}{4}\Bigl(\frac{y^4_l}{2}-\frac{35}{16} y^2_l + \frac{13}{8}\Bigr) ,\notag \\
   & \mathcal{H}^{(2)}_{i,l} = (y^2_l+2)^i Z^{(2)}_{\text{rest}}(y_l,\tau) = \frac{q^2 (y^2_l+2)^i}{4} \frac{3(y_l^2+2)}{16 (4 y_l^2+9)^2}
\end{align}
(the matrices $\overline{\mathcal{Z}}^{(s)}$ contain $\bar{q}$ in place of $q$).
$\mathcal{Z}^{(0)}$ is a Vandermonde matrix, hence its determinant is known and this implies that
\begin{align}
\label{eq:detMn_2inst}
    \det M^{(n)} = & \frac{(2\pi)^{n(n+1)}}{Z_{S^4}^{n+1} \, (n+1)!} \int_{\mathbb{R}^{n+1}} \prod_{k=0}^n \mathrm{d}y_k \, 4y^2_k \, e^{-4 \pi \text{Im}\tau \, y^2_k} \, \left|Z_{\text{1L}}(i y_k)\right|^2 \prod_{p<s} (y^2_p-y^2_s)^2 \times \notag \\
    & \times \underset{i,l=0,\dots,n}{\det}(\mathbb{I}+\sum_{s=1}^2 \mathcal{K}^{(s)}+(\mathcal{Z}^{(0)})^{-1} \mathcal{H}^{(2)})_{i,l} \underset{j,m=0,\dots,n}{\det}(\mathbb{I}+\sum_{s=1}^2 \overline{\mathcal{K}}^{(s)}+(\overline{\mathcal{Z}^{(0)}})^{-1} \overline{\mathcal{H}}^{(2)})_{j,m}
\end{align}
where the matrices $\mathcal{K}^{(s)}$ are defined as $\mathcal{K}^{(s)} = (\mathcal{Z}^{(0)})^{-1} \mathcal{Z}^{(s)}$.\\
Now we focus on one of the two determinants inside the integral. Expanding up to order $q^2$ in the computation of the determinant and thanks to the additivity of the trace, the following structure comes out
\begin{align} \label{eq:2inst_small_eigen}
    \det\Bigl[\mathbb{I}+\sum_{s=1}^2 \mathcal{K}^{(s)}+(\mathcal{Z}^{(0)})^{-1} \mathcal{H}^{(2)}\Bigr] &=  1 + \text{Tr}(\mathcal{K}^{(1)}) + \text{Tr}\Bigl[\mathcal{K}^{(2)}+(\mathcal{Z}^{(0)})^{-1} \mathcal{H}^{(2)}\Bigr] - \frac{1}{2} \text{Tr}(\mathcal{K}^{(1)^2}) + \frac{1}{2} \text{Tr}(\mathcal{K}^{(1)})^2 =\notag \\
    &= 1 + \underbrace{\text{Tr}(\mathcal{K}^{(1)})}_{\mathcal{O}(q)} + \underbrace{\text{Tr}(\mathcal{K}^{(2)}) - \frac{1}{2} \text{Tr}(\mathcal{K}^{(1)^2}) + \frac{1}{2} \text{Tr}(\mathcal{K}^{(1)})^2}_{\mathcal{O}(q^2)_{\text{pol}}} + \underbrace{\text{Tr}\Bigl[(\mathcal{Z}^{(0)})^{-1} \mathcal{H}^{(2)}\Bigr]}_{\mathcal{O}(q^2)_{\text{rest}}}
\end{align}
where the three contributions $\mathcal{O}(q^2)_{\text{pol}}$ derive from the polynomial part in $a$ of the instanton partition function, while the last one $\mathcal{O}(q^2)_{\text{rest}}$ from the remainder. We want to analyze the results deriving from these two kinds of contribution separately (it is possible thanks to the expansions we are going to consider in order to remain at order $q^2$, when taking the logarithm of the correlator).

\subsection{Polynomial part} \label{subsec:chap2_6_1}
It can be checked that the instantonic contributions coming from the polynomial part are 
\begin{align} 
    & \mathcal{O}(q) = \frac{q}{2} \Bigl(\frac{(n+1)(n-6)}{2} + \sum_{\alpha=0}^n y^2_\alpha\Bigr) \label{eq:1_inst_contr_traces}\\
    & \mathcal{O}(q^2)_{\text{pol}} = \frac{q^2}{8} \Bigl[\frac{2 n^4 - 20 n^3 + 35 n^2 + 83 n + 26}{8} + \Bigl(\frac{13}{8} + (n-6)(n+1)\Bigr)\sum_{\alpha=0}^n y^2_\alpha + \Bigl(\sum_{\alpha=0}^n y^2_\alpha\Bigr)^2 \Bigr]\label{eq:2_inst_contr_traces}
\end{align}
Performing the change of variables $x_i = y^2_i$ \footnote{Keep in mind that the integrands in terms of $y_i$ are even in these eigenvalues, hence each integral, that goes from $-\infty$ to $\infty$, is twice the integral from $0$ to $\infty$. For this reason the factor 4, repeated $n+1$ times, does not change} and expanding the products of the two determinants inside the integrals in \eqref{eq:detMn_2inst} up to $\mathcal{O}(q^2)$, the determinant of $M^{(n)}$ turns out to be
\begin{align}\label{eq:det_2inst__Mn}
    \det M^{(n)} = & \frac{4^{n+1} \, (2 \pi)^{n(n+1)}}{Z_{S^4}^{n+1}} \left\{X_n^{(0)} + \frac{q+\bar{q}}{2} \left(\frac{(n+1)(n-6)}{2} X_n^{(0)} + X_n^{(1)}\right) + \right. \notag \\
    & +\frac{(q+\bar{q})^2}{8} \left[ \frac{2n^4-20 n^3+35 n^2+83 n+26}{8} X_n^{(0)} + \left(\frac{13}{8} + (n-6)(n+1)\right) X_n^{(1)} + X_n^{(2)} \right]+ \notag \\
    & \left. + \frac{q \bar{q}}{4} \left(\frac{46+37 n-9 n^2}{8} X_n^{(0)} - \frac{13}{8} X_n^{(1)} \right) \right\}
\end{align}
where
\begin{equation}
X_n^{(p)} = \frac{1}{(n+1)!} \int_{\mathbb{R}^{n+1}_+} \prod_{k=0}^n \mathrm{d}x_k \, x^\frac{1}{2}_k \, e^{-4 \pi \text{Im}\tau \, x_k} \, \left|Z_{\text{1L}}(i \sqrt{x_k})\right|^2 \prod_{p<s}(x_p-x_s)^2 \left(\sum_{\alpha=0}^n x_\alpha\right)^p  \label{eq:X_n}\end{equation}
Before studying and evaluating the quantities $X_n^{(0)},X_n^{(1)},X_n^{(2)}$, recall that we are ultimately interested in the logarithm of the correlator. Therefore, taking the the logarithm of the determinant of the matrix $M^{(n)}$ \eqref{eq:det_2inst__Mn} and further expanding to $\mathcal{O}(q^2)$, we obtain
\begin{align}
\label{eq:log_detMn_2inst}
    \log \det M^{(n)} & = \log \left[\frac{4^{n+1} \, (2 \pi)^{n(n+1)}}{Z_{S^4}^{n+1}} \right] + \log \, X_n^{(0)} + \frac{q+\bar{q}}{2} \left(\frac{(n+1)(n-6)}{2} +  \frac{X_n^{(1)}}{X_n^{(0)}}\right) + \notag \\
    & + \frac{(q+\bar{q})^2}{8} \left[ \frac{2n^4-20 n^3+35 n^2+83 n+26}{8} + \left(\frac{13}{8} + (n-6)(n+1)\right) \frac{X_n^{(1)}}{X_n^{(0)}} + \frac{X_n^{(2)}}{X_n^{(0)}} \right] + \notag \\
    & + \frac{q \bar{q}}{4} \left(\frac{46+37 n-9 n^2}{8} - \frac{13}{8} \frac{X_n^{(1)}}{X_n^{(0)}} \right) - \frac{1}{2} \frac{(q+\overline{q})^2}{4} \left(\frac{(n+1)(n-6)}{2} +  \frac{X_n^{(1)}}{X_n^{(0)}}\right)^2 
\end{align}
At this point the logarithm of the correlator is the difference of $\log \det M^{(n)}$ and $\log \det M^{(n-1)}$ (adding $4 n \log2 = \log (16^n)$), hence it is time to study and evaluate the multivariate integral \eqref{eq:X_n} for a generic (non-negative) integer value of $p$.\\
For $p=0$ this quantity \eqref{eq:X_n} has already been studied in detail in \cite{Grassi:2019txd}, which we are going to quickly review in order to understand how to cope with $p=1,2$. The multivariate integral $X_n^{(0)}$ is equal to
\begin{equation}\label{eq:Xn_x}
    X_n^{(0)} = \frac{e^{12 (n+1) \log \gamma_G - (n+1) - \frac{(n+1)}{3} \log 2}}{(n+1)!} \int_{\mathbb{R}^{n+1}_+} \prod_{k=0}^n \mathrm{d}x_k \,  e^{-4 \pi \text{Im} \tau_R\, x_k} \, x^\frac{1}{2}_k \, (x_k+1)^\frac{1}{2} \, U(x_k) \prod_{p<s}(x_p-x_s)^2
\end{equation}
where $\text{Im}\tau_R = \text{Im}\tau + \frac{2 \log 2}{\pi} \equiv \frac{t}{4 \pi}$, $\gamma_G$ is the Gleisher constant and the function $U(x)$ is defined as
\begin{equation}
    U(x) = \frac{e^{8 x \, \log 2} \left|Z_{\text{1L}}(i \sqrt{x_k})\right|^2}{(x_k+1)^{\frac{1}{2}}} e^{-12 \log \gamma_G + 1 + \frac{\log 2}{3}}
\end{equation}
which is bounded and it tends to 1 for $x \to \infty$. In order to proceed, we apply the change of variables $z_k = t x_k$ in such a way that
\begin{equation}\label{eq:Xn_z}
     X_n^{(0)} = \frac{e^{12 (n+1) \log \gamma_G - (n+1) - \frac{(n+1)}{3} \log 2}}{t^{(n+1)(n+2)} \, (n+1)!} \int_{\mathbb{R}^{n+1}_+} \prod_{k=0}^n \mathrm{d}z_k \,  e^{-z_k} \, z^\frac{1}{2}_k \, (z_k+t)^\frac{1}{2} \, U(z_k;t) \prod_{p<s}(z_p-z_s)^2
\end{equation}
In \cite{Grassi:2019txd} the multivariate integral written here is studied using the Coulomb gas formalism, hence what we are going to expose is valid only at the \textbf{perturbative} level. The relevant feature to stress now is that this quantity is independent on $t$ (always at the perturbative level), indeed it turns out that (details can be seen in the Appendix A of \cite{Grassi:2019txd}) 
\begin{equation}\label{eq:final_expr_Xn}
    X_n^{(0)} = \frac{e^{12 N \log \gamma_G - N - \frac{N}{3} \log 2}}{t^{(n+1)(n+2)}} e^{N^2 \log N - \frac{3}{2} N^2 + N \log N + (\log(2 \pi)-1) N + \frac{7}{48} \log N + \mathcal{O}(1)} 
\end{equation}
with $N=n+1$. So, at the perturbative level, $X_n$ depends on $t$ only as $t^{-(n+1)(n+2)}$.\\
Regarding the integral $X_n^{(p)}$ for $p>0$, we can follow the same steps that led to \eqref{eq:Xn_x} for $X_n^{(0)}$. This implies that \footnote{I thank Guillaume Dubach for the suggestion at the second line.}
\begin{align}\label{eq:Yn_x}
    X_n^{(p)} & = \frac{e^{12 (n+1) \log \gamma_G - (n+1) - \frac{(n+1)}{3} \log 2}}{(n+1)!} \int_{\mathbb{R}^{n+1}_+} \prod_{k=0}^n \mathrm{d}x_k \,  e^{-t\, x_k} \, x^\frac{1}{2}_k \, (x_k+1)^\frac{1}{2} \, U(x_k) \prod_{p<s}(x_p-x_s)^2 \left(\sum_{\alpha=0}^n x_\alpha\right)^p = \notag \\
    & = (-1)^p \frac{e^{12 (n+1) \log \gamma_G - (n+1) - \frac{(n+1)}{3} \log 2}}{(n+1)!} \int_{\mathbb{R}^{n+1}_+} \partial_t^p \left[ \prod_{k=0}^n \mathrm{d}x_k \,  e^{-t\, x_k} \, x^\frac{1}{2}_k \, (x_k+1)^\frac{1}{2} \, U(x_k) \prod_{p<s}(x_p-x_s)^2 \right]
\end{align}
The derivative can be brought outside the integral thanks to the theorem of differentiation under the integral sign. In this way
\begin{equation} \label{eq:general_p_Xn}
    X_n^{(p)} = (-1)^p \, \partial_t^p X_n^{(0)}
\end{equation}
Now, if we simply use the perturbative expression for $X_n^{(0)}$ \eqref{eq:final_expr_Xn} \footnote{We are assuming that only the perturbative part of $X_n^{(0)}$ determines the perturbative expression of $X_n^{(p)}$ for $p>0$; in other words the derivative with respect to $t$ does not transform a non-perturbative term in $n$ of $X_n^{(0)}$ in a perturbative one for $X_n^{(p)}$. This assumption is supported a posteriori from the results that will be obtained later on.}, we find (restricting here to $p=1,2$ and coming back to the expression for generic $p$ at the end of Subsection \ref{subsec:chap2_7_1})
\begin{equation}
\label{eq:Yn_explicit_pert}
    X_n^{(1)} = \frac{(n+1)(n+2)}{t} X_n^{(0)} , \qquad X_n^{(2)} = \left[\frac{(n+1)(n+2)}{t^2} + \frac{(n+1)^2 (n+2)^2}{t^2}\right] X_n^{(0)}
\end{equation}
Obviously, in $X_n^{(0)}$ there are also non perturbative contributions in $n$ that have not been examined in \cite{Grassi:2019txd}, differently from \cite{Hellerman:2021duh}; for this reason we see that the instanton we are considering gives a certain contribution also at the non-perturbative level.\\
We stress that the numerator of the first term of $X_n^{(2)}$ comes from the first derivative of $X_n^{(0)}$, the numerator of the second term from the second derivative. The role of this quartic term in $n$, which should not be present at the end for $\log \det M^{(n)}$ \footnote{It would give rise to terms proportional to $n^2$ in the logarithm of the correlator, which are not expected from the EFT prediction in \eqref{eq:original_exp}}, is indeed the one of cancelling the quartic term coming from $- \frac{1}{2} \frac{(q+\overline{q})^2}{4} \left(\frac{(n+1)(n-6)}{2} +  \frac{X_n^{(1)}}{X_n^{(0)}}\right)^2$ in the expansion of the logarithm of $\det M^{(n)}$ in \eqref{eq:log_detMn_2inst}. Indeed, replacing \eqref{eq:Yn_explicit_pert} in \eqref{eq:log_detMn_2inst}, we obtain (always at the perturbative level)
\begin{align}
    \log \det M^{(n)} = & \log \left[\frac{4^{n+1} (2 \pi)^{n(n+1)}}{Z_{S^4}^{n+1}} \right] + \log \, X_n^{(0)} + \frac{q+\bar{q}}{2} \left(\frac{(n+1)(n-6)}{2} + \frac{(n+1)(n+2)}{t}\right) + \notag \\
    & + \frac{(q+\bar{q})^2}{8} \left[\frac{2n^4-20 n^3+35 n^2+83 n+26}{8} + \left(\frac{13}{8}+ (n-6)(n+1)\right) \frac{(n+1)(n+2)}{t} + \right. \notag \\
    & \left. + \frac{\textcolor{red}{(n+1)^2 (n+2)^2}+(n+1)(n+2)}{t^2} \right] + \frac{q \bar{q}}{4} \left(\frac{46+37 n-9n^2}{8} - \frac{13}{8} \frac{(n+1)(n+2)}{t}\right) + \notag \\
    & - \frac{1}{2} \frac{(q+\bar{q})^2}{8} \left(\frac{(n+1)(n-6)}{2} + \frac{\textcolor{blue}{(n+1)(n+2)}}{t}\right)^2
\end{align}
As it can be seen from a direct computation, the quartic term of before $(n+1)^2 (n+2)^2$ (and highlighted in red) cancels with the one coming from the square in the last line (look at the object highlighted in blue). In the end we find
\begin{align} \label{eq:final_log_detMn}
    \log \det M^{(n)} = & \log \left[\frac{4^{n+1} (2 \pi)^{n(n+1)}}{Z_{S^4}^{n+1}} \right] + \log \, X_n^{(0)} + \frac{q+\bar{q}}{2} \left(\frac{(n+1)(n-6)}{2} + \frac{(n+1)(n+2)}{t}\right) + \notag \\
    & + \frac{q^2+\bar{q}^2}{8} \left(\frac{(9n-46)(n+1)}{8} + \frac{13 (n+1)(n+2)}{8 t}\right) +\frac{(q+\bar{q})^2}{8} \frac{(n+1)(n+2)}{t^2}
\end{align}
Going to the logarithm of the correlator, we get \footnote{To facilitate comparison to the existing literature, we find it convenient to give the instanton expansion of the sum $\log G_{2n}(\tau,\bar{\tau})+\log Z_{S^4}(\tau,\bar{\tau})$, thus removing the $n$-independent contribution of the $S^4$ partition function. In this way, the coefficient of $n^0$ will be $B(\tau, \bar{\tau})$ and not $\tilde{B}(\tau,\bar{\tau})$, as it will be clear in the following Section.}
\begin{empheq}[box=\fbox]{align} \label{eq:final_expression_2inst_large_vev}
   \log (Z_{S^4}(\tau, \bar{\tau}) \, G_{2n}(\tau,\bar{\tau})) & =  \log\left(\frac{X_n^{(0)}}{X_{n-1}^{(0)}}\right) + \log \left(4 (2 \pi)^{2 n}\right) + 4 n \log 2 + \left[\frac{q+\overline{q}}{2} (n-3) + \right.\notag \\
   & \left. + \frac{q^2+\overline{q}^2}{8}\left(\frac{9 n - 23}{4} + \frac{13}{4} \frac{n+1}{t}\right) \right] + 2 (n+1) \left(\frac{q + \overline{q}}{2 t} + \frac{1}{2} \frac{(q + \overline{q})^2}{4 t^2}\right)
\end{empheq}
The 0-instanton sector is exactly the same as in \cite{Grassi:2019txd} and with this final result we conclude that, at least from the polynomial part in $a$ of the partition function, instantons only affect the coefficients $A(\tau,\bar{\tau})$ and $\tilde{B}(\tau,\bar{\tau})$ of the perturbative large-$n$ expansion, i.e. the terms of order $n^1$ and $n^0$, respectively \footnote{We remark that the non-perturbative terms, which we are neglecting throughout the present study, are expected to receive instanton corrections, affecting their $\tau$ dependence}.\\
Before proceeding to the contribution of $\mathcal{O}(q^2)_{\text{rest}}$ in \eqref{eq:2inst_small_eigen}, it is interesting to notice that we can rewrite \eqref{eq:log_detMn_2inst} in terms of universal quantities, that are the \textbf{cumulants}. This gives an elegant explanation of the elimination of all the terms that have powers in $n$ larger than 2 in the logarithm of $\det M^{(n)}$, that we saw before. We will see in Subsection \ref{subsec:chap2_7_1} that this simple structure persists beyond 2 instantons. \\
In \eqref{eq:log_detMn_2inst} the following ratios appear
\begin{align}
& \frac{X_n^{(1)}}{X_n^{(0)}} = \frac{\int_{\mathbb{R}^{n+1}_+} \prod_{k=0}^n \mathrm{d}x_k \, x^\frac{1}{2}_k \, e^{-4 \pi \text{Im}\tau \, x_k} \, Z_{\text{1L}}(i \sqrt{x_k}) \prod_{p<s}(x_p-x_s)^2 \left(\sum_{\alpha=0}^n x_\alpha\right)}{\int_{\mathbb{R}^{n+1}_+} \prod_{k=0}^n \mathrm{d}x_k \, x^\frac{1}{2}_k \, e^{-4 \pi \text{Im}\tau \, x_k} \, Z_{\text{1L}}(i \sqrt{x_k}) \prod_{p<s}(x_p-x_s)^2} \equiv \langle \text{Tr}\chi \rangle \\
& \frac{X_n^{(2)}}{X_n^{(0)}} = \frac{\int_{\mathbb{R}^{n+1}_+} \prod_{k=0}^n \mathrm{d}x_k \, x^\frac{1}{2}_k \, e^{-4 \pi \text{Im}\tau \, x_k} \, Z_{\text{1L}}(i \sqrt{x_k}) \prod_{p<s}(x_p-x_s)^2 \left(\sum_{\alpha=0}^n x_\alpha\right)^2}{\int_{\mathbb{R}^{n+1}_+} \prod_{k=0}^n \mathrm{d}x_k \, x^\frac{1}{2}_k \, e^{-4 \pi \text{Im}\tau \, x_k} \, Z_{\text{1L}}(i \sqrt{x_k}) \prod_{p<s}(x_p-x_s)^2} \equiv \langle (\text{Tr} \chi)^2 \rangle 
\end{align}
where we indicate as $\chi$ the matrix whose eigenvalues are the $x_i$'s. These are the moments of $(\text{Tr} \chi)^p$ with respect to the matrix ensemble defined by $X_n^{(0)}$ \footnote{Here we explicitly considered $p=1,2$, but the same is valid in terms of generic non-negative integer $p$, i.e. $\langle (\text{Tr} \chi)^p\rangle = X_n^{(p)}/X_n^{(0)}$. \label{footnote:general_moment}}. The $p$-point cumulants $\langle (\text{Tr} \chi)^p \rangle_c$ are those core quantities in which $\langle (\text{Tr} \chi)^p \rangle$ decomposes according to the partitions of the power $p$ \cite{Eynard:2015aea}. Up to 2 instantons we get \footnote{We are considering the case in which the insertions of the cumulants and of the moments are equal}
\begin{align}
    & \langle \text{Tr}\chi \rangle = \langle \text{Tr}\chi \rangle_c \notag \\ 
    & \langle (\text{Tr}\chi)^2 \rangle = \langle (\text{Tr}\chi)^2 \rangle_c + \langle \text{Tr}\chi \rangle_c^2 
\end{align}
By inverting these relations we can find the cumulants in terms of the moments
\begin{align}
    & \langle \text{Tr}\chi \rangle_c = \langle \text{Tr}\chi \rangle \notag \\ 
    & \langle (\text{Tr}\chi)^2 \rangle_c = \langle (\text{Tr}\chi)^2 \rangle - \langle \text{Tr}\chi \rangle^2
\end{align}
Using the results shown above at the perturbative level \eqref{eq:Yn_explicit_pert}, we can obtain the following expressions for the moments
\begin{align}
    & \langle \text{Tr}\chi\rangle = \frac{(n+1) (n+2)}{t} \notag \\
    & \langle (\text{Tr}\chi)^2\rangle = \frac{(n+1)(n+2)[(n+1)(n+2)+1]}{t^2}
\end{align}
while the cumulants are
\begin{align}
    & \langle \text{Tr}\chi\rangle_c = \langle \text{Tr}\chi\rangle = \frac{(n+1) (n+2)}{t} \notag \\
    & \langle (\text{Tr}\chi)^2\rangle_c = \frac{(n+1)(n+2)[(n+1)(n+2)+1]-(n+1)^2 (n+2)^2}{t^2} = \frac{(n+1)(n+2)}{t^2} = \frac{\langle \text{Tr}\chi\rangle_c}{t} \notag 
\end{align}
As we can see, the moment $\langle (\text{Tr}\chi)^p\rangle$ is a function of $n$ of order $2 p$ (as it can be obtained from \eqref{eq:general_p_Xn} together with \eqref{eq:final_expr_Xn}. By the way, we are going to come back on it more explicitly at the end of Subsection \ref{subsec:chap2_7_1}), while the cumulants are always quadratic in $n$. This is important because this implies that the polynomial part of the instanton partition function does not contribute to the perturbative structure of the correlator. This is due to the fact that the term quadratic in $n$ in $\log \det M^{(n)}$ flows into $A$ of $\log G_{2n}$, the one linear in $n$ flows in $\tilde{B}$ and the constant one cancels, as already seen above.\\
For this reason, it is important that $\log \det M^{(n)}$ is expressed in terms only of cumulants. In particular, \eqref{eq:log_detMn_2inst} becomes
\begin{align} \label{eq:intermediate_log_detMn_cum}
    \log \det M^{(n)} & = \log \left[\frac{4^{n+1} (2 \pi)^{n(n+1)}}{Z_{S^4}^{n+1}}\right] + \log \, X_n^{(0)} + \frac{q+\bar{q}}{2} \left(\frac{(n+1)(n-6)}{2} +  \langle \text{Tr}\chi \rangle_c \right) + \notag \\
    & + \frac{(q+\bar{q})^2}{8} \left[ \frac{2n^4-20 n^3+35 n^2+83 n+26}{8} + \left(\frac{13}{8} + (n-6)(n+1)\right) \langle \text{Tr}\chi \rangle_c + \langle (\text{Tr}\chi)^2 \rangle \right] + \notag \\
    & + \frac{q \bar{q}}{4} \left(\frac{46+37 n-9 n^2}{8} - \frac{13}{8} \langle \text{Tr}\chi \rangle_c \right) - \frac{1}{2} \frac{(q+\overline{q})^2}{4} \left(\frac{(n+1)(n-6)}{2} +  \langle \text{Tr}\chi \rangle_c\right)^2 
\end{align}
but we can see that $\frac{(q+\bar{q})^2}{8}\langle (\text{Tr}\chi)^2 \rangle$ combines with $-\frac{(q+\bar{q})^2}{8}\langle \text{Tr}\chi \rangle^2$ in order to give rise to the cumulant, which is quadratic in $n$. This kind of cancellation, which produces only cumulants in $\log \det M^{(n)}$, is the one we discussed before for the cancellation of all terms in $n$ of order larger than 2.\\
In few words, due to the fact that the moments appear only through their cumulants inside $\log \det M^{(n)}$, all terms containing more than 2 powers of $n$ drop out, in accordance with the EFT prediction. In Section \ref{sec:chap2_7}, by going beyond 2 instantons, we will see that the moment $\langle (\text{Tr}\chi)^p \rangle$ is a polynomial in $n$ of degree $2 p$, as already said before, whereas the corresponding cumulant $\langle (\text{Tr}\chi)^p \rangle_c$ is only quadratic in $n$, independently of $p$. At the level of the present, order-by-order analysis, this appears as a miracle. But we will see in the all-order argument of Subsection \ref{subsec:chap2_7_3} that is has a clean geometric origin.\\
At this point we can write \eqref{eq:intermediate_log_detMn_cum} in the following way
\begin{empheq}[box=\fbox]{align} \label{eq:final_logdetMn_2inst}
\log \det M^{(n)} = &\log\left[\frac{4^{n+1} (2 \pi)^{n(n+1)}}{Z_{S^4}^{n+1}}\right] + \log\, X_n^{(0)} + \left[\frac{q+\overline{q}}{2} \frac{(n-6)(n+1)}{2} + \right. \notag \\
& \left. + \frac{q^2+\overline{q}^2}{8}\left(\frac{(n+1)(9 n - 46)}{8} + \frac{13}{8} \frac{\langle \text{Tr}\zeta\rangle_c}{t}\right) \right] + \sum_{k=1}^2 \frac{1}{k!} \left(\frac{q+\overline{q}}{2 t}\right)^k \langle (\text{Tr}\zeta)^k \rangle_c
\end{empheq}
where we introduce the matrix $\zeta$ in such a way that
\begin{equation}
   \langle (\text{Tr}\chi)^k \rangle_c =\frac{\langle (\text{Tr}\zeta)^k \rangle_c}{t^k}
\end{equation}

\subsection{Non-polynomial part} \label{subsec:chap2_6_2}
We can now move on to examine the contribution of $Z^{(2)}_{\text{rest}}$, which we recall to be, after the application of the Andrèief identity
\begin{equation}
   \text{Tr}\Bigl[(\mathcal{Z}^{(0)})^{-1} \mathcal{H}^{(2)}\Bigr]
\end{equation}
with the matrices $\mathcal{Z}^{(0)}$ and $\mathcal{H}^{(2)}$ defined in \eqref{A-matrices}. Notice that we can study this contribution separately with respect to the others because, thanks to the linearity of the multivariate integrals and then to the expansion of the logarithm, this gives terms that sum to the result for $\log \det M^{(n)}$ found in the previous Subsection, as \eqref{eq:final_log_detMn}.\\
We notice that the elements of the matrix $\mathcal{H}^{(2)}$ in \eqref{A-matrices} are analytical in $y_l \in [0,+\infty), \forall l=0,\dots,n$. For this reason $1/(9+4 y^2_l)^2$ can be seen as a sum of a geometric series
\begin{align}
\mathcal{H}^{(2)}_{i,l} & = \frac{q^2 (y^2_l+2)^i}{4} \frac{3 (y^2_l+2)}{16(9+4 y^2_l)^2} = \frac{q^2 \, (y^2_l+2)^{i+1}}{64*27} \frac{1}{1-\left(-\frac{16 y^4_l}{81} - \frac{8 y^2_l}{9}\right)} = \frac{q^2 \, (y^2_l+2)^{i+1}}{1728} \sum_{k=0}^\infty (-1)^k \left(\frac{16}{81} y^4_l + \frac{8}{9} y^2_l\right)^k = \notag \\
& =  \frac{q^2 \, (y^2_l+2)^{i+1}}{1728} \sum_{k=0}^\infty \sum_{s=0}^k (-1)^k
\begin{pmatrix}
k \\
s
\end{pmatrix}
\left(\frac{16}{81}\right)^k \left(\frac{9}{2}\right)^s (y^2_l)^{2k- s}
&
\end{align}
We can write this last series as a power series in $y_l^2$. In particular we set $2k-s = r$; we notice that $k$ is a free index going from 0 to $\infty$ and so, by fixing it to a specific integer value, we get that $s \in [0,k]$, from which $r \in [k,2k]$. By the way, now we want to write a series in which $r$ is free to go from 0 to $\infty$ and, by fixing it to a given value, $k$ has to vary within a finite interval. At the end, it turns out that
\begin{equation}
    \mathcal{H}^{(2)}_{i,l} = \frac{q^2 \, (y^2_l+2)^{i+1}}{1728} \sum_{r=0}^\infty c_r \, (y_l^2)^r = \frac{q^2}{1728} \sum_{r=0}^\infty (\beta_r)_{i,l}
\end{equation}
where the matrix $\beta_r$ has elements $(\beta_r)_{i,l} = c_r (y^2_l+2)^{i+1}(y_l^2)^r$ and the coefficients $c_r$ are
\begin{equation}
    c_r = \sum_{k=\lfloor \frac{r+1}{2} \rfloor}^r (-1)^k 
    \begin{pmatrix}
        k \\
        2k-r
\end{pmatrix}
    2^{2k+r} \, \,9^{-r}
\end{equation}
with $c_0=1$. In this way, going to $\text{Tr}\Bigl[(\mathcal{Z}^{(0)})^{-1} \mathcal{H}^{(2)}\Bigr]$, we can study the contribution from each of the addends of $\mathcal{H}^{(2)}$, with the goal of determining their behavior in terms of $n$. The first terms are
\begin{align*}
& \frac{q^2}{1728} \, \text{Tr}[(\mathcal{Z}^{(0)})^{-1} \beta_0] = \frac{q^2}{1728} \left[(n+2)(n+1) + \sum_{\alpha=0}^n y^2_\alpha\right] \\
& \frac{q^2}{1728} \, \text{Tr}[(\mathcal{Z}^{(0)})^{-1} \beta_1] = \frac{q^2}{1728} c_1 \left[\frac{2}{3}(n+2)(n+1) n + 2(n+1) \sum_{\alpha=0}^n y^2_\alpha + \sum_{\alpha=0}^n y^4_\alpha\right] \\
& \frac{q^2}{1728} \, \text{Tr}[(\mathcal{Z}^{(0)})^{-1} \beta_2] = \frac{q^2}{1728} c_2 \left[\frac{(n+2)(n+1) n (n-1)}{3}+ 2(n+1) n \sum_{\alpha=0}^n y^2_\alpha + 2 (n+1) \sum_{\alpha=0}^n y^4_\alpha+  2 \sum_{i<j}^n y^2_i y^2_j + \sum_{\alpha=0}^n y^6_\alpha\right]
\end{align*}
\begin{align*}
& \frac{q^2}{1728} \, \text{Tr}[(\mathcal{Z}^{(0)})^{-1} \beta_3] = \frac{q^2}{1728} c_3 \left[\frac{2}{15} (n+2)(n+1) n (n-1) (n-2) + \frac{4}{3}(n+1) n (n-1) \sum_{\alpha=0}^n y^2_\alpha + 2 (n+1) n \sum_{\alpha=0}^n y^4_\alpha + \right. \\
& \left.+ 4 \, n \sum_{i<j}^n y^2_i y^2_j + 2 (n+1) \sum_{\alpha=0}^n y^6_\alpha + 2 \sum_{i,j=0 \, (i\ne j)}^n y^2_i y^4_j + \sum_{\alpha=0}^n y^8_\alpha\right]
\end{align*}
As one can see, at each value of $r$ there appear different functions of $n$ multiplying different polynomials of the $y_\alpha^2$'s. We name them $f_r^{(p_\ell)}(n)$, where $\{p_\ell\}_\ell$ label all the different functions of $n$ that multiply the various polynomials of degree $p$ showing up at level $r$. For example: there is obviously a single polynomial of degree $0$, $P_0(y_\alpha^2)=1$, and $f_0^{(0)}(n)=(n+2)(n+1), f_1^{(0)}(n)=\frac{2}{3}(n+2)(n+1)n$, and so on; there is still a single polynomial of degree 1, $P_1(y_\alpha^2)=\sum_\alpha y_\alpha^2$, and $f_0^{(1)}(n)=1, f_1^{(1)}(n)=2(n+1)$, and so on. There are, instead, two distinct polynomials of degree $2$, the sum of squares of the $y^2_\alpha$'s, $P_{2_1}(y_\alpha^2)=\sum_\alpha y_\alpha^4$, and the symmetric degree-$2$ polynomial of the $y^2_\alpha$'s, $P_{2_2}(y_\alpha^2)=\sum_{\alpha<\beta} y_\alpha^2 y_\beta^2$: here we have $f_0^{(2_1)}(n)=0,f_1^{(2_1)}(n)=1,f_2^{(2_1)}(n)=2(n+1)$, and so on, whereas $f_0^{(2_2)}(n)=f_1^{(2_2)}(n)=0,f_2^{(2_2)}(n)=2$, and so on. \\
Hence we can write the contribution of $\beta_r$ as
\begin{align}
\frac{q^2}{1728} \, \text{Tr}[(\mathcal{Z}^{(0)})^{-1} \beta_r] = \frac{q^2}{1728} c_r & \left[f_r^{(0)}(n) + f_{r}^{(1)}(n) \sum_{\alpha=0}^n y^2_\alpha + f_{r}^{(2_1)}(n) \sum_{\alpha=0}^n y^4_\alpha + f_{r}^{(2_2)}(n) \, \, P_{2_2}(y^2_\alpha) + \right. \notag \\
& \left. + f_{r}^{(3_1)}(n) \sum_{\alpha=0}^n y^6_\alpha + \dots \right]
\end{align}
At this point the entire non-polynomial contribution to \eqref{eq:detMn_2inst} can be written as
\begin{align} \label{eq:intermediate_expression_non_pert}
    \text{Tr}[(\mathcal{Z}^{(0)})^{-1} \mathcal{H}^{(2)}] & = \frac{q^2}{1728} \left[\sum_{r=0}^\infty c_r \, f_{r}^{(0)}(n) + \left(\sum_{r=0}^\infty c_r \, f_{r}^{(1)}(n)\right) \sum_{\alpha=0}^n y^2_\alpha + \left( \sum_{r=0}^\infty c_r \, f_{r}^{(2_1)}(n) \right) \sum_{\alpha=0}^n y^4_\alpha + \right. \notag \\
    & \left. + \left( \sum_{r=0}^\infty c_r \, f_{r}^{(2_2)}(n) \right) \, \, P_{2_2}(y^2_\alpha) + \left(\sum_{r=0}^\infty c_r \, f_{r}^{(3_1)}(n)  \right)\sum_{\alpha=0}^n y^6_\alpha + \dots \right]
\end{align}
We are going to perform our analysis up to the contribution of the function $f_{r}^{(2_2)}(n)$, but the procedure we are going to discuss can be applied in the similar way to the other terms.\\
From the cases highlighted above, we find that the functions $f_{r}^{(0)}(n), f_{r}^{(1)}(n), f_{r}^{(2_1)}(n)$ and $f_{r}^{(2_2)}(n)$ are \footnote{The functions $f_r^{(p_\ell)}(n)$ not explicitly written are meant to be zero.}
\begin{align}
f_r^{(0)}(n)&=2^{r+1}\begin{pmatrix}
        n+2 \\
        r+2
\end{pmatrix} \qquad 0\leq r\leq n \,,\\
f_r^{(1)}(n)&= 2^r\begin{pmatrix}
        n+1 \\
        r
\end{pmatrix}\qquad\quad 0\leq r\leq n+1\,, \\
f_r^{(2_1)}(n)&= 2^{r-1}\begin{pmatrix}
        n+1 \\
        r-1
\end{pmatrix}\qquad 1\leq r\leq n+2\,, \\
f_r^{(2_2)}(n)&= 2^{r-1}\begin{pmatrix}
        n \\
        r-2
\end{pmatrix}\qquad 2\leq r\leq n+2\,.
\end{align}
From these expressions we clearly see that the extrema for the summations in $r$ change from $0$ to $\infty$ in \eqref{eq:intermediate_expression_non_pert}, depending on the function $f_{r}^{(p_\ell)}(n)$ we are considering. In particular
\begin{align} \label{eq:intermediate_expression_non_pert_2}
    \text{Tr}[(\mathcal{Z}^{(0)})^{-1} \mathcal{H}^{(2)}] & = \frac{q^2}{1728} \left[\sum_{r=0}^n c_r \, f_{r}^{(0)}(n) + \left(\sum_{r=0}^{n+1} c_r \, f_{r}^{(1)}(n)\right) \sum_{\alpha=0}^n y^2_\alpha + \left( \sum_{r=1}^{n+2} c_r \, f_{r}^{(2_1)}(n) \right) \sum_{\alpha=0}^n y^4_\alpha + \right. \notag \\
& \left. +\left( \sum_{r=2}^{n+2} c_r \, f_{r}^{(2_2)}(n) \right) \, \, P_{2_2}(y^2_\alpha) + \dots \right]
\end{align}
Coming back to the multivariate integrals, since the trace we are computing goes as $\mathcal{O}(q^2)$ and we are working at 2 instantons, then, keeping in account also the contribution of the anti-instanton sector, we get
\begin{align}
 \det M^{(n)}_{\text{rest}} = & \frac{(2\pi)^{n(n+1)}}{Z_{S^4}^{n+1} (n+1)!} \int_{\mathbb{R}^{n+1}} \prod_{k=0}^n \mathrm{d}y_k \, 4y^2_k \, e^{-4 \pi \text{Im}\tau \, y^2_k} \, \left|Z_{\text{1L}}(i y_k)\right|^2 \prod_{p<s} (y^2_p-y^2_s)^2 \frac{q^2 + \overline{q}^2}{1728} \left[ \sum_{r=0}^n c_r \, f_{r}^{(0)}(n) + \right. \notag \\
 & \left. \left(\sum_{r=0}^{n+1} c_r \, f_{r}^{(1)}(n)\right) \sum_{\alpha=0}^n y^2_\alpha + \left(\sum_{r=1}^{n+2} c_r \, f_{r}^{(2_1)}(n)\right) \sum_{\alpha=0}^n y^4_\alpha + \left(\sum_{r=2}^{n+2} c_r \, f_{r}^{(2_2)}(n)\right) \, P_{2_2}(y^2_k) + \dots\right]
\end{align}
Changing the variables $x_i = y^2_i$, we obtain
\begin{align}
 \det M^{(n)}_{\text{rest}} & = \frac{4^{n+1} \, (2\pi)^{n(n+1)}}{Z_{S^4}^{n+1}} \frac{q^2 + \overline{q}^2}{1728} \left[ \left(\sum_{r=0}^n c_r \, f_{r}^{(0)}(n) \right) X_n^{(0)} + \left(\sum_{r=0}^{n+1} c_r \, f_{r}^{(1)}(n)\right) X_n^{(1)}  + \left(\sum_{r=1}^{n+2} c_r \, f_{r}^{(2_1)}(n)\right) Z_{1,n}  + \right. \notag \\
 & \left. + \left(\sum_{r=2}^{n+2} c_r \, f_{r}^{(2_2)}(n)\right) \, Z_{2,n} + \dots\right]
\end{align}
where
\begin{align}
& Z_{1,n} = \frac{1}{(n+1)!}\int_{\mathbb{R}^{n+1}_+} \prod_{k=0}^n \mathrm{d}x_k \, x^\frac{1}{2}_k \, e^{-4 \pi \text{Im}\tau \, x_k} \, \left|Z_{\text{1L}}(i \sqrt{x_k})\right|^2 \prod_{p<s} (x_p-x_s)^2 \left(\sum_{\alpha=0}^n x^2_\alpha\right)  \equiv \langle \text{Tr} (\chi^2) \rangle \,  X_n^{(0)} \\
& Z_{2,n} =\frac{1}{(n+1)!}\int_{\mathbb{R}^{n+1}_+} \prod_{k=0}^n \mathrm{d}x_k \, x^\frac{1}{2}_k \, e^{-4 \pi \text{Im}\tau \, x_k} \, \left|Z_{\text{1L}}(i \sqrt{x_k})\right|^2 \prod_{p<s} (x_p-x_s)^2 \left(P_{2_2}(x_k)\right)  \equiv \langle P_{2_2}(x_0,\dots,x_n) \rangle \, X_n^{(0)}
\end{align}
Recalling that $\det M^{(n)}_{\text{rest}}$ is added to the determinant \eqref{eq:det_2inst__Mn} and then, in taking the logarithm and expanding it up to $\mathcal{O}(q^2)$, we have to divide by $X_n^{(0)}$, we get (always at the perturbative level)
\begin{align} \label{eq:final_log_det_2inst_complete}
   \log \det M^{(n)} = \eqref{eq:final_log_detMn} + \frac{q^2 + \overline{q}^2}{1728} & \left[\sum_{r=0}^n c_r \, f_{r}^{(0)}(n) + \left(\sum_{r=0}^{n+1} c_r \, f_{r}^{(1)}(n)\right) \frac{X_n^{(1)}}{X_n^{(0)}} + \left(\sum_{r=1}^{n+2} c_r \, f_{r}^{(2_1)}(n)\right) \frac{Z_{1,n}}{X_n^{(0)}} + \right. \notag \\
   & \left. + \left(\sum_{r=2}^{n+2} c_r \, f_{r}^{(2_2)}(n)\right) \, \frac{Z_{2,n}}{X_n^{(0)}} + \dots\right]
\end{align}
For what seen in the previous Subsection in \eqref{eq:Yn_explicit_pert}, going to the variables $z_k = t x_k$, we have that $X_n^{(1)}/X_n^{(0)} \simeq t^{-1}, Z_{n,1}/X_n^{(0)} \simeq Z_{n,2}/X_n^{(0)} \simeq t^{-2}$. Thanks to this fact, we can parametrize the contribution to the logarithm of the determinant of the $M^{(n)}$ matrix from $Z^{(2)}_{\text{rest}}$ as a power series in $t^{-1}$
\begin{equation}
    \log \det M^{(n)} = \eqref{eq:final_log_detMn} +\frac{q^2 + \overline{q}^2}{1728} \sum_{p=0}^\infty F^{(p)}(n) \, t^{-p}
\end{equation}
At this point we want to determine the large-$n$ behaviour of the functions $F^{(0)}(n), F^{(1)}(n)$ and $F^{(2)}(n)$ in terms of $n$.\\
For what concerns $F^{(0)}(n)$, we get \footnote{Since we are interested in the large charge limit $n \to \infty$, in the following we are going to consider, for simplicity, the upper extrema of the summations in \eqref{eq:final_log_det_2inst_complete} as an even number for the computation of the $F^{(p)}(n)$ functions, i.e. $n = 2m$. In the case of $n$ odd, the same asymptotics that is shown is also found.}
\begin{equation}
    F^{(0)}(n=2m) = \sum_{r=0}^n c_r \, f_{r}^{(0)}(n)  = \frac{81}{32} - \frac{17}{32} 81^{-m} - \frac{m}{2} 81^{-m}
\end{equation}
We see that $F^{(0)}(m)$ tends to a constant for $m \to \infty$, which does not play any role in the computation of the logarithm of the correlator (since it cancels out when computing the difference with $\log \det M^{(n-1)}$), and it goes to that constant in an exponentially fast way, hence it is a non-perturbative contribution to the correlator. So this piece does not contribute to the asymptotic expansion of $\log(Z_{S^4} \, G_{2n})$.\\ 
For what concerns $F^{(1)}(n)$, we have to isolate from $X_n^{(1)} /X_n^{(0)}$ the $t$-independent part and since this term goes as $t^{-1}$, this is $t \frac{X_n^{(1)}}{X_n^{(0)}}$, which is a polynomial of the second order in $n$, hence
\begin{equation}
    F^{(1)}(n=2m) = \left(\sum_{r=0}^{n+1} c_r \, f_{r}^{(1)}(n)\right) \left(t \frac{X_n^{(1)}}{X_n^{(0)}}\right)  = \left(- \frac{7}{9} 81^{-m} - \frac{16 \, m}{9} 81^{-m}\right) (2m+1)(2m+2)
\end{equation}
This shows that also $F^{(1)}(m)$ gives rise only to non-pertubative corrections to the correlator.\\
$F^{(2)}(n)$ contains two contributions and analyzing them independently, we have
\begin{align}
    & F^{(2_1)}(n=2m) = \left(\sum_{r=1}^{n+2} c_r \, f_{r}^{(2_1)}(n)\right) \left(t^2 \frac{Z_{1,n}}{X_n^{(0)}} \right) = \left(\frac{8}{27} 81^{-m} + \frac{64 \, m}{81} 81^{-m} \right) \mathcal{P}_{2_1}(m) \\
    & F^{(2_2)}(n=2m) = \left(\sum_{r=2}^{n+2} c_r \, f_{r}^{(2_2)}(n)\right)\left(t^2 \frac{Z_{2,n}}{X_n^{(0)}} \right) = \left(\frac{32}{27} 81^{-m} - \frac{512 \, m}{81} 81^{-m} \right) \mathcal{P}_{2_2}(m)
\end{align}
where $\mathcal{P}_{2_1}(m)$ and $\mathcal{P}_{2_2}(m)$ are quartic polynomials in $m$ \footnote{Strictly speaking, these objects are polynomials if we work with the Wishart-Laguerre ensemble emerging from the large-eigenvalue limit of $|Z_{\text{1L}}(i a)|^2$ \cite{Grassi:2019txd,Livan:2011uvs}. Nevertheless, we argue that, in the ensemble defined by \eqref{eq:X_n}, they all grow no faster than a power for large $n$.}. Since $F^{(2)}(n)$ is the sum of these two functions, it clearly exhibits a non-perturbative behaviour at large $m$.\\
Due to the fact that the structure of the functions $f_r^{(p_l)}(n)$ is always the same, we expect that this behaviour will not change beyond the quadratic order in $1/t$, although we have not verified it explicitly.\\
To summarize, we have shown that the non-polynomial part of the $2$-instanton correction contributes only exponentially suppressed terms in the large-$n$ expansion of the logarithm of the ($Z_{S^4}$-rescaled) correlator. Therefore, up to $2$ instantons, its perturbative asymptotic expansion is fully captured by \eqref{eq:final_expression_2inst_large_vev}, i.e. by the polynomial parts of the instanton corrections, or equivalently by the large-VEV regime of the free energy.\\
The procedure we have outlined can be equally well applied beyond $2$ instantons, separating the polynomial part of the corrections from the non-polynomial remainder. We argue that, by doing so, one will arrive at the same conclusion. Motivated by this conjecture, in the next Section we will first push the analysis up to $4$ instantons, but focusing on the polynomial parts only: we will compute the large-$n$ expansion of the logarithm of the ($Z_{S^4}$-rescaled) two-point function and find perfect match with earlier results in the literature. This provides non-trivial support to our claim that $\mathcal{F}_0$ and $\mathcal{F}_1$ alone fully encode the large-charge expansion at the perturbative level. Armed with this evidence, we will perform an all-instanton computation in the large-VEV regime. Besides confirming our claim, this will provide an independent exact derivation of the coefficient $A$ and, at the same time, the first explicit derivation of the scheme-independent coefficient $\tilde{B}$,\footnote{We recall that in \cite{Hellerman:2021yqz} $\tilde{B}$ is expressed as $\tilde{B}=B-\log Z_{S^4}$ and, by specifying a renormalization scheme for the partition function, only $B$ is computed, whereas $\log Z_{S^4}$ is left implicit.} albeit only through its large-VEV contribution.

\section{The large-VEV sector} \label{sec:chap2_7}
In this Section we will concentrate on the polynomial terms in the instanton partition function and isolate the coupling dependence that flows into $A$ and $B$ from the one affecting the non-perturbative terms. This will allow us to push the computation of $A$ and $B$ up to $4$ instantons in Subsection \ref{subsec:chap2_7_1}. In Subsection \ref{subsec:chap2_7_2} we will show that our results perfectly agree with those of \cite{Hellerman:2020sqj,Hellerman:2021yqz}, which have been derived with different methods. Finally, in Subsection \ref{subsec:chap2_7_3}, we will prove at arbitrary instanton order (i.e. exactly in $\tau$) that the large-VEV sector, consisting solely of the contributions of $\mathcal{F}_0$ and $\mathcal{F}_1$ to the free energy, not only reproduces the correct coefficients of the $n\log n$ and $\log n$ terms, in agreement with \cite{Grassi:2019txd}, but also affects only the $\tau$ dependence of the coefficient $A(\tau,\bar{\tau})$ in the large-charge expansion \eqref{eq:log_corr_general} (with $d=2$).

\subsection{Four-instanton calculation}\label{subsec:chap2_7_1}
The polynomial truncation of the instanton partition function, which amounts to the instanton corrections to $\mathcal{F}_0$ and $\mathcal{F}_1$ only, reads
\begin{equation}
    Z_{\text{inst}} \simeq 1 + \frac{q}{2} \, (a^2-3) + \frac{q^2}{4}\Bigl(\frac{a^4}{2}-\frac{35}{16} a^2 + \frac{13}{8}\Bigr) + \frac{q^3}{8}\Bigl(\frac{a^6}{6} - \frac{11}{16}a^4 + \frac{7}{48} a^2 + \frac{3}{8} \Bigr) +\frac{q^4}{16} \Bigl(\frac{a^8}{24} - \frac{3}{32} a^6 - \frac{517}{1536} a^4 +\frac{285}{2048}a^2 + \frac{141}{512} \Bigr)
\end{equation}
Following the same steps of the case at 2 instantons discussed in the previous Section (computing \\
$\partial_\tau^i \partial_{\bar{\tau}}^j (e^{-4 \pi \, \text{Im} \tau \, a^2} |Z_{\text{inst}}|^2)$ and applying Andrèief identity), we arrive at the following expression of $\det M^{(n)}$ (equivalent to \eqref{eq:first_expr_det} but without the matrix $\mathcal{H}^{(2)}$)
\begin{equation}\label{eq:det_4inst}
    \det M^{(n)} = \frac{(2\pi)^{n(n+1)}}{Z_{S^4}^{n+1} \, (n+1)!} \int_{\mathbb{R}^{n+1}} \prod_{k=0}^n \mathrm{d}y_k \, 4y^2_k \, e^{-4 \pi \text{Im}\tau \, y^2_k} \, \left|Z_{\text{1L}}(i y_k)\right|^2 \underset{i,l=0,\dots,n}{\det}((\sum_{s=0}^4 \mathcal{Z}^{(s)})_{i,l}) \underset{j,m=0,\dots,n}{\det}((\sum_{s=0}^4 \overline{\mathcal{Z}}^{(s)})_{j,m})
\end{equation}
where the matrices $\mathcal{Z}^{(0)},\mathcal{Z}^{(1)},\mathcal{Z}^{(2)}$ are the ones defined in \eqref{A-matrices}, while
\begin{align}
& \mathcal{Z}^{(3)}_{i,l} = \frac{q^3 (y^2_l+3)^i}{8}\Bigl(\frac{y^6_l}{6}-\frac{11}{16} y^4_l + \frac{7}{48} y^2_l + \frac{3}{8}\Bigr), \notag \\ 
& \mathcal{Z}^{(4)}_{i,l} = \frac{q^4 (y^2_l+4)^i}{16}\Bigl(\frac{y^8_l}{24}-\frac{3}{32} y^6_l - \frac{517}{1536} y^4_l + \frac{285}{2048} y^2_l + \frac{141}{512}\Bigr)
\end{align}
Computing explicitly the determinant of the Vandermonde matrix using Binet's theorem, we arrive at
\begin{align}
\label{eq:detMn_2inst}
    \det M^{(n)} =  \frac{(2\pi)^{n(n+1)}}{Z_{S^4}^{n+1} \, (n+1)!} \int_{\mathbb{R}^{n+1}} & \prod_{k=0}^n \mathrm{d}y_k \, 4y^2_k \, e^{-4 \pi \text{Im}\tau \, y^2_k} \, \left|Z_{\text{1L}}(i y_k)\right|^2 \prod_{p<s} (y^2_p-y^2_s)^2 \underset{i,l=0,\dots,n}{\det}(\mathbb{I}+\sum_{s=1}^4 \mathcal{K}^{(s)})_{i,l} \times \notag \\
    & \times \underset{j,m=0,\dots,n}{\det}(\mathbb{I}+\sum_{s=1}^2 \overline{\mathcal{K}}^{(s)})_{j,m}
\end{align}
and focusing on one of the two determinants, we obtain
$$\begin{aligned}
 \det(\mathbb{I}+\sum_{s=1}^4 \mathcal{K}^{(s)}) & = 1 + \overbrace{\text{Tr}(\mathcal{K}^{(1)})}^{\mathcal{O}(q)} +  \overbrace{\text{Tr}(\mathcal{K}^{(2)})-\frac{1}{2}\text{Tr}(\mathcal{K}^{(1)^2}) + \frac{1}{2}\text{Tr}(\mathcal{K}^{(1)})^2}^{\mathcal{O}(q^2)_{\text{pol}}} + \\
 & \left.
    \begin{aligned}
        & + \text{Tr}(\mathcal{K}^{(3)}) - \text{Tr}(\mathcal{K}^{(1)} \mathcal{K}^{(2)}) + \frac{1}{2} \text{Tr}(\mathcal{K}^{(1)^3}) + \\
        & + \text{Tr}(\mathcal{K}^{(1)}) \text{Tr}(\mathcal{K}^{(2)}) - \frac{1}{2} \text{Tr}(\mathcal{K}^{(1)}) \text{Tr}(\mathcal{K}^{(1)^2}) + \\
        & + \frac{1}{6} \text{Tr}(\mathcal{K}^{(1)})^3 + 
    \end{aligned}
    \right\} \quad \mathcal{O}(q^3)_{\text{pol}} \\
& \left.
    \begin{aligned}
        & + \text{Tr}(\mathcal{K}^{(4)}) - \frac{1}{2}\text{Tr}(\mathcal{K}^{(2)^2}) - \text{Tr}(\mathcal{K}^{(1)}\mathcal{K}^{(3)}) + \text{Tr}(\mathcal{K}^{(1)^2}\mathcal{K}^{(2)}) - \frac{1}{4}\text{Tr}(\mathcal{K}^{(1)^4}) + \\
        & + \frac{1}{8}\text{Tr}(\mathcal{K}^{(1)^2})^2 + \frac{1}{3} \text{Tr}(\mathcal{K}^{(1)}) \text{Tr}(\mathcal{K}^{(1)^3}) - \text{Tr}(\mathcal{K}^{(1)}) \text{Tr}(\mathcal{K}^{(1)} \mathcal{K}^{(2)}) + \\
        & - \frac{1}{2} \text{Tr}(\mathcal{K}^{(1)^2}) \text{Tr}(\mathcal{K}^{(2)}) + \frac{1}{2} \text{Tr}(\mathcal{K}^{(2)})^2 + \text{Tr}(\mathcal{K}^{(1)})\text{Tr}(\mathcal{K}^{(3)}) + \frac{1}{2} \text{Tr}(\mathcal{K}^{(1)})^2 \text{Tr}(\mathcal{K}^{(2)}) + \\ 
        & - \frac{1}{4} \text{Tr}(\mathcal{K}^{(1)})^2 \text{Tr}(\mathcal{K}^{(1)^2}) + \frac{1}{24} \text{Tr}(\mathcal{K}^{(1)})^4
    \end{aligned}
    \right\} \quad \mathcal{O}(q^4)_{\text{pol}}
\end{aligned}$$
The first line has already been studied in the previous Section, hence we move on to $\mathcal{O}(q^3)_{\text{pol}}$ and $\mathcal{O}(q^4)_{\text{pol}}$
\begin{align}\label{orderq3}
    \mathcal{O}(q^3)_{\text{pol}} & = \frac{q^3}{48} \Bigl[\frac{2 n^6 - 30n^5 + 141 n^4 - 136 n^3 - 353 n^2 - 8 n + 36}{16} + \Bigl. \frac{12 n^4-120 n^3 + 249 n^2 + 303 n + 14}{16} \sum_{\alpha=0}^n y^2_\alpha + \Bigr. \notag \\
    &  + \frac{3}{2} \Bigl(\frac{13}{4} + (n-6)(n+1)\Bigr) \Bigl(\sum_{\alpha=0}^n y^2_\alpha\Bigr)^2 + \Bigl(\sum_{\alpha=0}^n y^2_\alpha\Bigr)^3 \Bigr]
\end{align}
\begin{align} \label{orderq4}
    \mathcal{O}(q^4)_{\text{pol}} = & \frac{q^4}{384} \Bigl[\frac{(n+1)(8 n^7 - 168n^6 + 1392 n^5 - 5560 n^4 + 10174 n^3 -5618 n^2 - 1269 n+846)}{128} + \Bigr. \notag \\ 
    &+ \frac{128 n^6 - 1920 n^5 + 9648 n^4 - 14944 n^3 - 8729 n^2 + 10664 n + 855}{256} \sum_{\alpha=0}^n y^2_\alpha + \notag \\
    & \Bigl. + \Bigl(\frac{3}{2}n^4 - 15 n^3 + 36 n^2 + \frac{27}{2} n - \frac{517}{64}\Bigr) \Bigl(\sum_{\alpha=0}^n y^2_\alpha\Bigr)^2 + \Bigl(2 n^2 - 10 n - \frac{9}{4}\Bigr)\Bigl(\sum_{\alpha=0}^n y^2_\alpha\Bigr)^3 + \Bigl(\sum_{\alpha=0}^n y^2_\alpha\Bigr)^4\Bigr]
\end{align}
Performing the usual change $x_i = y^2_i$ and expanding the products of the two determinants inside the integrals up to $\mathcal{O}(q^4)$, the determinant of $M^{(n)}$ turns out to be
\begin{align}\label{determinant_final_4inst}
  \det M^{(n)} & = \frac{4^{n+1} (2 \pi)^{n(n+1)}}{Z_{S^4}^{n+1}}\Bigl[X_n^{(0)} + \frac{q+\overline{q}}{2} \textcolor{red}{a_n} + \frac{(q+\overline{q})^2}{8} \textcolor{red}{b_n} + \frac{q \overline{q}}{4} \frac{46+37n-9n^2}{8} X_n^{(0)} - \frac{q \overline{q}}{4} \frac{13}{8} X_n^{(1)} + \frac{(q+\overline{q})^3}{48} \textcolor{red}{c_n} + \Bigr.\notag \\
  & - \frac{q \overline{q}^2+q^2\overline{q}}{16} \frac{9 n^4 - 82 n^3 + 123 n^2 + 310 n + 96}{8} X_n^{(0)} + \frac{q \overline{q}^2+q^2\overline{q}}{16} \frac{124+139n-31n^2}{8} X_n^{(1)} + \notag \\
  & -\frac{q \overline{q}^2+q^2\overline{q}}{16} \frac{26}{8} X_n^{(2)} + \frac{(q+\overline{q})^4}{384} \textcolor{red}{d_n} - \frac{q \overline{q}^3+q^3\overline{q}}{96} \frac{3(176 n^2-816n-455)}{64} X_n^{(2)}-\frac{q \overline{q}^3+q^3\overline{q}}{96} \frac{39}{8} X_n^{(3)} + \notag\\
  & - \frac{q \overline{q}^3+q^3\overline{q}}{96} \frac{3(n+1)(36 n^5 - 544 n^4 + 2774 n^3 - 4686 n^2 - 535 n + 570}{128} X_n^{(0)} + \notag\\
  & - \frac{q \overline{q}^3+q^3\overline{q}}{96} \frac{3(392 n^4 - 3664 n^3 + 6960 n^2 + 8632 n + 317)}{256} X_n^{(1)} + \notag \\
  & - \frac{q^2 \overline{q}^2}{64} \frac{(n+1)(144n^5-2176n^4+10772n^3-15756n^2-8549n-506)}{128} X_n^{(0)} + \notag \\
  & - \frac{q^2 \overline{q}^2}{64} \frac{1568 n^4-14656 n^3+25968 n^2 + 42224 n + 8135}{256} X_n^{(1)} + \notag \\
  & - \Bigl. \frac{q^2 \overline{q}^2}{64} \frac{352 n^2 - 1632 n - 1079}{32} X_n^{(2)} - \frac{q^2 \overline{q}^2}{64} \frac{13}{2} X_n^{(3)} \Bigr]
\end{align}
where in red we have highlighted new objects
\begin{align}
    & a_n = \int_{\mathbb{R}^{n+1}_+} \mathrm{d}\mu \frac{\eqref{eq:1_inst_contr_traces}}{\frac{q}{2}}=  \frac{(n+1)(n-6)}{2} X_n^{(0)} + X_n^{(1)} \\
    & b_n = \int_{\mathbb{R}^{n+1}_+} \mathrm{d}\mu \frac{\eqref{eq:2_inst_contr_traces}}{\frac{q^2}{8}} = \frac{2 n^4 - 20 n^3 + 35 n^2 + 83 n + 26}{8} X_n^{(0)} + \Bigl(\frac{13}{8} + (n-6)(n+1)\Bigr) X_n^{(1)} + X_n^{(2)} 
\end{align}
\begin{align}
    c_n &= \int_{\mathbb{R}^{n+1}_+} \mathrm{d}\mu \frac{\eqref{orderq3}}{\frac{q^3}{48}} = \frac{2 n^6-30 n^5 + 141 n^4 - 136 n^3 - 353 n^2 - 8n +36}{16} X_n^{(0)} + \notag \\
    & + \frac{12 n^4-120 n^3 + 249 n^2 + 303 n + 14}{16} X_n^{(1)} + \frac{3}{2} \Bigl(\frac{13}{4} + (n-6)(n+1)\Bigr) X_n^{(2)} + X_n^{(3)} 
\end{align}
\begin{align}
    d_n &= \int_{\mathbb{R}^{n+1}_+} \mathrm{d}\mu \frac{\eqref{orderq4}}{\frac{q^4}{384}} = \frac{(n+1)(8 n^7 - 168n^6 + 1392 n^5 - 5560 n^4 + 10174 n^3 -5618 n^2 - 1269 n+846)}{128} X_n^{(0)} + \notag \\
    & + \frac{128 n^6 - 1920 n^5 + 9648 n^4 - 14944 n^3 - 8729 n^2 + 10664 n + 855}{256} X_n^{(1)} +\notag \\
    & + \Bigl(\frac{3}{2} n^4-15 n^3 + 36 n^2 + \frac{27}{2} n - \frac{517}{64}\Bigr) X_n^{(2)} + \left(2 n^2 - 10 n - \frac{9}{4}\right) X_n^{(3)} + X_n^{(4)}
\end{align}
where we have introduced the measure $\mathrm{d}\mu = \prod_{k=0}^n \mathrm{d}x_k \, x^\frac{1}{2}_k \, e^{-4 \pi \text{Im}\tau \, x_k} \, \left|Z_{\text{1L}}(i \sqrt{x_k})\right| \prod_{p<s} (x_p-x_s)^2$ for simplicity. From all of this we see that there are two new objects to be computed, that are $X_n^{(3)}$ and $X_n^{(4)}$. Using \eqref{eq:general_p_Xn} for $p=3,4$ we obtain that
\begin{equation}
    X_n^{(3)} = - \partial_t^3 X_n^{(0)} \qquad X_n^{(4)} = \partial_t^4 X_n^{(0)}
\end{equation}
and, using the perturbative expression of $X_n^{(0)}$ \eqref{eq:final_expr_Xn}, we get
\begin{align}
   & X_n^{(3)} = \frac{(n+1)(n+2)[(n+1)(n+2)+1][(n+1)(n+2)+2]}{t^3} X_n^{(0)}\notag \\
   & X_n^{(4)} =  \frac{(n+1)(n+2)[(n+1)(n+2)+1][(n+1)(n+2)+2][(n+1)(n+2)+3]}{t^4} X_n^{(0)}
\end{align}
This implies that (recall what said in footnote \ref{footnote:general_moment}: $\langle (\text{Tr}\chi)^p \rangle = X_n^{(p)}/X_n^{(0)}$)
\begin{align}
    & \langle (\text{Tr} \chi)^3 \rangle=\frac{(n+1)(n+2)[(n+1)(n+2)+1][(n+1)(n+2)+2]}{t^3} \notag \\
    & \langle (\text{Tr} \chi)^4 \rangle=\frac{(n+1)(n+2)[(n+1)(n+2)+1][(n+1)(n+2)+2][(n+1)(n+2)+3]}{t^4}
\end{align}
In light of the fact that we expressed the results at 2 instantons in terms of cumulants, we would like to do the same also here, if possible, and hence it is useful to see the relations between moments and cumulants for 3 and 4 insertions. It turns out that
\begin{align}
    & \langle (\text{Tr}\chi)^3\rangle = \langle (\text{Tr}\chi)^3 \rangle_c + 3 \langle (\text{Tr}\chi)^2 \rangle_c \langle \text{Tr}\chi \rangle_c + \langle \text{Tr}\chi\rangle_c^3 \notag \\
    & \langle (\text{Tr}\chi)^4\rangle = \langle (\text{Tr}\chi)^4 \rangle_c + 4 \langle (\text{Tr}\chi)^3 \rangle_c \langle \text{Tr}\chi\rangle_c + 6 \langle (\text{Tr}\chi)^2 \rangle_c \langle \text{Tr}\chi\rangle_c^2 + 3 \langle (\text{Tr}\chi)^2\rangle_c^2 + \langle \text{Tr}\chi \rangle_c^4
\end{align}
By inverting these relations we can find the cumulants in terms of the moments
\begin{align} \label{cumvsvevs2}
    & \langle (\text{Tr}\chi)^3\rangle_c = \langle (\text{Tr}\chi)^3 \rangle - 3 \langle (\text{Tr}\chi)^2 \rangle \langle \text{Tr}\chi \rangle + 2 \langle \text{Tr}\chi\rangle^3 \notag \\
    & \langle (\text{Tr}\chi)^4\rangle_c = \langle (\text{Tr}\chi)^4 \rangle - 4 \langle (\text{Tr}\chi)^3 \rangle \langle \text{Tr}\chi\rangle + 12 \langle (\text{Tr}\chi)^2 \rangle \langle \text{Tr}\chi\rangle^2 - 3 \langle (\text{Tr}\chi)^2\rangle^2 - 6 \langle \text{Tr}\chi \rangle^4
\end{align}
At this point we are ready to analyse the contributions of the instantons (always at the perturbative level) after having expanded the logarithm of \eqref{determinant_final_4inst} up to $\mathcal{O}(q^4)$. We are going to do it separately and we are going to report also the 1 and 2 instantons results, which have to be the same with respect the ones obtained in the previous Section
\subsection*{1 instanton}
Clearly here we only have one class of contribution
\begin{equation}
    \boxed{\frac{q+\overline{q}}{2} \frac{a_n}{X_n^{(0)}} = \frac{q+\overline{q}}{2} \left(\frac{(n-6)(n+1)}{2} + \frac{\langle \text{Tr}\zeta \rangle_c}{t}\right)}
\end{equation}
as already obtained in the previous Section in \eqref{eq:final_logdetMn_2inst}.
\subsection*{2 instantons}
Here we have two kinds of contributions: the one that contains $(q+\overline{q})^2$ and the one with $q \overline{q}$.
\begin{itemize}
    \item \textcolor{red}{$(q+\overline{q})^2$}
    \begin{align*}
    \frac{(q+\overline{q})^2}{8} \left[ \frac{b_n}{X_n^{(0)}} - \left(\frac{a_n}{X_n^{(0)}}\right)^2\right] & = \frac{q^2+\overline{q}^2}{8}\left[\frac{(n+1)(9 n - 46)}{8} + \frac{13}{8} \frac{\langle \text{Tr}\zeta\rangle_c}{t}\right] + \frac{q\overline{q}}{4}\left[ \frac{(n+1)(9 n - 46)}{8} +  \frac{13}{8} \frac{\langle \text{Tr}\zeta\rangle_c}{t}\right] + \\
    & + \frac{1}{2} \left(\frac{q+\overline{q}}{2 t}\right)^2 \langle (\text{Tr}\zeta)^2 \rangle_c
    \end{align*}
    \item $\textcolor{red}{q \overline{q}}$
    \begin{equation*}
        \frac{q \overline{q}}{4} \left[- \frac{(n+1)(9 n - 46)}{8} - \frac{13}{8} \frac{\langle \text{Tr}\zeta\rangle_c}{t}\right]
    \end{equation*}
\end{itemize}
Summing these two ingredients we find that the total 2 instantons contribution is
\begin{equation}
    \boxed{\frac{q^2+\overline{q}^2}{8}\left[\frac{(n+1)(9 n - 46)}{8} + \frac{13}{8} \frac{\langle \text{Tr}\zeta\rangle_c}{t}\right] + \frac{1}{2} \left(\frac{q+\overline{q}}{2 t}\right)^2 \langle (\text{Tr}\zeta)^2 \rangle_c}
\end{equation}
which is the same as the one reported in the second line of \eqref{eq:final_logdetMn_2inst}.
\subsection*{3 instantons}
Here we have two kinds of contributions
\begin{itemize}
    \item $\textcolor{red}{(q+\overline{q})^3}$
    \begin{align*}
    &\frac{(q+\overline{q})^3}{48} \left[ \frac{c_n}{X_n^{(0)}} - 3 \frac{a_n b_n}{\left(X_n^{(0)}\right)^2} + 2 \left(\frac{a_n}{X_n^{(0)}}\right)^3\right] = \frac{q^3+\overline{q}^3}{48}\left[\frac{19 n^2-71 n-90}{4} + \frac{23}{4} \frac{\langle \text{Tr}\zeta \rangle_c}{t} +  \frac{39}{8} \frac{\langle (\text{Tr}\zeta)^2\rangle_c}{t^2} \right] + \\
    & + \frac{q^2\overline{q}+q\overline{q}^2}{16}\left[ \frac{19 n^2-71 n-90}{4} + \frac{23}{4} \frac{\langle \text{Tr}\zeta\rangle_c}{t} +\frac{39}{8} \frac{(\langle \text{Tr}\zeta)^2\rangle_c}{t^2}\right] + \frac{1}{6} \left(\frac{q+\overline{q}}{2 t}\right)^3 \langle (\text{Tr}\zeta)^3 \rangle_c
    \end{align*}
    \item $\textcolor{red}{q^2\overline{q}+q\overline{q}^2}$
    \begin{align*}
        & \frac{q^2\overline{q}+q\overline{q}^2}{16} \left[\frac{-9 n^4 + 82 n^3 - 123 n^2 - 310 n -96}{8} + \frac{124+139 n - 31 n^2}{8} \frac{\langle \text{Tr}\zeta\rangle_c}{t} - \frac{13}{4} \frac{\langle (\text{Tr}\zeta)^2 \rangle_c}{t^2}+ \right. \\
        & \left.  - \frac{46+37 n - 9n^2}{4} \left(\frac{(n-6)(n+1)}{2} + \frac{\langle \text{Tr}\zeta\rangle_c}{t}\right) + \frac{13}{4} \frac{\langle \text{Tr}\zeta\rangle_c}{t} \left(\frac{(n-6)(n+1)}{2} + \frac{\langle \text{Tr}\zeta\rangle_c}{t}\right)\right]
    \end{align*}
\end{itemize}
Summing these two ingredients we find that the total 3 instantons contribution is
\begin{empheq}[box=\fbox]{align*}
  & \frac{q^3+\overline{q}^3}{48}\left[\frac{19 n^2-71 n-90}{4} + \frac{23}{4} \frac{\langle \text{Tr}\zeta\rangle_c}{t} +  \frac{39}{8} \frac{\langle (\text{Tr}\zeta)^2\rangle_c}{t^2}\right] + \frac{1}{6} \left(\frac{q+\overline{q}}{2 t}\right)^3 \langle (\text{Tr}\zeta)^3 \rangle_c + \notag\\
  & \frac{q^2 \overline{q} + q \overline{q}^2}{16} \frac{13}{8} \frac{\langle (\text{Tr}\zeta)^2 \rangle_c}{t^2}
\end{empheq}
\subsection*{4 instantons}
Here we have three kinds of contributions
\begin{itemize}
    \item $\textcolor{red}{(q+\overline{q})^4}$
    \begin{align*}
    &\frac{(q+\overline{q})^4}{384} \left[ \frac{d_n}{X_n^{(0)}} - 3 \left(\frac{b_n}{X_n^{(0)}}\right)^2 -4 \frac{a_n c_n}{\left(X_n^{(0)}\right)^2} + 12 \left(\frac{a_n}{X_n^{(0)}}\right)^2 \frac{b_n}{X_n^{(0)}}- 6 \left(\frac{a_n}{X_n^{(0)}}\right)^4\right] =\\
    & \frac{q^4+\overline{q}^4}{384}\left[\frac{3(1257 n^2 - 4421 n-5678)}{128} + \frac{8103}{256} \frac{\langle \text{Tr}\zeta\rangle_c}{t} + \frac{1979}{64} \frac{\langle (\text{Tr}\zeta)^2\rangle_c}{t^2} + \frac{39}{4} \frac{\langle (\text{Tr}\zeta)^3\rangle_c}{t^3} \right] + \\
    &+ \frac{q^3\overline{q}+q\overline{q}^3}{96}\left[ \frac{3(1257 n^2 - 4421 n-5678)}{128} + \frac{8103}{256} \frac{\langle \text{Tr}\zeta\rangle_c}{t} +\frac{1979}{64} \frac{\langle (\text{Tr}\zeta)^2\rangle_c}{t^2} + \frac{39}{4} \frac{\langle (\text{Tr}\zeta)^3\rangle_c}{t^3}\right] + \\
    & +\frac{q^2 \overline{q}^2}{64} \left[\frac{3(1257 n^2 - 4421 n-5678)}{128} + \frac{8103}{256} \frac{\langle \text{Tr}\zeta\rangle_c}{t} +\frac{1979}{64} \frac{\langle (\text{Tr}\zeta)^2\rangle_c}{t^2} + \frac{39}{4} \frac{\langle (\text{Tr}\zeta)^3\rangle_c}{t^3}\right] + \\
    &+ \frac{1}{24} \left(\frac{q+\overline{q}}{2 t}\right)^4 \langle (\text{Tr}\zeta)^4\rangle_c
    \end{align*}
    \item $\textcolor{red}{q^3\overline{q}+q\overline{q}^3}$ and $\textcolor{red}{q^2\overline{q}^2}$. These two classes have very long expressions, which are not so useful to report here. As expected, many terms simplify with the corresponding ones coming from the previous $(q+\overline{q})^4$ family;
\end{itemize}
Summing all the three ingredients we find that the total 4 instantons contribution is
\begin{empheq}[box=\fbox]{align*}
   & \frac{q^4+\overline{q}^4}{384} \left[\frac{3(1257 n^2 - 4421 n-5678)}{128} + \frac{8103}{256} \frac{\langle \text{Tr}\zeta\rangle_c}{t} + \frac{1979}{64} \frac{\langle (\text{Tr}\zeta)^2\rangle_c}{t^2} + \frac{39}{4} \frac{\langle (\text{Tr}\zeta)^3\rangle_c}{t^3} \right]\\
   & + \frac{1}{24} \left(\frac{q+\overline{q}}{2 t}\right)^4 \langle (\text{Tr}\zeta)^4\rangle_c + \frac{q^3 \overline{q} + q \overline{q}^3}{96} \left(\frac{23}{4} \frac{\langle (\text{Tr}\zeta)^2 \rangle_c}{t^2} + \frac{39}{8} \frac{\langle (\text{Tr}\zeta)^3 \rangle_c}{t^3}\right) + \\
   & + \frac{q^2 \overline{q}^2}{64} \left(\frac{169}{64} \frac{\langle (\text{Tr}\zeta)^2 \rangle_c}{t^2} + \frac{13}{4} \frac{\langle (\text{Tr}\zeta)^3 \rangle_c}{t^3}\right)
\end{empheq}
From all these computations, we find that
\begin{empheq}[box=\fbox]{align}
   \log \det M^{(n)} & = \log\left[\frac{4^{n+1} (2 \pi)^{n(n+1)}}{Z_{S^4}^{n+1}}\right] + \log \, X_n^{(0)} + \left[\frac{q+\overline{q}}{2} \frac{(n-6)(n+1)}{2} +  \right. \notag \\
   & + \frac{q^2+\overline{q}^2}{8}\left(\frac{(n+1)(9 n - 46)}{8} + \frac{13}{8} \frac{\langle \text{Tr}\zeta\rangle_c}{t}\right)+ \frac{q^3+\overline{q}^3}{48}\left(\frac{19 n^2 - 71 n - 90}{4} + \right. \notag \\
   & \left. + \frac{23}{4} \frac{\langle \text{Tr}\zeta\rangle_c}{t} + \frac{39}{8} \frac{\langle (\text{Tr}\zeta)^2\rangle_c}{t^2} \right) + \frac{q^4+\overline{q}^4}{384} \left(\frac{3 (1257 n^2 - 4421 n - 5678)}{128} \right.+ \notag \\
   & \left. \left. + \frac{8103}{256} \frac{\langle \text{Tr}\zeta\rangle_c}{t} + \frac{1979}{64} \frac{\langle (\text{Tr}\zeta)^2\rangle_c}{t^2} + \frac{39}{4} \frac{\langle (\text{Tr}\zeta)^3\rangle_c}{t^3}\right)\right] + \sum_{k=1}^4 \frac{1}{k!} \left(\frac{q+\overline{q}}{2 t}\right)^k \langle (\text{Tr}\zeta)^k \rangle_c + \notag \\
   & + \frac{q^2 \overline{q} + q \overline{q}^2}{16} \frac{13}{8} \frac{\langle (\text{Tr}\zeta)^2 \rangle_c}{t^2} + \frac{q^3 \overline{q} + q \overline{q}^3}{96} \left(\frac{23}{4} \frac{\langle (\text{Tr}\zeta)^2 \rangle_c}{t^2} + \frac{39}{8} \frac{\langle (\text{Tr}\zeta)^3 \rangle_c}{t^3}\right) + \notag \\
   & + \frac{q^2 \overline{q}^2}{64} \left(\frac{169}{64} \frac{\langle (\text{Tr}\zeta)^2 \rangle_c}{t^2} + \frac{13}{4} \frac{\langle (\text{Tr}\zeta)^3 \rangle_c}{t^3}\right)
\end{empheq}
The above formulae confirm a key fact that we anticipated in Subsection \ref{subsec:chap2_6_1}: only the cumulants appear in the final result. Recall that only the specific combination of expectation values appearing in the cumulants guarantees that, no matter how high the power of $\text{Tr}\chi$ is, the power of $n$ that survives in the final result is at most 2, compatibly with the EFT predictions. In order to see it, it is useful to derive the expression of $\langle (\text{Tr} \chi)^p \rangle_c$ for generic positive integer $p$.\\
Since, at the perturbative level, the dependence of $X_n^{(0)}$ on $t$ is $X_n^{(0)} \simeq t^{-(n+1)(n+2)}$, then, by using \eqref{eq:general_p_Xn}, we get
\begin{equation}
    X_n^{(p)} =  \, \left(\prod_{l=0}^{p-1} [(n+1)(n+2)+l] \right) \frac{X_n^{(0)}}{t^p}
\end{equation}
Thanks to these relations, we find
\begin{equation}
    \boxed{\langle (\text{Tr}\chi)^p\rangle \equiv \frac{X_n^{(p)}}{X_n^{(0)}} = \frac{\prod_{l=0}^{p-1} [(n+1)(n+2)+l] \,\,X_n^{(0)}}{t^p \,X_n^{(0)}} = \frac{\prod_{l=0}^{p-1} [(n+1)(n+2)+l]}{t^p}}
\end{equation}
from which 
\begin{equation}
 \boxed{\langle (\text{Tr}\zeta)^p \rangle = \prod_{l=0}^{p-1} [(n+1)(n+2)+l]}
\end{equation}
At this point, using \eqref{cumvsvevs2}, we can evaluate the cumulants. Explicitly we can do it up to 4 points, i.e.
\begin{align}
    & \langle \text{Tr}\zeta\rangle_c = \langle \text{Tr}\zeta\rangle = (n+1) (n+2) \notag \\
    & \langle (\text{Tr}\zeta)^2\rangle_c = (n+1)(n+2)((n+1)(n+2)+1)-(n+1)^2 (n+2)^2 = (n+1)(n+2) = \langle \text{Tr}\zeta\rangle_c \notag \\
    & \langle (\text{Tr}\zeta)^3\rangle_c = 2 (n+1)(n+2) \notag \\
    & \langle (\text{Tr}\zeta)^4\rangle_c = 6 (n+1)(n+2)
\end{align}
and we propose the following expression for generic positive integer $p$
\begin{equation}
\label{eq:general_cumulants}
    \boxed{\langle (\text{Tr}\zeta)^p\rangle_c = (p-1)! \, \, (n+1)(n+2)}
\end{equation}
which is quadratic in $n$. This phenomenon, which at this stage may appear somewhat miraculous, will find a natural explanation in our general argument presented in Subsection \ref{subsec:chap2_7_3}.	\\
At this point we can write the perturbative large-$n$ expansion of the logarithm of the ($Z_{S^4}$-rescaled) two-point function
\begin{empheq}[box=\fbox]{align} \label{eq:log_corr_4inst}
   \log \left(Z_{S^4}(\tau,\bar{\tau} G_{2n}(\tau, \bar{\tau}) \right) & =  \log\left(\frac{X_n^{(0)}}{X_{n-1}^{(0)}}\right) + \log \left(4 (2 \pi)^{2 n}\right) + 4 n \log 2 + \left[\frac{q+\overline{q}}{2} (n-3) + \right.\notag \\
   & \left. + \frac{q^2+\overline{q}^2}{8}\left(\frac{9 n - 23}{4} + \frac{13}{4} \frac{n+1}{t}\right) \right]  + \frac{q^3+\overline{q}^3}{48}\left(\frac{19 n - 45}{2} + \frac{23}{2} \frac{n+1}{t} + \right. \notag \\
   & \left.+  \frac{39}{4} \frac{n+1}{t^2}\right) + \frac{q^4+\overline{q}^4}{384} \left(\frac{3 (1257 n - 2839)}{64}  + \frac{8103}{128} \frac{n+1}{t} + \right. \notag \\
   & \left. \left. + \frac{1979}{32} \frac{n+1}{t^2} + 39 \frac{n+1}{t^3} \right)\right]+ 2 (n+1) \sum_{k=1}^4 \frac{1}{k!} \left(\frac{q+\overline{q}}{2 t}\right)^k (k-1)! + \notag \\
   & + \frac{q^2 \overline{q} + q \overline{q}^2}{16} \frac{13}{4} \frac{n+1}{t^2} + \frac{q^3 \overline{q} + q \overline{q}^3}{96} \left(\frac{23}{2} \frac{n+1}{t^2} + \frac{39}{2} \frac{n+1}{t^3}\right) + \notag \\
   & + \frac{q^2 \overline{q}^2}{64} \left(\frac{169}{32} \frac{n+1}{t^2} + 13 \frac{n+1}{t^3}\right)
\end{empheq}
This final expression clearly shows that instanton contributions to the large-$n$ asymptotic expansion of the logarithm of the correlator are either exponentially suppressed in $n$, or limited to the sole coefficients of $n^1$ and $n^0$. Such a structure persists at any instanton order, as we will show in closed form in Subsection \ref{subsec:chap2_7_3}, and it is in perfect agreement with the expectations of the EFT.

\subsection{The coefficients $A$ and $B$} \label{subsec:chap2_7_2}
In the final result of the previous Subsection, \eqref{eq:log_corr_4inst}, we can extract the expressions, corrected up to 4 instantons, for the coefficients $A$ (the term of order $n^1$) and for the coefficient $B$ (the scheme-dependent term of order $n^0$). In fact, the $\tau$-dependence of these coefficients is known exactly: it has been derived in \cite{Hellerman:2020sqj} and \cite{Hellerman:2021yqz} by employing different methods, such as S-duality and EFT techniques. In this Subsection, we will verify that these exact quantities, when expanded to the fourth order, reproduce our results.

\subsubsection{\boldsymbol{$A(\tau,\bar{\tau})$}}
The authors of \cite{Hellerman:2020sqj} derived the following exact expression for the coefficients $A$ as a function of the UV coupling $\tau$ (which is the one we have considered: recall that $t = 4 \pi \text{Im}\tau + 8 \log2$)
\begin{equation} \label{eq:A_Hellerman}
    A(\tau,\bar{\tau}) = \log \, F'(\tau) + \log \, \overline{F'(\tau)} - \log \left[-4 \left(F(\tau)-\overline{F(\tau)}\right)^2\right]
\end{equation}
where $F(\tau)$ is the inverse of the modular lambda function, which gives the UV coupling in terms of the IR one $\sigma$ (as called in that reference, but we indicated it as $2 \tau_\text{R}$), $q = \lambda(\sigma)$ (look at \eqref{eq:F0_expression_coupling})
\begin{equation} \label{eq:F_tau_Hellerman}
    F(\tau) = 2 \tau + \frac{4 i}{\pi} \log(2) - \frac{i}{\pi} \left( \frac{q}{2} + \frac{13}{64} q^2 + \frac{23}{192} q^3 + \frac{2701}{32768} q^4 + \mathcal{O}(q^5)\right)
\end{equation}
We will provide an independent derivation of formula \eqref{eq:A_Hellerman} in Subsection \ref{subsec:chap2_7_3}. For now, let us perform its instanton expansion and compare it with the coefficient of $n^1$ in \eqref{eq:log_corr_4inst}. Denoting by $A^{(k)}(\tau, \bar{\tau})$ the $k$th instanton contribution and recalling that $t \equiv 4 \pi \text{Im} \tau + 8 \log 2$, we find the following

\begin{align}\label{Eq:As}
A^{(0)}(\tau,\bar\tau)=&2\log\frac{\pi}{t}\,,\nonumber\\
A^{(1)}(\tau,\bar\tau)=&\frac{q+\overline{q}}{2}\left(1+\frac{2}{t}\right)\,,\nonumber\\
A^{(2)}(\tau,\bar\tau)=&\frac{q^2+\overline{q}^2}{32}\left(9+\frac{13}{t}\right)+\left(\frac{q+\overline{q}}{2t}\right)^2\,,\nonumber\\
A^{(3)}(\tau,\bar\tau)=&\frac{q^3+\overline{q}^3}{96}\left(19+\frac{23}{t}+\frac{39}{2t^2}\right)+q\overline{q}(q+\overline{q})\frac{13}{64t^2}+\frac{2}{3}\left(\frac{q+\overline{q}}{2t}\right)^3\,,\nonumber\\
A^{(4)}(\tau,\bar\tau)=&\frac{q^4+\overline{q}^4}{8192}\left(1257+\frac{2701}{2t}+\frac{3958}{3t^2}+\frac{832}{t^3}\right)+\frac{q\overline{q}(q^2+\overline{q}^2)}{64}\left(\frac{23}{3t^2}+\frac{13}{t^3}\right)\nonumber\\
&+\,\frac{q^2\overline{q}^2}{64}\left(\frac{169}{32t^2}+\frac{13}{t^3}\right) +\frac{1}{2}\left(\frac{q+\overline{q}}{2t}\right)^4\,.
\end{align}
It is immediate to verify that these expressions are in perfect agreement with the results in Subsection \ref{subsec:chap2_7_1}.\footnote{Note that, to extract the $0$-instanton contribution to $A(\tau,\bar\tau)$ from \eqref{eq:log_corr_4inst}, we have to subtract from the same instanton sector contribution the quantity $4\log2+2\log2-2$, which comes from the explicit $4\log2$ and from the expansion of $\Gamma\left(2n+\frac{5}{2}\right)$ in \eqref{eq:log_corr_general} with $d=2$.}

\subsubsection{$\boldsymbol{B(\tau,\bar{\tau})}$}
For the coefficient of $n^0$, the authors of \cite{Hellerman:2021yqz} introduced the quantity $\tilde{B}(\tau, \bar{\tau}) = B(\tau,\bar{\tau}) - \log Z_{S^4}$ \footnote{Actually, the authors used for the second quantity the name $\log Z_{\text{Pestun-Nekrasov}}$, but it turns out that our $Z_{S^4}$ in \eqref{eq:part_func_4sphere} is exactly $Z_{\text{Pestun-Nekrasov}}$. Let us study, under the weak coupling limit, our $Z_{S^4}$ in \eqref{eq:part_func_4sphere}. Since we are in the limit $\text{Im}\tau \to \infty$, we can neglect all the instantons ($Z_{\text{inst}}=1$) and then the largest part of the contribution to the partition function comes from the limit $a \to 0$, under which the 1-loop partition function has the following behaviour 
\begin{equation*}
    Z_{\text{1L}}(i a) \underset{a \to 0}{=} 1 - 12 \, \zeta(3) \, a^4 + 40 \, \zeta(5) \, a^6 + \mathcal{O}(a^8)
\end{equation*}
In this way \eqref{eq:part_func_4sphere} becomes
\begin{equation*}
    Z_{S^4}(\tau, \bar{\tau}) = \frac{1}{4 \pi (\text{Im}\tau)^\frac{3}{2}} \left[1 - \frac{45 \, \zeta(3)}{16 \pi^2 \, (\text{Im} \tau)^2 } + \frac{525 \, \zeta(5)}{64 \pi^3 \, (\text{Im}\tau )^3} + \mathcal{O}\left[(\text{Im} \tau)^{-4}\right]\right] + \mathcal{O}(q)
\end{equation*}
and it coincides with (B.25) of \cite{Hellerman:2021yqz}}, free of the counterterm ambiguities of the path integral that plague the partition function. Then they computed $B(\tau,\bar\tau)$ in a scheme where the $S^4$ partition function is the one used by AGT \cite{Alday:2009aq}, whose instanton contributions do not contain the $U(1)$ factor of the Nekrasov sum \cite{Nekrasov:2002qd}. Translated to the Pestun-Nekrasov scheme \cite{Pestun:2007rz}, their expression reads

\begin{align} \label{eq:B_Hellerman}
B(\tau,\bar{\tau}) = \tilde{B}(\tau, \bar{\tau}) + \log Z_{S^4} & = 12 \log \gamma_G - 1 - \frac{9}{2} \log 2 - \frac{3}{2} \log \pi + \frac{1}{3}( \log q + \log \bar{q}) + \frac{4}{3} (\log(1-q) + \log(1-\bar{q})) + \notag \\
   &- 2 \log\left[\frac{F(\tau)-\overline{F(\tau)}}{2 i}\right] - 4 \left[\log \eta\left(F(\tau)\right) + \log \overline{\eta \left(F(\tau)\right)}\right]
\end{align}
where $\eta$ is the Dedekind eta function, $Z_{S^4}$ and $F(\tau)$ are defined in \eqref{eq:part_func_4sphere} and \eqref{eq:F_tau_Hellerman}, respectively.\\
Let us now perform the instanton expansion of \eqref{eq:B_Hellerman} and compare it with the coefficient of $n^0$ in \eqref{eq:log_corr_4inst}. Denoting by $B^{(k)}(\tau,\bar{\tau})$ the $k$th instanton contribution and recalling that $t \equiv 4 \pi \text{Im} \tau + 8 \log 2$, we find the following 

\begin{align}\label{Eq:Bs}
B^{(0)}(\tau,\bar\tau)=& 12 \log \gamma_G  + \frac{1}{6}\log 2 + \frac{1}{2}\log \pi - 2 \log t -1 \,,\nonumber\\
B^{(1)}(\tau,\bar\tau)=&\frac{q+\overline{q}}{2}\left(-3 + \frac{2}{t}\right) \,,\nonumber\\
B^{(2)}(\tau,\bar\tau)=&\frac{q^2+\overline{q}^2}{32}\left(-23+\frac{13}{t}\right)+\left(\frac{q+\overline{q}}{2t}\right)^2\,,\nonumber\\
B^{(3)}(\tau,\bar\tau)=&\frac{q^3+\overline{q}^3}{96}\left(-45+\frac{23}{t}+\frac{39}{2t^2}\right)+q\overline{q}(q+\overline{q})\frac{13}{64t^2}+\frac{2}{3}\left(\frac{q+\overline{q}}{2t}\right)^3\,,\nonumber\\
B^{(4)}(\tau,\bar\tau)=&\frac{q^4+\overline{q}^4}{8192}\left(-2839+\frac{2701}{2t}+\frac{3958}{3t^2}+\frac{832}{t^3}\right)+\frac{q\overline{q}(q^2+\overline{q}^2)}{64}\left(\frac{23}{3t^2}+\frac{13}{t^3}\right)\nonumber\\
&+\,\frac{q^2\overline{q}^2}{64}\left(\frac{169}{32t^2}+\frac{13}{t^3}\right) +\frac{1}{2}\left(\frac{q+\overline{q}}{2t}\right)^4\,.
\end{align}
It is immediate to verify that these expressions are in perfect agreement with the results in Section \ref{subsec:chap2_7_1}.\footnote{Note that, to extract the $0$-instanton contribution to $B(\tau,\bar\tau)$ from \eqref{eq:log_corr_4inst}, we have to subtract from the same instanton sector contribution the quantity $\frac{5}{2}\log2+\frac{1}{2}\log\pi$, which comes from the expansion of $\Gamma\left(2n+\frac{5}{2}\right)$ in \eqref{eq:log_corr_general} with $d=2$.}

\subsection{All-order derivation}\label{subsec:chap2_7_3}
Motivated by the order-by-order analysis don in the previous Sections, we now truncate the free energy \eqref{eq:genus_expansion_prep} and just retain the terms $g=0$ and $g=1$, which are the leading ones in the large-VEV regime. What makes this truncation particularly useful is that exact expressions for both $\mathcal{F}_0$ and $\mathcal{F}_1$ are known. Starting from $\mathcal{F}_0$, recall from \eqref{eq:F0_expression_coupling} that it can be written as
\begin{equation}
\mathcal{F}_0 = 2 \pi i \tau_{\text{R}} = 2 \pi i \, F(\tau)
\end{equation}
Notice that $F(\tau)$ encodes the only data of the Seiberg-Witten theory of SU(2) SQCD with four fundamental hypermultiplets which cannot be fixed by dimensional analysis.\\
As for $\mathcal{F}_1$, we wrote its explicit expression in \eqref{eq:u_plane_formula_F1}, which can be extracted from the geometry of the Seiberg-Witten curve \cite{Moore:1997pc} (see also \cite{Manschot:2019pog}) and that we report here for simplicity:
\begin{equation}
\mathcal{F}_1(u) = -\frac{1}{2} \log \left(\frac{\partial a}{\partial u} \right) + \frac{1}{12} \log \Delta(u)
\end{equation}
where $\Delta(u)\propto u^6$ is the physical discriminant of the Seiberg-Witten curve. We specifiy that this expression of the $\mathcal{F}_1$ component naturally incorporates also the contribution of the Vandermonde determinant in the 4-sphere partition function in \eqref{eq:expression_ZS4_genus} (the factor $4 a^2$). We need to express this $\mathcal{F}_1(u)$ in terms of $a$. From the Matone's relation, we obtain
\begin{equation}
u = - q \frac{\partial \mathcal{F}_0}{\partial q} = - \frac{1}{2 \pi i} \frac{\partial \mathcal{F}_0}{\partial \tau} = - F'(\tau)
\end{equation}
In this way, neglecting irrelevant additive numerical constants \footnote{These constants correspond to rescaling the partition function by pure numbers, which does not affect any scheme-independent quantity, and hence any observable.}, we arrive at
\begin{equation}
\mathcal{F}_1 = \frac{3}{2} \log \, a + \log \, F'(\tau) + \text{const}
\end{equation}
With the above quantities, we can write the truncated version of the 4-sphere partition function \eqref{eq:expression_ZS4_genus}, which takes the form
\begin{equation}
Z_{S^4} |_{g=0,1} = \int_{\mathbb{R}} \, \mathrm{d}a \, e^{2 \pi i a^2 \left(F(\tau) - \overline{F(\tau)}\right)} \, e^{3 \log a + \log \, F'(\tau) + \log \, \overline{F(\tau)} + \text{const}} \propto \frac{\left|F'(\tau)\right|^2}{\left(F(\tau)-\overline{F(\tau)}\right)^2}\;=\;\partial_\tau\partial_{\bar\tau}\log\left(F(\tau)-\overline{F(\tau)}\right)
\end{equation}
where again, in the second step, we neglected irrelevant multiplicative numerical constants. Note that the large-VEV truncation crucially makes the integral in $a$ trivial to perform, yielding simply a function of $\tau$. The latter, in turn, can be interpreted geometrically as the pull-back through $F(\tau)$ of the hyperbolic metric on the IR upper-half-plane, whose K\"ahler potential is $\log{\rm Im}F(\tau)$, as shown in the last equality. This peculiarity translates into remarkable features of the corresponding matrix of derivatives \eqref{eq:correlation_function_S4}. In particular, its determinant can be written as 
\begin{equation} \label{eq:det_M_trunc}
\det M^{(n)}(\tau,\bar\tau)|_{g=0,1} = h_{n+1}\left(\frac{\left|F'(\tau)\right|}{F(\tau)-\overline{F(\tau)}}\right)^{n(n+1)}\,,
\end{equation}
where $h_{n+1}$ is a number satisfying the following recursion relation
\begin{equation}\label{eq:recurs}
h_{n+1}=(-1)^n\,n!\,(n+1)!\,h_n\,,\qquad \quad h_1\equiv1\,.
\end{equation}
Equation \eqref{eq:det_M_trunc} shows that, in computing the determinant, all derivatives of $F(\tau)$ higher than the first one drop out! Therefore, the result ends up being proportional to a power of the partition function itself. Moreover, such a power is only quadratic in $n$, because it originates from taking $\sum_i^n i$ holomorphic derivatives of the K\"ahler potential and just as many anti-holomorphic ones. This explains why in our order-by-order analysis all powers of $n$ higher than two cancel out in the cumulants.\footnote{Recall that, from the matrix-model perspective, each $t$-derivative produces two powers of $n$, as is manifest from \eqref{eq:Xn_z}.}\\
Using \eqref{eq:final_correlator}, \eqref{eq:det_M_trunc}, and \eqref{eq:recurs}, the logarithm of the two-point function $G_{2n}(\tau, \bar{\tau})$ reads
\begin{align} \label{eq:G_2n_exact}
\log G_{2n}(\tau,\bar{\tau}) & = \log\left(\Gamma(n+1) \Gamma(n+2)\right)\nonumber\\
&+ n\left\{4\log2+\log F'(\tau) + \log\overline{F'(\tau)}-\log\left[-\left(F(\tau)-\overline{F(\tau)}\right)^2\right]\right\}
\end{align}
Note that this expression is also exact in $n$, because here, as opposed to the previous Sections (see e.g. \eqref{eq:log_corr_4inst}), we did not include the full one-loop partition function, but only its large-VEV portion. Comparing \eqref{eq:G_2n_exact} with the large-charge expansion \eqref{eq:log_corr_general} with $d=2$, which defines the coefficients $A(\tau,\bar{\tau})$ and $\tilde{B}(\tau,\bar{\tau})$, we see that the former coincides with the exact expression \eqref{eq:A_Hellerman} found in \cite{Hellerman:2020sqj}, whereas the latter turns out to be independent of $\tau$:
\begin{equation}\label{eq:tildeB_exact}
\tilde{B}(\tau,\bar\tau)|_{g=0,1}=\frac{1}{2}\log\frac{\pi}{8}
\end{equation}
It is important to remark, however, that, while $A(\tau,\bar\tau)$ receives no contributions from small VEVs, \eqref{eq:tildeB_exact} only gives the large-VEV piece of $\tilde{B}(\tau,\bar\tau)$. Thus, we conclude that the non-trivial $\tau$ dependence of $\tilde{B}(\tau,\bar\tau)$ originates exclusively from the small-VEV sector of the partition function.

\section{Outlook on Argyres-Douglas theories}\label{sec:chap2_8}
Given the very limited role that instantons play at large charge in the scale-invariant SU(2) SQCD, we would be tempted to speculate that the same holds in the asymptotically-free cases (one, two, and three fundamental flavors), which, at special values in their (extended) Coulomb branch, we saw that they flow to the three rank-one Argyres-Douglas SCFTs (see e.g.~\cite{Argyres:1995xn}). If this were literally the case, the one-loop Nekrasov partition function of the gauge theory would be sufficient to read off the universal behavior at large charge of the correlator of the corresponding Argyres-Douglas theory, $\log \langle \mathcal{O}^n(0) \, \overline{\mathcal{O}}^n (\infty)\rangle \underset{n \to \infty}{\simeq} \log \Gamma (d n+\alpha+1)+ \dots$, which, expressing $\alpha$ in terms of $d$, becomes\footnote{Recall that, in the context of gauge theory, $d$ emerges as the rational number controlling the $a$ dependence of the VEV of the Coulomb-branch operator $\mathcal{O}$ around the Argyres-Douglas point $a_*$, i.e. $\langle\mathcal{O}\rangle \sim(a-a_*)^d$.}
\begin{equation} \label{eq:G_2n_AD}
G_{2n}|_{\rm universal} \underset{n \to \infty}{\simeq} \Gamma\left(dn+\frac{3d-1}{2}\right)\,.
\end{equation}
 
However, such a circumstance is unlikely to happen because, as opposed to the conformal case, gauge-theory instantons are decisive to determine how the large-radius approximation of the free energy, and in particular $\mathcal{F}_1$, depends on $a$ at strong coupling \cite{Moore:1997pc}.\footnote{For example, in the case of one flavor, $\mathcal{F}_1$ tends to a constant in the weakly-coupled region $a\to\infty$, whereas it goes like $\frac{3}{10}\log(a-a_*)$ around the Argyres-Douglas point $a_*$. In the conformal case, instead, the instanton sum only contributes an $a$-independent addend in $\mathcal{F}_1$.} The large-radius free energy, in turn, enters prominently the leading behavior of the correlator at large charge. Indeed,  by truncating the free energy to just $\mathcal{F}_0$ and $\mathcal{F}_1$, analogously to what we did in Subsection \ref{subsec:chap2_7_3},\footnote{For $d=2$, we are actually neglecting instantons and focusing only on the tree-level and one-loop parts of $\mathcal{F}_0$ and $\mathcal{F}_1$. In this case, taking derivatives w.r.t.~the coupling corresponds to simply inserting powers of the Coulomb-branch operator, whose normalization here we are careless about. Recall what already said in Section \ref{sec:non_lag_theories}.} one gets for the matrix \eqref{eq:correlation_function_S4} \cite{Bissi:2021rei}
\begin{equation} \label{eq:Mn_AD}
M^{(n)}_{i j} \propto \Gamma\left(\frac{d}{2}(i+j+3)-1\right)\,,
\end{equation}
where we neglected all numerical factors in front, as well as the normalization by the partition function, which do not alter the universal behavior of the correlator at large charge. It is possible to show, numerically, that the ratio of determinants \eqref{eq:final_correlator}, with $M^{(n)}$ given by \eqref{eq:Mn_AD}, has the same leading universal behavior of \eqref{eq:G_2n_AD}, i.e. Exp$[(dn+\frac{3}{2}d-1)\log n]$, while it fails to reproduce the subleading terms $\mathcal{O}(e^{1/n})$.

The situation at small radius is radically different. Here, the instanton partition function can be organized as a power series in the small quantity $\Lambda R$, where $\Lambda$ is the dynamically-generated scale of the gauge theory and $R$ is the sphere radius
\begin{equation}
Z_{\rm inst}=1+\sum_{k=1}^\infty v_k(a,m,R) (\Lambda R)^k\,,
\end{equation}
where $v_k(a,m,R)$ are dimensionless functions, regular in the limit $R\to0$. As a consequence, there is no instanton contribution to the leading $R^0$ order.\footnote{Instead, in the conformal case, the instanton sum yields again an $a$-independent function of $\tau$ at order $R^0$, which however, as we learned in Section \ref{sec:chap2_6}, produces only exponentially-suppressed contributions in the large-$n$ $\log(Z_{S^4} G_{2n})$.} Concerning the one-loop partition function, it turns out that all three asymptotically-free gauge theories are such that $Z_{\rm 1L}=1+\mathcal{O}(R^2)$ for $R\to0$. This implies that, around the Argyres-Douglas point, the reduced partition function of the gauge theory $\hat{Z}\equiv e^{-\mathcal{F}_0 \, a^2 R^2}Z$ goes to a constant in the limit $R\to0$.\footnote{This can be independently confirmed by analyzing the partition function of Argyres-Douglas theories via the holomorphic anomaly equation \cite{Fucito:2023txg}.}

The above discussion leads us to an educated guess for a function \emph{interpolating} between the large-radius and the small-radius regimes of the partition function
\begin{equation} \label{eq:guess}
|\hat{Z}(a)|^2|_{g=0,1,\infty} \sim  (a^2 + {\rm const})^{\frac{3}{2}(d-1)}\qquad\; d\neq2\,,
\end{equation}
where $g=\infty$ indicates the leading contribution at small radius. In the same spirit as what was done in \cite{Grassi:2019txd} for the conformal case,\footnote{Note that in the conformal case, $d=2$, we have the Vandermonde term $a^2$ at small radius, which changes \eqref{eq:guess} to $|\hat{Z}(a)|^2|_{g=0,1,\infty}\sim a^2 \sqrt{a^2 + t}$ (see \eqref{eq:Xn_z}).} it would be very interesting to verify whether \eqref{eq:guess} leads to two-point functions whose universal behavior at large charge fully matches the EFT prediction \eqref{eq:G_2n_AD}. We expect such a behavior to be completely independent of the specific value of the constant appearing in \eqref{eq:guess}. We leave this question to future study.\\

Regarding the non-universal behavior in the perturbative large-charge expansion, i.e. the coefficients $A$ and $\tilde{B}$, we can actually use 
\eqref{eq:Mn_AD} to compute them for other rank-one SCFTs, such as the three Argyres-Douglas theories \cite{Argyres:1995xn}, $\mathcal{H}_0,\mathcal{H}_1,\mathcal{H}_2$, for which $d=6/5,4/3,3/2$ respectively, and the three Minahan-Nemeschansky theories \cite{Minahan:1996cj}, $E_6,E_7,E_8$, for which $d=3,4,6$ respectively. All of them share the feature of being isolated models, namely they do not have exactly-marginal parameters. Therefore, for these SCFTs, the coefficients $A$ and $\tilde{B}$ are pure numbers. The results of the present paper for the SQCD case ($d=2$) leads us to argue that what we will find for the coefficient $A$ in the other SCFTs is the complete result (albeit dependent on the normalization of the Coulomb-branch operator), whereas we expect that the result for $\tilde{B}$ will be modified by the inclusion of small VEVs.

To set up the computation, we take
\begin{equation}
Z_{S^4}=\int_\mathbb{R}{\rm d}a \,e^{-a^2}a^{3(d-1)}\,,
\end{equation}
where we used that $\mathcal{F}_1=\frac{3}{2}(d-1)\log a$, modulo an irrelevant additive constant. Then we fix a convenient normalization of the Coulomb-branch operator, such that $u=a^d$, and consider the matrix\footnote{For $d=2$, this normalization corresponds to fixing the loop-renormalized coupling to the value $t=2\pi$.}
\begin{equation}\label{eq:matricesAD}
M^{(n)}_{ij} = \frac{1}{Z_{S^4}}\int_\mathbb{R}{\rm d}a\, a^{d(i+j)}\,e^{-a^2}a^{3(d-1)}\;=\;\frac{\Gamma\left(\frac{d}{2}(i+j+3)-1\right)}{\Gamma\left(\frac{3d}{2}-1\right)}\,.
\end{equation}
For $d\neq2$, to the best of our knowledge, the ratio of determinants \eqref{eq:final_correlator} for the matrices \eqref{eq:matricesAD} can only be computed numerically. As mentioned above, such a ratio shares with the EFT prediction \eqref{eq:G_2n_AD} the same coefficients of $n\log n$ and $\log n$, differing only for terms that vanish for $n\to\infty$. Therefore, we can compute $A$ using the following sequence\footnote{We ignore the factor $16^n$ in \eqref{eq:final_correlator}.}
\begin{equation} \label{eq:A_sequence}
   A = \lim_{n \to \infty} A_n\,, \qquad  A_n = \frac{\log (G_{2n}) - \log \Gamma \left(d n + \frac{3(d-1)}{2} \right)}{n}\,,
\end{equation}
and, using the obtained value, we can compute $\tilde{B}$ using the following sequence

\begin{equation} \label{eq:B_sequence}
   \tilde{B} = \lim_{n \to \infty} \tilde{B}_n\,, \qquad   \tilde{B}_n = \log (G_{2n}) - \log \Gamma \left(d n + \frac{3(d-1)}{2} \right)- A \, n \,.
\end{equation}

A useful algorithm that accelerates the convergence is the so-called ``Richardson extrapolation'',\footnote{We thank Alba Grassi for suggesting this method.} which is defined as follows. Let $\{a_n\}_{n \ge 1}$ be a sequence, its Richardson extrapolation of order $N$ is
\begin{equation}
    R_m^N[\{a_n\}_n] = \sum_{k=0}^N \frac{(-1)^{N+k} \, (k+m)^N \, a_{k+m}}{k! (N-k)!}\,.
\end{equation}
Once numerical stability is reached to a chosen precision, by changing the parameters $(N,m)$, one approximates $\lim_{n}a_n$ with $R_m^N[\{a_n\}_n]$. Here we set $N=10$. Then we notice that, if we increase $d$, we should also increase $m$ in order to reach numerical stability. In particular, for the three Minahan-Nemeschansky models, the choice $m=200$ allows us to have stability for the first $9$ digits for the coefficient $A$ and $7$ digits for the coefficient $\tilde{B}$.\footnote{Note that the precision for $\tilde{B}$ is lower because it depends on the numerical value of $A$, which is itself subjected to an error.} In contrast, for the three Argyres-Douglas theories, choosing $m=10$ already ensures stability to the same level of approximation. In Table \ref{Tab:A_B} we report our approximation for $A$ and $\tilde{B}$ up to the $7$th decimal digit.
\begin{table}[H]
\centering
\begin{tabular}{|c|c|c|c|c|c|c|}
\hline
 & $\mathcal{H}_0$ & $\mathcal{H}_1$ & $\mathcal{H}_2$ & $E_6$ & $E_7$ & $E_8$ \\ \hline
$A$    & $-1.6904874$    & $-1.6298583$ & $-1.5605388$ & $-1.1373603$ & $-0.9662579$ & $-0.7444107$  \\ \hline
$\tilde{B}$    & $0.1507129$    & $0.1020099$  & $0.0044435$ & $-2.0144218$ & $-4.1600920$ & $-9.6973840$   \\ \hline
\end{tabular}
\caption{Values of $A$ and $\tilde{B}$ for the rank-one Argyres-Douglas and Minahan-Nemeschansky theories, extracted via the Richardson extrapolation.}\label{Tab:A_B}
\end{table}

As a crosscheck, we perform this numerical computation also for the SQCD case, i.e.~for $d=2$, where we have analytic expressions, recalling that \eqref{eq:matricesAD} corresponds to the regime of large VEVs and perturbative in the gauge coupling. We find $A=-1.3862944$ and $\tilde{B}=-0.4673558$. The former approximates $A=-2\log2$, which is indeed the $0$-instanton value of $A$ displayed in \eqref{Eq:As}, if we set $t=2\pi$ to account for the peculiar normalization of the Coulomb-branch operator we have chosen here. The latter approximates $\tilde{B}=\tfrac12\log\tfrac\pi8$, which is indeed the coupling-independent, large-VEV value of $\tilde{B}$ shown in \eqref{eq:tildeB_exact}.
\chapter{Stability under Charged Perturbations and Super radiance of Topological Stars}\label{chap_3}
The present Chapter is the first in which we apply the techniques we developed in Chapter 1 for $\mathcal{N}=2$ supersymmetric gauge theories to the realm of gravity. The objects whose dynamical properties we want to investigate are the so called \textit{fuzzballs} \cite{Mathur:2005zp,Bena:2022rna,Bena:2022ldq,Mathur:2024ify,Lunin:2001jy, Mathur:2009hf,Bambi:2025wjx,Lunin:2001fv}, according to which black holes should be thought as the superposition of quantum states, some of which admit a classical description as smooth and horizonless gravity solutions with the same conserved quantities (mass, charge and angular momentum) of the black hole. Remarkably, these geometries emit radiation preserving quantum information, solving one of the most important puzzles that come out when putting together General Relativity with Quantum Mechanics. In particular we focus on the recently discovered \textit{Topological Star} geometry of Bah and Heidmann \cite{Bah:2020ogh, Bah:2020pdz, Heidmann:2021cms, Bah:2021owp, Heidmann:2022zyd}. The importance of this geometry relies on the fact that it is characterized by two parameters in such a way that, depending on their relative values, a Topological Star behaves as a black  (i.e. with an event horizon) or an horizonless object. In this context, following the new results of \cite{Cipriani:2024ygw}, first we send a massive scalar charged perturbation (in the guise of a string) against the Topological Star and see how it reacts. This information is codified in the Quasi-Normal Modes (QNMs), that are complex frequencies $\omega_{\text{QNM}}$ that enter the response through the exponential factor $e^{- i \omega_{\text{QNM}} t}$. Using several methods, including the one derived from the SW curves analyzed in Chapter \ref{chap_1}, we see that the imaginary part of $\omega_{\text{QNM}}$ is negative, which is a signal of the stability of this object under that kind of perturbation. The other phenomenon that we investigate is the \textit{superradiance}: it consists in the amplification of the wave scattered off the geometry under consideration, producing a decrease of the energy and charge of the central object. This takes place only when the central object admits an horizon. We find that, when the Topological Star behaves as a horizonless geometry, the superradiance is not present, while when it is a black object, then such phenomenon happens and everything is in agreement with the information paradox.\\
The organization of this Chapter is the following. In Section \ref{sec:intro} we give an introduction to the concepts of Fuzzballs and Topological Stars (TSs), showing how they allow to solve the puzzles cited before. In Section \ref{Top star} we recall the basic elements of the TS solution and discuss the coupling of its electric field with electrically charged strings. In Section \ref{quantumSW} we review the theory of quantum SW curves for $\mathcal{N} = 2$ SYM with gauge group SU(2), adapting it to the case of a TS background. Section \ref{Leaver} illustrates another method to compute QNMs for the TS, the so called Leaver method. The frequencies of the charged QNMs of a TS calculated with these two methods are contained in Section \ref{TopstarQNM} and compared with those arising from a numerical integration. Section \ref{Charge superradiance} investigates charge superradiance for the TS in the regime of a black object. We compute the near-extremal near-superradiant QNMs and the prompt ring-down modes and finally the amplification factor is defined and corrected with the instantonic contributions. The outcome is compared with a numerical prediction.

\section{Introduction to Fuzzballs and Topological Star (TS)} \label{sec:intro}
Black holes (BH) are nowadays regarded as the ideal playground to focus on when attempting to solve the lasting clash between Quantum Mechanics (QM) and General Relativity (GR). Indeed, when we consider the quantum theory of black holes, there are mainly three puzzles that spontaneously arise and that are strongly entangled. We are going to list and analyze them, in such a way it is clearer how String Theory resolves all of them through the introduction of the Fuzzball paradigm. We mainly focus on \cite{Mathur:2005zp,Bena:2022rna,Bena:2022ldq,Mathur:2024ify}, but other interesting and useful references are \cite{Lunin:2001jy, Mathur:2009hf,Bambi:2025wjx}.

\subsection{The puzzles}
At the classical level (i.e. in the original formulation of GR by Einstein) it is largely known that BHs are objects whose gravity is so strong that nothing, not even light, matter and any other kind of information, can escape from the so called event horizon, which is a surface of no return that surrounds a given BH. For this reason, the horizon region is a vacuum, since everything close to that surface is moved to the center of the BH, which is at its core. This implies that, not only the exterior of a BH (i.e. the region outside the horizon) is independent from how the BH formed and how it originated, but also that it is completely determined by three parameters, which are mass, charge and angular momentum. It is commonly said that "BHs have no hair" to refer to this feature, which states a BH uniqueness.\\
Later on, it was discovered by Bekenstein \cite{PhysRevD.7.949} that BHs are akin to thermodynamical systems, in the sense that they obey laws similar to the ones of Thermodynamics. For this reason they possess an associated entropy-like quantity in the form of the event horizon’s area, since classical computations in GR show that such area never decreases during any process. exactly as the entropy of a thermodynamical system.\\

Troubles arise when this classical BH physics in GR is combined with QM. producing several puzzles that we want to recall here.
\begin{enumerate}
\item \textbf{The entropy puzzle}: in \cite{PhysRevD.7.2333} Bekenstein made more explicit the relation between the entropy of a BH with its horizon's area, using QM. In particular
\begin{equation} \label{eq:bek_entropy}
S_{\text{Bek}} = \frac{c^3}{\hbar} \frac{A}{4 G}
\end{equation}
with $A$ the area of the BH's horizon. We should also recall that for a normal system described by statistical mechanics the entropy is related to the number of microstates of that system; in the present case $S_{\text{Bek}} = \log N_{\text{micro}}$, where $N_{\text{micro}}$ is such number of microstates. Due to the BH uniqueness we cited above, the natural question that arises is what the entropy \eqref{eq:bek_entropy} is counting. For example, the BH at the core of the Milky Way should have about $e^{10^{90}}$ microstates, while, from the outside, the "BH have no hair" statement implies that its states is unique. This produces that huge problem that $e^{10^{90}} \ne 1$. So, we can summarize this puzzle by asking: \textbf{does the entropy \eqref{eq:bek_entropy} correspond to a count of BH microstates as for a normal system described by statistical mechanics?}
\item \textbf{The information paradox}: this is one of the biggest puzzles and its origin simply relies on the fact that gravity is attractive. In order to understand it, it might be useful to start with an easy situation described by Special Relativity (SR), simply hinting at the end of this how GR calculation corrects the final prediction of SR and seeing how problems arise. Consider a BH with mass $M$ and a test mass $m$ at a distance $r$ from the centre of the BH. The total energy of the probe is given by its intrinsic energy $m c^2$ and the negative potential energy, hence we can schematically write
\begin{equation}
E \simeq m c^2 - \frac{G M m}{r} 
\end{equation}
It is clear that for distances smaller than $G M/c^2$ the total energy of the probe is negative. A GR computation leads to the slightly more precise result that this total energy is negative at distances
\begin{equation}
r < \frac{2 G M}{c^2} \equiv R_s
\end{equation}
where $R_s$ is the famous Schwarzschild readius, i.e. the horizon radius. Notice that the energy that we are considering is the one measured from infinity.\\
Until this point nothing strange takes place and this simply because we have considered a classical treatment of the situation. If we insert QM, there is a problem that spontaneously arises. We said before that the horizon region is a vacuum, but, when looking at it from a quantum point of view, this always contains particle anti-particle fluctuations. Thanks to this fact, Hawking realized that the situation described above leads to an instability of the quantum vacuum \cite{Hawking1975}. For example, an electron-positron pair with energy $\Delta E$ can emerge from the vacuum and then disappear after a time $\Delta t$ in such a way that $\Delta E \Delta t \lesssim \hbar$. If we consider these fluctuations close to the horizon of a BH, one member of the pair can be inside the horizon, hence having negative energy, and the other outside, with positive energy. The total energy of the pair is hence equal to $\Delta E = 0$. The uncertainty relation of above leads to $\Delta t \lesssim \infty$, which means that the pair does not annihilate. Consequently, the outer member of the pair goes to infinity as "Hawking radiation", while the inner member, carrying negative energy, lowers the mass of the BH by a corresponding amount. This means that the BH slowly evaporates, with a radiation described by a temperature $T_{\text{Haw}} = 1/(4 \pi R_s) = 1/(8 \pi G M)$ (we have set here $c=1$ for simplicity).\\

At a first glance, the presence of Hawking radiation seemed to be consistent with Bekenstein entropy. Indeed, the thermodynamical relation $T \, \mathrm{d}S = \mathrm{d}E$ demands that BHs should have a temperature different from 0, which is what was found by Hawking. In particular, the principle of detailed balance requires that if BHs absorb (which they do), then they must also radiate, but this is only a quantum effect (indeed the entropy is inversely proportional to $\hbar$). After this connection, nonetheless Hawking realized that the process by which the radiation is produced leads to a serious contradiction \cite{PhysRevD.14.2460,Hawking1982}. Let us see it. The two members of the pair are in an entangled state and the precise value of the entanglement is not relevant, so we can just model the entangled state as
\begin{equation}
\ket{\psi}_{\text{pair}} = \frac{1}{\sqrt{2}} \left(\ket{0}_o \ket{0}_i + \ket{1}_o \ket{1}_i\right)
\end{equation}
where $o$ refers to the outer member of the pair, $i$ to the inner one. Here we are assuming that $\{\ket{0}_o, \ket{1}_o\}$ and $\{\ket{0}_i, \ket{1}_i\}$ are orthonormal states for the subsystem formed by the outer and inner particles, respectively.\\
Now, given a system with subsystems $A$ and $B$ whose states $\{\ket{u_i}_A\}$ and $\{\ket{v_i}_B\}$ (with $i=1,\dots,m$) form an orthonormal basis, then the entanglement entropy for the state $\ket{\Psi} = \sum_{i=1}^m \alpha_i \ket{u_i}_A \ket{v_i}_B$ is
\begin{equation} \label{eq:ent_entropy}
S_{\text{ent}}(\ket{\Psi}) = - \sum_{i=1}^m \alpha_i^2 \log (\alpha_i^2)
\end{equation}
Applying this definition to the present case, we see that $S_{\text{ent}}(\ket{\psi}_{\text{pair}}) = \log(2)$ and hence each emission step increases the entanglement entropy of the radiation with the remaining BH by $\log(2)$. The main problem of this takes place near the endpoint of evaporation. Indeed, assume (as Hawking did) that the BH completely evaporates. Then the total final radiation is entangled, but there is nothing that it is entangled with. Consequently the entire system cannot be described by a wavefunction, but only by a density matrix. In other words, it is impossible to reconstruct the interior state of a BH (apart from mass, charge and angular momentum) from the exterior data and thus from the final state of the Hawking radiation. The evaporation process cannot be represented through a unitary transformation of states in a Hilbert space. Summarizing, BH evaporation, as predicted by GR and QM, is inconsistent with the foundational postulate of unitarity of QM. This is called the BH information paradox.\\
This problem is so huge that, a possible explanation of the entropy puzzle we explained before, remains still present. In fact, for what concerns the previous puzzle, one may think that the difference between the number of microstates of the BH with the one coming from the BH uniqueness could be found by looking at a Planck sized neighborhood of the singularity and these differences are thus not visible in the classical description. After all, the matter that made up the hole disappeared into the singularity. But, if the information about the microstates resides near the singularity as we are now supposing, then the outgoing radiation is insensitive to the details of the microstate and when the hole has evaporated away we cannot recover the information contained in the matter which went into the hole. This, once again, violates the unitarity of QM. The robustness of this paradox relies on the fact that it uses very few assumptions about the physics relevant to the hole. Indeed we have assumed that quantum gravity effects are confined to a small length scale (like the Planck length) and that the curvature scales at the horizon are much larger than this length for large BHs. In few words, it seems that the precise theory of quantum gravity is irrelevant to the process of Hawking radiation and thus for the resolution of the paradox.\\

One solution that was given in order to solve this serious problem, but that at the end turned out to fail in its goal, is the effect of \textit{small corrections}. Hawking did a leading-order computation using quantum fields on curved space and it is certainly possible that small quantum gravity effects could yield a small correction to the state of the created pair. In particular, at the step $k$ of the emission, we can write
\begin{equation} \label{eq:small_corrections}
\ket{\psi}_{\text{pair},k} = \frac{1}{\sqrt{2}} \left(\ket{0}_{o_k} \ket{0}_{i_k} + \ket{1}_{o_k} \ket{1}_{i_k}\right) + \ket{\delta \psi_k} \,, \qquad |\ket{\delta \psi_k}| < \epsilon
\end{equation}
with $\epsilon \ll 1$. The number of quanta produced by the BH is
\begin{equation}
N_{\text{pairs}} \simeq S_{\text{Bek}} \simeq \left(\frac{M}{m_p}\right)^2
\end{equation}
with $m_p$ the mass of the pair and typically $N_{\text{pairs}} \gg 1$. Perhaps the small corrections in \eqref{eq:small_corrections} introduce subtle correlations among the large number of radiated quanta $o_k$ in such a way that at the end of the evaporation the radiation state is not entangled anymore with the remaining BH. In this way, the paradox is solved: Hawking's leading order entanglement $S_{\text{ent}} = N_{\text{pairs}} \log(2)$ would be removed by the large number of subleading correction terms.\\
The fact is that in \cite{Mathur:2009hf} it was shown that such small corrections \eqref{eq:small_corrections} cannot resolve the problem. Assuming that
\begin{itemize}
\item  the correction to each pair is small as in \eqref{eq:small_corrections};
\item once a quantum of the radiation $o_k$ has reached a very large distance from the BH, any further changes to the BHs state are not relevant to the quantum itself;
\item the vacuum is unique
\end{itemize}
then the entanglement entropy at step $N$ will keep rising as
\begin{equation}
S_{\text{ent}}(N+1) > \underbrace{S_{\text{ent}}(N) + \log(2)}_{\text{see below} \, \eqref{eq:ent_entropy}} - 2 \epsilon
\end{equation}
This small corrections theorem states that small modifications to semi-classical dynamics can never resolve the paradox, since we need order-\textit{unity} modifications.\\
We can summarize this puzzle by asking: \textbf{how should we resolve the puzzle of monotonically rising entanglement?}
\item \textbf{The breakdown of the semiclassical approximation}: consider a shell of mass $M$ that is collapsing to form a BH. As we know and as we already said, the curvature $\mathcal{R}$ is low at the horizon scale, in particular
\begin{equation}
\mathcal{R} \simeq \frac{1}{(GM)^2} \ll \frac{1}{l^2_p}
\end{equation}
where $l_p$ is the Planck scale. For this reason one expects that the usual low energy semiclassical dynamics holds and the shell passes smoothly through the horizon radius. The question is: if the shell goes through the horizon, how will its information ever come out? Points on the shell that are inside the horizon cannot send signals to the horizon or at the exterior of it. Even though new physics can occur at the singularity, how can this modify the pair creation at the horizon and resolve Hawking's paradox?\\
The puzzle can be schematically formulated in the following way: \textbf{what can lead to a breakdown of semiclassical dynamics at the horizon where curvature are low?}
\end{enumerate}

\subsection{Resolution: the Fuzzball proposal}
In the years following Hawking's original paper, there have been many attempts at resolving the puzzles we listed in the previous Subsection, in particular the information paradox. Through all these years it turned out that working purely within QM and GR could lead only to a dead point. Only a radical modification to the standard knowledge was needed. In this regard, String Theory gives a very relevant contribution to the resolution of the previous problems. Indeed, the first real advance came with the works of Sen \cite{Sen:1995in} and Strominger and Vafa \cite{Strominger:1996sh}. They showed that a string theoretic description of a BH has the capacity to count its microstates. This was demonstrated in a particular limit: the BHs are supersymmetric (which implies that their Hawking temperature is 0, as it can be shown) and the gravitational coupling $G$ vanishes. Their mayor result was that the states of the string system have an entropy that exactly matches the horizon areas of the BH at finite $G$.\\

\begin{figure}
    \centering
    \includegraphics[width=0.6\linewidth]{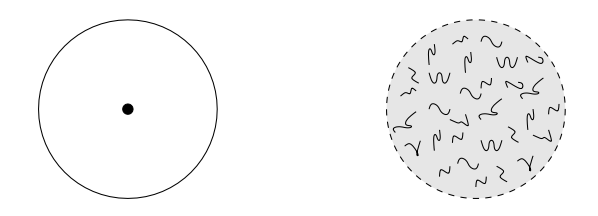}
    \caption{Schematic representation of a BH (on the left) and of a fuzzball (on the right), highlighting the main difference in their structure: in the former case we have a singularity (the black dot) surrounded by an horizon (the black solid line) and this gives rise to the information paradox; in the latter, we have a smooth and horizonless geometry without the presence of a singularity, (the dashed black line is the surface of the fuzzball and the spacetime is smooth in reaching it). Image taken from \cite{Mathur:2005zp}.}\label{fig:BH_vs_fuzzball}
  \end{figure}

However, in order to resolve the information paradox, it is not sufficient to count the microstates, indeed one must determine also their structure and understand it at strong coupling, where the stringy bound state is expected to describe a BH. Without entering in all the details (which go beyond the scope of the present Chapter and of the entire thesis), we simply say that all the results obtained in this scenario suggest that in String Theory one never gets the traditional BH with a vacuum horizon, instead one obtains a horizon-sized \textit{fuzzball}, which radiates from its surface like a normal body, evading in this way the information paradox. This gives rise to the \textit{fuzzball proposal} \cite{Lunin:2001fv}: singular and horizon-full BHs should be thought as the superposition of quantum states, some of which admit a classical description as smooth and horizonless gravity solutions with the same conserved quantities (mass, charge and angular momentum) of the BH. Remarkably, these geometries emit radiation preserving quantum information. From this explanation we understand how fuzzballs manage to describe the structure and dynamics of BH states. According to \cite{Bena:2022rna}, we should distinguish the notions of \textit{microstate geometries} and \textit{fuzzballs}, in particular:
\begin{itemize}
\item \textit{microstate geometries} are smooth horizonless solutions of the Supergravity theory, which is the low energy limit of String Theory. These solutions can have the same mass, charge and angular momentum as a BH;
\item \textit{fuzzballs} are the general class of objects in which the BH interior is supplanted by some new, quantum, horizon-scale structure having itself no horizon.
\end{itemize}
From these definitions, it is possible to show that microstate geometries are examples of fuzzballs, but not all fuzzballs need to be microstate geometries. In particular, the latter are coherent states within the ensemble of all BH microstates, which are fuzzballs. It is interesting to say that in String Theory it has been possible to construct these horizonless objects (resolving the information paradox) because it has many degrees of freedom. This goes under the name of \textit{fuzzball principle} \cite{Bena:2022rna}, according to which \textbf{horizons and singularities only appear if one tries to describe gravity using a theory that has too few degrees of freedom to resolve the physics}.\\ 
A schematic picture that represents the difference between the BH interior as suggested by GR and the fuzzball is shown in Figure \ref{fig:BH_vs_fuzzball}. From it we easily see that while the classical hole has all its mass at the core of spacetime, a fuzzball is a microstate which is a horizon-size quantum object but without an horizon, radiating from its surface like a normal body.
\\
  
We have realized how this paradigm solves the puzzle of the information paradox explained in the previous Subsection. For what concerns the entropy puzzle,  we simply say that the Bekenstein entropy corresponds to a count of these microstates of a given BH. The explanation for the resolution of the breakdown of the semiclassical approximation is more subtle and we limit ourselves to give a hint of how it is solved. Details can be found in \cite{Mathur:2024ify}.\\
It is natural to expect that semiclassical dynamics breaks down when curvatures exceed the Planck scale, i.e. $\mathcal{R} \gg l_p^{-2}$. But, as we noted above, this is not sufficient to resolve the information paradox because we need order unity corrections to the dynamics at the horizon. It is possible to show that there exists another way of breaking the semiclassical approximation, due to an effect that can be traced back to the large value of the Bekenstein entropy \eqref{eq:bek_entropy}. Indeed fuzzballs are bound states of the elementary excitations in String Theory. The existence of a bound state in a field theory gives rise to consequences to the correlation functions computed on the vacuum, as notoriously known. This effect is usually small in a normal QFT with a limited set of bound states. But the correlations created by fuzzballs are significant because:
\begin{itemize}
\item they have an extended structure (for what said before);
\item they have a very high degeneracy, given by the Bekenstein entropy, which increases with the size of the BH (it is related to the horizon area)
\end{itemize}
The relevant aspect is that the existence of these correlations created by the fuzzballs in the gravitational vacuum leads exactly to a breakdown of the semiclassical approximation.\\

The identification of some (a)typical microstate geometries has been rather successful for supersymmetric BHs \cite{Giusto:2004id, Giusto:2004ip, Giusto:2011fy, Turton:2012ny, Bena:2015bea, Bena:2016agb, Bena:2016ypk, Bena:2017xbt, Bena:2017upb, Bianchi:2016bgx, Bianchi:2017bxl}. This task has been more challenging for non-BPS or even non-extremal BHs (which is the ultimate goal if one aims to find microstate geometries for astrophysical BHs). For instance, JMaRT solitons \cite{Jejjala:2005yu}, the first known non-BPS smooth horizonless geometries, are not viable microstate geometries: a BH with the same charges would be over-rotating, thus its horizon would collapse leading to a naked singularity. Moreover, due to the presence of an ergoregion without horizon, JMaRT solitons are both affected by an ergoregion instability \cite{Cardoso:2005gj} and a charge instability \cite{Bianchi:2023rlt}. Very recent developments in the fuzzball programme for non-BPS geometries also include microstrata \cite{Ganchev:2021pgs, Ganchev:2023sth} and Schwarzschild-like topological solitons \cite{Bah:2022yji}.\\

The study of BHs is not only appealing for theoretical reasons: since a few years ago, new observational techniques have been yielding outstanding results. LIGO and Virgo interferometers \cite{LIGOScientific:2016aoc, LIGOScientific:2016lio} have been able to detect a signal of gravitational waves coming from the late phase (ring-down) of the merger of two BHs. Moreover, the Event Horizon Telescope \cite{EventHorizonTelescope:2019dse} has provided the first direct images of the light ring and the accretion disk of a BH. Given these results, one might hope to directly test novel ideas in BH physics or even Quantum Gravity, through direct observations. In particular, these may allow to discriminate between BHs and fuzzball geometries due to their different dynamical properties. In addition, the fuzzball conjecture could be tested as opposed to alternative models \cite{Cardoso:2016rao, Cardoso:2017cqb, Cardoso:2019rvt}, collectively called Exotic Compact Objects (ECO). A non-exhaustive list of the latter also includes boson stars \cite{Schunck:2003kk}, gravastars \cite{Mazur:2001fv, Danielsson:2017riq}, firewalls \cite{Almheiri:2012rt} and wormholes \cite{Solodukhin:2005qy} (also see \cite{Balasubramanian:2022gmo, Balasubramanian:2022lnw, Climent:2024trz} for recent progress).\\

The dynamical aspects of gravitational objects can be examined by probing them with particles, waves or strings \cite{Bianchi:2017sds}. BHs and fuzzballs can be discriminated by identifying observables with different behaviours for the two types of geometries: the shape of the light rings \cite{Bianchi:2018kzy, Bianchi:2020des, Bianchi:2020yzr}, the spectrum of Quasi-Normal Modes (QNMs) \cite{Chowdhury:2007jx, Chakrabarty:2019ujg, Bena:2020yii, Ikeda:2021uvc}, the Tidal Love Number (TLN) \cite{Consoli:2022eey, Bianchi:2023sfs, DiRusso:2024hmd} and the multi-polar structure \cite{Bena:2020see, Bianchi:2020bxa, Bena:2020uup, Bianchi:2020miz, Mayerson:2020tpn, Bah:2021jno}. Superradiance \cite{Brito:2015oca} is another significant phenomenon in this context. It consists in the amplification of wave perturbations scattering off BHs, which thereby lose energy, charge and angular momentum. Superradiance has been introduced and mostly studied \cite{Press:1972zz, Teukolsky:1974yv} for rotating BHs, where it can be considered as the wave counterpart of the Penrose process; nonetheless it has been also highlighted to occur for static charged BHs \cite{DiMenza:2014vpa, Benone:2015bst, Baake:2016oku}. Remarkably, superradiance can only take place for black objects and not for fuzzball geometries which, instead, preserve information.

\subsection{Topological Stars (TSs) and motivations for studying them}
From what just explained, String Theory has developed several constructions of horizonless smooth microstate geometries but all of them have in common two key ingredients: having \textit{extra compact dimensions}, that can degenerate in some points of spacetime, defining its end, and possessing \textit{electromagnetic fluxes} which avoid the collapse of the structure due to its gravitational structure. At the level of terminology, we call bubbles the non-trivial smooth topological cycles that are generated by the degeneracy of the extra dimensions at different loci of spacetime and they are supported by the aforementioned fluxes as a replacement for the horizon. The construction techniques in this regard are quite involved, since they require to turn on various degrees of freedom, giving rise to numerous gauge fields and scalars.\\
In this Chapter we want to consider the simplest framework to work with, which is Einstein-Maxwell theories with one extra compact dimension in addition to the four ones. In particular we focus on a recently discovered smooth horizonless non-BPS geometry called Topological Star (TS) \cite{Bah:2020ogh, Bah:2020pdz, Heidmann:2021cms, Bah:2021owp, Heidmann:2022zyd}. This is a non-supersymmetric, static and spherically symmetric solution of Einstein-Maxwell theory in 5 dimensions, whose 4-dimensional reduction is a BH solution of Einstein-Maxwell-Dilaton theory. A TS is characterized by a non-trivial topology and its Gregory-Laflamme instability \cite{Gregory:1993vy} is avoided by turning on electromagnetic fluxes. Specifically, the magnetic field of a TS is sourced by a magnetic monopole, whereas the electric field is sourced by a string wound around the compact direction. These gauge fields have not to be considered as the usual gauge fields in Electromagnetism under reduction to four dimensions, but more as the descendants of other fields that come out from the low-energy limit of Quantum Gravity. A double Wick rotation or an analytical continuation of the parameters \cite{Bah:2021irr} maps the TS to a black object with the same mass and charges, as we are about to say and we are going to deeply see in the next Section. This is called Black String (BS) and is a two-charge non-rotating instance of a BH in the Cvetic-Youm class \cite{Cvetic:1996xz, Cvetic:1997uw}.\\

\begin{figure}[H]
    \centering
    \includegraphics[width=0.62\linewidth]{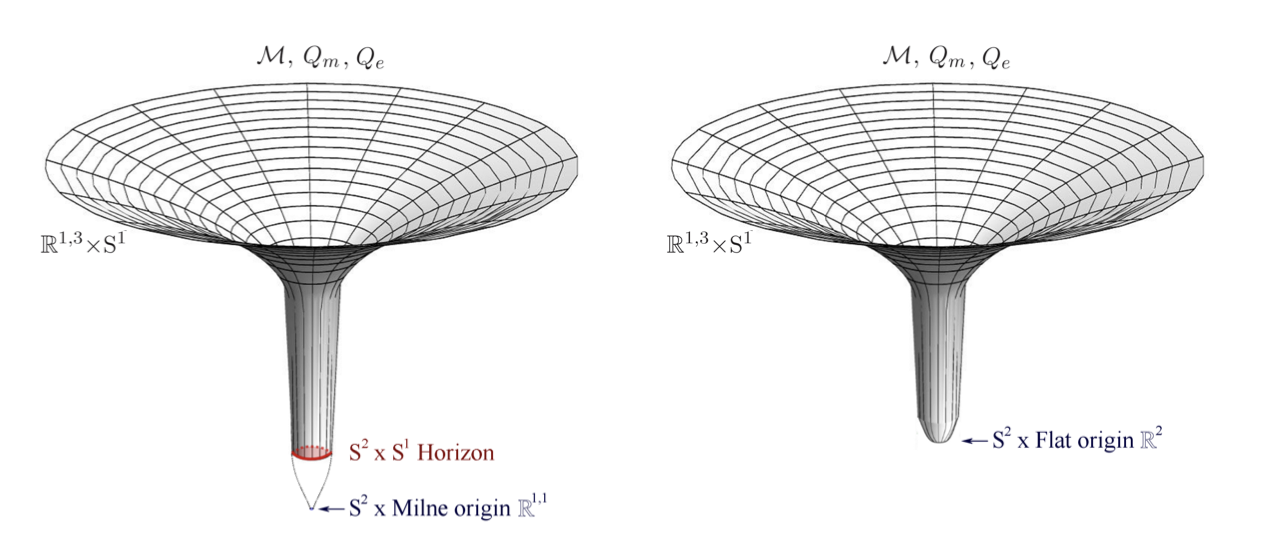}
    \caption{Representation of the two types of solutions of 5-dimensional Einstein-Maxwell theory. $\mathcal{M}$ is their 4-dimensional mass, while $(Q_m,Q_e)$ are the 4-dimensional conserved magnetic and electric charges associated to the fluxes that prevent the gravitational collapse of the structure. Image taken from \cite{Bah:2020pdz}.}\label{fig:BS_TS}
  \end{figure}

Now we want to give a quick presentation of TS and BS, passing to analyze these two phases in more detail in the next Section. The general solutions that can describe one of the two previous regimes superpose a Schwarzschild string, which is a $S^1$ fibration over a 4-dimensional Schwarzschild BH, and a static Bubble of Nothing, which is a Euclidean Schwarzschild solution with a time direction. The former has a horizon where the timelike Killing vector $\partial_t$ shrinks (as usual), while the latter has a smooth bubble (denoted also a \textit{bolt}) where the spacelike Killing vector $\partial_y$ vanishes, where we have defined $y$ the coordinate of the extra compact dimension. These two solutions are related by the double Wick rotation $(t, y) \to (i y, i t)$. For this reason the final class of two-charge solutions, obtained from superposing the two previous solutions into only one, is invariant under the double Wick rotation. As we are going to see, depending on which of the horizon or the bolt comes first, the solution is either a two-charge BS or a two-charge smooth bubble, which is a TS. Under the reduction to four dimensions (called \textit{Kaluza-Klein reduction}), the BSs correspond to charged non-rotating BHs, while the TSs give rise to a naked singularity due to the degeneracy of the extra dimension. A schematic representation of these two objects is shown in Figure \ref{fig:BS_TS}
As we are going to say, the BS has a bubble in the interior of the horizon and it hides its curvature singularity. The topology at the bubble is a 2-sphere $S^2$ sitting at the origin of a smooth Milne space \cite{Milne1936} \footnote{The metric of a Milne space is a particular case of the metric of Friedmann, Lemaitre, Robertson and Walker but without matter, radiation and cosmological constant.}. For what concerns the TSs, the solutions recall the BSs but they are smooth since the horizon is placed inside the bolt, which delimits the spacetime signaling its end. The bolt is described by a 2-sphere $S^2$ at the origin of a flat $\mathbb{R}^2$. The regularity at the bolt, as we are going to see, constrains the overall size in a way that it depends on the size of the extra dimension. The only way to get around this problem for phenomenological reasons is to add a conical defect, which is not considered as a singularity in String Theory as we are going simply to say in the next Section, without entering details.\\

The purpose of the investigation of this Chapter is to explore the dynamical behaviour of TS and BS under charged linear scalar perturbations. Indeed the question about the stability of this geometry without SUSY is quite basic. A single TS is a caricature of a microstate geometry for astrophysical BHs, in the sense that even if they enjoy all the properties of microstate geometries, at the level of phenomenological features they are quite far to reproduce the ones that have been observed: for example, they are `spherical spacetime mirror' for photons \cite{Heidmann:2022ehn}, then the possible values of charges and masses that are expected do not cover a range so large for phenomenological observations and, ultimately, it is a configuration that is never 4-dimensional, consequently if these microstates are correct, we will need evidences of a fifth dimension \footnote{I would like to thank Massimo Bianchi, Giuseppe Dibitetto and Alejandro Ruiperez for clarification on this issue.}. Nevertheless, trying to make more of its properties closer to the phenomenological requests helps also to bring to light a richer microstructure of this object and, because of the large symmetry of the its metric, the dynamics of particles or waves in this background is separable. Relevant information about their dynamical features can be easily extracted: TSs, then, represent very useful tools to learn or to test the typical behavior expected for more general smooth horizonless geometries.\\

The response of TS under linear scalar massive or massless perturbations has been already studied in \cite{Heidmann:2022ehn, Heidmann:2023ojf, Bianchi:2023sfs}. In particular, photon geodesics, QNMs and Tidal Love Numbers (TLNs) in the TS background have been analyzed, allowing to argue for the linear stability of TSs. Moreover, being their dynamical TLN real, TSs have been proved to undergo deformations only and not dissipation, consistently with the absence of a horizon. The study of the stability of the same geometry under electromagnetic and gravitational perturbations is examined in \cite{Bena:2024hoh}, while the generalization to the rotating case for this geometry is investigated in \cite{Bianchi:2025uis,Heidmann:2025pbb}, where the solution is closer to one of phenomenological interest since this kind of geometry is asymptotically flat. Many other properties of this spacetime, as the propagation of scalar perturbations, the computation of self-force and radiative losses, scalar waves from unbound orbits, the characterization of geodesic deviations and the scattering angle using a self-force approach, together with an exact resummation of the renormalized angular momentum in the eikonal limit, are reported in \cite{Bianchi:2024vmi,Bianchi:2024rod,DiRusso:2025lip,Bini:2025qyn,Bianchi:2025aei,Bini:2025ltr}. The cases of nonradial stability and extreme mass ratios around TS are analyzed in \cite{Dima:2024cok,Dima:2025zot,Melis:2025iaw}.\\

In \cite{Heidmann:2022ehn, Heidmann:2023ojf, Bianchi:2023sfs}, the probes have been chosen to be chargeless or at most to have Kaluza-Klein momentum along the compact direction. In the present Chapter, instead, we investigate the possibility of instabilities when the probes are \textit{charged} under the electric field of a TS, in analogy with the charge instability pointed out for JMaRT geometries \cite{Bianchi:2023rlt}. The electric gauge field of a TS is a 2-form, sourced by an electrically charged string, wound around the compact direction, so, differently from the previous treatments, probes are not point-like particles. These could be identified as some of the winding modes of TS's electric string, after they slide off the bolt during the discharge mechanism.\\ 

We will make use of the correspondence between BH/fuzzball perturbation theory and qSW curves of $\mathcal{N} = 2$ SYM theory with gauge group SU(2) \cite{Aminov:2020yma}, already analyzed in Chapter \ref{chap_1}. Besides providing a further tool to compute QNMs \cite{Bianchi:2021xpr, Bianchi:2021mft, Bianchi:2022qph, Bianchi:2023rlt, Bianchi:2023sfs} this correspondence has allowed to devise a definition for the dynamical TLN \cite{Consoli:2022eey, Bianchi:2023sfs, DiRusso:2024hmd}, by taking into account instantonic corrections. In all these developments, a prominent role has been played by the Heun connection formulae \cite{Bonelli:2021uvf, Bonelli:2022ten}, recently found exploiting the Alday-Gaiotto-Tachikawa (AGT) duality \cite{Alday:2009aq}. In particular, the radial equation for waves in TS/BS background is a Confluent Heun Equation (CHE), which can be mapped to the CHE arising from a qSW curve with three massive hypermultiplets, as already showed in Chapter \ref{chap_1}. Alternative semi-analytical procedures for the computation of QNMs are based on integrability techniques \cite{Fioravanti:2021dce, Fioravanti:2020udo, Rossi:2022mwx}. From the QNM spectrum, we infer that TSs are stable also when probed with scalar \textit{charged} perturbations. Indeed, only modes with negative imaginary part have been found. \\

Since we are considering here static geometries, TSs and BSs do not give rise to rotation superradiance. We investigate, instead, the possibility of charge superradiance. No amplitude increases have been discovered for TSs at any frequencies. This is consistent with the fact that information is preserved by fuzzball geometries. By contrast, BSs can be well-suited to examine charge superradiance. We show the presence of an enhancement of wave amplitudes in a certain range of frequencies for BSs, through the analysis of the amplification factor. On grounds of the BH/fuzzball perturbation theory - qSW curves correspondence, we have added the instantonic corrections to the definition of this observable. The results have been compared with a corresponding numerical integration. The absence of mirror mechanisms, though, excludes the possibility of superradiant instabilities \cite{Press:1972zz}.

\section{Charged scalar perturbations in a Topological Star background} \label{Top star}
In this Section we present the solution of a TS. We probe this geometry with strings that are charged with respect to the electric field of the TS itself. Accordingly, we devise a shift of variables in the scalar wave equation, that takes into account the stringy nature and the charge of the probes.

\subsection{The background}
TSs are smooth horizonless static solutions of the Einstein-Maxwell theory in five dimensions, described by the action \cite{Bah:2020ogh, Bah:2020pdz}
\be
S_5=\int \mathrm{d}^5x \, \sqrt{-\det(g)}\left({\frac{R}{2\k_5^2}}-{\frac{1}{2}}\bigl|F^{(m)}\bigr|^2-{\frac{1}{2}}\bigl|F^{(e)}\bigr|^2\right)
\ee
where $\k_5$ is the five-dimensional Einstein gravitational constant, $F^{(m)}$ and $F^{(e)}$ are the magnetic 2-form and electric 3-form field strengths respectively. The corresponding equations of motion are\footnote{We set $\varepsilon_{01234} = +1$.}
\be \label{eqmotion}
\mathrm{d} \star F^{(m)}= \mathrm{d} \star F^{(e)}=0\quad,\quad R_{\mu\nu}=\k_5^2\left(T_{\mu\nu}-{\frac{1}{3}}g_{\mu \nu}{T_\a}^\a\right)
\ee
where here $\star$ is the Hodge star operator in five dimensions and $T_{\mu\nu}$ is the stress energy tensor
\be
T_{\mu\nu}={F^{(m)}}_{\m\a}{{F^{(m)}}_\nu}^\a-{\frac{1}{4}}g_{\mu\nu}{F^{(m)}}_{\a\b}{F^{(m)}}^{\a\b}+{\frac{1}{2}}\Bigg[{F^{(e)}}_{\mu\a\b}{{F^{(e)}}_\n}^{\a\b}-{\frac{1}{6}}g_{\mu\nu}{F^{(e)}}_{\a\b\g}{F^{(e)}}^{\a\b\g}\Bigg]
\ee
The metric of a TS is given by
\be \label{eq:top}
\mathrm{d}s^2=- f_s(r) \, \mathrm{d}t^2 + {\frac{\mathrm{d}r^2}{f_b (r) f_s (r)}} + r^2 \,(\mathrm{d}\q^2 + \sin^2\q \, \, \mathrm{d} \phi^2) + f_b(r) \, \mathrm{d}y^2
\ee
where
\be \label{eq:warping}
f_s(r) = 1-\frac{r_s}{r}  , \quad f_b(r) = 1-\frac{r_b}{r}
\ee
In \eqref{eq:top}, $t$ and $(r, \q, \phi)$ are the time and the spherical coordinates of the 4-dimensional spacetime, while $y$ is a further compact coordinate such that $y \sim y + 2 \pi R_y$. The metric of the present TS is then static, asymptotically $\mathbb{R}^{1, 3} \times S^1_y$ and spherically symmetric in 4 dimensions\footnote{As already said in the previous Section, this gravity solution can be seen as the harmonic superposition of a Schwarzschild String (4-dimensional Schwarzschild BH $\times S^1_y$) and a static Bubble of Nothing (a Euclidean Schwarzschild solution with a time direction). Introduced to describe an instability of the Kaluza-Klein vacuum \cite{Witten:1981gj}, the Bubble of Nothing has been recently studied to understand the stability of non-SUSY vacua \cite{Ooguri:2017njy,  GarciaEtxebarria:2020xsr, Dibitetto:2020csn, Bomans:2021ara} and in the framework of the Cobordism Conjecture \cite{Delgado:2023uqk}.}.\\

Three singularities can be identified for \eqref{eq:top}: 
\begin{itemize}
\item $r=0$ is a full-fledged curvature singularity; 
\item $r =r_s$ is a horizon, being such that the component $g_{tt}$ of the metric changes sign (as hinted at the end of the previous Section); 
\item $r = r_b$ is the bolt singularity, where the compact $y$-circle degenerates to a point. 
\end{itemize}
In correspondence of the latter singularity, the spacetime smoothly terminates at a cap. According to the location of the singularities we can distinguish two kinds of solutions:
\begin{itemize}
\item TS for $r_b > r_s > 0$: the horizon and the singularity are pushed out of the geometry, which is then smooth and horizonless; 
\item BS for $r_s \geq r_b>0$: the spacetime is horizon-full. 
\end{itemize}
The two different regimes are linked by the double analytical continuation (the double Wick rotation we cited above)
$$
(t, y) \leftrightarrow (i t, i y) \quad \mathrm{and} \quad r_s \leftrightarrow r_b
$$
This agrees with what said at the end of the previous Section, in the sense that if we start with a regime that has a horizon (the BS), where the timelike Killing vector $\partial_t$ shrinks, then the other regime (the TS), which is the symmetric counterpart of the previous one, has a spacelike Killing vector $\partial_y$ that shrinks at the bolt.\\

The presence of electromagnetic fluxes prevents the gravitational collapse of the solution. The expressions of the electric and magnetic field strengths are strongly constrained by the spherical and double Wick rotation symmetries and they turn out to be
\be \label{eq:fieldstrength}
F^{(e)}=\frac{Q}{r^2} \, \mathrm{d}r \wedge \mathrm{d}t\wedge \mathrm{d}y\quad,\quad F^{(m)}=P\sin\q \, \mathrm{d} \q \wedge \mathrm{d}\phi
\ee
The equations of motion \eqref{eqmotion} are solved if
\be \label{eq:TopCharge}
P^2+Q^2=\frac{3 r_b r_s}{2\k_5^2}
\ee
The source for the field strength $F^{(e)}$ is a charged line wound along the compact direction $y$. Indeed this is the object that naturally couples to the 2-form gauge potential
\be \label{eq:gaugepotential}
B^{(e)} = -\frac{Q}{r} \, \mathrm{d} t \wedge \mathrm{d} y
\ee
In order to understand the nature of the parameter $Q$ in \eqref{eq:fieldstrength} and \eqref{eq:gaugepotential}, in Appendix \ref{Charged wire} we investigate the case of an infinitely extended and uniformly charged wire. The final result is that $Q$ can be interpreted as the linear charge density of this wire and this implies, in particular units discussed in the Appendix \ref{appmeas}, that $[Q] = [L]$, consistently with \eqref{eq:TopCharge}.\\ 

The absence of Gregory-Laflamme \cite{Gregory:1993vy} and thermodynamical instabilities for the TS further requires that (the analogous condition for the BS arises after $r_b \leftrightarrow r_s$) \cite{Bah:2021irr} 
\be
r_s < r_b < 2 r_s
\ee
Finally, \eqref{eq:top} admits a 4-dimensional reduction to the Einstein-Maxwell-Dilaton theory. This allows to deduce the corresponding ADM mass and the conserved electric and magnetic charges $Q_e$ and $Q_m$ \cite{Bah:2020pdz}
\be
\label{ADM mass}
\mathcal{M} = \frac{2 \pi}{\kappa_4^2} (2 r_s + r_b) \, , \qquad Q_m = \frac{P}{e} \, , \qquad Q_e = \frac{Q}{e}
\ee
with $e$ defined in such a way that $e^2 = (2 \pi R_y)^{-1}$. The first expression for the mass holds both for the TS and for the BS. Remarkably, TSs are non-BPS fuzzball geometries, since extremality is obtained for $r_b = r_s$ (hence just for one particular instance of BS).\\

\begin{figure}[H]
    \centering
    \includegraphics[width=0.6\linewidth]{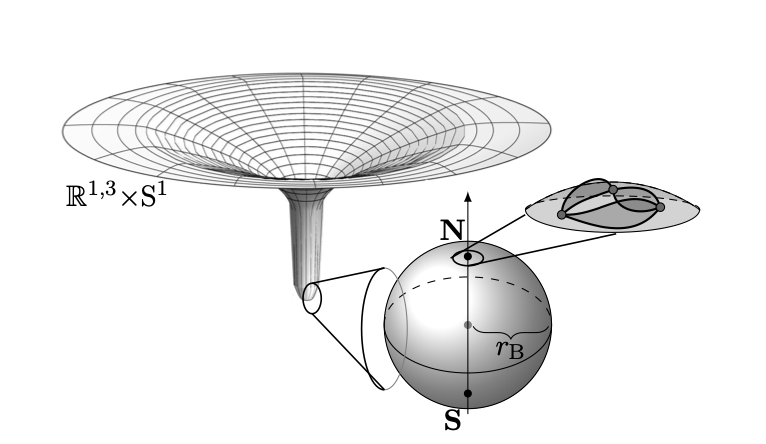}
    \caption{Schematic picture of the TS, zooming the situation near the bolt. From this we see that a generic TS is made of two different scales: we have a large scale $r_b$, that corresponds to the size of the large $S^2$ bubble of order $k R_y$, and then a small scale $r_{\text{bubbles}} \ll r_b$ where the small bubbles are present and that make the geometry entirely smooth. Image taken from \cite{Bah:2020pdz}.}\label{fig:TS_details}
  \end{figure}

We said that, for the TS regime ($r_b > r_s$), the request for regularity at the bolt constrains the overall size of our object to the size of the extra dimension. Let us see it, following the discussion of \cite{Bah:2020pdz}. The region near $r=r_b$ is best described by the local radial coordinate
\be
\rho^2 \equiv \frac{4 (r-r_b)}{r_b-r_s}
\ee
and then taking the limit $\rho \to 0$. With this procedure the metric \eqref{eq:top} becomes
\be
\mathrm{d}s^2 \simeq - \frac{r_b-r_s}{r_b} \mathrm{d} t^2 + r^2_b \left(\mathrm{d} \rho^2 + \frac{r_b-r_s}{4 r^3_b} \rho^2 \mathrm{d} y^2 + \mathrm{d} \theta^2 + \sin^2 \theta \mathrm{d} \phi^2 \right)
\ee
From this we see that the $(\theta,\phi)$ subspace describes a 2-sphere $S^2$ of radius $r_b$. For what concerns the subspace parametrized by $(\rho,y)$, this corresponds to a smooth flat space $\mathbb{R}^2$ only if we correctly match the size of the compact dimension to $r_s$ and $r_b$. Indeed, writing the metric of the 2-dimensional Euclidean space in polar coordinates $(R_{\mathbb{R}^2},\psi)$, with $\psi \sim \psi + 2 \pi$, and imposing  
$$\mathrm{d} R_{\mathbb{R}^2}^2 + R_{\mathbb{R}^2} \mathrm{d} \psi^2 = r^2_b \mathrm{d} \rho^2 + \frac{r_b-r_s}{4 r^3_b} (r_b \rho)^2 \mathrm{d} y^2 $$
we arrive at $\mathrm{d} \psi^2 = \frac{r_b-r_s}{4 r^3_b} \mathrm{d} y^2$. Keeping in mind that the periodicity of the $y$ dimension is $2 \pi R_y$ and the one of $\psi$ is $2 \pi$, then we obtain that
\be \label{eq:Ry}
R^2_y = \frac{4 r^3_b}{r_b-r_s}
\ee 
Thanks to this condition, the topology at the coordinate singularity corresponds to a bolt, i.e. to a smooth $S^2$ bubble sitting at an origin of $\mathbb{R}^2$. It is possible to show that also the gauge fields are regular at this locus. Now, we can recast \eqref{eq:Ry} in the following way
\be
r_s = r_b \left(1-\frac{4 r^2_b}{R^2_y}\right)
\ee
Since $r_s$ and $r_b$ must have the same sign for what showed above, this gives an important bound to the size of the bubble
\be
r^2_b \leq \frac{R^2_y}{4}
\ee
In other words, the TS is at best the size of the extra dimension. In order to have an astrophysical bubble, one needs to have a large extra dimension which is incompatible with the real world. One possibility to solve this problem is assuming that the local metric around the bubble has a conical defect, meaning that it has the topology of a smooth $\mathbb{Z}_k$ quotient of $\mathbb{R}^2 \times S^2$. In fact, in this way \eqref{eq:Ry} becomes
\be
R^2_y = \frac{4 r^3_b}{k^2 (r_b-r_s)} \Leftrightarrow r_s = r_b \left(1 - \frac{4 r^2_b}{k^2 R^2_y}\right)
\ee
Taking $k$ large allows for the construction of astrophysical bubbles with a size much larger that the extra dimension.\\
The parameter $k$ has important implication onto the classical degrees of freedom that constitute the TS. While in String Theory spacetimes with conical singularities can always be smoothed, in Einstein-Maxwell theory it is not so obvious. Without entering details, which instead can be found in \cite{Bah:2020pdz}, we simply say that the physical interpretation of these conical singularities is that they arise as two Kaluza-Klein monopoles of charge $k$ and $-k$ at north and south poles of the bolt (corresponding, respectively, to $\theta=0$ and $\theta = \pi$) and that each one can be replaced by $k-1$ smooth bubbles without conical defects in these poles. This is depicted in Figure \ref{fig:TS_details}

With an analysis similar to the one just followed, we can study the topology of our spacetime near the horizon and the bubble in the case of the BS regime. We consider only the non-extremal case, i.e. $r_s > r_b$. The extremal one can be found in \cite{Bah:2020pdz}. The first coordinate singularity is at $r=r_s$ and the topology can be found by considering the metric with coordinate
$$\rho^2 \equiv \frac{4 (r-r_s)}{r_s-r_b} r^2_s$$
and taking the limit $\rho \to 0$. The metric \eqref{eq:top} reduces to
\be
\mathrm{d}s^2 \simeq - \frac{r_s-r_b}{4 r^3_s} \rho^2 \mathrm{d} t^2 + \mathrm{d} \rho^2 + r^2_s \left(\mathrm{d} \theta^2 + \sin^2 \theta \mathrm{d} \phi^2 \right) + \frac{r_s-r_b}{r_s} \mathrm{d} y^2
\ee
From it we see that the horizon has the topology of a $S^2 \times \,S^1$ and the radii of the $S^2$ and S$^1$ are $r_s$ and $\sqrt{\frac{r_s-r_b}{r_s}}R_y$ respectively.\\
Opposite to the TS, the second coordinate singularity, $r_b$, is part of the spacetime and it is inside the BS. As it is notoriously known, we should consider the Kruskal coordinates to extend the metric in the whole interior, but since the local metric at $r=r_b$  (which is the focus of our interest) is the same as the one with spherical coordinates $(r,\theta,\phi)$, we follow this alternative path. For this reason, we define the coordinate
\be
\rho^2 \equiv \frac{4 (r-r_b)}{r_s-r_b}
\ee
and for $\rho \to 0$ the metric \eqref{eq:top} becomes
\be
\mathrm{d}s^2 \simeq - \frac{r_s-r_b}{r_b} \mathrm{d} t^2 + r^2_b \left(-\mathrm{d} \rho^2 + \frac{r_s-r_b}{4 r^3_b} \rho^2 \mathrm{d} y^2 + \mathrm{d} \theta^2 + \sin^2 \theta \mathrm{d} \phi^2 \right)
\ee
We have the usual $S^2$ component described by $(\theta,\phi)$, but now, since $r$ is a timelike coordinate (we are beyond the horizon), the $(\rho,y)$ subspace does not correspond to $\mathbb{R}^2$ as done above for the TS, but it can be shown to describe a quotient of $\mathbb{R}^{1,1}$ by a boost. This space is called \textit{Misner space}. In order to understand it, we consider the $2 \pi$-periodic angle $\varphi$ as $y = \varphi R_y$ and then we perform the following transformation
\be
T \equiv -\rho \cosh(\gamma \varphi) \, , \qquad R \equiv - \rho \sinh(\gamma \varphi) \qquad \text{with} \, \, \, \gamma^2 \equiv \frac{r_s-r_b}{4 r^3_b} R^2_y
\ee
In this way the metric of the $(\rho,y)$ subspace reduces to
\be
-\mathrm{d} \rho^2 + \frac{r_s-r_b}{4 r^3_b} \rho^2 \mathrm{d} y^2 = - \mathrm{d} T^2 + \mathrm{d} R^2
\ee
From the definitions of $T$ and $R$, the $(\rho,\varphi)$ subspace describes a lower cone in the two-dimensional $(T,R)$ Minkowski space and the apex of the cone corresponds to the coordinate singularity where the $y$-circle reduces to a point. Hence it is a Milne region of a Misner space, as hinted at the end of the previous Section.

\subsection{Stringy probes in the Topological Star background} \label{Stringy}
\begin{figure}[H]
    \centering
    \includegraphics[width=0.6\linewidth]{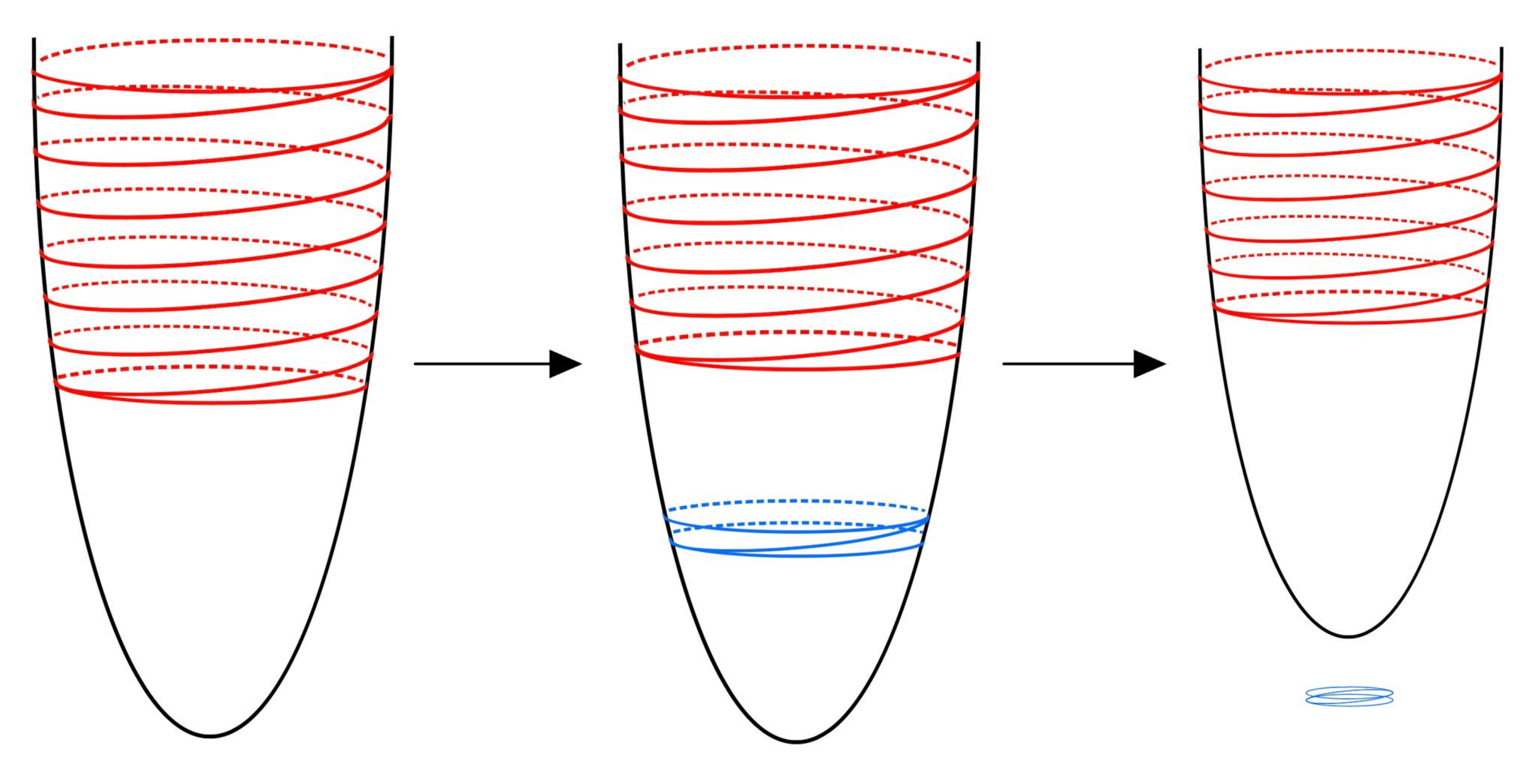}
    \caption{Representation of the (electric) discharge mechanism for a TS, from which the probe string (in blue) originates. The red line is the source for the electric field strength $F^{(e)}$.}\label{fig:discharge}
  \end{figure}

We want to investigate the response of a TS background to (electric) charged probes. Both the source and the probes of the gauge potential \eqref{eq:gaugepotential} are strings wound along the compact direction\footnote{Stringy probes have been also helpful to explore the phenomenon of tidal scrambling in fuzzball geometries \cite{Martinec:2020cml, Ceplak:2021kgl}.}. We expect the probe strings to originate from the discharge mechanism of the background. In particular, the TS geometry smoothly caps off at $r = r_b$ hence the winding modes of the source can slip off the bolt singularity with no obstacles. This is depicted in Figure \ref{fig:discharge}. However, we work in the probe limit.

Let us focus on the instance of rigid probes, which propagate without oscillations. Under this assumption, the embedding coordinates of their worldsheet in the TS geometry are 
\be \label{eq:embedding}
X^M (\tau, \sigma) = (X^\mu (\tau), \, Y = y_0 + n \sigma R_y) \quad, \quad \sigma \in [0, 2\pi]  
\ee
The spacetime indices are split as $M = (\mu, y)$, with $\mu = (t,r,\theta,\phi)$. ($\tau$, $\sigma$) are the (dimensionless) worldsheet coordinates of the string, $n$ is an integer and corresponds to the winding number.\\

The dynamics of the string in the TS background is governed by the non-linear sigma model \cite{Callan:1985ia}
\be \label{eq:sigmamodel}
S  = \frac{1}{4 \pi \alpha'} \int_\Sigma \mathrm{d}\sigma \mathrm{d}\tau \sqrt{-h} \Big [ G_{MN} (X) \partial_a X^M \partial_b X^N h^{ab} + B_{MN} (X) \partial_a X^M \partial_b X^N \varepsilon^{ab} \Big ]
\ee
In this action, $\Sigma$ represents the worldsheet of the string, $h^{ab} = \mathrm{diag}(-+)$ is the worldsheet metric, $\varepsilon_{ab}$ is the 2-dimensional antisymmetric tensor (such that $\varepsilon_{\t \s} = 1$). Moreover $G_{MN}$ and $B_{MN}$ are the metric \eqref{eq:top} and the gauge potential \eqref{eq:gaugepotential} of the background (the dilaton is missing in the TS solution). 
The conjugate momenta to the spacetime coordinates in \eqref{eq:sigmamodel} read 
\be \label{eq:momentum}
P_M = \frac{\partial L}{\partial \dot{X}^M}
\ee
where $\dot{X}^M = \partial_\t X^M$ and the Lagrangian $L = L\bigl [X^M, \partial_a X^M \bigr]$ is such that 
$$
S = \int \mathrm{d} \t \, L
$$
The effect of the gauge field on the expression of the conjugate momenta can be studied by splitting the Lagrangian as 
\be
L = L_G + L_B
\ee
with 
$$
L_G = \frac{1}{4 \pi \alpha'} \int_0^{2 \pi} \mathrm{d}\sigma \sqrt{-h} \, G_{MN} (X) \partial_a X^M \partial_b X^N h^{ab}
$$
$$
L_B = \frac{1}{4 \pi \alpha'} \int_0^{2 \pi} \mathrm{d}\sigma \sqrt{-h} \, B_{MN} (X) \partial_a X^M \partial_b X^N \varepsilon^{ab} 
$$
We expect the latter term to shift $P_M = \frac{\partial L_G}{\partial \dot{X}^M}$, i.e. the momentum in absence of the coupling with the electric field. In particular, the gauge potential \eqref{eq:gaugepotential} and the embedding coordinates \eqref{eq:embedding} imply that 
\be \label{eq:LB}
L_B = \frac{n R_y}{\alpha'} \frac{Q}{r} \dot{X}^t
\ee
where the integral in the coordinate $\sigma$ has been easily performed since no dependence on $\sigma$ is left in the integrand. Being \eqref{eq:LB} only dependent on $\dot{X}^t$ (and not on the other $\dot{X}$s) all conjugate momenta are left invariant by the coupling with the gauge field, except
\be
P_t \rightarrow P_t + \frac{q Q}{r}
\ee
or equivalently in terms of the energy $E$ ($P_t = -E$)
\be \label{eq:Eshift}
E \rightarrow E - \frac{q Q}{r}
\ee
In these expressions we have implicitly used the definition  
\be \label{eq:qdefinition}
q = \frac{n R_y}{\alpha'}
\ee

The presence of a compact direction also leads to a shift in the mass of the probe (compared to point-like particles in 4 dimensions)
\be \label{eq:massshift}
\mu^2 \rightarrow \mu^2 + f_b(r) q^2 + \frac{p^2}{f_b (r)} 
\ee
The warping factor \eqref{eq:warping} $f_b = 1 - \frac{r_b}{r}$ arises from the metric \eqref{eq:top}: indeed, the mass of the string depends on the actual length of the compactification radius which is modulated by the $g_{yy}$ component of the metric. In \eqref{eq:massshift}, besides \eqref{eq:qdefinition}, we have made use of the definition 
\be
p = \frac{\hbar m}{R_y}
\ee
where $m$ is the integer number representing Kaluza-Klein momentum units.

\subsection{Charged scalars in the Topological Star background}
At the linear level, scalar perturbations decouple from the vector and tensor modes. Therefore, the wave equation for a massive scalar in the background of a TS is simply the Klein-Gordon equation\footnote{Keep in mind that, according to the measure units of Appendix \ref{appmeas}, the mass parameter in this equation is actually $\tilde{\mu}^2 = \frac{\mu^2}{\hbar^2}$. This also applies to the momentum $p$ appearing in \eqref{eq:waveeq}: it should be $\tilde{p} = p/\hbar$. In the following we will omit the tilde in both instances.}
\be
\Box \Phi = \mu^2 \Phi
\ee
The radial and angular dynamics can be separated by imposing the ansatz
\be \label{eq:waveeq}
\Phi=e^{i(-\omega t+p y+m_\phi\phi)}R(r)S(\q)
\ee
In this expression, the exponential contains the conserved conjugate momenta: for instance $\omega$ is linked to the energy introduced in \eqref{eq:Eshift}, through the definition $E = \hbar \omega$.\\The solutions $S(\q)$ of the angular equation are the spherical harmonics in $S^2$. The radial equation, instead, reads
\be
\label{eqradial}
(r-r_b)(r-r_s)R''(r)+(2r-r_b-r_s)R'(r)+\Bigg[r^2\left(\frac{r\omega^2}{r-r_s}-\frac{r p^2}{r-r_b}-\mu^2\right)-\ell(\ell+1)\Bigg]R(r)=0
\ee

The charged scalar modes analyzed in this Chapter can be seen as scalar excitations of the stringy probes of Subsection \ref{Stringy}. The coupling to the electric field and the string nature of the probe can be taken into account by replacing the frequency and the mass according to \eqref{eq:Eshift} and \eqref{eq:massshift}. Thus, in the radial equation we perform the following shifts
\be\label{shift}
\omega\rightarrow \omega-\frac{q Q}{r} \quad,\quad \mu^2\rightarrow \mu^2+q^2 \left(1-\frac{r_b}{r}\right) + \frac{p^2} {(1 - \frac{r_b}{r})}
\ee
After these substitutions, the radial equation \eqref{eqradial} can be recast in the canonical (Schr\"odinger) form 
\be
\label{eq:eqradialcan}
\psi''(r)+Q_W(r) \psi(r)=0
\ee
by introducing the function $\psi(r)$, related to the original one by (look at Chapter \ref{chap_1} for details)
\be
R(r)=\frac{\psi(r)}{\sqrt{(r-r_s)(r-r_b)}}
\ee
In our instance, the characteristic $Q-$function $Q_W(r)$ appearing in \eqref{eq:eqradialcan} is
\begin{align}
\label{eq:Qgrav1}
Q_W(r)=&\frac{A_4 r^4+A_3 r^3+A_2 r^2+ A_1 r+A_0}{4(r-r_b)^2(r-r_s)^2}\\\nn
A_4=&-4(\omega^2-\m^2-p^2-q^2)\\\nn
A_3=&4(p^2 r_s + q^2 (2 r_b + r_s) + (r_b + r_s)\m^2 - 2 q Q\omega - r_b\omega^2)\\\nn
A_2=&4(\ell(\ell+1)+q^2(-Q^2+r_b^2+2r_br_s)+r_br_s\m^2-2q Qr_b\omega)\\\nn
A_1=&4(\ell(\ell+1)(r_b+r_s)+q^2r_b(r_br_s-Q^2))\\\nn
A_0=&r_b^2-2(1+2\ell(\ell+1))r_br_s+r_s^2
\end{align}
From the expression \eqref{eq:Qgrav1} of the $Q$-function, we infer that the radial dynamics of charged scalar perturbations in the TS background is described by a Confluent Heun Equation (CHE), with two regular singularities at $r=r_b$ and $r=r_s$ and an irregular one at $r= \infty$.



\section{Quantum Seiberg-Witten curves for \texorpdfstring{$\mathcal{N}=2$}{N=2} SYM with flavours} \label{quantumSW}
In this Section, we will make use of the correspondence between the ODEs coming from BH/fuzzball perturbation theory and quantum Seiberg - Witten (qSW) curves of $\mathcal{N} = 2$ SYM theory with gauge group SU(2) \cite{Aminov:2020yma}. In particular in this Chapter the map concernes the Confluent Heun equation (CHE) and the qSW curve with $N_f=3$. The other scenarios with all the confluences of the Heun equation and the link to the corresponding qSW curve are collected in \cite{Bianchi:2022wku} and have been reported also at the end of Chapter \ref{chap_1}.

Following the notation of \cite{Bianchi:2021mft}, we provide a short overview of how to construct the quantum Seiberg-Witten (qSW) curves for $\mathcal{N}=2$ theories with an SU(2) gauge group, as already done in Chapter \ref{chap_1}. Subsequently, we recall how to compute the quantum periods $(a, a_D)$, which are instrumental in deriving exact quantization conditions for the QNM frequencies and characterize the response of the geometry to tidal deformations.
\subsection{Quantum Seiberg-Witten curves}
The classical SW curves for SU(2) gauge theory with $N_f=3$ fundamental hypermultiplets with masses $m_i$ in flat space are \footnote{With respect to Chapter \ref{chap_1}, here we are using the symbol $\Lambda$ in place of $q= e^{2 \pi i \tau}$, which instead has been used in \eqref{eq:qdefinition}.}
\be
\Lambda y^2 P_L(x)+y P_0(x)+P_R(x)=0
\ee
with
\be\label{classSW}
P_L(x) {=} x{-}m_3, \qquad P_0(x) {=} x^2{-}u{+}\Lambda \left(x{+}\frac{1}{2}{-}m_1{-}m_2{-}m_3\right), \qquad P_R(x) {=} (x{-}m_1) (x{-}m_2)
\ee
In these definitions $\Lambda=e^{2\pi i \tau}$, $\tau$ is the gauge coupling and $u$ is the Coulomb branch (CB) modulus. In presence of a non trivial $\Omega-$background, $\epsilon_1=1$, $\epsilon_2=0$ (in units of $\hbar$), the dynamics of the gauge theory is described by the quantum curve which can be obtained by the classical one by promoting the variables $x$ and $y$ to operators satisfying the commutation relation
\be
[\hat{x},\hat{y}]= \hat{y}
\ee
The qSW curve follows from the classical one \eqref{classSW} 
\be
\label{eqnSWorig}
\left[\Lambda \hat{y}^{1/2}  P_L(\hat{x})\hat{y}^{1/2} +  P_0(\hat{x}) + \hat{y}^{-1/2}P_R (\hat{x})\hat{y}^{-1/2}\right] f(z)=0
\ee
This gauge theory can be understood in terms of D4-branes forming pairs stretched between two {NS5}-branes or between an NS5 and infinity, positioned both to the left and right. The locations of the zeros of $P_{L,R}(x)$, $P_0(x)$ determine the positions of the D4-branes in these three regions, serving as parameters for the masses and defining the CB (look at the Hanany-Witten setup in Section \ref{sec:hanany_witten} of Chapter \ref{chap_1}).

Equation \eqref{eqnSWorig} can be put in normal form by redefining the wave function as 
\be
f(z) = e^{-\frac{q z}{2}} z^{\frac{m_1+m_2}{2}} (1+z)^{-\frac{1+m_1+m_2}{2}} \Psi(z)
\ee
in such a way that \eqref{eqnSWorig} becomes
\be
\label{eqnSWcan}
\Psi''(z) + Q_{12}(z) \Psi(z) = 0
\ee
with
\be
\label{QSW12}
Q_{12}(z) {=}{-}\frac{\L^2}{4}{+}\frac{1{-}(m_1{-}m_2)^2}{4z^2}{+}\frac{1{-}(m_1{+}m_2)^2}{4(1{+}z)^2}{-}\frac{m_3\Lambda}{z}{-}\frac{1{-}2m_1^2{-}2m_2^2{+}4u{+}2\Lambda(m_1{+}m_2{-}1)}{4z(1{+}z)}
\ee

\subsection{Gauge/gravity dictionaries}
The identification between $Q_W$ \eqref{eq:Qgrav1} and $Q_{12}$ \eqref{QSW12} allows us to construct the dictionary between the parameters of the gravity solution and of the gauge theory side. 
For the TS we choose
\be
z = \frac{r-r_b}{r_b-r_s}
\ee
in such a way that $r= \infty \leftrightarrow z = \infty$, $r=r_b \leftrightarrow z=0$ and $r=r_s \leftrightarrow z=-1$. This is suitable for the present case because the values $z<0$ on the gravity side are not physical, exactly like for the case with $r<r_b$. 
By doing this change of variables, and setting $\mu=p=0$ for simplicity, we get the following dictionary
\begin{align}
\label{dictionary ts}
&\hspace{1.5cm} m_1 =m_2= -\frac{(q Q-r_s\omega)}{\sqrt{\frac{r_b}{r_s}-1}},\quad  m_3=\frac{-q^2 r_s - 2 q Q \omega + (r_b+ 2 r_s) \omega^2}{2i \sqrt{\omega^2-q^2}} \\
& u = \left(\ell{+}\frac{1}{2} \right)^2{-}q^2[Q^2{+}(r_b{-}r_s)r_s]{+}4 q Q r_s \omega {-} 3 r_s^2 \omega^2 {-}i\sqrt{\omega^2{-}q^2}(r_b{-}r_s)\left(1{+}2\frac{q Q{-}r_s\omega}{\sqrt{\frac{r_b}{r_s}{-}1}}\right) \notag\\ 
&\hspace{5cm}\Lambda= -2 \text{i} (r_b-r_s) \sqrt{\omega^2-q^2} \notag 
\end{align}

We choose to impose $p=\mu=0$ also for the BS. In this instance
\be
z=\frac{r-r_s}{r_s-r_b}
\ee
so that the dictionary is the following
  \be\label{dictBS}
m_1=-m_2=\frac{i(r_s\omega-q Q)}{\sqrt{1-\frac{r_b}{r_s}}},\quad m_3=\frac{-q^2 r_s -2q Q\omega+(r_b+2r_s)\omega^2}{2i\sqrt{\omega^2-q^2}}
  \ee
 $$ 
  u=\left(\ell+\frac{1}{2}\right)^2-q^2Q^2+2qQ(r_b+r_s)\omega-\omega^2(r_b^2+r_b r_s+r_s^2)+i (r_b-r_s)\sqrt{\omega^2-q^2}
$$
 $$\Lambda=-2i(r_s-r_b)\sqrt{\omega^2-q^2}$$

\subsection{Quantum Seiberg-Witten periods}
The dynamics of the gauge theory is codified in terms of a single holomorphic function which is the qSW period $a$. In order to compute it, we have to Fourier transform \eqref{eqnSWorig} bringing it in the form of a difference equation \cite{Poghossian:2010pn,Fucito:2011pn, Poghosyan:2020zzg}, whose integrability condition can be written as a continued fraction
\be\label{diffeqSW}
P_0(a)=\frac{\Lambda M(a+1)}{P_0(a+1)-\frac{\Lambda M(a+2)}{P_0(a+2)-...}}+\frac{\Lambda M(a)}{P_0(a-1)-\frac{\Lambda M(a-1)}{P_0(a-2)-...}}
\ee
with
\be
M(x)= P_L\left(x-\frac{1}{2}\right)P_R\left(x-\frac{1}{2}\right)
\ee
Equation \eqref{diffeqSW} with \eqref{classSW} can be solved for $u(a,\Lambda,m_i)$ order by order in $\Lambda$, getting
\begin{align}
\label{uvsa}
u(a,\Lambda,m_i) = & a^2 + \Lambda \left[\frac{1}{2}(1-m_1-m_2-m_3) - \frac{2 m_1 m_2 m_3}{4 a^2 - 1}\right] + \frac{\Lambda^2}{128 (a^2-1)} \left[4 a^2 -5 + \notag \right.\\
&\left.+ 4(m^2_1+m^2_2+m^2_3) - \frac{48 (m_1^2 m_2^2 + m_2^2 m_3^2 + m_1^2 m_3^2)}{4a^2 -1} + \frac{64 m^2_1 m^2_2 m^2_3(20 a^2+7)}{(4 a^2 -1)^3} \right]+\notag\\&+\dots
\end{align}
We can invert this series in order to obtain 
\begin{align}
\label{avsu}
a(u,\Lambda,m_i) =& \sqrt{u} + \frac{\Lambda}{4 \sqrt{u}} \left(\frac{4 m_1 m_2 m_3}{4u-1} + m_1 + m_2 + m_3 - 1\right) + \\\nn
&-\frac{\Lambda ^2}{256 \sqrt{u}} \left(\frac{1024 m_1^2 m_2^2 m_3^2}{(4
   u-1)^3}-\frac{256 m_1 m_2 \left(m_1 \left(m_2 m_3-2\right)-2
   \left(m_2+m_3-1\right)\right) m_3}{(1-4 u)^2}+\right.\\\nn
   &\left.+\frac{8
   \left(m_2+m_3+m_1 \left(1-4 m_2
   m_3\right)-1\right){}^2}{u}+\frac{\left(4 m_1^2-1\right)
   \left(4 m_2^2-1\right) \left(4 m_3^2-1\right)}{u-1}+\right.\\\nn
   &\left.+\frac{64
   \left(\left(\left(1{-}12 m_3^2\right) m_2^2{+}4 m_3
   m_2{+}m_3^2\right) m_1^2{+}4 m_2 m_3 \left(m_2{+}m_3{-}1\right)
   m_1{+}m_2^2 m_3^2\right)}{4 u{-}1}{+}4\right)+\\&{+}\dots
\end{align}
The Nekrasov-Shatashvili (NS) prepotential $\mathcal{F}_{\text{NS}}(a,\Lambda)$ \footnote{We do not explicitly write the dependence from the masses, as instead done in Chapter \ref{chap_1}, for simplicity.} is obtained by integrating the quantum version of the Matone relation \cite{Matone:1995rx,Flume:2004rp}
\be\label{matonerel}
u=-\Lambda \frac{\partial \mathcal{F}_{\text{NS}}(a,\Lambda)}{\partial \Lambda}
\ee
Furthermore, we have to add a $\Lambda$-independent constant obtained from the one-loop prepotential, so that
\be
\mathcal{F}_{\text{NS}}(a,\Lambda)=\mathcal{F}_{\text{tree}}(a,\Lambda)+\mathcal{F}_{1\text{-L}}(a)+\mathcal{F}_{\text{inst}}(a,\Lambda)
\ee
where
\begin{align}
\label{Prepotential}
& \mathcal{F}_{\text{tree}}(a,\Lambda) = -a^2 \log \Lambda \notag \\
& \mathcal{F}_{\text{inst}}(a,\Lambda) = \Lambda \frac{1-m_2-m_3+4 a^2 (m_1 + m_2 + m_3 -1) + m_1 (4 m_2 m_3 - 1)}{2(4a^2-1)} + \dots\\\nn
&\frac{\partial \mathcal{F}_{1\text{-L}}(a)}{\partial a} = \log \left[\frac{\Gamma^2(1+2a)}{\Gamma^2(1-2a)} \prod_{i=1}^3 \frac{\Gamma\left(\frac{1}{2}+m_i-a\right)}{\Gamma\left(\frac{1}{2}+m_i+a\right)}\right]
\end{align}
Finally, the $a_D-$period is given by
\be\label{addef}
a_D=-\frac{1}{2\pi i} \frac{\partial \mathcal{F}_{\text{NS}}}{\partial a}
\ee
QNM frequencies are obtained by imposing the exact quantization condition \cite{Bonelli:2022ten,Consoli:2022eey}:
\be\label{condquantqnm}
a_D=n\quad,\quad n=0,1,2,...
\ee
once expressed in terms of the gravity variables via the dictionary. The latter is solved in $\omega$ in the complex plane.

\section{Leaver method} \label{Leaver}

In this Section, we describe a semi-analytical method used to solve the radial differential equation involving a continued fraction, introduced by Leaver \cite{Leaver:1985ax, Leaver:1990zz}. This method has been applied in \cite{Bianchi:2023sfs, Heidmann:2023ojf} to the study of QNMs for neutral scalar waves in TS geometry. 

For generality, let us implement this technique on the (1,2)-qSW curve \eqref{eqnSWcan}-\eqref{QSW12}, once we redefine the canonical function as
\be
\Psi(z)=\sqrt{z(1+z)}R(z)
\ee
so that the equation becomes
\begin{equation}\label{12qswN}
z(1+z)R''(z)+(1+2z)R'(z)+\Big[-\frac{\Lambda^2z^2}{4}+\frac{\Lambda(\Lambda+4m_3)z}{4}-\frac{(m_1-m_2)^2}{4z}+\frac{(m_1+m_2)^2}{4(1+z)}+
\end{equation}
$$+\frac{1}{4}(1-4u+2\Lambda(1-m_1-m_2-2m_3))\Big]R(z)=0$$
The coordinate $z$ is the coordinate that originates at the cap in the case where the geometry is smooth $(r_b>r_s)$, while it originates at the horizon in the case of a BS
\be
z = \begin{cases}
\frac{r - r_b}{r_b - r_s} & \quad \mathrm{if} \ r_b>r_s \\
\frac{r - r_s}{r_s - r_b} & \quad \mathrm{if} \ r_s > r_b
\end{cases}
\ee
Once the appropriate gauge/gravity dictionary is chosen, the Frobenius exponents for the regular singular points are chosen in such a way that at $z=0$ they encode the information that the solution is regular in the case of smooth geometry or that the wave is ingoing in the case of a BS, while at infinity, where the singularity is irregular, the wave is outgoing in both cases. The other Frobenius exponent at $z=-1$ is chosen such that the recurrence relation is three-termed. Therefore, the correct ansatz is
\begin{equation}\label{leava}
    R_L(z)=e^{-\frac{\Lambda z}{2}}z^{-\frac{m_1-m_2}{2}}(1+z)^{\frac{m_1-m_2-2(1+m_3)}{2}}\sum_{n=0}^\infty c_n\left(\frac{z}{1+z}\right)^n
\end{equation}
If we plug \eqref{leava} in \eqref{12qswN}, we obtain a three terms recursion relation of the form
\be
\alpha_{n}c_{n+1}+\beta_n c_n+\gamma_n c_{n-1}=0\quad,\quad n\ge 0
\ee where the coefficients are
\begin{align}
\label{coeff Leaver}
\alpha_n=&(1+n)(1+n-m_1+m_2)\\\nn
\beta_n=&{-}\left(m_3{+}n\right) (\Lambda {+}2 n){-}m_2 \left(\Lambda {+}m_3{+}2
   n{+}1\right){+}m_1 \left(m_2{+}m_3{+}2 n{+}1\right){-}m_3{-}2
   n{-}u{-}\frac{3}{4}\\\nn
\gamma_n=&(n+m_2+m_3)(n-m_1+m_3)
\end{align}
and where we chose $c_{-1}=0$. The QNMs associated to the overtone $n$ are obtained by truncating the recursion and solving numerically the continuous fraction
\be
\label{fraction Leaver}
\beta_n =\frac{\alpha_n\gamma_n}{\beta_{n-1}-\frac{\alpha_{n-2}\gamma_{n-1}}{\beta_{n-2}-\dots}}+\frac{\alpha_n\gamma_{n+1}}{\beta_{n+1}-\frac{\alpha_{n+1}\gamma_{n+2}}{\beta_{n+2}-\dots}}
\ee
viewed as an equation for the QNMs $\omega$.

The two methods exposed in Section \ref{quantumSW} and in the current one can be applied in both cases of TSs ($r_b>r_s$) and BSs ($r_s>r_b$). 

\section{Charged QNMs of Topological Star} \label{TopstarQNM}
In this Section we analyze the case of the TSs. We compute the QNMs with the qSW technique and the Leaver method; furthermore, we obtain the same quantities also with a direct integration and then we compare the results with these different methods.
\subsection{Computation of the QNMs}
\begin{figure}[H]
    \centering
    \includegraphics[width=0.5\textwidth]{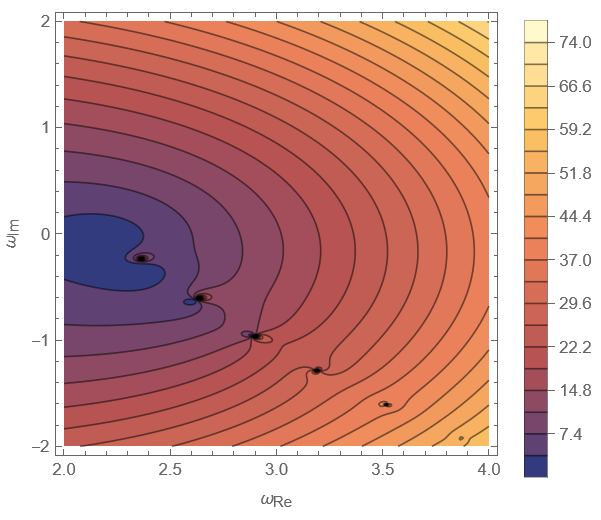} \\
    \includegraphics[width=0.3\textwidth]{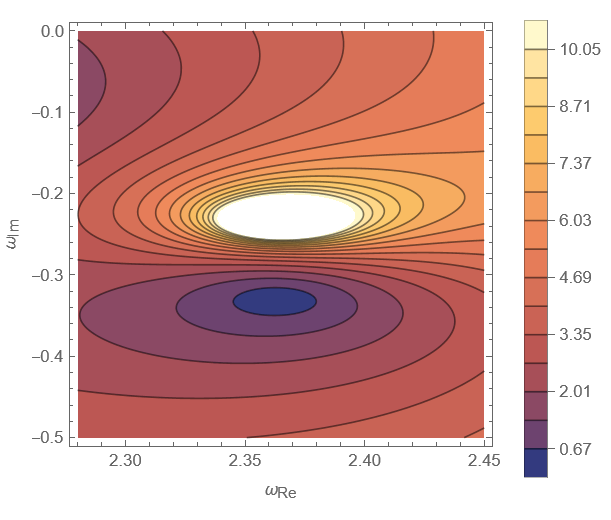}
    \includegraphics[width=0.3\textwidth]{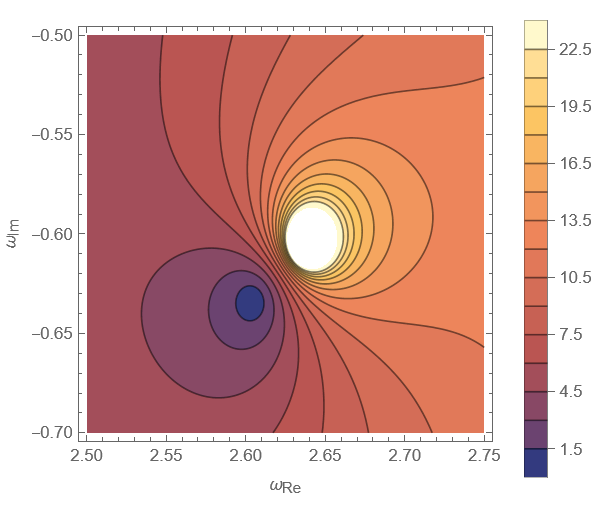}
    \includegraphics[width=0.3\textwidth]{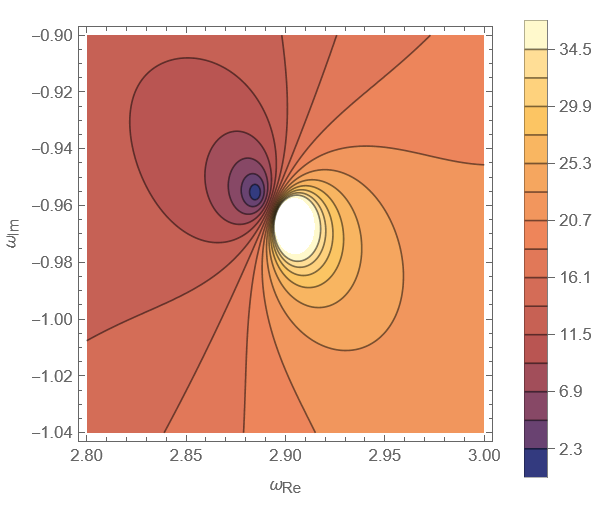}
    \label{fig:immagine1}
    \caption{Plots of the modulus of the Leaver difference equation \eqref{fraction Leaver} in the complex $\omega$-plane for $r_b=1$, $r_s=0.8$, $\ell=2$, $Q=2$, $q=0.5$, $p=0$, $\m=0$. The three plots at the bottom are zoomed-in snapshots of the regions around the first three zeros of the Leaver function.}\label{plotL}
\end{figure}

The dictionary \eqref{dictionary ts} is the fundamental ingredient required for implementing the Leaver method \eqref{coeff Leaver}, \eqref{fraction Leaver} and the SW quantization condition \eqref{Prepotential}, \eqref{addef}, \eqref{condquantqnm}. As demonstrated in Figure \ref{plotL} and Tables \ref{Tabb1}, \ref{Tabb2} and \ref{Tabb3}, only stable modes are present. In the last column, by $SW_s$ we mean that the number of instantonic corrections used in the SW procedure is equal to $s$.\\

The resulting QNM frequencies have also been validated against those obtained using a direct integration method, which involves matching numerical solutions in different integration regions. Starting from the cap, in accordance with the solution $(\ref{RCAPts})$, we impose a regular boundary condition and integrate the equation up to a specified extraction point. From this point, following the solution $(\ref{RINF})$, we impose outgoing waves and continue integration until reaching infinity. The two solutions can be matched by requiring the vanishing of the Wronskian, which then becomes an eigenvalue equation for the QNM frequencies. In Table \ref{Tabb2}, the column associated to the numerical integration is empty because this algorithm fails to be predictive when the QNMs are highly damped, that is when their imaginary parts are close to 1. Indeed in this scenario we need a huge number of subleading corrections at the cap and at infinity to estimate the QNMs with high imaginary parts.\\

As it can be seen, the results of the QNMs corresponding to the different methods are in agreement. We expect that a larger number of instantonic corrections in the qSW method will produce estimates closer to the ones provided by numerical integration and Leaver method.

\begin{table}[H]
\centering
\begin{tabular}{|c|c|c|c|}
\hline
$\ell$ & Numerical integration & Leaver                & $SW_4$ \\ \hline
$0$    & $1.34819 - 0.218003 \, i$    & $1.34819 - 0.218003 \, i$  & $1.33277 - 0.261032 \, i$   \\ \hline
$1$    & $1.73838 - 0.126107 \, i$    & $1.73838 - 0.126107 \, i$  & $1.72053 - 0.117156 \, i$   \\ \hline
$2$    & $2.18231 - 0.0797013 \,  i$    & $2.18231 - 0.0797013 \, i$ & $2.17005 - 0.0762405 \, i$   \\ \hline
$3$    & $2.63417 - 0.0517219 \,  i$    & $2.63417 - 0.0517219 \, i$ & $2.6258 - 0.060272 \, i$   \\ \hline
\end{tabular}
\caption{TS QNMs for different values of $\ell$ and fixed overtone $n=0$ and $r_b=1$, $r_s=0.8$, $Q=2$, $q=0.5$, $p=\mu=0$.}\label{Tabb1}
\end{table}
\begin{table}[H]
\centering
\begin{tabular}{|c|c|c|c|}
\hline
$n$ & Numerical integration & Leaver                & $SW_2$ \\ \hline
$1$    & $-$    & $2.36322 - 0.330499 \, i$  & $2.24082 - 0.293755 \, i$   \\ \hline
$2$    & $-$    & $2.60406 - 0.633878 \, i$  & $2.83614 - 0.700827 \, i$   \\ \hline
$3$    & $-$    & $2.88498 - 0.955453 \, i$  & $2.89907 - 0.926271 \, i$   \\ \hline
\end{tabular}
\caption{TS QNMs for fixed $\ell=2$, $r_b=1$, $r_s=0.8$, $Q=2$, $q=0.5$, $p=\mu=0 $ and for various overtone numbers.}\label{Tabb2}
\end{table}
\begin{table}[H]
\centering
\begin{tabular}{|c|c|c|c|}
\hline
$n$ & Numerical integration & Leaver                & $SW_4$ \\ \hline
$0$    & $5.81013 - 0.000491187 \,i$    & $5.81013 - 0.000491187\, i$  & $5.81895 - 0.0139528 \,i$   \\ \hline
$1$    & $5.97759 - 0.0411436 \,i$    & $5.97759 - 0.0411436 \,i$  & $5.96178 - 0.0271379\, i$   \\ \hline
$2$    & $6.13468 - 0.17184\, i$    & $6.13468 - 0.17184\, i$  & $6.12144 - 0.14779\, i$   \\ \hline
$3$    & $-$    & $6.31023 - 0.343684 \, i$  & $6.30019 - 0.385102\, i$   \\ \hline
\end{tabular}
\caption{TS QNMs for fixed $\ell=10$, $r_b=1$, $r_s=0.8$, $Q=2$, $q=0.5$, $p=\mu=0 $ and for various overtone numbers.}\label{Tabb3}
\end{table}

\section{Charged QNMs and Superradiance of Black String} \label{Charge superradiance}
In this Section we move to examine the BSs. We explain why BSs can produce charge superradiance and we also show that in TSs case, instead, this phenomenon cannot take place. At this point, we can introduce the \textit{near extremal, near superradiant limit} together with the \textit{probe limit}, arriving at computing the QNMs with the usual two methods of the qSW and Leaver and now also with the matching asymptotic expansion. Prompt Ring Down modes are expected as well and indeed they are computed, in addition to the two persistent methods, with a WKB approximation. At the end of this Section, the amplification factor of superradiance for BSs is computed, keeping in account also the first instantonic corrections.    
 
\subsection{Charge superradiance}
The TS and the BS are static (in the present discussion) and charged geometries and hence they can only give rise to charge superradiance. Focusing on the BS, we have to determine first of all the behaviour of the radial wave function $\Psi(z)$ at the boundaries of the spacetime under consideration, that is $z=0$ and $z=\infty$. By solving \eqref{eqnSWcan} with $Q_{12}$ given by \eqref{QSW12} adapted at the two boundaries in exam, we get
\be
\Psi(z) \simeq 
\begin{cases}
 \mathcal{O} z^{\frac{1}{2} - i \, \, \frac{q Q - \omega r_s}{\sqrt{1-\frac{r_b}{r_s}}}} +  \mathcal{T} z^{\frac{1}{2} + i \, \, \frac{q Q - \omega r_s}{\sqrt{1-\frac{r_b}{r_s}}}}, \qquad\qquad\qquad z=0 \\
 \mathcal{R} e^{ i (r_s-r_b) \, z \, \,\sqrt{\omega^2-q^2}} +  \mathcal{I} e^{ -i (r_s-r_b) \, z \, \,\sqrt{\omega^2-q^2}}, \quad z \to +\infty\\
\end{cases}
\ee
where, following the convention of \cite{Brito:2015oca}, $\mathcal{I}$ and $\mathcal{R}$ are, respectively, the amplitudes of  incident and reflected waves (defined at spatial infinity $z = \infty$), while $\mathcal{T}$ and $\mathcal{O}$ are the amplitudes of transmitted and outgoing waves across the boundary $z=0$ (the horizon, in this geometry).\\
The probability currents associated to the two asymptotic solutions are
\begin{align}
& J_0 = \frac{2 i (-q Q + r_s \omega)}{\sqrt{1-\frac{r_b}{r_s}}} (|\mathcal{O}|^2-|\mathcal{T}|^2) \\
& J_\infty = 2 \, i \, (r_b-r_s) \sqrt{\omega^2-q^2} (|\mathcal{I}|^2-|\mathcal{R}|^2)
\end{align}
and they must be equal due to the conservation and this implies
\be
\label{incident wave amplitude}
|\mathcal{I}|^2 = |\mathcal{R}|^2 + \frac{(q Q-r_s \omega) \sqrt{r_s}}{(r_s-r_b)^{\frac{3}{2}} \sqrt{\omega^2-q^2}} (|\mathcal{O}|^2 - |\mathcal{T}|^2) 
\ee
Notice that if $\omega < \omega_{SR} = \frac{q Q}{r_s}$ then the wave is superradiantly amplified ($|\mathcal{I}|^2 < |\mathcal{R}|^2$) \footnote{For a BH the boundary conditions at the horizon involve only ingoing waves, so that $\mathcal{O}=0$.}. For this reason $\omega_{SR}$ will be denoted as \textit{superradiance threshold}.\\
This result is in complete agreement with the one we could obtain by using the classical laws of BH thermodynamics, as firstly understood by Bekenstein \cite{PhysRevD.7.949}. Indeed, the reduction from 5 to 4 dimensions for BS gives rise to a Reissner-Nordstr\"om BH with radius of the horizon equal to $r_s$. The first law of thermodynamics for this kind of BH reads
\be
\label{1st law BH}
\delta M = \frac{T_H}{4} \delta A_H + \Phi_H \delta \mathcal{Q} 
\ee
where $T_H$, $A_H$, $\Phi_H$ are the temperature, the area and the electric potential at the horizon, hence $\Phi_H = \mathcal{Q}/r_s$. If we send a monochromatic wave with frequency $\omega$ and charge $q$, the variation of the mass of the background in terms of its electric charge is
\be
\delta M = \delta \mathcal{Q} \frac{\omega}{q}
\ee
and so, plugging in \eqref{1st law BH}, we get
\be
\delta \mathcal{Q} = \frac{T_H q}{4} \frac{\delta A_H}{\omega - q \Phi_H}
\ee
In order to have charge superradiance ($\delta \mathcal{Q} < 0$), since $\delta A_H \geq 0$ for the second law of BH thermodynamics, we must have $\omega - q \Phi_H < 0$. This is exactly what we found out in \eqref{incident wave amplitude}.\\

It is worth noticing that in the case of TS $(r_b>r_s)$, it is not possible to have superradiant amplification. This circumstance arises from the fact that smooth geometries resolve the information paradox and therefore the amplitude of the incident wave on our compact object will be equal to the reflected one. Once the regularity is imposed at the cap of the geometry $r=r_b$, the amplification curve obtained by numerically integrating the radial wave equation is zero. This will be shown in Figure \ref{plotAmpli} together with the amplification factor of BS.

\subsection{Near extremal and near superradiant modes}
In this Subsection we will analytically point out some peculiar modes, referred as \textit{near extremal, near superradiant modes}, that will be computed by exploiting some matching asymptotic expansion technique.\\
Firstly notice that in the near extremal limit ($r_s \to r_b$) the theory decouples ($\Lambda \to 0$) and, in order to avoid divergences arising in the dictionary \eqref{dictBS}, we must fix $\omega = \omega_{SR}$.\\ 

Let us now apply the technique previously cited.
The differential equation \eqref{eqnSWcan} with $Q_{12}$ in \eqref{QSW12} in the decoupling limit $\Lambda=0$ can be solved in terms of hypergeometric functions (this point will be crucial in Chapter \ref{chap_5} for a different problem). The solution that guarantees ingoing waves at the horizon $z=0$ is
\be
\Psi(z)=z^{\frac{1}{2}(1-m_1+m_2)}(1+z)^{\frac{1}{2}(1+m_1+m_2)}{}_2F_1\Big[\frac{1}{2}+m_2-\sqrt{u},\frac{1}{2}+m_2+\sqrt{u},1-m_1+m_2,-z\Big]
\ee
The behavior of the previous solution at infinity is
\be\label{sol1}
\Psi(z)\sim z^{\frac{1}{2}-\sqrt{u}}\left(1+{z^{2\sqrt{u}}\Gamma(\frac{1}{2}-m_1-\sqrt{u})\Gamma(\frac{1}{2}+m_2-\sqrt{u})\Gamma(2\sqrt{u})\over \Gamma(\frac{1}{2}-m_1+\sqrt{u})\Gamma(\frac{1}{2}+m_2+\sqrt{u})\Gamma(-2\sqrt{u})}\right)
\ee
At $z=\infty$ the differential equation becomes
\be
\Psi''(z)+\left(-{m_3\Lambda\over z}-{\Lambda^2\over4}+{1-4u\over 4z^2}\right)\Psi(z)=0
\ee
whose general solution is
\be\label{solinfpm}
\Psi(z)=\sum_{\alpha=\pm}A_\alpha e^{{\alpha\Lambda z\over2}}(z\Lambda)^{{1\over2}{+}\sqrt{u}}U\Big[{1\over2}{-}\alpha m_3{+}\sqrt{u},1{+}2\sqrt{u},{-}\alpha\Lambda z\Big]
\ee
In order to have only outgoing waves at infinity, we are forced to set $A_+=0$ (and $A_-=1$ without loss of generality). After that we can expand the previous \eqref{solinfpm} solution at the horizon
\be\label{sol2}
\Psi(z)\sim z^{{1\over2}-\sqrt{u}}\left(1+{z^{2\sqrt{u}}\Lambda^{2\sqrt{u}}\Gamma({1\over2}+m_3+\sqrt{u})\Gamma(-2\sqrt{u})\over \Gamma\left(\frac{1}{2}+m_3-\sqrt{u}\right)\Gamma(2\sqrt{u})}\right)
\ee
The two solutions \eqref{sol1} and \eqref{sol2} match if
\be\label{1loop}
\Lambda^{2\sqrt{u}}-{\Gamma(1+2\sqrt{u})^2\Gamma({1\over2}-m_1-\sqrt{u})\Gamma\left(\frac{1}{2}+m_2-\sqrt{u}\right)\Gamma\left(\frac{1}{2}+m_3-\sqrt{u}\right)\over \Gamma(1-2\sqrt{u})^2\Gamma\left(\frac{1}{2}-m_1+\sqrt{u}\right)\Gamma\left({1\over2}+m_2+\sqrt{u}\right)\Gamma\left(\frac{1}{2}+m_3+\sqrt{u}\right) }=0
\ee
Since the first term in \eqref{1loop} is going to zero, in order to satisfy the equation we require that one of the $\Gamma-$functions in the denominator of the second term must diverge. The correct choice is
\be
{1\over2}+m_2+\sqrt{u}=-n\quad, \quad n=0,1,2,...
\ee
Using the dictionary \eqref{dictBS} and replacing $\omega=\omega_{SR}-\nu \sqrt{r_s-r_b}$, we can find the correction to the superradiant frequency by expanding for small $r_s-r_b$:
\be
\label{omega match}
\omega_{match}={q Q\over r_s}-i{\ell+1+n\over r_s^{3/2}}\sqrt{r_s-r_b}
\ee
The observant reader will have noticed that the matching condition \eqref{1loop} coincides with the quantization of the cycle $a_D$ as written in \eqref{condquantqnm}, where only the tree-level and the 1-loop contributions are present. So, starting from \eqref{1loop}, in order to consider the istantonic corrections, let us replace $\sqrt{u}=a$ with the $a$-cycle defined as \eqref{avsu} and we also add the instantonic contribution of the prepotential \eqref{Prepotential}, getting at the end
\be
\label{SW freq}
\Lambda^{2a}-{\Gamma(1+2a)^2\Gamma({1\over2}-m_1-a)\Gamma(\frac{1}{2}+m_2-a)\Gamma(\frac{1}{2}+m_3-a)\over \Gamma(1-2a)^2\Gamma(\frac{1}{2}-m_1+a)\Gamma({1\over2}+m_2+a)\Gamma(\frac{1}{2}+m_3+a) }e^{\partial_a F_{\text{inst}}(\Lambda)}=0
\ee 
The results obtained with the matching asymptotic expansion \eqref{omega match}, with the SW quantization \eqref{SW freq} and with the Leaver continued fraction \eqref{coeff Leaver},\eqref{fraction Leaver}, using the dictionary \eqref{dictBS}, are displayed in Tables \ref{Tab 2 BS} and \ref{Tab 3 BS}.

\begin{table}[H]
\centering
\begin{tabular}{|c|c|c|c|}
\hline
$r_b$   & $\omega_{match}$               & Leaver                 & $SW_4$                 \\ \hline
$0.9$   & $1 - 0.948683 \, i$    & $1.51407 - 1.10179 \, i$  & $1.51404 - 1.10182 \, i$  \\ \hline
$0.95$  & $1 - 0.67082 \, i$     & $1.06253 - 0.7356 \, i$   & $1.06253 - 0.7356 \, i$   \\ \hline
$0.99$  & $1 - 0.3 \, i$        & $1.01209 - 0.305512 \, i$ & $1.01209 - 0.305512 \, i$ \\ \hline
$0.999$ & $1 - 0.0948683 \, i$ & $1.0012 - 0.0950405 \, i$ & $1.0012 - 0.0950407 \, i$ \\ \hline
\end{tabular}
\caption{Near extremal near superradiant modes corresponding to the first overtone number $n=0$ for $r_s=1$, $\ell=2$, $q=0.5$, $Q=2$, $\mu=p=0$.}
\label{Tab 2 BS}
\end{table}

\begin{table}[H]
\centering
\begin{tabular}{|c|c|c|c|}
\hline
$r_b$   & $\omega_{match}$               & Leaver                 & $SW_4$                 \\ \hline
$0.9$   & $1 - 1.26491 \, i$    & $1.4123 - 1.39 \, i$  & $1.41232 - 1.38995 \, i$  \\ \hline
$0.95$  & $1 - 0.894427 \, i$     & $1.07965 - 1.00857 \, i$   & $1.07965 - 1.00857 \, i$   \\ \hline
$0.99$  & $1 - 0.4 \, i$        & $1.01595 - 0.409877 \, i$ & $1.01596 - 0.409876 \, i$ \\ \hline
$0.999$ & $1 - 0.126491 \, i$ & $1.0016 - 0.126801 \, i$ & $1.0016 - 0.126801 \, i$ \\ \hline
\end{tabular}
\caption{Near extremal near superradiant modes corresponding to the second overtone number $n=1$ for $r_s=1$, $\ell=2$, $q=0.5$, $Q=2$, $\mu=p=0$.}
\label{Tab 3 BS}
\end{table}

\subsection{Prompt Ring Down modes}

As extensively shown in \cite{Bianchi:2022qph,Bianchi:2023sfs,DiRusso:2024hmd}, in general, within smooth and horizonless solutions, one can distinguish between unstable and metastable modes, corresponding to frequencies near the maximum or minimum of the effective potential, respectively. Unstable frequencies are also referred to as prompt ring down modes, bearing in mind the analogy with BH mergers. Given the shape of the effective potential of the BS, which exhibits only one unstable light ring, we expect such modes to be present in the spectrum of QNMs and we aim to provide a WKB estimate of their frequency values. Exploiting the results shown in \cite{Bianchi:2021mft}, prompt ring down frequencies must satisfy the Bohr-Sommerfeld quantization condition
\be
\int_{z_-}^{z_+}\sqrt{Q(z)}dz=\pi\left(n+\frac{1}{2}\right)
\ee
with $n$ the overtone number and $z_{\pm}$ the two turning points. In the semiclassical limit the turning points collide and the integral can be approximated as
\be
\int_{z_-}^{z_+}\sqrt{Q(z)}dz\simeq \int_{z_-}^{z_+}\sqrt{Q(z_c)+\frac{Q''(z_c)}{2}(z-z_c)^2}dz\simeq\frac{i\pi Q(z_c)}{\sqrt{2Q''(z_c)}}
\ee
where $z_c\in[z_-,z_+]$ the critical point $Q'(z_c)=0$. So we are left with
\be
\frac{Q(z_c)}{\sqrt{2Q''(z_c)}}=-i\left(n+\frac{1}{2}\right)
\ee
The equation can be solved perturbatively by substituting the information that $\omega=\omega_{\text{R}}+i \omega_{\text{I}}$ and expanding for small $\omega_{\text{I}}$. It is easy to see that $\omega_{\text{R}}$ is simply the critical frequency at the unstable photon sphere. The results are shown in Tables \ref{tabb6} and \ref{tabb7} and are compared with the results provided by the Leaver method and SW quantization.
\begin{table}[H]
\centering
\begin{tabular}{|c|c|c|c|}
\hline
$\ell$ & WKB                   & Leaver               & $SW_4$                    \\ \hline
$0$    & $0.884096 - 0.402014 \, i$ & $1.04019 - 0.582305 \, i$ & $1.0416 - 0.571598 \, i$ \\ \hline
$1$    & $1.31439 - 0.144228 \, i$  & $1.36621 - 0.13088 \, i$  & $1.37756 - 0.138392 \, i$  \\ \hline
$2$    & $1.68218 - 0.13711 \, i$   & $1.71123 - 0.133916 \, i$ & $1.71388 - 0.108494 \, i$  \\ \hline
$3$    & $2.05574 - 0.135935 \, i$  & $2.07588 - 0.134619 \, i$ & $2.06643 - 0.135559 \, i$  \\ \hline
$4$    & $2.43315 - 0.135379 \, i$  & $2.44854 - 0.134684 \, i$ & $2.44221 - 0.138805 \, i$  \\ \hline
$5$    & $2.81279 - 0.134979 \, i$  & $2.82523 - 0.134556 \, i$ & $2.82053 - 0.143357 \, i$  \\ \hline
$6$    & $3.19382 - 0.134657 \, i$  & $3.20425 - 0.134375 \, i$ & $3.20389 - 0.143435 \, i$  \\ \hline
$7$    & $3.57575 - 0.134387 \, i$  & $3.58472 - 0.134186 \, i$ & $3.58429 - 0.144436 \, i$  \\ \hline
$8$    & $3.9583 - 0.134157 \, i$   & $3.96617 - 0.134008 \, i$ & $3.9698 - 0.145597 \, i$   \\ \hline
$9$    & $4.34129 - 0.13396 \, i$   & $4.34831 - 0.133845 \, i$ & $4.35221 - 0.144408 \, i$  \\ \hline
$10$   & $4.72462 - 0.133788 \, i$  & $4.73094 - 0.133697 \, i$ & $4.73747 - 0.146576 \, i$  \\ \hline
\end{tabular}
\caption{Prompt ring down modes for the first overtone number $n=0$ and $r_s=1$, $r_b=0.8$, $q=0.5$, $Q=2$, $p=\mu=0$ for various $\ell$.}\label{tabb6}
\end{table}
\begin{table}[H]
\centering
\begin{tabular}{|c|c|c|c|}
\hline
$r_b$ & $WKB$                  & Leaver               & $SW_4$                   \\ \hline
$0.5$ & $1.67435 - 0.180019 \, i$ & $1.72041 - 0.170672 \, i$ & $1.70821 - 0.156853 \, i$ \\ \hline
$0.6$ & $1.67723 - 0.166767 \, i$ & $1.71766 - 0.159428 \, i$ & $1.7079 - 0.144709 \, i$  \\ \hline
$0.7$ & $1.67991 - 0.152547 \, i$ & $1.71465 - 0.147257 \, i$ & $1.70806 - 0.130406 \, i$ \\ \hline
$0.8$ & $1.68218 - 0.13711 \, i$  & $1.71123 - 0.133916 \, i$ & $1.71388 - 0.108494 \, i$ \\ \hline
$0.9$ & $1.6836 - 0.120151 \, i$  & $1.70704 - 0.119066 \, i$ & $1.71739 - 0.121633 \, i$ \\ \hline
\end{tabular}
\caption{Prompt ring down modes for the first overtone number $n=0$ and $r_s=1$, $\ell=2$, $q=0.5$, $Q=2$, $p=\mu=0$ for various $r_b$.}\label{tabb7}
\end{table}

\subsection{Amplification factor}
The general solution at infinity \eqref{solinfpm} has to be matched with \eqref{sol1}, encoding ingoing boundary conditions at the horizon. This happens if 
\begin{align}
A_{+}{=}&\frac{\Gamma \left(\frac{1}{2}{-}m_3{-}\sqrt{u}\right) \Gamma \left(\frac{1}{2}{-}m_3{+}\sqrt{u}\right)}{ \Gamma
   \left(\frac{1}{2}{-}m_3{+}\sqrt{u}\right) \Gamma \left(\frac{1}{2} + m_3{-}\sqrt{u}\right){-}  ({-}1)^{2 \sqrt{u}}\Gamma \left(\frac{1}{2} {-} m_3{-}\sqrt{u}\right) \Gamma
   \left(\frac{1}{2} {+} m_3{+}\sqrt{u}\right)}\times\\\nn
   &\Bigg[\frac{ \Gamma \left(2 \sqrt{u}\right) \Gamma \left(\frac{1}{2}{-}m_1{-}\sqrt{u}\right) \Gamma \left(\frac{1}{2}{+}m_2{-}\sqrt{u}\right)
   \Gamma \left(\frac{1}{2}{+}m_3{-}\sqrt{u}\right)}{\Gamma \left({-}2 \sqrt{u}\right)}+\\\nn
   &{-}\frac{\Lambda
   ^{2 \sqrt{u}} \Gamma \left({-}2 \sqrt{u}\right)  \Gamma \left(\frac{1}{2}{-}m_1{+}\sqrt{u}\right) \Gamma \left(\frac{1}{2}{+}m_2{+}\sqrt{u}\right) \Gamma
   \left(\frac{1}{2}{+}m_3{+}\sqrt{u}\right)}{\Gamma \left(2 \sqrt{u}\right)}\Bigg]
\end{align}
\begin{align}
A_-{=}&\frac{\Gamma \left(\frac{1}{2}{+}m_3{-}\sqrt{u}\right) \Gamma \left(\frac{1}{2}{+}m_3{+}\sqrt{u}\right)}{ \Gamma
   \left(\frac{1}{2}{-}m_3{+}\sqrt{u}\right) \Gamma \left(\frac{1}{2}{+}m_3{-}\sqrt{u}\right){-} ({-}1)^{2 \sqrt{u}} \Gamma \left(\frac{1}{2}{-}m_3{-}\sqrt{u}\right) \Gamma
   \left(\frac{1}{2}{+}m_3{+}\sqrt{u}\right)}\times\\\nn
   &\Bigg[\frac{ \Gamma \left(-2 \sqrt{u}\right) \Lambda ^{2 \sqrt{u}} \Gamma \left(\frac{1}{2}{-}m_1+\sqrt{u}\right) \Gamma
   \left(\frac{1}{2} {+} m_2+\sqrt{u}\right) \Gamma \left(\frac{1}{2}{-}m_3{+}\sqrt{u}\right)}{\Gamma \left(2 \sqrt{u}\right)}+\\\nn
   &-\frac{\Gamma \left(2
   \sqrt{u}\right) \Gamma \left(\frac{1}{2}-m_1-\sqrt{u}\right) \Gamma \left(\frac{1}{2}+m_2-\sqrt{u}\right) \Gamma
   \left(\frac{1}{2}-m_3-\sqrt{u}\right)}{\Gamma \left(-2 \sqrt{u}\right)}\Bigg]
\end{align}
At infinity, the behaviour of \eqref{solinfpm} is
\be
\Psi(z) \sim Y_+  e^\frac{\Lambda z}{2} z^{m_3} {+} Y_- e^{-\frac{\Lambda z}{2}} z^{- m_3}
\ee

\be
Y_+ = - (-1)^{\frac{1}{2}+ m_3 - \sqrt{u}}\Lambda^{m_3} A_+ \quad,\quad Y_-= \Lambda^{-m_3}A_-
\ee
The amplification factor is defined as \cite{Brito:2015oca, Consoli:2022eey} 
\be
Z =\Bigl| \frac{Y_+}{Y_-}  \Bigr|^2 -1
\ee
Introducing the quantity 
\be \label{eq:Qgrav}
Q_{\rm{grav}} = \frac{qQ - \omega r_s}{\sqrt{1- \frac{r_b}{r_s}}} 
\ee
in the near-decoupling limit ($\L \ll 1$), the dictionary \eqref{dictBS} becomes 
\be\label{approxdic}
m_1 = - m_2 = -\text{i} Q_{\text{grav}}, \qquad m_3 = \frac{q}{2} \sqrt{r_s^2 - Q^2}, \qquad u = \left(\ell + \frac{1}{2}\right)^2 
\ee
In order to reproduce results known in literature \cite{Consoli:2022eey,Starobinskil:1974nkd}, we consider the approximation $m_3 \simeq 0$, which is consistent with $q \simeq 0$ \footnote{This parameter can be small since it is proportional to $R_y$.}, called \textit{probe limit}.\\

In these approximations, the amplification factor can be written as
\begin{equation}
   \label{Z-Sigma}
   Z \sim  \left|\frac{1- \Sigma}{1 + \Sigma}\right|^2-1 \sim -4 \text{Re}\Sigma 
\end{equation}
where
\begin{equation}\label{sigma0}
\Sigma=\frac{\Lambda^{2\sqrt{u}}\Gamma\left({1\over2}+\sqrt{u}\right)\Gamma\left({1\over 2}-m_1+\sqrt{u}\right)\Gamma\left({1\over2}+m_2+\sqrt{u}\right)\Gamma\left(-2\sqrt{u}\right)^2}{\Gamma\left({1\over2}-\sqrt{u}\right)\Gamma\left({1\over 2}-m_1-\sqrt{u}\right)\Gamma\left({1\over2}+m_2-\sqrt{u}\right)\Gamma\left(2\sqrt{u}\right)^2}
\end{equation}
Plugging in the previous equation \eqref{sigma0} the approximated dictionary \eqref{approxdic} and expanding for small $Q_{grav}$ we obtain
\be
\Sigma=\frac{i(-1)^\ell 2^{-1-4\ell} Q_{grav}\Lambda^{2\ell+1}\Gamma\left(1+\ell\right)^4}{ \Gamma(2+2\ell)^2\Bigl[\left({\frac{1}{2}}\right)_\ell\Bigr]^2}
\ee
\begin{figure}
    \centering
    \includegraphics[width=0.8\linewidth]{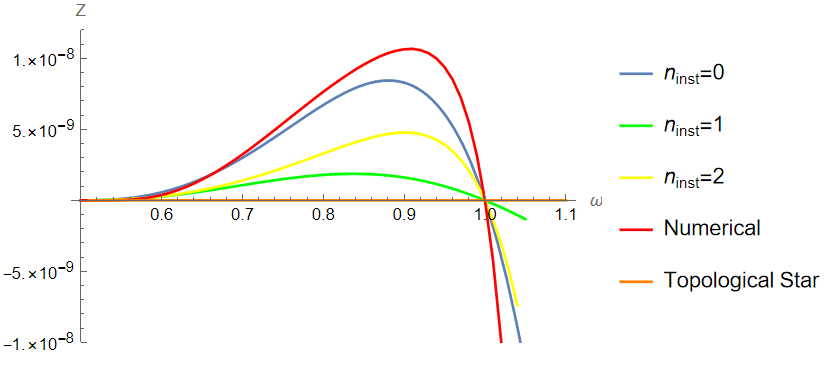}
    \caption{Comparison among different BS amplification factors: the red line represents the curve obtained by numerical integration, while blue, yellow and green curves represent \eqref{Z-Sigma} with \eqref{amplA} and with 0, 1 and 2 instantonic corrections, respectively. The chosen parameters are $r_s=1$, $r_b=0.8$, $\ell=2$, $q=0.5$, $Q=2$, $p=\mu=0$. The orange line refers to the amplification factor of TS with the same parameters as BS with $r_b$ and $r_s$ exchanged.}\label{plotAmpli}
  \end{figure}
In order to take the instantonic corrections into account we have to perform the replacement $\ell\rightarrow a-\frac{1}{2}$:
\be\label{amplA}
\Sigma=\frac{\pi ^2 (-1)^a 2^{1-8 a} Q_{grav} \Lambda ^{2 a} \Gamma
   \left(a+\frac{1}{2}\right)^2}{\Gamma (a)^2 \Gamma (a+1)^2}
\ee
Unfortunately, as it is shown in Figure \ref{plotAmpli}, the first instantonic corrections do not get closer to the numerical curve (see Appendix \ref{appA}), stating that the convergence would require a larger number of instantonic corrections. In the same Figure, we display the amplifaction factor for the TS and it is clear that it is zero within the numerical precision of the software employed in the calculation (Wolfram Mathematica).\\
The superradiance here observed for the charged non-rotating BS is not linked to the superradiant instability highlighted in \cite{Press:1972zz, Cardoso:2005gj, Cardoso:2005vk}: indeed, the BS lacks the reflection mechanism needed to trap and enhance the outgoing perturbation.\\

Concluding, exploiting several methods, and the common ones in all the cases analyzed are the correspondence with the qSW curves for $\mathcal{N}=2$ SYM with SU(2) gauge group and $N_f=(1,2)$ fundamental hypermultiplets and the Leaver method, we computed the QNMs for the TSs and the BSs under the perturbation of a charged scalar probe, in the guise of a stringy object winding around the compact dimension. The origin of this study originates from the suggestion of \cite{Bianchi:2023sfs}. All the results that come out show that the imaginary part of the frequencies associated to these modes is negative and furthermore the different computation methods give compatible results. This confirms the stability of the TS solution with respect to linearized \textit{charged} and scalar perturbations. Despite the TS and the JMaRT soliton are non-BPS smooth horizonless geometries, the former seems not to be affected by the charged linear instability. Then, as argued in \cite{Bianchi:2023rlt}, the peculiar behavior of the JMaRT ergoregion and the charge instability should not to be attributed to non-extremality, but to the over-rotation.\\

For what rotational superradiance is concerned, as exposed in \cite{Brito:2015oca}, it occurs for geometries with a horizon inside an ergoregion. Our focus, instead, has been on static backgrounds which can only give rise to a charge superradiance - very much as Reissner - Nordstr\"om \cite{Brito:2015oca, DiMenza:2014vpa, Benone:2015bst, Baake:2016oku} - due to the presence of a horizon and a charged probe with respect to the electromagnetic fluxes of the solutions\footnote{We are aware of the fact that superradiance can also take place for horizonless geometries \cite{Brito:2015oca}, leading to superradiant instabilities. Nonetheless, as highlighted in this paper, this does not happen for the TSs.}. Our results confirm charge superradiance for the BSs; the TSs, instead, do not superradiate, as expected for fuzzball geometries which conserve information.\\

We have also had the opportunity to study the Tidal Love Numbers of the TSs and the BSs in the case of charged probes. Focusing on the former, we find that these numbers exhibit the same peculiarities of the ones for the case of neutral probes as analyzed in \cite{DiRusso:2024hmd}, that is there are poles in correspondence of metastable QNMs.\\ 

The relevance of the results of this Chapter is that the stability of the considered geometries only for linear scalar perturbations has been established. This gave rise to another study, carried out in \cite{Bena:2024hoh}, concerning the stability of modes of higher spins (e.g. vector bosons and gravitons). The final result is that the TS is stable also in this scenario.
\chapter{Machine Learning Method for Trajectories of a Particle in a Fuzzball Geometry}\label{chap_4}
This Chapter keeps on with the study of gravitational systems, but without using $\mathcal{N}=2$ SUSY for the determination of several aspects in gravity. In particular, while in the latter we considered a central object (a Topological Star) and we studied its dynamical properties by probing it with a charged string, now we analyze the situation in which we have a particle moving inside the geometry generated by another central object, which is a particular type of fuzzball: the D1-D5 circular fuzzball, formed by the intersection of a D1 and a D5 branes. This gives rise to a 10-dimensional object and we study the motion of a massless neutral (with respect to the gauge fields defined in this geometry) particle in the approximation in which we neglect any emission of gravitational waves. The original aspect, that follows the analysis of \cite{Cipriani:2025ini}, is that the trajectories (in particular the geodesics) are obtained using a method developed in the Machine Learning (ML) scenario, the Hamiltonian Neural Network (HNN), a particular example of a Physics-Informed Neural Network (PINN) that exploits the hamiltonian nature of the motion in order to solve the dynamical equations. The final goal is comparing the performance of the HNN against those obtained by standard numerical integrators, in particular Euler and Runge-Kutta methods. It turns out that not only the outcomes of the HNN have the same level of accuracy of the numerical integrators in three different cases of interest, but also that in the critical one, i.e. when the particle moves along bound orbits, the HNN outperforms the two standard numerical integrators.\\
This Chapter is hence organized as follows. In Section \ref{sec:introduction}, we give a general overview of the contemporary and growing use of PINNs and HNNs in many different areas of Physics and a summary of the results we are going to show in this Chapter. In Section \ref{sec:HNN}, we introduce the HNN-based strategy for solving a generic set of Hamiltonian equations. In Section \ref{sec:fuzzball}, we review the theoretical background of the D1-D5 fuzzball and its geodesics. In Section \ref{results:planar}, we present the numerical results for planar geodesics, while the results for the non-planar ones are discussed in Section \ref{results:non_planar}.

\section{\label{sec:introduction} General overview and summary of the results}

Physics-Informed Neural Networks (PINNs) are a widely used tool in today's Machine Learning (ML) landscape. They consist of Neural Networks (NNs) that, during the training phase, learn to solve the differential equations governing the physical laws of a system in a model-independent way. When these differential equations correspond to Hamilton equations of motion, we refer to them as Hamiltonian Neural Networks (HNNs). The HNN paradigm was introduced in \cite{HNN} and we closely follow the strategy proposed in \cite{Harvard}. The key advantages of HNNs over standard numerical integrators can be summarized as follows:

\begin{itemize}
	\item the predicted solution is analytical in time and not limited to a discrete set of time steps;
	\item conservation laws, symmetries, constraints and prior knowledge of the system can be easily incorporated at the level of the architecture and of the loss function to improve the predictability of the HNN;
	\item the minimization process of the loss function occurs under the constraint that the solution satisfies the system of equations at all times simultaneously and independently, thus avoiding any iterative mechanism.
\end{itemize}

The last point is crucial for systems whose (effective) potential exhibits unstable critical points, for which small changes of the initial conditions lead to completely different behaviors of the solutions, similar to what happens in chaotic dynamical systems. As it is well known, any numerical integrator operating iteratively, building the solution based on the result of the previous time step, accumulates errors and inevitably loses accuracy over long time scales in such situations. HNNs have the potential to overcome this issue.\\

While HNNs have been applied in various physical contexts\footnote{Due to the rapidly increasing number of ML applications, the literature on this topic is vast. An incomplete selection of applications includes \cite{HNN, Crammer:2020,Lagaris_1998,DiHan:2021,deveney:2021,raissi:2017,Wu:2022,Cornell:2022enn,Jain:2022dsh,Luna:2022rql,Luna:2024spo,Geng:2024zzw,Louppe:2017ipp,Nabian:2020,Hashemi:2024azx,Lagaris_1998}.}, their diffusion in high-energy physics remains limited. In \cite{Cipriani:2025ini}, this technique was applied for the first time in the field of String Theory, where ML has only begun to gain attraction in recent years \cite{Cole:2019enn,Cole:2021nnt,MacFadden:2024him,Lanza:2023vee,Lanza:2024mqp,Deen:2020dlf,PhysRevD.111.086007,Ashmore:2019wzb,Jejjala:2020wcc,Larfors:2021pbb,Berglund:2022gvm,Berglund:2023ztk,Anderson:2023viv,Hendi:2024yin,Mirjanic:2024gek,Seong:2023njx,Choi:2023rqg,Seong:2024wkt,He:2024jao}. Specifically, we focus on using HNNs to determine the non-critical and unstable critical geodesics of a massless particle moving in a D1-D5 circular fuzzball geometry by solving the associated equations of motion \footnote{The scattering properties, scalar perturbations and tidal deformations of this physical system has already been extensively studied in \cite{Bianchi:2017sds,Bianchi:2022qph,DiRusso:2024hmd}}. It serves as a well-suited test case, given that the exact trajectory, to which we will refer to as the \textit{ground truth}, can be explicitly computed. Thus, the results presented here do not add new physical insights into the system under study; rather, they are aimed to assess the feasibility and performance of the HNNs compared to standard numerical methods in a controlled setting. The present investigation, as clearly demonstrated by the numerical results, shows that the HNN performs comparably to numerical integrators for non-critical trajectories and surpasses them for unstable critical trajectories, predicting accurate solutions even over long time scales.\\
This result suggests that HNNs, and PINNs in general, may offer a robust and reliable solution for numerically determining unstable solutions, not only in String Theory but in all systems where the underlying physics gives rise to complex dynamics. This one, together with other features discussed later on, is one of the reasons that favor these Machine Learning-based algorithms over standard integrators, even when their performances appear comparable. Moreover, the promising outcomes  pave the way for a systematic study of more intricate geometries, where the absence of a ground truth or the lack of physical information necessitates reliable and flexible tools for analysis. HNNs and PINNs are ideal candidates in this regard. 

\section{Hamiltonian Neural Networks}\label{sec:HNN}
In this Section we provide an introduction to HNNs by summarizing the basic concepts. We refer mostly to \cite{Harvard,HNN}, from which we have taken inspiration for this work.\\
At the level of notation, we denote by $\textbf{q}=(q^1,...,q^n)$  and $\textbf{p}=(p_1,...,p_n)$, respectively, the generalized space coordinates and conjugate momenta, with $ n \geq 1$ the number of degrees of freedom. In this way, the Hamilton equations, satisfied by the Hamiltonian $\mathcal{H}(\textbf{q},\textbf{p})$, are \footnote{The notation $\dot{x}$ denotes the time derivative.}

\begin{flalign}\label{eq:generic_hamilton_equations}
\dot{\textbf{q}} = \dfrac{\partial\mathcal{H}}{\partial\textbf{p}}, 
\qquad
\dot{\textbf{p}} = -\dfrac{\partial\mathcal{H}}{\partial\textbf{q}}.
\end{flalign}

As it is well known, along the the dynamics of the system the Hamiltonian is conserved, so we can define the conserved quantity $\mathrm{E}=\mathcal{H}(\textbf{q},\textbf{p})$. In the following we are going to use occasionally the definition of the vector $\textbf{z}\in \mathbb{R}^{2n}$ as
\begin{equation}\label{eq:sympl_coordinates}
    \textbf{z}=(\textbf{q},\textbf{p})^{T}
\end{equation}

\subsection{HNN}\label{sec:HNN_Presentation}
It is time to present the strategy based on  Hamiltonian Neural Networks (HNNs) to solve \eqref{eq:generic_hamilton_equations} for a generic system. We will then specialize to the case of the D1-D5 fuzzball geometry whose details are provided in Section \ref{sec:fuzzball}. In order to start appreciating the differences between ML and standard numerical integrators, let us recall the widely used first order semi-implicit Euler method to evaluate a trajectory in the time range $[0,T]$.  This method respects the symplectic structure of the  Hamiltonian system. In particular it automatically conserves a quantity which is the original Hamiltonian shifted by a term proportional to the discretization time step. In fact, the algorithm requires splitting the time interval in $N+1$ points equally spaced with time step $\Delta t$ and such that $T=N\Delta t$.  The solution is then computed  in correspondence of the discrete times $t_j=j\Delta t$, with $j=0,\dots, N$, by updating the generalized coordinates and the corresponding momenta at time step $j+1$ from those at time step $j$ \footnote{The notation $\widetilde{x}$ denotes a quantity obtained with a numerical integrator.} in the following way

\begin{flalign}\label{eq:Euler}
    \widetilde{q}_k(t_{j+1})
    &=\widetilde{q}_k(t_j)+\Delta t\dfrac {\partial\mathcal{H}\big(\widetilde{\textbf{q}}(t_j),\widetilde{\textbf{p}}(t_j)\big)}{\partial p_k}, \nonumber \\[4pt]
    \widetilde{p}_k(t_{j+1})&=\widetilde{p}_k(t_j)-\Delta t\dfrac {\partial\mathcal{H}\big(\widetilde{\textbf{q}}(t_{j+1}),\widetilde{\textbf{p}}(t_j)\big)}{\partial q_k}.
\end{flalign}

\begin{figure}[H]
    \centering
    \includegraphics[scale=0.5]{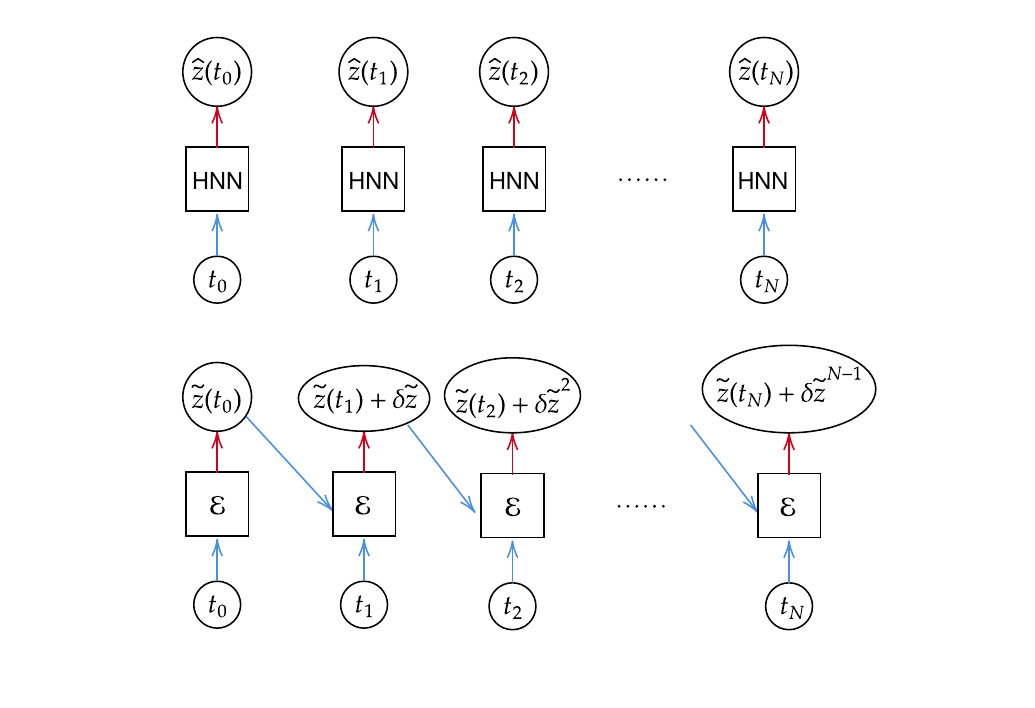}
    \caption{
    Representation of the algorithmic flow for a Hamiltonian Neural Network (HNN, on top) and for the first-order semi-implicit Euler method ($\mathcal{E}$, on bottom).  Blue arrows refer to the inputs at each step, while the red ones to the outputs. As it can be seen, the former produces an output individually for each time without using the information at the previous step, while the latter works iteratively accumulating errors.}
    \label{fig:Euler_vs_HNN}
\end{figure}

By employing this iterative method, it is inevitable to accumulate errors through all the input time series, making the prediction of trajectories, especially at large times, usually affected by large discretization errors.  In practice, for particularly complex systems accurate solutions are found only in  the limit of $\Delta t \mapsto 0$ (i.e. $N\mapsto \infty$) where the method is expected to preserve the real Hamiltonian. On the contrary, an HNN provides an analytical solution independently for any input time. This methodological difference is schematically sketched in Figure~\ref{fig:Euler_vs_HNN}. 

\begin{figure}[H]
\centering
    \includegraphics[scale=0.67]{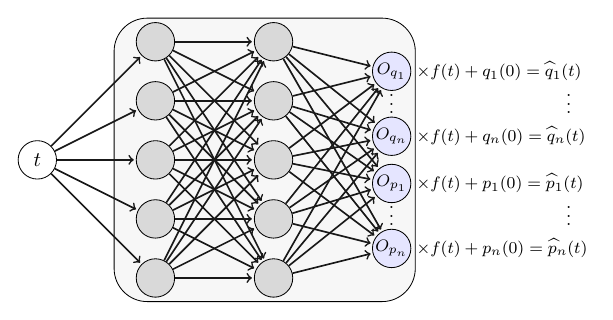}
    \caption{\label{fig:HNN} 
The neural network architecture consists in one input node for the time $t$, an arbitrary number of hidden layers and an output layer with a number of nodes (colored light blue) equal to the number of non-trivial Hamilton equations that constitute the system. The network outputs are multiplied by $f(t)=1-e^{-t}$ and added to the initial conditions. Notice that $f(0)=0$ is such that the initial conditions are exaclty satisfied by the predicted solution.
    } 
\end{figure}

Recalling the definition \eqref{eq:sympl_coordinates}, we can write the HNN solution as \footnote{The notation $\widehat{x}$ denotes a quantity predicted by a neural network.}

\begin{flalign} \label{eq:z_hat}
    \mathbf{\widehat{z}}(t,\textbf{w})=\mathbf{z}_{0}+f(t)\mathbf{O}(t,\textbf{w}),
\end{flalign}

where

\begin{flalign}
\mathbf{O}(t,\textbf{w})=\big(O_{q_1}(t,\textbf{w}),\cdots,O_{q_n}(t,\textbf{w}),O_{p_1}(t,\textbf{w}),\cdots,O_{p_n}(t,\textbf{w}) \big)^T
\end{flalign}

is the output of the NN whose architecture is designed to be a $\mathbb{R}\mapsto \mathbb{R}^{2n}$ map, with the time $t$ being the input, and $\mathbf{z}_0\in \mathbb{R}^{2n}$ is the vector of the initial conditions at $t=0$. \footnote{In case in which some momenta are conserved, due to the fact that the corresponding coordinates are cyclic, then we will not put the nodes associated to them. This is exactly what we mean by non-trivial Hamilton equations in the caption of Figure\ref{fig:HNN}} The functions $O_{q_k}$ and $O_{p_k}$, with $k=1,\dots,n$, represent single output nodes assuming values in $\mathbb{R}$. The single variables in \eqref{eq:z_hat} are therefore obtained according to

\begin{flalign}\label{eq:qp_hat}
\widehat{q}_k(t,\textbf{w})= q_k(0)+f(t)O_{q_k}(t,\textbf{w})\,,   
\nonumber\\[6pt]
\widehat{p}_k(t,\textbf{w})= p_k(0)+f(t)O_{p_k}(t,\textbf{w})\,,
\end{flalign}

See also Figure~\ref{fig:HNN} to visualize the network diagram. 

The symbol $\textbf{w}$ represents the collection of the network weights. We will specify the network architecture in the results' Section, since it is not necessary for the purpose of explaining the method. Let us only clarify that the choice of the network architecture has to be such that the expression $\widehat{\textbf{z}}(t,\textbf{w})$ is analytical and differentiable with respect to the time for any set of weights $\textbf{w}$. We explicitly write in this Section the symbol $\textbf{w}$ for the sake of clarity, but we will omit it further on in the text, leaving it understood that whenever a variable is marked by the symbol $\;\widehat{}\;$ it will be dependent on the architecture and weights. The function $f(t)$, which does not contain trainable parameters, is chosen in such a way it satisfies $f(0)=0$ and it is introduced to automatically enforce the initial conditions\footnote{We remark that this might be also achieved by including explicitly the conditions at the level of the loss function. The advantage of the definitions \eqref{eq:qp_hat} is to automatically restrict the space of possible solutions by accelerating and stabilizing the minimization of the loss function.}. In this work we make the same choice as in \cite{Harvard} (where the impact of different choices of $f(t)$ is also illustrated) and define the bounded function 
 
\begin{flalign}\label{eq:ft}
    f(t)=1-e^{-t}.
\end{flalign}

We now explain in detail how the solutions to \eqref{eq:generic_hamilton_equations}, parametrized by \eqref{eq:qp_hat}, can be obtained through the minimization of a  specifically-designed loss function. Let us consider again the time interval $[0,T]$ sampled with $N+1$ points at constant time step $\Delta t$ and indicate $t_j=j\Delta t$ the time after $j$ steps. We define the loss function

\begin{equation} \label{eq:loss_dyn}
    L_\mathrm{dyn}(\textbf{w}) = \sum_{k=1}^{n} \Bigg\{\dfrac{1}{N} \sum_{j=0}^{N} \gamma_{q_k}\Bigg[\left.\Bigg(\dot{\widehat{q}}_k-\widehat{\pdv{\mathcal{H}}{p_{k}}}\Bigg)\right|_{t_j}\Bigg]^2 +\dfrac{1}{N} \sum_{j=0}^{N}\gamma_{p_k}\Bigg[\left.\Bigg(\dot{\widehat{p}}_k+\widehat{\pdv{\mathcal{H}}{q_k}}\Bigg)\right|_{t_j} \Bigg]^2\Bigg\},
\end{equation}

where

\begin{equation}\label{eq:q_dot}
    \left. \dot{\widehat{q}}_k\right|_{t_j} = \left.\dv{f(t)}{t}\right|_{t_j}\cdot O_{q_k}(t_j,\textbf{w})+f(t_j)\cdot \left.\dv{O_{q_k}(t,\textbf{w})}{t}\right|_{t_j}, \qquad\qquad \left. \widehat{\pdv{\mathcal{H}}{p_k}}\right|_{t_j}=\left.\pdv{\mathcal{H}}{p_k}\right|_{\widehat{\textbf{z}}(t_j,\textbf{w})}.
\end{equation}

Similarly one can obtain the definitions for $\left.\dot{\widehat{p}}_k\right|_{t_j}$ and $\left.\widehat{\pdv{\mathcal{H}}{q_k}}\right|_{t_j}$ in \eqref{eq:loss_dyn} by replacing $\widehat{q}_k$ with $\widehat{p}_k$ and viceversa. The factors $\gamma_{q_k}$ and $\gamma_{p_k}$ can be chosen to weight a variable more or less relative to the others. Throughout most of the Chapter, we will simply use $\gamma_{q_k}=\gamma_{p_k}=1$.\\
\eqref{eq:q_dot} makes manifest the dependence of $L_\mathrm{dyn}(\textbf{w})$ on the network weights and on the way $f(t)$ enters the loss function. The subscript \quote{dyn} refers to the fact that,  for a set of weights such that $L_\mathrm{dyn}(\textbf{w})=0$, the network predictions \eqref{eq:qp_hat} would exactly satisfy the dynamics expressed by the Hamilton equations~(\ref{eq:generic_hamilton_equations}) at the discrete times $t_j$.  The possibility to actually meet the condition  $L_\mathrm{dyn}(\textbf{w})=0$ is supported by the so-called Universal Approximation Theorem \cite{hornik1989multilayer,Goodfellow-et-al-2016}, which states that a neural network large enough is able to approximate any continuous function with arbitrary accuracy. In practice this possibility is spoiled by the fact that a neural network has a finite number of neurons and then, as customary in the ML field, we can at most search for the set of weights that make the loss function as small as possible, that is we want to solve the equation
\begin{flalign}\label{eq:minimization}
 \pdv{L_\mathrm{dyn}(\textbf{w})}{\textbf{w}}=0   
\end{flalign}
and find an approximation to the solution. We will call \textit{training} the process of minimization of the loss function, but it has to be emphasised that the training is not data-driven but rather equation-driven. The meaning is that the only information to let the NN fulfill the desired task is codified in the Hamilton equations, which are analytically known after defining the Hamiltonian, and no assumption on the true trajectory is required. From this perspective the HNNs alone offer a model independent solution to the problem and additional informations, e.g. the knowledge of the solution at given times due to experiments or different methods, can be easily incorporated in the training process to improve the accuracy of the predicted solution.\\

\begin{figure}
\includegraphics[scale=0.5]{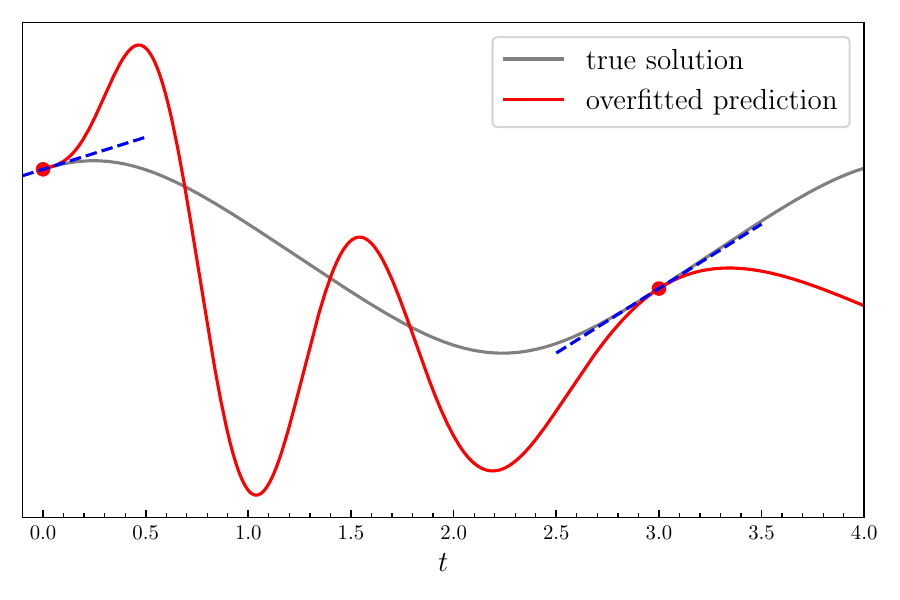}
\caption{\label{fig:overfitting} 
Demonstrative example of overfitting for a HNN. In this case the dynamical loss is minimized only for $t=0$ and $t=3$ (red points), in correspondence of which the predicted solution is accurate, while it rapidly deviates from the true one otherwise. }
\end{figure}

Let us stress that \eqref{eq:qp_hat} are continuous and differentiable in the variable $t$ and then, after the training, they represent an \textit{analytical} solution to the Hamilton equations, providing a significant advantage over numerical integrators, where the solution is only known at predefined discrete time points. However, a natural concern is the reliability of the prediction for times different from the $t_j$ entering $L_\mathrm{dyn}$. As for other typical ML tasks (as regression ones), also HNNs can suffer from over-fitting, by providing an accurate solution for $t=t_j$ but considerably different from the ground truth for $t$ in between $t_j$ and $t_{j+1}$. A demonstrative example of the problem is shown in Figure~\ref{fig:overfitting}. This problem is expected to be particularly severe as $\Delta t$ increases ($N\to 0$), since the solution is less constrained in between two consecutive points, and vanish in the limit $\Delta t \to 0$ ($N\to \infty$). A possibility to overcome the problem is to sample the interval $[0,T]$ differently at the beginning of each epoch, for instance by adding gaussian noise to a given initial set of times $t_j$, or to choose $N$ large and check a posteriori that the loss function, computed for a set of times different from $t_j$, keeps at the same level as the one minimized during the training. We are going to follow the latter strategy. Of course, there is no chance that the prediction could be accurate for $t$ outside the interval $[0,T]$, since the network has not been instructed on the dynamics in that regime.

Another great advantage of the procedure outlined above is the possibility to extend the definition of the loss function by including terms accounting for conservation of physical quantities. Particularly important is the term imposing the energy conservation, already introduced in \cite{Harvard}, that can be defined as

\begin{flalign}\label{eq:loss_energy}
    L_\mathrm{energy}(\textbf{w})=\dfrac{1}{N} \sum_{j=0}^{N}\big[\mathcal{H}\big(\widehat{\textbf{z}}(t_j,\textbf{w})\big)-E_0\big]^{2},  
\end{flalign}

where $E_0=\mathcal{H}(\textbf{z}_0)$ is the value of the energy to be conserved along the trajectory. We then define the total loss function that has to be minimized as a convex linear combination of $L_\mathrm{dyn}$ and $L_\mathrm{energy}$,

\begin{flalign}\label{eq:total_loss}
    L(\lambda,\textbf{w})=(1-\lambda)L_\mathrm{dyn}(\textbf{w})+\lambda L_\mathrm{energy}(\textbf{w}),   
\end{flalign}

where we have introduced the trade-off parameter $\lambda\in~[0,1]$. The practical advantage of introducing the energy conservation term in the loss function is making the convergence towards the correct solution faster and the training procedure more stable. Moreover, setting $\lambda>0$ serves also as regulator to select a unique solution in those situations where $L_\mathrm{dyn}$ might admit multiple ones. We will show in fact that setting $\lambda >0$ dramatically affects the performance of the neural network in some cases. Similarly one can introduce other terms enforcing the predicted trajectories to conserve physical quantities.\\
However, the loss function \eqref{eq:total_loss} is not particularly helpful in comparing the performances of the neural network at varying $\lambda$ since $\lambda$ itself, especially when close to 0 or 1, alters the order of magnitude of $L(\lambda,\textbf{w})$ at a given  $\textbf{w}$. During the training we therefore also monitor individually $L_\mathrm{dyn}(\textbf{w})$ and $L_\mathrm{energy}(\textbf{w})$ in order to have a measure of the goodness of the solution independently from $\lambda$.\\
Further regularization terms limiting the space of the solutions according to additional knowledge about the system can be introduced as well to improve the accuracy of the network. We will come back to it later in the results' Section since this is case-dependent.\\

We conclude this Section by observing that we have specialized (\ref{eq:loss_dyn}) for the case of the Hamilton equations, but in general one can write a proper loss function to search for the solution of any system of ordinary differential equations. For instance, the geometry described in the next Section enjoys also the separability of the motion and it allows the derivation of a different, but physically equivalent, set of equations of motion. This alternative point of view is analyzed in Appendix E of \cite{Cipriani:2025ini}, with the aim of highlighting differences between the usage of HNN.
The scope of applicability of PINNs is therefore extremely wide and such technology is a valuable tool that can complement standard numerical integrators and, at times, replace them.

\section{\label{sec:fuzzball}D1-D5 fuzzball geometry}
The physical configuration we consider in this Chapter is the geodesic motion (i.e. the free fall trajectory) of a scalar massless neutral particle inside the geometry of a D1-D5 circular fuzzball. The fuzzball's notion was already defined at the beginning of the previous Chapter, in order to see how it solves the information-loss paradox \cite{Mathur:2009hf} and allows also to give a microscopic interpretation in terms of number of microstates to the entropy even in General Relativity, enforcing the connection of BHs to thermodynamical systems \cite{PhysRevD.7.949}. We also said that there are many ways to construct these smooth-horizonless geometries, but since we moved then to examine the Topological Star solution which is not a fuzzball, now we can simply describe what kind of construction we consider in this Chapter.\\
We need the already mentioned D$p$-branes (look at Chapter \ref{chap_1}). Indeed these objects carry mass and charge and, consequently, are subjected to the action of forces. One can combine a number of different branes in order to form a stable gravitational object, with a given mass and charge. For the scope of this Chapter we are interested in the gravitational object obtained by combining D1 and D5 branes and which has a circular profile.\\

Before proceeding to see all the details of the dynamics, we need to recall some basic ingredients with the aim of setting the formalism in the gravity context. It is well known that the free fall of a neutral particle inside a certain $d$-dimensional spacetime with metric $g_{\mu \nu}$ and coordinates $x^{\mu}$ ($\mu = 0, \dots, d-1$) happens along the geodesics, which are solutions of differential equations known as \textit{geodesic equations}. On the other hand, it is also well known that the trajectories in the physical space are such that the action is minimized; this can be done by solving the \textit{Euler-Lagrange equations}

\begin{flalign} \label{E-L Eqns}
\frac{\mathrm{d}}{d\mathrm{s}}\frac{\partial \mathcal{L}}{\partial \dot{x}^\mu}=\frac{\partial \mathcal{L}}{\partial x^\mu},
\end{flalign}

where $s$ is the affine parameter of the geodesic of the particle and $\mathcal{L}$ its Lagrangian. From all of this we understand that the geodesic equations are equivalent to the Euler-Lagrange ones and it can be shown that it is evident if the Lagrangian has the following form

\begin{flalign} \label{Lagrangian}
\mathcal{L}=\frac{1}{2} g_{\mu\nu}\dot{x}^\mu\dot{x}^\nu\quad,\quad \dot{x}^\mu=\frac{\mathrm{d} x^\mu}{\mathrm{d}s}.
\end{flalign}

In order to move to the  Hamiltonian viewpoint, we have to compute the canonical conjugate momenta

\begin{flalign} \label{eq:Momenta}
P_\mu= \frac{\partial \mathcal{L}}{\partial \dot{x}^\mu}
\end{flalign}

and the  Hamiltonian $\mathcal{H}$ is obtained through a Legendre transformation of the Lagrangian

\begin{flalign}\label{eq:Hamiltonian}
    \mathcal{H} = P_\mu \dot{x}^\mu (P) - \mathcal{L}(x,\dot{x}(P)) = \frac{1}{2} g^{\mu \nu} P_\mu P_\nu.
\end{flalign}

Notice that what we have called here $x$ and $P$ correspond to $\mathbf{q}$ and $\mathbf{p}$ introduced at the beginning of Section \ref{sec:HNN}, but it is the typical notation in the gravity context. Futhermore, we consider as time the affine parameter $s$ in place of the coordinate time $t$ used in Section \ref{sec:HNN_Presentation}. This implies no conceptual and practical differences in the ML strategy proposed. The mass shell condition for a particle of mass $\mu_0$ reads \cite{Bianchi:2017sds,Chervonyi:2013eja} $\mathcal{H}= - \frac{1}{2}\mu_0^2$. Since we are interested in massless geodesics, we obtain the following important condition: $\mathcal{H}=0$. Finally, we recall that, if the metric components and thus the Lagrangian do not depend on a specific coordinate $x^\mu$ (which is said to be cyclic), then the corresponding momentum $P_\mu$ is conserved, as can be read from the Euler-Lagrange equations \ref{E-L Eqns}.

\subsection{Geometry}\label{sec:Geometry}
Now we study the properties of the motion of a massless particle in the geometry of the D1-D5 circular fuzzball. 
The spacetime metric generated by this object is

\begin{flalign} \label{eq:metric}
\dd s^2 =&H^{-1}\left[-(\dd t+\omega_\phi \dd \phi)^2+(\dd z+\omega_\psi \dd \psi)^2\right] +H\Bigg[(\rho^2+a_f^2\cos^2\theta)\left(\frac{\dd\rho^2}{\rho^2+a_f^2} +\dd\theta^2\right)
\nonumber\\[6pt]
&+\rho^2\cos^2\theta \dd\psi^2+(\rho^2+a_f^2)\sin^2\theta \dd\phi^2\Bigg] +\left(\frac{H_1}{H_5}\right)^{1/2}\dd z_{1,2}^2+\left(\frac{H_5}{H_1}\right)^{1/2}\dd z_{3,4}^2
\end{flalign}

where

\begin{flalign} \label{eq:def_params}
H  =\sqrt{H_1H_5} \,, \quad H_i =1+\frac{L_i^2}{\rho^2+a_f^2\cos^2\theta}\,, \quad \omega_\phi&=\frac{a_fL_1L_5\sin^2\theta}{\rho^2+a_f^2\cos^2\theta}\,, \quad \omega_\psi=\frac{a_fL_1L_5\cos^2\theta}{\rho^2+a_f^2\cos^2\theta}.
\end{flalign}

$a_f$ is the radius of the circular profile, $L_1$ and $L_5$ are the charges of D1 and D5 branes respectively \footnote{Notice that $L_1$ and $L_5$ have dimensions of a length, while in measure units with $c=G=\frac{1}{4 \pi \epsilon_0}=1$, the charge has dimension of a squared length. Consequently the real charge is related to $L_i$ via multiplication by a dimensional parameter.}. These are the parameters of the geometry that we are considering and it is also rotating. It has to be said that the complete solution describing the fuzzball does not contain only the metric \eqref{eq:metric}, but also other fields, which are not relevant for the dynamics of the neutral probe we are taking into consideration and for this reason we have not written them. This spacetime is 10-dimensional \footnote{For $L_1=L_5=0$ (i.e. $H_1=H_5=1$ and $\omega_\phi = \omega_\psi = 0$) and under a suitable coordinate change the metric \eqref{eq:metric} describes flat spacetime.}, but four out of the ten coordinates, $z_I$ with $I=1,\dots,4$, parametrize compact directions and for this reason we can safely set their corresponding momenta $P_{z_I}$ equal to 0 and neglect these coordinates. The coordinates $(\rho,\theta,\phi,\psi)$ are called \textit{oblate spheroidal coordinates} and they are related to the standard cartesian ones $(x_1,x_2,x_3,x_4)$ through the relations

\begin{equation}\label{eq:spheroidal_coordinates}
    x_1 = \sqrt{\rho^2 + a_f^2} \sin\theta \cos \phi \qquad x_2 =  \sqrt{\rho^2 + a_f^2} \sin\theta \sin \phi  \qquad x_3 = \rho \cos \theta \cos \psi \qquad x_4 = \rho \cos \theta \sin \psi.
\end{equation}

Notice that the metric components do not depend on $t$, $\phi$, $\psi$ and $z$, consequently the associated momenta $P_t = - E$, $P_\phi = J_\phi$, $P_\psi=J_\psi$ and $P_z$ are conserved. Following \eqref{Lagrangian}, \eqref{eq:Momenta} and \eqref{eq:Hamiltonian}, we can arrive at the expression of the  Hamiltonian of a particle moving inside this geometry, which is

\begin{flalign}\label{eq:Ham_tot}
    \mathcal{H} = & \frac{1}{2 \left(\rho^2 +a_f^2\cos^2\theta\right)H}\Bigg[P_\rho^2 (\rho^2+a_f^2) + P_\theta^2 + \frac{(J_\psi a_f - P_z L_1 L_5)^2}{\rho^2} + \frac{J_\psi^2}{\cos^2\theta} + \frac{J^2_\phi}{\sin^2\theta} - \frac{(J_\phi a_f - E L_1 L_5)^2}{\rho^2 + a_f^2} \nonumber\\[6pt]
    &- (E^2-P_z^2)(\rho^2+a_f^2+L_1^2+L_5^2) + (E^2-P_z^2) a_f^2 \sin^2\theta \Bigg].
\end{flalign} 

From this, we can compute the derivatives, as required by Hamilton equations, that should then be used inside the loss function as explained in the previous Section \ref{sec:HNN}.\\

Before proceeding with this, we have to pay attention to a particular feature of this geometry, that will constitute another aspect to probe the goodness of the HNN, that is the \textit{separability} of the dynamics along the radial $\rho$ and the angular $\theta$ directions. Indeed, from the previous expression of the Hamiltonian, since it has to be 0 for the massless particle, we can neglect the overall factor and, by examining the terms within the square brackets, we notice that, by introducing the \textit{Carter constant} $K^2$ \cite{Chandrasekhar:1984siy}, we can distinguish the motion along the direction parametrized by $\rho$ from the one along the direction parametrized by $\theta$. In particular

\begin{equation}\label{eq:QR}
P_\rho^2 = Q_R(\rho) =(E^2-P_z^2) \left(1+\frac{L_1^2+L_5^2}{\rho^2+a_f^2}\right)+\frac{(J_\phi a_f {-} E L_1L_5)^2}{(\rho^2{+}a_f^2)^2}{-}\frac{(J_\psi a_f {-} P_z L_1L_5)^2}{\rho^2(\rho^2{+}a_f^2)} -\frac{K^2}{\rho^2{+}a_f^2}
\end{equation}
and
\begin{flalign}\label{eq:QA}
P_\theta^2{=}&Q_A(\theta){=}K^2{-}\frac{J_\psi^2}{\cos^2\theta}{-}\frac{J_\phi^2}{\sin^2\theta}{-}(E^2{-}P_z^2) a_f^2\sin^2\theta .
\end{flalign}

Thanks to this, it is possible to integrate these two equations, getting expressions in terms of elliptic integrals. We are not going to do this in the most general case, since we are interested in the planar and a specific non planar motions. In the former situation we report also the procedure that shows how to integrate these equations, getting the ground truth that will constitute the reference point with respect to which we compare the results coming from the HNN and the standard numerical integrator.\\

A common peculiarity of BHs and
other compact gravitating objects is the presence of \quote{photon-spheres} or light-rings (light-halos for rotating objects), that are stable or unstable bound orbits of particles separating the asymptotically flat region from the horizon (or the inner region for fuzzballs) \cite{Bianchi:2020yzr,Bianchi:2021yqs, Bianchi:2022wku}. In the following we will denote them as \textit{critical geodesics}. The simultaneous vanishing of the radial momentum $P_\rho^2(\rho) \equiv Q_R(\rho)$ and its derivative
\begin{equation}\label{critcond}
Q_R(\rho_c,J_c)=Q'_R(\rho_c,J_c)=0
\end{equation}
individuate the critical radius $\rho_c$ and the corresponding critical angular momentum $J_c$.
The radial effective potential $-Q_R(\rho)$ depends also on all the other parameters of the geometry, hence fixing all of them we get a particular shape. In this case, $\rho_c$ can correspond both to a maximum or a minimum of $Q_R$ itself. In the former case the critical geodesic is unstable, in the latter it is stable. It is worth noticing that this identification is possible only if the geometry is separable. Determining the critical geodesics for non separable geometries is something not clear and the HNNs could help in this direction.\\ 
We now examine the planar and non-planar situations separately, determining also the critical geodesics.

\subsection{Planar case} \label{sec:Planar_Case}
Due to the axial (non fully spherical) symmetry, the geodesics are non planar in general. Some simplifications arise when the motion is in a fixed plane. In order to study this case, we should solve the Euler-Lagrange equation for $\theta$ coordinate (which is a second order differential equation) imposing as initial conditions $\theta(s_0) = \theta_0$ (the fixed value of the angle) and $\dot{\theta}(s_0)=0$, where $s_0$ is the value of the affine parameter at the beginning of the motion. In this way we obtain a Cauchy problem and if the only solution is $\theta(s)=\theta_0$, then the motion takes place only along the plane corresponding to $\theta = \theta_0$.\\
For the geometry we are studying, this happens if $\theta_0=~0,\pi/2$ \footnote{Actually these are specific cases of a more general class of geodesics, denoted as \textit{shear free} \cite{Bianchi:2022qph}.}, which correspond to the two equatorial planes. Since $\theta$ is constant, then $P_\theta=0$. Furthermore, in order to avoid divergences in \eqref{eq:QA} for $\theta=\pi/2$, we are forced to take $J_\psi = 0$, conversely $J_\phi=0$ for $\theta=0$. Furthermore, we set also $P_z=0$, which simplifies the treatment without altering the generality of final results. In the following we will consider only the case with $\theta_0 = \pi/2$.\\
The  Hamiltonian \eqref{eq:Ham_tot} becomes

\begin{equation}\label{eq:Ham_plan}
\mathcal{H} = \frac{1}{2 \rho^2 H(\theta=\pi/2)}\Bigg[P_\rho^2 (\rho^2+a_f^2) + J^2_\phi  - \frac{(J_\phi a_f - E L_1 L_5)^2}{\rho^2 + a_f^2} - E^2 (\rho^2+a_f^2+L_1^2+L_5^2)+ E^2 a_f^2 \Bigg].
\end{equation} 

and its derivatives are reported in Appendix A of \cite{Cipriani:2025ini}.\\

For what concerns the separability, the equation \eqref{eq:QA} with $P_\theta = 0, \theta=\pi/2, J_\psi=0, P_z=0$ can be solved with the Carter constant

\begin{equation}
 K^2 = J_\phi^2+E^2 a_f^2,
\end{equation}

which plugged into the radial equation \eqref{eq:QR}, gives

\begin{equation}\label{eq:Prho_Plan}
    P_\rho^2 = E^2 \left(1+\frac{L_1^2+L_5^2}{\rho^2+a_f^2}\right) + \frac{(J_\phi a_f - E L_1 L_5)^2}{(\rho^2+a_f^2)^2} - \frac{J_\phi^2+a_f^2 E^2}{\rho^2+a_f^2}.
\end{equation}

We introduce now the impact parameter $b$
\begin{flalign} \label{impact param}
    b = \frac{J_\phi}{E}
\end{flalign}

In this way these equations can be expressed in terms of this parameter. Using the expressions of the momenta $P_i$ in terms of the derivatives of the coordinates, $\dot{x}^i$, from \eqref{eq:Momenta} and inverting all these relations, we can find an equation for the derivative of $\rho$ and $\phi$ with respect to the coordinate time $t$. At the end we get

\begin{flalign}\label{eq:diffeq}
\frac{\mathrm{d} \rho(t)}{\mathrm{d} t} = \frac{\dot{\rho}}{\dot{t}} = &\pm \frac{(\rho^2+a_f^2)\sqrt{L_1^2 (L_5^2+\rho^2)+\rho^2 (L_5^2+\rho^2-b^2)+ a_f^2 (L_1^2+L_5^2+\rho^2) - 2 a_f b L_1 L_5}}{(L_1^2+\rho^2)(L_5^2+\rho^2)+a_f^2(L_1^2+L_5^2+\rho^2)-a_f b L_1 L_5},
\nonumber\\[6pt]
\frac{\mathrm{d} \phi(t)}{\mathrm{d} t} = \frac{\dot{\phi}}{\dot{t}} = &\frac{b \rho^2 + a_f L_1 L_5}{(L_1^2+\rho^2)(L_5^2+\rho^2)+a_f^2(L_1^2+L_5^2+\rho^2)-a_f b L_1 L_5}.
\end{flalign}

Details can be found in Appendix \ref{APP:separability}. In the first equation of \eqref{eq:diffeq}, the negative sign refers to the phase in which the particle is moving closer to the central object, the positive sign refers to the opposite situation.\\
We can find the critical geodesics by imposing the vanishing of \eqref{eq:Prho_Plan} and its radial derivative. Introducing the impact parameter $b$, we have two equations in the variables $\rho,b$ and we can solve them in terms of the parameters $L_1,L_5,a_f$. Keeping in mind that $\rho$ and $b$ are non negative, then we obtain the final results

\begin{equation}\label{eq:stable_ph}
\rho_{c,1} = 0 \qquad b_{c,1} = \frac{a_f^2 (L_1^2+L_5^2)+L_1^2 L_5^2}{2 a_f L_1 L_5}\,,
\end{equation}

and

\begin{equation}\label{eq:unstable_ph}
\rho_{c,2} = \sqrt{L_1 L_5 - a_f (L_1+L_5)} \qquad b_{c,2} = L_1+L_5-a_f \,.
\end{equation}

\begin{figure}[t]
  \centering
  \includegraphics[scale=0.8]{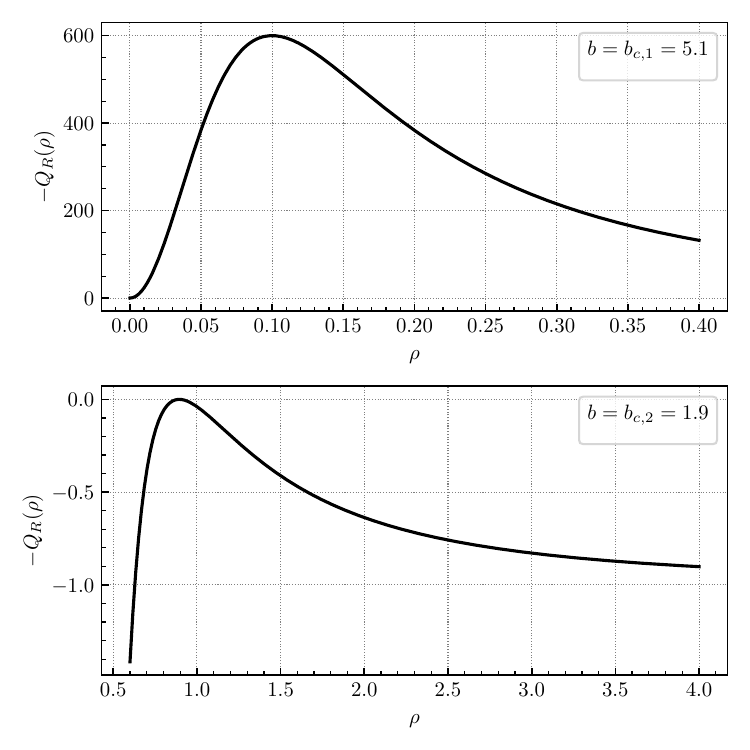}
  \caption{
  D1-D5 circular fuzzball radial effective potential in the equatorial plane $\theta=\pi/2$ for $L_1 = L_5 = 1$ and $a_f=0.1$. \emph{Top-panel}:  at $b_c = b_{c,1} = 5.1$, the stable photon-sphere corresponds to $\rho_c = \rho_{c,1}=0$. \emph{Bottom-panel}:  
  at $b_c = b_{c,2} = 1.9$, the unstable photon-sphere corresponds to $\rho_c = \rho_{c,2}=\sqrt{0.8}= 0.8944$.
  }
  \label{fig:plotQ}
\end{figure}

Figure~\ref{fig:plotQ} shows the plot of the effective radial potential $-P^2_\rho=-Q_R(\rho)$ for these two cases by setting $L_1=L_5=1$ and $a_f=0.1$. As it can be seen, $\rho_{c,1}$ describes a stable photon-sphere (top-panel), while $\rho_{c,2}$ describes an unstable one (bottom-panel). By plugging $\theta = \pi/2$ and $\rho=0$ in \eqref{eq:spheroidal_coordinates} we obtain

\begin{equation}
 x_1 = a_f \cos \phi \qquad x_2 =  a_f \sin \phi \qquad x_3 = x_4 = 0
\end{equation}

Hence when $\rho = 0$ we get on the plane $(x_1,x_2)$ a circle of radius $a_f$ and this is the stable geodesic \footnote{In the $\theta = 0$ plane described by the coordinates $(x_3,x_4)$ (since $x_1=x_2=0$), the fuzzball instead corresponds to a point located at $x_3=x_4=0$.}. In the following we will indicate $(x_1,x_2)$ as the usual $(x,y)$ and only focus on the orbit with radius $\rho_{c} = \sqrt{L_1 L_5 - a_f (L_1+L_5)}$ and impact parameter equal to $b_c = L_1+L_5-a_f$. From the above discussion, it is clear that, when $b<b_c$ the particle, coming from infinity, arrives at the compact object and then it bounces, while for $b>b_c$ the particle reaches a minimum value of the radius and then it moves away again. The former case will be denoted as \textit{sub-critical}, while the latter as \textit{over-critical}.\\
Another way of formulating the difference of these situations from the first equation of \eqref{eq:diffeq} is the following:
\begin{itemize}
    \item when $b<b_c$, the radicand does not admit real zeros in $\rho$. It implies that $\mathrm{d} \rho / \mathrm{d} t$ never vanishes, meaning that the distance always decreases (with minus sign) or always increases (with plus sign). If the particle comes from infinity, then it reaches $\rho = 0$ and then it bounces;
    \item when $b=b_c$, the radicand has two real and positive zeros that are coincident and they correspond to $\rho_c$;
    \item when $b>b_c$, the radicand has two distinct real and positive zeros, hence we have two critical radii $\rho_{-}$ and $\rho_{+}$, with $\rho_{-} < \rho_{+}$. If the particle comes from infinity, it reaches first $\rho_{+}$ and after this it goes away again.
\end{itemize}
It has to be noticed that, in this case, we can rewrite the equatorial geodesics in terms of elliptic integrals using \eqref{eq:diffeq}. This constitutes the ground truth in the comparison of the results of the HNN and the numerical integrators. Details of the determination of the ground truth can be found in Appendix \ref{APP:groundtruth}. 
In the following we will set, in the planar case, $L_1 = L_5$ without loss of generality, much simplifying the involved equations.

\subsection{Non-planar case}\label{Non-Planar Case}
Now we want to analyze the motion that takes place not only in the equatorial plane. In particular we set
\begin{equation}
    P_z = J_\psi = 0
\end{equation}
and we fix $\psi = 0$. It corresponds to a family of 3-dimensional geodesics in the Cartesian space through the relations outlined in \eqref{eq:spheroidal_coordinates}, that is

\begin{equation}
\label{eq:change_coord_3d}
x = \sqrt{\rho^2 + a_f^2} \sin\theta \cos \phi \qquad y = \sqrt{\rho^2 + a_f^2} \sin \theta \sin \phi \qquad z = \rho \cos \theta
\end{equation}

where we have denoted $x_3$ as $z$. The general  Hamiltonian \eqref{eq:Ham_tot} now reads

\begin{flalign}\label{eq:Ham_Non-Plan}
    \mathcal{H} & = \frac{1}{2 \left(\rho^2 +a_f^2\cos^2\theta\right)H}\Bigg[P_\rho^2 (\rho^2+a_f^2)  + P_\theta^2 + \frac{J^2_\phi}{\sin^2\theta} - \frac{(J_\phi a_f - E L_1 L_5)^2}{\rho^2 + a_f^2} - E^2(\rho^2+a_f^2+L_1^2+L_5^2-a_f^2 \sin^2\theta) \Bigg].
\end{flalign} 

The derivatives with respect to the variables that define the Hamilton equations are always reported in Appendix A of \cite{Cipriani:2025ini}.\\

We discuss now how to find the critical geodesics, that is the only case we study with the HNN since it is the most physically relevant one for observational reasons. The momenta \eqref{eq:QR} and \eqref{eq:QA} now become

\begin{flalign}
P_\rho^2  = E^2 \left(1+\frac{L_1^2+L_5^2}{\rho^2+a_f^2}\right) + \frac{(J_\phi a_f - E L_1L_5)^2}{(\rho^2+a_f^2)^2}-\frac{K^2}{\rho^2+a_f^2} \label{eq:QR_Non Plan} 
\end{flalign}
and
\begin{flalign} \label{eq:QA_Non_Plan} 
P_\theta^2  = K^2-\frac{J_\phi^2}{\sin^2\theta}- E^2 a_f^2 \sin^2\theta,
\end{flalign}

respectively.  For $P_\theta\propto \dot{\theta}\neq 0$ the angular dynamics is non-trivial. Let us rewrite \eqref{eq:QR_Non Plan} as follows,

\begin{flalign}
\frac{(\rho^2+a_f^2)^2P_\rho^2}{E^2}&=\mathcal{R}(\rho)=\rho^4+A\rho^2+B\,,
\nonumber\\[6pt]
A&=2a_f^2+L_1^2+L_5^2-b^2\,,
\nonumber\\[6pt]
B&=a_f^2(\zeta^2-b^2+a_f^2+L_1^2+L_5^2)\,,
\end{flalign}

where

\begin{flalign} \label{eq:def_zeta_bs}
\zeta=b_\phi-\frac{L_1L_5}{a_f},\quad b_\phi=\frac{J_\phi}{E},\quad b=\frac{K}{E}\,.
\end{flalign}

The critical conditions are $\mathcal{R}(\rho_c)=\mathcal{R}'(\rho_c)=0$ and these correspond to two equations in terms of the variables $\rho_c, \zeta_c$ and $b_c$. We can solve them in terms of $\zeta_c$ and $b_c$, which become functions of $\rho_c$, that is

\begin{flalign}\label{eq:critreg}
\zeta_c^2=\frac{(\rho_c^2+a_f^2)^2}{a_f^2} \, \, , \qquad b_c^2=&2\rho_c^2+2a_f^2+L_1^2+L_5^2.
\end{flalign}

In order to fix $\rho_c$ we must work on the condition $P^2_\theta \geq 0$. The details of all steps that have to be done in order to arrive at the determination of $\rho_c$ and the examination of the corresponding critical geodesics are reported in Appendix \ref{APP:nonplangeo}. Indeed, as it is shown there, these critical geodesics generically wrap around spheroidal zones. There is only one case in which the critical geodesics wrap around an oblate spheroid and we will study only this last scenario. The critical radius has the following expression

\begin{equation} \label{eq:crit_radius}
\rho_c=\sqrt{L_1L_5-a_f^2} \, \, , \qquad  0<a_f\leq \sqrt{L_1L_5}.
\end{equation}

These geodesics are \textit{counterrotating} with respect to the fuzzball, i.e. they rotate in the opposite sense with respect to the one of the central object. From \eqref{eq:crit_radius} we can obtain the expressions of $\zeta_c$ and $b_c$, which are

\begin{align} \label{eq:final_params_nonplan}
\zeta_c = -\frac{L_1 L_5}{a_f} \leftrightarrow b_{\phi,c}=0 \, \, , \qquad  \qquad b_c = L_1+L_5.
\end{align}

The reason why we have the negative sign of \eqref{eq:critreg} for $\zeta_c$ is explained in Appendix \ref{APP:nonplangeo}.

\section{\label{results:planar}Results for planar geodesics}

In this Section we present in detail the implementation of the HNNs, the strategies and the numerical results for the planar geodesics in the $\theta=\pi/2$ plane and in the three scenarios, over-critical ($b>b_c$), critical ($b=b_c$) and sub-critical ($b<b_c$), illustrated in Section \ref{sec:Planar_Case}. The Hamilton equations that describe the geodesics are given by the following system of coupled differential equations,
\begin{align}
\dot \rho = \pdv{\mathcal{H}}{P_\rho} \quad , \quad \dot P_\rho  =-\pdv{\mathcal{H}}{\rho} \quad , \quad \dot \phi = \pdv{\mathcal{H}}{J_\phi}.
\end{align}
The derivatives on the right-hand side are always explicitly written in Appendix A of \cite{Cipriani:2025ini}.\\
For all the three cases,  we fix the geometry parameters to $a_f=0.1$, $L_1=L_5=L=1$ and $E=1$. With this choice, the critical values of the impact parameter \eqref{eq:unstable_ph} and of the critical radius are
\begin{equation}\label{eq:planar_crit_params_values}
  b_\mathrm{c}=1.9, \qquad \rho_c = \sqrt{0.8} = 0.8944.
\end{equation}
In the remainder of this Section we outline the general strategy, while the case-dependent details and the associated results will be presented separately in the following three subsections.\\

The neural networks are explicitly designed to map the input time $s$ to the three-dimesional output \\
$(O_\rho(s),O_{P_\rho}(s),O_\phi(s))$, from which the solutions are built according to 
\begin{align}
\widehat{\rho}(s) = \rho_0  +f(s)O_\rho(s) \qquad , \qquad \widehat{P_\rho}(s) = P_{\rho,0}+f(s)O_{P_\rho}(s) \qquad , \qquad \widehat{\phi}(s) = \phi_0 +f(s)O_{\phi}(s).
\end{align}

 Let us stress that the role of the time $t$ in Section \ref{sec:HNN_Presentation} is now played by the affine parameter $s$ and this  does not entail any difference neither in the strategy nor in the implementation of the neural networks. Indeed, Section \ref{sec:HNN_Presentation} can be re-read just replacing $t$ with $s$.\\
 Our implementation of the neural networks\footnote{A demonstrative notebook is provided in \cite{notebook}. The complete code is available upon request.} relies on the libraries Keras \cite{keras} and TensorFlow \cite{tensorflow}. The trainings are performed by employing the Adam optimizer of \cite{Adam}.  In all the cases the weights are initialized by drawing random numbers from a normal distribution with center $0$ and width $0.05$.\\
The loss function we minimize is the combination given in \eqref{eq:total_loss} and during the training we monitor individually also $L_\mathrm{dyn}$ and $L_\mathrm{energy}$, which are defined  respectively in \eqref{eq:loss_dyn} and \eqref{eq:loss_energy}. Unless otherwise specified, we set $\gamma_\rho=~\gamma_{P_\rho}=~\gamma_\phi=1$ inside the loss function \eqref{eq:loss_dyn}.
As usually happens when training a neural network, the loss function fluctuates continuously over the epochs. Therefore, we save the updated weights at the end of each epoch and, at the end of the training, we consider the set of weights corresponding to the lowest value reached by the loss function as the optimal set for the trajectory prediction.\\
If not otherwise specified, the learning rate $\eta$ is defined as function of the number of the epoch according to the scheduler function
\begin{flalign}\label{eq:learning_rate_scheduler}
\eta(\mathrm{epoch}) =  \eta_f + \frac{\eta_i-\eta_f}{1+\exp\big[-(\eta_c-\mathrm{epoch})/\sigma_\eta \big]},
\end{flalign}
with $\eta_i=8\times 10^{-4}$, $\eta_f=\eta_i/10$, $\eta_c=10^5$ and $\sigma_\eta=20\times10^3$. It can be easily seen that \eqref{eq:learning_rate_scheduler} is a decreasing function such that $\eta\sim \eta_i$ for $\mathrm{epoch}\ll \eta_c$ and $\eta\sim\eta_f$ for $\mathrm{epoch}\to \infty$. Setting a decreasing learning rate during the training is a standard technique to improve the convergence towards the minimum of the loss function and the choice of the aforementioned parameters has been done after a careful investigation.\\
In addition, for each case we perform several trainings by changing $\lambda$ and the neural network architecture and judge the best choice according to the lowest loss function. We will show an example of this study for the over-critical case and for the other cases (including the non-planar motion in Section \ref{results:non_planar}) we will only quote the configuration that according to our analysis is the one that leads to the best performance.\\

As explained in Section \ref{sec:Planar_Case} from \eqref{eq:spheroidal_coordinates}, the solution in the $x-y$ Cartesian plane (recall that now $\theta=\pi/2$) is given by

\begin{flalign}\label{eq:cartesian_map}
x =\sqrt{\rho^2+a_f^2}\cos \phi \qquad , \qquad y =\sqrt{\rho^2+a_f^2}\sin \phi
\end{flalign}

For the three cases we set the  initial value of the radius $\rho_0$ to 10. The initial values of the variables $P_{\rho}$ and $\phi$ depend instead on the choice of the value of the impact parameter $b$. On the Cartesian plane, the impact parameter $b$ is identified with $y_0$ and therefore, the initial value of the angular variable is given by inverting the second equation of 
 \eqref{eq:cartesian_map}, that is

\begin{equation}\label{eq:plan_phi0}
    \phi_0 = \arcsin{\left(\dfrac{b}{\sqrt{\rho_0^2+a_{f}^{2}}}\right)}.
\end{equation}

The value of $P_{\rho,0}$ can be obtained by solving $\mathcal{H}(0)=0$ in (\ref{eq:Ham_plan}).\\
In each case we set the interval of discrete times in correspondence of which the loss function \eqref{eq:total_loss} has to be minimized by splitting the interval $\mathcal{T}_\mathrm{train}=[0,T]$ in $N$ equally spaced points. We keep this set fixed during the entire training process. We then construct a second interval of discrete times, denoted by $\mathcal{T}_\mathrm{val}$, by adding 5 additional equally spaced points between two consecutive times in $\mathcal{T}_\mathrm{train}$ and excluding the times in $\mathcal{T}_\mathrm{train}$. The interval $\mathcal{T}_\mathrm{val}$ thus contains $5\times(N-1)$ points which are not used during the training. For all the trainings performed in this work we have explicitly checked that, at the end of the training, the loss function evaluated for the set $\mathcal{T}_\mathrm{val}$ is compatible with that actually minimized within 5\%. This is a numerical check that the over-fitting of the solution has not occurred. Finally, we show in the plots the results predicted for the joint interval $\mathcal{T}_\mathrm{train}\cup\mathcal{T}_\mathrm{val}$ which contains $N+5\times(N-1)$ points in total. \\

In order to assess the performance of the neural networks we compare our results with those obtained by other numerical integrators. In particular we consider the first-order semi-implicit Euler integrator \eqref{eq:Euler}, that, as it is well known, preserves the symplectic structure of  Hamiltonian systems by conserving the energy up to a shift term proportional to the discretization step, and the more advanced Runge-Kutta method of order 5(4), \texttt{RK45} in short, which is not symplectic though. To implement the latter we use the SciPy package \texttt{solve\_ivp}. The package (see the documentation at \cite{solve_ivp}) allows conveniently to set the precision of the solution $y(t)$ in terms of absolute and relative tolerances, determined by the parameters $a_\mathrm{tol}$ and $r_\mathrm{tol}$ respectively, by keeping the local error estimates smaller than $a_\mathrm{tol}+ r_\mathrm{tol} |y(t)|$.  We have set in all cases the relative and absolute tolerances to the minimum possible values $r_\mathrm{tol}=10^{-15}$ and $a_\mathrm{tol}=10^{-15}$, respectively.\\
We evaluate the numerical solution obtained from \texttt{RK45} in correspondence of the interval $\mathcal{T}_\mathrm{train}\cup\mathcal{T}_\mathrm{val}$ as well. The numerical solution provided by the Euler method is instead strongly dependent on the discretization step $\Delta s$, which regulates the update of solution in \eqref{eq:Euler}. We consider two time steps, $\Delta s=T/N$ which corresponds to the same step of $\mathcal{T}_\mathrm{train}$ and $\Delta s=T/(100N)$, which is instead 100 times finer. In the discussion of the results we will refer to these different solutions by Euler with $1\times N$ and $100\times N$ points respectively.\\

As explained in Appendix~\ref{APP:groundtruth}, we can compute exactly $\phi$ as function of $\rho$ and then, by plugging these values in \eqref{eq:cartesian_map}, we get the exact trajectory in the Cartesian plane. We call this trajectory \quote{ground truth} as customary in ML studies. The comparison with the ground truth only allows for a qualitative check of the correctness of the predicted geodesics. In order to have a quantitative measure of the error committed by the neural network and the numerical integrators, we can compare the exact function $\phi(\widehat{\rho})$ with $\widehat{\phi}(\widehat{\rho})$ and $\phi(\tilde{\rho})$ with $\tilde{\phi}(\tilde{\rho})$. The same comparison can be carried out also for $P_\rho$, whose true value can be found as function of $\rho$ by solving 
$\mathcal{H}(s)=0$ as explicitly done in \eqref{eq:Prho_exact_overcriical}-\eqref{eq:Prho_exact_subcritical}. 

We pass now to examine separately the three regimes.

\begin{figure}[t!]
\includegraphics[scale=0.8]{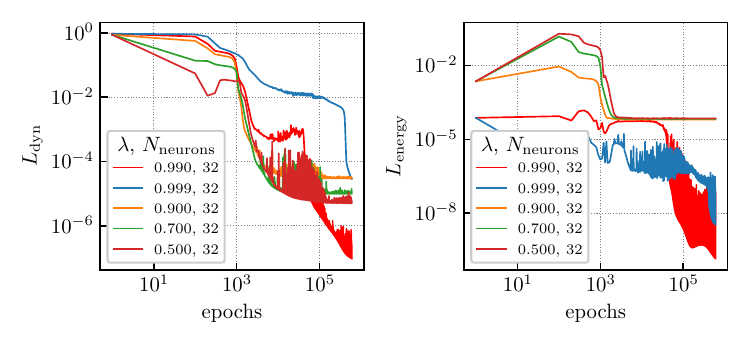}
\includegraphics[scale=0.8]{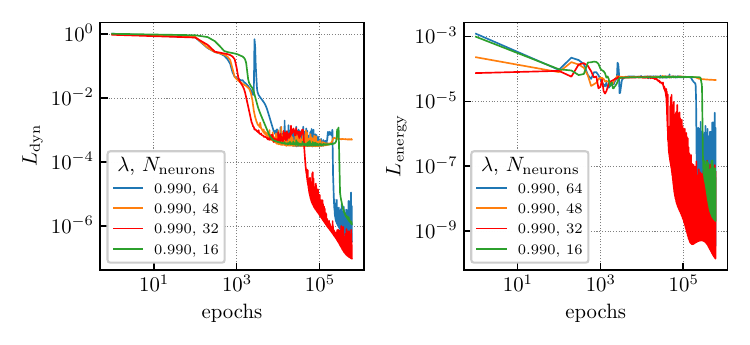}
\caption{
\label{fig:hamilton_overcritical_tuning} 
Tuning of the parameters for the over-critical case with $b=1.91$. In all the tests the neural network has 2 hidden layers and $\tanh(x)$ as activation function. \emph{Top-panels}: $L_\mathrm{dyn}$ (left) and $L_\mathrm{energy}$ (right) at fixed number of 32 neurons and changing $\lambda$ from 0.5 to 0.99. \emph{Bottom-panels}: the same as the top-panel but at fixed $\lambda=0.99$ and varying the number of neurons from 16 to 64. The four plots show that the best performance, in terms of minimization of the loss function, is given by the choice $\lambda=0.99$ and $N_\mathrm{neurons}=32$, corresponding to the red color.
}
\end{figure}

\subsection{\label{sec:planar_overcritical}Over-critical case}
In this Subsection we discuss the results for the over-critical case and set $b=1.91$. The initial conditions are therefore given by

\begin{flalign}\label{eq:init_val_over}
\phi_0 = 0.1922 \qquad , \qquad \mathcal{H}(0) = 0 \qquad , \qquad P_{\rho,0} = -0.9917.
\end{flalign}

\begin{figure}[H]
\centering
\subfloat[][ \footnotesize{Loss functions for the over-critical case with $b=1.91$. The training has been performed with $\lambda=0.99$, $N_\mathrm{neurons}=32$, 2 hidden layers and $\tanh(x)$ as activation function.
} \label{fig:Hamilton_overcritical_loss}]
   {\includegraphics[scale=0.65]{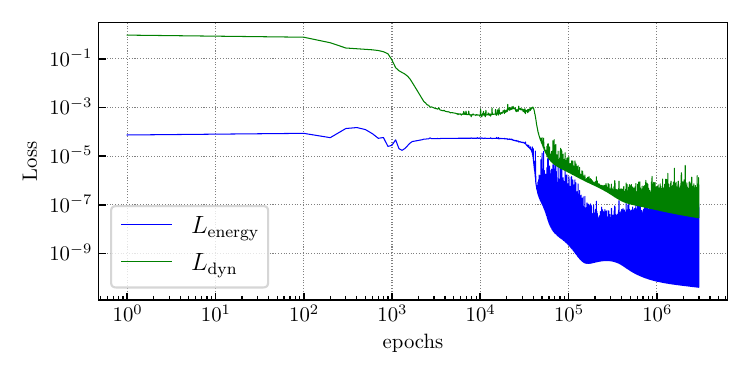}} \quad
\subfloat[][Energy conservation over the time in the over-critical case with $b=1.91$  and $N=400$ for different methods. The conserved value of the energy $\mathcal{H}(0)=0$ is represented by the dotted black line. \label{fig:hamilton_overcritical_H} ]
   {\includegraphics[scale=0.65]{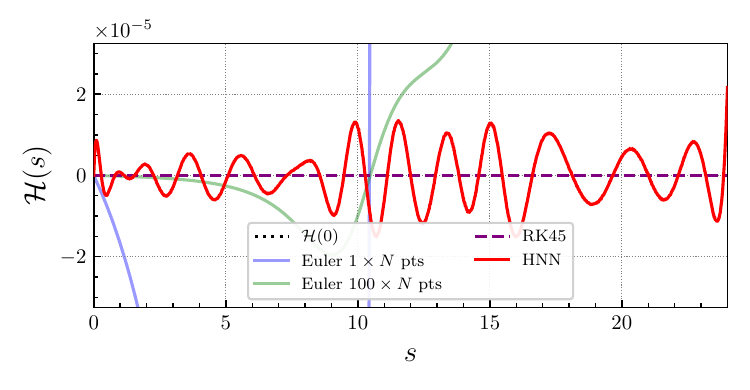}}
   \caption{Loss functions (on the left) and energy conservation (on the right) for the over-critical case.}
\label{fig:subfig1}
\end{figure}

To find the optimal setup, we have repeated the training several times by varying the trade-off parameter $\lambda$ and the number  of neurons $N_\mathrm{neurons}$  while keeping fixed the number of hidden layers to 2 and the activation function $\sigma(x)$ to the hyperbolic tangent
\begin{equation}\label{eq:act_fun}
    \sigma(x) = \tanh(x)
\end{equation}
For $\lambda $ we have considered the values 0.5, 0.7, 0.9, 0.99 and 0.999 while for $N_\mathrm{neurons}$  we have considered the values 16, 32, 48 and 64. 
The interval of time $\mathcal{T}_\mathrm{train}$ is given by splitting the interval $[0,24]$ in $N=400$ points. The neural network is trained with $600\times 10^3$ epochs for each combination of parameters. The loss functions $L_\mathrm{dyn}$ and $L_\mathrm{energy}$ are displayed in Figure~\ref{fig:hamilton_overcritical_tuning}, where the top-panel shows the dependence on $\lambda$ at fixed $N_\mathrm{neurons}=32$, while the bottom-panel shows the dependence on $N_\mathrm{neurons}$ at fixed $\lambda=0.99$. The study shows a strong dependence upon both $\lambda$ and $N_\mathrm{neurons}$. In particular, at fixed $N_\mathrm{neurons}=32$, the loss functions get stuck in a plateau for $\lambda\le 0.9$ highlighting that, indeed, the addition of the conservation term $L_\mathrm{energy}$ favors a faster and better minimization of the total loss function. We have observed no significant improvement by going from 2 to 3 hidden layers. \\
The study shows that the best performance is achieved by the combination $\lambda=0.99$ and $N_\mathrm{neurons}=32$. We have then trained the neural network with such combination up to $3\times 10^6$ epochs and we have used it to predict the geodesic. The corresponding loss functions are shown in Figure~\ref{fig:Hamilton_overcritical_loss}.

Figure~\ref{fig:hamilton_overcritical_H} shows the function $\mathcal{H}(s)$ for the predicted geodesic (labeled by \quote{HNN} and represented by the solid red line), for the Runge-Kutta method (purple dashed line) and for the Euler method with respectively $N$ (blue) and $100\times N$ (green) points. The plot shows that the HNN conserves the energy very well on average with fluctuations between $-2\times 10^{-5}$ and $2\times 10^{-5}$. The Euler methods with both $N$ and $100\times N$ points exhibit larger violation of the energy that increases with the time $s$. The \texttt{RK45} integrator is remarkably good with violations from $\mathcal{H}(0)=0$ of order $10^{-9}$.\\

\begin{figure}
\includegraphics[scale=0.7]{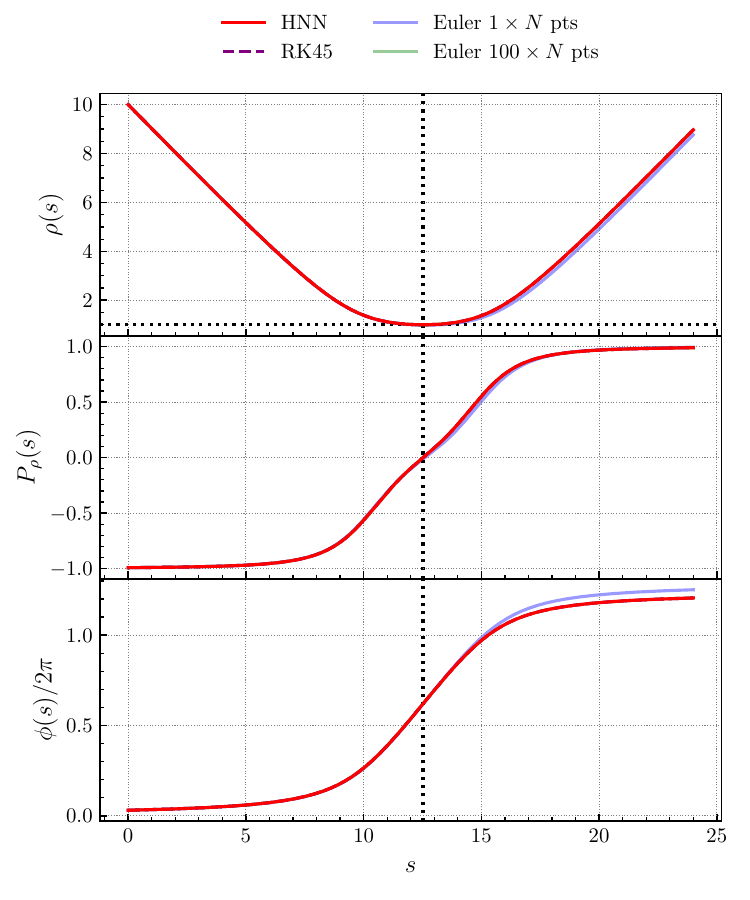}
\caption{
\label{fig:Hamilton_overcritical_variables}
From top to bottom: $\rho(s)$, $P_\rho(s)$ and $\phi(s)/2\pi$ for different methods in the over-critical case with $b=1.91$ and $N=400$. The vertical dotted line in the three panels represents $\widehat{s}_+=12.521$, that is the time at which the minimum of $\rho$ is reached and with respect to which the motion is symmetric. The motions to the left and to the right of $\widehat{s}_+$ are respectively incoming and outgoing. The horizontal dotted line in the top-panel represents $\rho_+=1.000138$.
}
\end{figure}

Before showing the results for $\rho(s)$, $P_\rho(s)$ and $\phi(s)$, let us discuss their expectations based on the physical arguments of Section \ref{sec:Planar_Case}. At the beginning of the motion, the particle approaches the fuzzball (we refer to this phase as \quote{incoming motion}), and consequently $\rho$ decreases, until it reaches the minimum value $\rho_+=1.000138$, which is the largest root of the radicand in \eqref{eq:diffeq}. Afterward, the particle moves away from the fuzzball (we refer to this phase as \quote{outgoing motion}), and $\rho$ starts increasing again. We denote by $s_+$ the time at which the variable $\rho$ attains its minimum $\rho_+$. Due to the symmetry of the system under $\rho\mapsto -\rho$, the function $\rho(s)$ must be symmetric with respect to $s_+$.   On the other hand, $P_\rho$ starts from the negative value $P_{\rho, 0}$, given in \eqref{eq:init_val_over}, and increases over time until it reaches zero at $s_+$, eventually approaching a plateau at infinite time. Similarly, $\phi$ starts from the initial value $\phi_0$ written in \eqref{eq:init_val_over}, then increases and asymptotically reaches a plateau at infinite time.\\
This qualitative description of the motion is numerically confirmed by the functions $\rho(s)$, $P_\rho(s)$ and $\phi(s)/2\pi$ that are shown in  Figure~\ref{fig:Hamilton_overcritical_variables} for different methods. The three dynamical variables behave qualitatively as expected and there is no qualitative difference between the prediction of the HNN, \texttt{RK45} and the Euler method with $100\times N$ points, while Euler with $N$ points visibly deviates from the other. From the numerical solution of the HNN we find that the minimum value of $\rho$ is  $\widehat{\rho}_+=1.000153$, 0.002\% away of $\rho_+$ , at time  $\widehat{s}_+=12.521$.\\

In order to have a quantitative measure of the quality of the solutions predicted by the HNN we consider the quantities
\begin{flalign}\label{eq:Prho_error}
\delta P_\rho(\hat\rho)&=\widehat{P_\rho}(\hat{\rho})-P_\rho(\hat{\rho}),
\end{flalign}
and
\begin{flalign}\label{eq:phi_error}
\delta \phi(\hat\rho)&=\widehat{\phi}(\hat{\rho})-\phi(\hat{\rho}).
\end{flalign}

\begin{figure}[H]
\centering
\subfloat[][ \footnotesize{Error on the function $\phi(\rho)$ for different methods and for the incoming motion (\emph{top-panel}) and the outgoing motion (\emph{bottom-panel)} in the over-critical case with $b=1.91$. The vertical dotted line represents $\rho_+=1.000138.$
} \label{fig:Hamilton_overcritical_phi_error}]
   {\includegraphics[scale=0.65]{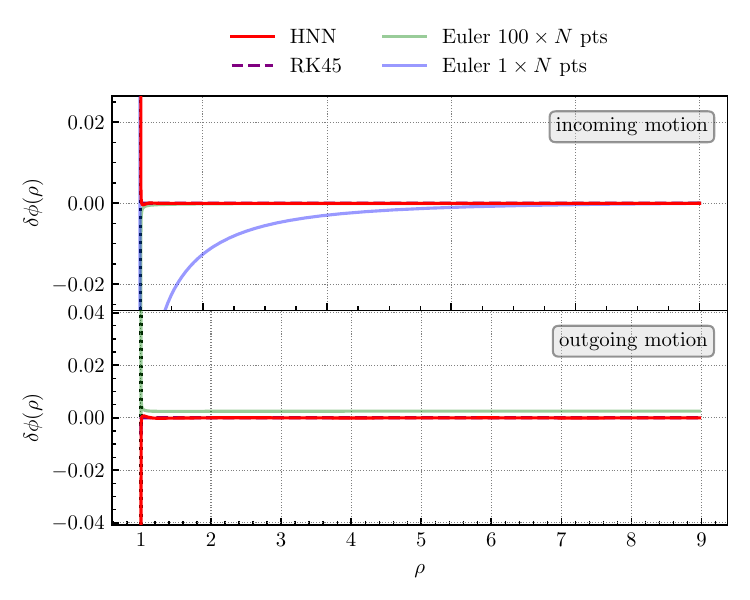}} \quad
\subfloat[][Error on the variable $P_\rho(\rho)$  for different methods and for the incoming motion (\emph{top-panel}) and the outgoing motion (\emph{bottom-panel)} in the over-critical case with $b=1.91$.  The vertical dotted line represents $\rho_+=1.000138.$
\label{fig:Hamilton_overcritical_Prho_error} ]
   {\includegraphics[scale=0.65]{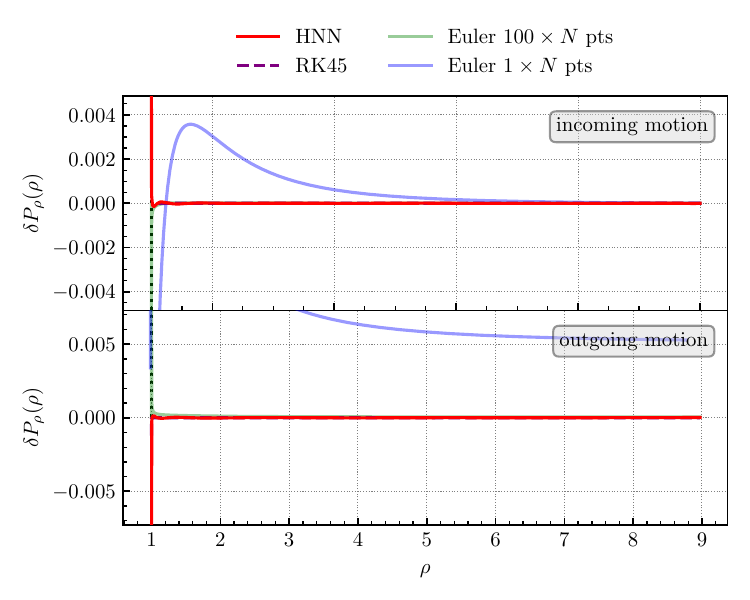}}
   \caption{Errors on $\phi(\rho)$ (on the left) and $P_\rho(\rho)$ (on the right) for the different methods.}
\label{fig:subfig2}
\end{figure}

Equivalent quantities can be considered also for the solutions obtained from the numerical integrators. As anticipated before, the functions $P_\rho(\rho)$ and $\phi(\rho)$ can be computed exactly and therefore \eqref{eq:Prho_error} and \eqref{eq:phi_error} are good indicators to quantify the error made by a given method. For this reason we will refer to these quantities as \quote{errors} henceforth. The functions  $\phi(\rho)$ and $P_\rho(\rho)$ are obtained as explained in Appendix~\ref{APP:groundtruth}.\\
The errors of the HNN and the numerical integrators are shown in Figure~\ref{fig:Hamilton_overcritical_phi_error} and Figure~\ref{fig:Hamilton_overcritical_Prho_error} for the variable  $\phi$ and $P_\rho$ respectively. Both the figures show the error in the incoming motion (top-panel) and outgoing motion (bottom-panel). As it can be seen, the error of the HNN on the variable $\phi$ oscillates uniformly around 0 with an order of magnitude of $10^{-4}$ and with a spike occurring for $\rho$ close to $\rho_+$. An analogous picture is found for the variable $P_\rho$ but with an order of magnitude of $10^{-5}$. The small scale of the errors (compare with $P_\rho(\rho)$ and $\phi(\rho)$ in Figure~\ref{fig:phi_exact} and Figure~\ref{fig:Prho_exact} in Appendix \ref{APP:groundtruth}) is a reassuring evidence that the Hamilton equations have been solved by the HNN with a high accuracy.  The Euler method with both $N$ and $100\times N$ points has larger errors than HNN. \texttt{RK45} is instead the most precise, with errors that fluctuate on the scale $10^{-9}$ and $10^{-10}$ for $\phi$ and $P_\rho$ respectively.

\begin{figure}[H]
\centering
\subfloat[][ \footnotesize{Geodesics in the Cartesian plane obtained from different methods in the over-critical case with $b=1.91$ and $N=400$. The ground truth is represented by the solid black line, while the small circle at the center (0,0) has radius $a_f$ and represents the fuzzball. The black arrow indicates the incoming direction of the particle.
} \label{fig:Hamilton_overcritical_cartesian}]
   {\includegraphics[scale=0.7]{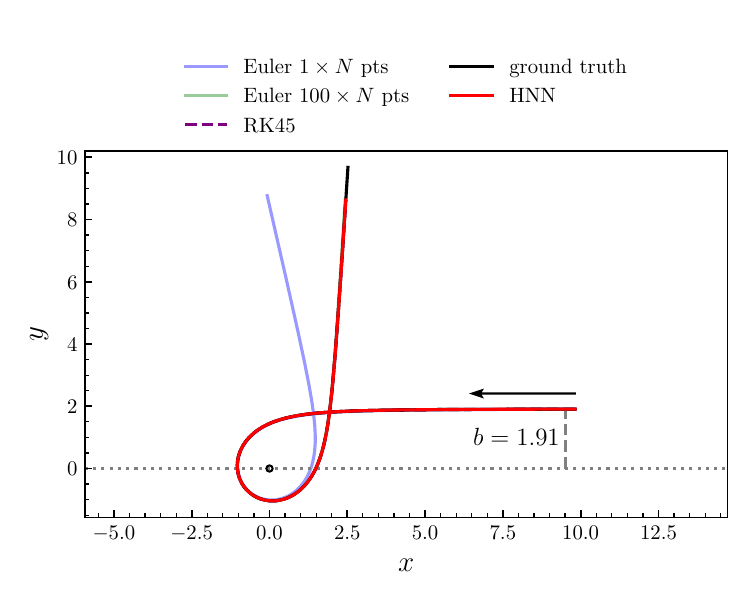}} \quad
\subfloat[][Geodesics in the Cartesian plane predicted by the HNN for different values of the impact parameter $b$.
\label{fig:Hamilton_overcritical_multib} ]
   {\includegraphics[scale=0.75]{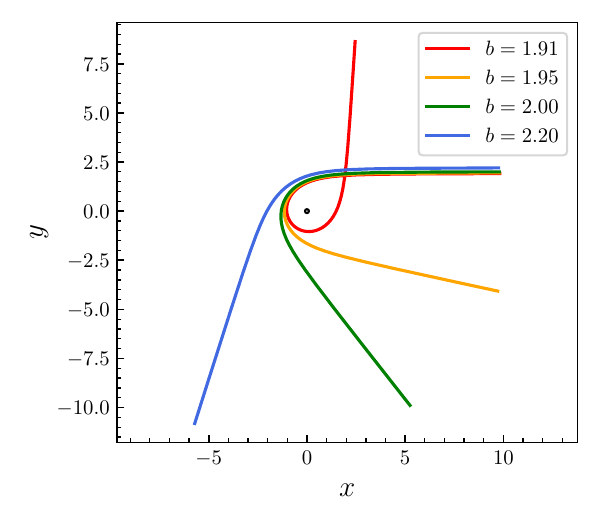}}
   \caption{Geodesics for the over-critical case.}
\label{fig:subfig3}
\end{figure}

We eventually project the motion obtained for the different methods  to the Cartesian plane by using \eqref{eq:cartesian_map} and compare to the ground truth. The geodesics in the Cartesian plane are shown in Figure~\ref{fig:Hamilton_overcritical_cartesian}. The HNN and \texttt{RK45} have no visible differences with respect to the ground truth, showing that they are both efficient integrators of the Hamilton equation of motions. Concerning the Euler method, for the same number of points as those used to train the neural network the trajectory shows a clear deviation from the ground truth over long timescales and a reliable result is obtained only by considering a number of points (time step) larger (smaller) by a factor 100. 

We conclude this subsection by employing the HNN to compute the geodesics for impact parameters $b=1.95$, $2.0$ and $2.2$. The three cases have been obtained by considering the same setup as $b=1.91$ discussed above, that is $\lambda=0.99$, $N_\mathrm{neurons}=32$, $N=400$, $\mathcal{T}_\mathrm{train}=[0,24]$ and $3\times10^6$ epochs. For all the three cases we have computed the errors and checked that the quality of the solution is at the same level as the case with $b=1.91$. In Figure~\ref{fig:Hamilton_overcritical_multib} we show the final geodesics predicted by the HNN in the Cartesian plane.  As it can be seen from the plot, the scattering angle at which the particle moves away from the fuzzball increases as the impact parameter increases and this is expected by the theory (in fact, as the impact parameter increases, the particle is less affected by the gravitational field of the fuzzball, and therefore its trajectory is less bended).

\subsection{\label{sec:planar_critical}Critical case}
In this Subsection we present the critical case obtained by setting $b=b_c=1.9$. The initial conditions now read

\begin{flalign}\label{eq:init_val_critical}
\phi_0 = 0.1912 \qquad , \qquad \mathcal{H}(0) = 0 \qquad , \qquad P_{\rho,0} = -0.9919.
\end{flalign}

\begin{figure}[H]
\centering
\subfloat[][ \footnotesize{Loss functions for the critical case with $b=b_c=1.9$. The training has been performed with $\lambda=0.99$, $N_\mathrm{neurons}=48$, 2 hidden layers and $\tanh(x)$ as activation function.
} \label{fig:Hamilton_critical_loss}]
   {\includegraphics[scale=0.65]{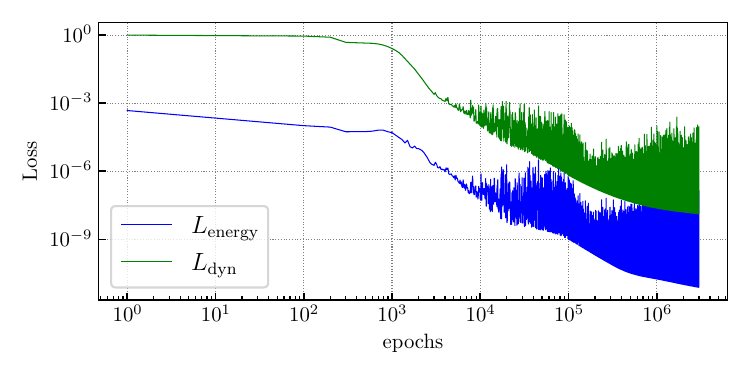}} \quad
\subfloat[][Energy conservation over the time in the critical case with $b=b_c=1.9$  and $N=800$ for different methods. The conserved value of the energy $\mathcal{H}(0)=0$ is represented by the dotted black line.  
\label{fig:Hamilton_critical_H} ]
   {\includegraphics[scale=0.65]{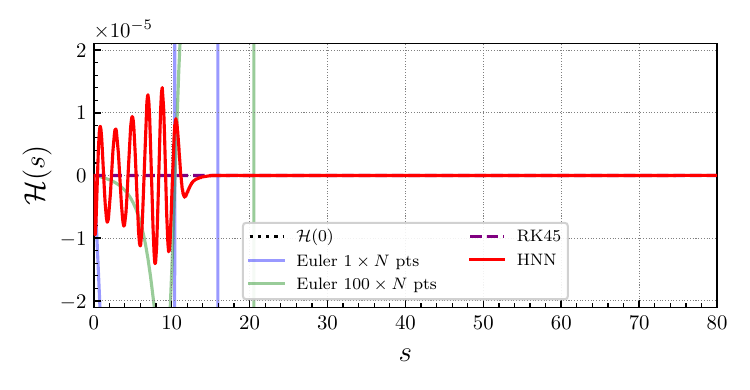}}
   \caption{Loss functions (on the left) and energy conservation (on the right) for the critical case.}
\label{fig:subfig4}
\end{figure}

For this case we have considered the time interval $\mathcal{T}_\mathrm{train}=[0,80]$ sampled with $N=800$ points. The reason behind such a long time interval is to probe the stability of the trajectory on the critical radius circumference over long timescales. We have repeated the tuning procedure discussed in the over-critical case and found that the optimal setup is given by $\lambda=0.99$ and $N_\mathrm{neurons}=48$. The number of hidden layers is 2 and the activation function is given by \eqref{eq:act_fun}. The neural network is trained with $3\times 10^6$ epochs and the corresponding loss functions $L_\mathrm{dyn}$ and $L_\mathrm{energy}$ are shown in Figure~\ref{fig:Hamilton_critical_loss}.\\
In Figure~\ref{fig:Hamilton_critical_H} we show the function $\mathcal{H}(s)$ for the different methods. The energy is very well conserved on average by the HNN with oscillations of order $10^{-5}$ for $s$ smaller than $\simeq 14$ and of order $10^{-9}$ for larger times. The Euler method with $N$ and $100\times N$ points does not conserve the energy from the very beginning. Particular attention should be paid to the \texttt{RK45}, which conserves the energy on a scale of $10^{-13}$. This information seems to indicate the \texttt{RK45} method as the best for trajectory prediction. However, as we will see shortly, the exact energy conservation by a non-symplectic method does not automatically imply an exact solution of the equations of motion.

\begin{figure}[H]
\centering
\subfloat[][ \footnotesize{From top to bottom: $\rho(s)$, $P_\rho(s)$ and $\phi(s)/2\pi$ for different methods in the critical case with $b=b_c=1.9$ and $N=800$. The horizontal black dotted line in the top panel represents the critical radius $\rho_c=0.8944$, on which the trajectory must settle. The trajectory predicted by the HNN is the only one among the different methods to correctly reproduce this feature.
} \label{fig:Hamilton_critical_variables}]
   {\includegraphics[scale=0.65]{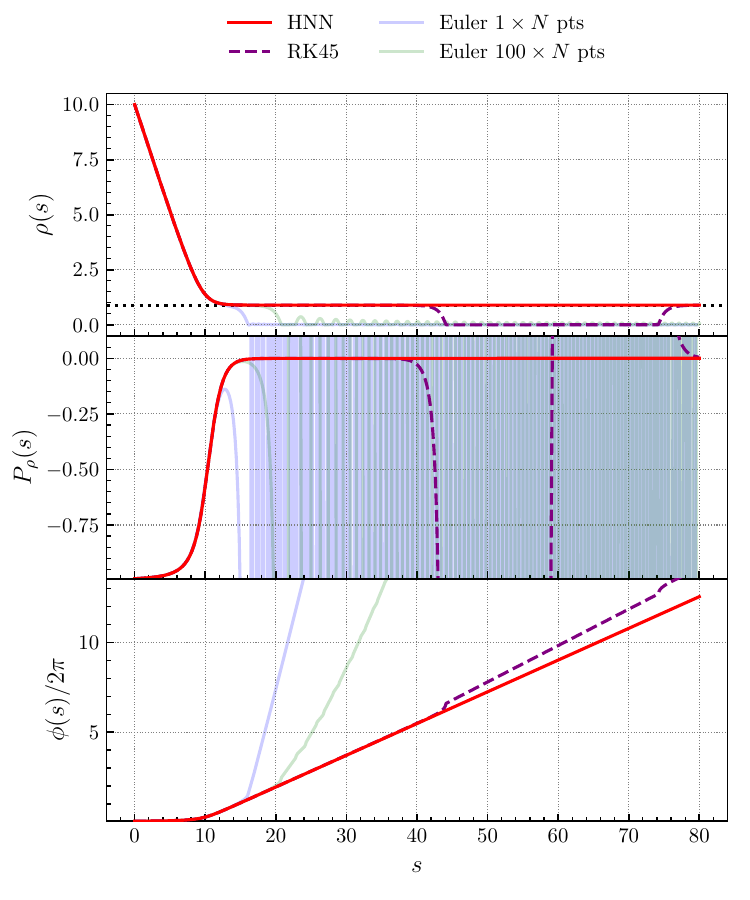}} \quad
\subfloat[][Stability of the trajectory predicted by the HNN over the long timescale.  The vertical dotted line is drawn at $s\simeq 22$, the time at which approximately the trajectory reaches the critical circumference.   \emph{Top-panel}: difference between $\widehat{\rho}(s)$ and $\rho_c=0.8944$. \emph{Bottom-panel}:  predicted $\widehat{P_\rho}(s)$.\label{fig:Hamilton_critical_differences} ]
   {\includegraphics[scale=0.65]{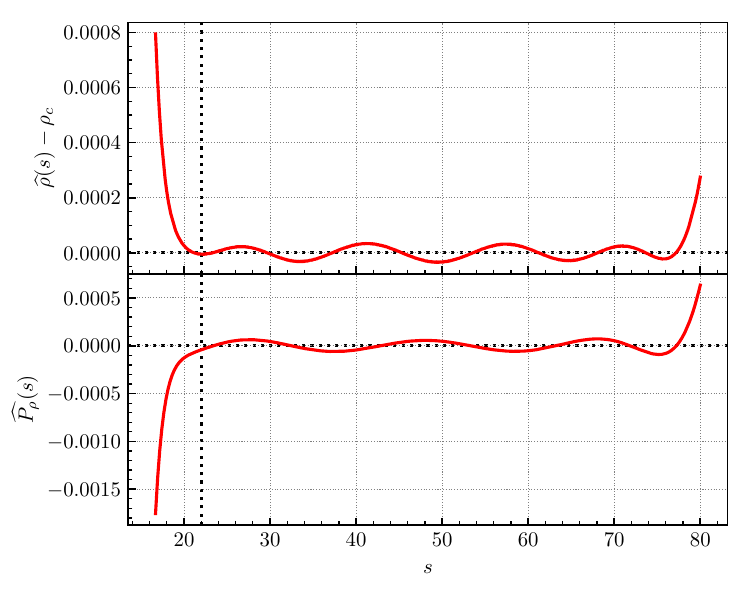}}
   \caption{Predicted variables from the different methods (on the left) and the stability of the solution obtained from the HNN (on the right) in the critical case.}
\label{fig:subfig5}
\end{figure}

Concerning the physics of the motion, in this case we only have an incoming phase in which the radial coordinate $\rho(s)$ decreases from $\rho_0$ up to reach asymptotically the value $\rho_c=0.8944$. 
 Correspondingly, the radial momentum $P_\rho(s)$ increases from the negative value $P_{\rho,0}$ given in \eqref{eq:init_val_critical}  up to freezing at 0. The function $\phi(s)$ is instead expected to have a linear grow as the particle stabilizes rotating on the critical geodesic.\\
Figure~\ref{fig:Hamilton_critical_variables} shows the functions $\rho(s)$, $P_\rho(s)$ and $\phi(s)/2\pi$ for HNN and the numerical integrators. As it can be seen, the HNN is the only method that correctly reproduces the expected dynamics of the motion. The Euler method with both $N$ and $100\times N$ points starts failing around $s\simeq 15$ and exhibit a badly oscillating behavior. The  \texttt{RK45},  despite exactly conserving energy, fails for $s>44$. In particular $\rho(s)$ for both Euler and \texttt{RK45} approaches touches 0, which corresponds to a situation in which the particle trajectory has collapsed onto the sphere of radius $a_f$ and it is physically impossible for this impact parameter. The variable $\phi(s)$ predicted by the HNN exhibits the correct linear grow for times larger  than $s\simeq 22$, which we can consider as the time at which the trajectory stabilizes on the critical circumference. The angle $\phi(s)$ is plotted normalized by $2\pi$ in order to give an idea of the number of windings of the trajectory around the fuzzball. From time $s = 22$ to time $s = 80$, the trajectory completes approximately 10 full windings and remains stable throughout.\\
Since the numerical integrators clearly fail, it does not make sense to compare the error of the HNN with the error of the integrators. Rather, to assess the stability of the solution predicted by the neural network, we show in the Figure~\ref{fig:Hamilton_critical_differences} the quantities  $\widehat{\rho}(s)-\rho_c$ and $\widehat{P_\rho}(s)$ for large times. Both quantities oscillate around 0 on a remarkably small scale. A small deviation from 0 is observed for $s\simeq 76$. It is possible that it could simply be reduced by increasing the number of epochs. As it can be seen in the Figure~\ref{fig:Hamilton_critical_loss}, in fact, the loss functions have not yet reached their minimum.\\
At this point, we emphasize that although the \texttt{RK45} method is very accurate and the Euler method is symplectic, the HNN is the only one that correctly integrates the equations of motion, outperforming both methods in terms of efficiency and accuracy. The HNN ability to predict a stable trajectory over long timescales is due to the introduction of the energy conservation term, which penalizes solutions that violate this constraint. The ability to enforce constraints within the ML framework proves to be an extremely beneficial factor in this case, enabling accurate results where standard methods fail.\\

\begin{figure}
    \centering
    \includegraphics[scale=0.65]{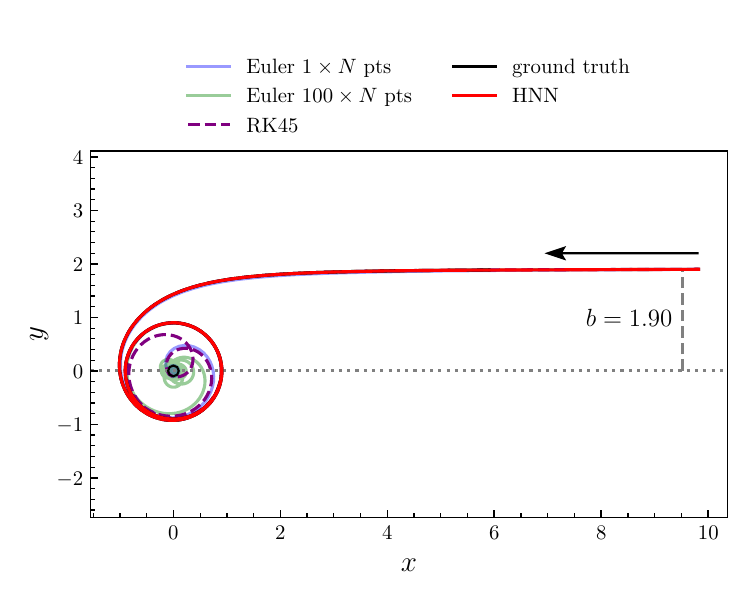}
    \caption{
    \label{fig:Hamilton_critical_cartesian}
Geodesic in the Cartesian plane obtained from different methods in the critical case with $b=b_c=1.9$ and $N=800$. The ground truth is represented by the solid black line, while the small circle at the center (0,0) has radius $a_f$ and it represents the fuzzball. The black arrow indicates the incoming direction of the particle.
    }
\end{figure}

We then project the  motion to the Cartesian plane by means of (\ref{eq:cartesian_map}) and the result is shown in Figure~\ref{fig:Hamilton_critical_cartesian} together with the ground truth. As it can be appreciated, the geodesic predicted by the HNN remains stable along the circumference of radius $\sqrt{\rho_c^2+a_f^2}$, while all the other methods predict a geodesic that collapses onto the fuzzball.

\subsection{Sub-critical case} \label{subsec:sub-critical}

In this Subsection we present the results for the last regime, that is the sub-critical one, and we study the situation with $b=1.89$. The initial conditions are

\begin{flalign}\label{eq:init_val_subcritical}
\phi_0 = 0.1901 \qquad , \qquad \mathcal{H}(0) = 0 \qquad , \qquad P_{\rho,0} = -0.9921.
\end{flalign}

This case is particularly challenging from the HNN perspective, as the underlying physics involves an impact of the particle against the fuzzball, hence the radius $\rho$ decreases from $\rho_0$ down to zero (this can be also deduced from the first of \eqref{eq:diffeq}, since the radicand never vanishes). As a result, the conjugate momentum $P_\rho$ increases in absolute value until it reaches the value
\begin{equation} \label{minimal_prho_sub_critical}
    \overline{P_{\rho}} \equiv P_\rho(\rho=0)= -80.1249,
\end{equation}
where we have used $P_\rho(\rho)$ in \eqref{eq:Prho_exact_subcritical}. The problem arises from the fact that the decrease from $P_{\rho,0}$ to $\overline{P_\rho}$ is very rapid in a small interval of values of $\rho$, which are close to 0, and this is shown in the right-panel of Figure~\ref{fig:Prho_exact}. The same behavior is observed in the time domain, as it can be  seen beforehand in the central panel of Figure~\ref{fig:Hamilton_subcritical_variables}. From this perspective, the differential equation governing the dynamics of $P_\rho$ is a stiff equation and, as well known, stiff equations can be a challenge to numerical methods, HNN included. Indeed, we did not manage to obtain accurate solutions by using the same strategy adopted in the past two Subsections. Therefore, in the following we propose two alternative approaches to overcome this difficulty and get a reliable prediction for the geodesic.

\subsubsection{\label{sec:splitting}Approach 1: splitting the time interval}

\begin{figure}
    \centering
    \includegraphics[scale=0.65]{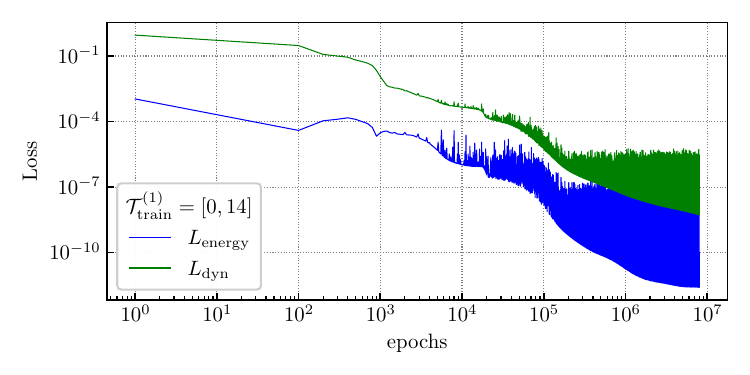}
    \includegraphics[scale=0.65]{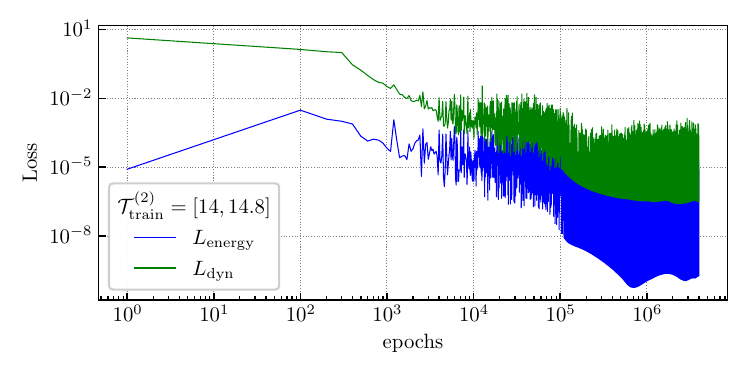}
    \includegraphics[scale=0.65]{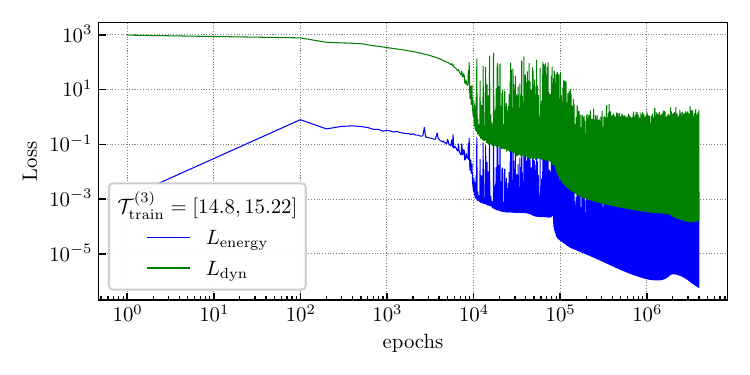}
    \caption{
    \label{fig:Hamilton_subcritical_losses}
    Loss functions for the sub-critical case with $b=1.89$. From top to bottom the losses refer to the training intervals $\mathcal{T}_\mathrm{train}=[0,14]$, $[14,14.8]$ and $[14.8,15.22]$.
    }
\end{figure}

In the first approach we split the time interval over which we want to compute the solution in such a way to separate the stiff regime of $P_\rho$ from that in which its decrease is smoother. This can be easily achieved by dividing the total time interval into consecutive sub-intervals and we will refer to this as \quote{time-splitting approach}. For each sub-interval, a neural network is trained from scratch using as initial conditions the values of the predicted variables at the final time of the previous interval. We consider the following three time intervals, $\mathcal{T}_\mathrm{train}^{(1)}=[0,14]$, for which the initial conditions are given in    \eqref{eq:init_val_subcritical}, $\mathcal{T}_\mathrm{train}^{(2)}=[14,14.8]$, for which the initial conditions are $\rho_0=0.6552$, $P_{\rho,0}=-0.8982$ and $\phi_0=5.8452$, and finally
$\mathcal{T}_\mathrm{train}^{(3)}=[14.8,15.22]$, for which the initial conditions are $\rho_0=0.3014$, $P_{\rho,0}=-7.0688$ and $\phi_0=7.2790$. All the three intervals are divided into $800$ points for a total number of $N=2400$ points. Notice that, at the practical level, the time interval to feed to the neural network always starts at time 0, that is $\mathcal{T}_\mathrm{train}^{(2)}$ and $\mathcal{T}_\mathrm{train}^{(3)}$ actually are $[0,0.8]$ and $[0,0.42]$. Setting as initial conditions the values predicted at the end of the previous interval is what guarantees that the predictions, on the sub-intervals, glued together eventually correspond to the solution in the overall interval $\mathcal{T}_\mathrm{train}=[0,15.22]$. \\
Figure~\ref{fig:Hamilton_subcritical_losses} shows the loss functions for the three time intervals. The numbers of epochs, respectively, are $8 \times 10^6$, $4\times 10^6$ and $4\times 10^6$. The neural networks in all three cases consist of 3 hidden layers with 64 neurons per each. $\lambda=0.99$ again and the activation function always is $\tanh{(x)}$. \\

\begin{figure}[H]
\includegraphics[scale=0.7]{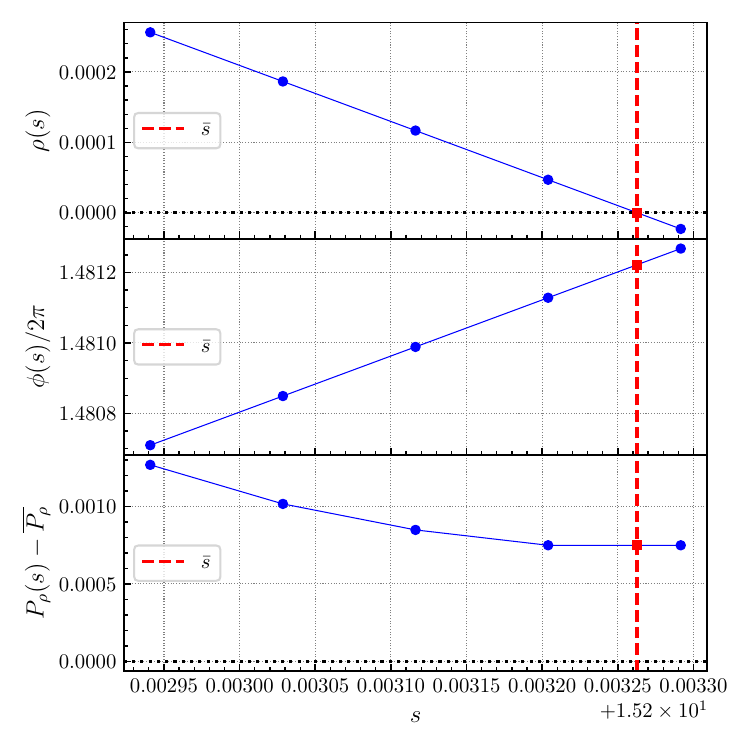}
\caption{
\label{fig:prediction_interpolation}
From top to bottom: interpolation of $\rho(s)$, $\phi(s)/2\pi$ and $P_\rho(s)-\overline{P_\rho}$ at the time $\bar{s}$. The plot refers to the HNN with the time-splitting approach for which $\bar{s}=15.2032$ (vertical dashed line). In the top and bottom panel we highlight the value 0 with a horizontal dotted line.
}
\end{figure}

The final time $s=15.22$ has been chosen such that $\rho$ reaches and slightly exceeds the value 0. We call $\bar{s}$ the  time at which $\rho(\bar{s})=0$. The motion in the interval $[0,\bar{s}]$ corresponds to the incoming phase, in which the particle approaches the fuzzball and then bounces it, and, thanks to the symmetry $\rho\mapsto -\rho$ of the system, the outgoing phase can be found by reflecting the incoming motion with respect to $\bar{s}$. Figure~\ref{fig:prediction_interpolation} shows the procedure we have employed to determine $\bar{s}$ in the case of HNN. In the top-panel we determine $\bar{s}$ by calculating the time at which the curve $\rho(s)$ intersects the line $\rho=0$ (assuming a linear ansatz due to the small range of times we are focusing on). The resulting time is $\bar{s}=15.2032$ and it is marked by the red vertical dashed line. The fact that the mathematical solution to the system predicts $\rho <0$ for $s>\bar{s}$ is irrelevant to the physics since the geodesics in the Cartesian plane (see \eqref{eq:cartesian_map}) depends on $\rho^2$.  We then interpolate $\phi(s)/2\pi$ and $P_\rho(s)$ at the time $\bar{s}$ (middle- and bottom-panels) and discard all the points for which $s>\bar{s}$.  The procedure is repeated independently also for the numerical integrators.

\begin{figure}[H]
\centering
\subfloat[][ \footnotesize{Energy conservation for the sub-critical case with $b=1.89$  for the different methods and by following the time-splitting approach. The conserved value of the energy $\mathcal{H}(0)=0$ is represented by the dotted black line.
} \label{fig:Hamilton_subcritical_H}]
   {\includegraphics[scale=0.65]{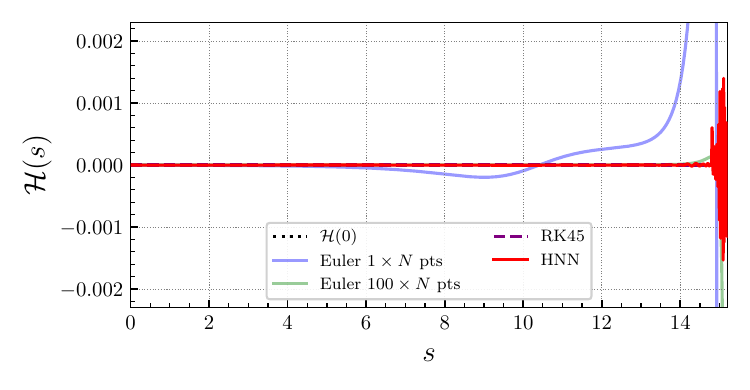}} \quad
\subfloat[][ \footnotesize{From top to bottom: $\rho(s)$, $P_\rho(s)$ and $\phi(s)/2\pi$ for the different methods in the sub-critical case with $b=1.89$ and $N=2400$ obtained with the time-splitting approach. The vertical dotted line in the three panels represents $\bar{s}$ for the HNN. In the middle-panel we show also a zoom in the region $s\in[15.185,\bar{s}]$.
} \label{fig:Hamilton_subcritical_variables}]
   {\includegraphics[scale=0.65]{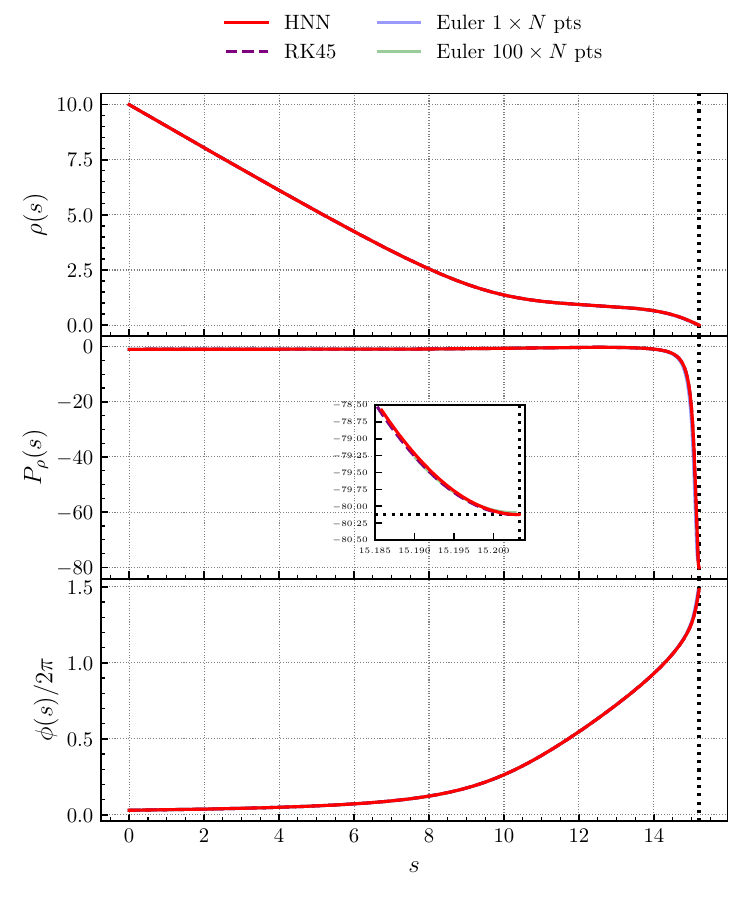}}
   \caption{Energy conservation (on the left) and predicted variables from the different methods (on the right) in the sub-critical case using the time-splitting approach.}
\label{fig:subfig6}
\end{figure}

The conservation of the energy is shown in Figure~\ref{fig:Hamilton_subcritical_H} for the different methods. The HNN conserves very well the energy over the time and an increasing of the oscillations up to $\pm 0.001$ is observed as $s\to \bar{s}$. The Euler method has larger deviations both for $N$ and $100\times N$ points while the \texttt{RK45} method conserves the energy much better compared to the others.\\
The dynamical variables $\rho(s)$, $P_\rho(s)$ and $\phi(s)/2\pi$ obtained from the different methods are shown in Figure~\ref{fig:Hamilton_subcritical_variables}. No visible differences are observed among them. \\

\begin{figure}
\centering
\subfloat[][Errors on the variables $\phi(\rho)$ and $P_\rho(\rho)$ in the sub-critical case with $b=1.89$ obtained from the time-splitting approach. \label{fig:Hamilton_subcritical_errors} ]
   {\includegraphics[scale=0.63]{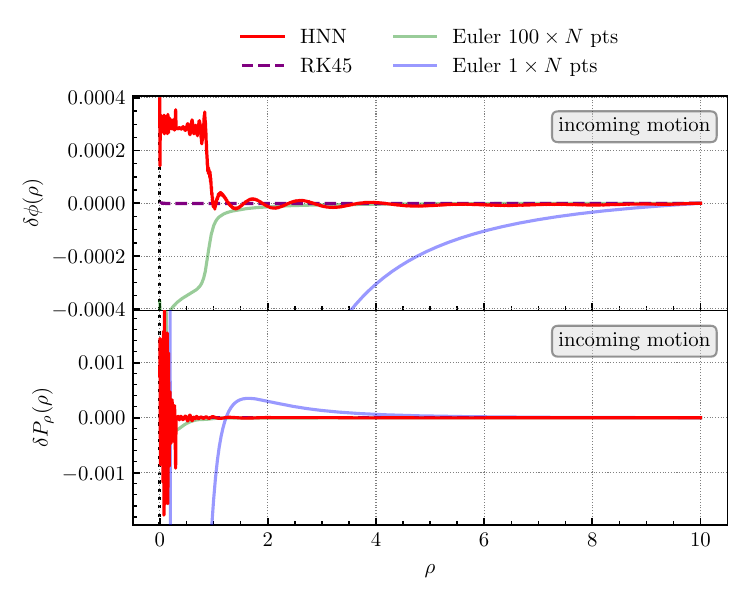}} \quad
\subfloat[][Geodesic in the Cartesian plane obtained from the different methods in the sub-critical case with $b=1.89$ and by applying the time-splitting approach. The ground truth is represented by the solid black line, while the small circle at center (0,0) has radius $a_f$ and it represents the fuzzball. The black arrow indicates the incoming direction of the particle. \label{fig:Hamilton_subcritical_cartesian} ]
   {\includegraphics[scale=0.63]{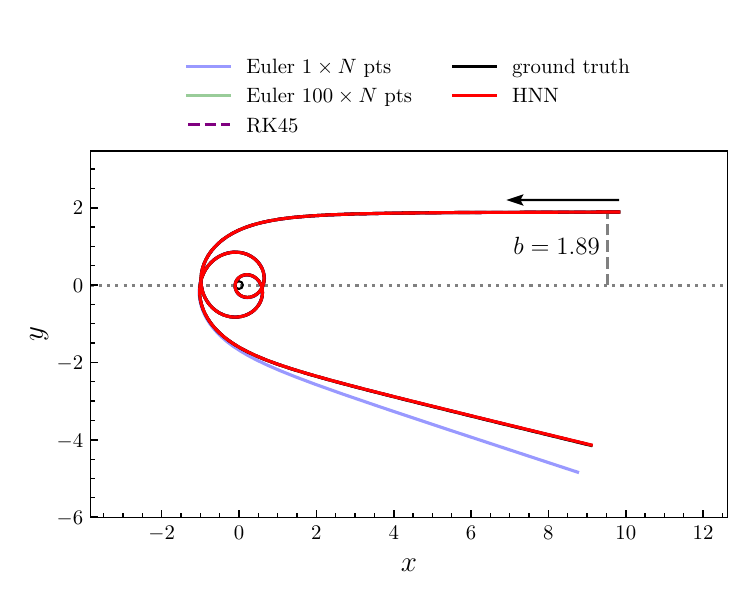}}
   \caption{Errors on the variables $\phi(\rho)$ and $P_\rho(\rho)$ (on the left) and the final geodesic in the Cartesian plane (on the right) in the sub-critical case using the time-splitting approach from the different methods.}
\label{fig:subfig7}
\end{figure}

To quantify the goodness of the solutions, we still rely on the errors as defined in \eqref{eq:phi_error} and \eqref{eq:Prho_error}, where now for the exact $P_\rho(\rho)$ we use \eqref{eq:Prho_exact_subcritical}. The errors are plotted in Figure~\ref{fig:Hamilton_subcritical_errors} for the different methods in the incoming motion (the outgoing motion would be identical, since we obtain it from reflection of the first one). The errors made by the HNN tend to increase for both the variables as $\rho \to 0$, but remain very small compared to the scale of the variables in question, showing that the HNN is able to solve the system with high accuracy. The Euler method with both $N$ and $100\times N$ points show larger deviations, while \texttt{RK45} is very accurate.\\
Figure~\ref{fig:Hamilton_subcritical_cartesian} shows the complete (incoming and ougoing phases) geodesic in the Cartesian plane obtained from the different methods. HNN, Euler with $100\times N$ points and \texttt{RK45} are qualitatively overlapping with the ground truth, while Euler with $N$ points (blue line) deviates significantly from it during the outgoing phase.

\subsubsection{Approach 2: preconditioning of the HNN output}

The second approach we propose consists in providing an offset to the network's output associated with the problematic variable, in this case $P_\rho$, relieving the network from having to learn a particularly stiff behavior. In order to understand better this concept, let us consider the following redefinition 
\begin{flalign}\label{eq:z_hat_precond}
    \widehat{P_{\rho}}(s,\textbf{w}) = P_{\rho,0}+f(s)\big[O_{P_\rho}(s,    \textbf{w})+h(s)\big],
\end{flalign}
where we have introduced the function $h(s)$, representing the new offset with respect to which the output $O_{P_\rho}$ has to be trained to solve the Hamilton equation for $P_\rho(s)$. The function $h(s)$ is chosen arbitrarily and does not depend on the network weights. The idea is to incorporate into $h(s)$ the complexity of the solution $P_\rho(s)$, in such a way that the residual function $O_{P_\rho}(s,\textbf{w})$, that has to be learned by the HNN, is close to zero or has a much smoother behavior. Let us emphasize that minimizing the loss function leads to finding the solution for $P_{\rho}(s)$ regardless of the definition of $h(s)$. If $h(s)$ is completely different from the expected solution, the offset will be automatically corrected by the output $O_{P_\rho}(s,\textbf{w})$. From this perspective, the strategy remains model-independent. We will refer to this approach as \quote{preconditioning} of the HNN output.\\
In order to capture the rapid decrease of $P_{\rho}(s)$ near $\overline{s}$, we have considered the following Breit-Wigner-type parametrization for $h(s)$
\begin{flalign} \label{eq:fit_ansatz}
    h(s)= \dfrac{\big(\overline{P_\rho} -P_{\rho,0}\big)\,\Gamma^{2}}{\Gamma^{2}+(s-\bar{s})^2}.
\end{flalign}

\begin{figure}[H]
    \centering
    \includegraphics[scale=0.8]{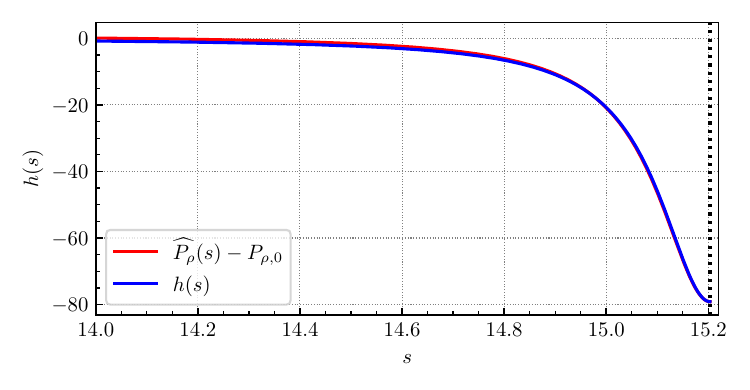}
    \caption{ Result of the fit (blue line) compared to the fitted quantity (red line). In this case $\widehat{P_\rho}(s)$ is the prediction of the HNN obtained from the time-splitting approach in Section \ref{sec:splitting}.
    \label{fig:h}
    }
\end{figure}

With the aim of minimizing the amount of prior information, we have considered the accurate prediction $\widehat{P_\rho}(s)$ obtained from the time-splitting procedure of the previous Subsection, then $\bar s =15.2032$ and fitted the quantity  $\widehat{P_\rho}(s)-P_{\rho,0}$ in the time range $s\in[14,\bar s]$ with the fit ans\"atz \eqref{eq:fit_ansatz}, getting $\Gamma = 0.1213$. In the fitting procedure we have neglected the effect of the function $f(s)$ since we are interested in parameterizing the solution at $s\gg 1$, for which $f(s)\simeq 1$. The result of the fit is drawn in blue in Figure~\ref{fig:h}. As it can be seen, the so-defined $h(s)$ function provides an excellent parametrization for $\widehat{P_\rho}(s)-P_{\rho,0}$ (in red) for times close to $\bar s$. Let us say that in general one can rely on other source of information, if available, such as the result of the numerical integrators, or one can design ansätze by leveraging theoretical knowledge of the system.\\
With this setup, we have trained a neural network made of 3 hidden layers and 64 neurons per each for $30\times 10^6$ epochs.  $\eta_i$, appearing in \eqref{eq:learning_rate_scheduler}, is set to $4\times 10^{-4}$. The time interval is $\mathcal{T}_\mathrm{train}=[0,15.22]$ and it is divided in $N=800$ points. 
In addition we have set $\gamma_{P_\rho}=0.01$ (recall the definition of the dynamics loss function in \eqref{eq:loss_dyn}) in order to give more importance to the dynamical equations for $\rho(s)$ and $\phi(s)$, since the one for $P_\rho(s)$ is basically already solved.\\

\begin{figure}[H]
\centering
\subfloat[][ \footnotesize{Loss functions for the sub-critical case with $b=1.89$ by using the preconditioning approach.
} \label{fig:prec_loss}]
   {\includegraphics[scale=0.65]{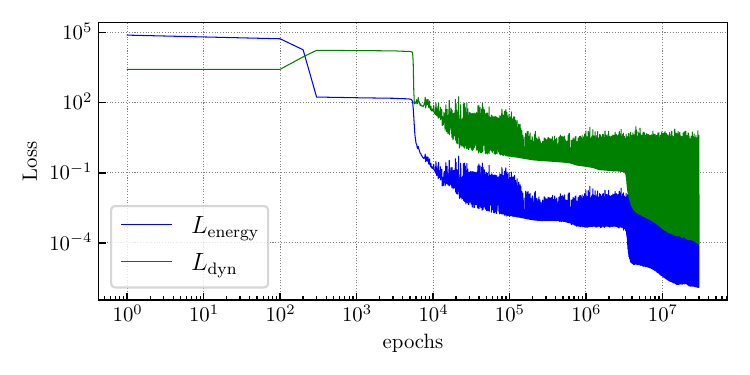}} \quad
\subfloat[][Energy conservation for the sub-critical case with $b=1.89$  for the different methods and by following the preconditioning approach. The conserved value of the energy $\mathcal{H}(0)=0$ is represented by the dotted black line.
\label{fig:prec_H} ]
   {\includegraphics[scale=0.65]{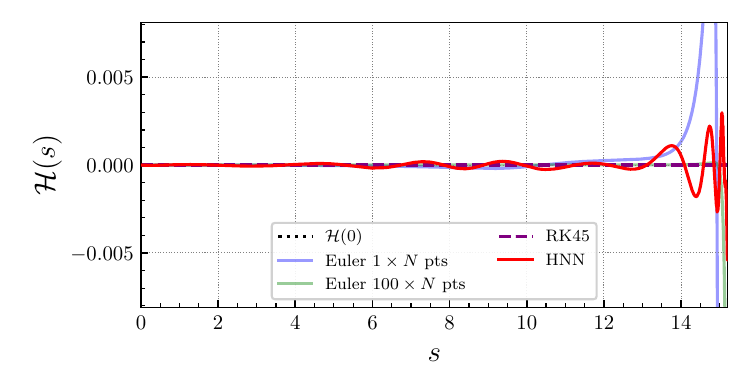}}
   \caption{Loss functions (on the left) and energy conservation (on the right) for the sub-critical case using the preconditioning approach.}
\label{fig:subfig7}
\end{figure}

The loss functions are shown in Figure~\ref{fig:prec_loss} and are at a similar level as those corresponding to the third interval in the time-splitting approach, which are displayed in the bottom-panel of Figure~\ref{fig:Hamilton_subcritical_losses}. \\
The results from the numerical integrators are taken from the previous Subsection and we will not comment them here again. For the HNN predictions we have repeated the interpolation procedure illustrated in Figure~\ref{fig:prediction_interpolation} to determine $\bar s$ (which still remains at the value $15.2032$) and consequently to discard the points for which $s>\bar s$. This procedure is also needed to obtain the outgoing trajectory by reflection of the incoming one. \\
As shown in Figure~\ref{fig:prec_H}, the energy is well conserved on average and, compared to the time-splitting method (see Figure~\ref{fig:Hamilton_subcritical_H}), the deviations from $\mathcal{H}(0)=0$ are slightly larger and reach 0.005 in absolute value for $s\to \bar {s}$. It should be noted that both $L_\mathrm{dyn}$ and $L_\mathrm{energy}$ have not yet reached a plateau, meaning that the accuracy of the solution, and then the conservation of the energy, can be further improved by training the network for more epochs.\\

\begin{figure}[H]
\centering
\subfloat[][ \footnotesize{From top to bottom: $\rho(s)$, $P_\rho(s)$ and $\phi(s)/2\pi$ for the different methods in the sub-critical case with $b=1.89$ and $N=800$ obtained with the preconditioning approach. The vertical dotted line in the three panels represents $\bar{s}$ for the HNN. In the middle-panel we show also a zoom in the region $s\in[15.185,\bar{s}]$.
} \label{fig:prec_variables}]
   {\includegraphics[scale=0.65]{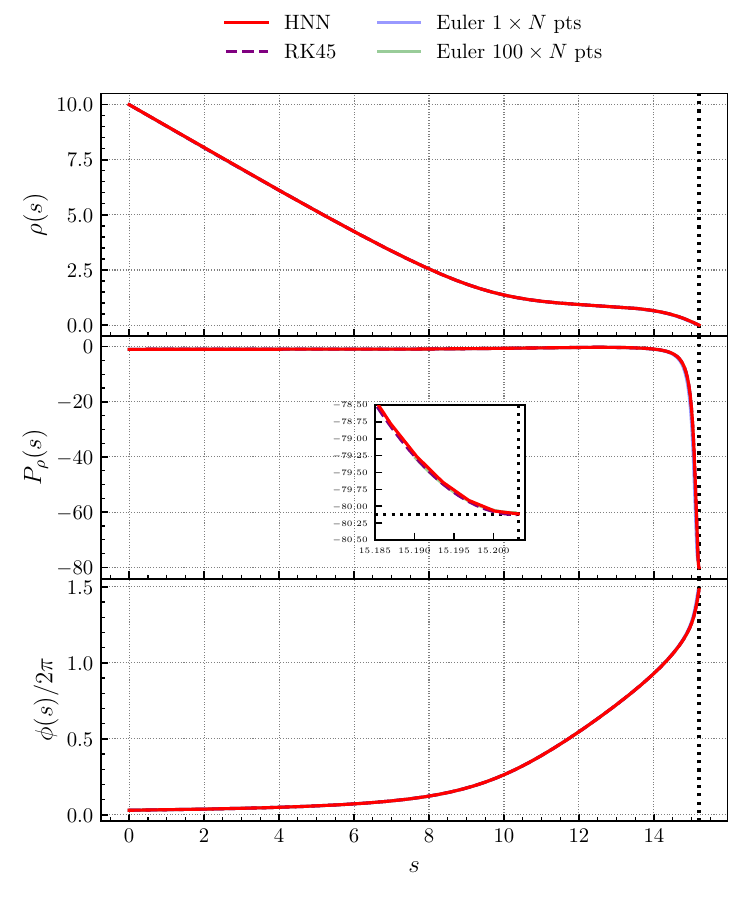}} \quad
\subfloat[][Errors on the variables $\phi(\rho)$ and $P_\rho(\rho)$ in the sub-critical case with $b=1.89$ obtained from the preconditioning approach. \label{fig:prec_errors} ]
   {\includegraphics[scale=0.65]{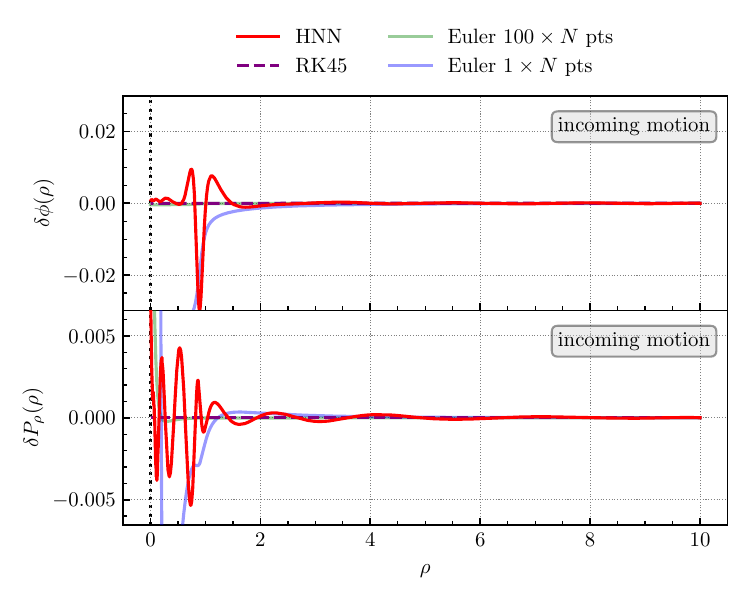}}
   \caption{Predicted variables from the different methods (on the left) and the errors on the variables $\phi(\rho)$ and $P_\rho(\rho)$ (on the right) in the sub-critical case using the preconditioning approach.}
\label{fig:subfig8}
\end{figure}

The predicted variables $\rho(s)$, $P_\rho(s)$ and $\phi(s)/2\pi$ are shown in Figure~\ref{fig:prec_variables} from top to bottom. Analogously to what observed in the time-splitting approach (Figure~\ref{fig:Hamilton_subcritical_variables}), there is no visually appreciable difference between the different methods.\\
In order to make a quantitative statement, we plot the errors $\delta \phi(\rho)$ and $\delta P_\rho( \rho)$ in, respectively, the top- and bottom-panel of Figure~\ref{fig:prec_errors}. For $\phi(\rho)$ we observe errors that are about 100 times larger than than the ones shown in the top panel of Figure~\ref{fig:Hamilton_subcritical_errors}, while for $P_\rho$ we see that the error increases by about a factor of 5. The worse performance compared to before can be improved by increasing the number of epochs, but we emphasize that the error scale remains very small.

\begin{figure}[H]
    \centering
    \includegraphics[scale=0.7]{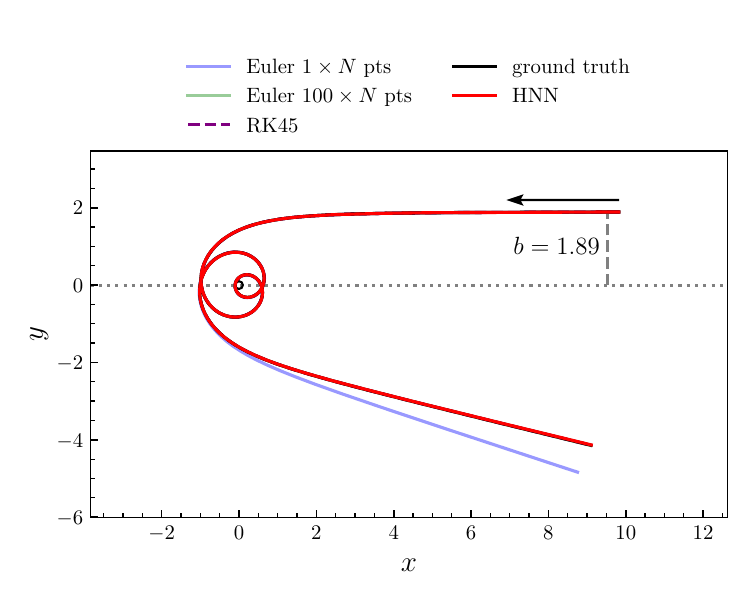}
    \caption{
    \label{fig:prec_cartesian}
Geodesic in the Cartesian plane obtained from different methods in the sub-critical case with $b=1.89$ and by applying the preconditioning approach. The ground truth is represented by the solid black line, while the small circle at the center (0,0) has radius $a_f$ and represents the fuzzball. The black arrow indicates the incoming direction of the particle.}
\end{figure}

Finally, in Figure~\ref{fig:prec_cartesian} we show the predicted geodesic in the Cartesian plane compared to the other methods and to the ground truth. As we can see, the solution predicted by the HNN is still very accurate and in agreement with the ground truth. The strategy proposed here effectively allows incorporating information about the solutions during the training phase while preserving model independence and facilitating the minimization process of the loss function.

\section{\label{results:non_planar}Results for non-planar geodesics}

In this Section we present the results for the non-planar case which is discussed in Section \ref{Non-Planar Case} and whose dynamics is determined by the following five ($J_\phi$ is not considered since it is constant) Hamilton equations

\begin{align}\label{eq:Hamilton_nonplanar}
   \dot \rho = \pdv{\mathcal{H}}{P_\rho} \quad , \quad \dot P_\rho  =-\pdv{\mathcal{H}}{\rho} \quad , \quad \dot \phi = \pdv{\mathcal{H}}{J_\phi} \quad , \quad \dot \theta = \pdv{\mathcal{H}}{P_\theta} \quad , \quad \dot P_\theta = -\pdv{\mathcal{H}}{\theta}. 
\end{align}

The derivatives on the right-hand side are explicitly written in Appendix A of \cite{Cipriani:2025ini}. From there one can immediately appreciate the extreme complexity of such set of equations. Indeed, the non-planar case presents a much greater challenge compared to the planar one, due to the higher number of coupled dynamical variables and the increased complexity of the dynamics.\\ 
As already hinted in Section \ref{Non-Planar Case}, our analysis of non-planar geodesics has been limited to the critical ones, as they are crucial from both an observational and theoretical perspective. Their determination relies on the separability of motion.\\
The geometry parameters that we fix in this analysis are $L_1=0.5$, $L_5=1$, $a_f=0.2$, $E=1$ (from Appendix \ref{APP:nonplangeo} it is clear that $L_1$ and $L_5$ cannot be equal). With this choice, the critical parameters $\zeta_c$, $b_c$ and $\rho_c$ assume the values

\begin{flalign}\label{eq:crit_params_values}
 \zeta_c = -2.5 \qquad , \qquad b_c = 1.5 \qquad , \qquad \rho_c = \sqrt{0.46} = 0.6782,
\end{flalign}
while $b_{\phi,c}=0$ (see Appendix~\ref{APP:nonplangeo}).\\

In order to accommodate the Hamilton equations \eqref{eq:Hamilton_nonplanar}, we need a neural network with 5 output nodes, from which the solutions are assembled according to

\begin{flalign}
\widehat{\rho}(s) = \rho_0  +f(s) & O_\rho(s) \qquad , \qquad \widehat{P_\rho}(s) = P_{\rho,0}+f(s)O_{P_\rho}(s) \qquad , \qquad \widehat{\phi}(s) = \phi_0 +f(s)O_{\phi}(s),
\nonumber\\[6pt]
& \widehat{P_\theta}(s) = P_{\theta,0} +f(s)O_{P_\theta}(s) \qquad , \qquad \widehat{\theta}(s) = \theta_0 +f(s)O_{\theta}(s).    
\end{flalign}

We have chosen the following initial conditions

\begin{equation}\label{eq:init_cond_nonplan1}
  \rho_0 = 3, \qquad \theta_0 = 0.9, \qquad \phi_0 = 1.
\end{equation}

The value $P_{\theta,0}$ is obtained from \eqref{eq:QA_Non_Plan}, in particular we have (recall that $E=1$)
\begin{equation}
    \frac{P^2_{\theta,0}}{E^2} = P^2_{\theta,0} = b^2_c - a_f^2 \sin^2 \theta_0
\end{equation}

from which
\begin{flalign}\label{eq:init_cond_nonplan2}
    P_{\theta,0} = 1.4918.
\end{flalign}

The positive sign of the solution determines the initial direction of the trajectory. Finally, $P_{\rho,0}$ is obtained by requiring that the  Hamiltonian in \eqref{eq:Ham_Non-Plan} vanishes at $s=0$. Using \eqref{eq:crit_params_values}, \eqref{eq:init_cond_nonplan1} and \eqref{eq:init_cond_nonplan2}, we gain
\begin{equation}
P_{\rho,0} = -0.9447.
\end{equation}

Again, we will compare the prediction of the HNN with the \texttt{RK45} method and with the symplectic first-order Euler method. However, in this case, we are unable to derive a ground truth for the geodesic in the Cartesian space and, therefore, we cannot quantify the errors introduced by the different integration methods. To assess the quality of a solution, we can only rely on the expected theoretical behavior of the variables. \\
Since we have chosen the parameters to select the critical trajectory, $\rho(s)$ is expected to decrease from $\rho_0$  until it reaches and settles at the value $\rho_c$.  Consequently, $P_\rho(s)$ is expected to increase from the negative $P_{\rho,0}$ to 0 and remain at that value. Concerning $\theta(s)$ and $\phi(s)$, we expect that, when the particle enters the critical regime with $\rho(s)\simeq \rho_c$ and it wraps around the oblate spheroid \eqref{eq:oblate_spheroid}, they have a  linear grow in time. Consequently, $P_\theta(s)$ in the critical regime should have an oscillating trend with constant period.\\
Out of completeness, we project the oblate spheroid \eqref{eq:oblate_spheroid} onto the three planes $x-y$ ($\theta = \pi/2$), $x-z$ ($\phi = 0$) and $y-z$ ($\phi = \pi/2$). To do this we rely on \eqref{eq:change_coord_3d}, which maps the spherical coordinates $\rho$, $\theta$ and $\phi$ to the Cartesian ones $x$, $y$ and $z$. It turns out that the profile of the geodesic in the plane $x-y$ is given by the circumference

\begin{flalign}
\label{eq:ellipse_xy}
& x^2 + y^2 = \rho^2_c+a_f^2 \qquad x-y \quad \text{plane}\,,
\end{flalign}

and, in the other two planes, by the two following ellipses 

\begin{flalign}
\label{eq:ellipse_xz}
& \frac{x^2}{\rho^2_c+a_f^2} + \frac{z^2}{\rho^2_c} = 1 \qquad x-z \quad \text{plane}\,,
\\[6pt]
\label{eq:ellipse_yz}
& \frac{y^2}{\rho^2_c+a_f^2} + \frac{z^2}{\rho^2_c} = 1 \qquad y-z \quad \text{plane}.
\end{flalign}

This concludes our a priori knowledge of the geodesic, based on physical principles. \\

We now discuss the training phase of the neural network. We have thoroughly investigated various parameter choices and architectures. Below, we present the strategy that, according to our analysis, has proven to be the most efficient. As training time interval we have set the range $\mathcal{T}_\mathrm{train}=[0,60]$ equally sampled with $N=2400$ points. Also in this case, after the training, we have validated the loss function on the interval $\mathcal{T}_\mathrm{val}$ to check the no over-fitting has occurred.\\

The first difference with respect to the planar case is the choice of the neural network architecture. As anticipated, the variable $P_\theta(s)$ is expected to have a periodic trend, in contrast to the monotonic trend of the others. For this reason we have found convenient to consider a two-block architecture with a common input node given by the time. The first block consists in 2 hidden layers with 64 neurons per each and the output $O_{P_\theta}$. For this first block we have used a sinusoidal activation function since, as remarked in \cite{Harvard}, using trigonometric functions is particularly convenient when the output is expected to have a periodic behavior. The second block consists in 2 hidden layers with 128 neurons per each, hyperbolic-tangent activation function and the remaining four outputs  $O_\rho$, $O_{P_\rho}$, $O_\phi$ and $O_\theta$.\\
The second difference is the addition of a penalty term to the loss function  \eqref{eq:total_loss}.  This contribution is introduced to penalize the  solutions such that $P_\rho(s)>0$ and takes the form

\begin{flalign}
    L_{\text{penalty}}(\textbf{w}) = \sum_{s \in \mathcal{S}_p} \bigg[\max(\widehat{P_\rho}(s,\mathbf{w}),0)\bigg]^2 
\end{flalign}

where $\mathcal{S}_p=\{8,20,30,40,50,60\}$. The total loss function actually minimized is therefore

\begin{flalign}
    L(\lambda,\textbf{w}) = (1-\lambda)L_\mathrm{dyn}(\textbf{w})+\lambda L_\mathrm{energy}(\textbf{w})+ L_\mathrm{penalty}(\textbf{w}).
\end{flalign}

The penalty term is activated only for $\widehat{P_\rho}(s,\textbf{w})>0$, while it does not contribute to the loss function when, correcltly, $\widehat{P_\rho}(s,\textbf{w})<0$. We have observed that without this additional term, the HNN tends to predict unstable trajectories. Let us stress that we are using an information from the physics that can be deduced without any prior knowledge of the actual solution. This is a clear example of how various types of information about the solutions can be fruitfully incorporated into the loss function to restrict the solution space and improve the network's predictability. The training is performed by using the Adam optimizer and we monitor individually both $L_\mathrm{dyn}$ and $L_\mathrm{energy}$.\\



\begin{figure}
\centering
\subfloat[][ \footnotesize{Loss functions for the non-planar case.
} \label{fig:nonplanar_loss}]
   {\includegraphics[scale=0.65]{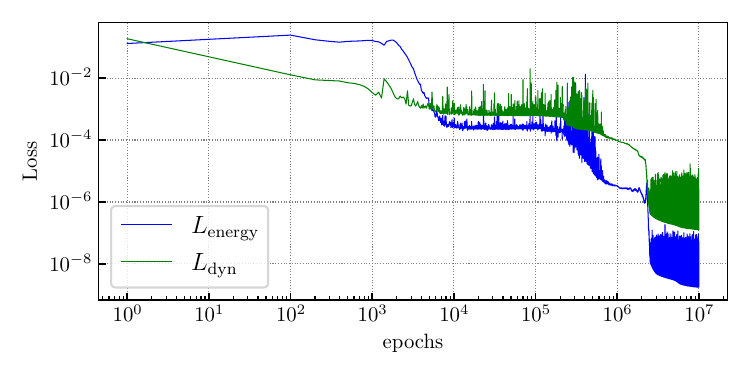}} \quad
\subfloat[][Energy conservation over the time for the non-planar case. The conserved value of the energy $\mathcal{H}(0)=0$ is represented by the dotted black line. 
\label{fig:nonplanar_H} ]
   {\includegraphics[scale=0.65]{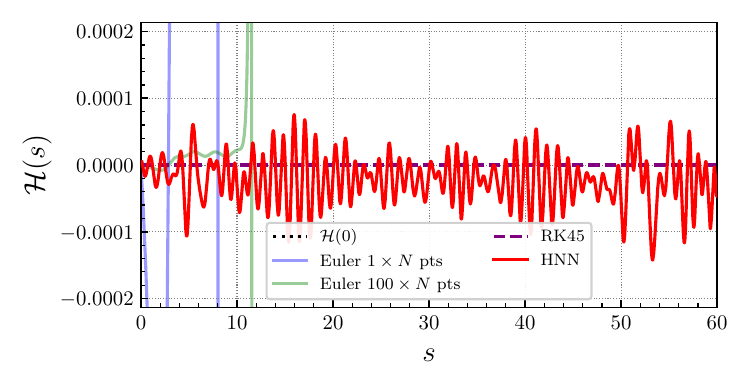}}
   \caption{Loss functions (on the left) and energy conservation (on the right) for the non-planar motion analyzed in the text.}
\label{fig:subfig9}
\end{figure}

\begin{figure}
\centering
\subfloat[][ \footnotesize{From top to bottom: $\rho(s)$, $P_\rho(s)$, $\phi(s)/2\pi$, $P_\theta(s)$ and $\theta(s)/\pi$ for the different methods in the non-planar case. The horizontal black dotted line in the top panel represents the critical radius $\rho_c=0.6782$, on which the trajectory must settle. The trajectory predicted by the HNN is the only one among the different methods to correctly reproduce this feature.}\label{fig:nonplanar_variables} ]
   {\includegraphics[scale=0.82]{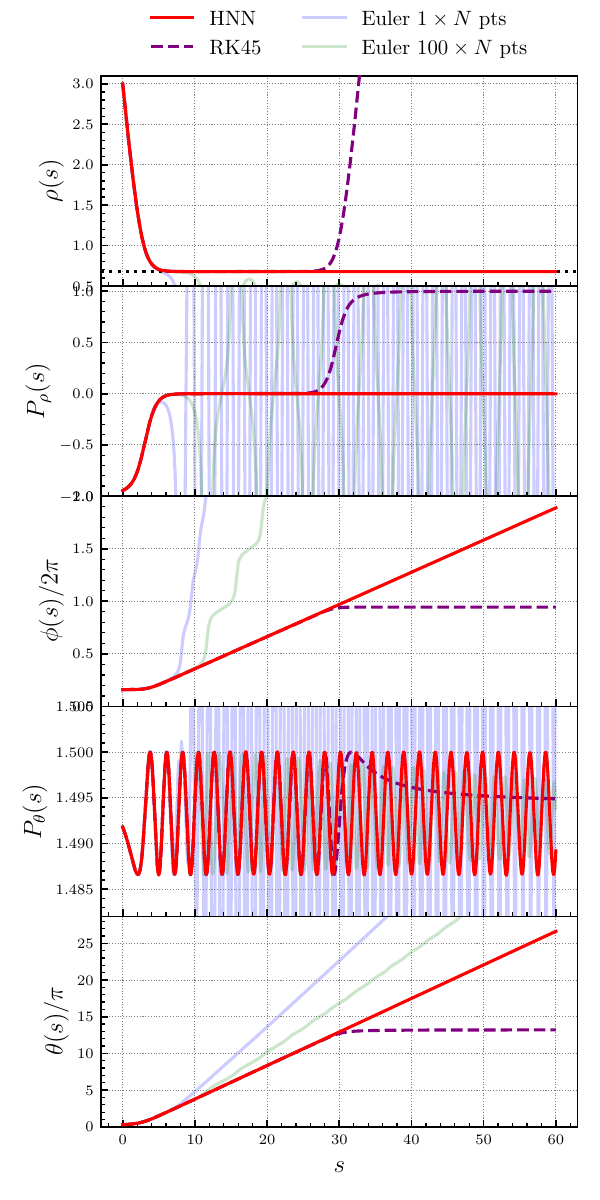}} \quad
\subfloat[][Geodesics in the non-planar case for the different methods, projected onto the, respectively from top to bottom, $x-y$, $x-z$ and $y-z$ Cartesian plane.  The  solid black line represents the fuzzball surface. The dashed light-gray  lines represent the circumference and ellipses defined in  \eqref{eq:ellipse_xy}-\eqref{eq:ellipse_yz}. The black arrow indicates the incoming direction of the particle.\label{fig:nonplanar_cartesian} ]
   {\includegraphics[scale=0.67]{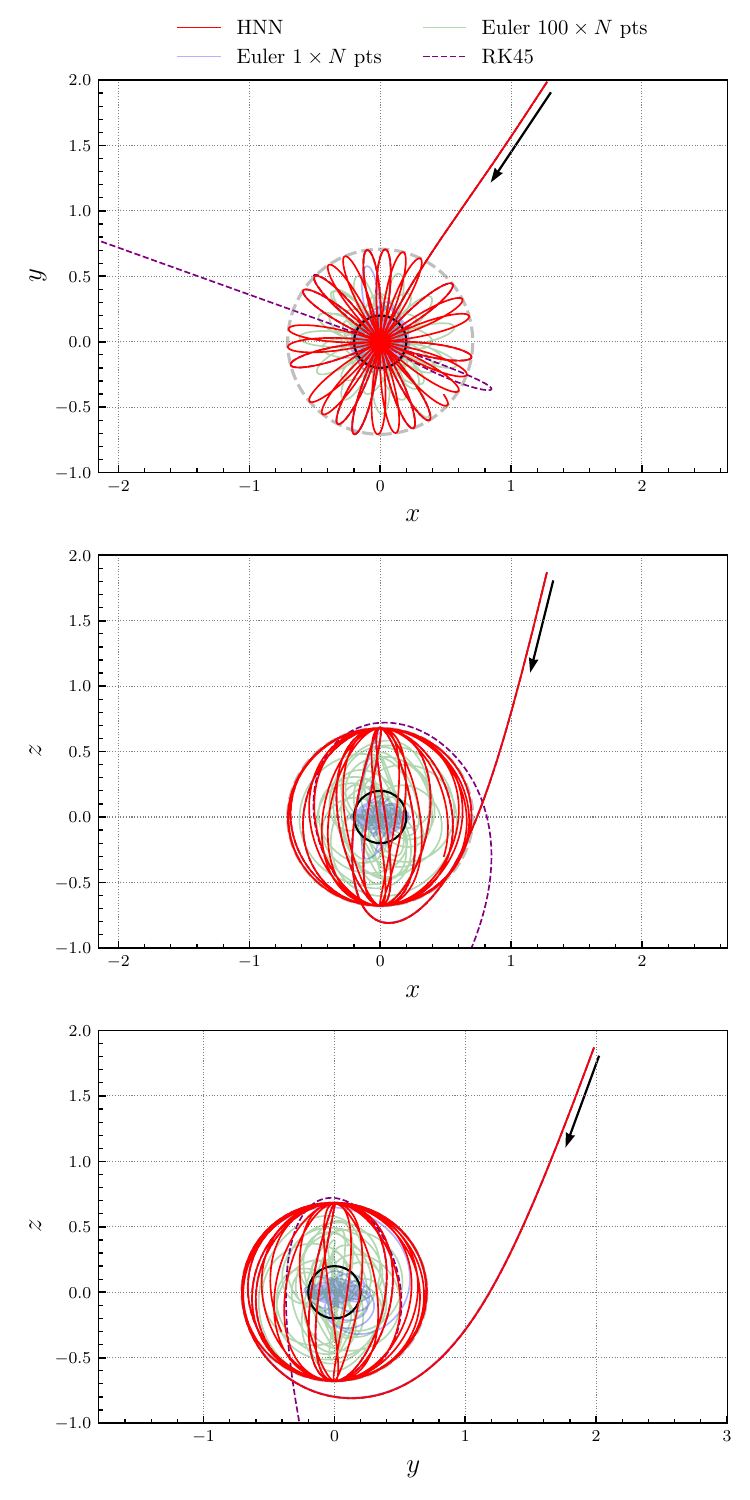}}
   \caption{Final outcomes and predicted trajectories from the different methods in the non-planar case.}
\label{fig:subfig}
\end{figure}

Concerning the remaining algorithmic parameters, the ones entering $\eta(\mathrm{epoch})$ in \eqref{eq:learning_rate_scheduler} are $\eta_i=8\times 10^{-8}$, $\eta_f=\eta_i/50$, $\eta_c=500\times 10^3$ and $\sigma_\eta=100\times 10^3$.  In this case, it turns out that the small value $\lambda=0.1$ gives better performances compared to $\lambda = 0.99$ used in all the cases of the planar study. We interpret this result in light of the fact that the dynamics are much more complex and, therefore, should be prioritized over energy conservation during the minimization process.\\
Having fixed the setup, we have trained the neural network with $10\times 10^6$ epochs and the loss functions are shown in Figure~\ref{fig:nonplanar_loss}. The plot shows a good performance which is comparable to those obtained in the planar case.\\ 
In Figure~\ref{fig:nonplanar_H} we show the energy $\mathcal{H}(s)$ for the different methods. The HNN conserves the energy on average with small fluctuations between $-0.0001$ and $+0.0001$. The Euler methods starts failing in conserving the energy already at small times. The \texttt{RK45} method appears to conserve energy with great accuracy, but, as we have already seen in the planar critical case in Section \ref{sec:planar_critical}, this does not necessarily mean that its numerical solution is the correct one. \\

We realize this by looking at the predicted $\widehat{\rho}(s)$, $\widehat{P_\rho}(s)$, $\widehat{\phi}(s)/2 \pi$, $\widehat{P_\theta}(s)$ and $\widehat{\theta}(s)/\pi$, which are shown, from top to bottom in the given order, in Figure~\ref{fig:nonplanar_variables}. As it can be seen, the different methods yield very different solutions and the one predicted by the HNN is the only one which is in fully agreement with the theoretical expectations discussed above. The critical regime for which $\rho(s)\simeq \rho_c$ is reached around $s=6$ and it is maintained stable for all subsequent times. The prediction of $P_\theta(s)$ as an oscillating function is remarkably good and such a result is due to the use of a trigonometric activation function in the architecture block responsible for the output associated with $P_{\theta}$. The Euler method with both $N$ and $100\times N$ points fails in remaining in critical regime and even predicts $\rho(s)\simeq 0$, which corresponds to a particle which has collapsed onto the fuzzball.  The \texttt{RK45} method is in agreement with the HNN up to $s\simeq 28$ and after this $\rho(s)$ starts increasing, while the angular variables flatten, showing a particle that moves away from the critical trajectory as if it were in a situation analogous to the over-critical case presented in Section \ref{sec:planar_overcritical}.\\

We now convert the spherical coordinates to the Cartesian ones by using \eqref{eq:change_coord_3d}. In Figure~\ref{fig:nonplanar_cartesian} we show the trajectories from the different methods on the $x-y$, $x-z$ and $y-z$ planes, respectively from top to bottom. We also display the curves (colored in gray) given in \eqref{eq:ellipse_xy}, \eqref{eq:ellipse_xz}  and \eqref{eq:ellipse_yz}, which represent the profile of the oblate spheroid. We can appreciate how the geodesic predicted by the HNN, once entered in the critical regime,  exactly describes the profile given by the aforementioned curves in the three Cartesian planes. With this observation we can conclude that the HNN is the only method that, in this challenging situation due to the instability of the critical solution,  solves the Hamilton equations with high accuracy and allows for a solid and stable numerical study of the critical non-planar geodesic of a massless particle immersed in the D1-D5 circular fuzzball. This is the main result of this Chapter and paves the way for future studies on more complex geometries. For instance, as a future outlook a detailed study of geodesic motion in the so-called D1-D5-P (three-charge) fuzzball would be highly interesting. This is a nontrivial generalization of the present D1-D5 (two-charge) fuzzball with an additional Kaluza-Klein (KK) charge. The main complication arises from the fact that, in general, this geometry does not allow for a separation of radial and angular dynamics via the introduction of a Carter-like constant; separation is possible only on equatorial planes \cite{Bianchi:2018kzy}. Furthermore, studying the geodesic motion in this new scenario will help to identify the photon spheres, which serve as a polar star in analyzing the Quasi-Normal Mode (QNM) spectrum \cite{Bianchi:2023rlt,Bianchi:2023sfs,Cipriani:2024ygw}.\\

\begin{figure}
    \centering
    \includegraphics[scale=0.8]{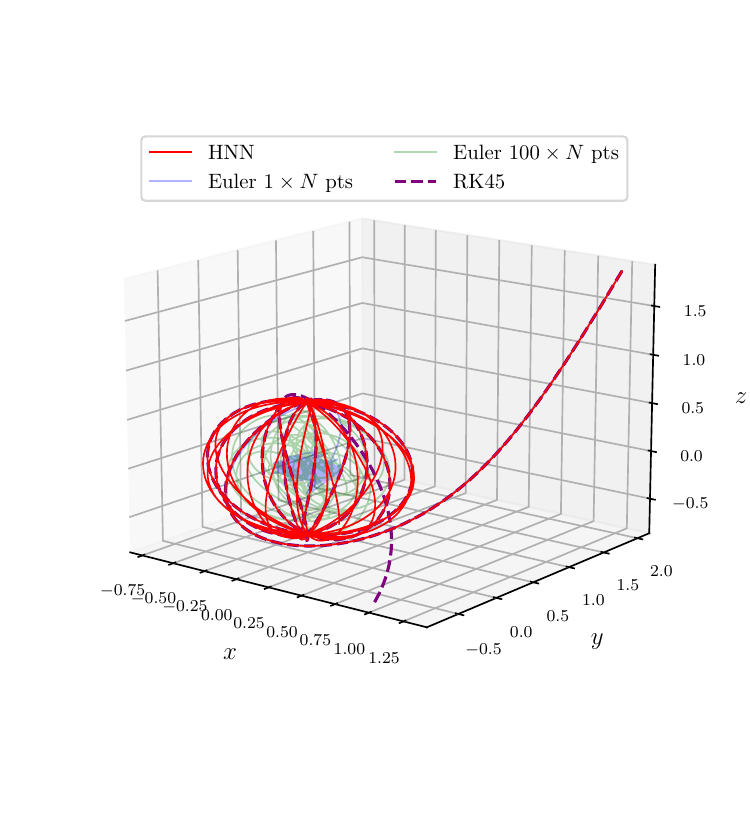}
    \caption{
    \label{fig:nonplanar_3D}
 Geodesic in the 3D Cartesian space for different methods in the non-planar geometry.
    }
\end{figure}

We conclude by showing the geodesics in the 3D Cartesian space in Figure~\ref{fig:nonplanar_3D}. While it does not add anything new to the previous discussion, it still looks beautiful and quite promising.

\chapter{Resummation of Post-Minkowskian and Post-Newtonian Gravitational Waveforms}\label{chap_5}
In Chapter \ref{chap_3} we saw how to probe the dynamical properties of a central object by sending against it a specific kind of perturbation. In Chapter \ref{chap_4}, as a step forward, we determined the trajectory of a massless particle moving inside a specific spacetime, in the approximation of neglecting the emission of gravitational waves. The present final Chapter is thus devoted to analyze the process of emission of gravitational waves when we have an object moving around another one, making an additional step forward. We begin with the simplest situation, i.e. a binary system consisting of a light particle with mass $\mu$ moving on an arbitrary trajectory around a Schwarzschild BH with mass $M$, with $\mu \ll M$ (\textit{probe limit}). We are interested on the inspiral phase, characterized by distances and velocities of the lighter body from the heavier one that are respectively very large ($r \gg 2M$) and very small ($\omega r \ll 1$). The final quantity we focus on is the luminosity, i.e. the amount of energy per unit of time emitted by the system towards infinity.\\
Our investigation method, as already done in Chapter \ref{chap_3}, relies on the correspondence between the equations describing the gravitational processes we are interested on and the SW curves. Indeed the propagation of gravitational waves is described by the so called \textit{Teukolsky equation}, which, in the case of a Schwarzschild background, is a Confluent Heun Equation (CHE). As already seen in Chapters \ref{chap_1} and \ref{chap_3}, a CHE can be viewed as the quantum version of the SW curve. In this scenario, this curve governs the low energy dynamics of a $\mathcal{N}=2$ supersymmetric gauge theory with gauge group SU$(2)^2$ and couplings $z = 2M/r$ and $y = 2 i \omega r$. Notice how the inspiral phase corresponds precisely to the weak coupling limit of this gauge theory and since the solution of the CHE (and so of the Teukolsky one) is mapped to the instanton partition function of the same theory, we derive the solution of the CHE from first principles. The aforementioned correspondence is also called \textit{SW correspondence}.\\
The main result of this Chapter, based on \cite{Cipriani:2025ikx}, is the so called \textit{hypergeometric representation} of the Confluent Heun function, which allows to express the homogeneous solutions of the Teukolsky equation in terms of a single hypergeometric function and its first derivative. From them we can determine the inhomogenous solution and then the luminosity. After the verification that the new method provides good results, as it can be seen from the comparison against numerical outcomes, we then move to estimate the benefit of using it in the framework of the factorization of the waveform, as already done in \cite{Damour:2008gu,Ivanov:2025ozg}. This kind of analysis is deeply performed in \cite{Cipriani:2026xmx} and in this Chapter we are going to present the most important aspects, even the extension to binary systems with comparable masses.\\
The plan of the Chapter is the following: in Section \ref{sec:intro_Chap5} we present the scenario in which our study is placed; in Section \ref{sec2:Teukolsky} we quickly move to recall some basic notions in order to arrive at the Teukolsky equation, whose properties (in terms of nature of the singularities) are examined; in Section \ref{sec2} we derive the PM and PN expansions of the confluent Heun function, that is directly related to the homogeneous ingoing and outgoing solutions of the Teukolsky equation. In Section \ref{sec3} we apply the results of Section \ref{sec2} to determine the inhomogenous solution of the same equation, with the aim of studying the emitted gravitational waveforms. In Section \ref{sec4} we test our results against those obtained via the MST method and numerical integration. Section \ref{sec:factorization} contains the treatment of the factorization of the waveforms and also the extension of our results to binary systems with comparable masses.

\section{General scenario} \label{sec:intro_Chap5}
 Recent advances in the sensitivity of the experimental apparatus for detecting gravitational waves (GWs) necessitate increasingly precise theoretical computations. The next generation of space-based experiments, such as the Laser Interferometer Space Antenna (LISA), promises to revolutionize the GW detection by significantly enhancing sensitivity and exploring entirely new wavelength ranges compared to ground-based interferometers \cite{LISA:2022kgy}. LISA will detect long-wavelength GWs originating from sources like supermassive black hole mergers, extreme mass-ratio inspirals (EMRIs), galactic binaries, and the Stochastic Gravitational Wave Background—sources largely inaccessible to ground-based observatories.
The gravitational signals from these sources are diverse. For instance, GWs from supermassive black hole mergers will produce strong signals with large signal-to-noise ratios, while those from EMRIs, though weaker, will carry exceptionally detailed information about the gravitational sources.
Indeed,  EMRIs waveforms, spanning thousands of cycles during their inspiral phase, provide a unique opportunity to probe fundamental questions about the nature of gravitational sources, including their compactness, the presence or absence of horizons, light rings, echoes \cite{Cardoso:2019rvt},
and the validity of Einstein's theory of gravity and its possible extensions \cite{Gair:2012nm,Yunes:2013dva}.\\

Two main analytical tools have been developed for studying the problem we just described: the multipole expansion \cite{RevModPhys.52.299,Damour:1982wm,Blanchet:2013haa} and perturbation theory \cite{1973ApJ...185..635T,PhysRevD.47.1497,PhysRevD.49.4016,Poisson:1994yf,Shibata:1994jx,Tagoshi:1994sm,Tanaka:1996lfd,Mano:1996vt,Mino:1997bx,Suzuki:1998vy,Fujita:2010xj,Fujita:2011zk}. The latter, valid in the probe limit of the binary, the so called EMRI systems, is the focus of this Chapter. In this framework, GWs are computed by solving the Einstein's equations to linear order in the perturbations induced by the motion of a lighter body orbiting around a very massive black hole (BH). The radial and angular propagations of the waves are governed by ordinary differential equations of the confluent Heun type. Analytic solutions of these  equations are not known, but it is possible to write them as an expansion in the Newton constant  (Post-Minkowskian, or PM, approximation) or in the orbital velocity (Post-Newtonian, or PN, approximation).\\
In the case of circular orbits, the PM and PN approximations are linked and characterised by a single expansion parameter $v$ (the tangential velocity).
 The current state-of-the-art for the emitted energy in perturbation theory  is the 22nd PN (order $v^{44}$) computation in \cite{Fujita:2012cm}. The expansion  is known to be converging slowly \cite{PhysRevD.47.1497,PhysRevD.47.1511} due to the presence of logarithmic, leading to an infinite tower of tails and tails of tails non-local interactions    \cite{Damour:2008gu,Damour:2009kr,Buonanno:2009zt,Damour:2012ky}, and transcendental terms.
In \cite{Aminov:2020yma,Bianchi:2021xpr,Bianchi:2021mft,Bonelli:2021uvf,Bonelli:2022ten,Consoli:2022eey,Bautista:2023sdf,Aminov:2023jve,Ikeda:2021uvc,Fucito:2023afe} novel techniques, capitalizing on recent advances in quantum field theories, have been exploited to  derive an exact formula for the  Heun connection  matrix and alternative formulations of the waveforms, in the probe limit, were proposed in terms of the instanton partition function of a SU(2)$^2$ quiver gauge theory. Following this path the infinite tower of logarithmic terms have been resummed into exponentials.\\

More recently, multipole methods \cite{Bini:2023fiz,Bini:2024ijq,Bini:2024rsy}, scattering amplitude techniques \cite{Jakobsen:2021smu,Mougiakakos:2021ckm,Cristofoli:2021vyo,DiVecchia:2023frv,Brandhuber:2023hhy,Herderschee:2023fxh,Elkhidir:2023dco,Georgoudis:2023lgf,DeAngelis:2023lvf,Brandhuber:2023hhl,Georgoudis:2023eke,Georgoudis:2023ozp,Georgoudis:2024pdz,Brunello:2024ibk,Alessio:2024onn} and black hole perturbation theory \cite{Fucito:2024wlg} have been applied to the study of gravitational scattering processes, fostering a fruitful interplay and mutual cross-checking among these complementary approaches. Within the quantum field theory framework, the PM series describes the expansion in loops (weak coupling) of the scattering amplitude, keeping finite the relative velocity of the binary.
Using perturbation theory, we will show that at any fixed order in the gravitational coupling, the entire series of PN corrections in the probe limit of the gravitational waveform can be  resummed and expressed in terms of a single hypergeometric function and its first derivative. Analogous results hold for the PN expansion itself, where the exact dependence on the  gravitational coupling corrections  is determined at any given order in the small-velocity limit.\\
The PM and PN series provide novel hypergeometric representations of the confluent Heun function along the lines of the well studied MST solutions \cite{Mano:1996vt}. The main difference between the two is that the MST ans\"atz involves an infinite sum of hypergeometric functions, while the solution we put forward here is written in terms of a single hypergeometric function and its first derivative. The crucial advantage of this representation is that it provides a simple mathematical proof of the Heun connection formulae originally found in \cite{Aminov:2020yma,Bianchi:2021xpr, Bonelli:2021uvf,Bonelli:2022ten}. The connection matrix coefficients are given as ratios of gamma functions with arguments depending on a single
(but highly non-trivial) function codifying the holonomies of the Confluent Heun function around the irregular singular point.  This function, known as the "renormalized angular momentum" in the MST context \cite{PhysRevD.49.4016, Poisson:1994yf,Shibata:1994jx,Tagoshi:1994sm,Tanaka:1996lfd,Mano:1996vt,Mino:1997bx,Suzuki:1998vy,Fujita:2010xj,Fujita:2011zk},
     can be related to the so called Seiberg-Witten period of a ${\cal N}=2$ supersymmetric gauge theory \cite{Consoli:2022eey,Bautista:2023sdf,Aminov:2023jve,Ikeda:2021uvc,Fucito:2023afe} or
  the "anomalous dimension" in the effective field theory description of the gravitational wave \cite{Goldberger:2009qd,Trestini:2023wwg,Almeida:2021jyt,Edison:2023qvg,Ivanov:2025ozg}.
    The Heun connection formula has been recently applied in a variety of physical contexts, including the study of QNMs for gravity backgrounds (BHs, fuzzballs, D-branes), correlators in thermal field theories, holography, cosmological perturbations and scattering amplitudes in gravity \cite{Bianchi:2021xpr,Bianchi:2021mft,Fioravanti:2021dce,Bianchi:2022wku,Dodelson:2022eiz,Dodelson:2022yvn,Imaizumi:2022qbi,Fioravanti:2022bqf,Imaizumi:2022dgj,Bianchi:2022qph,Dodelson:2023vrw,Bianchi:2023rlt,Bianchi:2023sfs,Giusto:2023awo,Aminov:2023jve,BarraganAmado:2023apy,Lei:2023mqx,BarraganAmado:2023wxt,Fucito:2023afe,Bautista:2023sdf,Ge:2024jdx,Cipriani:2024ygw,Arnaudo:2024rhv,Bianchi:2024vmi,Bianchi:2024mlq,DiRusso:2024hmd,Bianchi:2024rod,Aminov:2024aan,BarraganAmado:2024tfu,Bautista:2024emt,Liu:2024eut,Silva:2025khf,Aprile:2025hlt,Bhatta:2025kil}.\\

To keep the material self-contained, in this Chapter we present a derivation of the Heun solutions (and its confluences) from first principles  with no too much explicit reference to supersymmetric gauge theory or localization, which sit in the background as inspiration to some of our ans\"atze, that nonetheless could be also accepted as such. Finally, we remark that even if the main motivation for resumming PM and PN series relies on the study of open orbits, our hypergeometric representations provide an algorithm which is very efficient when implemented in a Mathematica code to describe the PN expansions of bounded orbits. Indeed, here we compute the gravitational waveform and the energy flux, in the case of circular orbits, up to 30 PN orders. This because the present method is based on a recursive scheme in determining the final solution. The results are tested against the computation at 22 PN orders available in literature \cite{Fujita:2012cm}  and  the numerical methods from the  Black Hole Perturbation Toolkit (BHTK) \cite{BHPToolkit} based on the MST approach.\\
After that we have tested the goodness of the hypergeometric representations, we compare these results with the ones already existing in literature \cite{Damour:2008gu,Ivanov:2025ozg} in terms of factorizing the waveform in the frequency domain. The primary goal is clearly seeing how we resum all the tail and transcendental contributions in a way which is new with respect to the two existing versions; this allows to enhance the convergence properties of the PN expansion of the waveforms, which produces better results in the evaluation of the total energy flux.

\section{Determination and properties of Teukolsky equation} \label{sec2:Teukolsky}
In this Section we start by recalling all the ingredients that are needed in order to determine the equation that governs the propagation of BH perturbations and, in particular, of GWs, the so called \textit{Teukolsky equation}, whose explicit derivation is reported in Appendix \ref{app:Teuk}. After having written this equation, we are going to study the nature of its singularities together with the behavior of the solution close to these ones in the homogeneous case, in order to pave the way to the new analysis carried out in \cite{Cipriani:2025ikx} for solving it. We are going to use the mostly plus signature for a 4-dimensional spacetime $(-+++)$.

\subsection{The Newman-Penrose formalism}
A useful way to arrive at the determination of the aforementioned equation is considering the components of the \textit{Weyl tensor} and then projecting it along a null tetrad. This formalism was first proposed by Newman and Penrose \cite{Newman:1961qr} and we are going to quickly follow the presentation of \cite{Maggiore:2018sht}, with slight different choices as variables' names and normalizations. In particular we need to recall the notions of null tetrads and the Weyl tensor.\\

Given a 4-dimensional spacetime with metric $g_{\mu \nu}$, a Newman-Penrose null tetrad is a set of four linearly independent four-vectors
\be
q^\mu_{a} \equiv \left(l^\mu, n^\mu, m^\mu, \bar{m}^\mu\right)
\ee
where $a=1,\dots,4$ labels the four-vector. In addition, $l^\mu$ and $n^\mu$ have to be real, while $m^\mu$ is complex, with $\bar{m}^\mu$ the complex conjugate of $m^\mu$. These four-vectors are chosen to be null with respect to the metric $g_{\mu \nu}$, i.e.
\be
g_{\mu \nu} l^\mu l^\nu = g_{\mu \nu} n^\mu n^\nu = g_{\mu \nu} m^\mu m^\nu = g_{\mu \nu} \bar{m}^\mu \bar{m}^\nu = 0
\ee
Furthermore,
\be
g_{\mu \nu} m^\mu \bar{m}^\nu = 1 \qquad, \qquad g_{\mu \nu} l^\mu n^\nu = -1
\ee
In other words
\be \label{eq:orthogonality_relations}
q_a \cdot q_b \equiv q_a^\mu g_{\mu \nu} q^\nu_b = \begin{pmatrix}
0 & -1 & 0 & 0 \\
-1 & 0 & 0 & 0 \\
0 & 0 & 0 & 1 \\
0 & 0 & 1 & 1
\end{pmatrix}
\ee
From all of this we learn that the metric can be written in the following way in terms of these four-vectors
\be
g^{\mu \nu} = m^\mu \bar{m}^\nu + m^\nu \bar{m}^\mu - l^\mu q^\nu - l^\nu q^\mu 
\ee
For example, by using this relation and \eqref{eq:orthogonality_relations}, in can be shown that $g^{\mu \nu}l_\nu = l^\mu$ and similarly for all the other vectors. Later we will see that any four-vector or tensor can be projected onto the null tetrad, by contracting the spacetime indices. For example, for what concerns the Riemann tensor we obtain $R_{\mu \nu \rho \sigma} q^\mu_a q^\nu_b q^\rho_c q^\sigma_d$.\\

The basic quantities of the Newman-Penrose formalism are some projections of the Weyl tensor onto the null tetrad and for this reason it could be useful to recall the main properties of this tensor. In the same 4-dimensional spacetime, the Weyl tensor is defined as
\be
C_{\mu \nu \rho \sigma} = R_{\mu \nu \rho \sigma} - \frac{1}{2} \left(g_{\mu \rho} R_{\nu \sigma} - g_{\mu \sigma} R_{\nu \rho} - g_{\nu \rho} R_{\mu \sigma} + g_{\nu \sigma} R_{\mu \rho}\right) + \frac{1}{6} R \left(g_{\mu \rho} g_{\nu \sigma} - g_{\mu \sigma} g_{\nu \rho}\right)
\ee
where the Riemann and Ricci tensors, together with the curvature scalar, are
\begin{align}
& R^{\mu}_{\, \nu \alpha \beta} = \partial_\alpha \Gamma^\mu_{\nu \beta} - \partial_\beta \Gamma^\mu_{\nu \alpha} +  \Gamma^\mu_{\rho \alpha} \Gamma^\rho_{\beta \nu} - \Gamma^\mu_{\rho \beta} \Gamma^\rho_{\alpha \nu} \notag \\
& R_{\mu \nu} = R^{\alpha}_{\, \mu \alpha \nu} \qquad , \qquad R = g^{\mu \nu} R_{\mu \nu}
\end{align}
with the Christoffel's symbols given by
\be
\Gamma^\mu_{\alpha \beta} = \frac{1}{2} g^{\mu \nu} \left(\partial_\alpha g_{\nu \beta} + \partial_\beta g_{\nu \alpha} - \partial_\nu g_{\alpha \beta}\right)
\ee
Notice that $C_{\mu \nu \rho \sigma}$ vanishes with a contraction of any pair of its indices, hence it is the trace-free part of the Riemann tensor. From its expression, it can be seen that the Weyl tensor enjoys the same symmetries as the Riemann tensor, i.e. antisymmetry with respect to the exchange of the first or of the second pair of indices and symmetry with respect to the exchange of the first pair with the second one. Furthermore, also the cyclicity with respect to the last three indices
\be
C_{\mu \nu \rho \sigma} + C_{\mu \sigma \nu \rho} + C_{\mu \rho \sigma \nu} = 0
\ee
All these properties, together with the vanishing of all its traces, reduce the number of independent components of such tensor to 10, in four dimensions.\\

At this point, projecting the Weyl tensor onto a given null tetrad allows us to define several Newman-Penrose quantities, also known as Weyl scalars. We are interested, in this Chapter, only to the following one
\be \label{eq:psi_4}
\Psi_4 = C_{\mu \nu \rho \sigma} n^\mu \bar{m}^\nu n^\rho \bar{m}^\sigma 
\ee
since this will be useful for the determination of the gravitational waveform, which is the quantity we are going to be mainly interested on during this Chapter, since from it we can define the total energy carried by the GW and measured at infinity.

\subsection{Schwarzschild background geometry and its linear perturbations}
As we said in the previous Section, we want to analyze the GW emitted by a particle, with mass $\mu$, moving around a Schwarzschild BH. Hence we want to recall the ingredients of this geometry that will be important later. The metric is, in the usual Schwarzschild coordinates $(t,r,\theta,\phi)$
\be \label{eq:Schwarzschild_metric}
\mathrm{d} s^2 = -f(r) \mathrm{d} t^2 + \frac{\mathrm{d} r^2}{f(r)} + r^2 \mathrm{d} \theta^2 + r^2 \sin^2\theta \mathrm{d} \phi^2
\ee
where $f(r) = 1- \frac{2 M}{r}$ and $M = G M_{\text{BH}}$, where $M_{\text{BH}}$ is the mass of the BH. We will mainly consider $ \mu \ll M_{\text{BH}}$, except in the last Section where we see how our method (born in BH Perturbation Theory) can be potentially extended to the case of a binary system with comparable mass.\\
The null tetrad that we consider is the \textit{Kinnersley tetrad}, which, in the Schwarzschild components is \footnote{Observe that these expressions match with (12.314)-(12.315)-(12-316) of \cite{Maggiore:2018sht}, but there we have to set the spin of Kerr BH equal to 0.}
\begin{align} \label{eq:kinnersly_tetrad}
& l^\mu = \frac{1}{f(r)} \partial_t + \partial_r = \left(\frac{1}{f(r)}, 1, 0, 0\right)^T \qquad , \qquad n^\mu = \frac{1}{2} \left(\partial_t - f(r) \partial_r \right) = \left(\frac{1}{2}, -\frac{f(r)}{2}, 0, 0\right)^T \notag \\
& m^\mu = \frac{1}{\sqrt{2} r}\left( \partial_\theta + \frac{i}{\sin \theta}\partial_\phi \right)= \left(0, 0, \frac{1}{\sqrt{2} r}, \frac{i}{\sqrt{2} r \sin \theta}\right)^T \qquad , \qquad \bar{m}^\mu = \left(0, 0, \frac{1}{\sqrt{2} r}, -\frac{i}{\sqrt{2} r \sin \theta}\right)^T
\end{align}
It can be seen that this tetrad satisfies the conditions \eqref{eq:orthogonality_relations} and also that
\be
R_{\mu \nu \alpha \beta} n^\mu \bar{m}^\nu n^\alpha = R_{\mu \nu \alpha \beta} n^\mu \bar{m}^\nu \bar{m}^\beta = 0
\ee
This implies that, on Schwarzschild spacetime, the scalar $\Psi_4$ is identically vanishing.\\

Let us pass to consider the perturbation of the Schwarzschild's metric, generated by the probe with mass $\mu \ll M_{\text{BH}}$. In this way the quantity $\Psi_4$ will not be trivial anymore and our goal is determining it. In general, we indicate the full metric as
\be
\textbf{g}_{\alpha \beta} = g_{\alpha \beta} + h_{\alpha \beta}
\ee
where $g_{\alpha \beta}$ is the background Scwarzschild's metric \eqref{eq:Schwarzschild_metric} and $h_{\alpha \beta}$ is the perturbation. Similarly, we have also the perturbations to the Riemann and Weyl tensors, in such a way that
\be
\textbf{R}_{\alpha \beta \gamma \delta} = R_{\alpha \beta \gamma \delta} + \delta R_{\alpha \beta \gamma \delta} + \dots \qquad \textbf{C}_{\alpha \beta \gamma \delta} = C_{\alpha \beta \gamma \delta} + \delta C_{\alpha \beta \gamma \delta} + \dots 
\ee
We are considering all the quantities corrected by the perturbation and by the mass $\mu$ of the probe only at the linear order. In principle, we should introduce a similar notation also for the tetrad and then for the Newman-Penrose scalars, but since $\Psi_4 = 0$, then we simply have
\be
\mathbf{\Psi}_4 = \delta C_{\mu \nu \rho \sigma} n^\mu \bar{m}^\nu n^\rho \bar{m}^\sigma + \dots
\ee
where $l^\mu, n^\mu, m^\mu, \bar{m}^\mu$ are the vectors of the tetrad of background.\\
The Einstein tensor is given by
\be
\textbf{G}_{\alpha \beta} = \textbf{R}_{\alpha \beta} - \frac{1}{2} \textbf{g}_{\alpha \beta} \textbf{R}
\ee
from which we write the following Einstein equations
\be
\textbf{G}_{\alpha \beta} = 8 \pi G \textbf{T}_{\alpha \beta}
\ee
Since the only source is the light particle, the components of the (full) stress energy tensor are
\be \label{eq:stress_components_original}
\textbf{T}^{\alpha \beta}(x) = \frac{\mu}{\sqrt{- \det \textbf{g}(x)}} \int \frac{\delta^{(4)} (x-x(\lambda) \dot{x}^\alpha(\lambda) \dot{x}^\beta (\lambda)}{\sqrt{- \dot{x}^\rho(\lambda) \textbf{g}_{\rho \sigma}(x(\lambda)) \dot{x}^\sigma(\lambda)}} \mathrm{d} \lambda
\ee
At the leading order in the perturbation induced by the light object the Einstein equations reduce to (recall that the background Scwarzschild metric is solution of the same equations in the vacuum)
\be \label{eq:linearized_Einstein_equations}
\delta G_{\alpha \beta} = 8 \pi G T_{\alpha \beta}
\ee
where the indices of $T$ are raised and lowered with the background metric. In order to understand what are $T^{\alpha \beta}$, since we are restricting to the linear order in the perturbation $\mu$ and due to the fact that \eqref{eq:stress_components_original} contains $\mu$ in front of the integral, we can approximate the full metric $\textbf{g}$ appearing there with the background one. At this point, thanks to the invariance of the integral under reparametrization of the affine parameter $\lambda$, we can fix it to the proper time of the particle with respect the background, $\lambda = \tau$. In this way we obtain that
\be \label{eq:relation_proper_time}
\dot{x}^\mu (\tau) g_{\mu \nu} (x(\tau)) \dot{x}^\nu (\tau) = -1
\ee
In this way
\be \label{eq:stress_components_appr}
T^{\alpha \beta}(x) = \frac{\mu}{\sqrt{- \det g(x)}} \int \delta^{(4)} (x-x(\tau)) \dot{x}^\alpha (\tau) \dot{x}^\beta (\tau) \mathrm{d} \tau 
\ee

\subsection{Teukolsky equation and its properties} \label{sec:general_Teuk_discussions}
In order to arrive to the Teukolsky equation, we need to decompose the Weyl scalar $\mathbf{\Psi}_4$ in modes
\be  \label{eq:mode_dec_psi4}
r^4 \mathbf{\Psi}_4 = \int \frac{\mathrm{d} \omega}{2 \pi} \underbrace{\sum_{\ell =2}^\infty \sum_{m = 1}^\ell}_{\equiv \sum_{\ell, m}} e^{-i \omega t} R_{\ell m \omega} (r) Y^{\ell m}_{-2} (\theta, \phi)
\ee
$\omega$ is the wavefrequency, $(t,r,\theta,\phi) \equiv x$ are the generic spacetime coordinates, eventually associated to the observer's ones, and $Y^{\ell m}_{-2}$ are the $-2$-spin-weighted spherical harmonics
\begin{equation*}
Y^{\ell m}_s (\theta) = e^{i m \phi} \, \sin^{2 \ell} \left(\frac{\theta}{2}\right) \sqrt{\frac{(2 \ell +1)(\ell-m)! (\ell+m)!}{4 \pi (\ell-s)! (\ell+s)!}} \sum_{r=0}^{\ell-s} (-)^{\ell+m+r-s} \begin{pmatrix}
\ell -s \\
r
\end{pmatrix} \begin{pmatrix}
\ell + s \\
r+s-m
\end{pmatrix} \cot^{2 r+s-m} \left(\frac{\theta}{2}\right)
\end{equation*}
All the passages that lead to the Teukolsky equation are reported in Appendix \ref{app:Teuk} and, specializing to the case of gravitational perturbations, we have the following ordinary second order differential equation
\be \label{eq:teuk}
{\Delta (r)^2 } \frac{\mathrm{d}}{\mathrm{d}r} \left[ \frac{1}{\Delta (r)} R_{\ell m \omega}'(r) \right]  + \left(\frac{\omega^2 r^4 + 4 i (r - M)  \omega r^2}{\Delta (r)} - 8 i \omega r - (\ell + 2)(\ell - 1) \right)R_{\ell m \omega}(r)  = T_{\ell m \omega}(r)
\ee
where $\Delta(r) = r(r-2M)$ and the source is defined as
\be
T_{\ell m \omega}(r) = 2 \pi G \int \mathrm{d} t \, \sin \theta \, \mathrm{d} \theta \, \mathrm{d} \phi \, e^{i \omega t} \, Y^{l m *}_{-2} \, \mathcal{E}[T]
\ee
(the reason of it is clear from Appendix \ref{app:Teuk}). $T$ is \eqref{eq:stress_components_appr} and $\mathcal{E}$ is an operator whose action on a symmetric tensor $X_{\alpha \beta}$ is given by
\begin{align*}
\mathcal{E} [X] & = 2 r^4 L_{-1} L_0 X_{n n} + \sqrt{2} r^3 f(r)^2 J_+ \left(\frac{r^2}{f(r)} L_{-1} X_{n \bar{m}}\right) + \sqrt{2} r f(r)^2 J_+ \left(\frac{r^4}{f(r)} L_{-1} X_{n \bar{m}}\right) + \notag \\
& + r f(r)^2 J_+ \left[r^4 J_+ \left(r X_{\bar{m} \bar{m}}\right)\right]
\end{align*}
where
\begin{equation*}
X_{n n} = X_{\alpha \beta} \, n^\alpha n^\beta \quad , \quad X_{n \bar{m}} = X_{\alpha \beta} \, n^\alpha \bar{m}^\beta \quad , \quad X_{\bar{m} \bar{m}} = X_{\alpha \beta} \, \bar{m}^\alpha \bar{m}^\beta
\end{equation*}
and the differential operators $L_s$ and $J_\pm$ are defined as (in the usual Schwarzschild coordinates)
\begin{equation*}
L_s = \partial_\theta - \frac{i}{\sin \theta} \partial_\phi + s \cot \theta \qquad , \qquad J_\pm = \partial_r \mp \frac{1}{f(r)} \partial_t 
\end{equation*}

A way for finding the solution of the Teukolsky equation \eqref{eq:teuk} is defining a suitable Green function. In particular, by denoting as $R_{\text{in},\omega}(r)$ and $R_{\text{out},\omega}(r)$ the two solutions of the \textit{homogeneous} equation satisfying ingoing and outgoing boundary conditions at the horizon and at infinity respectively (we are going to analyze them in the following Section), we can construct a constant with respect to $r$ out of the Wronskian between these two solutions in the following way
\be \label{eq:constant_W}
W = \frac{R_{\text{in},\omega}(r) \, \partial_r \, R_{\text{out},\omega}(r) - R_{\text{out},\omega}(r) \, \partial_r \, R_{\text{in},\omega}(r)}{\Delta(r)}
\ee
($\partial_r W =0$ as it follows using the equation \eqref{eq:teuk} itself). At this point, the function
\be \label{eq:Green_func}
G_\omega (r,r')=\frac{1}{W} \left\{
\begin{array}{ccc}
R_{\text{in},\omega} (r')   R_{\text{out},\omega} (r)  & ~~~~~~~&  r'< r\\
R_{\text{in},\omega} (r)   R_{\text{out}, \omega} (r')  & ~~~~~~~ &   r< r' \\
\end{array}
\right.
\ee
satisfies the following relation
\be
\mathcal{O}^\ell_{-2 \omega}(r) \left[G_{\omega}(r,r')\right] = \Delta(r')^2 \delta(r-r')
\ee
and so it is a Green function. In this way a solution of the equation we are studying is
\be
R_{\ell m \omega} (r) = \int_{2M}^\infty G_\omega (r,r') \, \Delta(r')^{-2} T_{\ell m \omega} (r')
\ee
From this explanation we see that we have first of all to see how to compute the solutions of the homogeneous equation. This is done in the following Section, but now it is quite useful to inspect some properties of the homogeneous Teukolsky equation

\subsubsection{Singularities and behaviors of the homogeneous solution} \label{sec:asymptotic_behaviors}
We indicate the general solution of the homogeneous equation, without specifying the appropriate boundary condition, as $R_{\ell m \omega}^{\text{hom}}(r)$.\\
Given an ordinary linear differential equation of $n$-th order in the complex plane
\be
f^{(n)} (z) + \sum_{i=0}^{n-1} t_i(z) f^{(i)}(z) = 0
\ee
with $t_i(z)$ meromorphic functions, the Frobenius method can be applied to find solutions as power series times $(z-z_0)^\delta$ near any given $z_0$ in the complex plane where $\delta$ is not necessarily an integer. This presents no difficulty for $z_0$ an ordinary point or also a \textit{regular singular point}, which means that the function $t_{n-i}(z)$ for each $i$ has a pole of order at most $i$ at $z_0$. In this way the Frobenius method furnishes $n$ independent solutions for $\delta$ near $z_0$. If this is not the situation, $z_0$ is said to be an \textit{irregular singular point} and the behavior of the solution of the equation near this kind of points has to be determined in a different way.\\
Specializing to a second order differential equation, the point $z=z_0$ is a regular singular point if $t_1(z)$ and $t_0(z)$ have, respectively, a pole in $z_0$ of order at most 1 and 2. In other words, writing the equation as
\be \label{eq:general_equation_Frobenius}
f''(z) + \frac{p(z)}{z-z_0} f'(z) + \frac{q(z)}{(z-z_0)^2} f(z) = 0
\ee
the functions $p(z)$ and $q(z)$ have to be analytic in $z=z_0$. Assuming this is the case, the Frobenius method leads to the following equation for the index $\delta$
\be \label{eq:indicial_equation}
\delta (\delta-1) + p(z_0) \delta + q(z_0) = 0
\ee
and it is called \textit{indicial equation}, which is, as expected, of the second order in $\delta$.\\

Let us apply all this explanation to our Teukolsky equation \eqref{eq:teuk}. Writing it in the form \eqref{eq:general_equation_Frobenius} we have
\be
R_{\ell m \omega}^{\text{hom}''}(r) + \frac{2 (M-r)}{r(r-2M)} R_{\ell m \omega}^{\text{hom}'}(r) + \frac{\left[2 M \left(\ell^2+\ell-2+6 i \omega r \right) + r \left(2-\ell-\ell^2-4 i \omega r + \omega^2 r^2 \right)\right]}{r(r-2M)^2} R_{\ell m \omega}^{\text{hom}}(r) = 0
\ee
and we clearly see that $r_0=0,2M$ are regular singular points. Since $r=0$ is inside the horizon, we do not care about it and so we apply the Frobenius method near the point $r=2M$. In reference to \eqref{eq:general_equation_Frobenius}, the functions $p(r)$ and $q(r)$ are
\be
p(r) = \frac{2 (M-r)}{r} \quad , \quad q(r) = \frac{\left[2 M \left(\ell^2+\ell-2+6 i \omega r \right) + r \left(2-\ell-\ell^2-4 i \omega r + \omega^2 r^2 \right)\right]}{r}
\ee
and hence the indicial equation gives rise to
\be
\delta(\delta-1) + p(2M) \delta + q(2M) = \delta(\delta-1) - \delta + 4 M \omega (i + M \omega) = 0 \Rightarrow \delta= \{2 i M \omega, 2 - 2 i M \omega\}
\ee
In this way the behavior of the general solution of the homogeneous equation near $r=2M$ is
\be \label{eq:behavior_horizon}
R_{\ell m \omega}^{\text{hom}}(r) \underset{r \to 2 M}{\simeq} A_{\text{hor}} (r-2 M)^{2-2 i M \omega} + B_{\text{hor}} (r-2M)^{2 i M \omega}
\ee
where $A_{\text{hor}},B_{\text{hor}}$ are constants.\\
In order to determine the nature of the point $r=\infty$, we change variable $r = w^{-1}$ in \eqref{eq:teuk}, getting at the end
\be
R_{\ell m \omega}^{\text{hom}''}(w) + \frac{4-6 M w}{w (1-2 M w)} R_{\ell m \omega}^{\text{hom}'}(w) + \frac{\left[2 M w^3 \left(\ell^2+\ell-2 \right) - w^2 \left(-2+\ell+\ell^2-12 i M \omega\right) - 4 i w \omega + \omega^2 \right]}{w^4(1-2 M w)^2} R_{\ell m \omega}^{\text{hom}}(w) = 0
\ee
and we easily read that $t=0$ is an irregular singular point. The way that we follow here in order to determine the behavior of the solution near $r=\infty$ is applying the WKB approximation, already quickly encountered in Chapter \ref{chap_3} and that now we want to recall, in such a way that we can compute the aforementioned behavior.\\
Given a second-order homogeneous differential equation in the Schr\"odinger form \cite{Bender:1999box}
\be
\epsilon^2 \frac{\mathrm{d}^2 g(z)}{\mathrm{d} z^2} + Q(z) g(z) = 0
\ee
with $\epsilon \ll 1$ a small parameter, the WKB method consists in looking a solution in the form
\be
g(z) = e^{\frac{1}{\eta} \sum_{n=0}^\infty \eta^n S_n(z)}
\ee
where $\eta \to 0$ (called \textit{semiclassical limit}; this parameter can be thought as $\hbar$ in Quantum Mechanics). In this way plugging this ansatz in the previous equation we get
\begin{align}
& \epsilon^2 \left[\frac{1}{\eta^2} \left(\sum_{n=0}^\infty \eta^n S'_n \right)^2 + \frac{1}{\eta} \sum_{n=0}^\infty \eta^n S''_n \right]= - Q(z) \Leftrightarrow \notag \\
& \frac{\epsilon^2}{\eta^2} S'^2_0 + \frac{\epsilon^2}{\eta} \left(2 S'_0 S'_1 + S''_0 \right) + \mathcal{O}\left(\epsilon^2 \eta^0\right) = -Q(z)
\end{align}
In the limit of $\eta \to 0$ the dominant term is 
\be
\frac{\epsilon^2}{\eta^2} S'^2_0 \simeq - Q(z)
\ee
This implies that $\eta$ is proportional to $\epsilon$ and setting them equal, we obtain that, at order $\epsilon^0$ (the leading one)
\be \label{eq:WKB_condition}
S'^2_0 = -Q(z) \Rightarrow S_0 = \pm i \int \sqrt{Q(z')} \mathrm{d} z'
\ee
If then we pass to the first-order correction in $\epsilon$ we get
\be
2 S'_0 S'_1 + S''_0 = 0
\ee
By the way, in the following we are going to be interested only on the leading order behavior of our solution, hence we determine only $S_0$ according to \eqref{eq:WKB_condition}.\\
Coming back to our Teukolsky equation, we need first of all to put in the canonical form as described in Chapter \ref{chap_1}, so we introduce the function $\varphi(r)$ such that
\be \label{eq:R_vs_varphi}
R_{\ell m \omega}^{\text{hom}}(r) = \sqrt{r} \sqrt{r-2M} \varphi(r)
\ee
and the equation becomes
\be
\varphi''(r) + \underbrace{\frac{2 M r \left(\ell + \ell^2 + 6 i \omega r\right) + r^2 \left(-\ell -\ell^2 + (\omega r)^2 - 4 i \omega r \right)}{r^2 (r-2M)^2}}_{Q(r)} \varphi(r) = 0
\ee
Since we are interested to the limit of $r \to \infty$, then we expand $\sqrt{Q(r)}$ for large $r$, obtaining
\be
\sqrt{Q(r)} \underset{r \to \infty}{\simeq} \omega + \frac{2 (M \omega-i)}{r} + \mathcal{O}\left(r^{-2}\right)
\ee
in such a way that, from \eqref{eq:WKB_condition}
\be
S_0(r) = \pm i \left(\omega r + 2 (M \omega - i ) \log (r) \right) \Rightarrow \varphi(r) \underset{r \to \infty}{\simeq} e^{\pm i (\omega r + 2 (M \omega -i) \log (r)} = r^{\pm 2} e^{\pm i \omega (r + 2 M \log(r))}
\ee
We introduce the tortoise coordinate $r_* = r + 2 M \log \left(\frac{r}{2M}-1\right)$ and, being in the limit $r \to \infty$, we have $r_* \simeq r + 2M \log(r) - 2 M \log(2M)$ and the last term is a constant. So we can write
\be
\varphi(r) \underset{r \to \infty}{\simeq} r^{\pm 2} e^{\pm i \omega r_*}
\ee
We keep in mind that we want the behavior of $R_{\ell m \omega}^{\text{hom}}(r)$, related to $\varphi(r)$ through \eqref{eq:R_vs_varphi}. This relation, in the limit of $r \to \infty$, implies that $R_{\ell m \omega}^{\text{hom}}(r) \underset{r \to \infty}{\simeq} r \varphi(r)$, leading to the following behavior of the general solution of the homogeneous equation near $r=\infty$

\be \label{eq:behavior_infinity}
R_{\ell m \omega}^{\text{hom}}(r) \underset{r \to \infty}{\simeq} A_\infty r^3 e^{i \omega r_*} + \frac{B_\infty}{r} e^{- i \omega r_*}
\ee
where $A_\infty, B_\infty$ are other constants.

\section{The homogeneous equation}\label{sec2}
In this Section we show a new method to solve the homogeneous Teukolsky equation, based on the results of \cite{Cipriani:2025ikx}. Having realized that our equation admits two regular singular points and one irregular one, it can be recast as a Confluent Heun equation (CHE), according to what has been explained at the end of Chapter \ref{chap_1}. In order to see how to do it, let us recall the more general CHE defined in mathematical literature.\\
It is an ODE of the second order of the form
\be
 \text{HeunC}''(t)+ \left( \frac{\beta_0}{t}+\frac{\beta_1}{t-1} +\lambda\right) \text{HeunC}'(t) + \frac{\alpha_1 t -\alpha_0}{t(t-1)} \text{HeunC}(t)  =0 \label{CHE0}
\ee
with two regular singularities at $t=0,1$ and an irregular one at infinity. $\{\alpha_0,\alpha_1,\beta_0,\beta_1,\lambda\}$ are complex parameters that specify the equation. A basis of solutions is given by
 \beaq
&&  {\rm HeunC} [\alpha_0,\alpha_1,\beta_0,\beta_1 ,\lambda, t] \\
 &&  t^{1-\beta_0 } \text{HeunC}[ \alpha_0+(1-\beta_0)(\lambda-\beta_1),\alpha_1 +\lambda(1-\beta_0 ) ,2-\beta_0 ,\beta_1 ,\lambda ,t] \nn
\eeaq
  with ${\rm HeunC}$ the confluent Heun function.

\subsection{Confluent Heun-Gauge-Gravity dictionaries}
As already said many times, the CHE can be viewed as the quantum version of the SW curve governing the dynamics of a $\mathcal{N}=2$ supersymmetric SU(2) gauge theory with three flavours of masses $m_i$,  transforming in the fundamental representation of the gauge group \cite{Bianchi:2021mft,Bianchi:2021xpr} living on a curved Nekrasov-Shatashvili background \cite{Nekrasov:2009rc}. The quantum curve can be written in the form\footnote{Equation (\ref{CHE}) can be recast in the normal Schr\"odinger like form
   \be
\Psi''(z) + Q_{\rm gauge}(z)\Psi =0 \qquad , \qquad G_1(z)=e^{{\frac{x}{2 z}}} z^{m_3-1}(1-z)^{m_1+m_2-1\over 2} \Psi(z) \label{normal}
\ee
 with  	
 \be
 Q_{\rm gauge} = {-}\frac{x^2}{4 z^4}{+}\frac{x m_3}{z^3}{+}\frac{1{-} \left(m_1{-}m_2\right)^2}{4(z{-}1)z}+\frac{ 1{-}
 		\left(m_1{+}m_2\right){}^2}{4(z-1)^2 z}  +\frac{u{-}\ft{1}{4}{+}\ft{x}{2}
 		\left(m_1{+}m_2{-}1\right)}{(z{-}1)
 		z^2} \label{qgauge}
\ee
   }
\be \label{CHE}
G_1''(z) + G_1'(z) \left(\frac{1 - m_1 - m_2}{z - 1} - \frac{2 \left(m_3 - 1\right)}{z} + \frac{x}{z^2}\right) + G_1(z) \left(\frac{\frac{1}{4} - u + m_3(m_3 - 1)}{z^2} + \frac{u + \frac{3}{4} + \delta}{(z - 1) z}\right)=0
 \ee
with $\delta = m_1 m_2 + m_1 m_3 + m_2 m_3 - m_1 - m_2 - m_3$, $x$ the gauge coupling and $u$ the Coulomb branch parameter. Notice that the complex parameters $\{\alpha_0,\alpha_1,\beta_0,\beta_1,\lambda\}$ that characterize \eqref{CHE0} correspond now to $\{m_1, m_2, m_3, u, x\}$. In particular, \eqref{CHE0} and \eqref{CHE} are related by the Confluent Heun-Gauge dictionary
  \beaq
  G_1(z)  &=& z^{-\frac{\alpha_1}{\lambda}}   \text{HeunC}\left(z^{-1}\right) \qquad , \qquad   t = z^{-1} \qquad, \qquad \lambda = -x \nn\\
   \alpha _0 &=& u-\frac{1}{4} +m_1(1-m_1) +(m_1+m_3-1) x  \qquad , \qquad \alpha_1= \left(m_1+m_3-1\right) x \nn\\
     \beta _0&=& 1 -m_1+m_2 \qquad , \qquad \beta_1=1 -m_1-m_2
  \eeaq
As we can see, the advantage of this dictionary, and hence of working with the equation \eqref{CHE}, is that the actual variable $z$ varies inside a finite interval: since we need the irregular singular point $t=\infty$ and the following regular one which is $t=1$, then in this way $z \in (0,1)$.\\
In order to understand how to recast the homogeneous Teukolsky equation \eqref{eq:teuk} (with $T_{\ell m \omega}(r)=0$) into \eqref{CHE}, we need first of all to find the suitable change of variable between $r$ and $z$ that preserves the nature of the singularities. It turns out to be
\be
r = \frac{2M}{z}
\ee
At this point, applying this change of variable to the homogeneous version of \eqref{eq:teuk} and writing in the canonical form the final equation, we obtain a given characteristic function $Q_{\text{Teuk}}(\omega, M, \ell, s; z)$ (we are keeping the spin generic now, so look at \eqref{eq:teuk_general_s}; at the end we set $s=-2$). Comparing it with $Q_{\text{CHE}}(m_1, m_2, m_3, x, u; z)$ that comes out when writing in the Schr\"odinger form \eqref{CHE}, we obtain the following Gauge-Gravity dictionary

\begin{align}
& x=4 i M \omega \quad , \quad m_1 =  2 i M \omega = \frac{x}{2} \quad ,\quad  m_2=m_3^*= s + 2 i M \omega = s + \frac{x}{2} \notag \\
& u= \left(\ell + \frac{1}{2}\right)^2 + 2 i M \omega - 4 i M \omega s - 4 M^2 \omega^2 = \left(\ell + \frac{1}{2} \right)^2 + \frac{x}{2} - x s + \frac{x^2}{4}
\end{align}
and the relation between the two solutions $R_{\ell m \omega}^{\text{hom}}(z)$ and $G_1(z)$ is
\be
R_{\ell m \omega}^{\text{hom}}(z) = e^{-\frac{x}{2z}} (1-z)^{-\frac{s+m_1+m_2}{2}} \left(\frac{x}{z}\right)^{m_3-1-s} G_1(z)
\ee
Specializing to the case with $s=-2$ we have

\begin{align}
R_{\ell m \omega}^{\text{hom}}(z) &=  e^{-\frac{x}{2z}} (1-z)^{1-\frac{m_1+m_2}{2}} \left(\frac{x}{z}\right)^{1+m_3} G_1(z)  \qquad , \quad z= \frac{2M}{r}  \qquad, \qquad x=4 i M \omega \label{rdict} \\
m_1 &=  \frac{x}{2} \quad ,\quad  m_2=m_3^*=-2+  \frac{x}{2}   \quad , \quad  u=\left(\ell+\frac{1}{2}\right)^2+\frac{5}{2}x +\frac{x^2}{4}\nn
\end{align}

So, once we have determined $G_1(z)$, we also have the homogeneous solution of the Teukolsky equation and for this reason we are going to analyze the former.\\
In light of the SW correspondence we explained in Section \ref{sec:intro_Chap5}, there are two limits under which the CHE \eqref{CHE} admits simple solutions in terms of hypergeometric functions. These correspond in taking or $z \ll1$ keeping $x/z$ finite or the opposite, i.e. $x/z \ll 1$ with $z$ finite. While the latter limit can be directly applied to equation \eqref{CHE} simply by setting $x=0$ (at the leading order), the former is more subtle. A easier way to study this limit is writing the \textit{same} equation \eqref{CHE} in terms of the variable
\be
y \equiv \frac{x}{z} = 2 i \omega r
\ee
in such a way that the leading order behavior of the limit $z \ll 1$ with $y$ finite corresponds in setting $x=0$ in the following equation
\be \label{eq:CHE_y}
 G_0''(y) + G_0'(y) \left(\frac{1- m_1 - m_2}{y - x} + \frac{ m_1 + m_2 + 2m_3 - 1}{y} - 1\right) + G_0(y) \left(\frac{u+\frac{3}{4} + \delta}{(x-y) y} + \frac{1 + \delta + m_3(m_3 - 1)}{y^2}\right)=0
 \ee
The relation between the functions $G_0(y)$ and $G_1(x/y)$ is
\be
G_1\left(\frac{x}{y}\right) = g(x) G_0(y)
\ee
with $g(x)$ an arbitrary constant that will be determined later.\\
 We will use both descriptions (\ref{CHE}) and (\ref{eq:CHE_y})  of the CHE. Notice that the two limits cited before both imply that $x$ has to be small and it is called the soft (or large wavelength) limit, which will be always assumed in all our procedure. In the gravity context, the two limits correspond to the PM (when $z=2M/r \ll 1$) and PN (when $y = 2 i \omega r \ll 1$) expansions of the Confluent Heun solution, so we will refer to $G_0$ and $G_1$ as the PM and PN representations of the confluent Heun function respectively.
In particular, sending $x\to 0$,  \eqref{eq:CHE_y} and (\ref{CHE}) reduce to hypergeometric equations with solutions (at the leading order in $x$, i.e. $x^0$)
 \beaq \label{eq:leading_order_G0_G1}
{\rm PM} : && \qquad G^0_\alpha(y)   \underset{x\to 0\atop y~{\rm finite}}{\approx}  H^0_\alpha(y)  \nn\\
{\rm PN} : && \qquad G^1_\alpha(z) \underset{x\to 0\atop z~{\rm finite}}{\approx}  H^1_\alpha(z)   \label{lim12}
 \eeaq
   with $\alpha=\pm$ labelling the two independent solutions,
\beaq \label{h012}
H^0_\alpha(y) &=& y^{\frac{1}{2}-\alpha a-m_3} \,
   _1F_1\left(\ft{1}{2}-\alpha a-m_3;1-2\alpha
   a;y\right) \nn\\
H^1_\alpha(z) &=& z^{-\frac{1}{2}+\alpha a+m_3} \,
   _2F_1\left(\frac{1}{2}+\alpha a-m_1,\frac{1}{2}+\alpha a-m_2;1+2\alpha
   a;z\right)
\eeaq
and  $a \underset{x\to 0}{\approx} \sqrt{u}$.  On the other hand, at order $k$ in $x$, \footnote{This is the only expansion that will appear in our formulae. With an abuse of notation we will define our results a PN expansion at order $k$ even if other contributions to our final formulae will be given in closed form.} we will look for solutions of the form
 \beaq
{\rm PM}_k : && \qquad G^0_\alpha(y)  = y^{\frac{1}{2}-\alpha a-m_3 } \sum_{i=0}^k \sum_{j=0}^\infty  c^{PM}_{ij}( \alpha a) x^i  y^{j-i} \nn  \label{defsigma}\\
{\rm PN}_k : && \qquad G^1_\alpha(z)  = z^{-\frac{1}{2}+\alpha a+m_3 }\sum_{j=0}^k\sum_{i=0}^\infty c^{PN}_{ij}( \alpha a) z^{i-j} x^{j}
 \eeaq
 where $a(u)$ is a function of $u$. Writing
\be \label{eq:u_vs_a_recursion}
u = a^2+\sum_{i=1}^k u_i(a) x^i
\ee
one can determine the coefficients $c^{PM}_{ij}(a)$, $c^{PN}_{ij}(a)$ and $u_i(a)$  by solving the CHE equations  order by order in $x$. If we set $k=0$ and then we apply this procedure, at the end we find that $G^0_\alpha$ and $G^1_\alpha$ reduce to \eqref{eq:leading_order_G0_G1}.\\
We notice that $a$ codifies the monodromy of the solutions around the irregular singular point $z=0$ ($y=\infty$). Moreover,  since $G^1_\alpha(z)$ and $ G^0_\alpha(\ft{x}{z})$ are solutions of the same equation and have the same monodromy around the irregular singular point, they should be proportional to each other
\be
G^1_\alpha(z) = g_{\alpha}(x) G^0_\alpha(\ft{x}{z})
\ee
There is a region where the two limits overlap and it corresponds to the fact that $x \ll z \ll 1$, in such a way that also $y \ll 1$. This is called \textit{near zone} and the form of the solutions as the series \eqref{defsigma} is natural within such a region. The main result of this Section, as we are going to show, is that the series PM$_k$  and PN$_k$  in \eqref{defsigma} can be exactly resummed and written in terms of a single hypergeometric function and its first derivative even in the regions where one between $z$ or $y$ is not small anymore. This will provide us with a first principle proof of the connection formulae for the Confluent Heun functions, where the whole non-triviality of the CHE is codified into the single function $a(u)$.\\

For concreteness, as already said above, we will focus on the domain $z\in (0, 1)$. The general solution $G_1(z)=\sum_\alpha c_\alpha G^1_\alpha(z)$  behaves near the boundaries as
 \beaq
  G_1(z) &\underset{z\to 0}{\approx}&  B_{-} (1+\ldots) +B_{+}   z^{ 2 m_3}  e^{ x\over z}  (1+\ldots) \nn\\
  G_1(z) &\underset{z\to 1}{\approx}&  D_{-}(1+\ldots)  +D_{+} (1-z)^{m_1+m_2} (1+\ldots) \, \, \, \, \, \, \, \, \,  \label{gbcgen0}
\eeaq
where the dots stand for subleading terms in these limits. Specifying boundary conditions corresponds to fix two linear combinations out of the four coefficients $B_{\pm}$,  $D_{\pm}$. We will derive connection formulae relating the coefficients $B_\pm$ to  $D_\pm$ and viceversa.

\subsection{PM approximation: exterior region}

In this Subsection we consider the limit $x\ll 1$ keeping $y$ finite (so also $z \ll 1$). This corresponds to the PM approximation and we refer to this region as the \textit{exterior region}: the reason of it is that according to the dictionary \eqref{rdict}, $z \ll 1 \Leftrightarrow r \gg 2M$. For what already said, the solutions that are naturally defined here are the $G^0_\alpha(y)$ and the goal is giving a way of resumming the entire series in the variable $y$ (that assumes generic values) in the first line of \eqref{defsigma} at a fixed order $k$ in $x$. Clearly, by increasing $k$ the solution becomes more appropriate also in the opposite region that we are going to discuss later, i.e. $x \ll 1$ keeping $z$ finite.\\
For what previously discussed, at leading order in $x$ we take
  \be
  G^0_\alpha(y)=H^0_\alpha(y)
 \ee
   For higher orders in $x$, proceeding by trial and error, we come to the following ans\"atz
\be
    G^0_\alpha(y) =     P_{0}(y) \, H^0_\alpha(y) {+} \widehat{P}_{0}(y) \,  y H^{0}_\alpha{}'(y)   \label{g00}
\ee
where the prime stands for derivation with respect the argument and
 \beaq
P_{0 }( y)  &=& 1+\sum_{i=1}^k   \sum_{j=0}^{i-1}   c_{ij} \, x^i y^{j-i}  \nn\\  \widehat{P}_{0 }(y)   &=& \sum_{i=1}^k   \sum_{j=0}^{i-1}   \widehat{c}_{ij} \, x^i y^{j-i}   \label{pp0bis}
\eeaq
are polynomials of order $k$ in $x$\footnote{For $k=1$ M.Billò, M.L.Frau and A.Lerda first obtained \eqref{g00} resumming the contributions of the Young tableaux entering the partition function of the supersymmetric gauge theory.}. We plug this ans\"atz into \eqref{eq:CHE_y} and use the differential equation to express higher derivatives of $H^0_\alpha$ in terms of $H^0_\alpha$ and its first derivative themselves. Therefore one finds a system of algebraic linear equations that determines the coefficients $c_{ij}$, $\widehat{c}_{ij}$  and $u_i$, whose complete forms are given in \eqref{csol}. The first non trivial coefficient is $c_{10}$ that involves, in the recursive scheme shown in \eqref{csol}, the coefficient $c_{00}$ that does not enter our ans\"atz, hence we can set $c_{00}=1$. By doing it, one finds, for the first terms in the recursion
\beaq
 c_{10} &=& u_1=\frac{1}{2}(1-m_1-m_2-m_3)  -{2 m_1 m_2 m_3\over 4a^2-1} \nn\\
 \widehat{c}_{10}& =&\partial_{m_3} u_1=-\frac{1}{2}   -{2 m_1 m_2 \over 4a^2-1}   \label{cus}
 \eeaq
  It is easy to check that $c_{ij}$, $\widehat{c}_{ij}$ are invariant under $a\to -a$. This is true at any PM order, so all the dependence on $\alpha$ of the solution is codified in $H^0_\alpha$.
  The results (\ref{cus}) are special cases of more general relations that determine the coefficients $c_{i,i-1}$, $\widehat{c}_{i,i-1}$ in terms of $u_i$
  \beaq
  & c_{i,i-1} = u_i  \, , \nn\\
  & \gamma_m \equiv 1+\sum_{i=1}^\infty  \widehat{c}_{i,i-1}x^i = {\rm exp}\left\{  \sum_{i=1}^\infty {\partial_{m_3} u_i\over i} x^i \right\}  \quad \label{dk1f}
  \eeaq
 The combination in the exponential will be denoted in the following as
  \be
   \mathcal{F}_{\text{inst}} (a,x) \equiv  -\sum_{i=1}^\infty {u_i(a)\over i} x^i \label{finst}
  \ee
The subscript "inst" reminds the reader that this quantity is interpreted as the instanton prepotential in the gauge theory description of the CHE. Indeed in the limit $z\to 0$, where one of the two gauge couplings is turned off, the quiver theory reduces to a  ${\cal N}=2$ supersymmetric SU(2) gauge theory with three massive  fundamental flavours with prepotential ${\cal F}_{\rm inst}$. Indeed, from \eqref{eq:u_vs_a_recursion} and integrating the quantum Matone relation, we get
\be
\mathcal{F}_{\text{tree}} + \mathcal{F}_{\text{inst}} = - a^2 \log x - \sum_{i=1}^\infty \int u_i(a) x^{i-1} \mathrm{d} x
\ee
Now, the coefficients $u_i(a)$ depend only from $a$ and the masses $m_1,m_2,m_3$. Only if we use the dictionary \eqref{rdict} the dependence from $x$ comes out, but since we are following here a complete general treatment, we should not care about it, or at least, if we are want to plug the dictionary, then the $x$ that appears inside the masses should be treated differently with respect to the integration variable. After having performed the integration, then we can restore its meaning to the one of the coupling. By the way, all of this implies that we can bring outside the integral all the $u_i(a)$ and in this way we obtain
\be \label{eq:identification_Ftree_Finst}
\mathcal{F}_{\text{tree}} + \mathcal{F}_{\text{inst}} = - a^2 \log x - \sum_{i=1}^\infty \frac{u_i(a)}{i} x^i
\ee
as written before. The identification of $\mathcal{F}_{\text{1-L}}$ will be seen later.\\

We notice that  \eqref{g00} is exact in $y$  at the $k$-th PM order, so we can extrapolate it to the region of large $y$ using the standard hypergeometric transformation rule
 \beaq
H^0_\alpha (y) = \sum_{\alpha'=\pm}  B_{\alpha \alpha'}  \widetilde{H}^{0}_{\alpha'}(y)
\label{braiding0}\eeaq
 with
 \be
   \widetilde{H}^0_{\alpha}(y) =y^{-m_3(1+\alpha ) }
e^{ y (1+\alpha)\over 2} {}_2 F_0 \left(  \frac{1}{2}-a+\alpha  m_3,\frac{1}{2}+a+\alpha
   m_3,\frac{\alpha }{y}\right)
   \ee
and
  \be
  B_{\alpha \alpha'} =e^{\frac{{\rm i} \pi}{2}  \left(1-\alpha' \right) \left({1\over 2}  -\alpha a - m_3 \right)}   \frac{\Gamma (1-2 \alpha a  ) }{\Gamma
   \left(\frac{1}{2}-\alpha a  -\alpha' m_3 \right)}
  \ee
the braiding matrix elements. We can therefore introduce an equivalent basis of solutions, naturally defined in the \textit{far region}
  \be
  \widetilde{G}^0_{\alpha}(y) =   P_{0}(y)   \widetilde{H}^{0}_{\alpha}(y)   {+} \widehat{P}_{0}(y)   y \widetilde{H}^{0}_{\alpha}{}'(y)
\ee
 related to the previous one by the connection formulae
\be\label{conn_form_B}
 G^0_\alpha(y) =   \sum_{\alpha'=\pm}  B_{\alpha \alpha'}  \widetilde{G}^{0}_{\alpha'} (y)
\ee
All of this highlights the main difference of our hypergeometric representation for the Confluent Heun function with respect to the one of MST \cite{Mano:1996vt}: the latter involves an infinite bilateral sum of hypergeometric functions, while the former only a single hypergeometric function and its first derivative. This allows to provide a simple mathematical proof of the Heun connection formulae. By the way, this is only a conceptual aspect, indeed, at the practical level, we are going to show that all our final results for the energy fluxes will be compatible with the ones obtained from MST. It is due to the fact that, thanks to the contiguous relations of the hypergeometric functions, the bilateral series of MST can be written in terms of an hypergeometric function and its first derivative and all the infinite summations are shifted inside the present polynomials \footnote{I am very thankful to Thibeault Damour for discussion on this point and to Hasmik Poghosyan and Rubik Poghossian for having directly shown it.}.

\subsection{PN approximation: interior region}

In this Subsection we consider the limit $x\ll 1$ keeping $z$ finite (so also $y \ll 1$). This corresponds to the PN approximation and we refer to this region as the \textit{interior region}. The solutions naturally defined here are the $G^1_\alpha(z)$ and so the goal now is providing a way of resumming the entire series in the variable $z$ in the second line of \eqref{defsigma} at a fixed order $k$ in $x$. The analysis proceeds {\it mutatis mutandis} following the same steps of the previous Subsection, exchanging the roles of $y$ and $z$.\\
At leading order in $x$ we  take
  \be
  G^1_\alpha(z)=H^1_\alpha(z)
 \ee
 and higher orders  follow from the ans\"atz
\be
    G^1_\alpha(z) =     P_{1}(z) \, H^1_\alpha(z) {+} \widehat{P}_{1}(z) \,  z H^{1}_\alpha{}'(z)   \label{g0}
\ee
where the prime stands for derivation and
 \beaq
P_{1 }(z) & = &1+\sum_{j=1}^k   \sum_{i=0}^{j-1}   d_{ij} \, z^{i-j} x^j  \nn\\ \widehat{P}_{1 }(z)   &=& \sum_{j=1}^k   \sum_{i=0}^{j}   \widehat{d}_{ij} \, z^{i-j} x^j \label{pp1bis}
\eeaq
As before, plugging this ans\"atz into \eqref{CHE} and imposing the satisfaction of the differential equation, we arrive at a recursive scheme for $d_{ij}, \widehat{d}_{ij}$ and $u_i$, which is not reported since it will not be useful for the scope of this Chapter, but its structure is similar to the one of \eqref{csol}. So, setting $d_{00}=1$, one finds that the first non-trivial coefficients are
\beaq
 d_{01} =\frac{1}{2}+ \frac{2 m_3\left(1-m_3\right) }{4 a^2-1} , \quad
 \widehat{d}_{01} =- \widehat{d}_{11}=\frac{  2 m_3}{4 a^2-1}
  \eeaq
  We notice that the results are invariant under $a\to -a$. This is true at any PN order, so all the dependence on $\alpha$ of the solution is codified in $H^1_\alpha$.\\

Proceeding as before, we can extrapolate the PN representation to the region $z\simeq 1$ using the hypergeometric transformation rule
   \beaq
  H^1_\alpha(z) = \sum_{\alpha'=\pm}  F_{\alpha\alpha' } \widetilde{H}^1_{\alpha'}(z)
\label{fusion0}\eeaq
with
  \be
\widetilde{H}^{1}_{\alpha}(z)  = z^{a{+}m_3{-}\frac{1}{2}} (1{-}z)^{\frac{1}{2} \left(m_1{+}m_2\right)
   \left(1{+}\alpha \right)} \, _2F_1\left(\frac{1}{2}{+}a{+}\alpha m_1, \frac{1}{2}{+}a{+}\alpha m_2 ;1{+}\alpha\left(m_1{+}m_2\right) ;1{-}z\right)
\ee
and
\be
   F_{\alpha\alpha'}  = \frac{\Gamma (1{+}2\alpha a ) \Gamma
   \left(-\alpha' \left(m_1+m_2\right) \right)}{\Gamma \left( \ft12+\alpha a  -\alpha' m_1  \right) \Gamma \left( \ft12 +\alpha a  -\alpha' m_2  \right)}
\ee
the fusion matrix elements. An equivalent  basis of solutions defined in the \textit{near horizon region} is therefore given by
\be
   \widetilde{G}^1_{\alpha}(z) =   P_{1}(z) \, \widetilde{H}^{ 1}_{\alpha}(z)   {+} \widehat{P}_{1}(z) \,  z\, \widetilde{H}^{ 1 }_{\alpha}{}'(z)
\ee
 satisfying the connection formulae
\be\label{conn_form_F}
   G^1_\alpha(z) = \sum_{\alpha'=\pm}  F_{\alpha\alpha' } \widetilde{G}^1_{\alpha'}(z)
\ee

 \subsection{Summary picture and determination of the asymptotic coefficients}

  The functions $G^0_\alpha(\ft{x}{z})$ and $G^1_\alpha(z)$ are solutions of the same differential equation and have the  same monodromy around $z=0$, so they should be proportional to each other. We denote by $g_\alpha(x)$ their ratios
 \begin{equation}
  g_{\alpha}(x)= \frac{G^1_\alpha(z)}{G^0_\alpha \left(\frac{x}{z}\right)} =x^{\alpha a + m_3 - \frac{1}{2}} \left[ 1 + x \left(\frac{1}{4} - \frac{m_3 \left(m_1 + m_2 + 2 m_3 - 2\right) - m_1 m_2}{4 a^2-1} + \frac{8 \, \alpha \, a \, m_1 m_2 m_3}{\left(4 a^2 - 1\right)^2} \right)+\ldots  \right ] \label{galpha}
\end{equation}
Expanding for small $z$ and $y$ one can check explicitly that this ratio depends only on $x=z y$. Furthermore one finds
\be
 \gamma(x) ={g_+(x)\over g_-(x) }=   x^{2a} e^{-\partial_a {\cal F}_{\rm inst}(a,x) }   \label{gamma2}
\ee
with ${\cal F}_{\rm inst}(a,x)$ defined in (\ref{finst}).  Collecting all the different pieces together, we can draw the following summary pictiure

\begin{figure}[H]
    \centering
    \includegraphics[width=1 \linewidth]{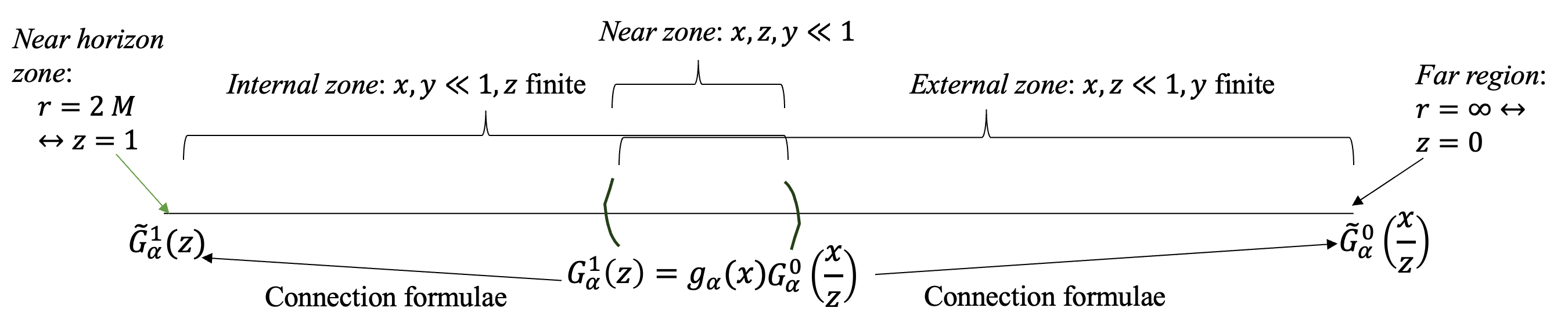}
    \caption{Summary picture that represents the four equivalent bases of solutions we discussed so far and how they are related. They suitably cover different patches of the interval $z \in (0,1) \Leftrightarrow r \in (2M,\infty)$.}
  \end{figure}
  
where we see that we have at our disposal four different bases of solutions, so that the general solution can be written as
\beaq
G_1(z) &=&  \sum_\alpha c_\alpha g_\alpha(x)  G^0_\alpha(\ft{x}{z} )  +\ldots  = \sum_{\alpha,\alpha'}  c_\alpha g_\alpha(x) B_{\alpha \alpha'}  \widetilde{G}^0_{\alpha'}(\ft{x}{z} )   +\ldots\nn\\
&=&  \sum_\alpha c_\alpha   G^1_\alpha(z )+\ldots  = \sum_{\alpha,\alpha'}  c_\alpha  F_{\alpha \alpha'}  \widetilde{G}^1_{\alpha'}(z )+\ldots  \label{g4}
\eeaq
The first line corresponds to the PM approximation where $x$ is taken small keeping $y$ finite. The second equality in this line is obtained using the connection formulae  \eqref{conn_form_B}. Similarly, the second line corresponds to the PN approximation where  $x$ is taken small keeping $z$ finite and the right hand side follows from \eqref{conn_form_F}. Finally dots stand for terms of order $x^{k+1}$ in the corresponding limits.\\

In order to apply all this procedure to the homogeneous Teukolsky equation, we need to find an expression for the asymptotic coefficients of $G_1(z)$ appearing in \eqref{gbcgen0}. This will be important when we impose ingoing and outgoing boundary conditions to the solution $R_{\ell m \omega}^{\text{hom}}(z)$. For what concerns the coefficients $B_\pm$, we need to go to $z \simeq 0$, hence we use the first line of \eqref{g4} and the connection formulae \eqref{conn_form_B}. Exploiting the fact that the leading order behavior of $\widetilde{G}^0_{\alpha'}(\ft{x}{z})$ is
\be
\widetilde{G}^0_{\alpha'}\left(\frac{x}{z}\right) \underset{z \to 0}{\simeq} \left(\frac{x}{z}\right)^{-m_3(1+\alpha')} e^{\frac{x}{z} \frac{1+\alpha'}{2}} e^{-\frac{1+\alpha'}{2} \partial_{m_3} \mathcal{F}_{\text{inst}}}
\ee 

we obtain
\be
G_1(z ) \underset{z\to 0}{\simeq} \sum_{\alpha,\alpha'}  c_\alpha g_\alpha(x) B_{\alpha \alpha'}   \left(\frac{x}{z}\right)^{-m_3 (1+\alpha')}  e^{ \frac{x (1+\alpha')}{2 z}}  e^{ - \frac{ 1+\alpha'}{2} \partial_{m_3} {\cal F}_{\rm inst}   }
\ee
Comparing with the first line of \eqref{gbcgen0} gives us the asymptotic coefficients
\be \label{bdcoef_B}
B_{\alpha'} = \sum_{\alpha}  c_\alpha g_\alpha(x) B_{\alpha \alpha'}  x^{-m_3 (1+\alpha' )}  e^{- \frac{ 1+\alpha'}{2} \partial_{m_3} {\cal F}_{\text{inst}}} 
\ee

For what concerns the remaining coefficients $D_\pm$, we have to go to $z \simeq 1$, so we now use the second line of \eqref{g4} and the connection formulae \eqref{conn_form_F}. Knowing that, at the leading order
\be
\widetilde{G}^1_{\alpha'} (z) \underset{z \to 1}{\simeq} (1-z)^{\frac{1+\alpha'}{2} (m_1+m_2)} h_{\alpha'}
\ee
with
\be
h_{\alpha'}=    P_{1}(1) \,    {+} {1{+}\alpha'\over 2 }(m_1{+}m_2)   \widehat{P}_{1}{}'(1) \,
\ee
we have
\be
G_1(z) \underset{z\to 1}{\simeq} \sum_{\alpha,\alpha'}  c_\alpha  F_{\alpha \alpha'} (1 - z)^{\frac{1 + \alpha'}{2}(m_1 + m_2)} \, h_{\alpha'}\label{ginf}
\ee
Comparing with the second line of \eqref{gbcgen0} provides the asymptotic coefficients 
\be \label{bdcoef_D}
D_{\alpha'}  =   \sum_\alpha c_\alpha  F_{\alpha \alpha'}  h_{\alpha'} 
\ee

Before coming back to the gravity side, we need to say something about how to compute $a$, in such a way that our solutions of the CHE are completely determined. In principle, from the relation \eqref{eq:u_vs_a_recursion}, since now we know all the coefficients $u_i(a)$ at the order we wish and using the fact that $u = \left(\ell+\frac{1}{2}\right)^2 + \frac{5}{2} x + \frac{x^2}{4}$ from the dictionary \eqref{rdict}, we can invert this series order by order in $x$. Doing it at high PM order can be time consuming for a Mathematica code. There is an alternative, more direct way to find the function $a(u)$ based on the connection between this quantity and the quantum SW period of a SU(2) theory with three fundamental flavors, as already discussed in Chapters \ref{chap_1} and \ref{chap_3}. Indeed, as already seen, the quantum SW period $a$ is known to satisfy the continuous fraction equation \cite{Poghosyan:2020zzg}
 \beaq
 \label{fractionequality}
 \frac{ x M(a+1)}{P(a+1) -\frac{ x M(a+2)}{P(a+2)-\ldots
 }}
 +\frac{ x M(a)}{P(a-1)-\frac{ x M(a-1)}{P(a-2)-\ldots
 }}-P(a)=0
 \eeaq
 where now
 \beaq
 P(a)&=&a^2+a x+2x -\frac{3}{4}x^2-\left(\ell+\frac{1}{2}\right)^2  , \nn\\
 M(a)&=& \left( a-\frac{x}{2}-\frac{1}{2} \right) \left(a-\frac{x}{2}+\frac{3}{2} \right) \left(a+\frac{x}{2}+\frac{3}{2} \right)
 \eeaq
 
 Equation \eqref{fractionequality} can be solved for $a(\ell)$ order by order in $x$. For example for $\ell=2,3$ one finds
 \beaq\label{acycleexp}
 a(2) &=&\frac{5}{2}+\frac{107 x^2}{840}-\frac{1695233 x^4}{148176000}+\frac{76720109901233 x^6}{30764716012800000}-\frac{71638806585865707261481 x^8}{99644321084605000704000000}+\ldots \nn\\
 a(3) &=&\frac{7}{2}+\frac{13 x^2}{168}-\frac{10921 x^4}{4346496}+\frac{95353832269 x^6}{493385539046400}-\frac{23627105510827613 x^8}{1148838359958761472000}+\ldots
 \eeaq
From these expressions we also see how $a$ is related to the renormalized angular momentum in the MST context \cite{PhysRevD.49.4016, Poisson:1994yf,Shibata:1994jx,Tagoshi:1994sm,Tanaka:1996lfd,Mano:1996vt,Mino:1997bx,Suzuki:1998vy,Fujita:2010xj,Fujita:2011zk} and to the anomalous dimension in the effective field theory description of GWs \cite{Goldberger:2009qd,Trestini:2023wwg,Almeida:2021jyt,Edison:2023qvg,Ivanov:2025ozg}, as already said in the Introduction.

\subsection{Homogeneous Teukolsky solutions}\label{sec3}

We are now ready to pass to analyze the solutions of the homogeneous Teukolsky equation, imposing in particular the boundary conditions we mentioned above.\\
Recalling the relation in \eqref{rdict} between $R_{\ell m \omega}^{\text{hom}}(z)$ and $G_1(z)$
\be
R_{\ell m \omega}^{\text{hom}} (z) = e^{-\frac{x}{2z}} (1-z)^{1-\frac{m_1+m_2}{2}} \left(\frac{x}{z}\right)^{1+m_3} G_1(z)
\ee
and using the asymptotic behaviors of $G_1(z)$ as written in \eqref{gbcgen0}, we have

\be \label{eq:asymptotic_behavior_general}
R_{\ell m \omega}^{\text{hom}} (z)  \simeq 
\left\{
\begin{array}{lll}
(x^{1+m_3} B_-) e^{-\frac{x}{2z}} z^{-1-m_3} + (x^{1+m_3} B_+) e^{\frac{x}{2z}} z^{m_3-1} &~~~~~~~   &\quad z \to 0   \\
\left(e^{-\frac{x}{2}} x^{1+m_3} D_- \right) (1-z)^{1-\frac{m_1+m_2}{2}} + \left(e^{-\frac{x}{2}} x^{1+m_3} D_+ \right) (1-z)^{1 + \frac{m_1+m_2}{2}}  &~~~~~~~   &\quad z \to 1
\end{array}
\right. 
\ee
\\
When considering the ingoing boundary condition at the horizon and outgoing boundary condition at infinity, the function $R_{\ell m \omega}^{\text{hom}}(z)$ becomes $R_{\text{in},\omega}(z)$ and $R_{\text{out},\omega}(z)$ respectively; inside their asymptotic behaviors written in \eqref{eq:asymptotic_behavior_general}, for the expressions of $B_\pm$ and $D_\pm$ in \eqref{bdcoef_B} and \eqref{bdcoef_D} we replace $c_\alpha$ with $c_\alpha^{\text{in,out}}$. In this way we define the asymptotic coefficients

 \begin{align} \label{bdcoef2}
 B^{\text{in,out}}_{\alpha'} & =  x^{1+m_3} \sum_{\alpha}  c^{\text{in,out}}_\alpha g_\alpha(x) B_{\alpha \alpha'}  x^{ -m_3(1+\alpha')}    e^{ -\frac{ 1+\alpha'}{2} \partial_{m_3} \mathcal{F}_{\text{inst}}} =  \sum_{\alpha}  c^{\text{in,out}}_\alpha g_\alpha(x) B_{\alpha \alpha'}  x^{1- \alpha' m_3}    e^{-\frac{ 1+\alpha'}{2} \partial_{m_3} \mathcal{F}_{\text{inst}}}
       \nn\\
 D^{\rm in,out}_{\alpha'} &= e^{-\frac{x}{2}} x^{1+m_3} \sum_\alpha c^{\text{in,out}}_\alpha  F_{\alpha \alpha'}  h_{\alpha'}  = \sum_\alpha c^{\text{in,out}}_\alpha  F_{\alpha \alpha'}  h_{\alpha'}  e^{-\frac{x}{2} }  \,   x^{1+m_3}  
 \end{align}
 
By using the dictionary \eqref{rdict}, we find

\beaq
R_{\text{in},\omega} (r)  & \simeq &
\left\{
\begin{array}{lll}
B^{\rm in}_-   e^{-i \omega r}  \left(\frac{2M}{r}\right)^{1+2 i M \omega} + B^{\rm in}_+ e^{i \omega r}  \left(\frac{2M}{r}\right)^{-3-2i M \omega } &~~~~~~~   &\quad r \to \infty   \\
D^{\rm in}_- 	 (1-\frac{2M}{r})^{2-2i M \omega} + D^{\text{in}}_+ \left(1-\frac{2M}{r}\right)^{2 i M \omega}   &~~~~~~~   &\quad r \to 2 M    \\
\end{array}
\right.  \nn\\
R_{\text{out},\omega} (r)  & \simeq &
\left\{
\begin{array}{lll}
B^{\rm out}_-   e^{-i \omega r}  \left(\frac{2M}{r}\right)^{1+2 i M \omega} + B^{\rm out}_+ e^{i \omega r}  \left(\frac{2M}{r}\right)^{-3-2i M \omega } &~~~~~~~   &\quad r \to \infty   \\
D^{\rm out}_- 	 (1-\frac{2M}{r})^{2-2i M \omega} + D^{\text{out}}_+ \left(1-\frac{2M}{r}\right)^{2 i M \omega}   &~~~~~~~   &\quad r \to 2 M    \\
\end{array}
\right.
\eeaq
\\
First of all we notice that these two asymptotic behaviors exactly match with the ones determined using the indicial equation for $r \simeq 2M$ and the WKB method for $r \simeq \infty$ in Section \ref{sec:asymptotic_behaviors} (provided to suitably redefine the constant factors). In order to impose the ingoing boundary condition at the horizon for $R_{\text{in}}$ and the outgoing one at infinity for $R_{\text{out}}$, we have to impose
\be \label{eq:boundary_conditions_Rin_Rout}
D^{\text{in}}_+ = B^{\text{out}}_- = 0
\ee
which will allow us to determine (even if not completely) the constants $c^{\text{in,out}}_\alpha$, as we are going to see. In this way we finally have
\beaq  \label{bcpsi}
R_{\text{in},\omega} (r)  & \simeq &
\left\{
\begin{array}{lll}
B^{\rm in}_-   e^{-i \omega r}  \left(\frac{2M}{r}\right)^{1+2 i M \omega} + B^{\rm in}_+ e^{i \omega r}  \left(\frac{2M}{r}\right)^{-3-2i M \omega } &~~~~~~~   &\quad r \to \infty   \\
D^{\rm in}_- (1-\frac{2M}{r})^{2-2i M \omega}  &~~~~~~~   &\quad r \to 2 M    \\
\end{array}
\right.  \nn\\
R_{\text{out},\omega} (r)  & \simeq &
\left\{
\begin{array}{lll}
B^{\rm out}_+ e^{i \omega r}  \left(\frac{2M}{r}\right)^{-3-2i M \omega } &~~~~~~~   &\quad r \to \infty   \\
D^{\rm out}_- 	 (1-\frac{2M}{r})^{2-2i M \omega} + D^{\text{out}}_+ \left(1-\frac{2M}{r}\right)^{2 i M \omega}   &~~~~~~~   &\quad r \to 2 M    \\
\end{array}
\right.
\eeaq
\\
In light of what said for the inhomogeneous solution in Section \ref{sec:general_Teuk_discussions}, we will need to evaluate the Green function \eqref{eq:Green_func}, defined in terms of the constant $W$ built out of the Wronskian in \eqref{eq:constant_W}. Using the behaviors in \eqref{bcpsi}, we can determine this quantity for $r \to \infty$. In this way we get

\be
W= \frac{R_{\text{in},\omega} (r) \, \partial_r \, R_{\text{out},\omega} (r)- \partial_r \, R_{\text{in},\omega} (r) \, R_{\text{out},\omega} (r)}{\Delta(r)} = \frac{i \omega}{2 M^2}  B^{\text{in}}_- B^{\text{out}}_+
\label{ww}
\ee

Since we are mainly interested in the physics at large distances $z \ll 1$, it is convenient to work with the PM representation; for this reason, in the following, we use as independent variables the two dimensionless combinations
\be
  y = 2  i \omega r  \qquad , \qquad    x= 4 i \omega M  \label{ekappa}
\ee
and then we consider as independent solutions of the Teukolsky equation the functions 
\be
g_\alpha(x) R_\alpha(y) = g_\alpha (x) e^{-\frac{y}{2}} \left(1- \frac{x}{y}\right)^{2-\frac{x}{2}}  y^{- 1 - \frac{x}{2}}   G^0_{\alpha}(y)
   \label{rka}
\ee
in such a way that the incoming and outgoing solutions can be written as linear combinations of these ones
\be \label{eq:general_Rin_Rout}
  R_{\text{in,out},\omega}(y)    =     \sum_{\alpha} c^{\text{in,out}}_\alpha g_\alpha(x)  R_{\alpha}(y)
\ee

We are focused on the PN expansion of the solutions, since we want to describe binary systems in the limit where the distance between the two objects is large and velocities small. For example, for circular orbits with tangential velocity $v$, this corresponds to the limit where $v$ is small,  $y \sim v$ and $x\sim v^3$. Therefore the PN expansion can be written as
\be \label{PNexpRa}
 R_\alpha (y) = y^{\frac{3}{2}-\alpha a} \sum_{n=0}^\infty  R_n(\alpha a, y)
\ee
with
\beaq
 &&  R_0(a,y)= 1 \nn\\
&& R_1(a,y)=\frac{2 y }{1-2 a}\nn\\
&& R_2(a,y)=\frac{(17-2 a) y ^2}{ 16(2 a-1)(a-1)}+\frac{(2 a-3) x}{4 y }\nn\\
 && R_3(a,y)=
  \frac{(7-2 a) y ^3}{8(2a-1)(2a-3)(a-1)}+\frac{(11-6 a) x}{4(2
   a-1)} \label{rfinaln} \nn\\
  &&R_4(a,y)=  \frac{\left(4 a^2{-}72 a{+}163\right) y ^4}{512 (a{-}2) (a{-}1) (2 a{-}3) (2
   a{-}1)}\nn\\
   &&+\frac{\left({-}4 a^2{+}48 a{-}119\right) y  x}{64 (2 a{-}1)(a{-}1)}{+}\frac{\left(8 a^3{-}4 a^2{-}18 a{+}9\right) x^2}{ 64y ^2 (a{+}1) }
\eeaq
and
\be
   a(\ell)=\ell+\frac{1}{2}  +\frac{\left(15 \ell^4+30 \ell^3+28 \ell^2+13 \ell+24\right)  x^2}{8 \ell (\ell+1) (2 \ell+1) \left(4 \ell^2+4 \ell-3\right)}+\ldots \label{als}
\ee

\subsubsection{The incoming solution} \label{sec:incoming_solution}

 The incoming solution is obtained by requiring no outgoing waves at the horizon, i.e. by setting $D^{\text{in}}_+=0$. According to \eqref{bdcoef2}, such condition leads to
\be
  \sum_{\alpha=\pm}  c^{\rm in}_\alpha F_{\alpha +}   =0 \label{dp0}
\ee
This determines the ratio  of the two coefficients to be
\be
\ {c^{\rm in}_+\over  c^{\rm in}_- } =- { F_{-+}   \over F_{++}   } =
\frac{\Gamma({-} 2 a) \Gamma \left(  \ft12{-} m_1    {+}a \right)
\Gamma \left(   \ft12{-} m_2 {+}a \right) }{\Gamma(2 a) \Gamma \left(   \ft12{-}m_1{-}a  \right)
\Gamma \left( \ft12{-} m_2 {-} a  \right)}   \label{cinup}
\ee
We stress that only the ratio  $c^{\rm in}_+/c^{\rm in}_-$ has been fixed, leaving undetermined the overall normalization $c^{\rm in}_-$. This dependence however cancels out in the Green function after dividing by the factor $W$ \eqref{ww}. It is convenient therefore to define a ratio which does not depend on overall normalizations
\be  \label{rgg}
 \mathfrak{R}_{\text{in},\omega} (y)=  \frac{R_{\text{in},\omega} (y)}{x  B^{\rm in}_-} =   C_{\rm in}   \left[  R_-(y)  - \gamma  \frac{F_{-+}}{F_{++}}  R_+(y)  \right] 
\ee
with
\be
C_{\rm in}  = \frac{x^{\frac{x}{2}}}{B_{--}} \left( 1-  \gamma \frac{F_{-+}  B_{+-} }{ F_{++}B_{--} } \right)^{-1} \underset{x \to 0}{\simeq}(-1)^{ \ell+1} \frac{\Gamma \left(\ell - 1\right)}{\Gamma (2  \ell + 2 )}  +\ldots
\label{Cin}
\ee
In deriving this expression we have used the relations \eqref{eq:general_Rin_Rout} for $R_{\text{in},\omega}(y)$, \eqref{gamma2} for the ratio $g_+/g_-$, \eqref{bdcoef2} for $B^{\text{in}}_-$ and \eqref{cinup} for $c^{\text{in}}_+/c^{\text{in}}_-$. We stressed in \eqref{eq:identification_Ftree_Finst} that $\gamma(x) = x^{2 a} e^{-\partial_a \mathcal{F}_{\text{inst}}}$ reproduces the exponential of the derivative with respect to $a$ of the tree level and instantonic parts of the prepotential, in particular $\gamma = e^{-\partial_a \left(\mathcal{F}_{\text{tree}} + \mathcal{F}_{\text{inst}}\right)}$. For what concerns the 1-loop sector, we notice that the ratio $(F_{-+} B_{+-}) / (F_{++} B_{--})$ reproduces $e^{-\partial_a \mathcal{F}_{\text{1-L}}}$. In this way it is interesting to notice that
\be
e^{2 \pi i a_D} = e^{-\partial_a \left(\mathcal{F}_{\text{tree}}+\mathcal{F}_{\text{1-L}}+\mathcal{F}_{\text{inst}}\right)} = \gamma \, \frac{B_{+-} F_{-+}}{B_{--} F_{++}}
\ee
We saw in Chapter \ref{chap_3} that the quantization of the $a_D$ period, $a_D = n$, leads to the determination of the QNMs frequencies, solving the previous equation with respect to $\omega$. Using the relation just written, we realize that such condition brings to
\be
1 = \gamma \, \frac{B_{+-} F_{-+}}{B_{--} F_{++}} \Rightarrow B^{\text{in}}_- = 0
\ee
This implies the vanishing of the Wronskian, as expected when determining of the QNMs.\\
We notice that in the PN limit $ R_\alpha (y) \sim  y^{{3\over 2}-\alpha a}$ and $\gamma \sim x^{2a}$, so the $\gamma R_+$-contribution in (\ref{rgg}) is suppressed by an extra factor $(x/y)^{2a}$ with respect to  $R_-$. The incoming solution is therefore dominated by the $R_-$ component. Using (\ref{als}) one finds the following PN expansion
{
	\beaq
	&&   R_-(y)=y^{a+{3\over 2} } \left[  1{+}\frac{ y }{1{+}\ell} {+} \left( \frac{(9{+}\ell) y
		^2}{8 (1{+}\ell) (3{+}2\ell)} {-} \frac{(\ell{+}2) x}{2 y }\right){+} \left(\frac{(\ell{+}4) y ^3}{8 (\ell{+}1) (\ell{+}2) (2 \ell{+}3)} {-}\frac{(7{+}3 \ell) x}{4
		(\ell+1)}\right)  \right. \nn\\
	&& \left. +
	\left(\frac{\left(50{+}19 \ell{+}\ell^2\right) y
		^4}{128 (\ell{+}1) (\ell{+}2) (2 \ell{+}3) (2 \ell{+}5)} {+} \frac{(\ell^2-1)   ( \ell+2) x^2}{4 (2\ell-1) y ^2}-\frac{(\ell+4)
		(\ell+9) x y }{16 (\ell+1) (2 \ell+3)}\right) +\ldots \right]
	\eeaq
}
It is important to observe that at higher orders both $R_\pm(y)$ exhibit  poles in the limit where $\ell \to \mathbb{Z}$, but one can check that they cancel against each other in the combination \eqref{rgg}. For example, for $\ell=2$ the poles  in the two components first appear at order $v^{16}$.

\subsubsection{The outgoing solution}

Outgoing boundary condition corresponds to setting $B^{\rm out}_-=0$,  that according to (\ref{bdcoef2}) leads to
\be
{c_-^{\rm out} \over c_+^{\rm out}}  = -\frac{\gamma {B_{+-}}}{B_{--}}
\ee
Following the same passages of the incoming solution, in order to get rid of the undetermined normalization $c^{\text{out}}_+$ we consider the ratio of $R_{\text{out}}(y)$ with the rest of $W$ that has not been taken into consideration for $R_{\text{in}}(y)$, as seen in the previous Subsection. For this reason we define
\be
   \mathfrak{R}_{\text{out},\omega} (y)=(2M)^3 \frac{R_{\rm out} (y)}{ B^{\rm up}_+} =    C_{\rm up}    \left[  R_+(y)   -  \frac{B_{+-}}{B_{--}}  R_-(y)  \right] \label{rupfinal}
\ee
with
\beaq
  C_{\rm up}&=&   \frac{ (2M)^3  e^{\partial_{m_3} \mathcal{F}_{\rm inst} } }{x^{3+\frac{x}{2}} B_{++} }     \left( 1-\frac{B_{- +} B_{+-}}{B_{+ +}B_{- -}} \right)^{-1}  = \frac{(2M)^3}{x^{3+\frac{x}{2}}}   \frac{\Gamma(2-\ell)}{\Gamma(-2\ell)}   +\ldots
\eeaq
We notice that the leading term on the right hand side is finite in the limit where $\ell$ becomes an integer with $\ell \geq 2$. The same result is obtained by setting $\ell  \to \mathbb{N}$ from the very beginning and sending then $x \to 0$.
In the PN limit, the solution is dominated by the $R_+(y)$ component with PN expansion
\beaq
&& R_+(y)=  y^{{3\over 2}-a}  \left[  1{-}\frac{ y }{\ell}{+} \left(\frac{(\ell-1) x}{2 y }{-}\frac{(\ell{-}8) y
	^2}{8 l (2\ell{-}1)}\right){+} \left(-\frac{(3\ell{-}4) x}{4
	\ell}{+}\frac{(\ell{-}3) y^3}{8 (\ell{-}1) \ell (2\ell{-}1)}\right) \right. \nn\\
&& \left. +
\left(\frac{( \ell{-}1) \ell (\ell{+}2) x^2}{4 (2\ell{+}3) y ^2}-\frac{(\ell{-}8) (\ell{-}3)
	x y }{16 \ell (2\ell{-}1)}+\frac{\left(32{-}17 \ell{+}\ell^2\right) y ^4}{128
	(\ell{-}1) \ell (2\ell{-}3) (2\ell{-}1)}\right) \right]
\eeaq

It is important to observe that again the poles in $R_\pm(y)$  in the limit where $\ell \to \mathbb{Z}$ cancel against each other in the combination \eqref{rupfinal}. For $\ell=2$ the poles in the two components first appear at order $v^{7}$.\\

Using the completely determined normalized functions $\mathfrak{R}_{\text{in}}(y)$ in \eqref{rgg} and $\mathfrak{R}_{\text{out}}(y)$ in \eqref{rupfinal}, the Green function becomes
\be \label{eq:Green_func_frak}
G_\omega (r,r')= \left\{
\begin{array}{ccc}
\mathfrak{R}_{\text{in},\omega} (r') \,  \mathfrak{R}_{\text{out},\omega} (r)  & ~~~~~~~&  r'< r\\
\mathfrak{R}_{\text{in},\omega} (r)  \, \mathfrak{R}_{\text{out}, \omega} (r')  & ~~~~~~~ &   r< r' \\
\end{array}
\right.
\ee

\section{The inhomogeneous equation: waveform and luminosity}
We are now ready to pass to examine the inhomogeneous solution, which we rewrite here for convenience
\be
R_{\ell m \omega} (r) = \int_{2M}^\infty G_{\omega}(r,r') \frac{T_{\ell m \omega}(r')}{\Delta(r')^2} \mathrm{d} r'
\ee
Since we are aimed at determining the waveform and the totale energy flux measured by an observer located at infinity, we consider the limit $r \to \infty$, which leads to
\be
R_{\ell m \omega}(r) \underset{r \to \infty}{\simeq} \mathfrak{R}_{\text{out},\omega}(r) \int_{2M}^\infty \frac{\mathfrak{R}_{\text{in},\omega}(r') \, T_{\ell m \omega}(r')}{\Delta (r')^2} \mathrm{d} r'
\ee
Using the asymptotic behavior of $\mathfrak{R}_{\text{out},\omega}(r) \underset{r \to \infty}{\simeq} r^3 e^{i \omega r_*}$ (as it can be seen from \eqref{bcpsi}), we can write
\be \label{eq:asymptotic_behavior_inhom}
R_{\ell m \omega} (r) \underset{r \to \infty}{\simeq} r^3 e^{i \omega r_*} Z_{\ell m \omega}^\infty
\ee
where we have defined the quantity
\be \label{eq:Z_lmomega}
Z_{\ell m \omega}^\infty \equiv \int_{2 M}^\infty \frac{\mathfrak{R}_{\text{in},\omega}(r') \, T_{\ell m \omega}(r')}{\Delta(r')^2} \mathrm{d} r'
\ee
Plugging this trend \eqref{eq:asymptotic_behavior_inhom} inside the harmonic decomposition of the $\mathbf{\Psi}_4$ scalar \eqref{eq:mode_dec_psi4}, we get
\be \label{eq:psi_4_asymptotic}
\mathbf{\Psi}_4 \underset{r \to \infty}{\simeq} \frac{1}{r} \int \frac{\mathrm{d} \omega}{2 \pi} \sum_{\ell,m} e^{- i \omega (t-r_*)} \, Z_{\ell m \omega}^\infty \, Y^{\ell m}_{-2} (\theta, \phi)
\ee

\subsection{Relation between the waveform and the Weyl scalar $\mathbf{\Psi}_4$ at infinity}
In order to get the relation between the waveform emitted by the binary system and the Weyl scalar under examination at infinity, we need to introduce the \textit{retarded time} $u$ defined as
\be \label{eq:retarded_time}
t = u + r_*(r)
\ee
We consider at the moment the retarded coordinates $(u,r,\theta,\phi)$ and we impose the Bondi gauge, according to which
\be
h_{rr} = h_{r A} = 0 \qquad , \qquad \gamma^{AB}h_{AB} = 0 \Rightarrow h_{\phi \phi} = -\sin^2\theta h_{\theta \theta}
\ee
where $\gamma^{AB}$ is the inverse metric defined on the 2-sphere parametrized by the angular coordinates $\theta,\phi$ and the indices $A,B$ run along these coordinates. At this point, when we take in consideration the limit $r \to \infty$ with $u,\theta,\phi$ fixed, we impose the standard falloffs as $r \to \infty$
\be \label{eq:falloffs}
h_{AB}(u,r,\theta,\phi) \simeq r h^{(-1)}_{AB}(u,\theta,\phi) + \ldots
\ee
(all the other non-vanishing components are at most $\mathcal{O}(r^0)$). At the end of the computation it turns out that
\be
\mathbf{\Psi}_4 \simeq -\frac{1}{2r} \left(\ddot{h}_{\theta \theta}^{(-1)} - \frac{i}{\sin \theta} \ddot{h}_{\theta \phi}^{(-1)} \right)
\ee
where the double dots refer to derivatives with respect to $u$ or $t$, since when $r,\theta,\phi$ are fixed these two operations coincide, as it can be seen from \eqref{eq:retarded_time}. Passing to the standard components
\be
h_+ = \frac{1}{2} \left(\hat{e}^\mu_\theta h_{\mu \nu} \hat{e}^\nu_\theta - \hat{e}^\mu_\phi h_{\mu \nu} \hat{e}^\nu_\phi\right) \qquad h_\times = e^\mu_\theta h_{\mu \nu} \hat{e}^\nu_\phi
\ee
with
\be
\hat{e}_\theta = \frac{1}{r} \partial_\theta \qquad \hat{e}_\phi = \frac{1}{r \sin \theta} \partial_\phi
\ee
we obtain that
\begin{align}
& h_{\theta \theta} = r^2 h_+  \Rightarrow \ddot{h}_{\theta \theta} = r^2 \ddot{h}_+ \notag \\
& h_{\theta \phi} = r^2 \sin \theta h_\times \Rightarrow \ddot{h}_{\theta \phi} = r^2 \sin \theta \ddot{h}_\times
\end{align}
From the falloffs \eqref{eq:falloffs} we see that
\be
\ddot{h}^{(-1)}_{\theta \theta} \simeq r \ddot{h}_+ \qquad \ddot{h}^{(-1)}_{\theta \phi} \simeq r \sin \theta \ddot{h}_\times
\ee
In this way \footnote{We have to say that in \cite{Maggiore:2018sht} we read that $\mathbf{\Psi}_{4,\text{Maggiore}} \simeq -\ddot{h}$, in the same limit of $r \to \infty$. The reason of it relies on a different choice of the vectors of the tetrad. In particular $l^\mu_{\text{here}} = \frac{l^\mu_{\text{Maggiore}}}{\sqrt{2}}$ and $n^\mu_{\text{here}} = \frac{q^\mu_{\text{Maggiore}}}{\sqrt{2}}$. In this way $\mathbf{\Psi}_{4,\text{here}} = \frac{1}{2} \mathbf{\Psi}_{4,\text{Maggiore}}$.}
\be
\mathbf{\Psi}_4 \simeq -\frac{1}{2} \left(\ddot{h}_+ - i \ddot{h}_\times \right) \equiv -\frac{1}{2} \ddot{h}
\ee
From the asymptotic behavior of $\mathbf{\Psi}_4$ for $r \to \infty$ in \eqref{eq:psi_4_asymptotic} and integrating with respect to time twice, we get at the end
\be \label{eq:waveform_infinity}
h(t,r,\theta,\phi) \underset{r \to \infty}{\simeq} \frac{1}{r} \int \frac{\mathrm{d}\omega}{2 \pi} \sum_{\ell,m} e^{- i \omega (t-r_*)} \frac{2 Z_{\ell m \omega}^\infty}{\omega^2} \, Y^{\ell m}_{-2} (\theta,\phi)
\ee
We indicate the $(\ell,m)$ component of the waveform in the frequency domain as
\be \label{eq:h_lm_generic}
h_{\ell m \omega} \equiv \frac{2}{\omega^2} Z_{\ell m \omega}^\infty = \frac{2}{\omega^2} \int_{2 M}^\infty \frac{\mathfrak{R}_{\text{in},\omega}(r') T_{\ell m \omega}(r')}{\Delta (r')^2} \mathrm{d} r'
\ee
From all of this we see that, in order to determine the waveform, we need to know the quantity $Z_{\ell m \omega}^\infty$.

\subsection{Final result for the source term $Z_{\ell m \omega}^\infty$}\label{sec:source_term}
All the detailed passages for the explicit computation of $Z_{\ell m \omega}^\infty$ are reported in Appendix \ref{app:Z_lm_inf}. Here we limit ourselves to express the logical steps and write the final expression. The first one is writing the quantity $\mathcal{E}[T]/\Delta^2$, that enters $Z_{\ell m \omega}^\infty$ via $T_{\ell m \omega}/\Delta^2$, in a more convenient way using the commutation property of the operator $J_+$ with a given power of $r$. At that point, integrating by parts several times and using some properties of the operators $L_s$, we get an expression for $Z_{\ell m \omega}^\infty$ which can be made more explicit only by invoking the stress-energy tensor components in \eqref{eq:stress_components_appr}. After this, we are left by considering the motion of the particle along a given trajectory. Using the Hamiltonian formalism for studying the geodesics, together with some identities of the application of the operators $L_s$ on the spherical harmonics, we arrive at
\be \label{eq:final_Z_generic}
Z_{\ell m \omega}^\infty = G \int \frac{\mathrm{d} t}{r(t)^2} \, e^{i \omega t - i m \phi(t)} \sum_{j=0}^2 b^j_{\ell m} \mathcal{L}^j \left[\mathfrak{R}_{\text{in},\omega}(r)\right]
\ee
where we have defined
\be
b^0_{\ell m} = \pi \sqrt{(\ell^2-1) \ell (\ell+2)} \, S^{\ell m}_0 \left(\frac{\pi}{2}\right)\, \, , \quad b^1_{\ell m} = 2 \pi \sqrt{(\ell-1)(\ell+2)} \, S^{\ell m}_{-1} \left(\frac{\pi}{2}\right) \, \, , \qquad b^2_{\ell m} = 2 \pi \, S^{\ell m}_{-2} \left(\frac{\pi}{2}\right)
\ee
and
\begin{align} \label{eq:operators_Lj}
& \mathcal{L}^0 = \frac{E}{f} \left(1+\frac{r'(t)}{f}\right)^2 \, \, , \qquad \mathcal{L}^1 = \frac{i J}{r} \left(1+\frac{r'(t)}{f}\right) \left(2-r \, \partial_r + \frac{i \omega r}{f}\right) \notag \\
& \mathcal{L}^2 =  \frac{f \, J^2}{E \, r^2} \left[\frac{1}{f^2} \left(i M \omega + \frac{\omega^2 r^2}{2}- i \omega r\right) + \left(\frac{i \omega r}{f} + 1\right)r \, \partial_r - \frac{1}{2} r^2 \, \partial_r^2\right]
\end{align}

\subsubsection{Restriction to the circular case}
In the following we are going to compute physical observables along circular orbit, hence it is useful to see what are the implications of this on the formulae we derived so far.\\
The first one is that $r(t) \equiv r_\Omega$, with $r_\Omega$ the radius of the orbit, and so $r'(t) =0$. Furthermore, we also have $\phi(t) = \Omega \, t$, with $\Omega$ the orbital frequency. This produces a huge simplification in the expression of $Z_{\ell m \omega}^\infty$ in \eqref{eq:final_Z_generic}. In fact the only dependence on the time $t$ is in $\int \mathrm{d} t \, e^{i t (\omega - m \Omega)} = 2 \pi \delta(\omega - m \Omega)$. In this way, setting for simplicity $G=1$
\be \label{eq:Z_lm_circular}
Z_{\ell m \omega}^\infty = \frac{1}{r_\Omega^2} \sum_{j=0}^2 b^j_{\ell m} \mathcal{L}^j \left[\mathfrak{R}_{\text{in},\omega} (r)\right]|_{r=r_\Omega} \, \, 2 \pi \delta(\omega - m \Omega)
\ee
where, inside the operators $\mathcal{L}^j$, we have to set $r'=0$ and, after having performed all the derivatives, we impose $r=r_\Omega$. At this point the $(\ell,m)$ component of the waveform \eqref{eq:h_lm_generic} becomes
\be
h_{\ell m \omega} = \frac{2}{\omega^2} \frac{1}{r_\Omega^2} \sum_{j=0}^2 b^j_{\ell m} \mathcal{L}^j \left[\mathfrak{R}_{\text{in},\omega} (r)\right]|_{r=r_\Omega} \, \,2 \pi \delta(\omega - m \Omega)
\ee
implying that the waveform in the time domain \eqref{eq:waveform_infinity} simply becomes
\be
h(t,R,\theta,\phi) \underset{R \to \infty}{\simeq} \frac{1}{R} \sum_{\ell,m} e^{-i m \Omega (t-R_*)}\, Y^{\ell m}_{-2} (\theta,\phi) \, \frac{2}{(m \Omega)^2 \, r^2_\Omega}\sum_{j=0}^2 b^j_{\ell m} \mathcal{L}^j \left[\mathfrak{R}_{\text{in},\omega} (r)\right]|_{r=r_\Omega, \omega = m \Omega}
\ee
where we have changed the distance of the observer $r$ appearing in \eqref{eq:waveform_infinity} with $R$ in order to not confuse with the radius of the orbit.\\
From all of this we realize that we can set from the beginning $\omega = m \Omega$ in the circular case. This allows us to rewrite $Z_{\ell m \omega}^\infty$ in \eqref{eq:Z_lm_circular} in a more convenient way, together with another constraint coming from the circular motion. Indeed the third Kepler's law implies that
\be \label{eq:third_Kepler}
r_0 = M^{\frac{1}{3}} \Omega^{-\frac{2}{3}} = \frac{M}{v^2}
\ee 
where $v$ is the tangential velocity of the probe. Thanks to this it can be proven that the factor $J^2 f(r_0)/(E r^2_0)$ that appears inside $\mathcal{L}^2$ is simply $J \, \Omega$. So, rearranging all the terms inside the operators $\mathcal{L}^j$ in \eqref{eq:operators_Lj} in such a way that they multiply $\mathfrak{R}_{\text{in},\omega}$, $r \, \partial_r \mathfrak{R}_{\text{in},\omega}$ and $r^2 \, \partial_r^2 \mathfrak{R}_{\text{in},\omega}$, we can write
\be \label{eq:Z_lm_simplified_circular}
Z_{\ell m \omega}^{\infty} = \frac{1}{r_\Omega^2} \left[A_0 \, \mathfrak{R}_{\text{in},\omega}(r) + A_1 r \, \partial_r \mathfrak{R}_{\text{in},\omega}(r) + A_2 ^2 \, \partial_r^2 \mathfrak{R}_{\text{in},\omega}(r)\right]|_{r=r_0}
\ee
with
\begin{align} \label{eq:A_coefficients}
& A_0 = \frac{E}{f(r)} b^0_{\ell m} + \frac{2 i J}{r} \left(1 + \frac{i m \Omega r}{2 \, f(r)}\right) b^1_{\ell m} - J \Omega \frac{i m \Omega r}{f(r)^2} \left(1 - \frac{M}{r} + \frac{i m\Omega r}{2}\right) b^2_{\ell m} \notag \\
& A_1 = J \left[- \frac{i}{r} b^1_{\ell m} + \Omega \left(1 + \frac{i m \Omega r}{f(r)}\right) b^2_{\ell m}\right] \quad , \quad A_2 = -\frac{1}{2} J \Omega \, b^2_{\ell m}
\end{align}
This will be the relevant formula that we are going to use in our computations.\\
The other simplification that comes out from the circular motion is when we pass to examine the PN expansion. Indeed, thanks to the third Kepler's law \eqref{eq:third_Kepler}, we firstly see that PM and PN expansions concide
\be
1 \underset{\text{PM}}{\gg} z = \frac{2 M}{r_0} = 2 v^2 \underset{\text{PN}}{\ll} 1
\ee
and this produces the following dependence on $v$ of the following quantities
\be \label{eq:scaling_v}
\omega = m \Omega = \frac{m v^3}{M} \qquad , \qquad E = \frac{\mu (1-2v^2)}{\sqrt{1-3 v^2}} \qquad , \qquad J =\frac{\mu M}{v \sqrt{1- 3 v^2}}
\ee
In the following we will consider all these quantities in unity of the mass of the probe.

\subsection{The luminosity}
At this point we can define what is the observable that we are going to compute, in order to see if the new hypergeometric representation of the Confluent Heun function explained in Section \ref{sec2} produces results that are good and that deserve to be further analyzed. It is the luminosity, defined as the amount of energy per unit of time emitted by the system towards infinity. For this reason it is also called \textit{total energy flux} and it is given by the formula (see \cite{Mino:1997bx} and references therein)
\be \label{eq:total_flux}
F_{\text{tot}} = \sum_{\ell=2}^\infty \sum_{m=1}^\ell F_{\ell m} = \sum_{\ell=2}^\infty \sum_{m=1}^\ell \frac{\left|Z_{\ell m \omega}^\infty\right|^2}{2 \pi \omega^2} = \sum_{\ell=2}^\infty \sum_{m=1}^\ell  \frac{\omega^2}{8 \pi} |h_{\ell m \omega}|^2 \equiv F_{22}^N \sum_{\ell=2}^\infty \sum_{m=1}^\ell \eta_{\ell m}  
\ee
where $F_{22}^{\text{N}}$ is the Newtonian (i.e. quadropolar) contribution, that turns out to be \footnote{If we want to restore the dependence from $\mu$, we have $F_{22}^N = \frac{32}{5} \left(\frac{\mu}{M}\right)^2 v^{10}$}
\be \label{eq:Newton_contribution}
F_{22}^N = \frac{32 \, v^{10}}{5 M^2}
\ee
as it will be clear in Section \ref{sec:factorization}.

\section{Results of the homogeneous solutions and the energy flux at 30PN}\label{sec4}

In this Section we use perturbation theory to compute the homogeneous solutions $\mathfrak{R}_{\text{in},\omega}(r)$ and $\mathfrak{R}_{\text{out},\omega}(r)$ for given values of the parameters characterizing them and at some distances, together with the energy flux produced by a circular EMRI binary system at order 30PN, i.e. $ v^{60} $, at the leading order of the probe limit. The results will be compared against the 22PN results available in the literature \cite{Fujita:2012cm} and the numerical computations using the package \textit{Teukolsky} of the Black Hole Perturbation Toolkit (BHTK) \cite{BHPToolkit} based on the MST method \cite{PhysRevD.49.4016,Poisson:1994yf,Shibata:1994jx,Tagoshi:1994sm,Tanaka:1996lfd,Mano:1996vt,Mino:1997bx,Suzuki:1998vy,Fujita:2010xj,Fujita:2011zk} and on a numerical integration of the differential equation. We stress the fact that by  30PN we mean that the polynomials $P_0$, $\widehat{P}_0$ and the exponent $a$, defining the solution, are truncated at order 30PN, but the hypergeometric functions and the $c^{\rm in}_\alpha(a)$ coefficients specifying the solution are kept exact, so an infinite number of terms beyond 30PN are taken into account. In particular, the infinite tower of tail (log divergences) and transcendental ($\pi$, Euler constant, \ldots) contributions are resummed into exponentials and Gamma functions. The observables we consider are  the $(\ell,m)$ harmonic modes of the incoming solution $\mathfrak{R}_{\text{in}}$, i.e. $\eta_{\ell m}$ defined in \eqref{eq:total_flux}, and the total energy flux radiated to infinity by the system.\\

In the present analysis we consider distances $r$ larger than the innermost stable circular orbit (ISCO) radius  $r_{\text{ISCO}} = 6 M$, so, keeping in mind that $v^2=M/r$, the domain of variability of the tangential velocity is $v\in \left[ 0,\frac{1}{\sqrt{6}}\right] $. We find that the first 12 digits of 30PN results are stable  for any choice of $v$ in this domain, so we will always display this number of digits.  The numerical computations and those with the MST method are performed  setting in the BHTK Mathematica files: AccuracyGoal = 1000, PrecisionGoal = 70, WorkingPrecision = 80,  SetPrecision=80.  With this choice of parameters both numerical and MST results agree to a 12 digits precision. In our Tables these results will be denoted as BHTK.

\subsection{The necessary ingredients}
In this Subsection we collect all the ingredients needed in the computations of the incoming, outgoing solutions at order $k=30$ in the hypergeometric PM representation. We write\footnote{ Here we rewrite the second terms in the right hand sides of \eqref{rgg}, \eqref{rupfinal},  as the images under $a \to -a$ of the first terms,  using the transformation rules
    $R_- \to R_+$, $\gamma \to \gamma^{-1}$,  $F_{-\alpha'} \leftrightarrow  F_{+ \alpha'}$ and $B_{-\alpha'} \leftrightarrow  B_{+ \alpha'}$. }
\beaq
   \mathfrak{R}_{\text{in},\omega} (y) &=&      C_{\rm in}    R_-(y) + (a \to -a)  \nn\\
    \mathfrak{R}_{\text{out},\omega} (y) & =&     C_{\rm up}    R_+(y) + (a \to -a) \label{rinrup}
\eeaq
with, explicity using the dictionary for the masses \eqref{rdict} \footnote{As already said in Section \ref{sec:incoming_solution}, we notice that  $\gamma_{\rm tot} = \gamma  \ft{F_{-+}  B_{+-} }{ F_{++}B_{--}}  = e^{2\pi {\rm i} a_D}= e^{-\partial_a  ({\cal F}_{\rm tree}+{\cal F}_{\rm 1-L}+{\cal F}_{\rm inst}) } $ collects the  tree level, 1-loop and instanton contributions to the dual SW period $a_D$.}
\beaq \label{collection1}
R_\alpha (y) &=& e^{- {y\over 2  } } \left(1- \ft{x}{y}\right)^{ 2 {-}  {x\over 2}    }  y^{ {-}1{-}{x\over 2}}    \left[  P_{0}( y) \, H^0_\alpha(y) {+} \widehat{P}_{0}( y) \,  y H^{0}_\alpha{}'(y)   \right] \nn\\
H^0_\alpha(y) &=& y ^{\frac{5}{2} +\frac{x}{2}-\alpha a } \, _1F_1\left( \ft52 +\ft{x}{2}- \alpha a   ;1-2 \alpha a;y \right)\nn\\
C_{\rm in}  &=& x^{x\over 2 } e^{ -\frac{{\rm i} \pi}{2}(5+2a+x) }  {    \Gamma(-\ft32+a-\ft{x}{2} )\over \Gamma(1+2a) }
\left( 1- \gamma_{\rm tot}  \right)^{-1}  \nn\\
C_{\rm up}  &=&    { (2M)^3    \over    x^{3+\frac{x}{2}}   \, \gamma_m }   {    \Gamma( \ft52-a+\ft{x}{2} )\over \Gamma(1-2a) }   \left( 1-e^{-2\pi {\rm i} a} \, \ft{\cos \ft{\pi}{2}(x+2a)}{\cos \ft{\pi}{2}(x-2a)}  \right)^{-1}\nn\\
 \gamma_m  &=&1+\sum_{i=1}^\infty  \widehat{c}_{i,i-1} x^i   \quad , \quad  \gamma =x^{2a} e^{-\partial_a {\cal F}_{\rm inst} } \nn\\
 \gamma_{\rm tot} &=& \gamma \, e^{-2\pi {\rm i} a} \,\frac{\Gamma ({-}2 a)^2 \Gamma \left({-}\frac{3}{2}{+}a{-}\frac{x}{2}\right) \Gamma
		\left(\frac{1}{2}{+}a{-}\frac{x}{2}\right) \Gamma
		\left(\frac{5}{2}{+}a{-}\frac{x}{2}\right)}{\Gamma (2 a)^2 \Gamma
		\left(-\frac{3}{2}{-}a{-}\frac{x}{2}\right) \Gamma \left(\frac{1}{2}{-}a{-}\frac{x}{2}\right)
		\Gamma \left(\frac{5}{2}{-}a{-}\frac{x}{2}\right)}
\eeaq
    The coefficients $c_{ij}, \widehat{c}_{ij}$ in the polynomials
  \be
P_{0 }( y)  = 1+\sum_{i=1}^k   \sum_{j=0}^{i-1}   c_{ij} \, x^i y^{j-i}  \quad , \quad  \widehat{P}_{0 }( y)   = \sum_{i=1}^k   \sum_{j=0}^{i-1}   \widehat{c}_{ij} \, x^i y^{j-i} \\  \label{pp0bis2}
\ee
 are determined by the recursion relations
 \beaq
 c_{ij}&=& \left[  A_1 \,c_{i,j-1}+A_2\, c_{i-1,j}+ A_3\,  c_{i-1,j-1}+\widehat{A}_{1} \widehat{c}_{i,j-1}+\widehat{A}_{2}
 \widehat{c}_{i-1,j} +\widehat{A}_{3} \widehat{c}_{i-1,j-1} \right. \nn\\
 && \left.    +  \sum _{s=1}^{k}
 c_{s,s-1} (C_1  \,  c_{i-s,j-s} + \widehat{C}_1\,  \widehat{c}_{i-s,j-s}) \right]  \left[ 8 \Delta_{ij} (\Delta_{ij}^2 -4 a^2)  \right]^{-1}  \nn\\
 \widehat{c}_{ij}&=& \left[  B_1 \,c_{i,j-1}+B_2\, c_{i-1,j}+ B_3\,  c_{i-1,j-1}+\widehat{B}_{1} \widehat{c}_{i,j-1}+\widehat{B}_{2}
 \widehat{c}_{i-1,j} +\widehat{B}_{3} \widehat{c}_{i-1,j-1} \right. \nn\\
 && \left.    +  \sum _{s=1}^{k}
 c_{s,s-1} (C_2  \,  c_{i-s,j-s} + \widehat{C}_2\,  \widehat{c}_{i-s,j-s}) \right]  \left[ 8 \Delta_{ij} (\Delta_{ij}^2 -4 a^2)  \right]^{-1} \nn\\
 u_i  &=& c_{i,i-1}   \label{csol}
 \eeaq
starting from  $c_{0i}=\delta_{i0}$,  $\widehat{c}_{0i}=0$, with
\beaq
A_1 &=&  -8 \left(\Delta _{ij}+1\right) \left(\Delta _{ij}-x-5\right) \quad, \quad \widehat{A}_{1}= 4 \left(4 a^2-(x+5)^2\right) \Delta
_{i,j} \nn\\
A_3 &=&  8 \Delta _{ij}
\left(\Delta _{ij}-x-5\right) \quad ,\quad \widehat{A}_{3}= 4 \left(4 a^2-(x+5)^2\right) \left(-\Delta
_{i,j}+x-2\right)\nn\\
A_2&=&  2\left(3 x^2{+}6 x{-}12 a^2{-}13\right) \Delta _{ij}{+}8 \Delta _{ij}^3{-}8 (x{-}1)
\Delta _{ij}^2{+}8 a^2 (x{-}7){-}2( x^3{+}3 x^2{-}13 x{-}15), \nn\\
\widehat{A}_{2}&=&  \left(4 a^2-(x+5)^2\right) \left(-2 (x-1) \Delta _{ij}+4 a^2+x^2-2
x-3\right) \nn\\
B_1&=&  -16 \left(\Delta _{ij}+1\right) \quad ,\quad \widehat{B}_{1}= 8 \Delta _{ij}
\left(\Delta _{ij}+x+5\right) 	\quad, \quad \widehat{B}_{3}= 8 \left(-\Delta_{i,j}^2-7 \Delta _{ij}+x^2+3 x-10\right) \nn\\
B_2&=&  4 \left(-2 (x-1) \Delta_{i,j}+4 a^2+x^2-2 x-3\right) \quad , \quad  B_3= 16 \Delta _{ij} \nn\\
\widehat{B}_{2} &=&  2 \left(-\Delta _{ij}+x+1\right) \left(-4 \Delta _{ij}^2-8 \Delta _{ij}+12 a^2+x^2+2 x-15\right)  \nn\\
C_1 &=&  8 \left(\Delta _{ij}-x-5\right) \quad , \quad \widehat{C}_1= 16 a^2-4(x+5)^2 \quad , \quad C_2=  16  \quad , \quad  \widehat{C}_2=8 \left(\Delta_{ij}+x+5\right)
\eeaq
and $\Delta_{ij}=i-j$. At the end of the computation, we set $y = 2 i \omega r$ and $x = 4 i M \omega$.

\subsection{Final results}

In this Subsection we use our formulae expressed at order 30PM in order to compute the homogeneous solutions, the $(\ell, m)$ harmonic modes $\eta_{\ell m}$ and the total energy flux. We set $M=1$ and we start with the evaluation of the two functions $\mathfrak{R}_{\text{in},\omega}(r)$ and $\mathfrak{R}_{\text{out},\omega}(r)$ for $\ell=2$ and $\omega=0.1$.\\
The results obtained with our method are compared in Table \ref{trinrup} against those derived with the MST method and numerical integration using the BHTK \cite{BHPToolkit} code. In all the subsequent Tables, the displayed digits are those that are insensitive to the contribution of terms of order higher than $v^{60}$ in $P_0(y)$ and $\widehat{P}_0(y)$. The agreement is up to 10 and 14 digits even for small values of $r$,  where the PM approximation is less reliable.\\

\begin{table*}
\centering			
\begin{tabular}{ccccc}
\hline
 $ \mathfrak{R}_{\text{in},0.1}$   & {\small $r=5$ } &  {\small  $r=100$ } \\ \hline
{\small HYP}& {\small -0.004086908784 -{\rm i} 0.000826769654} & {\small 252.949132499007 -{\rm i} 211.392731187846}  \\
{\small BHTK}& {\small -0.004086908784 -{\rm i} 0.000826769654} & {\small 252.949132499009 -{\rm i} 211.392731187848} \\ \hline
$ \mathfrak{R}_{\text{out},0.1}$  &  {\small $r=5$} &  {\small  $r=100$ } \\ \hline
{\small HYP}& {\small 2632.048432 -{\rm i} 840.387207} & {\small  -15205.329855167 -{\rm i} 989985.766975648 }\\
{\small BHTK}& {\small  2632.048432 -{\rm i} 840.387207} &  {\small -15205.329855168 -{\rm i} 989985.766975651 }\\ \hline
\end{tabular}
\caption{Comparison between the 30PM representation of the homogeneous solutions $\mathfrak{R}_{\text{in},\omega}(r)$ and $\mathfrak{R}_{\text{out},\omega}(r)$ and MST/Numerical results for $\omega=0.1, \ell=2$. }\label{trinrup}
\end{table*}

\begin{figure}[H]
    \centering
    \includegraphics[width=0.3\textwidth]{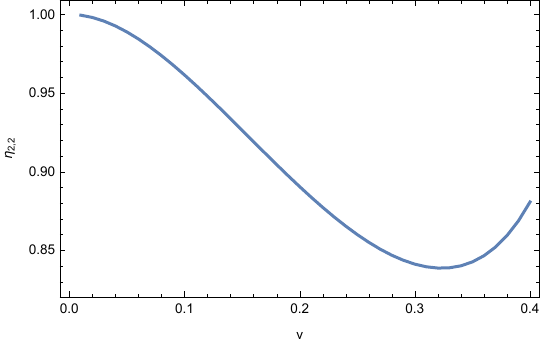}
    \includegraphics[width=0.3\textwidth]{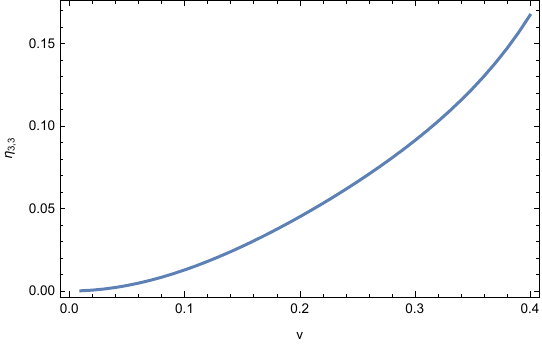}
    \includegraphics[width=0.3\textwidth]{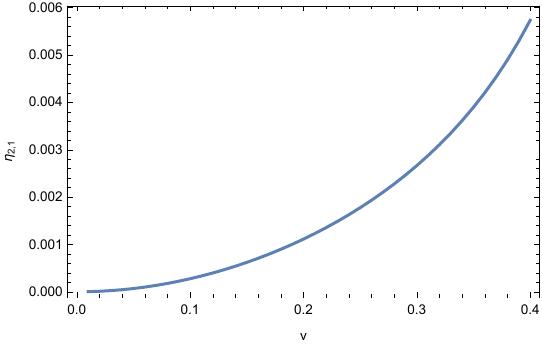}
    \caption{Luminosity vs velocity. We display the $v$-dependence of the ratio between $\eta_{\ell m}$ and the Newton luminosity $F_{22}^N$ \eqref{eq:Newton_contribution} for the three dominating modes. a: (2,2), b: (3,3), c: (2,1).}
    \label{plotlum}
\end{figure}

Then we pass to the harmonic modes, in order then to arrive at the luminosity. In Figure \ref{plotlum} we display the $v$-dependence of the luminosity for the $(\ell,m)=(2,2)$, $(3,3)$, $(2,1)$ harmonic modes computed from \eqref{rgg} at order $k=30$  in the PN approximation.

The convergence speed of the results  for some $(\ell,m)$ modes is showed in Tables \ref{Tconvergence} and \ref{Tconvergence2}.  The digits in bold style underline the  agreement between computations based on the PM hypergeometric representations and the MST/numerical results. As expected, the speed of convergence varies with $v$: in the case of the $(2,2)$ mode, for $v=0.1$ an agreement at 12 digits is already reached at 10PM order, while for $v=0.4$, the contributions at order 30PM  are needed to reach a precision at all the considered digits. A similar behavior is also found for higher $(\ell,m)$ modes. The last line of each sub-table shows the results coming from the  22PN computation in \cite{Fujita:2012cm}.  We observe that, opposite to what happens if one uses the 22PN expansion, our results agree against those obtained by  the MST/Numerical methods for the first 10-12 digits independently of the $\ell$ mode. The difference is evident in the case of large $\ell$, see for example the results for the $(10,1)$ mode in Table \ref{Tconvergence2}.

\begin{table*}
\centering
\begin{tabular}{|cccc|}
\hline
 $k$ & $\eta_{2,2}(0.1) $ &   $\eta_{2,2}(0.3)$ & $\eta_{2,2}(0.4)$ \\ \hline
5 & \textbf{0.9614837055}13    & \textbf{0.8411}81503167   & \textbf{0.8807}96957664 \\
 10 & \textbf{0.961483705502}  & \textbf{0.84117617}8745 & \textbf{0.8807}13162076 \\
 15 & \textbf{0.961483705502}    & \textbf{0.84117617132}8 & \textbf{0.8807054}81610 \\
 20 & \textbf{0.961483705502}   & \textbf{0.841176171325}& \textbf{0.88070541}2127 \\
 25 & \textbf{0.961483705502}    & \textbf{0.841176171325}  & \textbf{0.8807054116}30 \\
 30 & \textbf{0.961483705502}    & \textbf{0.841176171325 } & \textbf{0.880705411627} \\ \hline
 BHTK& \textbf{0.961483705502}   & \textbf{0.841176171325} & \textbf{0.880705411627} \\  \hline
 22PN& \textbf{0.961483705502}   & \textbf{0.84117617132}6  & \textbf{0.880705}608981  \\ \hline \hline
  $k$ & $  \eta_{2,1}(0.1) $ &   $  \eta_{2,1}(0.3)$ & $  \eta_{2,1}(0.4)$ \\ \hline
  5 & \textbf{0.0002761309188}56    & \textbf{0.0026641}1244976   & \textbf{0.0057}2740466492 \\
 10 & \textbf{0.000276130918862}  & \textbf{0.002664146}79182 & \textbf{0.005730}06836479 \\
 15 & \textbf{0.000276130918862}    & \textbf{0.002664146863}48 & \textbf{0.00573016}433854 \\
 20 & \textbf{0.000276130918862}   & \textbf{0.00266414686353} & \textbf{0.0057301655}1089 \\
 25 & \textbf{0.000276130918862}    & \textbf{0.00266414686353}  & \textbf{0.005730165520}70 \\
 30 & \textbf{0.000276130918862}    & \textbf{0.00266414686353} & \textbf{0.00573016552076} \\  \hline
 BHTK& \textbf{0.000276130918862}   & \textbf{0.00266414686353} & \textbf{0.00573016552076} \\ \hline
 22PN& \textbf{0.000276130918862}   & \textbf{0.00266414686353} & \textbf{0.005730165}04468 \\ \hline \hline
  $k$ & $  \eta_{3,3}(0.1) $ &   $  \eta_{3,3}(0.3)$ & $  \eta_{3,3}(0.4)$ \\ \hline
  5 & \textbf{0.0127391414224}    & \textbf{0.091120}2006109   & \textbf{0.16}6992708659 \\
 10 & \textbf{0.0127391414224}  & \textbf{0.0911203195}830 & \textbf{0.167029}390871\\
 15 & \textbf{0.0127391414224}    & \textbf{0.091120319524}5 & \textbf{0.1670292}60059 \\
 20 & \textbf{0.0127391414224}   & \textbf{0.0911203195244} & \textbf{0.1670292582}39 \\
 25 & \textbf{0.0127391414224}    & \textbf{0.0911203195244}  & \textbf{0.167029258222} \\
 30 & \textbf{0.0127391414224}    & \textbf{0.0911203195244} & \textbf{0.167029258222} \\  \hline
 BHTK&\textbf{0.0127391414224}    & \textbf{0.0911203195244}  & \textbf{0.167029258222} \\ \hline
 22PN&\textbf{0.0127391414224}    & \textbf{0.09112031952}62  & \textbf{0.167029}722596  \\ \hline \hline
 $k$ & $ 10^6\times \eta_{3,2}(0.1) $ &   $ 10^4\times \eta_{3,2}(0.3)$ & $ 10^3\times  \eta_{3,2}(0.4)$ \\ \hline
  5 & \textbf{7.695632933} & \textbf{5.99470}001058   & \textbf{2.2415}1427554 \\
 10 & \textbf{7.695632933} & \textbf{5.9947011}1376 & \textbf{2.24157}466375\\
 15 & \textbf{7.695632933}   & \textbf{5.99470112089} & \textbf{2.2415769}1351 \\
 20 & \textbf{7.695632933}   & \textbf{5.99470112089} & \textbf{2.24157695}298\\
 25 & \textbf{7.695632933}    & \textbf{5.99470112089}  & \textbf{2.24157695337} \\
 30 & \textbf{7.695632933}    & \textbf{5.99470112089} & \textbf{2.24157695337} \\ \hline
 BHTK& \textbf{7.695632933}   & \textbf{5.99470112089} & \textbf{2.24157695337} \\ \hline
 22PN& \textbf{7.695632933}    & \textbf{5.9947011208}8  & \textbf{2.241576}33248\ \\ \hline \hline
 $k$ & $10^6 \times \eta_{3,1}(0.1) $ &   $10^6 \times \eta_{3,1}(0.3)$ & $10^5 \times \eta_{3,1}(0.4)$ \\ \hline
5 & \textbf{1.1829878889} & \textbf{8.212183}6172   & \textbf{1.328}87186107 \\
 10 & \textbf{1.1829878889} & \textbf{8.21218325}69 & \textbf{1.32886}658816\\
 15 & \textbf{1.1829878889}   & \textbf{8.2121832518} & \textbf{1.3288658}4761 \\
 20 & \textbf{1.1829878889}   & \textbf{8.2121832518}& \textbf{1.328865835}99\\
 25 & \textbf{1.1829878889}    & \textbf{8.2121832518}  & \textbf{1.32886583589} \\
 30 & \textbf{1.1829878889}    & \textbf{8.2121832518} & \textbf{1.32886583589} \\ \hline
 BHTK& \textbf{1.1829878889}   & \textbf{8.2121832518} & \textbf{1.32886583589} \\ \hline
 22PN& \textbf{1.1829878889}   & \textbf{8.212183251}7  & \textbf{1.328865}77283 \\ \hline \hline
\end{tabular}
\caption{Convergence of the PM$_k$ results for $\eta_{\ell,m}$, with $\ell=2,3, 1 \leq m \leq \ell$, normalized with the Newtonian luminosity $F_{22}^N$ \eqref{eq:Newton_contribution}.  The results obtained using the MST/ NUM methods and the 22PN computation in \cite{Fujita:2012cm} are displayed in the last two lines of each sub-table.}
\label{Tconvergence}
\end{table*}

\begin{table*}
\centering
\begin{tabular}{|cccc|}
\hline
 $k$ & $10^{39} \times \eta_{10,1}(0.1) $ &   $10^{31} \times \eta_{10,1}(0.3)$ & $10^{29} \times \eta_{10,1}(0.4)$ \\ \hline
5 & \textbf{9.351820404}3 & \textbf{8.6461}6460005   & \textbf{4.13}218882098\\
10 & \textbf{9.3518204044} & \textbf{8.64619838996} & \textbf{4.13187842}500\\
15 & \textbf{9.3518204044}   & \textbf{8.64619838996} & \textbf{4.13187842498}\\
20 & \textbf{9.3518204044}   & \textbf{8.64619838996}& \textbf{4.13187842498}\\
25 & \textbf{9.3518204044}    & \textbf{8.64619838996}  & \textbf{4.13187842498} \\
30 & \textbf{9.3518204044}   & \textbf{8.64619838996} & \textbf{4.13187842498} \\ \hline
BHTK& \textbf{9.3518204044}   & \textbf{8.64619838996} & \textbf{4.13187842498} \\ \hline
22PN  & \textbf{9.3518204044}    & \textbf{8.64619}786463  & \textbf{4.1}2989060363 \\ \hline
\end{tabular}
\caption{Convergence of the PM$_k$ results for $\eta_{10,1}$ normalized by the Newtonian luminosity $F_{22}^N$ \eqref{eq:Newton_contribution}. The results obtained using the MST/ NUM methods and the 22PN computation in \cite{Fujita:2012cm} are displayed in the last two lines.}
\label{Tconvergence2}
\end{table*}

Finally in Table \ref{comparison dE/dt} we compare our 30PN results for the total energy flux \eqref{eq:total_flux} against those obtained by MST/Numerical methods and  by the 22PN computation in \cite{Fujita:2012cm}. We notice that in the whole range of $v$ the results obtained with the 30PN hypergeometric representations for the total luminosity $F_{\text{tot}}$ agree with those obtained by the numerical integration and with the MST method up to 11-13 digits.

\begin{table*}
\centering
\begin{tabular}{|cccc|}
\hline
 & $v=0.1$   & $v=0.3$ & $v=0.4$ \\ \hline
HYP & 0.974719282291879  & 0.950003411484245 & 1.10991197121 \\ \hline
BHTK & 0.974719282291999   & 0.950003411484254 & 1.10991197120 \\ \hline
22PN  & 0.974719282291877   & 0.950003411477562 & 1.10990619336 \\ \hline
\end{tabular}
\caption{Total luminosity $F_{\text{tot}}$.}\label{comparison dE/dt}
\end{table*}

\section{Factorization of the waveform} \label{sec:factorization}
From these results, we have checked the goodness of the hypergeometric representation of the Confluent Heun function and for this reason it is quite important to frame such outcomes with the ones that already exist in the literature. In particular we want to compare our new resummation of tail and transcendental contributions with two existing proposals, the one of \cite{Damour:2008gu} and that will be denoted as DIN and then of \cite{Ivanov:2025ozg}, that will be indicated as ILPZ. The exposition of the main results in this Chapter will deal only with the fundamental aspects, but all the details can be consulted in \cite{Cipriani:2026xmx}. We are going to consider at the moment only the test-mass limit. The possible extension to binary systems with equal masses will be analyzed later.\\

We start with a brief overview of the proposals of the previous two references. They focus on the harmonic components of the waveform $h_{\ell m \omega}$ (in the following we are going to drop the index $\omega$) and the authors of \cite{Damour:2008gu} proposed for the first time the following factorized form for such quantity
\be \label{eq:fact_DIN}
h_{\ell m} = h_{\ell m}^{(N,\epsilon)} \, \hat{S}^{(\epsilon)}_{\text{eff}} \, T^{\text{DIN}}_{\ell m} \, \left(\rho_{\ell m}^{\text{DIN}}\right)^\ell \, e^{i \delta_{\ell m}^{\text{DIN}}}
\ee
where $\epsilon$ denotes the parity of the multipolar waveform, i.e. even for mass-generated multipoles (for which $\epsilon=0$) and odd for current-generated ones (for which $\epsilon = 1$). In the circular case, $\epsilon$ is equal to the parity of $\ell +m$, so $\epsilon = (\ell + m) \, \, \text{mod} \, 2$. $h_{\ell m}^{(N,\epsilon)}$ is the Newtonian multipolar waveform, whose expression in the probe limit is
\be
h_{\ell m}^{(N,\epsilon)} = n^{(\epsilon)}_{\ell m} \, v^{\ell + \epsilon} \, Y^{\ell-\epsilon, -m}_0 \, \left(\frac{\pi}{2}, \phi \right)
\ee
with
\be
n^{(0)}_{\ell m} = (i m)^\ell \, \frac{8 \pi}{(2 \ell +1) !!} \sqrt{\frac{(\ell+1)(\ell+2)}{\ell (\ell-1)}} \quad , \quad n^{(1)}_{\ell m} = -(i m)^\ell \, \frac{16 \pi \, i}{(2 \ell +1) !!} \sqrt{\frac{(2 \ell+1)(\ell+2) (\ell^2-m^2)}{(2 \ell -1) (\ell+1) \ell (\ell-1)}}
\ee

For what concerns $\hat{S}^{(\epsilon)}_{\text{eff}}$, it is the dimensionless effective source term and it depends on the parity of the waveform. In the even case, since the leading order source of gravitational radiation is given by the energy density, this quantity is the Effective One Body Hamiltonian per unit mass $\mu$, $\hat{H}_{\text{eff}}$. Restricting along circular orbits and at the leading order of the probe limit, $\hat{H}_{\text{eff}}$ reduces to the usual conserved energy of a test-mass $\mu$ in a Schwarzschild background of mass $M$. In other words, in reference to \eqref{eq:scaling_v}, we have
\be
\hat{S}^{(0)}_{\text{eff}} = \hat{H}_{\text{eff}} = \frac{E}{\mu} = \frac{1-2 v^2}{\sqrt{1-3v^2}}
\ee
In the odd case we have to consider the angular momentum of the system and so, always along circular trajectories and at the leading order of the test-mass limit, we have, in reference to \eqref{eq:scaling_v}
\be
\hat{S}^{(1)}_{\text{eff}} = \frac{J \, v}{\mu \, M} = \frac{1}{\sqrt{1-3v^2}}
\ee
Passing to $T_{\ell m}^{\text{DIN}}$, it is a resummed version of an infinite numbers of \textit{leading} logarithms entering the tail effects and for this reason it is referred as "tail effect" (it includes also transcendental contributions, not only the leading logarithms). Its explicit expression is
\be
T_{\ell m}^{\text{DIN}} = \frac{\Gamma \left(\ell + 1 - 2 i \, H_{\text{real}} \omega\right)}{\Gamma (\ell+1)} \, e^{\pi H_{\text{real}} \omega} \, e^{2 i H_{\text{real}} \, \omega \log \left(2 \omega r_0 \right)}
\ee
where $r_0 = 2 M$ \footnote{In \cite{Pan:2010hz} it is highlighted that $r_0$ has to be set to $2 M/\sqrt{e}$ in order to restore the correct PN results. In our case we will see that this aspect will be automatically satisfied. \label{footnote:r0}} and $H_{\text{real}}$ is the real energy of the system. The dependence on the quantum number $m$ arises from the fact that, along circular orbits, $\omega = m \Omega$. Defining the symmetric mass ratio for a binary system with masses $m_1,m_2$ as $\nu = m_1 m_2/(m_1+m_2)^2$, it turns out that
\be
H_{\text{real}} = M \sqrt{1 + 2 \nu \left(\hat{H}_{\text{eff}} -1 \right)}
\ee 
If $m_1 \equiv M \gg m_2 \equiv \mu$, then we see that $\nu \simeq \mu/M$ and at the leading order of the probe limit $\nu \to 0$ we clearly see that $H_{\text{real}} = M$. For this reason in this Section, except for the case of binary system with comparable masses, we write
\be \label{eq:tail_DIN}
T_{\ell m}^{\text{DIN}} = \frac{\Gamma \left(\ell + 1 - 2 i M \omega\right)}{\Gamma (\ell+1)} \, e^{\pi M \omega} \, e^{2 i M \omega \log \left(2 \omega r_0 \right)}
\ee
The complex quantity $(\rho_{\ell m}^{\text{DIN}})^\ell \, e^{i \delta_{\ell m}^{\text{DIN}}}$ is the remainder, which is divided in phase and amplitude. For this reason $f_{\ell m}^{\text{DIN}} = (\rho_{\ell m}^{\text{DIN}})^\ell$ is a real quantity which is PN expanded. The fact of considering as building block for the amplitude of the remainder $\rho_{\ell m}^{\text{DIN}}$ rather than $f_{\ell m}^{\text{DIN}}$ is that, if $\rho_{\ell m}^{\text{DIN}} = 1 + A v^2 + \ldots$, then it turns out that $f_{\ell m}^{\text{DIN}} = 1 + \ell \, A v^2 + \ldots$; in other words, setting $f_{\ell m}^{\text{DIN}} = (\rho_{\ell m}^{\text{DIN}})^\ell$ is an additional resummation, which produces better convergence properties of the PN expansion.\\

Passing to the new proposal of \cite{Ivanov:2025ozg}, the authors provide a tail factor that resums all the \textit{universal} logarithmic terms using EFT notions. They test it at 4PN accuracy but in \cite{Cipriani:2026xmx} we show that sub-leading logs are reabsorbed up to 7PN order included. In order to understand the ingredients that enter their tail factor, we have to quickly explain an important concept they point out. We already said that the quantum period $a$ is related to the renormalized angular momentum that we call now $\hat{\ell}$ \footnote{In the standard MST perturbation theory literature it is called $\nu$, but we do not want to confuse it with the symmetric mass ratio introduced before.} as
\be
\hat{\ell} = a- \frac{1}{2}
\ee
On the EFT side, \cite{Ivanov:2025ozg} showed that $\hat{\ell}$ precisely determines the anomalous dimension of BH multipole moments, that is called (restoring at the moment the dependence from $G$)
\be \label{eq:hat_l}
\gamma_{\ell m}^{\text{BH}} = \hat{\ell} (G M \omega, \hat{S}) - \ell
\ee
where $\hat{S} \equiv S/M^2$ is the dimensionless spin of the Kerr black hole. From the computation of $a$, then the knowledge of $\gamma_{\ell m}^{\text{BH}}$ follows. \cite{Ivanov:2025ozg} pointed out that from the anomalous dimension of BHs one can compute an universal anomalous dimension $\gamma_{\ell m}^{\text{univ}}$ that is valid for a general systems (e.g., like the case of a binary with comparable masses) that is obtained suitably expanding the BH anomalous dimension around $\hat{S} = 0$ and then replacing the mass of the black hole, $M$, with the energy of the system $H_{\text{real}}$ and the dimensionless black hole spin, $\hat{S}$, with $\mathcal{J} = J/H_{\text{real}}^2$, where $J$ is the angular momentum of the system. The universal dimension is then given by
\be \label{eq:univ_anom_dim}
\gamma_{\ell m}^{\text{univ}} = \left[\gamma_{\ell m}^{\text{BH}} (G H_{\text{real}} \, \omega , 0 ) + \partial_{\hat{S}} \gamma_{\ell m}^{\text{BH}} (G H_{\text{real}} \, \omega, 0) \mathcal{J}\right]_{G^{2 \ell +1}}
\ee
where $[\ldots]_{G^n}$ denotes an expansion through order $n$. Following in spirit the proposal of  \cite{Ivanov:2025ozg} we will then use the so computed universal anomalous dimension to define the quantity
\be \label{eq:hat_hat_l}
\hat{\hat{\ell}} = \ell + \gamma^{\text{univ}}_{\ell m}
\ee
Now, for the test-mass case in the Schwarzschild scenario, we clearly see that $\gamma_{\ell m}^{\text{univ}} = \gamma_{\ell m}^{\text{BH}}$, implying that $\hat{\hat{\ell}} = \hat{\ell}$. When passing to the comparable mass case, this equality is not true and the more subtle passage from $a$ inside the new tail factor to $\hat{\hat{\ell}}$ will be quickly explained in Section \ref{sec:comparable_mass}. Details can be found in \cite{Cipriani:2026xmx}.\\
At this point we are ready to write the proposal of \cite{Ivanov:2025ozg} for the factorization of the waveform. The structure is the same as \eqref{eq:fact_DIN}, in particular

\be \label{eq:fact_ILPZ}
h_{\ell m} = h_{\ell m}^{(N,\epsilon)} \, \hat{S}^{(\epsilon)}_{\text{eff}} \, T^{\text{ILPZ}}_{\ell m} \, \left(\rho_{\ell m}^{\text{ILPZ}}\right)^\ell \, e^{i \delta_{\ell m}^{\text{ILPZ}}}
\ee
From the reference, we can find the tail factor written in the following form \footnote{The argument of the $\log$ function at exponent reported in the reference is $2 \omega r_{\text{orb}}$, but it is a typo, as it can be seen from the final PN expansion of the waveforms.}
\be \label{eq:tail_ILPZ_radiative}
T^{\text{ILPZ}}_{\ell m} = \frac{\Gamma \left(\hat{\hat{\ell}}+1-2 i M \omega \right)}{\Gamma \left(\hat{\hat{\ell}}+1\right)} \, e^{\pi M \omega} \, (\omega r_{\text{orb}})^{\hat{\hat{\ell}}-\ell} \, e^{2 i M \omega \log \left(2 m \phi_0^{\text{ILPZ}}\right)} \, e^{i \frac{\pi}{2}(\ell - \hat{\hat{\ell}})}
\ee
where $\phi_0^{\text{ILPZ}} = e^{11/12-\gamma_E}/4$. As for the DIN factorization, the dependence on the quantum number $m$ arises from the fact that, along circular orbits, $\omega = m \Omega$. $\left(\rho_{\ell m}^{\text{ILPZ}}\right)^\ell \, e^{i \delta_{\ell m}^{\text{ILPZ}}}$ is the remainder, expanded at 4PN in the same reference and we are going to expand it further. $r_{\text{orb}}$ is the orbital scale that comes out from the running of the RG evolution of the multipoles. The EFT description does not allow to fix the exact coefficient that relates this quantity with the renormalization energy scale and for this reason all the transcendental contributions are not entirely captured according to this proposal. By the way, the authors set $\omega r_{\text{orb}}$ to the following expression (in part by hand in order to match with \cite{Faye:2014fra})
\be
\omega \, r_{\text{orb}} = \frac{(M \Omega)^{\frac{1}{3}}}{\frac{1}{4} e^{\frac{11}{12} - \gamma_E}} = \frac{x^{\frac{1}{2}}}{\frac{1}{4} e^{\frac{11}{12} - \gamma_E}}
\ee
It must be said that the expression of the tail factor in \eqref{eq:tail_ILPZ_radiative} is valid in the so called \textit{radiative coordinates}, while the form of the DIN tail factor in \eqref{eq:tail_DIN} is written in the \textit{Schwarzschild coordinates}. In order to pass from the former to the latter, we have to multiply by the factor $e^{2 i M \omega \log(\Omega/\Omega_0)}$, where
\begin{equation}
\Omega_0 = \frac{e^{\frac{11}{12}-\gamma_E}}{4 r_0} = \frac{e^{\frac{17}{12}-\gamma_E}}{8 M}
\end{equation}
and here we have used the fact that $r_0 = 2 M/\sqrt{e}$. The reason for this transformation is explained in \cite{Cipriani:2026xmx}. So, at the end we find
\be
T^{\text{ILPZ, Sch}}_{\ell m} = \frac{\Gamma \left(\hat{\hat{\ell}}+1-2 i M \omega \right)}{\Gamma \left(\hat{\hat{\ell}}+1\right)} \, e^{\pi M \omega} \, (\omega r_{\text{orb}})^{\hat{\hat{\ell}}-\ell} \, e^{2 i M \omega \log \left(2 \omega r_0 \right)} \, e^{i \frac{\pi}{2}(\ell - \hat{\hat{\ell}})}
\ee
Since in our construction the form of the tail is automatically in the Schwarzschild coordinates, we will work in these ones and for this reason we are going to omit the specification Sch for the tail factor.

\subsection{New factorization}
In this context we want to see the factorization of the waveform that comes out using the new hypergeometric representation we discussed in the previous Sections. The relevant differences are the following: the $\rho_{\ell m}$ functions obtained with the two previous methods contain, starting from a given PN order that depends on $\ell$, some logarithms and transcendental terms, while instead the new functions that we are going to develop contain only hypergeometric functions and so, after being PN expanded, are polynomials in $v^2$. The other novelty concerns the fact that now we have a new tail factor.\\
In order to understand the origin of the new factorization, we have to recall the structure of $\mathfrak{R}_{\text{in},\omega}(r)$, from which we can infer the body of the waveform through \eqref{eq:Z_lm_simplified_circular}. From the equation \eqref{rgg}, we express all the ingredients that appear there in terms of the gravity variables

\begin{align}
\mathfrak{R}_{\text{in},\omega}(r) & = \frac{(4 i \hat{\omega})^{2 i \hat{\omega}} \, e^{- i \pi \left(\frac{5}{2} +a + 2 i \hat{\omega} \right)} \Gamma \left(a - 2 i \hat{\omega} - \frac{3}{2}\right)}{\Gamma \left(1+2 a \right)} \frac{1}{1- (4 i \hat{\omega})^{2a} \gamma_{\text{inst}} \frac{e^{- 2 i a \pi} \Gamma(1-2a)^2 \Gamma \left(\frac{5}{2} + a - 2 i \hat{\omega} \right) \Gamma \left(\frac{1}{2} + a - 2 i \hat{\omega} \right) \Gamma \left(-\frac{3}{2} + a - 2 i \hat{\omega} \right)}{\Gamma(1+2a)^2 \Gamma \left(\frac{5}{2} - a - 2 i \hat{\omega} \right) \Gamma \left(\frac{1}{2} - a - 2 i \hat{\omega} \right) \Gamma \left(-\frac{3}{2} - a - 2 i \hat{\omega} \right)}} \times \notag \\
&  \times \left[R_-(r) - (4 i \hat{\omega})^{2a} \gamma_{\text{inst}} \frac{\Gamma(1-2a) \Gamma \left(\frac{5}{2} + a - 2 i \hat{\omega}\right) \Gamma \left(\frac{1}{2} + a - 2 i \hat{\omega} \right)}{\Gamma(1+2a) \Gamma \left(\frac{5}{2} - a - 2 i \hat{\omega} \right) \Gamma \left(\frac{1}{2} - a - 2 i \hat{\omega} \right)} R_+(r) \right] 
\end{align}
where we have introduced the quantity $\hat{\omega} \equiv M \omega$ for simplicity. We have to recall that the functions $R_{\pm}(r)$ contain, respectively, the factors $(2 i \omega r)^{-1-2 i \hat{\omega}} G^0_{\pm}(r)\simeq (2 i \omega r)^{-1-2 i \hat{\omega}} \,(2 i \omega r)^{\frac{5}{2}\mp a + 2 i \hat{\omega}} = (2 i \omega r)^{\frac{3}{2} \mp a}$ (see \eqref{defsigma} with $m_3 = -2-2 i \hat{\omega}$). These factors, once PN expanded, generate logarithmic and transcendental contributions. It is thus useful to separate them from the rest and define the functions \textit{without} it as
\be
\mathcal{S}_\alpha (r) \equiv (2 i \omega r)^{-\left(\frac{3}{2} - \alpha a \right)} R_{\alpha}(r)
\ee
in such a way that they contain only hypergeometric functions. At this point we collect the factor $(2 i \omega r)^{\frac{3}{2} + a}$ in front of the square brackets, so to recast the equation above as
\begin{align}
\label{eq:R_in_fin_explicit}
& \mathfrak{R}_{\text{in},\omega}(r) = \notag \\
& = \textcolor{red}{\overbrace{(4 \hat{\omega})^{2 i \hat{\omega}} (2 \omega r)^{\frac{3}{2}+a}}^{T^{1,\text{log}}}} \textcolor{blue}{\overbrace{\frac{ \, e^{\pi \hat{\omega}} e^{- i \frac{\pi}{2} \left(\frac{7}{2} +a\right)} \Gamma \left(a - 2 i \hat{\omega} - \frac{3}{2}\right)}{\Gamma \left(1+2 a \right)}}^{T^{1,\text{trnsc}}}} \frac{1}{1-  \gamma_{\text{inst}}\textcolor{red}{\underbrace{(4 \hat{\omega})^{2a}}_{T^{2,\text{log}}}} \textcolor{blue}{\tiny{\underbrace{\frac{e^{-i a \pi} \Gamma(1-2a)^2 \Gamma \left(\frac{5}{2} + a - 2 i \hat{\omega} \right) \Gamma \left(\frac{1}{2} + a - 2 i \hat{\omega}\right) \Gamma \left(-\frac{3}{2} + a - 2 i \hat{\omega} \right)}{\Gamma(1+2a)^2 \Gamma \left(\frac{5}{2} - a - 2 i \hat{\omega} \right) \Gamma \left(\frac{1}{2} - a - 2 i \hat{\omega} \right) \Gamma \left(-\frac{3}{2} - a - 2 i \hat{\omega} \right)}}_{T^{2,\text{trnsc}}}}}}
 \notag \\
& \times \left[\mathcal{S}_-(r) - \gamma_{\text{inst}}\textcolor{red}{\overbrace{\left(\frac{2 M}{r}\right)^{2a}}^{\tilde{T}^{\text{log}}}}  \textcolor{blue}{\overbrace{\frac{\Gamma(1-2a) \Gamma \left(\frac{5}{2} + a - 2 i \hat{\omega} \right) \Gamma \left(\frac{1}{2} + a - 2 i \hat{\omega} \right)}{\Gamma(1+2a) \Gamma \left(\frac{5}{2} - a - 2 i \hat{\omega} \right) \Gamma \left(\frac{1}{2} - a - 2 i \hat{\omega} \right)}}^{\tilde{T}^{\text{trnsc}}}} \mathcal{S}_+(r) \right]
\end{align}
It is important to highlight what the various pieces of this function bring once PN-expanded within \eqref{eq:Z_lm_simplified_circular}. In particular, it is possible to show that the terms highlighted in red give rise to all logarithmic terms (from which the apex log). By contrast, those in blue only produce transcendental numbers, i.e. those terms with $(\pi,\gamma_E)$ etc (from which the apex trnsc). Finally, the parts in black give only polynomials with rational coefficients (notice that $\gamma_{\text{inst}}= 1 + \mathcal{O}((M \omega)^2)$). \\
The structure of \eqref{eq:R_in_fin_explicit} remains unchanged under the action of the $r$-derivatives. For this reason, we propose to write the waveform in the following factorized form that is structurally analogous to the factorization of the previous two methods but improves it thanks to the new analytic knowledge
\be
h_{\ell m} =  h_{\ell m}^{(N,\epsilon)} \, \hat{S}^{(\epsilon)}_{\text{eff}} \, T^{\text{New}}_{\ell m} \, \left(\rho_{\ell m}^{\text{New}}\right)^\ell \, e^{i \delta_{\ell m}^{\text{New}}} \left(1 + \tilde{T}_{\ell m} \, \tilde{f}_{\ell m} e^{i \tilde{\delta}_{\ell m}}\right)
\ee
where
\begin{align} \label{eq:external_tail_new}
& T^{\rm New}_{\ell m} = \notag \\
 & = \left(-\frac{e^{- \mathrm{i} \ell \pi} (2 \mathrm{i} \omega r)^{2+\ell} \Gamma(\ell-1)}{\Gamma(2 \ell +2)}\right)^{-1} \frac{\textcolor{red}{T_{\ell m}^{1,\text{log}}} \, \, \textcolor{blue}{T_{\ell m}^{1,\text{trnsc}}}}{1- \gamma_{\text{inst}} \textcolor{red}{T_{\ell m}^{2,\text{log}}} \, \, \textcolor{blue}{T_{\ell m}^{2,\text{trnsc}}}} = \textcolor{blue}{\frac{\Gamma \left(\hat{\ell} - 1 - 2 i \hat{\omega} \right)}{\Gamma \left(2 \hat{\ell} + 2\right)} e^{\pi \hat{\omega}}} \textcolor{red}{(2 \omega r)^{\hat{\ell} - \ell} e^{2 i \hat{\omega} \log \left(4 \hat{\omega} \right)}} \textcolor{blue}{e^{\frac{i \pi}{2} \left(\ell - \hat{\ell} \right)}} \times \notag \\
& \times \frac{\Gamma \left(2 \ell + 2\right)}{\Gamma \left(\ell -1\right)}  \frac{1}{1- \textcolor{red}{(4 \hat{\omega})^{2\hat{\ell} +1}} \gamma_{\text{inst}} \textcolor{blue}{\frac{e^{-i \left(\hat{\ell} + \frac{1}{2}\right) \pi} \Gamma(-2\hat{\ell})^2 \Gamma \left(\hat{\ell} + 3 - 2 i \hat{\omega} \right) \Gamma \left(\hat{\ell} + 1 - 2 i \hat{\omega} \right) \Gamma \left(\hat{\ell} - 1 - 2 i \hat{\omega} \right)}{\Gamma(2\hat{\ell}+2)^2 \Gamma \left(2 - \hat{\ell} - 2 i \hat{\omega} \right) \Gamma \left(-\hat{\ell} - 2 i \hat{\omega} \right) \Gamma \left(-2 - \hat{\ell} - 2 i \hat{\omega} \right)}}}
\end{align}
\be \label{eq:internal_tail}
\tilde{T}_{\ell m} = - \gamma_{\text{inst}} \textcolor{red}{\tilde{T}_{\ell m}^{\text{log}}} \textcolor{blue}{\tilde{T}_{\ell m}^{\text{trnsc}}}  =  - \gamma_{\text{inst}} \textcolor{red}{\left(\frac{2M}{r}\right)^{2\hat{\ell}+1}} \, \textcolor{blue} {\frac{\Gamma(-2\hat{\ell}) \Gamma \left(\hat{\ell} + 3 - 2 i \hat{\omega} \right) \Gamma \left(\hat{\ell} + 1 - 2 i \hat{\omega} \right)}{\Gamma(2 \hat{\ell} + 2) \Gamma \left(2 - \hat{\ell} - 2 i \hat{\omega} \right) \Gamma \left(-\hat{\ell} - 2 i \hat{\omega} \right)}}
\ee
Notice that the last tail was not taken in consideration in the two previous factorizations and for this reason we have not put the apex New on it. For what concerns the remainders, we are able to furnish expressions of $\rho_{\ell m}^{\text{New}}$ and $\tilde{f}_{\ell m}$, which contain only hypergeometric function and so produce only polynomial after being PN expanded. In particular it turns out that
\be
(\rho_{\ell m}^{\text{New}})^\ell = \left|\frac{1}{\hat{S}^{(\epsilon)}_{\text{eff}} \, b^{\epsilon}_{\ell m} \, \left[- i \ell \left(\frac{J}{\mu r}\right)_{\text{Lead}}\right]^\epsilon} \, \sum_{n=0}^2 A_n \, \mathcal{R}^-_n\right| \qquad , \qquad \tilde{f}_{\ell m} = \left| \left(-\frac{\ell+1}{\ell}\right)^{-\epsilon} \frac{\sum_{n=0}^2 A_n \mathcal{R}^+_n}{\sum_{n=0}^2 A_n \mathcal{R}^-_n} \right|
\ee
where the factors $A_n$ are the ones defined in \eqref{eq:A_coefficients}, then
\be
\mathcal{R}^\alpha_n = (2 i \omega r)^{-\left(\frac{3}{2} - \alpha a\right)} \, r^n \, \partial_r^n R_\alpha(r) |_{r=r_0, \omega = m \Omega, a = \hat{\ell} + \frac{1}{2}}
\ee
and by $(J/(\mu r))_{\text{Lead}}$ we mean that we have to consider the leading order term when this quantity is PN expanded, i.e. setting $r=M/v^2$ it turns out that $(J/(\mu r))_{\text{Lead}} = v$.\\
If we want to PN expand all the ingredients entering the waveform $h_{\ell m}$ in order to compare with the results known in literature, we have to replace $\omega r = m v$ and $M \omega \equiv \hat{\omega} = m v^3$, even inside $\hat{\ell}$. By the way, as we are going to show in Section \ref{sec:results_resummation}, we are going to PN expand only the functions $\rho_{\ell m}^{\text{New}}$ and $\tilde{f}_{\ell m}$, keeping the tail factors \eqref{eq:external_tail_new}-\eqref{eq:internal_tail} in their closed form and with this we evaluate the total energy flux.

\subsection{Comparison of the tail factors of the three proposals}
Before proceeding with the comparison of the final results, it is quite useful to report the tail factors of the three methods $T_{\ell m}^{\text{DIN}}, T_{\ell m}^{\text{ILPZ}}$ and $T_{\ell m}^{\text{New}}$ together, in such a way that we can easily see their similarities and differences. We do not take into consideration the tail $\tilde{T}_{\ell m}$ since there is not its counterpart in the two previous descriptions.\\
Using the same colors for the log and transcendental contributions, we have
\begin{align}
& T^{\rm DIN}_{\ell m} = \textcolor{blue}{\frac{\Gamma \left(\ell + 1 - 2 i M \omega \right)}{\Gamma \left(\ell+1\right)} e^{\pi M \omega}} \textcolor{red}{e^{2 i M \omega \log \left(2 \omega r_0 \right)}} \\
& T^{\rm ILPZ}_{\ell m} = \textcolor{blue}{\frac{\Gamma \left(\hat{\hat{\ell}} + 1 - 2 i M \omega \right)}{\Gamma \left(\hat{\hat{\ell}} + 1\right)} e^{\pi M \omega}} \textcolor{red}{(\omega r_{\rm orb})^{\hat{\hat{\ell}} - \ell} e^{2 i M \omega \log \left(2 \omega r_0\right)}} \textcolor{blue}{e^{\frac{i \pi}{2} \left(\ell - \hat{\hat{\ell}} \right)}}
\end{align} 
\begin{align} \label{eq:tail_new_summary}
T^{\rm New}_{\ell m} = & \textcolor{blue}{\frac{\Gamma \left(\hat{\ell} - 1 - 2 i M \omega \right)}{\Gamma \left(2 \hat{\ell} + 2\right)} e^{\pi M \omega}} \textcolor{red}{(2 \omega r)^{\hat{\ell} - \ell} e^{2 i M \omega \log \left(4 M \omega \right)}} \textcolor{blue}{e^{\frac{i \pi}{2} \left(\ell - \hat{\ell} \right)}} \frac{\Gamma \left(2 \ell + 2\right)}{\Gamma \left(\ell -1\right)}  \times \notag \\
& \times \frac{1}{1- \textcolor{red}{(4 M \omega)^{2\hat{\ell} +1}} \gamma_{\text{inst}} \textcolor{blue}{\frac{e^{-i  \left(\hat{\ell} + \frac{1}{2}\right) \pi} \Gamma(-2\hat{\ell})^2 \Gamma \left(\hat{\ell} + 3 - 2 i M \omega \right) \Gamma \left(\hat{\ell} + 1 - 2 i M \omega \right) \Gamma \left(\hat{\ell} - 1 - 2 i M \omega \right)}{\Gamma(2\hat{\ell}+2)^2 \Gamma \left(2 - \hat{\ell} - 2 i M \omega \right) \Gamma \left(-\hat{\ell} - 2 i M \omega \right) \Gamma \left(-2 - \hat{\ell} - 2 i M \omega \right)}}}
\end{align}
where $r_0 = 2M/\sqrt{e}$ for what discussed in the footnote~\ref{footnote:r0}. The factor that appears in the second line of $T_{\ell m}^{\text{New}}$ is not present in the other two, but it starts to contributes at higher PN order (in particular at 6PN) and for this reason we do not consider it in the comparison with $T_{\ell m}^{\text{DIN}}$ and $T_{\ell m}^{\text{ILPZ}}$. For the first factor, instead, we see that some of the terms that appear in it are the same as for the other two tails, but other ones are quite different. For example, while the arguments of the gamma functions in $T_{\ell m}^{\text{DIN}}$ and $T_{\ell m}^{\text{ILPZ}}$ have the same structure, with the only difference of the promotion of $\ell$ to $\hat{\hat{\ell}}$, instead the gamma functions that are present in the new tail are different. This will have an interesting consequence when we analyze the phases of the waveforms.\\
For the comparison between $T_{\ell m}^{\text{ILPZ}}$ and $T_{\ell m}^{\text{New}}$, the former is automatically defined with $\hat{\hat{\ell}}$ and the latter with $\hat{\ell}$ and these two quantities are not the same in general. For this reason, when discussing the extension to comparable mass case in Section \eqref{sec:comparable_mass}, the passage from $\hat{\ell}$ to $\hat{\hat{\ell}}$ is more subtle, but for the case we are analyzing now, i.e. test-mass limit in the Schwarzschild background, the two quantities are the same.\\
For what concerns the comparison between $T_{\ell m}^{\text{DIN}}$ and $T_{\ell m}^{\text{New}}$, we just want to focus the attention to the phase. For what motivated in footnote~\ref{footnote:r0}, we have already set $r_0=2M/\sqrt{e}$, but now let us consider the original value for this constant in the DIN formalism, i.e. $r_0 = 2M$. If we pass to compute the total phase of the waveform $h_{\ell m}$, it is possible to see that the final result obtained with the ingredients derived from the hypergeometric representation is compatible with the known PN results in literature, differently from the DIN case, in which we have to consider as $r_0$ the corrected quantity $r_0 = 2 M /\sqrt{e}$ (as noticed in \cite{Pan:2010hz}). For example, for $h_{22}$ it turns out that
\begin{align}
& \delta \left(T_{22}^{\text{New}}\right) = 4 \left(\gamma_E + 3 \log (2) + 3 \log (v) \right) v^3 + \frac{428 \pi}{105} v^6 + \left(\frac{1712 \pi^2}{315} - \frac{64}{3} \zeta(3) \right) + \mathcal{O}(v^{12}) \label{eq:phase_T22_new}\\
& \delta_{22}^{\text{New}} = -\frac{17}{3} v^3 - \frac{259}{81} v^9 + \mathcal{O} (v^{15}) \label{eq:phase_new} \\
& \delta \left(T_{22}^{\text{DIN}}\right) = \left(-\frac{11}{3} + 4 \gamma_E + 12 \log(2) + 12 \log(v) \right) v^3 + \left(24 - \frac{64}{3}  \zeta(3) \right) v^9 + \mathcal{O}(v^{15}) \label{eq:phase_T22_DIN} \\
& \delta_{22}^{\text{DIN}} = \frac{7}{3} v^3 + \frac{428 \pi}{105} v^6 + \left(\frac{1712 \pi^2}{315} - \frac{2203}{81}\right) v^9 + \mathcal{O}(v^{12}) \label{eq:phase_DIN}
\end{align}
Summing \eqref{eq:phase_T22_new} with \eqref{eq:phase_new} and \eqref{eq:phase_T22_DIN} with \eqref{eq:phase_DIN}, we obtain
\begin{align}
& \delta \left(T_{22}^{\text{New}}\right) + \delta_{22}^{\text{New}} = \left(-\frac{17}{3} + 4 \gamma_E + 12 \log(2 v) \right) + \frac{428 \pi}{105} v^6 + \left(\frac{1712 \pi^2}{315} - \frac{259}{81} - \frac{64}{3} \zeta(3) \right) v^9 + \mathcal{O}(v^{12}) \notag \\
& \delta \left(T_{22}^{\text{DIN}}\right) + \delta_{22}^{\text{DIN}} = \left(-\frac{11}{3} + 4 \gamma_E + 12 \log(2 v) \right) + \frac{428 \pi}{105} v^6 + \left(\frac{1712 \pi^2}{315} - \frac{259}{81} - \frac{64}{3} \zeta(3) \right) v^9 + \mathcal{O}(v^{12})
\end{align}
and the correct PN result in literature is the first. In the second the correction is reached if we set $r_0 = 2 M /\sqrt{e}$. Notice how, in the new factorization, all the $\pi$ and $\gamma_E$ contributions come only from the expansion of the phase of $T_{22}^{\text{New}}$, differently from the DIN prescription where some transcendental terms also come from the phase $\delta_{22}^{\text{DIN}}$. This pattern continues to hold also for all the other modes. \\
All other details in the comparison between these factorizations can be consulted in \cite{Cipriani:2026xmx}.

\subsection{Comparison of the final results between the three methods} \label{sec:results_resummation}
\begin{figure}
    \centering
    \includegraphics[width=0.49\textwidth]{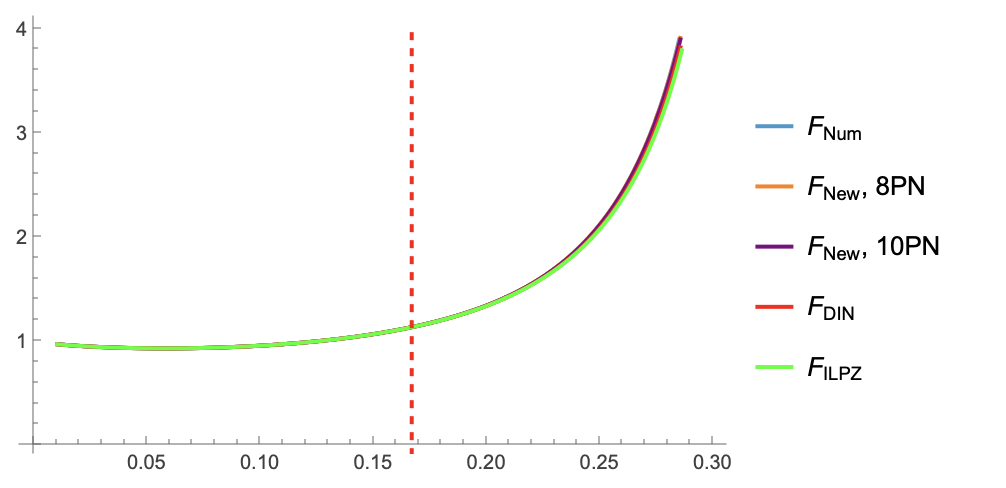}
    \includegraphics[width=0.49\textwidth]{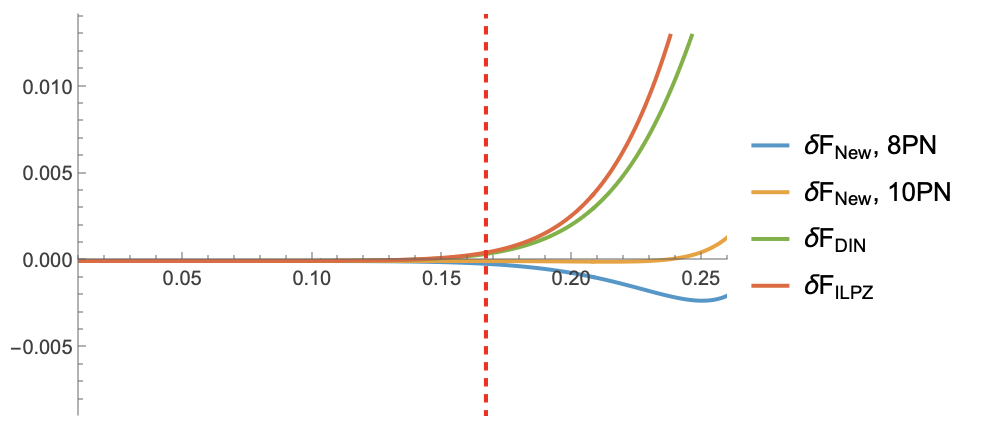}
    \includegraphics[width=0.49\textwidth]{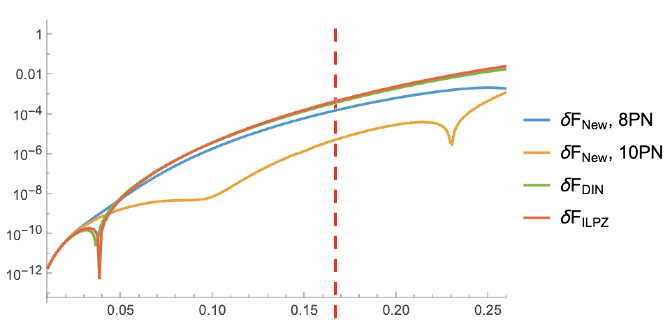}
    \caption{Total energy flux computed with the factorization of DIN and ILPZ, with the $\rho_{\ell m}$ functions expanded at 8PN, and with the new hypergeometric representation, with the $\rho_{\ell m}$ and $\tilde{f}_{\ell m}$ expanded at 8PN and 10PN, compared against the numerical solution. At the top-left we have the total luminosity, obtained by summing all the modes up to $\ell=8$, as a function of $X = (M \Omega)^{\frac{2}{3}}=v^2$. At the top-right we have the relative error of the four cases against the numerical solution and at the bottom we have the same relative error but plotted in the logarithmic scale. The dashed red line corresponds to the LSO, $X=\sqrt{\frac{1}{6}}$.}
    \label{plot:final_results_resummation}
\end{figure}

At this point we are left to see how our new resummation of all the logarithms and transcendental factors performs in the computation of the total energy flux, comparing it with the analogous outcomes coming from the other two prescriptions.\\
First of all we expand the $\rho_{\ell m}^{\text{New}}$ and $\tilde{f}_{\ell m}$, both at 8PN and at 10PN, becoming polynomials in $v^2$ and we leave the two tail factors $T_{\ell m}^{\text{New}}$ and $\tilde{T}_{\ell m}$ in their closed form. After this, we compute the luminosity summing all the modes up to $\ell = 8$ and then we repeat the same procedure also for the other two methods, where the $\rho_{\ell m}^{\text{DIN}}$ and $\rho_{\ell m}^{\text{ILPZ}}$ are expanded at 8PN, and we keep $T_{\ell m}^{\text{DIN}}$ and $T_{\ell m}^{\text{ILPZ}}$ in their closed form. All the final results are compared against the numerical solution, already known in literature. The reason why we consider, for our new method, both 8PN and 10PN is that we want, first of all, to check if our resummation procedure produces results that are better than the ones coming from the other two procedure. After having shown this, it is quite interesting to see if, increasing the PN order, the relative error against the numerical solution decreases or not. All of this is shown in Figure \ref{plot:final_results_resummation}. On the horizontal axis we have $X \equiv (M \Omega)^{\frac{2}{3}} = v^2$, where the last equality is valid only for circular case and for the third Kepler's law \eqref{eq:third_Kepler}.

As we can see, the result coming from our prescription gives the lowest errors. In particular, focusing on the last stable orbit (LSO) for simplicity, we have
\be
\left|\delta F_{\text{New}}^{\text{8PN}}\right|_{\text{LSO}} = 1.7 \cdot 10^{-4} \quad , \quad \left |\delta F_{\text{New}}^{\text{10PN}}\right|_{\text{LSO}} = 5.7 \cdot 10^{-6} \quad , \quad \left|\delta F_{\text{DIN}} \right|_{\text{LSO}} = 3.9 \cdot 10^{-4} \quad , \quad  \left|\delta F_{\text{ILPZ}} \right|_{\text{LSO}} = 2.3 \cdot 10^{-3}
\ee
If we leave the entire hypergeometric functions instead of their PN expansions for $\rho_{\ell m}^{\text{New}}$ and $\tilde{f}_{\ell m}$, exploiting the results of the previous Sections, we are able to evaluate the relative error at the LSO with $k=20$ and we find that
\be
\left| \delta F_{\text{Hyp}}^{k=20} \right|_{\text{LSO}} = 5.7 \cdot 10^{-7}
\ee 
Notice that we have normalized the total energy flux of each of the three methods by the Newtonian contribution $F_{22}^N$, which we can now directly evaluate thanks to the factorization we wrote before. Indeed it turns out that
\be
F_{22}^N = \frac{(2 \Omega)^2}{8 \pi} \left|h^{(N,0)}_{22}\right|^2 \underset{\Omega=\frac{v^3}{M}}{=} \frac{4 v^6}{8 \pi \, M^2} \frac{64}{5} \pi v^4 = \frac{32}{5} \frac{v^{10}}{M^2}
\ee
in agreement with \eqref{eq:Newton_contribution}.

\subsection{Extension to binary systems with comparable mass} \label{sec:comparable_mass}
In the light of the results in the test-mass limit, it is natural to apply the new method also to the comparable mass case. In principle, one would like to work with functions that are highly accurate in the test-mass limit (i.e., they are 10PN accurate) and then incorporate all the known PN information with the $\nu$ dependence. This information is currently limited to 4PN in the $(2,2)$ mode~\cite{Blanchet:2022vsm,Blanchet:2023sbv,Blanchet:2023bwj}  and up to 3PN in  the subdominant modes. Our goal is thus to start from the 4PN-expanded waveform and compute the corresponding $\nu$-corrections to the $\rho_{\ell m}$ in the PN sense.\\
Since we work at 4PN accuracy, we have to worry only about factoring out $\hat{S}_{\rm eff}^{(\epsilon)}$ and the external tail $T_{\ell m}^{\text{New}}$, since the internal one, $\tilde{T}_{\ell m}$ starts at 5PN order. Furthermore, for $T_{\ell m}^{\text{New}}$ we consider only the contributions coming from $T_{\ell m}^{1,\text{log}}$ and $T_{\ell m}^{1,\text{trnsc}}$ (the first line of \eqref{eq:tail_new_summary}), since the denominator $1-\gamma_{\text{inst}} \, T_{\ell m}^{2,\text{log}} \, T_{\ell m}^{2,\text{trnsc}}$ contributes from 6PN. The procedure to obtain the $\nu$-dependence in the $\rho_{\ell m}$'s is thus the standard one of \cite{Damour:2008gu}, but upgraded to the approach of \cite{Ivanov:2025ozg} because of the presence of the universal anomalous dimension. In practical terms, the technical steps are analogous to those of \cite{Ivanov:2025ozg}, the only difference residing in the fact that $T_{\ell m}^{\rm ILPZ}$ is replaced by $T_{\ell m}^{\text{New}}$. In particular, since $\nu\neq 0$, it is not true anymore that $\hat{\ell}$ in \eqref{eq:hat_l}, that naturally appears inside $T_{\ell m}^{\text{New}}$, is the same as $\hat{\hat{\ell}}$ in \eqref{eq:hat_hat_l}, which is present instead inside $T_{\ell m}^{\text{ILPZ}}$. The important step hence is to replace $\hat{\ell}$ with $\hat{\hat{\ell}}$ and the way we do it is the following: we compute the quantum period $a$ as described many times using the dictionary coming from the Kerr spacetime and then we keep only the terms that are linear in the spin, as it is evident from the definition of the universal anomalous dimension in \eqref{eq:univ_anom_dim}. For example, for $\ell= m = 2$ (which is the only case we are interested here, as we are going to explain) we obtain
\be \label{eq:a_Kerr}
\gamma_{2 2}^{\rm univ} = -\frac{214}{105} \epsilon^2 + \frac{4 \hat{S}}{3} \epsilon^3 - \frac{3390466}{1157625} \epsilon^4 + \frac{763726 \hat{S}}{99225} \epsilon^5 + \mathcal{O}(\epsilon^6)
\ee
where $\epsilon = M \omega$ and $\hat{S} \equiv S/M^2$ is the dimensionless spin of the Kerr black hole. Once done this, we have to apply the following replacements \footnote{Actually these have to be done only in the parts of all the tail contributions that are universal. It should be better understood and paves the way to future studies. I thank Julio Parra-Martinez for the discussion on this aspect.}
\be
M \to H_{\text{real}} \qquad , \qquad \hat{S} \to \mathcal{J} = \frac{J}{H_{\text{real}}^2}
\ee
where $H_{\text{real}}$ and $J$ are the total energy and angular momentum of the binary system, whose 5PN and 4PN accuracy expressions known in literature are respectively reported in Appendix \ref{app:energy_angularmom_4PN}. In this way the quantity $\gamma_{22}^{\text{univ}}$ written above is consistent with (24) of \cite{Ivanov:2025ozg}, setting $m=2$. The reason why the universal anomalous dimension is determined in the way explained before is that the universal tail contributions to the binary waveform are captured by the free-wave propagation in the Kerr background linearized in the spin sourced by the binary. The universality arises from the fact that this background is the universal part of the asymptotic metric of all compact gravitating sources \cite{Ivanov:2025ozg}. Implementing the usual factorization of the multipoles, one finds the $\nu$-dependent terms in the $\rho_{\ell m}^{\text{New}}$'s, that are listed in Appendix E of \cite{Cipriani:2026xmx}. The case of $\rho_{22}^{\text{New}}$ is also shown in \ref{app:energy_angularmom_4PN}.\\

A thorough evaluation of the performance of the new resummed waveform in the context of gravitational wave modeling purposes would require the complete implementation of the full waveform within, for example, \texttt{TEOBResumS-Dali} and assess the phasing performance with Numerical Relativity (NR) simulations. This analysis requires dedicated long work and as such is postponed to the future. Here, to get a first idea of the effects of the new resummation with respect to what implemented in \texttt{TEOBResumS-Dali} \cite{Nagar:2024oyk}, we consider the EOB adiabatic dynamics along circular orbits up to the LSO and evaluate $|\hat{h}^{(0)}_{22}(X)|$ on top of it, where by $\hat{h}_{\ell m}^{(\epsilon)}$ we mean the waveform without the Newtonian contribution
\be
h_{\ell m} = h_{\ell m}^{(N,\epsilon)} \, \hat{h}^{(\epsilon)}_{\ell m}
\ee
Let us specify the elements of the EOB dynamics. The effective Hamiltonian along circular orbits is now written as
\be
\hat{H}_{\rm eff}=\sqrt{A(u)(1+J^2(u) \, u^2)} \ ,
\ee
where $u=r^{-1}$, the $A(u)$ function (not to be confused with the coefficients $A_i$ in \eqref{eq:A_coefficients}) is taken from (6) of \cite{Nagar:2024oyk} and resummed  as in (8) therein. It depends on the effective 5PN function $a_6^c(\nu)$ that is informed by NR simulations as $a_6^c(\nu)=208.19\nu^2-318.26\nu+34.85$ (see (36) of \cite{Nagar:2024oyk} and related discussion). The circular orbits are determined by the condition $\partial_u\{A(u)[1+J^2(u) \, u^2]\}=0$, which leads to the following representation of the squared angular momentum 
\be
J^2(u) = -\dfrac{A'(u)}{(u^2 A(u))'} \ ,
\ee
where the prime stands for $\mathrm{d}/\mathrm{d}u$. From one of the Hamilton's equations we obtain the orbital frequency in the circular case as function of $u$
\be \label{eq:Omega_vs_u}
\Omega(u) = \dfrac{M A(u) J(u) u^2}{H_{\rm real}\hat{H}_{\rm eff}} \ ,
\ee
and from $\Omega(u)$ one obtains the frequency parameter $X=(M\Omega)^{2/3}$.\\ 

\begin{figure}[H]
    \centering
    \includegraphics[width=0.49\textwidth]{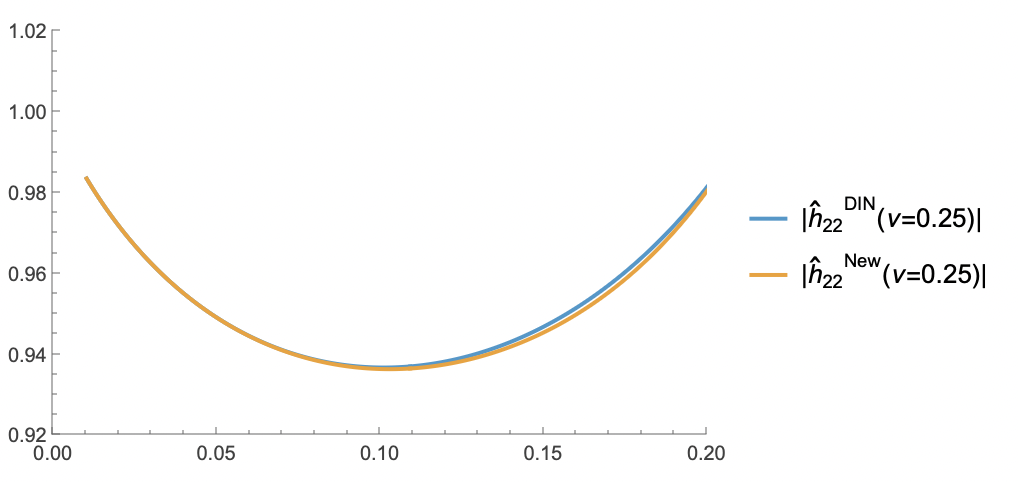}
    \includegraphics[width=0.49\textwidth]{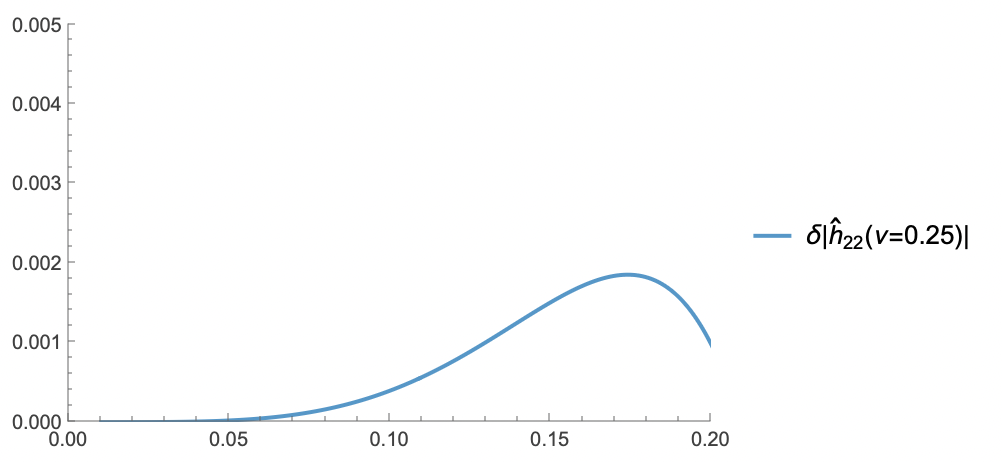}
    \caption{On the left we have $\left|\hat{h}_{22}^{\text{DIN}}\right|$ and $\left|\hat{h}_{22}^{\text{New}}\right|$ up to the LSO $X_{\text{LSO}} \simeq 0.20$ for an equal-masss binary system. On the right, we show the relative error of the latter against the former.}
    \label{plot:comparable_mass}
\end{figure}

In order to arrive at evaluating $|\hat{h}^{(0)}_{22}(X)|$, we specify a grid of numerical values of $X$ and from that we obtain $u$ solving numerically the equation above \eqref{eq:Omega_vs_u} for $u$ and then finally obtain $J(u)$ and $\hat{H}_{\rm eff}$. On this so defined adiabatic dynamics we compute $|\hat{h}_{22}(X)|$ in two ways starting from the 4PN analytical results. One the one hand, we use the prescription of \cite{Nagar:2024oyk}, i.e. the DIN tail factorization with a 4PN-accurate $\rho_{22}^{\rm DIN}$ function resummed as in (23) therein. On the other hand, we use our new factorization, including also all the factors that have been neglected in the determination of the PN expansion of the $\rho_{\ell m}^{\text{New}}$'s, i.e.
\be
\left|\hat{h}^{(0)}_{22}\right| = \left| \hat{H}_{\text{eff}} \, T_{22}^{\text{New}} \, (\rho_{22}^{\text{New}})^2\right| \, \left|1+ \tilde{T}_{22} \tilde{f}_{22}\right|
\ee
We are interested in determining the behavior of this quantity up to the LSO, defined by the condition $\partial_u\hat{H}_{\rm eff}=\partial_u^2\hat{H}_{\rm eff}=0$. We consider an equal-mass binary ($\nu = 1/4$), for which we have $u_{\rm LSO}=0.1956$ and $J_{\rm LSO}=10.7906$, that yield $M\Omega_{\rm LSO}=0.0845$ and $X_{\rm LSO}=0.1926$. The comparison is shown in Figure \ref{plot:comparable_mass}. It has to be said that, in our factorization, the quantum period replaced inside the tail factors is $a = \frac{5}{2} + \gamma_{22}^{\text{univ}}$, where for the last quantity we have used \eqref{eq:a_Kerr} up to order $\epsilon^4$ (with all the replacements we explained above).
It is interesting to note that our new factorization yields a waveform amplitude that is slightly, but nonegligibly, smaller than $|\hat{h}_{22}^{\rm DIN}|$. This will have consequences at the level of radiation reaction, since the fractional difference of $\sim 10^{-3}$ will gain an additional factor two. In conclusion, we can already predict that on \texttt{TEOBResumS-Dali} the straightforward change of the $\ell=m=2$ mode in radiation reaction will, by itself, yield a longer inspiral and a delayed plunge once all the other elements entering the EOB dynamics are kept fixed. All the other details of the discussion on this point can be seen in \cite{Cipriani:2026xmx}.\\

The study exposed in this Chapter is quite important for the various future outlooks it gives rise. For example, one of these is the application of the hypergeometric representation of the Confluent Heun function for the case of a particle moving along open trajectories, with the computation of the absorbed energy and angular momentum by the BH \cite{Cipriani:2026myb}. Another interesting future direction is the implementation of the same representation to study the Kerr BH, since also in this spacetime the propagation of GWs is governed by a CHE. A third possibility, as already hinted, is understanding whether, once that the new radiation reaction (and waveform) is implemented in the \texttt{TEOBResumS-Dali}, the model still retains the current flexibility that eventually allows it to make it highly faithful to NR simulations via calibration, as it is the case with DIN-like factorizations. Futhermore, the importance of our hypergeometric representation is that it can be also applied to physical systems governed by a Heun equation, as it is shown in Appendix \ref{sec5}, and it enlarges its range of applicability.

\chapter*{Conclusions}
\addcontentsline{toc}{chapter}{Conclusions}
\markboth{Conclusions}{}
The present Thesis has explored the versatility of the Seiberg–Witten (SW) curve as a tool to investigate a broad range of topics, from strongly coupled regimes in Quantum Field Theory (QFT) to applications in Gravity. The analysis confirms the remarkable effectiveness of this geometric framework in producing new results in High Energy Physics and Gravity.\\

In Chapter \ref{chap_2} we recalled how the SW curve allows to find AD theories as particular points of the moduli spaces of $\mathcal{N}=2$ SU(2) SQCD with $N_f=1,2,3$. Before turning to them, we first considered the case with $N_f=4$, since in the computation of the extremal correlator $G_{2n}$ in the large R-charge limit we have both a reference result from EFT \cite{Hellerman:2018xpi,Hellerman:2020sqj,Hellerman:2021yqz} and the possibility of carrying out the relevant perturbative computations explicitly. Following the analysis of \cite{Grassi:2019txd} in the language of random matrix theory, the main result of this part, reported in \cite{Cipriani:2026iws}, is the agreement with the EFT expectations for the perturbative behavior of $\log(G_{2n})$ and for the structure of the functions $A(\tau,\bar{\tau})$ and $B(\tau,\bar{\tau})$, now including instanton contributions for the first time. More precisely, working up to fourth instanton order in the Pestun-Nekrasov scheme, we perfectly reproduce the results of \cite{Hellerman:2020sqj,Hellerman:2021yqz} for $A(\tau,\bar{\tau})$ and $B(\tau, \bar{\tau})$, respectively, expanded at the corresponding instanton order, showing that they receive contributions exclusively from large VEVs. Furthermore, restricting to the large-VEV sector, we derive an all-instanton expression for $A(\tau,\bar{\tau})$ (which coincides with the exact expression of the same function in \cite{Hellerman:2020sqj}) and for the renormalization-scheme independent quantity $\tilde{B}(\tau, \bar{\tau}) = B(\tau,\bar{\tau}) - \log Z_{S^4} (\tau, \bar{\tau})$. The latter, remarkably, turns out to be completely independent of $\tau$. This leads us to the scheme-independent conclusion that the entire $\tau$ dependence of $\tilde{B}$ originates from small VEVs. One possible direction for future outlook is using direct localization computation along the lines of the one presented in this Chapter in order to check if instantons contribute to the negative powers of $n$ in the large-charge expansion of the logarithm of the correlator for higher-rank models, which is something verified in \cite{Brown:2023why,Paul:2023rka} using different techniques.\\
For the three AD theories $\mathcal{H}_0$, $\mathcal{H}_1$, and $\mathcal{H}_2$, we exposed a discussion that led to formulate an educated guess for a function interpolating between the two asymptotic regimes of large and small $aR$. Furthermore, we presented a numerical determination of the large-VEV contributions to the coefficients $A$ and $\tilde{B}$ for the three AD models of above together with the three rank-one MN theories. The evidence gathered in this Chapter strongly suggests that the value obtained for $A$ is exact, by contrast the numerical value of $\tilde{B}$ is expected to change once the small-VEV contribution is taken into account. In these models, no supersymmetric recursion relations are known to govern the correlators and it makes direct tests of the EFT predictions at large charge especially important. A more comprehensive study of this problem is left for future work.\\

Chapter \ref{chap_3} presented the first gravitational application of the SW curve, namely the study of the stability of the Topological Star (TS) under massive charged scalar perturbations. This is possible because the differential equation governing these perturbations is a Confluent Heun Equation (CHE), which can be mapped to the quantum SW curve of $\mathcal{N}=2$ SU(2) SQCD with $N_f=3$. Together with other methods which are the numerical integration, the Leaver one, the asymptotic matching condition and WKB, we computed several QNMs and, for all of them, we found not only compatibility between the different procedure, but also that the imaginary part of these complex frequencies is always negative, demonstrating that this gravitational object is stable under the kind of perturbation mentioned before. We then established that the TS exhibits charge superradiance when behaving as a Black String (i.e. when a horizon is present), whereas superradiance is absent in the smooth, horizonless phase, in accordance with expectations from the information-loss problem. The significance of these findings, obtained in \cite{Cipriani:2024ygw}, has already motivated further work, such as \cite{Bena:2024hoh}, where the stability of the same geometry under higher-spin perturbations (vector bosons and gravitons) has been investigated.\\

Chapter \ref{chap_4} reported the first-ever application of a Machine Learning algorithm, specifically a Hamiltonian Neural Network, in the realm of String Theory for what concerns the motion of a massless neutral particle inside a particular background, the one of a D1-D5 fuzzball. For this reason the SW formalism is set aside in this part of the Thesis. The results presented in \cite{Cipriani:2025ini} did not add new physical insights into the system under study; rather, they were aimed at assessing the feasibility and performance of the HNNs compared to standard numerical methods in a controlled setting. In particular our analysis has shown that, despite the indisputable complexity of the equations learned, HNNs are able to accurately reproduce the geodesics across all the mentioned cases, even over large time scales, whereas the considered numerical integrators (Euler and Runge-Kutta) fail, as expected, when computing unstable trajectories. This because they accumulate errors at each time step during the integration of the equations of motion. This result highlights another significant aspect that emerged from our analysis: the performance of the HNNs, due to the fact that the solution is obtained by simultaneously and independently considering all time steps, does not depend on the intrinsic stability or instability of the dynamics described by the Hamilton equations and it remains at the same level for the different cases. In contrast, the performance of the numerical integrators is highly case-dependent. As a future outlook, a detailed study of geodesic motion in the so-called D1-D5-P (three-charge) fuzzball would be highly interesting. This is a nontrivial generalization of the present D1-D5 (two- charge) fuzzball with an additional Kaluza-Klein (KK) charge. The main complication arises from the fact that, in general, this geometry does not allow for a separation of radial and angular dynamics. For this reason, no ground truth for comparison is present and hence a challenging aspect is finding a way to assess the quality of the prediction. For the HNNs, we can first rely on the loss function itself, which provides a quantitative measure of how well the predicted solution satisfies the equations of motion.\\

Chapter \ref{chap_5} returned to the SW methods to study the inspiral phase of a binary system consisting of a Schwarzschild black hole and an orbiting particle and the observable we have focused on was the energy flux carried by the GWs emitted by the system towards infinity. The first part of the Chapter, based on \cite{Cipriani:2025ikx}, introduced a new resummation method for the infinite series of PM and PN corrections to the waveform—at leading order in the probe limit and to a fixed order in the soft expansion. This new method is built upon the so called hypergeometric representation of the Confluent Heun function, that consists in expressing the solution of the CHE (corresponding to the Teukolsky equation that governs the propagation of GWs) as a combination of a hypergeometric function and its first derivative, with coefficients given by polynomials in $x=4 i M \omega$ truncated at a desired order $k$. Using this proposal, we computed the luminosity for circular orbits up to $k=30$ PN for various values of the tangential velocity. We stress that by $k$PN we mean that the aforementioned polynomials and the period $a$ defining the solution are truncated at order $k$PN, but the hypergeometric functions and the coefficients entering the new ans\"atze are kept exact. This implies that an infinite number of terms beyond $k$PN are taken into account, in particular all the tail and transcendental contributions are resummed into exponentials and gamma functions. For this reason, the agreement of the new results with the numerical ones is excellent, confirming the reliability of the new representation.\\
Building on this, we saw how to write in a factorized form the waveform with the new method. The main goal was investigating similarities and differences with respect to \cite{Damour:2008gu}, where the factorization was proposed for the first time and it resummed only the leading logarithms in the waveform, and to \cite{Ivanov:2025ozg}, which gives a more recent proposal in this regards and this resumms all the other logarithmic terms. A detailed analysis is presented in \cite{Cipriani:2026xmx} but the main points highlighted in the Chapter are that the new resummation is more comprehensive, since it also captures the transcendental contributions. Furthermore, the remainder functions $\rho_{\ell m}^{\text{New}}$ and $\tilde{f}_{\ell m}$, depending exclusively on hypergeometric functions, reduce to polynomials in the tangential velocity when PN-expanded, displaying improved convergence properties. All of this is enforced by the comparison with the numerical data, showing that the new method produces the lowest discrepancy with respect to the other two proposals. Also the investigation to binary systems with comparable masses is examined.\\
Future outlooks in this direction are diverse: application of the hypergeometric representation for the case of a particle moving along open trajectories, with the computation of the absorbed energy and angular momentum by the BH \cite{Cipriani:2026myb}; implementation of the same representation to study the Kerr BH. Finally we can apply employ our results to generate waveform templates for gravitational-wave data analysis.\\

In conclusion, this Thesis illustrates the predictive power and versatility of the Seiberg–Witten methods across a wide spectrum of physical regimes, and it opens several promising avenues for further research.

\appendix
\chapter{Convention for contraction between spinors}\label{app:convention}
Due to the importance spinors have in supersymmetry, it is relevant to stress how Lorentz group acts on them, outlining the conventions adopted in this thesis. In order to do this, we have to exploit the fact that SO$(1,3)$ and $\text{SU}(2) \otimes \text{SU}(2)$ have the same algebra and this can be used to classify finite dimensional representation of the Lorentz group; on the other hand, this equivalence is true only locally, while the relation that holds is \cite{Weinberg:1995mt}
$$\text{SO}(1,3) = \frac{\text{SL}(2,\mathbb{C})}{\mathbb{Z}_2}$$
which means that we can express the action of Lorentz group on spinors using also the representations of SL$(2,\mathbb{C})$. From now on, the position of indices will be crucial.\\
There are two main independent representations of SL$(2,\mathbb{C})$ group. Taken $N \in \text{SL}(2,\mathbb{C})$, we have
\begin{itemize}
\item \textit{fundamental representation}, for which
$$\psi_\alpha \rightarrow \psi'_\alpha = N_\alpha^{\,\, \beta} \psi_\beta \qquad \alpha,\beta = 1,2$$
\item \textit{complex conjugate representation}, for which
$$\overline{\psi}_{\dot{\alpha}} \rightarrow \overline{\psi}'_{\dot{\alpha}} = N^{* \, \, \dot{\beta}}_{\dot{\alpha}} \overline{\psi}_{\dot{\beta}} \qquad \dot{\alpha}, \dot{\beta} = \dot{1},\dot{2}$$
\end{itemize}
In other words, a spinor is a two complex component object $\psi = 
\begin{pmatrix}
\psi_1 \\
\psi_2 \\
\end{pmatrix}$
where $\psi_1$ and $\psi_2$ are complex Grassman numbers that transform in the two previous representations as we have just written. Undotted indices are used for the fundamental and all its depedent representations, dotted indices are used for the complex conjugate repesentation and for all the others that are equivalent. These two repersentations are not equivalent, in the sense that it does not exist a matrix $C$ such that
$$N = C N^* C^{-1}$$
There are two other representations that are important when we deal with spinors. In order to introduce them, it has to be noticed that SL$(2,\mathbb{C})$ has two invariant tensors, the metric tensor $\eta$, which lowers and raises spacetime indices, and the antisymmetric tensor $\epsilon$, which is used to raise and lower spinorial indices. Its expression in the two previous representations is
$$\epsilon_{\alpha\beta} = \epsilon_{\dot{\alpha}\dot{\beta}}= \begin{pmatrix}
0 & -1 \\
1 & 0
\end{pmatrix}   
\qquad \epsilon^{\alpha \beta} = \epsilon^{\dot{\alpha}\dot{\beta}} =
\begin{pmatrix}
0 & 1 \\
-1 & 0
\end{pmatrix}$$
At this point we define
$$\psi^{\alpha} = \epsilon^{\alpha\beta} \psi_{\beta}, \quad \psi_{\alpha} = \epsilon_{\alpha\beta}\psi^{\beta}, \quad \overline{\psi}_{\dot{\alpha}} = \epsilon_{\dot{\alpha}\dot{\beta}}\overline{\psi}^{\dot{\beta}}, \quad \overline{\psi}^{\dot{\alpha}} = \epsilon^{\dot{\alpha}\dot{\beta}}\overline{\psi}_{\dot{\beta}}$$
where there is the convention that adjacent indices are always contracted putting the epsilon tensor on the left.\\
It can be proved that
$$\psi'^{\alpha} = (N^{-1})_{\beta}^{\,\,\alpha}\psi^{\beta} = (N^{-1T})^{\alpha}_{\,\,\beta}\psi^\beta \qquad \overline{\psi}'^{\dot{\alpha}} = (N^{*-1})_{\dot{\beta}}^{\,\,\dot{\alpha}}\overline{\psi}^{\dot{\beta}} = (N^{*-1T})^{\dot{\alpha}}_{\,\,\dot{\beta}} \overline{\psi}^{\dot{\beta}}$$
and also that $N^{-1T}$ can be obtained from $N$ with the above similarity transformation using as matrix $C$ the epsilon tensor $\epsilon_{\alpha\beta}$; similarly $N^{*-1T}$ can be obtained by $N^*$ with a similarity transformation with matrix $\epsilon_{\dot{\alpha}\dot{\beta}}$. In other words this is telling us that the fundamental $\psi_\alpha$ (or $\overline{\psi}_{\dot{\alpha}}$) and antifundamental $\psi^\alpha$ (or $\overline{\psi}^{\dot{\alpha}}$) representations are equivalent and it is valid only because the structure is that one of SU$(2)$, for which fundamental and antifundamental are equivalent.\\
In this convention, lower undotted indices are row ones, while upper indices are column ones; dotted indices follow instead the opposite convention. All this implies that $(\psi_\alpha)^* = \overline{\psi}^{\dot{\alpha}}$, while under hermitian conjugation $\overline{\psi}_{\dot{\alpha}} = (\psi_\alpha)^\dagger$. Thanks to the isomorphism between SO$(1,3)$ and SL$(2,\mathbb{C})$, we see that $\psi_\alpha$ and $\overline{\psi}^{\dot{\alpha}}$ are representations of Lorentz group that we can label as
$$\psi_\alpha \equiv \Bigl(\frac{1}{2},0\Bigr) \qquad \overline{\psi}^{\dot{\alpha}} \equiv \Bigl(0,\frac{1}{2}\Bigr)$$ 
hence these are Weyl spinors.
Given the antisymmetric "metric" $\epsilon$ for spinors, we have to define the way indices are contracted in spinorial (scalar) products
\begin{subequations}
\begin{align}
& \psi \chi := \psi^\alpha \chi_\alpha \label{chipsi}\\
& \bar{\psi}\bar{\chi} := \bar{\psi}_{\dot{\alpha}} \bar{\chi}^{\dot{\alpha}} \label{barchibarpsi}
\end{align}
\end{subequations}
that is, in this covention, undotted indices are contracted from upper left to lower right, while dotted indices are contracted from lower left to upper right (nonetheless, this rule is not valid when we raise or lower indices with the $\epsilon$ tensor) and this rule is chosen in order to ensure that $(\psi\chi)^\dagger = \bar{\psi}\bar{\chi}$. In order to understand this, let us first show that \eqref{chipsi} and \eqref{barchibarpsi} are actual scalars
\begin{align*}
\begin{split}
& \psi \chi = \psi^\alpha \chi_\alpha = \epsilon^{\alpha\beta} \psi_\beta \chi_\alpha = \epsilon^{\beta\alpha} \psi_\alpha \chi_\beta = -\epsilon^{\alpha\beta} \psi_\alpha \chi_\beta = -\psi_\alpha \chi^\alpha = \chi^\alpha \psi_\alpha = \chi \psi \\
& \bar{\psi} \bar{\chi} = \bar{\psi}_{\dot{\alpha}} \bar{\chi}^{\dot{\alpha}} = \epsilon_{\dot{\alpha}\dot{\beta}} \bar{\psi}^{\dot{\beta}} \bar{\chi}^{\dot{\alpha}} = \epsilon_{\dot{\beta}\dot{\alpha}} \bar{\psi}^{\dot{\alpha}} \bar{\chi}^{\dot{\beta}} = -\epsilon_{\dot{\alpha} \dot{\beta}} \bar{\psi}^{\dot{\alpha}} \bar{\chi}^{\dot{\beta}} = -\bar{\psi}^{\dot{\alpha}} \bar{\chi}_{\dot{\alpha}} = \bar{\chi}_{\dot{\alpha}} \bar{\psi}^{\dot{\alpha}} = \bar{\chi} \bar{\psi}
\end{split}
\end{align*}
Under hermitian conjugation we get
$$(\psi \chi)^\dagger = (\psi^\alpha \chi_\alpha)^\dagger = \chi^\dagger_\alpha \psi^{\alpha \dagger} = \bar{\chi}_{\dot{\alpha}} \bar{\psi}^{\dot{\alpha}} = \bar{\chi} \bar{\psi} = \bar{\psi}\bar{\chi}$$
To conclude this Appendix, we see what is the role of the $\sigma$ matrices when they are coupled to spinors. In the proof that SL$(2,\mathbb{C})$ is the universal covering group of SO$(1,3)$, one uses the fact that under an element of SL$(2,\mathbb{C})$ the generic matrix $X = x^\mu \sigma_\mu$ transforms as $NXN^\dagger$ and the index structure of $N$ and $N^\dagger$, for what we have said right above, is $N_{\alpha}^{\,\,\beta}$ and $N^{* \dot{\beta}}_{\,\,\dot{\alpha}}$, so $\sigma_\mu$ naturally has one dotted and one undotted indices, being contracted with an undotted and a dotted spinor as
$$\psi \sigma^\mu \bar{\chi} \equiv \psi^\alpha \sigma^\mu_{\alpha\dot{\alpha}} \bar{\chi}^{\dot{\alpha}}$$
Raising the indices as we have shown above, it can be proved that
$$\bar{\sigma}^{\mu \dot{\alpha}\alpha} = \epsilon^{\dot{\alpha}\dot{\beta}} \epsilon^{\alpha \beta} \sigma^\mu_{\beta \dot{\beta}} = (\sigma_0,\sigma_i)$$
and, consequently
$$\bar{\psi} \bar{\sigma}^\mu \chi \equiv \bar{\psi}_{\dot{\alpha}} \bar{\sigma}^{\mu \dot{\alpha} \beta} \chi_\beta$$
\chapter{Instantons}\label{app:instantons}
This Appendix is devoted to the introduction of instantons, determining their main properties that are used with great importance in Chapters 1 and 2. We consider the non-supersymmetric case of Yang-Mills theory for simplicity, but the reasoning does not change when SUSY is present.\\
The YM action is 
\begin{equation} \label{eq:YM_action}
S_{YM} = \frac{1}{4 g^2} \int \mathrm{d}^4 x \text{Tr} \left(F_{\mu \nu} \widetilde{F}^{\mu \nu} \right)
\end{equation}
and the classical equations of motion satisfied by the gauge field are
\begin{equation} \label{eq:eqn_motion}
D_\mu F^{\mu \nu} = 0
\end{equation}
where $D_\mu$ is the usual covariant derivative $D_\mu = \partial_\mu - i [A_\mu,]$ and $F_{\mu \nu} = \partial_\mu A_\nu - \partial_\nu A_\mu - i[A_\mu, A_\nu]$. We want to find particular solutions of the equations \eqref{eq:eqn_motion}. In order to understand what we mean by particular, it is useful to recall some properties of the theta term. Its action is
\begin{equation}
\label{theta_term}
S_{\theta} = \frac{\theta}{32\pi^2} \int d^4 x \text{Tr} (F_{\mu\nu}\widetilde{F}^{\mu\nu})
\end{equation}
and it is a total derivative, since
$$\frac{1}{2} \int d^4 x \text{Tr}F_{\mu\nu}\widetilde{F}^{\mu\nu} = \int d^4x \epsilon^{\mu\nu\rho\sigma} \partial_\mu \text{Tr}\Bigl(A_\nu \partial_\rho A_\sigma + \frac{2}{3} A_\nu A_\rho A_\sigma \Bigr)$$
After a regularization consisting in computing the integral in a ball with finite radius, we send the volume of the ball to infinity and we expect that the integral vanishes. This is true only for abelian gauge theories, but not for non-abelian ones due to the presence of the so called \textit{extended field configurations}. One could ask how can it be possible that a field does not go to zero at infinity and the answer relies on the fact that the field tends to a pure gauge, which is similar to going to zero because it does not go to anything which is physical. In other words, the field goes to a configuration which is equivalent to the vacuum, but it is a pure gauge. \\
This aspect is important for the following reason. The core for any computation in quantum field theory is the partition function which has the following form after the analytical continuation in the Euclidean space 
$$Z[J] = \int \mathcal{D}[\Phi] e^{-\frac{S[\Phi]}{g^2} + \int d^4x J \Phi}$$ 
where $\Phi$ is a set of fields with source $J$, $S[\Phi]$ is the Euclidean action (real and bounded from below) and $g$ the dimensionless gauge coupling. In the semi-classical approximation ($g^2 \to 0$) the path integral, as we know very well, is dominated by configurations of fields that minimize the Euclidean action and so the expression of the integral is obtained by expanding around these configurations. One of these configurations are those ones in which the action vanishes, so these are perturbative vacua and minima of the classical potential; for what we have said just above for the presence of extended field configurations, there can exist other minima of the Euclidean action corresponding to a finite (meaning non-vanishing) value and these have to be considered. Notice that these configurations are not relevant in the weak coupling limit, indeed the leading semi-classical contribution of these configurations is $e^{-S/g^2}$.\\
These configurations, that are relevant for \eqref{theta_term}, are called \textit{instanton configurations}. Instantons are solutions of the equations of motion \eqref{eq:eqn_motion} obtained from the Euclidean action that approach pure gauge for $|x|\to \infty$. It can be shown that the integral in \eqref{theta_term} is an integer number $k$, dubbed as \textit{instanton number} or \textit{winding number}
\begin{equation}
\label{theta_term_2}
S_\theta = \frac{\theta}{32\pi^2} \int d^4x \text{Tr}F_{\mu\nu}\widetilde{F}^{\mu\nu} =  \theta k \qquad \text{with} \qquad k \in \mathbb{Z}
\end{equation}
This number is a topological quantity, in the sense that it does not change upon continuous deformations of the gauge field configuration. Moreover, since the action \eqref{theta_term_2} enters in the functional integral via the exponential map $e^{iS_\theta}$ (in Minkowsky space), then the $\theta$ term behaves as an angle, meaning that the following transformation
$$\theta \rightarrow \theta + 2\pi$$
is a symmetry of the theory.\\
Instantons have an intrinsic non-perturbative nature and in order to show this we need to find how they contribute to the path integral, so their action. This can be done noticing that the YM action \eqref{eq:YM_action} can be written as
\begin{equation}
S_{YM} = \frac{1}{8 g^2} \int \mathrm{d}^4 x \text{Tr} \left(F_{\mu \nu} \mp \widetilde{F}_{\mu \nu} \right)^2 \pm  \frac{1}{4 g^2} \int d^4 x \text{Tr} (F_{\mu\nu}\widetilde{F}^{\mu\nu})
\end{equation}
and since the first quantity is always non negative and the second integral is, thanks to \eqref{theta_term_2}
$$\pm  \frac{1}{4 g^2} \int d^4 x \text{Tr} (F_{\mu\nu}\widetilde{F}^{\mu\nu}) = \frac{8 \pi^2}{g^2} (\pm k) = \frac{8 \pi^2}{g^2} |k|$$
we gain that
\begin{equation} \label{eq:S_YM_inequality}
S_{YM} \geq \frac{8 \pi^2}{g^2} |k|
\end{equation}
The action is minimized when the bound is saturated and this occurs when
\begin{equation} \label{eq:inst_equations}
F_{\mu \nu} = \pm \widetilde{F}_{\mu \nu}
\end{equation}
which are the self dual (+) and anti-self dual (-) Yang Mills equations. Solutions to these equations \eqref{eq:inst_equations} are exactly the instantons (and this explains what said in Section \ref{sec:non_pert_corrections} regarding the fact that a (anti)-self dual field strength represents a (anti)-instanton) and their action is hence given by
\begin{equation} 
S_{\text{inst}} = \frac{8 \pi^2}{g^2} |k|
\end{equation}
At this point the instanton contributions to the path integral in the Euclidean sector are exponentially suppressed 
\begin{equation}
\label{instanton_contribution}
e^{-S_{\text{inst}}} = \Bigl(e^{-\frac{8 \pi^2}{g^2}}\Bigr)^{|k|}
\end{equation}
We see how this behavior has been obtained without any perturbative analysis, so the nature of instantons is non-perturbative. Notice that in Section \ref{sec:non_pert_corrections} it was implicitly assumed that $k$ was positive, as it can be seen from all the sums. The contribution of anti-instantons was codified in the coefficients $a_k$. The "completing the square" trick that we used before (bounding the action by a topological invariant quantity) is known as \textit{Bogomolnyi bound}, which is encountered other times in this thesis.\\
Finite action solutions have the physical interpretation of mediating quantum tunneling effects and instantons are such interpolating field configurations, in other words they interpolate between different vacua of the gauge theory. Futhermore, they are called in this way because their action density is localized at both a point in space and at an instant in (Euclidean) time. For further details, see \cite{Bertolini:2025jhz}, \cite{Muller-Kirsten:2012wla}, \cite{Shifman:1994ee} and \cite{Vandoren:2008xg}.
\chapter{Elements on Eletric-Magnetic duality}\label{app:EM_duality}
In this Appendix we embark on an in-depth study of Eletric-Magnetic duality, giving the main elements that are useful in the text to understand the analysis of Seiberg and Witten. We cannot give all the details for such ingredients and for this reason references will be given for the interested reader.\\
We know very well that the Maxwell equations in 3+1 dimensions in presence of only electric sources do not enjoy any symmetry that exchanges electric and magnetic fields (called eletric-magnetic duality). This is possible if the electric sources $\rho_e$ and $\textbf{J}_e$ are zero, but Dirac found out that it is possible to restore electric-magnetic duality in presence of sources if we add suitable magnetic source terms
\begin{align}
& \nabla \cdot \textbf{E} = \rho_e \qquad \nabla \cdot \textbf{B} = \rho_m \notag \\
& \nabla \times \textbf{B} - \frac{\partial \textbf{E}}{\partial t} = \textbf{J}_e \qquad \nabla \times \textbf{E} + \frac{\partial \textbf{B}}{\partial t} = \textbf{J}_m
\end{align} 
This allows, as we are going to see, also to explain the electric charge quantization. It is useful to rewrite these equations in terms of the abelian gauge field $A = A_\mu \mathrm{d} x^\mu$ using the language of forms. Hence we have the field strength $F = \mathrm{d} A = \frac{1}{2} F_{\mu \nu} \mathrm{d} x^\mu \mathrm{d} x^\nu$ and the equations $ \nabla \cdot \textbf{E} = \rho_e, \nabla \times \textbf{B} - \frac{\partial \textbf{E}}{\partial t} = \textbf{J}_e$ can be written as
\begin{equation} \label{eq:electric_Max_equation}
\mathrm{d} * F = * J_e
\end{equation}
and the equations $\nabla \cdot \textbf{B} = \rho_m,\nabla \times \textbf{E} + \frac{\partial \textbf{B}}{\partial t} = \textbf{J}_m
$ are equivalent to
 \begin{equation}\label{eq:magnetic_Max_equation}
\mathrm{d} F = * J_m
\end{equation}  
where $*$ denotes the Hodge operation in 4 dimensions \footnote{In general, given a differentiable manifold $M$ with $\text{dim}M = m$ and metric $g$, we denote as $\Omega^r(M)$ the space of smooth $r$-forms on $M$ and the Hodge operation * is a linear map $*: \Omega^r(M) \to \Omega^{m-r}(M)$ whose action on a basis vector of $\Omega^r(M)$ is defined by $$*(\mathrm{d} x^{\mu_1} \wedge \mathrm{d} x^{\mu_2} \wedge \dots \wedge \mathrm{d} x^{\mu_r} ) = \frac{\sqrt{|g|}}{(m-r)!} \epsilon^{\mu_1 \mu_2 \dots \mu_r}\,_{\nu_{r+1} \dots \nu_m} \mathrm{d}x^{\nu_{r+1}} \wedge \dots \wedge \mathrm{d} x^{\nu_m}$$}. Applying it to the field strength we get the dual 2-form
$$\widetilde{F} = \frac{1}{2} \epsilon_{\mu \nu \rho \sigma} F^{\rho \sigma} \mathrm{d} x^\mu \mathrm{d} x^\nu = - B^i \mathrm{d}x^0 \mathrm{d}x^i + \frac{1}{2} \epsilon_{ijk} E^i \mathrm{d}x^j \mathrm{d} x^k$$
where we have used the fact that $F=E^i \mathrm{d} x^0 \mathrm{d}x^i + \frac{1}{2} \epsilon_{i j k} B^i \mathrm{d} x^j \mathrm{d} x^k$. Since \eqref{eq:electric_Max_equation} and \eqref{eq:magnetic_Max_equation} involve respectively the electric and magnetic (1-)form currents, they are called \textit{electric} and \textit{magnetic Maxwell equations}.

\subsection*{Monopoles and Dyons: Dirac quantization condition and Schwinger-Swanziger equation}
We have just seen that, in order to have an exact symmetry between electric and magnetic field, the introduction of magnetic monopoles is needed. These are free magnetic charges that produce a magnetic field of the form
\begin{equation}
\textbf{B} = \frac{g \hat{\textbf{r}}}{4 \pi r^2} \Rightarrow \nabla \cdot \textbf{B} = g \delta(r)
\end{equation}
where $g$ is the magnetic charge of the monopole. Such statement is inconsistent when we couple Electromagnetism with Quantum Mechanics, because in this situation we can express the magnetic field as the curl of a non singular vector potential, which cannot be realized in the present case (the divergence of \textbf{B} is not zero). In order to solve this problem, Dirac introduced the so called \textit{Dirac string}: it is a solenoid of infinite length and infinitesimal diameter with one end connected to the magnetic monopole. The effect of this string would be the cancellation of the monopole contribution to the divergence equation. Consequently, the total magnetic field of the system monopole+string would admit a vector potential that is singular only along the string, but it is regular elsewhere. Since strings should be thought as unobservable objects, they cannot interact with particles; this implies that their position in space should be irrelevant. This suggests that the vector potential associated to strings laying in different positions in space should differ only by a gauge transformation. We cannot see the details of all of this, which instead can be consulted in \cite{Jackson:1998nia, Heras:2018uub, Harvey:1996ur}. The final result is that the fields derived from different string configurations are related by a gauge transformation with scalar function
\begin{equation}
\chi = \frac{g \phi}{2 \pi}
\end{equation}
where $\phi$ is the parameter of the gauge transformation (it codifies the turns around the string; indeed it is the azimuthal angle in the standard spherical coordinates). Now, if we couple an electric charge $e$ to the monopole magnetic field, under this gauge transformation the electron wavefunction $\psi$ transforms as
$$\psi \to e^{i e \chi} \psi = e^{i \frac{g e \phi}{2 \pi}} \psi$$
In order for this wavefunction not to be multivalued, we must impose that
\begin{equation} \label{eq:Dirac_quantization}
\frac{e g}{2 \pi} = m, \qquad m \in \mathbb{Z}
\end{equation}
This is the famous Dirac quantization condition, which leads to the conclusion that the existence of a single magnetic monopole would explain electric charge quantization. Furthermore, we can observe that if the electric charge is quantized in multiples of a fundamental unit $e$ ($q = n e$), the gauge parameter varies within the compact range $\chi \in [0, 2 \pi/e]$, indeed for $\chi = 0$ and $\chi = 2 \pi/e$ we get the same U$(1)$ transformation. We conclude that the existence of the monopole is strongly related to the presence of a U$(1)$ gauge group and viceversa, that is whenever we consider a theory with a U$(1)$ gauge group we will have magnetic monopoles.\\

Now we want to see the generalization of the Dirac quantization condition \eqref{eq:Dirac_quantization} to the presence of two dyons. A dyon is a particle that carries both electric and magnetic charge, denoted as $(q,g)$. We consider a reference frame in which a dyon of mass $m$ and electric and magnetic charges $(q_2,g_2)$ is fixed at the origin, while a second dyon of the same mass and charges $(q_1,g_1)$ moves with velocity $\mathbf{\dot{\textbf{r}}}$. The latter dyon is subjected to the Lorentz force due to both the electric and magnetic field generated by the former. Defining the second dyon's momentum $\mathbf{p} = m \dot{\textbf{r}}$ and with the typical analysis of classical mechanics, we can show that the quantity
\begin{equation}
\mathcal{J} = \mathbf{r} \times \mathbf{p} - \frac{q_1 g_2 - q_2 g_1}{4 \pi} \hat{\textbf{r}}
\end{equation}
is conserved and it is exactly the total angular momentum of the system of the two dyons (indeed it is conserved since the field generated by the first dyon is spherically symmetric). By projecting $\mathcal{J}$ along the radial direction we get a constant, which is $(q_1 g_2 - q_2 g_1)/(4 \pi)$, and imposing the quantization of this radial component we gain
\begin{equation} \label{eq:schwinger_equation}
\mathcal{J} \cdot \hat{\textbf{r}} = \frac{n}{2}, \, \, n \in \mathbb{N} \Rightarrow q_1 g_2 - q_2 g_1 = 2 \pi n
\end{equation}
and it is known Schwinger-Swanziger equation. It is invariant under the duality transformation that sends $q \to g$ and $g \to -q$.

\subsection*{Witten effect}
Witten effect \cite{Witten:1979ey} is an interesting phenomenon according to which in presence of a $\theta$ term the electric charge of a particle is shifted by a quantity proportional to the magnetic charge, becoming in this way a dyon. In order to show this, let us consider Electromagnetism coupled to an electric charge $n e$. Its action is the one of Yang-Mills of the Appendix \ref{app:instantons} (but now with gauge group U$(1)$) with the addition of a term that describes the coupling of the electromagnetic field with the particle itself, hence we can write the action in the language of forms in the following way
\begin{equation} \label{eq:action_form}
S = \frac{1}{2 e^2} \int F \wedge * F + n \int_L A_\mu \mathrm{d} x^\mu = \frac{1}{2 e^2} \int F \wedge * F + n \int A \wedge \delta_3(L)  
\end{equation}
where $\wedge$ is the usual wedge operations on differential forms. We have chosen to reabsorb the unity of electric charge $e$ in the definition of the gauge field $A$ in such a way that its covariant derivative simply gives $\partial_\mu - i A_\mu$. $L$ stands for the worldline of the particle and $\delta_3(L)$ is the delta function 3-form with support on $L$ \footnote{Consider for simplicity a 4-dimensional spacetime. For a positive integer $0 \leq p \leq 4$, it is possible to define a delta function $(4-p)$-form on a $p$-dimensional subspace $\Sigma_p$ of our spacetime as follows \begin{equation} \int_{\Sigma_p} \omega_p(x) = \int \omega_p(x) \wedge \delta_{4-p}(x;\Sigma_p) \end{equation} where $omega_p$ is a $p$-form field. Notice that the explicit form of the delta function is \begin{align} \delta_{4-p}(x;\Sigma_p) & = \delta_{\mu_1 \dots \delta_{4-p}}(x;\Sigma_p) \mathrm{d}x^{\mu_1} \wedge \dots \wedge \mathrm{d} x^{\mu_{4-p}} = \notag \\ & = \frac{\epsilon_{\mu_1 \dots \mu_{4-p} \nu_1 \dots \nu_p}}{p! (4-p)!} \mathrm{d}x^{\mu_1} \wedge \dots \wedge \mathrm{d} x^{\mu_{4-p}} \int_{\Sigma_p} \mathrm{d}y^{\nu_1} \wedge \dots \wedge \mathrm{d}y^{\nu_p} \delta^4(x-y) \end{align} where $(y^\mu)$ represents the coordinates of $\Sigma_p$ in the spacetime.}. At this point, varying the action and imposing stationarity we get
\begin{equation} \label{eq:electric_equation_notheta}
\frac{1}{e^2} \mathrm{d} * F = n \delta_3 (L)
\end{equation}
and the Bianchi identity is violated by the presence of the magnetic monopole charge $g$ (which is present thanks to the U$(1)$ gauge group)
\begin{equation} \label{eq:magnetic_equation}
\mathrm{d} F = e g \delta_3(\mathbf{r}) = 4 \pi m \delta_3(\mathbf{r})
\end{equation}
Notice that the RHS of this equation differs by a factor two from Dirac quantization condition \eqref{eq:Dirac_quantization} and this is due to the rescaling $e \to 2 e$. Indeed in Dirac's derivation of above we identified the gauge coupling $e$ with the charge of the electron, but now we are assuming that the fundamental electric charge corresponds to half the gauge coupling. Considering at this point only the spatial components we have the following equations for the divergence of electric and magnetic fields
\begin{equation} \label{eq:spatial_equations}
\frac{1}{e^2} \nabla \cdot \textbf{E} = n \delta_3(L) , \qquad \nabla \cdot \textbf{B} = 4 \pi m \delta_3(\textbf{r}) 
\end{equation}
Now, as we have already seen many times, when we treat a quantum system that couples electric and magnetic charges, we should consider the possibility of adding the $\theta$ term
\begin{equation}\label{eq:theta_term_form}
-\frac{\theta}{16 \pi^2} \int F \wedge F = - \frac{\theta}{32 \pi^2} \int \mathrm{d}^4 x F_{\mu\nu} \widetilde{F}^{\mu\nu}
\end{equation}
(look at \eqref{theta_term}, which, in the abelian case, does not have the trace). Considering the variation of the original action \eqref{eq:action_form} with the addition of the $\theta$ term, the equation \eqref{eq:electric_equation_notheta} becomes
\begin{equation} \label{eq:electric_equation}
\mathrm{d} \left(\frac{1}{e^2}* F - \frac{\theta}{8 \pi^2} F\right) = n \delta_3 (L)
\end{equation}
Now, integrating this equation over a three-dimensional space $\mathcal{M}_3$, we get
\begin{equation} \label{equation1}
\int_{\mathcal{M}_3} \mathrm{d} \left(\frac{1}{e^2}* F - \frac{\theta}{8 \pi^2} F\right) = n
\end{equation}
and, assuming that the boundary of $\mathcal{M}_3$ is a two-dimensional space that links the electric particle and the magnetic monopole, we get, from the LHS of this equation (thanks to Stokes' theorem)
$$\int_{\mathcal{M}_3} \mathrm{d} \left(\frac{1}{e^2}* F - \frac{\theta}{8 \pi^2} F\right) \underset{\text{for} \, \eqref{eq:magnetic_equation}}{=} \int_{\partial{\mathcal{M}_3}}  \frac{1}{e^2}* F - \frac{\theta}{2 \pi} m = \int_{\partial{\mathcal{M}_3}} \frac{1}{e^2} \textbf{E} \cdot \mathrm{d} \textbf{S} - \frac{\theta}{2 \pi} m$$
Now, the only integral on the RHS is the flux of the electric field and hence it is equal to the electric charge (divided by $e$ due to the normalization by $e^2$) and hence, plugging this last result in \eqref{equation1}, we obtain
\begin{equation}
\frac{q}{e} = n + \frac{\theta}{2 \pi} m = n + \frac{\theta}{8 \pi^2} g \, e \Rightarrow q = n e + \frac{\theta e^2}{8 \pi^2} g
\end{equation}
This concludes our proof: if the $\theta$ parameter is not 0, the electric charge acquires a magnetic one proportional to $\theta$, thus becoming a dyon.\\

For later purposes that are important for what treated in Chapter 1, we consider the following combination of the electric and magnetic charges (defined in terms of the doublet of integers $(n \, \, \,m)$)
\begin{equation}
q + i g = n e + \frac{\theta e m}{2 \pi} + i \frac{4 \pi m}{e} = e \left[n + m \left(\frac{\theta}{2 \pi} + \frac{4 \pi i}{e^2}\right)\right] = e (n + m \tau)
\end{equation}
where $\tau = \frac{4 \pi i}{e^2} + \frac{\theta}{2 \pi}$ is exactly the complexified coupling firstly introduced in \eqref{eq:tau_F_a} (where there the coupling constant is $g$, now instead is the fundamental unit charge $e$ since we are in QED).

\subsection*{Extended electric-magnetic duality}
When there is a non-zero $\theta$ term, the action is invariant under an infinite set of transformations, which is SL$(2,\mathbb{Z})$. In order to show this, we rewrite the action \eqref{eq:action_form}+\eqref{eq:theta_term_form} in terms of the coupling $\tau$
$$S = \frac{1}{16 \pi} \text{Im} \left[\tau (F+ i * F) \wedge *(F + i * F)\right]$$
It can be proved that this action is invariant under the following transformations
\begin{align} \label{eq:S_duality}
S : \, & F \to F_D = -\frac{4 \pi}{e^2} * F + \frac{\theta}{2 \pi} F \notag \\
& \tau \to \tau_D = -\frac{1}{\tau} \notag \\
&(n,m) \to (n_D, m_D) = (m,-n)
\end{align}
\begin{align} \label{eq:T_duality}
T : \, &F \to  F_D = F \notag \\
& \tau \to \tau_D = \tau +1 \notag \\
&(n,m) \to (n_D, m_D) = (n+m,m)
\end{align}
The inversion $S$ and the translation $T$ are the generators of SL$(2,\mathbb{Z})$. Notice that the $S$ transformation is the duality one cited above that leaves invariant the Schwinger-Swanziger equation \eqref{eq:schwinger_equation}. The previous group acts non linearly on the complexified coupling, in particular, given an element
\begin{equation*}
\begin{pmatrix}
a & b \\
c & d
\end{pmatrix} \in \text{SL}(2,\mathbb{Z})
\end{equation*}
its action on $\tau$ is
\begin{equation} \label{eq:tau_lin_frac}
\tau \to \tau' = \frac{a \tau + b}{c \tau + d}, \quad a,b,c,d \in \mathbb{Z}, \quad ad-bc = 1
\end{equation}
If we consider as element of the group $\mathbb{I}$ or $- \mathbb{I}$, the effect on $\tau$ is the same, hence the actual group acting on $\tau$ is $\text{PSL}(2, \mathbb{Z}) = \text{SL}(2,\mathbb{Z})/\mathbb{Z}_2$. Furthermore, $\tau$ assumes values only on the upper half $H^+$ of the complex plane, since $\text{Im} \tau = 4 \pi/e^2 >0$. Now, due to the fact that variables connected by elements of PSL$(2,\mathbb{Z})$ give the same physics, we can identify all the points on $H^+$ related by this group action by quotienting this space by PSL$(2,\mathbb{Z})$, getting
$$\frac{H^+}{\text{PSL}(2,\mathbb{Z})} = \left\{z \in \mathbb{C}: |z| \geq 1, -\frac{1}{2} \leq \text{Re}(z) \leq \frac{1}{2} \right\}$$
It is known as \textit{fundamental domain} of SL$(2,\mathbb{Z})$.

\subsection*{The Bogomol'nyi Bound: BPS states}
We have analyzed magnetic monopoles, but we have to say that the electromagnetic field coming out of a point to be considered a magnetic source cannot be a proper excitation of the theory, since its energy diverges in the IR
$$U_E = \frac{1}{2} \epsilon_0 \int \mathrm{d}^3 x \, E^2 = \frac{1}{2} \left(\frac{q}{4 \pi} \right)^2 \int_0^\infty 4 \pi r^2 \,\frac{\mathrm{d} r}{r^4} \underset{r \to 0} \to \infty$$
In Appendix \ref{app:instantons} we said that instead magnetic monopoles are specific kinds of solitons, that is finite energy field configurations that solve equations of motion with some specific boundary conditions. This can be shown when the U$(1)$ gauge group is embedded into a non abelian one. A typical example where this aspect can be proven is the SU$(2)$ Georgi-Glashow model \cite{Rajaraman:1982is, Alvarez-Gaume:1996ohl}. We are going to analyze only the main properties of it.\\
This is characterized by a Yang-Mills term coupled to a triplet of scalar fields transforming in the adjoint of SU$(2)$. The gauge bosons are denoted as $W^a_\mu$ and the scalars as $\phi^a$, with $a=1,2,3$ is the index of the adjoint representation. We introduce the following definitions for the covariant derivative and field strength
\begin{align}
&D_\mu \phi^a = \partial_\mu \phi^a + g f^{abf} W^b_\mu \phi^c \notag \\
& G^a_{\mu \nu} = \partial_\mu W^a_\nu - \partial_\nu W^a_\mu + g W^b_\mu W^c_\nu f^{abc}
\end{align}
where $g$ is the coupling (not to be confused with the magnetic charge, indicated in the following as $g_m$) and $f^{abc}$ are the structure constants. The Lagrangian of the model is
\begin{equation}
\mathcal{L} = -\frac{1}{4} G^a_{\mu \nu} G^{a \mu \nu} + \frac{1}{2} D_\mu \phi^a D^\mu \phi^a - V(\phi)
\end{equation}
where the scalar potential $V(\phi)$ is 
\begin{equation}
V(\phi) = -\frac{\lambda}{4} (\phi^a \phi^a -a^2)^2
\end{equation}
with  $\lambda >0$. The extremization of the action leads to the following equations of motion
\begin{align} \label{eq:GG_eqn_motion}
&\frac{\delta \mathcal{L}}{\delta W^\mu_a}=0 \Rightarrow D_\mu G^{a \mu \nu} = g f_{abc} (D^\nu \phi)^b \phi^c \notag \\
&\frac{\delta \mathcal{L}}{\delta \phi_a}=0 \Rightarrow D_\mu D^\mu \phi^a =  -\lambda \phi^a (\phi^b \phi^b - a^2)
\end{align}
knowing also that the field strength satisfies the Bianchi identity $D_\mu (* G^{\mu\nu}) = 0$.\\
The form of the scalar potential induces a Higgs mechanism. Indeed, considering a minimum energy solution of the theory, which means we impose
\begin{equation} \label{eq:classical_vacua_conditions}
\partial_\mu \phi^a = 0 \qquad W^a_\mu = 0 \qquad V(\phi)=0
\end{equation}
the potential is minimized whenever the scalar three-vector $\vec{\phi}$ satisfies
\begin{equation}
\phi^a \phi^a -a^2 = 0 \Rightarrow \left(\phi^1 \right)^2+\left(\phi^2 \right)^2 + \left(\phi^3 \right)^2 = a^2
\end{equation}
This is telling that the classical vacua of the theory cover a $S^2$ sphere with radius $a$. The expansion around a chosen vacuum breaks SU$(2)$ to U$(1)$. Indeed, the vacuum space is 2-dimensional and it is where we have the broken degrees of freedom (which are the Goldstone's bosons): since SU$(2)$ has 3 generators, then around a vacuum it breaks down to U$(1)$. The Goldstone bosons eventually give mass to one or more vector bosons of the theory. In this case only two of the three gauge vectors acquire mass and the one that remains massless is the gauge field associated to the U$(1)$ group. So, for a generic point $\vec{\phi}_{\text{vac}}$ of the vacuum manifold $S^2$, this gauge field is
\begin{equation}
A_\mu = \frac{\phi^a_{\text{vac}} W^a_{\mu}}{a}
\end{equation}
and the corresponding field strength is
\begin{equation} \label{eq:field_strength_U1}
F_{\mu \nu} = \frac{\phi^a_{\text{vac}} G^a_{\mu \nu}}{a}
\end{equation}
What we want to sketch is that this theory exhibits solutions for which the U$(1)$ field strength \eqref{eq:field_strength_U1} gives a magnetic monopole field with finite energy. In order to do so, we start by considering the energy functional of the theory
\begin{equation} \label{eq:energy_density}
\mathcal{E} = \int \mathrm{d}^3 x \left[ \frac{(E^a_i)^2 + (B^a_i)^2}{2} + \frac{(D_i \phi^a)^2 + (D_0 \phi^a)^2}{2} + V(\phi) \right]
\end{equation}
where $E^a_i = G^a_{0i}, B^a_i = \frac{1}{2} \epsilon_{ijk} G^{a j k}$. In order to have $\mathcal{E} < \infty$ we require that for $r \to \infty$ the integrand goes to 0 faster than $1/r^3$. Since classical vacua satisfy the conditions \eqref{eq:classical_vacua_conditions}, we realize that the energy configuration must reach a classical vacuum of the theory at each point of the boundary of the spatial domain of spacetime, placed at infinity. This is another $S^2$ sphere denoted as $S^2_{\text{phys}}$. Now, we call $\vec{\phi}_{\text{vac}}(\hat{r})$ the configuration at spatial infinity of the scalar triplet and since it satisfies $\phi^a_{\text{vac}} \phi^a_{\text{vac}} = a^2$ at every point of the boundary $S^2_{\text{phys}}$, we see that it defines the following mapping
$$\vec{\phi}_{\text{vac}} : S^2_{\text{phys}} \to S^2$$
where $S^2$ is the vacuum manifold introduced before. All these mappings are classified according to their winding number, which is a topological quantity which can be thought as the number of times $S^2_{\text{phys}}$ is covered as we go over the internal sphere $S^2$.\\
Let us now consider the following boundary conditions
\begin{equation}
D_\mu \phi^a_{\text{vac}} = \partial_\mu \phi^a_{\text{vac}} + g f^{abc} W^b_\mu \phi^c_{\text{vac}} = 0
\end{equation}
and we can solve them for $W^m_\mu$, getting
\begin{equation}
W^m_\mu = A_\mu \frac{\phi^m_{\text{vac}}}{a} - \frac{1}{g a^2} f^{mla} \phi^l_{\text{vac}} \partial_\mu \phi^a_{\text{vac}}
\end{equation}
Plugging this expression in \eqref{eq:field_strength_U1}, we get, for the U$(1)$ field strength
\begin{equation} \label{eq:final_Fmunu_GG}
F_{\mu \nu} = \partial_\mu A_\nu - \partial_\nu A_\mu -\frac{1}{g a^3} f^{mln} \phi^m_{\text{vac}} \partial_\mu \phi^l_{\text{vac}} \partial_\nu \phi^n_{\text{vac}}
\end{equation}
In order to illustrate how the topological winding number is related to the magnetic monopole charge $g_m$, one can follow the original approach of 't Hooft and Polyakov. Without entering in details, we directly give the final result, that is
\begin{equation}
g_m = \frac{4 \pi m}{g} \quad m \in \mathbb{Z}
\end{equation}
where $m$ is the winding number. Hence the magnetic monopole charge is proportional to the winding number of the scalar fields boundary configuration. Concluding, with certain boundary conditions, solitons may carry a magnetic charge.\\
At this point we are left to discuss a way of finding particular solutions to the model introduced in this Subsection. It consists in solving a set of first order differential equations, known as BPS equations. Let us rewrite the energy functional \eqref{eq:energy_density} in the following way
\begin{align} \label{eq:BPS_E}
\mathcal{E} & = \frac{1}{2} \int \mathrm{d}^3 x \left[(E^a_i - \sin \theta D_i \phi^a)^2 + (B^a_i - \cos \theta D_i \phi^a)^2 + (D_0 \phi^a)^2\right] + \int \mathrm{d}^3 x V(\phi) + \notag \\
& + \int \mathrm{d}^3 x (\sin\theta E^a_i D_i \phi^a + \cos \theta B^a_i D_i \phi^a)
\end{align}
where $\theta$ is a continuous parameter. Noticing that from the equations of motion \eqref{eq:GG_eqn_motion} and the Bianchi identity we have $D_i E^{ai} = g f^{abc} (D^0 \phi)^b \phi^c$ and $D_i B^i=0$, the last integral can be recast as
\begin{align}
& \int \mathrm{d}^3 x (\sin\theta E^a_i D_i \phi^a + \cos \theta B^a_i D_i \phi^a) = \int \mathrm{d}^3 x \partial^i (\sin\theta E^a_i \phi^a + \cos \theta B^a_i \phi^a) = \int_{S^2_{\text{phys}}} \mathrm{d}^2 s_i (\sin\theta E^a_i \phi^a_{\text{vac}} + \cos \theta B^a_i \phi^a_{\text{vac}}) = \notag \\
& = a \int_{S^2_{\text{phys}}} \mathrm{d}^2 s_i (\sin\theta E_i + \cos \theta B_i) = a (\sin \theta \, q + \cos \theta \, g_m)
\end{align}
where the second equality is due to Stokes' theorem and the quantities $E_i = \frac{E^a_i \phi^a_{\text{vac}}}{a}, B_i = \frac{B^a_i \phi^a_{\text{vac}}}{a}$ are the components of the U$(1)$ field strength $F_{\mu \nu}$ \eqref{eq:final_Fmunu_GG}. Now, since the first three integrals in \eqref{eq:BPS_E} are positive definite, then the following inequality is true
\begin{equation}
\mathcal{E} \geq a \, (\sin \theta \, q + \cos \theta \, g_m)
\end{equation}
and it is satisfied also when the RHS takes its maximum value, reached when $\cos \theta = \frac{g_m}{\sqrt{q^2 + g^2_m}}, \sin\theta = \frac{q}{\sqrt{q^2 + g^2_m}}$, that is
$$\mathcal{E} \geq a \sqrt{q^2 + g^2_m} = a \sqrt{q^2 + \left(\frac{4 \pi m}{g}\right)^2} = a g \, |n + \tau m| \Rightarrow$$
\begin{equation} \label{eq:BPS_final_result}
\mathcal{E} \geq a g \, |n+\tau m|
\end{equation}
The bound is saturated when $V(\phi)=0$ and
\begin{equation}
D_0 \phi^a = 0 \qquad E^a_i = \sin \theta D_i \phi^a \qquad B^a_i = \cos \theta D_i \phi^a
\end{equation}
The solutions to this set of equations minimize the energy functional $\mathcal{E}$ and extremize the action, hence they are solutions of the equations of motion. These are the BPS states, whose supersymmetric version is treated in Chapter 1 together with the mass formula used for determining the nature of singularities in the moduli space of SW theory. It can be shown that the monopole obtained from the proposal of 't Hooft and Polyakov satisfies the BPS equations listed right now and the same reasoning could be applied also to dyons, hence they are BPS states.
\chapter{Interplay between the factors of the Nekrasov partition function and the components of the genus expansion for $\mathcal{N}=2,\text{SU(2)}$ SQCD with $N_f=4$} \label{app:interplay}
In this Appendix we analyze the relation between the three components of the Nekrasov partition function \eqref{eq:expansion_Z_coupling} and the ones coming from the genus expansion of the prepotential \eqref{eq:genus_expansion_prep}. Putting together \eqref{eq:relation_Z_prep} with \eqref{eq:expansion_Z_coupling} we write
$$\mathcal{F} = \epsilon^2 \log Z_{\text{tree}} + \epsilon^2 \log Z_{\text{1-L}} + \epsilon^2 \log Z_{\text{inst}} = \mathcal{F}_{\text{tree}} + \mathcal{F}_{\text{1-L}} + \mathcal{F}_{\text{inst}}$$
Each of these three components admits a genus expansion (actually, as we are going to see, $\mathcal{F}_{\text{tree}}$ has only the $g=0$ component), which we recall to be an expansion for $a \gg \epsilon = R^{-1}$.\\
Let us start with the instantonic part. Let us recall, for the theory under consideration, the instanton partition function \eqref{eq:Z_inst_2_inst}, which is
\begin{equation}
Z_{\text{inst}}(a,\epsilon,q) = 1 + \frac{q}{2} \underbrace{\left(\frac{a^2}{\epsilon^2} -3\right)}_{\mathcal{Z}_1\left(\frac{a}{\epsilon}\right)} + \frac{q^2}{4} \underbrace{\frac{8 \frac{a^8}{\epsilon^8} + \frac{a^6}{\epsilon^6} - 91 \frac{a^4}{\epsilon^4} - 60 \frac{a^2}{\epsilon^2} +132}{\left(4 \frac{a^2}{\epsilon^2} + 9 \right)^2}}_{\mathcal{Z}_2\left(\frac{a}{\epsilon}\right)}+\dots
\end{equation}
At this point, we take the logarithm of this function and, expanding for small $q$, we obtain
$$\mathcal{F}_{\text{inst}} \simeq \epsilon^2 \left[ \frac{q}{2} \left(\frac{a^2}{\epsilon^2} - 3\right) + \frac{q^2}{4} \mathcal{Z}_2 - \frac{q^2}{4}\left(\frac{a^2}{\epsilon^2} - 3\right) \right]$$
From this expression we clearly understand the fact that, being $\mathcal{Z}_1$ a polynomial of the first order in $(a/\epsilon)^2$, it can only contribute to $\mathcal{F}_0$ and $\mathcal{F}_1$; instead, due to the fact that $\mathcal{Z}_2$ is a rational function in the same parameter, expanding it for $a/\epsilon \gg 1$, it will give rise also to negative powers in such parameter, implying that it contributes also to all the others $\mathcal{F}_{g \geq 2}$. In particular we get, up to the order $(a^2/\epsilon^2)^{-1}$
$$\mathcal{F}_{\text{inst}} \simeq a^2 \left(\frac{q}{2} + \frac{13}{64}q^2\right) + \epsilon^2 \left(-\frac{3}{2} q - \frac{23}{32} q^2\right) + \frac{\epsilon^4}{a^2} \frac{3 q^2}{1024} + \mathcal{O}\left(\frac{\epsilon^6}{a^4}\right)$$
from which we can read
\begin{align} \label{eq:genus_components_instantons}
& \mathcal{F}_{0, \text{1-inst}} = \frac{q}{2} \qquad \mathcal{F}_{1, \text{1-inst}} = -\frac{3}{2}q \notag \\
& \mathcal{F}_{0, \text{2-inst}} = \frac{13}{64} q^2 \qquad \mathcal{F}_{1, \text{2-inst}} = -\frac{23}{32} q^2 \qquad \mathcal{F}_{2, \text{2-inst}} = \frac{3}{1024} q^2 \dots
\end{align}
Notice how, if we send $\epsilon \to 0$, we find the SW prepotential up to 2 instantons showed in (B.29) of \cite{Alday:2009aq}. It is relevant to say that, since we have used the entire $\mathcal{Z}_2$, the single terms $\mathcal{F}_{g \geq 2, \text{2-inst}}$ are complete, they do not receive any correction from instantons with greater winding numbers.\\

Until this point we have performed first the expansion for small $q$ and then for $a/\epsilon$ very large. Due to the growing complexity of the functions $\mathcal{Z}_{n \geq 3} \left(\frac{a}{\epsilon}\right)$, it could be more convenient to exchange the order of the two limits, but in doing it we have to pay attention at what happens to the components of the prepotential. We have, always at two instantons
\begin{align*}
& \mathcal{Z}_1 \left(\frac{a}{\epsilon}\right)  = \frac{a^2}{\epsilon^2} -3 \\
& \mathcal{Z}_2 \left(\frac{a}{\epsilon}\right)  \simeq \frac{a^4}{2 \epsilon^4} -\frac{35}{16} \frac{a^2}{\epsilon^2} + \frac{13}{8} + \frac{3}{256} \frac{\epsilon^2}{a^2} + \mathcal{O} \left(\frac{\epsilon^4}{a^4}\right) 
\end{align*}
There are several things we can do. For example, if we keep the highest powers in $a/\epsilon$ in each of these two functions, we find $\mathcal{F}_{\text{inst}} \simeq \frac{q}{2} a^2$ and this continues to hold if we pick the same class of terms for greater instanton numbers. This because the maximum power in the expansion of $\mathcal{Z}_k$ is $\frac{1}{k!} \left(\frac{a}{\epsilon}\right)^{2k}$. This choice corresponds to approximate the instanton partition function to $Z_{\text{inst}} \simeq e^{\frac{q}{2} \frac{a^2}{\epsilon^2}}$, meaning that we are selecting only the 1-instanton contribution to the component $\mathcal{F}_0$ of the prepotential.\\
If we now insert the term in $a^2/\epsilon^2$ of $\mathcal{Z}_2$, we obtain that
$$\mathcal{F}_\text{inst} \simeq a^2 \left(\frac{q}{2} - \frac{35}{64} q^2\right)$$
As we can see, now the contribution of 2 instantons to $\mathcal{F}_0$ comes out, but it is not the correct one, as reported in \eqref{eq:genus_components_instantons}. This because we have neglected the constant term in $\mathcal{Z}_1$ which, from the expansion of the logarithm, contributes to the last value of $\mathcal{F}_{0,\text{2-inst}}$ just discussed in order to give the correct one. In particular, by including the $-3$ term of $\mathcal{Z}_1$, we obtain
$$\mathcal{F}_\text{inst} \simeq a^2 \left(\frac{q}{2} + \frac{13}{64} q^2\right) + \epsilon^2 \left(-\frac{3}{2} q - \frac{9}{8} q^2\right)$$
In this way, the contribution of the 1 instanton to $\mathcal{F}_1$ enters in the game and it is the correct one, but the same cannot be said for $\mathcal{F}_{1,\text{2-inst}}$ and that is because we have not included the constant term of $\mathcal{Z}_2$. By doing so we finally arrive at
$$\mathcal{F}_\text{inst} \simeq a^2 \left(\frac{q}{2} + \frac{13}{64} q^2\right) + \epsilon^2 \left(-\frac{3}{2} q - \frac{23}{32} q^2\right)$$
This reasoning is valid for any number of instantons we consider, so the take-home message is that if we expand the function $\mathcal{Z}_k$ for $a/\epsilon \gg 1$ and we consider all the terms up to the constant one, then all these will flow inside the $\mathcal{F}_0$ and $\mathcal{F}_1$ components of the instanton prepotential.\\
If we now consider also the term going as $\epsilon^2/a^2$ of $\mathcal{Z}_2$, we obtain
$$\mathcal{F}_\text{inst} \simeq a^2 \left(\frac{q}{2} + \frac{13}{64} q^2\right) + \epsilon^2 \left(-\frac{3}{2} q - \frac{23}{32} q^2\right) + \frac{\epsilon^4}{a^2} \frac{3 q^2}{1024}$$
and we directly reproduce the correct term $\mathcal{F}_{2,\text{2-inst}}$ shown in \eqref{eq:genus_components_instantons}. If we insert the term of order $\epsilon^4/a^4$ of the same function, we will directly reproduce the correct term $\mathcal{F}_{3, \text{2-inst}}$ and so on for all the others $\mathcal{F}_{g \geq 3, \text{2-inst}}$. Notice that it is valid only because $\mathcal{Z}_1$ is polynomial and so it does not contribute to these components of the prepotential. If we want to evaluate the term $\mathcal{F}_{2, \text{3-inst}}$ we will need both the contributions of 2 and 3 instantons.\\
For what concerns the tree level part, we know that
$$Z_{\text{tree}} = e^{2 \pi i \tau \frac{a^2}{\epsilon^2}} \Rightarrow \mathcal{F}_{\text{tree}} = \epsilon^2 \log Z_{\text{tree}} = 2 \pi i a^2$$
and hence all the components $\mathcal{F}_{g \geq 1}$ terms do not enter $\mathcal{F}_{\text{tree}}$.\\
Passing to the 1-loop part, it can be seen that the behaviour of $Z_{\text{1-L}} (a,\epsilon)$ as shown in \eqref{eq:Z_1_loop} for $a/\epsilon \gg 1$ is
\begin{equation} \label{eq:Z_1L_expansion}
Z_{\text{1-L}} \simeq e^{-8 \frac{a^2}{\epsilon^2} \log 2 + \beta} \, \, \frac{a}{\epsilon} + \mathcal{O} \left(\frac{\epsilon^2}{a^2}\right)
\end{equation}
where $\beta$ is a number (in particular $\beta = 12 \log \gamma_G - 1 - \frac{\log 2}{3}$). At this order we see that
$$\mathcal{F}_{\text{1-L}} \simeq -8 \log 2 \, a^2 + \epsilon^2 \left(\beta + \log \left(\frac{a}{\epsilon}\right)\right)$$
meaning that it contributes only to the first two components of the genus expansion of the prepotential. All the higher order terms in \eqref{eq:Z_1L_expansion} flow into $\mathcal{F}_{g \geq 2, \text{1-L}}$.\\
The final message of this Appendix is that, if one wants to work with only $\mathcal{F}_0$ and $\mathcal{F}_1$ terms of the genus expansion of the prepotential in the evaluation of the extremal correlator (as it is done in \cite{Grassi:2019txd}), it is sufficient to consider the expansion \eqref{eq:Z_1L_expansion} for the 1-loop part and the expansion of the functions $\mathcal{Z}_k$ for the instanton sector up to the constant one in $a/\epsilon$.
\chapter{Electric field of a charged wire} \label{Charged wire}
Consider an infinitely extended and uniformly charged wire along the $y$ direction in $\mathbb{R}^{1,D-1}$. The metric of this space is
\be \label{flatmetric}
\mathrm{d}s^2 = -\mathrm{d}t^2 + \mathrm{d}y^2 + \mathrm{d}r^2 + r^2 \mathrm{d}\Omega_{D-3}^2
\ee
where $\mathrm{d}\Omega_{D-3}^2$ is the metric of a $(D-3)$ - sphere of unitary radius. The transverse space $\mathcal{M}^{D-2}$ has the topology of $\mathbb{R}^+ \times \text{S}^{D-3}$. For a $(D-3)$-sphere embedded in a $(D-2)$-dimensional (flat) space, we introduce $D-3$ angles $\theta_1,\dots, \theta_{D-3}$ and only one of them is azimuthal ($\theta_{D-3} \in [0,2 \pi)$), while all the other ones are polar ($\theta_1,\dots, \theta_{D-4} \in [0,\pi)$). The induced metric is
\be
\label{induced metric}
\mathrm{d}\Omega^2_{D-3} = \sum_{j=1}^{D-3} \left(\prod_{i=1}^{j-1} \sin^2{\theta_i}\right) \mathrm{d}\theta_j^2
\ee 
In the transverse space the problem presents a radial symmetry, so the electric gauge potential sourced by the wire is
\be
B^{(e)} = -\Phi(r) \, \mathrm{d}t \wedge \mathrm{d}y
\ee
Hence, the electric field strength is
\be
F^{(e)} = \mathrm{d} B^{(e)} = E(r) \, \mathrm{d}r \wedge \mathrm{d}t \wedge \mathrm{d}y
\ee
The Hodge dual is
\be
\star F^{(e)} = - E(r) \, r^{D-3} \sqrt{|g_{\text{S}^{D-3}}|} \, \mathrm{d}\theta_1 \wedge \dots \wedge \mathrm{d}\theta_{D-3}
\ee
where $g_{\text{S}^{D-3}}$ is the determinant of \eqref{induced metric}. By Maxwell's equations\footnote{In this expression we have used that, for a generic $k$-form $A_k$ in a D-dimensional spacetime,
\be
\star \star A_k = (-)^s (-)^{k(D - k)} A_k
\ee
For \eqref{flatmetric}, $s=1$ and $\mathrm{d} \star F^{(e)}$ is a $(D-2)$-form.  
}, 
\be \label{Maxwell}
\mathrm{d}\star F^{(e)} = -\star j^{(e)}
\ee
and the 2-form electric current is $j^{(e)} = -\frac{\rho}{\epsilon_0} \mathrm{d}t \wedge \mathrm{d}y$, with $\rho$ the electric volume density, so 
\be
\star j^{(e)} = \frac{\rho}{\epsilon_0} \, r^{D-3} \, \sqrt{|g_{\text{S}^{D-3}}|} \, \mathrm{d}r \wedge \mathrm{d} \theta_1 \wedge \dots \wedge \mathrm{d} \theta_{D-3}
\ee
Due to the fact that $\rho$ is the electric volume density, then $\rho = \mathcal{Q}/(L_y V^{D-2})$, where $\mathcal{Q}$ is the electric charge along the length $L_y$ of the wire and $V^{D-2}$ is the volume of the transverse space. So $\rho = \lambda/V^{D-2}$, with $\lambda$ linear electric density, and consequently
\be
\label{d starFe}
\int_{\mathcal{M}^{D-2}} \mathrm{d} \star F^{(e)} = - \int_{\mathcal{M}^{D-2}}  \star j^{(e)} = - \frac{\lambda}{\epsilon_0 V^{D-2}} V^{D-2} = - \frac{\lambda}{\epsilon_0}
\ee
On the other side, thanks to Stokes' theorem, we can write
\be
\int_{\mathcal{M}^{D-2}} \mathrm{d} \star F^{(e)} = \int_{\text{S}^{D-3}} \star F^{(e)}
\ee
where we are assuming that the transverse space $\mathcal{M}^{D-2}$ has a ($D-3$)-sphere at the boundary at fixed $r$ embedded in it.
So
\be
\label{starFe}
\int_{\text{S}^{D-3}} \star F^{(e)} = - E(r) r^{D-3} \int_{\text{S}^{D-3}} \sqrt{|g_{S^{D-3}}|} \, \mathrm{d}\theta_1 \wedge \dots \wedge \mathrm{d}\theta_{D-3} = - V(\text{S}^{D-3}) E(r) r^{D-3}
\ee
Equalling \eqref{d starFe} with \eqref{starFe}, we get
\be
E(r) = \frac{\lambda}{V(S^{D-3}) \epsilon_0 r^{D-3}}
\ee
It is known that
\be
V(\text{S}^{D-3}) = \frac{2 \pi^{\frac{D-2}{2}}}{\Gamma\left(\frac{D-2}{2}\right)} 
\ee
and so
\be
\label{electric field}
E(r) = \frac{\lambda \Gamma\left(\frac{D-2}{2}\right)}{2 \epsilon_0 \pi^{\frac{D-2}{2}} r^{D-3}}
\ee
Notice that for $D=4$ we get the well known result of standard Electromagnetism in $1+3$ dimensions.\\
Comparing \eqref{electric field} with $Q/r^2$ of \eqref{eq:top} \footnote{We stress that, even if the probe has finite length while now we have considered an infinite length of the wire, the quantity $F^{(e)}$ is dimensionally the same.}, we realize that the parameter $Q$ has the dimensions of a linear charge density and in measure units such that
\be
\frac{1}{4 \pi \epsilon_0} = 1
\ee
they are the same.
\chapter{Measure Units} \label{appmeas}
We choose to work in measure units such that 
\be
c = G_5 = \frac{1}{4 \pi \varepsilon_0} = 1
\ee
In this framework, from Schwarzschild-Tangherlini metric \cite{Tangherlini:1963bw}, we find 
$$[M] = [L]^{D-3} = [L]^2$$ 
Consequently, the force has dimension $[F]=[M] \, [L]^{-1} = [L]$ and, from (\ref{electric field}), $[F] = [q] [E] = [q]^2 [L]^{-3}$, we get that the dimension of the electric charge is 
$$[q] = [L]^2$$
Finally, the dimension of the linear charge density is
$$[Q] = [q] [L]^{-1} = [L]$$
The fact that $[M] = [L]^2$ and $[Q] = [L]$ is consistent with the conditions \eqref{eq:TopCharge} and \eqref{ADM mass}. Notice that in these measure units it is inconsistent to set $\hbar=1$ and indeed, due to the fact that $[E] = [M]= [L]^2$, then 
$$[\hbar] = [E] [L] = [L]^3$$

Let us notice that according to these measure units, the tension of a string has dimension 
\be
[T] = [M] [L]^{-1} = [L]
\ee
hence the constant $\alpha'$ is such that
\be
[\alpha'] = [T]^{-1} = [L]^{-1}
\ee
In this way, the action \eqref{eq:sigmamodel} has the correct dimension $[S] = [\hbar] = [L]^3$. Moreover the parameter $q$ defined in \eqref{eq:qdefinition} has dimension
\be
[q] = [L]^2
\ee
as expected for a charge and consistently with \eqref{eq:massshift}.

\chapter{Numerical amplification factor} \label{appA}
In this Appendix, we briefly describe the numerical procedure implemented in Mathematica used to plot the amplification curve. The starting point consists in finding the leading and a sufficient number of subleading behaviors at infinity and at the cap. At infinity, the radial wave function behaves as
\begin{equation}\label{RINF}
    R_{\infty}(r)\simeq e^{i\epsilon \, r\sqrt{\omega^2-q^2}}r^{-1-\frac{i\epsilon(q^2r_s+2qQ\omega-(r_b+2r_s)\omega^2)}{2\sqrt{\omega^2-q^2}}}\sum_{n=0}^{N_{\infty}}d_n r^{-n}
\end{equation}
where $\epsilon=\pm1$ encodes both the behaviors at infinity and where we set $\mu=p=0$ for simplicity. 
The first coefficient is
\begin{align}
    \frac{d_1}{d_0}=&\frac{q^2 \left(2 r_b+r_s\right)-\omega ^2 r_b-2 q Q \omega }{4 \left(q^2-\omega ^2\right)}-\frac{1}{8 \epsilon 
   \left(\omega ^2-q^2\right)^{3/2}}\Big[i (4 q^3 Q \omega  \left(2 r_b+3
   r_s\right)+\\\nn
   &-2 \omega ^2 \left(q^2 \left(3 r_b r_s+2 r_b^2+6 r_s^2\right)+2 \ell  (\ell +1)\right)-4 q Q \omega ^3 \left(r_b+2
   r_s\right)+\omega ^4 \left(4 r_b r_s+3 r_b^2+8 r_s^2\right)+\\\nn
   &+q^4 \left(3 r_s^2-4 Q^2\right)+4 q^2 \ell  (\ell +1))\Big]\\\nn
 &  \dots
\end{align}
For a TS $(r_b>r_s)$ we have to impose regularity at the cap, so the correct ansatz is
\begin{equation}\label{RCAPts}
    R_c(r)\simeq \sum_{n=0}^{N_c}c_n(r-r_b)^n
\end{equation}
where
\begin{align}
    \frac{c_1}{c_0}=&-\frac{r_b(qQ-r_b\omega)^2-(r_b-r_s)\ell(\ell+1)}{(r_b-r_s)^2}\\\nn
    \frac{c_2}{c_0}=&\frac{1}{4(r_b-r_s)^4}\Big[3 \omega ^2 r_b^2 \left(2 q^2 Q^2 r_b^2+r_s
   \left(r_b-r_s\right)\right)-2 q Q \omega  r_b \left(r_b^2
   \left(2 q^2 Q^2+1\right)+r_b r_s-2 r_s^2\right)+\\\nn
   &{+}q^2
   \left(q^2 Q^4 r_b^2{+}Q^2 \left(r_b{-}r_s\right) \left(2
   r_b{+}r_s\right){+}r_b \left(r_b{-}r_s\right)^3\right){-}\ell^2
   \left(r_b{-}r_s\right) \left(r_b \left(2 \left(q Q{-}\omega 
   r_b\right)^2{+}1\right){-}r_s\right){+}\\\nn
   &{-}2 \ell 
   \left(r_b{-}r_s\right) \left(r_b \left(q Q{-}\omega 
   r_b\right)^2{+}r_b{-}r_s\right){-}4 q Q \omega ^3 r_b^5{+}\ell^4
   \left(r_b{-}r_s\right)^2{+}2 \ell ^3
   \left(r_b{-}r_s\right)^2{+}\omega ^4 r_b^6\Big]\\\nn
   &\dots
\end{align}
Otherwise for a BS the boundary conditions at the horizon require an ingoing wave
\begin{equation}\label{RCAPbs}
    R_H(r)=(r-r_s)^{\frac{i(qQ-r_s\omega)}{\sqrt{1-\frac{r_b}{r_s}}}}\sum_{n=0}^{N_H}\tilde{c}_n(r-r_s)^n
\end{equation}
where 
\begin{align}
    \frac{\tilde{c}_1}{\tilde{c}_0}=&\frac{1}{\left(r_b{-}r_s\right)^2 \left(i \sqrt{1{-}\frac{r_b}{r_s}}{-}2 q Q{+}2 \omega  r_s\right)}\Big[\left(r_b{-}r_s\right) \left(\omega  r_s{-}q Q\right){+}i \sqrt{1{-}\frac{r_b}{r_s}} (q^2 \left(Q^2 r_b{+}r_s \left(r_b{-}r_s\right)^2\right){+}\\\nn
    &{+}2
   q Q \omega  r_s \left(r_s-2 r_b\right)+\omega ^2 r_s^2 \left(3 r_b-2 r_s\right)+\ell  (\ell +1)
   \left(r_s-r_b\right))\Big]\\\nn
  &\dots 
\end{align}

The numerical integration is performed starting from the horizon (or the cap) and moving up to infinity. If we denote $A_{\text{in}}$ and $A_{\text{out}}$ as the zero-order coefficients $d_0$ of the asymptotically expanded wave at infinity \eqref{RINF}, respectively corresponding to $\epsilon=-1$ and $\epsilon=1$, and if we compare this function with its numerical counterpart along with their derivatives, we obtain expressions for $A_{\text{in}}$ and $A_{\text{out}}$. The final step consists in plotting the quantity $Z=\left|\frac{A_{\text{out}}}{A_{\text{in}}}\right|^2-1$ as a function of $\omega$.
\chapter{Separability} \label{APP:separability} 
In this Appendix we report the details that allow to determine the equations \eqref{eq:diffeq} of the separable case.\\
We start by computing the conjugate momenta using \eqref{eq:Momenta}, knowing the Lagrangian in the more general case (meaning keeping in count also the coordinates $z_I$, with $I=1,\dots,4$, that have been neglected throughout all the Chapter 4). We get the following results

\begin{equation} \label{momenta}
P_t=-\frac{\dot{t}+\omega_\phi \dot{\phi}}{H}=-E\quad,\quad P_z=\frac{\dot{z}+\omega_\psi \dot{\psi}}{H}\quad,\quad P_\rho=\frac{H(\rho^2+a_f^2\cos^2\theta)}{\rho^2+a_f^2}\dot{\rho}\quad,\quad P_\theta=H(\rho^2+a_f^2\cos^2\theta)\dot{\theta} 
\end{equation}
$$P_\phi=-\frac{\omega_\phi}{H}\dot{t}+\Bigg[H(\rho^2+a_f^2)\sin^2\theta-\frac{\omega_\phi^2}{H}\Bigg]\dot{\phi}=J_\phi \quad,\quad P_\psi=\frac{\omega_\psi}{H}\dot{z}+\left(H\rho^2\cos^2\theta+\frac{\omega_\psi^2}{H}\right)\dot{\psi}=J_\psi$$
$$P_{1,2}=\left(\frac{H_1}{H_5}\right)^{1/2}\dot{z}_{1,2}\quad,\quad P_{3,4}=\left(\frac{H_5}{H_1}\right)^{1/2}\dot{z}_{3,4}$$
where the only non conserved momenta are $P_\rho$ and $P_\theta$. By inverting the previous relations, we obtain
\begin{equation}\label{dotxinP}
\dot{t}=H E-\frac{\omega_\phi(J_\phi +E\omega_\phi)}{H(\rho^2+a_f^2)\sin^2\theta} \quad,\quad \dot{z}=H P_z-\frac{\omega_\psi(J_\psi-P_z\omega_\psi)}{H\rho^2\cos^2\theta} \quad,\quad \dot{\rho}=\frac{\rho^2+a_f^2} {H(\rho^2+a_f^2\cos^2\theta)}P_\rho
\end{equation}
$$\dot{\theta}=\frac{1}{H(\rho^2+a_f^2\cos^2\theta)}P_\theta \quad,\quad \dot{\phi}=\frac{J_\phi+E \omega_\phi}{H(\rho^2+a_f^2)\sin^2\theta} \quad,\quad \dot{\psi}=\frac{J_\psi-P_z \omega_\psi}{ H\rho^2\cos^2\theta}$$
$$\dot{z}_{1,2}=\left(\frac{H_5}{H_1}\right)^{1/2}P_{1,2}\quad,\quad \dot{z}_{3,4}=\left(\frac{H_1}{H_5}\right)^{1/2} P_{3,4}$$

In order to find \eqref{eq:diffeq}, we have to divide $\dot{\rho}$ by $\dot{t}$ and $\dot{\phi}$ by $\dot{t}$, where $\omega_{\phi}$ and $H$ are given in \eqref{eq:def_params}, $J_\phi = b E$ (from \eqref{impact param}) and $P^2_\rho$ is written in \eqref{eq:Prho_Plan}. By doing all these substitutions, the equations \eqref{eq:diffeq} are obtained.
\chapter{Derivation of the ground truth for the planar case} \label{APP:groundtruth} 
\begin{figure}[H]
\includegraphics[scale=0.8]{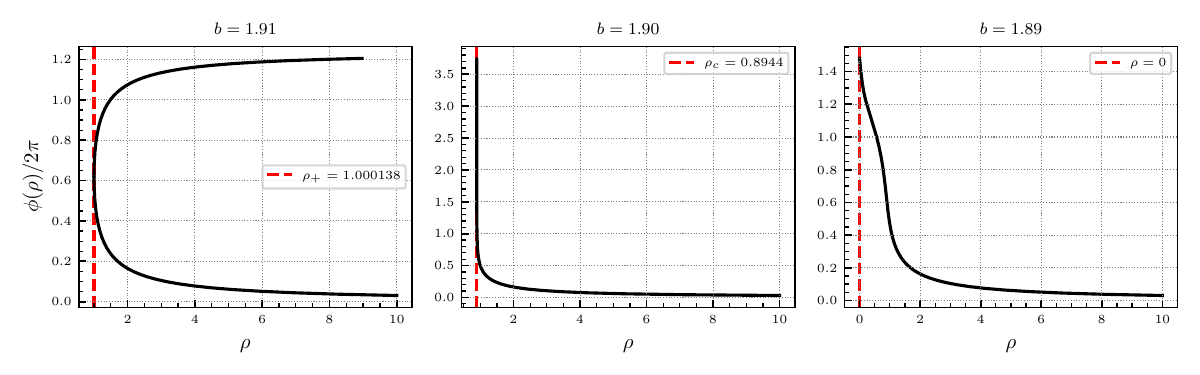}
\caption{
\label{fig:phi_exact}
Exact functions $\phi(\rho)/2\pi$ for the planar geometry. \emph{Left-panel}: over-critical case with $b=1.91$. The vertical dashed  line represents $\rho_+=1.000138$, which is the largest root of the radicand in \eqref{eq:diffeq}. \emph{Central-panel}: critical case with $b=b_c=1.9$. The vertical dashed  line is the critical radius $\rho_c=0.8944$, obtained from \eqref{eq:unstable_ph}. \emph{Right-panel}: sub-critical case with $b=1.89$. The vertical dashed line is drawn for $\rho=0$, below which the motion has no physical sense.
}
\end{figure}

As hinted in Section \ref{sec:Planar_Case}, the planar case can be studied by exactly integrating the equatorial geodesics.\\
The starting point is \eqref{eq:diffeq}, from which we get

\begin{equation} \label{eq:groundtruth}
\phi(\rho)-\phi_0=\pm \int_{\rho_0}^{\rho}\dd\rho'\frac{ \left(L_1 L_5 a_f+b \rho'^{2}\right)}{\left(\rho'^2+a_f^2\right) \sqrt{\rho'^4+\rho'^2
   \left(a_f^2-b^2+L_1^2+L_5^2\right)-2 b L_1 L_5 a_f+\left(L_1^2+L_5^2\right) a_f^2+L_1^2 L_5^2}},
\end{equation}

which holds for all the three regimes listed in Section \ref{sec:Planar_Case} (sub-critical, critical and over-critical). By numerically solving this integral, we can find the angle $\phi$ as a function of $\rho$ and, by using \eqref{eq:cartesian_map}, we can build the trajectory that we promote to ground truth in the comparison with the results obtained from the NN and the other numerical integrators. We have used extended-precision arithmetic to compute exactly the integral up to 32 digits. In the main text we have set the geometry parameters to $a_f=0.1$, $L_1=L_5=L=1$ and $E=1$. The functions $\phi(\rho)$, obtained by integrating \eqref{eq:groundtruth}, are shown in Figure~\ref{fig:phi_exact} for $b=1.91$ (over-critical regime), $b=b_c=1.9$ (critical regime) and $b=1.89$ (sub-critical regime).

In the over-critical regime,  the function $\phi(\rho)$ can be written in a closed form, which we now show for completeness. Indeed, when $b > b_c$ the shape of the potential $-Q_R$ is similar to the one in the bottom-panel of Figure~\ref{fig:plotQ} but shifted upwards and then intersecting the $\rho$-axis twice. This implies that the radicand inside the integral \eqref{eq:groundtruth}, proportional to $Q^2_R$, admits two distinct zeros. In terms of the variable $y$, defined in such a way that $\rho=a_f\sqrt{y}$, the integral becomes
\begin{flalign}\label{eqgeo1}
\phi-\phi_0=\frac{1}{2 a_f^2}\int_{\rho_0^2/a_f^2}^{\rho^2/a_f^2}\dd y\frac{(L_1L_5+a_f b y)}{(1+y)\sqrt{y(y-y_+)(y-y_-)}},
\end{flalign}
where $y_\pm=\rho_\pm^2/a_f^2$. In particular

\begin{flalign}
y_{\pm}=\frac{b^2-a_f^2-L_1^2-L_5^2\pm\sqrt{\left(a_f+b-L_1-L_5\right)
   \left(-a_f+b+L_1-L_5\right) \left(-a_f+b-L_1+L_5\right)
   \left(a_f+b+L_1+L_5\right)}}{2
   a_f^2}.
\end{flalign}
At the end of the day \eqref{eqgeo1} can be written as
\begin{align}
\phi-\phi_0=&\frac{1}{\rho _- a_f}\Big[L_1 L_5 \left(\Pi \left(-\frac{a_f^2}{\rho _-^2};\sin ^{-1}\left(\frac{\rho _-}{\rho }\right)|\frac{\rho _+^2}{\rho _-^2}\right)-\Pi
   \left(-\frac{a_f^2}{\rho _-^2};\sin ^{-1}\left(\frac{\rho _-}{\rho _0}\right)|\frac{\rho _+^2}{\rho _-^2}\right)-F\left(\sin
   ^{-1}\left(\frac{\rho _-}{\rho }\right)|\frac{\rho _+^2}{\rho _-^2}\right)+\right.\nn\\
   &\left.+F\left(\sin ^{-1}\left(\frac{\rho _-}{\rho _0}\right)|\frac{\rho
   _+^2}{\rho _-^2}\right)\right)-b a_f \left(\Pi \left(-\frac{a_f^2}{\rho _-^2};\sin ^{-1}\left(\frac{\rho _-}{\rho }\right)|\frac{\rho
   _+^2}{\rho _-^2}\right)-\Pi \left(-\frac{a_f^2}{\rho _-^2};\sin ^{-1}\left(\frac{\rho _-}{\rho _0}\right)|\frac{\rho _+^2}{\rho
   _-^2}\right)\right)\Big],
\end{align}

where $\Pi(n,\phi|m)$ is the incomplete elliptic integral of the third kind and $F(\phi|m)$ is the incomplete elliptic integral of the first kind.

\begin{figure*}[t]
\includegraphics[scale=0.8]{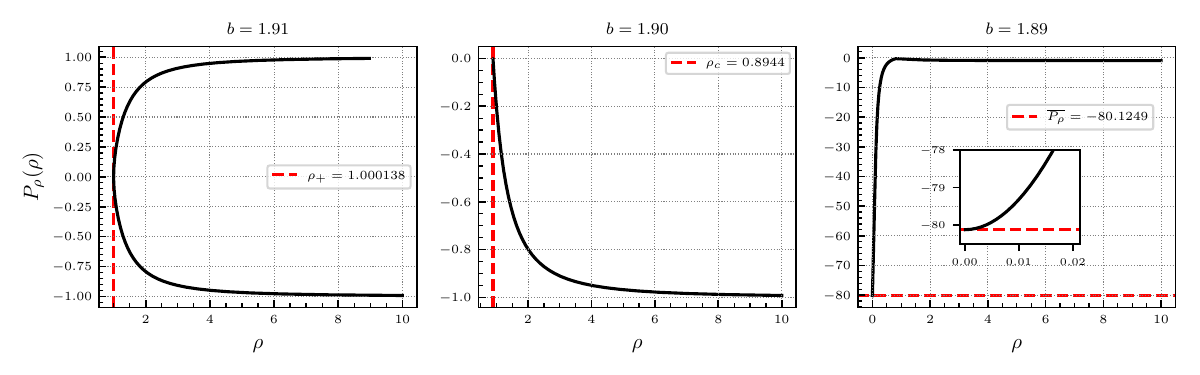}
\caption{
\label{fig:Prho_exact}
Exact functions $P_\rho(\rho)$ for the planar geometry. \emph{Left-panel}: over-critical case with $b=1.91$. The vertical dashed  line represents $\rho_+=1.000138$, which is the largest root of the radicand in \eqref{eq:diffeq}. \emph{Central-panel}: critical case with $b=b_c=1.9$. The vertical dashed  line is the critical radius $\rho_c=0.8944$, obtained from \eqref{eq:unstable_ph}. \emph{Right-panel}: sub-critical case with $b=1.89$. The horizontal dashed line represents $\overline{P_\rho}\equiv P_\rho(0)=-80.1249$, computed from \eqref{eq:Prho_exact_subcritical}. The small frame is a zoom in the region $\rho\in[0,0.02]$ and shows as $P_\rho(\rho)$ has a vanishing derivative for $\rho=0$.
}
\end{figure*}

The exact function $P_\rho(\rho)$ is easily obtained by solving the constraint $\mathcal{H}(\rho)=0$, with $\mathcal{H}$ the  Hamiltonian given in \eqref{eq:Ham_plan} at fixed choice of the parameters. For their values reported at the beginning of this Appendix we find
\begin{flalign}\label{eq:Prho_exact_overcriical}
  b=1.91\qquad   P_\rho (\rho) &= \pm \sqrt{\frac{6380 - 16381 \rho^2 + 10000 \rho^4}{1 + 200 \rho^2 + 10000 \rho^4}}\,,
\\[6pt] \label{eq:Prho_exact_critical}
b=1.90\qquad P_\rho (\rho) &= -\frac{20 (-4 + 5 \rho^2)}{1 + 100 \rho^2}\,,
\\[6pt] \label{eq:Prho_exact_subcritical}
     b=1.89\qquad P_{\rho}(\rho)&=-\sqrt{\dfrac{6420-15621\rho^{2}+10000\rho^{4}}{1+200\rho^{2}+10000\rho^{4}}}.
\end{flalign}
In the first equation the minus sign holds for the incoming phase, while the plus sign is valid for the outgoing phase. The three functions are plotted in Figure~\ref{fig:Prho_exact}

\chapter{Examination of the geometry of non-planar critical geodesics} \label{APP:nonplangeo}
In this Appendix we investigate in more details what has been said in Section \ref{Non-Planar Case} for the nature of the non-planar critical geodesics, showing that in general they wrap around a spheroidal zone.\\ 
We solved the critical conditions $\mathcal{R}(\rho_c) = \mathcal{R}'(\rho_c) = 0$ in terms of $\zeta_c$ and $b_c$, getting \eqref{eq:critreg}. In order to find the critical radius $\rho_c$, we must work on the condition $P_\theta^2\geq 0$. This can written as
\begin{equation} \label{pos Ptheta}
\beta^2-\frac{\beta_\phi^2}{\sin^2\theta}-\sin^2\theta\geq0 ,\qquad \text{with} \qquad \beta=\frac{b}{a_f},\beta_{\phi}=\frac{b_{\phi}}{a_f}.
\end{equation}
If we introduce $\sin^2\theta=\frac{1}{2}(1-X)$, where $X=\cos(2\theta)$, the solutions of the associated equation are
\begin{equation}
X_\pm=1-\beta^2\pm\sqrt{\beta^4-4\beta_\phi^2}.
\end{equation}
It is very easy to show that the discriminant $\beta^4-4\beta_\phi^2$ is always positive in the critical regime \eqref{eq:critreg}. From the definition of $X$, we can write down the following chain of inequalities
\be
-1\leq X_-\leq X_+\leq 1.
\ee
The condition $X_-\geq-1$ is always false while $X_+\leq 1$ is always verified, so that the first non trivial condition is $X_+\geq-1$, which yields
\begin{equation}
\beta^2-2-\sqrt{\beta^4-4\beta_\phi^2}\leq0 \leftrightarrow \sqrt{b^4-4 b^2_\phi a^2_f} \geq b^2-2 a_f^2.
\end{equation}
Since we want to study the critical geodesics, we should evaluate this relation in the critical regime \eqref{eq:critreg}. In doing so, we have to distinguish two cases depending on the sign of the right hand side. It turns out that if $b^2_c< 2 a_f^2$, then from \eqref{eq:critreg} $2 \rho_c^2 + L_1^2+L_5^2 < 0$ and it is never satisfied, so the only case to consider is when $b^2_c > 2 a_f^2$. This brings to the following condition
\begin{equation}
    b^2_c \geq b^2_{\phi,c}+a_f^2.
\end{equation}
When we substitute the expressions \eqref{eq:critreg}, we have to distinguish two cases according to the sign of $\zeta_c$, with the aim of solving with respect to $\rho_c$

$$a_f\zeta_c=\rho_c^2+a_f^2\Longrightarrow 0\leq\rho_c\leq\sqrt{a_f|L_1-L_5|-L_1L_5}\quad\text{for}\quad a_f\geq\frac{L_1L_5}{|L_1-L_5|}$$
\begin{equation}
    a_f\zeta_c=-(\rho_c^2+a_f^2)\Longrightarrow 
\begin{cases} \label{counter geo}
\sqrt{L_1L_5-a_f(L_1+L_5)}\leq\rho_c\leq \sqrt{L_1L_5+a_f(L_1+L_5)}, & 0<a_f\leq\frac{L_1L_5}{L_1+L_5} \\
0\leq\rho_c\leq \sqrt{a_f(L_1+L_5)-L_1 L_5}, & a_f>\frac{L_1L_5}{L_1+L_5}
\end{cases}
\end{equation}

The former case corresponds to a \textit{corotating} geodesic, the latter to a \textit{counterrotating} geodesic.\\
The inequality \eqref{pos Ptheta} has as solution $X_- \leq X \leq X_+$. Since $X_- < -1$ is always valid, then the first inequality is always satisfied. Hence it is left to solve $\cos(2 \theta) \leq X_+$, whose solution is (keeping in mind that $\theta \in [0,\pi]$)
\begin{flalign}\label{eq:interval_theta}
&\theta_{\text{bound}} \leq \theta \leq \pi-\theta_{\text{bound}} 
\nonumber\\[6pt]
&\text{with} \qquad \theta_{\text{bound}} = \frac{\arccos(X_+)}{2}.
\end{flalign}
Since $\phi$ varies from 0 to 2$\pi$, the geodesic describes a spheroidal zone. In order to understand what this is, we recall that the change of coordinates \eqref{eq:change_coord_3d} is a parametrization of an ellipsoid, at $\rho$ fixed, whose cartesian equation is
\begin{equation}\label{eq:oblate_spheroid}
    \frac{x^2}{\rho^2+a_f^2} + \frac{y^2}{\rho^2+a_f^2} + \frac{z^2}{\rho^2} = 1.
\end{equation}
By indicating in this context $a^2 = \rho^2+a_f^2, \, c^2 = \rho^2+a_f^2, \, d^2 = \rho^2$, then, if two of these quantities are equal, the ellipsoid is said to be a spheroid (or ellipsoid of revolution) and since $a=c>d$, then it is an oblate spheroid. As $\theta$ varies inside \eqref{eq:interval_theta}, then the surface is the one bounded by two planes that are parallel to the $x-y$ plane and intersect the oblate spheroid at $\theta=\theta_{\text{bound}}$ and $\theta = \pi-\theta_{\text{bound}}$. This region is called spheroid zone.\\ 

In the text we considered the case with $X_+ = 1$, corresponding to $\theta_{\text{bound}}=0$ and hence $\theta$ varies within its interval of definition, $\theta \in [0,\pi]$. For this reason such trajectory wraps around the complete oblate spheroid \eqref{eq:oblate_spheroid} with $\rho = \rho_c$. Furthermore, the condition $X_+ = 1$ brings to $\beta_{\phi,c} = 0 \leftrightarrow b_{\phi,c} = 0$ and so from the definition \eqref{eq:def_zeta_bs} we gain $\zeta_c = - \frac{L_1 L_5}{a_f}$. This explains the first condition of \eqref{eq:final_params_nonplan} and hence the fact that this geodesic is counterrotating. The critical radius can be found from $X_+ = 1$ expressed in terms of $\rho_c$ and then solving it with respect to $\rho_c$ or also from the first equation of \eqref{eq:critreg} by substituting $\zeta^2_c$ with $(L_1 L_5/a_f)^2$ and solving with respect to $\rho_c$ (which is, by definition, non negative). At the end we get
\begin{equation}
\rho_c=\sqrt{L_1L_5-a_f^2},\quad 0<a_f\leq \sqrt{L_1L_5}
\end{equation}
which is exactly \eqref{eq:crit_radius}. Notice that this condition has to satisfy \eqref{counter geo}. It turns out that $\sqrt{L_1 L_5} > \frac{L_1 L_5}{L_1+L_5}$ for every $L_1,L_5 >0$ and it can be shown that if $a_f$ satisfies the first or second condition of \eqref{counter geo}, then the corresponding critical radius $\rho_c$ falls within the associated interval.
\chapter{Derivation of Teukolsky equation for GWs} \label{app:Teuk}
This Appendix is devoted to the derivation of the Teukolsky equation that describes the radial propagation of GWs emitted by the probe moving around the Schwarzschild BH.\\
Given a symmetric tensor $X_{\alpha \beta}$, we consider the differential operator $\mathcal{E}$, whose action on the previous tensor is given by
\begin{align} \label{eq:E_operator}
\mathcal{E} [X] & = 2 r^4 L_{-1} L_0 X_{n n} + \sqrt{2} r^3 f(r)^2 J_+ \left(\frac{r^2}{f(r)} L_{-1} X_{n \bar{m}}\right) + \sqrt{2} r f(r)^2 J_+ \left(\frac{r^4}{f(r)} L_{-1} X_{n \bar{m}}\right) + \notag \\
& + r f(r)^2 J_+ \left[r^4 J_+ \left(r X_{\bar{m} \bar{m}}\right)\right]
\end{align}
where
\be
X_{n n} = X_{\alpha \beta} \, n^\alpha n^\beta \quad , \quad X_{n \bar{m}} = X_{\alpha \beta} \, n^\alpha \bar{m}^\beta \quad , \quad X_{\bar{m} \bar{m}} = X_{\alpha \beta} \, \bar{m}^\alpha \bar{m}^\beta
\ee
and the differential operators $L_s$ and $J_\pm$ are defined as (in the usual Schwarzschild coordinates)
\be \label{eq:L_J_operators}
L_s = \partial_\theta - \frac{i}{\sin \theta} \partial_\phi + s \cot \theta \qquad , \qquad J_\pm = \partial_r \mp \frac{1}{f(r)} \partial_t 
\ee
Taking inspiration from the LHS of the formula (4.7) in \cite{1973ApJ...185..635T} (setting the BH spin equal to 0), we can define the following operator $\Phi_s$
\begin{align} \label{eq:Phi_operator}
\Phi_s [\psi] & = \frac{r^4}{\Delta} \partial_t^2 \psi - \frac{1}{\sin^2 \theta} \partial_\phi^2 \psi - \Delta^{-s} \partial_r \left(\Delta^{s+1} \partial_r \psi \right) - \frac{1}{\sin \theta} \partial_\theta \left(\sin \theta \partial_\theta \psi \right) + \notag \\
& - 2 s \frac{i \cos \theta}{\sin^2 \theta} \partial_\phi \psi - 2 s \left(\frac{M r^2}{\Delta} - r \right) \partial_t \psi + \left(s^2 \cot^2 \theta -s \right) \psi
\end{align}
with $\Delta = r(r-2 M)$. Through an explicit computation it can be proved that, always at the linear level in the perturbation
\be
\mathcal{E} [\delta G] = - 4 \Phi_{-2} \left[r^4 \mathbf{\Psi}_4\right]
\ee
and it is a kinematical equality, the equations of motions are not used. Thanks to this, if we apply the operator $\mathcal{E}$ to both sides of the Einstein equation \eqref{eq:linearized_Einstein_equations}, we get
\be \label{eq:E_to_Einstein}
\Phi_{-2} \left[r^4 \mathbf{\Psi}_4 \right] = - 2 \pi G \mathcal{T} \qquad \text{with} \qquad \mathcal{T} \equiv \mathcal{E} [T]
\ee
Now we consider the following mode decomposition of the Weyl scalar $\mathbf{\Psi}_4$
\be \label{eq:mode_decomposition}
r^4 \mathbf{\Psi}_4 = \int \frac{\mathrm{d} \omega}{2 \pi} \sum_{\ell, m} e^{-i \omega t} R_{\ell m \omega} (r) Y^{\ell m}_{-2} (\theta, \phi)
\ee
where $(t,r,\theta,\phi)$ are generic spacetime coordinates, eventually associated to the observer's ones. These have not to be confused with the coordinates of the particle, as we are going to specify later. $Y^{\ell m}_{-2} (\theta, \phi) = e^{i m \phi} S^{\ell m}_{-2} (\theta)$ are the $-2$-spin-weighted spherical harmonics. Keeping the spin $s$ general, the functions $S^{\ell m}_s (\theta)$ satisfy the following differential equation
\be \label{eq:Slm_equation}
\frac{\partial_\theta \left(\sin \theta \partial_\theta S^{\ell m}_{s}(\theta)\right)}{\sin \theta} + \left[(\ell -s) (\ell +s +1) - \frac{m^2}{\sin^2\theta} - \frac{2 m s \cos \theta}{\sin^2 \theta} - s^2 \cot^2 \theta + s \right] S^{\ell m}_s = 0
\ee
and their expression is
\be
S^{\ell m}_s (\theta) = \sin^{2 \ell} \left(\frac{\theta}{2}\right) \sqrt{\frac{(2 \ell +1)(\ell-m)! (\ell+m)!}{4 \pi (\ell-s)! (\ell+s)!}} \sum_{r=0}^{\ell-s} (-)^{\ell+m+r-s} \begin{pmatrix}
\ell -s \\
r
\end{pmatrix} \begin{pmatrix}
\ell + s \\
r+s-m
\end{pmatrix} \cot^{2 r+s-m} \left(\frac{\theta}{2}\right)
\ee
By applying each term inside the operator $\Phi_s$ in \eqref{eq:Phi_operator} (with $s =-2$) to $r^4 \mathbf{\Psi}_4$ and by using the equation \eqref{eq:Slm_equation} for the part that involves $\sin^{-1} \theta \, \partial_\theta \left[\sin \theta \, \partial_\theta \left(r^4 \mathbf{\Psi}_4 \right) \right]$, we obtain
\be
\Phi_{-2} \left[r^4 \mathbf{\Psi}_4\right] = \int \frac{\mathrm{d} \omega}{2 \pi} \sum_{\ell,m} e^{- i \omega t} Y^{\ell m}_{-2} (\theta, \phi) \left(- \mathcal{O}^\ell_{-2 \omega} \left[R_{\ell m \omega} (r) \right]\right)
\ee
where we have defined the operator $\mathcal{O}^\ell_{s \omega}$ as
\begin{align} \label{eq:O_operator}
\mathcal{O}^\ell_{s \omega}(r) \left[R_{\ell m \omega} (r) \right] & = \Delta^{-s}(r) \partial_r \left[\Delta^{s+1}(r) \partial_r \left(R_{\ell m \omega} (r) \right)\right] + \left[\frac{(\omega r^2)^2 - 2 i s \omega r^2 (r-M)}{\Delta(r)} + \right. \notag \\
& \left. + 4 i s \omega r - (\ell -s) (\ell + s + 1) \right] R_{\ell m \omega}(r)
\end{align}
Thanks to the orthogonality condition of the spherical harmonics
\be
\int_{\theta = 0}^{\theta=\pi} \int_{\phi = 0}^{\phi =2 \pi} \sin \theta \, \mathrm{d} \theta \, \mathrm{d} \phi \, Y^{\ell m}_{-2}(\theta,\phi) \, Y^{l' m' *}_{-2}(\theta,\phi) = \delta_{\ell \ell'} \delta_{m m'}
\ee
we obtain that
\be
\int \mathrm{d} t \, \sin \theta \, \mathrm{d} \theta \, \mathrm{d} \phi \, e^{i \omega' t} \, Y^{l' m' *}_{-2}(\theta,\phi) \, \Phi_{-2} \left[r^4 \mathbf{\Psi}_4 \right] = - \mathcal{O}^{l'}_{-2 \omega'} \left[R_{\ell' m' \omega'} (r)\right]
\ee
Applying $\int \mathrm{d} t \, \sin \theta \, \mathrm{d} \theta \, \mathrm{d} \phi \, e^{i \omega t} \, Y^{l m *}_{-2}(\theta,\phi) $ to the both sides of \eqref{eq:E_to_Einstein} we get
\be \label{eq:operatorial_Teuk}
\mathcal{O}^\ell_{-2 \omega}(r) \left[R_{\ell m \omega} (r)\right] = T_{\ell m \omega} (r)
\ee
where 
\be \label{eq:components_Tlm}
T_{\ell m \omega}(r) \equiv 2 \pi G \int \mathrm{d} t \, \sin \theta \, \mathrm{d} \theta \, \mathrm{d} \phi \, e^{i \omega t} \, Y^{l m *}_{-2}(\theta,\phi) \, \mathcal{E}[T]
\ee
We want to stress that here we are integrating over all the possible values of the spacetime coordinates, or of the observer's coordinates (except over $r$, whose role will be relevant in the behavior of the solution of the equation we are determining). Indeed it is important to posses an expression for all $t,\theta,\phi$ (which are integrated out) because it is not known a priori the direction of the observer with respect to the orbit's plane of the binary system \footnote{I am grateful to Carlo Heissenberg for the discussion on this aspect.}. By the way, when we explicitly compute the source term, the fact that the support of the $\delta^{(4)}$ that appears inside the components of the stress energy tensor in \eqref{eq:stress_components_appr} is only along the particle's trajectory restricts the dependence on the trajectory itself, through the resolution of the integrals in $t,\theta,\phi$. This is analyzed in Section \ref{sec:source_term}.\\
\eqref{eq:operatorial_Teuk} is the Teukolsky equation shown in the main text in \eqref{eq:teuk}. It is possible also to keep the spin $s$ generic, obtaining
\be \label{eq:teuk_general_s}
\frac{1}{\Delta (r)^{s} } \frac{\mathrm{d}}{\mathrm{d} r} \left[ \Delta (r)^{s+1} R'_{\ell m \omega}(r) \right] + \left(\frac{\omega^2 r^4 - 2 i s (r - M)  \omega r^2  }{\Delta (r)} + 4 i s \omega r - (\ell - s)(\ell + s + 1) \right)R_{\ell m \omega}(r)  = T_{\ell m \omega} (r)
\ee
but now the components of the stress energy tensor on the RHS are different with respect to \eqref{eq:components_Tlm}. Indeed, the operator $\mathcal{E}$ involves up to two derivatives with respect to $r$ because we are now considering gravitational perturbations. For the case of scalar perturbations, no derivatives are involved inside $T_{\ell m \omega}$.
\chapter{Explicit computation of the source term $Z_{\ell m \omega}^\infty$} \label{app:Z_lm_inf}
In this Appendix we show all the detailed passages that lead to the explicit expression for the source term $Z_{\ell m \omega}^\infty$ that has been directly reported in \eqref{eq:final_Z_generic}. We recall that this quantity characterizes the waveform through the equation \eqref{eq:h_lm_generic}.\\
First of all we have to rewrite $T_{\ell m \omega}(r)$, which depends on the particle's trajectory.
\be
T_{\ell m \omega}(r) = 2 \pi G \int \mathrm{d} t \, \sin \theta \, \mathrm{d} \theta \, \mathrm{d} \phi \, e^{i \omega t} \, Y^{l m *}_{-2} \, \mathcal{E}[T]
\ee
Consider a generic function $g(r,t)$. The following relation is valid
\be
r^{\alpha} J_+(g) = J_+(r^\alpha \, g) - \alpha r^{\alpha-1} \, g \Leftrightarrow [r^\alpha,J_+] = -\alpha r^{\alpha-1}
\ee
with the operator $J_+$ defined in \eqref{eq:L_J_operators}. In this way, using the definition of the operator $\mathcal{E}$ in \eqref{eq:E_operator} and the property just exposed for $J_+$, we arrive at obtaining
\be
\frac{\mathcal{E}[T]}{\Delta(r)^2} = L_{-1} L_0 \left(\frac{2}{f(r)^2} \mathcal{T}_{nn}\right) + 2 \sqrt{2} L_{-1} \left[\frac{2 \mathcal{T}_{n \bar{m}}}{f(r)} + J_+ \left(\frac{r}{f(r)} \mathcal{T}_{n \bar{m}} \right)\right] + J^2_+ \left(r^2 \mathcal{T}_{\bar{m} \bar{m}}\right) + 2 J_+ \left(r \mathcal{T}_{\bar{m} \bar{m}}\right)
\ee
Given two generic functions of $\theta,\phi$, $\varphi(\theta,\phi)$ and $\psi(\theta,\phi)$, we introduce the short-hand notation
\be
\int_0^\pi \mathrm{d} \theta \sin \theta \int_0^{2\pi} \mathrm{d} \phi \varphi(\theta,\phi)^* \, \psi(\theta,\phi) \equiv (\varphi, \psi)
\ee
and, assuming that $(\varphi^* \, \psi)|_{\phi=2 \pi} -(\varphi^* \, \psi)|_{\phi=0} = 0$ and all the dependence from $\phi$ is of the form $e^{i m \phi}$ (as it will be the case), we obtain, integrating by parts 
\be \label{eq:property_L_integration}
\left(\varphi, L_{-s} \psi \right) = - \left(L^\dagger_{s+1} \varphi, \psi\right)
\ee
Coming back to our $Z_{\ell m \omega}^\infty$ in \eqref{eq:Z_lmomega}, we have
\begin{align}
Z_{\ell m \omega}^\infty & = \int_{2M}^\infty \mathrm{d} r \, \mathfrak{R}_{\text{in},\omega}(r) \, 2 \pi G \int \mathrm{d} t \, \sin \theta \mathrm{d} \theta \, \mathrm{d} \phi \, e^{i \omega t} Y^{\ell m *}_{-2} \frac{\mathcal{E}[T]}{\Delta(r)^2} = \notag \\
& = \int_{2M}^\infty \mathrm{d} r \, \mathfrak{R}_{\text{in},\omega}(r) \, 2 \pi G \int \mathrm{d} t \, \sin \theta \mathrm{d} \theta \, \mathrm{d} \phi \, e^{i \omega t} Y^{\ell m *}_{-2} \frac{2}{f(r)^2} L_{-1} L_0 \left(\mathcal{T}_{nn}\right) + \notag \\
& +  \int_{2M}^\infty \mathrm{d} r \, \mathfrak{R}_{\text{in},\omega}(r) \, 2 \pi G \int \mathrm{d} t \, \sin \theta \mathrm{d} \theta \, \mathrm{d} \phi \, e^{i \omega t} Y^{\ell m *}_{-2} 2 \sqrt{2} L_{-1} \left[\frac{2 \mathcal{T}_{n \bar{m}}}{f(r)} + J_+ \left( \frac{r}{f(r)} \mathcal{T}_{n \bar{m}} \right) \right] + \notag \\
& +  \int_{2M}^\infty \mathrm{d} r \, \mathfrak{R}_{\text{in},\omega}(r) \, 2 \pi G \int \mathrm{d} t \, \sin \theta \mathrm{d} \theta \, \mathrm{d} \phi \, e^{i \omega t} Y^{\ell m *}_{-2}  \left[J^2_+ \left(r^2 \mathcal{T}_{\bar{m} \bar{m}}\right) + 2 J_+ \left(r \mathcal{T}_{\bar{m} \bar{m}}\right)\right]
\end{align}
Now we want to bring $e^{i \omega t} \mathfrak{R}_{\text{in},\omega}(r)$ on the right: as long as we have a $L_s$ operator, which is multiplicative with respect to $t$ and $r$, we can simply do it, but when there is the operator $J_+$ we need to integrate by parts. Furthermore, applying twice the property \eqref{eq:property_L_integration}, we obtain
\be
\left(Y^{\ell m}_{-2}, L_{-1} L_0 (\mathcal{T}_{nn})\right) = \left(L_1^\dagger L_2^\dagger Y^{\ell m}_{-2}, \mathcal{T}_{nn}\right)
\ee
At the end we write
\begin{align}
Z_{\ell m \omega}^\infty = 2 \pi G \int_{2M}^{\infty} \mathrm{d} r \int_{-\infty}^{\infty} \mathrm{d} t & \left[\frac{2}{f(r)^2} \left(L_1^\dagger L_2^\dagger Y^{\ell m}_{-2}, \mathcal{T}_{nn}\right) - \frac{2 \sqrt{2}}{f(r)} \left(L_2^\dagger Y^{\ell m}_{-2}, \mathcal{T}_{n \bar{m}}\right) (2-r J_+) + \right. \notag \\
& \left. + \left(Y^{\ell m}_{-2}, \mathcal{T}_{\bar{m} \bar{m}}\right) \left(r^2 J^2_+ - 2 r J_+\right)\right] \, \mathfrak{R}_{\text{in},\omega}(r) \, e^{i \omega t}
\end{align}
Due to the fact that $L_s^\dagger$ and $J_+$ act on eigenfunctions of $\partial_\phi$ and $\partial_t$, we can substitute them with
\be
L_s^\dagger \to \overline{L}^\dagger_s = \partial_\theta - \frac{m}{\sin \theta} + s \cot \theta \qquad , \qquad J_+ \to \bar{J}_- = \partial_r - \frac{i \omega}{f(r)}
\ee
In this way
\begin{align}
Z_{\ell m \omega}^\infty = 2 \pi G \int_{2M}^{\infty} \mathrm{d} r \int_{-\infty}^{\infty} \mathrm{d} t \, e^{i \omega t} & \left[\frac{2}{f(r)^2} \left(\overline{L}_1^\dagger \overline{L}_2^\dagger Y^{\ell m}_{-2}, \mathcal{T}_{nn}\right) - \frac{2 \sqrt{2}}{f(r)} \left(\overline{L}_2^\dagger Y^{\ell m}_{-2}, \mathcal{T}_{n \bar{m}}\right) (2-r \bar{J}_-) + \right. \notag \\
& \left. + \left(Y^{\ell m}_{-2}, \mathcal{T}_{\bar{m} \bar{m}}\right) \left(r^2 \bar{J}^2_- - 2 r \bar{J}_-\right)\right] \, \mathfrak{R}_{\text{in},\omega}(r)
\end{align}
Using the fact that $Y^{\ell m}_{-2}(\theta,\phi) = e^{i m \phi} S^{\ell m}_{-2} (\theta)$, we have
\begin{align}
Z_{\ell m \omega}^\infty = 2 \pi G \int_{2M}^{\infty} \mathrm{d} r \int_{-\infty}^{\infty} \mathrm{d} t \, e^{i \omega t- i m \phi} & \left[\frac{2}{f(r)^2} \left(\overline{L}_1^\dagger \overline{L}_2^\dagger S^{\ell m}_{-2}, \mathcal{T}_{nn}\right) - \frac{2 \sqrt{2}}{f(r)} \left(\overline{L}_2^\dagger S^{\ell m}_{-2}, \mathcal{T}_{n \bar{m}}\right) (2-r \bar{J}_-) + \right. \notag \\
& \left. + \left(S^{\ell m}_{-2}, \mathcal{T}_{\bar{m} \bar{m}}\right) \left(r^2 \bar{J}^2_- - 2 r \bar{J}_-\right)\right] \, \mathfrak{R}_{\text{in},\omega}(r)
\end{align}
In order to write this quantity more explicitly, we have to use the components of the stress-energy tensor. Since $\det g = -r^4 \sin^2 \theta$, we write, from \eqref{eq:stress_components_appr}
\be
T^{\alpha \beta} = \mu \int \mathrm{d} \tau \, \frac{\delta^{(4)} (x-x(\tau))}{r^2 \sin \theta} \, u^\alpha(\tau) \, u^\beta (\tau)
\ee
and using this expression inside $Z_{\ell m \omega}^\infty$, we keep the integral in $\mathrm{d} \tau$ in front all the other ones and we solve the integrals in $\mathrm{d} t,\mathrm{d} r,\mathrm{d} \theta,\mathrm{d} \phi$ through the 4-dimensional delta function, that localizes all the coordinates along the trajectory $t(\tau), r(\tau), \theta(\tau), \phi(\tau)$. By doing so we have to notice that
\be
\mathcal{T}_{nn} = \int \mathrm{d} \tau \frac{\delta^{(4)}(x-x(\tau))}{r^2 \sin \theta} \, n_\alpha u^\alpha(\tau) \, u^\beta(\tau) n_\beta = \int \mathrm{d} \tau \frac{\delta^{(4)}(x-x(\tau))}{r^2 \sin \theta} \, (u \cdot n)^2
\ee
and the same is valid for $\mathcal{T}_{n \bar{m}}, \mathcal{T}_{\bar{m} \bar{m}}$, hence we get
\begin{align} \label{eq:Z_almost_finished}
Z_{\ell m \omega}^\infty = 2 \pi G \mu \int \frac{\mathrm{d} \tau}{r(\tau)^2} e^{i \omega t(\tau)- i m \phi(\tau)} & \left[ \frac{2}{f(r(\tau))^2} \left(\overline{L}_1^\dagger \overline{L}_2^\dagger \, S^{\ell m}_{-2}(\theta(\tau)) \right)^* \, (u \cdot n)^2 - \frac{2 \sqrt{2}}{f(r(\tau))} \left(\overline{L}_2^\dagger S^{\ell m}_{-2}(\theta(\tau))\right)^* (n \cdot u) (\bar{m} \cdot u) (2 + \right. \notag \\
& \left. -r(\tau) \bar{J}_-) + \left(S^{\ell m}_{-2}(\theta(\tau))\right)^* \, (u \cdot \bar{m})^2 \left(r^2(\tau) \bar{J}^2_- - 2 r(\tau) \bar{J}_-\right)\right] \, \mathfrak{R}_{\text{in},\omega}(r(\tau))
\end{align}
Now we consider the motion of the particle along the equatorial plane $\theta(\tau) \equiv \pi/2$ and, from the Lagrangian
\be
\mathcal{L} = \frac{1}{2} \mu g_{\mu \nu} \dot{x}^\mu \dot{x}^\nu \qquad \text{with} \, \, \, \dot{x}^\mu = \frac{\mathrm{d} x^\mu}{\mathrm{d} \tau}
\ee
we define the momenta $p_\mu = \partial \mathcal{L}/\partial \dot{x}^\mu$ and so $p_t$ and $p_\phi$ are conserved
\begin{align}
& p_t \equiv -E = - f(r(\tau)) \mu \dot{t} (\tau) \Rightarrow \frac{\mathrm{d} t}{\mathrm{d} \tau} = \frac{E}{\mu \, f(r(\tau))} \notag \\
& p_\phi \equiv J = \mu r^2(\tau) \dot{\phi}(\tau) \Rightarrow \frac{\mathrm{d} \phi}{\mathrm{d} \tau} = \frac{J}{\mu \, r^2(\tau)}
\end{align}
where $E$ is the energy of the particle and $J$ its angular momentum. Using the fact that $p_\theta(\tau)=0$ and the mass shell condition
\be
\mathcal{H} = \frac{1}{2} g^{\mu \nu} p_\mu p_\nu = -\frac{\mu^2}{2}
\ee
it is possible to obtain $\dot{r}(\tau)$, that we do not report since we will not interested on it. Thanks to this, we can compute
\be
u \cdot n = -\frac{1}{2} \left(\frac{E}{\mu} + \dot{r}(\tau)\right) \qquad u \cdot \bar{m} = -\frac{i}{\sqrt{2} \, r(\tau)} \frac{J}{\mu}
\ee
At this point we can come back to \eqref{eq:Z_almost_finished}, we have to use the following identities (valid also for generic $\theta$)
\be
\overline{L}_1^\dagger \overline{L}_2^\dagger \, S^{\ell m}_{-2} \left(\frac{\pi}{2}\right) = \sqrt{(\ell^2-1) \ell (\ell+2)} S^{\ell m}_{0} \left(\frac{\pi}{2}\right) \qquad \overline{L}_2^\dagger \, S^{\ell m}_{-2} \left(\frac{\pi}{2}\right) = -\sqrt{(\ell-1) (\ell+2)} S^{\ell m}_{-1} \left(\frac{\pi}{2}\right)
\ee
and then we choose as integration variable the coordinate time $t$, so we have to consider the following expressions
\be
\frac{\mathrm{d} \phi}{\mathrm{d} t} \equiv \phi'(t) = \frac{J}{E} \frac{f(r(t))}{r(t)^2} \quad , \quad \mathrm{d} \tau = \frac{f(r(t)) \mu}{E}  \mathrm{d} t \quad , \quad  \dot{r}(\tau) = \frac{E}{f(r(t)) \, \mu} r'(t)
\ee
where with the symbol prime we refer to derivation with respect to $t$. By applying all these operations, we finally gain
\be
Z_{\ell m \omega}^\infty = G \int \frac{\mathrm{d} t}{r(t)^2} \, e^{i \omega t - i m \phi(t)} \sum_{j=0}^2 b^j_{\ell m} \mathcal{L}^j \left[\mathfrak{R}_{\text{in},\omega}(r)\right]
\ee
where we have defined
\be
b^0_{\ell m} = \pi \sqrt{(\ell^2-1) \ell (\ell+2)} \, S^{\ell m}_0 \left(\frac{\pi}{2}\right)\, \, , \quad b^1_{\ell m} = 2 \pi \sqrt{(\ell-1)(\ell+2)} \, S^{\ell m}_{-1} \left(\frac{\pi}{2}\right) \, \, , \qquad b^2_{\ell m} = 2 \pi \, S^{\ell m}_{-2} \left(\frac{\pi}{2}\right)
\ee
and
\begin{align} 
& \mathcal{L}^0 = \frac{E}{f} \left(1+\frac{r'(t)}{f}\right)^2 \, \, , \qquad \mathcal{L}^1 = \frac{i J}{r} \left(1+\frac{r'(t)}{f}\right) \left(2-r \, \partial_r + \frac{i \omega r}{f}\right) \notag \\
& \mathcal{L}^2 =  \frac{f \, J^2}{E \, r^2} \left[\frac{1}{f^2} \left(i M \omega + \frac{\omega^2 r^2}{2}- i \omega r\right) + \left(\frac{i \omega r}{f} + 1\right)r \, \partial_r - \frac{1}{2} r^2 \, \partial_r^2\right]
\end{align}
\chapter{Energy and angular momentum for a binary system with comparable masses and expression of $\rho_{22}^{\text{New}}$.} \label{app:energy_angularmom_4PN}
The 5PN and 4PN expressions in terms of $X = (M \Omega)^{\frac{2}{3}}$ of the energy $H_{\text{real}}$ and the angular momentum $J$ of a binary system with comparable masses $m_1$ and $m_2$, whose symmetric mass ratio is $\nu = m_1 m_2/(m_1+m_2)^2$, are respectively
\begin{align}
\frac{H_{\text{real}}}{M} & = 1-\frac{1}{2} \nu  X \left\{1 + \left(-\frac{3}{4}-\frac{\nu}{12}\right) X + \left(-\frac{27}{8} + \frac{19}{8} \nu - \frac{\nu^2}{24} \right) X^2 + \left[-\frac{675}{64} + \left(\frac{34445}{576} - \frac{205}{96}\pi^2\right) \nu - \frac{155}{96} \nu^2 + \right. \right. \notag \\
& \left. \left. - \frac{35}{5184} \nu^3 \right] X^3 + \left[-\frac{3969}{128} + \left(-\frac{123671}{5760} + \frac{9037}{1536} \pi^2 + \frac{896}{15} \gamma_E + \frac{448}{15} \log(16 X)\right)\nu + \left(-\frac{498449}{3456} + \frac{3157}{576} \pi^2\right) \nu^2 +\right. \right. \notag \\
& \left. \left. + \frac{301}{1728} \nu^3 + \frac{77}{31104} \nu^4\right] X^4\right\}
\end{align}
\begin{align}
J & = \frac{\nu}{\sqrt{x}} \left\{1+ \left(\frac{3}{2} + \frac{\nu}{6}\right) X + \left(\frac{27}{8} - \frac{19}{8} \nu + \frac{\nu^2}{24}\right) X^2 + \left[\frac{135}{16} + \left(-\frac{6889}{144} + \frac{41}{24}\pi^2\right) \nu + \frac{31}{24} \nu^2 + \right. \right. \notag \\
& \left. \left. + \frac{7}{1296} \nu^3 \right] X^3 + \left[\frac{2835}{128} + \left(\frac{98869}{5760} - \frac{128}{3} \gamma_E - \frac{6455}{1536} \pi^2 - \frac{64}{3} \log(16 X)\right)\nu + \left(\frac{356035}{3456} - \frac{2255}{576} \pi^2\right) \nu^2 + \right. \right. \notag \\
& \left. \left. - \frac{215}{1728} \nu^3 - \frac{55}{31104} \nu^4\right] X^4\right\}
\end{align}
As already said in the text, we also want to report the expression of the quantity $\rho_{22}^{\text{New}}$ at 10 PN but with $\nu$-dependent terms at 4PN
\begin{align}
\rho_{22}^{\text{New}} & = 1+x \left(-\frac{43}{42}+\frac{55 \nu }{84}\right)+x^2 \left(-\frac{20555}{10584}-\frac{33025 \nu }{21168}+\frac{19583 \nu^2}{42336}\right)+x^3 \left[-\frac{4296031}{4889808}+\left(-\frac{48993925}{9779616}+\frac{41 \pi^2}{192}\right) \nu + \right. \notag \\
& \left.-\frac{6292061 \nu^2}{3259872}+\frac{10620745 \nu^3}{39118464}\right]+x^4 \left[\frac{9228174993589}{800950550400}+\left(\frac{10815863492353}{640760440320}-\frac{3485 \pi ^2}{5376}\right) \nu^2-\frac{2088847783 \nu^3}{11650189824} + \right. \notag \\
& \left. +\frac{70134663541 \nu ^4}{512608352256}+\nu \left(-\frac{2487107795131}{145627372800}+\frac{464 \gamma }{35}-\frac{9953\pi ^2}{21504}+\frac{928 \log (2)}{35}+\frac{232 \log (x)}{35}\right)\right]+ \notag \\
& -\frac{8938613036677 x^5}{2116091577600} -\frac{1060700697798333909671 x^6}{24231643979185843200} +\frac{3567168919606240724303840051 x^7}{43991338062012939037440000}+ \notag \\
& +\frac{8339316227220569285816625738049 x^8}{279101750471556914871889920000} -\frac{522338057689474511990262498143822507399 x^9}{857097472947610731676894961786880000} + \notag \\
&+\frac{1523513000214555169284583871085138536795675131x^{10}}{1333729377653777059562416250036563968000000} + \mathcal{O}(x^{11})
\end{align}
\chapter{The Heun equation}\label{sec5}
The generalization of the results in the main text of Chapter 5 to case of the non-Confluent Heun Equation is straightforward.
The Heun differential equation is defined as
\begin{equation}
	\label{Heun}
	G_1''(z) + \left( {\beta_0\over z} +{\beta_1\over z-1}+{\beta_2\over z-\lambda} \right) G_1'(z) +{ z \alpha_0 \alpha_1 - \sigma \over z (z-1)(z-\lambda)}  G_1(z)=0
\end{equation}
with  $\beta_2=1+\alpha_0+\alpha_1-\beta_0-\beta_1$ \footnote{This condition guarantees that the characteristic exponents that come out from the indicial equation for the regular singular point at infinity are $\alpha_0$ and $\alpha_1$.}.  The solutions of the Heun equation can be written as linear combinations of
\beaq
{\rm Heun}_-(z) &=& {\rm Heun}\left[\lambda,\sigma ,\alpha_0 ,\alpha_1 ,\beta _0,\beta _1,z\right]\\
{\rm Heun}_+(z) &=&  z^{1{-}\beta _0} {\rm Heun}\left[\lambda,\sigma {+}\left(1{-}\beta_0 \right) \left(  \beta_2{+}\lambda \beta_1  \right),1{+}\alpha_0 {-}\beta _0,1{+}\alpha_1 {-}\beta _0,2{-}\beta_0,\beta _1,z\right]\nn
\eeaq
where ${\rm Heun}\left[\lambda,\sigma ,\alpha_0 ,\alpha_1 ,\beta _0,\beta _1,z\right]$ is the Heun function which is normalized to one at $z=0$, so that
 \be
 {\rm Heun}_-(z) \underset{z \to 0}{\simeq}1+\ldots \quad , \quad {\rm Heun}_+(z)\underset{z \to 0}{\simeq}z^{1-\beta_0} (1+\ldots )
 \ee
The dots stand for corrections in $z$. The expansion coefficients of ${\rm Heun}_\alpha(z)$ can be obtained recursively, solving the equation  order by order in $z$.
The Heun differential equation can be related to the quantum SW curve of a ${\cal N}=2$ supersymmetric gauge theory with four fundamental hypermultiplets. To this aim, we write the six Heun parameters in terms of 4 masses $m_i$, a gauge coupling $x$ and a Coulomb branch parameter $u$, defined via the identifications\footnote{
Equation (\ref{Heun}) can be alternatively written in the normal form (\ref{normal}) with
\be
G_1(z) =  (1-z)^{-\frac{\beta _1}{2}} z^{-\frac{\beta _0}{2}}  (z-\lambda )^{-\frac{\beta_2}{2}} \, \Psi (z)
\ee
and
\beaq
Q_{\rm gauge} &=& \frac{\ft{1}{4}-
   \left(\ft{m_3+m_4}{2}\right){}^2}{(z-x)^2}+\frac{\ft{1}{4}-
   \left( \ft{m_3-m_4}{2}\right){}^2}{z^2}+\frac{\ft{1}{4}-
   \left(\ft{m_1+m_2}{2}\right){}^2}{(z-1)^2}+\frac{m_3^2+m_4^2+2 m_1
   m_2-1}{2 (z-1) z} \nn\\
 &&+  \frac{(1-x) \left( u+\ft{1}{4} -\frac{m_3^2}{2}-\frac{m_4^2}{2}\right)
    -\ft{x}{2}\left(1{-}m_1{-}m_2\right)
   \left(1{-}m_3{-}m_4\right)  }{
   (z-1) z (z-x)}
\eeaq
}
\beaq
  \alpha _0&=& 1{-}m_1{-}m_3 \quad ,\quad \alpha _1= 1{-}m_2{-}m_3\quad , \quad    \lambda =x  \\
  \beta_0&=& 1{-}m_3{+}m_4\quad ,\quad \beta _1=1 {-}m_1{-}m_2 \quad , \quad    \beta_2 =1-m_3-m_4  \nn\\
  \sigma &=&u (x-1)+\left(\ft{1}{2}-m_3\right){}^2+x \left(\ft34{-}m_1{-} m_2{-} m_3{+}m_1 m_2{+}m_1 m_3{+}m_2m_3  \right) \nn\eeaq
The mathematical problem we are interested in is looking for a solution in an interval, let us say   $z\in (0,1)$, satisfying some specific boundary conditions.
The general solution
\be
G_1(z)=\sum_\alpha B_\alpha \, {\rm Heun}_\alpha(z)
\ee
behaves, near the boundaries, as
\beaq
G_1(z) &\underset{z\to 1}{\simeq}&  D_{-}(1+\ldots)  +D_{+} (1{-}z)^{m_1+m_2} (1+\ldots)
  \nn\\
 G_1(z) &\underset{z\to 0}{\simeq}&  B_{-} (1+\ldots) +B_{+} z^{m_3-m_4} (1+\ldots)   \label{gbcgen}
\eeaq
Specifying boundary conditions corresponds to fix two linear combinations made out of the four coefficients $B_{\pm}$,  $D_{\pm}$.

\subsection{The hypergeometric representation}

 As for the confluent case, we start from the PM/PN ans\"atze
\be
   G^0_\alpha\left(\frac{x}{z}\right) g_\alpha(x)  =   G^1_\alpha(z )+\ldots   \label{gp0}
\ee
with
\beaq
 G^0_\alpha (y)  &=&      P_{0}( y) \, H^0_\alpha (y) {+} \widehat{P}_{0}( y) \,  y H^{0}_\alpha{}'(y)  \nn\\
  G^1_\alpha (z) &=&      P_{1}(z) \, H^1_\alpha (z) {+} \widehat{P}_{1}(z) \,  z  H^{1}_\alpha{}'(z )    \label{g0g1h} \nn\\
  H^0_\alpha(y) &=& y^{\frac{1}{2}- \alpha a  - m_3} \, _2F_1\left( \frac{1}{2}- \alpha a -m_3, \frac{1}{2}- \alpha a -m_4; 1-2  \alpha a;y\right) \nn\\
  H^1_\alpha(z) &=& z^{-\frac{1}{2}+  \alpha a  + m_3} \, _2F_1\left( \frac{1}{2}+ \alpha a -m_1,\frac{1}{2}+ \alpha a-m_2;1+2   \alpha a;z\right) \label{ghnonc}
\eeaq
and
\beaq
P_{0 }(y) & = &1+\sum_{i=1}^k   \sum_{j=0}^{i-1}   c_{ij} \, x^i y^{j-i}  \quad ,\quad \widehat{P}_{0 }(y)   = \sum_{i=1}^k   \sum_{j=0}^{i}   \widehat{c}_{ij} \, x^i y^{j-i} \nn\\ \\
P_{1 }(z)  &   = & 1+  \sum_{j=1}^k \sum_{i=0}^{j-1}    d_{ij}   \, z^{i-j} x^j  \quad ,\quad \widehat{P}_{1 }(z) = \sum_{j=1}^k \sum_{i=0}^{j}    \widehat{d}_{ij}   \, z^{i-j} x^j  \nn \label{pp2}
\eeaq
   We notice that now only the non-confluent hypergeometric functions are involved. As before, plugging the ans\"atze (\ref{gp0})  into (\ref{Heun}) and setting to zero the coefficients of $H^p_\alpha$ ($p=0,1$) and its derivatives, one finds a linear system of algebraic equations for the coefficients $c_{ij}$, $\widehat{c}_{ij}$, $d_{ij}$, $\widehat{d}_{ij}$.
  For example, for $k=1$, writing $u=a^2+u_1 x+\ldots$, one finds
\beaq
 c_{10} &=&
\frac{1}{2}(1-m_1-m_2-m_3) -\frac{2 m_1 m_2 m_3}{4 a^2-1}   \quad, \quad   \widehat{c}_{11} =  -\widehat{c}_{10}=\frac{1}{2}+ \frac{2 m_1 m_2}{4a^2-1}  \nn\\
 d_{01} &= &  \frac{2 \left(m_3-1\right) m_3 m_4}{4 a^2-1}-\frac{m_4}{2} \quad ,  \quad
 \widehat{d}_{11} = - \widehat{d}_{01}= \frac{2 m_3 m_4}{4 a^2-1}+\frac{1}{2} \nn\\
u_1 &=&  \frac{2 m_1 m_2 m_3 m_4}{4 a^2-1}+\frac{1}{8} \left(4
  a^2-1\right)+\frac{1}{2} \left( 1 - \sum_{i=1}^4 m_i+ \sum_{i<j}^4 m_i m_j   \right) \label{pp01s}
\eeaq
   As in the confluent case, we denote by
   \be
   g_\alpha(x) =\frac{G^1_\alpha(z)}{G^0_\alpha \left(\frac{x}{z}\right)}
   \ee
    the ratio of the two solutions.  Explicit computations show that, as in the confluent case, the ratio of the two components
   can be written as
  \be
\gamma(x) ={g_+(x)\over g_-(x) }=   x^{2a} e^{-\partial_a {\cal F}_{\rm inst}(a,x) } =x^{2 a} \left(1+ x \left(a -\frac{16 a m_1 m_2 m_3 m_4 }{\left(1-4 a^2 \right)^2} \right) + \ldots \right)  \label{gammaHeun}
\ee
with  ${\cal F}_{\rm inst} $ still is given by (\ref{finst}).
  The function  ${\cal F}_{\rm inst} $ corresponds now to the prepotential of a ${\cal N}=2$ supersymmetric SU(2) gauge theory with four flavours transforming in the fundamental representation of the previous gauge group.

\subsection{Connection formulae}

 As in the confluent case, connection formulae for the Heun function can be easily derived from the standard hypergeometric transformation rules of $H^p_\alpha$.
 We introduce the two extra sets of hypergeometric functions
{
	\beaq
	\widetilde{H}^{0}_{\alpha}(y) & = &  ( e^{ {-} i \pi }
   y)^{{-}\frac{1{+}\alpha}{2} m_{34} } \,
   _2F_1\left( \frac{1}{2}{-}a{-}m_3{+}\frac{1{+}\alpha}{2}
  m_{34}, \frac{1}{2}{+}a{-}m_3{+}\frac{1{+}\alpha}{2} m_{34}; 1{+}\alpha
  m_{34};\frac{1}{y}\right)   \label{hyp2}  \\
	\widetilde{H}^{1}_{\alpha}(z) & = & z^{a{+}m_3{-}\frac{1}{2}} (1{-}z)^{\frac{1+\alpha}{2} \left(m_1{+}m_2\right)} \, _2F_1\left( \frac{1}{2}{+}a{+}\alpha m_1,\frac{1}{2}{+}a{+}\alpha  m_2;1{+}\alpha \left(m_1+m_2\right);1{-}z\right)  \nn
	\eeaq
}
with $m_{34}=m_3-m_4$.
We notice that both $\widetilde{H}^0_{\alpha}(y)$ and $\widetilde{H}^1_{\alpha}(z)$ are invariant under the sign flipping $a \to -a$, due to hypergeometric identities.
The functions (\ref{hyp2}) are related to those introduced in (\ref{ghnonc}) by the standard hypergeometric relations
\be
H^0_\alpha (y) = \sum_{\alpha'=\pm}  B_{\alpha \alpha'}  \widetilde{H}^{0}_{\alpha'}(y) \quad , \quad
H^1_\alpha(z) = \sum_{\alpha'=\pm}  F_{\alpha \alpha' }  \widetilde{H}^{1}_{\alpha'}(z)
\label{g0con00}
\ee
with
\beaq
B_{\alpha \alpha'} &=&e^{-i \pi  \left( \alpha a
   +m_3-\frac{1}{2}\right)} \frac{\Gamma (1-2 \alpha a )  \Gamma \left( \alpha'\left(m_4-m_3\right) \right)}{\Gamma \left( \ft12 -\alpha a  - \alpha' m_3 \right)
    \Gamma \left( \ft12 -\alpha a + \alpha' m_4 \right)} \nn\\
F_{\alpha\alpha'}  &=&\frac{\Gamma (1+2 \alpha a ) \Gamma
   \left(-\alpha' \left(m_1+m_2\right) \right)}{\Gamma \left( \frac{1}{2} +\alpha a  - \alpha' m_1  \right) \Gamma \left( \frac{1}{2} +\alpha a  - \alpha' m_2 \right)}
\eeaq
Plugging \eqref{g0con00} into \eqref{g0g1h} one finds the connection formulae
\beaq
   G^0_\alpha(y) &=&   \sum_{\alpha'=\pm}  B_{\alpha \alpha'}  \widetilde{G}^0_{\alpha'} (y) ,\nn\\
   G^1_\alpha (z) &=& \sum_{\alpha'=\pm}  F_{\alpha \alpha'} \widetilde{G}^1_{\alpha'}(z)    \label{gp21}
\eeaq
with
\beaq
  \widetilde{G}^0_{\alpha}(y) &=&  P_{0}(y)   \widetilde{H}^{0}_{\alpha}(y)   {+} \widehat{P}_{0}(y)   y \widetilde{H}^{0}_{\alpha}{}'(y)   \nn\\
  \widetilde{G}^1_{\alpha}(z) &=&  P_{1}(z) \widetilde{H}^{ 1}_{\alpha}(z)   {+} \widehat{P}_{1}(z)   z \widetilde{H}^{ 1 }_{\alpha}{}'(z)  \label{gtildes}
\eeaq
describing the solution in the \textit{far region} and \textit{near horizon zone}.
The pairs $ G^0_\alpha(x/z)$, $ G^1_\alpha(z)$, $\widetilde{G}^0_{\alpha}(x/z)$, $ \widetilde{G}^0_{\alpha}(z) $ provide four different bases of solutions of the Heun equation.
Each basis covers a different patch of the interval $z\in (0,1)$. They are related to each other by the connection formulae (\ref{gp21}). Remarkably the Heun connection formulae (\ref{gp21}) take the familiar hypergeometric form, with the important difference that the argument of the gamma functions depends on the non-trivial function $a(u)$.
Finally, we notice that the Heun function corresponds to the minus component of the far region basis
\be
   {\rm Heun}_- (z) = \widetilde{G}^0_-(z)  \underset{z\to 0}{\simeq} 1+ \ldots \label{heunm}
\ee
Indeed in this limit $P_0\to 1$, $\widetilde{H}^0_- \to 1$ and (\ref{heunm}) follows from (\ref{gtildes}).
\backmatter
\cleardoublepage

\bibliographystyle{unsrt}      
\bibliography{bibALL} 

@article{Paul:2023rka,
    author = "Paul, Hynek and Perlmutter, Eric and Raj, Himanshu",
    title = "{Exact large charge in $ \mathcal{N} $ = 4 SYM and semiclassical string theory}",
    eprint = "2303.13207",
    archivePrefix = "arXiv",
    primaryClass = "hep-th",
    doi = "10.1007/JHEP08(2023)078",
    journal = "JHEP",
    volume = "08",
    pages = "078",
    year = "2023"
}

@article{Brown:2023why,
    author = "Brown, Augustus and Wen, Congkao and Xie, Haitian",
    title = "{Generating functions and large-charge expansion of integrated correlators in {\ensuremath{{N}}} = 4 supersymmetric Yang-Mills theory}",
    eprint = "2303.17570",
    archivePrefix = "arXiv",
    primaryClass = "hep-th",
    reportNumber = "QMUL-PH-23-03",
    doi = "10.1007/JHEP07(2023)129",
    journal = "JHEP",
    volume = "07",
    pages = "129",
    year = "2023"
}

@article{Minahan:1996cj,
    author = "Minahan, Joseph A. and Nemeschansky, Dennis",
    title = "{Superconformal fixed points with E(n) global symmetry}",
    eprint = "hep-th/9610076",
    archivePrefix = "arXiv",
    reportNumber = "USC-96-23",
    doi = "10.1016/S0550-3213(97)00039-4",
    journal = "Nucl. Phys. B",
    volume = "489",
    pages = "24--46",
    year = "1997"
}

@article{Livan:2011uvs,
    author = "Livan, Giacomo and Vivo, Pierpaolo",
    title = "{Moments of Wishart-Laguerre and Jacobi ensembles of random matrices: application to the quantum transport problem in chaotic cavities}",
    eprint = "1103.2638",
    archivePrefix = "arXiv",
    primaryClass = "cond-mat.stat-mech",
    doi = "10.5506/APhysPolB.42.1081",
    month = "3",
    year = "2011"
}

@article{Eynard:2015aea,
    author = "Eynard, Bertrand and Kimura, Taro and Ribault, Sylvain",
    title = "{Random matrices}",
    eprint = "1510.04430",
    archivePrefix = "arXiv",
    primaryClass = "math-ph",
    month = "10",
    year = "2015"
}

@article{Cipriani:2026myb,
    author = "Cipriani, Andrea and Fucito, Francesco and Heissenberg, Carlo and Morales, Jose Francisco and Russo, Rodolfo",
    title = "{''Waveforms'' at the Horizon}",
    eprint = "2602.05766",
    archivePrefix = "arXiv",
    primaryClass = "gr-qc",
    doi = "10.21468/SciPostPhys.21.1.003",
    journal = "SciPost Phys.",
    volume = "21",
    pages = "003",
    year = "2026"
}

@article{PhysRevLett.38.1440,
  title = {$\mathrm{CP}$ Conservation in the Presence of Pseudoparticles},
  author = {Peccei, R. D. and Quinn, Helen R.},
  journal = {Phys. Rev. Lett.},
  volume = {38},
  issue = {25},
  pages = {1440--1443},
  numpages = {0},
  year = {1977},
  month = {Jun},
  publisher = {American Physical Society},
  doi = {10.1103/PhysRevLett.38.1440},
  url = {https://link.aps.org/doi/10.1103/PhysRevLett.38.1440}
}

@article{PhysRevD.16.1791,
  title = {Constraints imposed by $\mathrm{CP}$ conservation in the presence of pseudoparticles},
  author = {Peccei, R. D. and Quinn, Helen R.},
  journal = {Phys. Rev. D},
  volume = {16},
  issue = {6},
  pages = {1791--1797},
  numpages = {0},
  year = {1977},
  month = {Sep},
  publisher = {American Physical Society},
  doi = {10.1103/PhysRevD.16.1791},
  url = {https://link.aps.org/doi/10.1103/PhysRevD.16.1791}
}

@article{Seiberg:1994rs,
    author = "Seiberg, N. and Witten, Edward",
    title = "{Electric - magnetic duality, monopole condensation, and confinement in N=2 supersymmetric Yang-Mills theory}",
    eprint = "hep-th/9407087",
    archivePrefix = "arXiv",
    reportNumber = "RU-94-52, IASSNS-HEP-94-43",
    doi = "10.1016/0550-3213(94)90124-4",
    journal = "Nucl. Phys. B",
    volume = "426",
    pages = "19--52",
    year = "1994",
    note = "[Erratum: Nucl.Phys.B 430, 485--486 (1994)]"
}

@article{Seiberg:1994aj,
    author = "Seiberg, N. and Witten, Edward",
    title = "{Monopoles, duality and chiral symmetry breaking in N=2 supersymmetric QCD}",
    eprint = "hep-th/9408099",
    archivePrefix = "arXiv",
    reportNumber = "RU-94-60, IASSNS-HEP-94-55",
    doi = "10.1016/0550-3213(94)90214-3",
    journal = "Nucl. Phys. B",
    volume = "431",
    pages = "484--550",
    year = "1994"
}

@article{PhysRev.159.1251,
  title = {All Possible Symmetries of the $S$ Matrix},
  author = {Coleman, Sidney and Mandula, Jeffrey},
  journal = {Phys. Rev.},
  volume = {159},
  issue = {5},
  pages = {1251--1256},
  numpages = {0},
  year = {1967},
  month = {Jul},
  publisher = {American Physical Society},
  doi = {10.1103/PhysRev.159.1251},
  url = {https://link.aps.org/doi/10.1103/PhysRev.159.1251}
}

@article{Haag:1974qh,
    author = "Haag, Rudolf and Lopuszanski, Jan T. and Sohnius, Martin",
    title = "{All Possible Generators of Supersymmetries of the s Matrix}",
    reportNumber = "Print-74-1630 (KARLSRUHE)",
    doi = "10.1016/0550-3213(75)90279-5",
    journal = "Nucl. Phys. B",
    volume = "88",
    pages = "257",
    year = "1975"
}

@book{Wess:1992cp,
    author = "Wess, J. and Bagger, J.",
    title = "{Supersymmetry and supergravity}",
    isbn = "978-0-691-02530-8",
    publisher = "Princeton University Press",
    address = "Princeton, NJ, USA",
    year = "1992"
}

@book{Bertolini:2025jhz,
    author = "Bertolini, Matteo",
    title = "{Supersymmetry - From the basics to exact results in gauge theories}",
    doi = "10.1142/14026",
    isbn = "978-981--980024-7, 978-981--980025-4, 978-981--980026-1",
    publisher = "World Scientific",
    month = "1",
    year = "2025"
}

@article{Quevedo:2010ui,
    author = "Quevedo, Fernando and Krippendorf, Sven and Schlotterer, Oliver",
    title = "{Cambridge Lectures on Supersymmetry and Extra Dimensions}",
    eprint = "1011.1491",
    archivePrefix = "arXiv",
    primaryClass = "hep-th",
    reportNumber = "DAMTP-2010-90, MPP-2010-143",
    month = "11",
    year = "2010"
}

@book{Terning:2006bq,
    author = "Terning, J.",
    title = "{Modern supersymmetry: Dynamics and duality}",
    doi = "10.1093/acprof:oso/9780198567639.001.0001",
    year = "2006"
}

@book{Cabibbo:Gauge,
    author = "Cabibbo, N. and Maiani, L. and Benhar, O.",
    title = "{An Introduction to Gauge Theories}",
    doi = "10.1201/9781315369723",
    year = "2017",
    publisher = "CRC Press"
}

@article{Alvarez-Gaume:1996ohl,
    author = "Alvarez-Gaume, Luis and Hassan, S. F.",
    title = "{Introduction to S duality in N=2 supersymmetric gauge theories: A Pedagogical review of the work of Seiberg and Witten}",
    eprint = "hep-th/9701069",
    archivePrefix = "arXiv",
    reportNumber = "CERN-TH-96-371",
    doi = "10.1002/prop.2190450302",
    journal = "Fortsch. Phys.",
    volume = "45",
    pages = "159--236",
    year = "1997"
}

@inproceedings{Bilal:1995hc,
    author = "Bilal, A.",
    title = "{Duality in N=2 SUSY SU(2) Yang-Mills theory: A Pedagogical introduction to the work of Seiberg and Witten}",
    booktitle = "{NATO Advanced Study Institute on Quantum Fields and Quantum Space Time}",
    eprint = "hep-th/9601007",
    archivePrefix = "arXiv",
    reportNumber = "LPTENS-95-53",
    pages = "21--43",
    year = "1997"
}

@book{Peskin:QFT,
    author = "Peskin, M. E.",
    title = "{An Introduction To Quantum Field Theory}",
    doi = "10.1201/9780429503559",
    year = "2018",
    publisher = "CRC Press"
}

@article{Fujikawa:1979ay,
    author = "Fujikawa, Kazuo",
    title = "{Path Integral Measure for Gauge Invariant Fermion Theories}",
    reportNumber = "INS-328",
    doi = "10.1103/PhysRevLett.42.1195",
    journal = "Phys. Rev. Lett.",
    volume = "42",
    pages = "1195--1198",
    year = "1979"
}

@article{Kapustin:2014lwa,
    author = "Kapustin, Anton and Thorngren, Ryan",
    title = "{Anomalies of discrete symmetries in three dimensions and group cohomology}",
    eprint = "1403.0617",
    archivePrefix = "arXiv",
    primaryClass = "hep-th",
    doi = "10.1103/PhysRevLett.112.231602",
    journal = "Phys. Rev. Lett.",
    volume = "112",
    number = "23",
    pages = "231602",
    year = "2014"
}

@article{Yamaguchi:2018xse,
    author = "Yamaguchi, Satoshi",
    title = "{{\textquoteright}t Hooft anomaly matching condition and chiral symmetry breaking without bilinear condensate}",
    eprint = "1811.09390",
    archivePrefix = "arXiv",
    primaryClass = "hep-th",
    reportNumber = "OU-HET-988",
    doi = "10.1007/JHEP01(2019)014",
    journal = "JHEP",
    volume = "01",
    pages = "014",
    year = "2019"
}

@article{Seiberg:1988ur,
    author = "Seiberg, Nathan",
    title = "{Supersymmetry and Nonperturbative beta Functions}",
    reportNumber = "IASSNS-HEP-88-5",
    doi = "10.1016/0370-2693(88)91265-8",
    journal = "Phys. Lett. B",
    volume = "206",
    pages = "75--80",
    year = "1988"
}

@book{Nakahara:2003nw,
    author = "Nakahara, M.",
    title = "{Geometry, topology and physics}",
    publisher = "Graduate Student Series in Physics",
    year = "2003"
}

@book{Griffiths:Algebraic_Geometry,
    author = "Griffiths, P. and Harris, J.",
    title = "{Principles of Algebraic Geometry}",
    volume = "19",
    year = "1978",
    publisher = "New York : Wiley"
}

@article{Hanany:1996ie,
    author = "Hanany, Amihay and Witten, Edward",
    title = "{Type IIB superstrings, BPS monopoles, and three-dimensional gauge dynamics}",
    eprint = "hep-th/9611230",
    archivePrefix = "arXiv",
    reportNumber = "IASSNS-HEP-96-121",
    doi = "10.1016/S0550-3213(97)00157-0",
    journal = "Nucl. Phys. B",
    volume = "492",
    pages = "152--190",
    year = "1997"
}

@article{Bianchi:2021mft,
    author = "Bianchi, Massimo and Consoli, Dario and Grillo, Alfredo and Morales, Jose Francisco",
    title = "{More on the SW-QNM correspondence}",
    eprint = "2109.09804",
    archivePrefix = "arXiv",
    primaryClass = "hep-th",
    doi = "10.1007/JHEP01(2022)024",
    journal = "JHEP",
    volume = "01",
    pages = "024",
    year = "2022"
}

@article{Bianchi:2024mlq,
    author = "Bianchi, Massimo and Dibitetto, Giuseppe and Morales, Jose Francisco",
    title = "{Gauge theory meets cosmology}",
    eprint = "2408.03243",
    archivePrefix = "arXiv",
    primaryClass = "hep-th",
    doi = "10.1088/1475-7516/2024/12/040",
    journal = "JCAP",
    volume = "12",
    pages = "040",
    year = "2024"
}

@article{Witten:1997sc,
    author = "Witten, Edward",
    title = "{Solutions of four-dimensional field theories via M-theory}",
    eprint = "hep-th/9703166",
    archivePrefix = "arXiv",
    reportNumber = "IASSNS-HEP-97-19",
    doi = "10.1201/9781482268737-38",
    journal = "Nucl. Phys. B",
    volume = "500",
    pages = "3--42",
    year = "1997"
}

@article{Nekrasov:2002qd,
    author = "Nekrasov, Nikita A.",
    title = "{Seiberg-Witten prepotential from instanton counting}",
    eprint = "hep-th/0206161",
    archivePrefix = "arXiv",
    reportNumber = "ITEP-TH-22-02, IHES-P-04-22",
    doi = "10.4310/ATMP.2003.v7.n5.a4",
    journal = "Adv. Theor. Math. Phys.",
    volume = "7",
    number = "5",
    pages = "831--864",
    year = "2003"
}

@article{Flume:2002az,
    author = "Flume, R. and Poghossian, R.",
    title = "{An Algorithm for the microscopic evaluation of the coefficients of the Seiberg-Witten prepotential}",
    eprint = "hep-th/0208176",
    archivePrefix = "arXiv",
    doi = "10.1142/S0217751X03013685",
    journal = "Int. J. Mod. Phys. A",
    volume = "18",
    pages = "2541",
    year = "2003"
}

@article{Bruzzo:2002xf,
    author = "Bruzzo, Ugo and Fucito, Francesco and Morales, Jose F. and Tanzini, Alessandro",
    title = "{Multiinstanton calculus and equivariant cohomology}",
    eprint = "hep-th/0211108",
    archivePrefix = "arXiv",
    reportNumber = "ROM2F-2002-27, SISSA-75-2002-FM, PAR-LPTHE-02-54, STR-02-042",
    doi = "10.1088/1126-6708/2003/05/054",
    journal = "JHEP",
    volume = "05",
    pages = "054",
    year = "2003"
}

@inproceedings{Nekrasov:2009rc,
    author = "Nekrasov, Nikita A. and Shatashvili, Samson L.",
    title = "{Quantization of Integrable Systems and Four Dimensional Gauge Theories}",
    booktitle = "{16th International Congress on Mathematical Physics}",
    eprint = "0908.4052",
    archivePrefix = "arXiv",
    primaryClass = "hep-th",
    reportNumber = "TCD-MATH-09-19, HMI-09-09, IHES-P-09-38",
    doi = "10.1142/9789814304634_0015",
    pages = "265--289",
    year = "2010"
}

@article{Poghossian:2010pn,
    author = "Poghossian, Rubik",
    title = "{Deforming SW curve}",
    eprint = "1006.4822",
    archivePrefix = "arXiv",
    primaryClass = "hep-th",
    reportNumber = "ROM2F-2010-10",
    doi = "10.1007/JHEP04(2011)033",
    journal = "JHEP",
    volume = "04",
    pages = "033",
    year = "2011"
}

@article{Fucito:2011pn,
    author = "Fucito, F. and Morales, J. F. and Pacifici, D. Ricci and Poghossian, R.",
    title = "{Gauge theories on $\Omega$-backgrounds from non commutative Seiberg-Witten curves}",
    eprint = "1103.4495",
    archivePrefix = "arXiv",
    primaryClass = "hep-th",
    reportNumber = "ROM2F-2011-03",
    doi = "10.1007/JHEP05(2011)098",
    journal = "JHEP",
    volume = "05",
    pages = "098",
    year = "2011"
}

@article{Zenkevich:2011zx,
    author = "Zenkevich, Yegor",
    title = "{Nekrasov prepotential with fundamental matter from the quantum spin chain}",
    eprint = "1103.4843",
    archivePrefix = "arXiv",
    primaryClass = "math-ph",
    reportNumber = "ITEP-TH-2-11",
    doi = "10.1016/j.physletb.2011.06.030",
    journal = "Phys. Lett. B",
    volume = "701",
    pages = "630--639",
    year = "2011"
}

@article{Poghosyan:2020zzg,
    author = "Poghosyan, Hasmik",
    title = "{Recursion relation for instanton counting for SU(2) $ \mathcal{N} $ = 2 SYM in NS limit of $\Omega$ background}",
    eprint = "2010.08498",
    archivePrefix = "arXiv",
    primaryClass = "hep-th",
    doi = "10.1007/JHEP05(2021)088",
    journal = "JHEP",
    volume = "05",
    pages = "088",
    year = "2021"
}

@article{Matone:1995rx,
    author = "Matone, Marco",
    title = "{Instantons and recursion relations in N=2 SUSY gauge theory}",
    eprint = "hep-th/9506102",
    archivePrefix = "arXiv",
    reportNumber = "DFPD-95-TH-25A",
    doi = "10.1016/0370-2693(95)00920-G",
    journal = "Phys. Lett. B",
    volume = "357",
    pages = "342--348",
    year = "1995"
}

@article{Flume:2004rp,
    author = "Flume, Rainald and Fucito, Francesco and Morales, Jose F. and Poghossian, Rubik",
    title = "{Matone's relation in the presence of gravitational couplings}",
    eprint = "hep-th/0403057",
    archivePrefix = "arXiv",
    reportNumber = "ROM2F-2004-05",
    doi = "10.1088/1126-6708/2004/04/008",
    journal = "JHEP",
    volume = "04",
    pages = "008",
    year = "2004"
}

@article{Bianchi:2022wku,
    author = "Bianchi, Massimo and Di Russo, Giorgio",
    title = "{Turning rotating D-branes and black holes inside out their photon-halo}",
    eprint = "2203.14900",
    archivePrefix = "arXiv",
    primaryClass = "hep-th",
    doi = "10.1103/PhysRevD.106.086009",
    journal = "Phys. Rev. D",
    volume = "106",
    number = "8",
    pages = "086009",
    year = "2022"
}

@article{Alday:2009aq,
    author = "Alday, Luis F. and Gaiotto, Davide and Tachikawa, Yuji",
    title = "{Liouville Correlation Functions from Four-dimensional Gauge Theories}",
    eprint = "0906.3219",
    archivePrefix = "arXiv",
    primaryClass = "hep-th",
    doi = "10.1007/s11005-010-0369-5",
    journal = "Lett. Math. Phys.",
    volume = "91",
    pages = "167--197",
    year = "2010"
}

@book{Weinberg:1995mt,
    author = "Weinberg, Steven",
    title = "{The Quantum theory of fields. Vol. 1: Foundations}",
    doi = "10.1017/CBO9781139644167",
    isbn = "978-0-521-67053-1, 978-0-511-25204-4",
    publisher = "Cambridge University Press",
    month = "6",
    year = "2005"
}

@book{Muller-Kirsten:2012wla,
    author = {M{\"u}ller-Kirsten, Harald J. W.},
    title = {{Introduction to Quantum Mechanics}: {Schr{\"o}dinger Equation and Path Integral}},
    doi = "10.1142/8428",
    isbn = "978-981-4397-73-5, 978-981-4397-74-2",
    publisher = "World Scientific",
    year = "2012"
}

@book{Shifman:1994ee,
    editor = "Shifman, Mikhail A.",
    title = "{Instantons in gauge theories}",
    publisher = "World Scientific",
    year = "1994"
}

@article{Vandoren:2008xg,
    author = "Vandoren, Stefan and van Nieuwenhuizen, Peter",
    title = "{Lectures on instantons}",
    eprint = "0802.1862",
    archivePrefix = "arXiv",
    primaryClass = "hep-th",
    reportNumber = "ITP-UU-08-05, SPIN-08-05",
    month = "2",
    year = "2008"
}

@book{Jackson:1998nia,
    author = "Jackson, John David",
    title = "{Classical Electrodynamics}",
    isbn = "978-0-471-30932-1",
    publisher = "Wiley",
    year = "1998"
}

@article{Heras:2018uub,
    author = "Heras, Ricardo",
    title = "{Dirac quantisation condition: a comprehensive review}",
    eprint = "1810.13403",
    archivePrefix = "arXiv",
    primaryClass = "physics.hist-ph",
    doi = "10.1080/00107514.2018.1527974",
    journal = "Contemp. Phys.",
    volume = "59",
    number = "4",
    pages = "331--355",
    year = "2018"
}

@inproceedings{Harvey:1996ur,
    author = "Harvey, Jeffrey A.",
    title = "{Magnetic monopoles, duality and supersymmetry}",
    booktitle = "{ICTP Summer School in High-energy Physics and Cosmology}",
    eprint = "hep-th/9603086",
    archivePrefix = "arXiv",
    reportNumber = "EFI-96-06",
    month = "3",
    year = "1996"
}

@article{Witten:1979ey,
    author = "Witten, Edward",
    title = "{Dyons of Charge e theta/2 pi}",
    reportNumber = "CERN-TH-2724",
    doi = "10.1016/0370-2693(79)90838-4",
    journal = "Phys. Lett. B",
    volume = "86",
    pages = "283--287",
    year = "1979"
}

@book{Rajaraman:1982is,
    author = "Rajaraman, R.",
    title = "{Solitons and Instantons: an Introduction to solitons and instantons in Quantum Field Theory}",
    year = "1982"
}

@article{Hellerman:2018xpi,
    author = "Hellerman, Simeon and Maeda, Shunsuke and Orlando, Domenico and Reffert, Susanne and Watanabe, Masataka",
    title = "{Universal correlation functions in rank 1 SCFTs}",
    eprint = "1804.01535",
    archivePrefix = "arXiv",
    primaryClass = "hep-th",
    reportNumber = "CALT-TH-2018-014, IPMU18-0059",
    doi = "10.1007/JHEP12(2019)047",
    journal = "JHEP",
    volume = "12",
    pages = "047",
    year = "2019"
}

@article{Hellerman:2020sqj,
    author = "Hellerman, Simeon and Maeda, Shunsuke and Orlando, Domenico and Reffert, Susanne and Watanabe, Masataka",
    title = "{S-duality and correlation functions at large R-charge}",
    eprint = "2005.03021",
    archivePrefix = "arXiv",
    primaryClass = "hep-th",
    doi = "10.1007/JHEP04(2021)287",
    journal = "JHEP",
    volume = "04",
    pages = "287",
    year = "2021"
}

@article{Hellerman:2021yqz,
    author = "Hellerman, Simeon and Orlando, Domenico",
    title = "{Large R-charge EFT correlators in N=2 SQCD}",
    eprint = "2103.05642",
    archivePrefix = "arXiv",
    primaryClass = "hep-th",
    month = "3",
    year = "2021"
}

@article{Grassi:2019txd,
    author = "Grassi, Alba and Komargodski, Zohar and Tizzano, Luigi",
    title = "{Extremal correlators and random matrix theory}",
    eprint = "1908.10306",
    archivePrefix = "arXiv",
    primaryClass = "hep-th",
    doi = "10.1007/JHEP04(2021)214",
    journal = "JHEP",
    volume = "04",
    pages = "214",
    year = "2021"
}

@article{Cipriani:2026iws,
    author = "Cipriani, Andrea and Savelli, Raffaele",
    title = "{On instantons at large charge}",
    eprint = "2602.19677",
    archivePrefix = "arXiv",
    primaryClass = "hep-th",
    doi = "10.1007/JHEP09(2026)195",
    journal = "JHEP",
    volume = "09",
    pages = "195",
    year = "2026"
}

@article{Argyres:1995jj,
    author = "Argyres, Philip C. and Douglas, Michael R.",
    title = "{New phenomena in SU(3) supersymmetric gauge theory}",
    eprint = "hep-th/9505062",
    archivePrefix = "arXiv",
    reportNumber = "IASSNS-HEP-95-31, RU-95-28",
    doi = "10.1016/0550-3213(95)00281-V",
    journal = "Nucl. Phys. B",
    volume = "448",
    pages = "93--126",
    year = "1995"
}

@article{Argyres:1995xn,
    author = "Argyres, Philip C. and Plesser, M. Ronen and Seiberg, Nathan and Witten, Edward",
    title = "{New N=2 superconformal field theories in four-dimensions}",
    eprint = "hep-th/9511154",
    archivePrefix = "arXiv",
    reportNumber = "RU-95-81, WIS-95-59-PH, IASSNS-HEP-95-95",
    doi = "10.1016/0550-3213(95)00671-0",
    journal = "Nucl. Phys. B",
    volume = "461",
    pages = "71--84",
    year = "1996"
}

@book{Birrell:1982ix,
    author = "Birrell, N. D. and Davies, P. C. W.",
    title = "{Quantum Fields in Curved Space}",
    doi = "10.1017/CBO9780511622632",
    isbn = "978-0-511-62263-2, 978-0-521-27858-4",
    publisher = "Cambridge University Press",
    address = "Cambridge, UK",
    series = "Cambridge Monographs on Mathematical Physics",
    year = "1982"
}

@book{DiFrancesco:1997nk,
    author = "Di Francesco, P. and Mathieu, P. and Senechal, D.",
    title = "{Conformal Field Theory}",
    doi = "10.1007/978-1-4612-2256-9",
    isbn = "978-0-387-94785-3, 978-1-4612-7475-9",
    publisher = "Springer-Verlag",
    address = "New York",
    series = "Graduate Texts in Contemporary Physics",
    year = "1997"
}

@book{Ammon:2015wua,
    author = "Ammon, Martin and Erdmenger, Johanna",
    title = "{Gauge/gravity duality}: {Foundations and applications}",
    doi = "10.1017/CBO9780511846373",
    isbn = "978-1-107-01034-5, 978-1-316-23594-2",
    publisher = "Cambridge University Press",
    address = "Cambridge",
    month = "4",
    year = "2015"
}

@article{Coleman:1973sx,
    author = "Coleman, Sidney R. and Gross, David J.",
    title = "{Price of asymptotic freedom}",
    doi = "10.1103/PhysRevLett.31.851",
    journal = "Phys. Rev. Lett.",
    volume = "31",
    pages = "851--854",
    year = "1973"
}

@article{Fucito:2023txg,
    author = "Fucito, Francesco and Grassi, Alba and Morales, Jose Francisco and Savelli, Raffaele",
    title = "{Partition functions of non-Lagrangian theories from the holomorphic anomaly}",
    eprint = "2306.05141",
    archivePrefix = "arXiv",
    primaryClass = "hep-th",
    reportNumber = "CERN-TH-2023-095",
    doi = "10.1007/JHEP07(2023)195",
    journal = "JHEP",
    volume = "07",
    pages = "195",
    year = "2023"
}

@article{Bissi:2021rei,
    author = "Bissi, Agnese and Fucito, Francesco and Manenti, Andrea and Morales, Jos{\'e} Francisco and Savelli, Raffaele",
    title = "{OPE coefficients in Argyres-Douglas theories}",
    eprint = "2112.11899",
    archivePrefix = "arXiv",
    primaryClass = "hep-th",
    reportNumber = "UUITP-65/21",
    doi = "10.1007/JHEP06(2022)085",
    journal = "JHEP",
    volume = "06",
    pages = "085",
    year = "2022"
}

@article{Papadodimas:2009eu,
    author = "Papadodimas, Kyriakos",
    title = "{Topological Anti-Topological Fusion in Four-Dimensional Superconformal Field Theories}",
    eprint = "0910.4963",
    archivePrefix = "arXiv",
    primaryClass = "hep-th",
    doi = "10.1007/JHEP08(2010)118",
    journal = "JHEP",
    volume = "08",
    pages = "118",
    year = "2010"
}

@article{Pestun:2007rz,
    author = "Pestun, Vasily",
    title = "{Localization of gauge theory on a four-sphere and supersymmetric Wilson loops}",
    eprint = "0712.2824",
    archivePrefix = "arXiv",
    primaryClass = "hep-th",
    reportNumber = "ITEP-TH-41-07, PUTP-2248",
    doi = "10.1007/s00220-012-1485-0",
    journal = "Commun. Math. Phys.",
    volume = "313",
    pages = "71--129",
    year = "2012"
}

@article{Gerchkovitz:2016gxx,
    author = "Gerchkovitz, Efrat and Gomis, Jaume and Ishtiaque, Nafiz and Karasik, Avner and Komargodski, Zohar and Pufu, Silviu S.",
    title = "{Correlation Functions of Coulomb Branch Operators}",
    eprint = "1602.05971",
    archivePrefix = "arXiv",
    primaryClass = "hep-th",
    reportNumber = "PUPT-2500",
    doi = "10.1007/JHEP01(2017)103",
    journal = "JHEP",
    volume = "01",
    pages = "103",
    year = "2017"
}

@article{Moore:1997dj,
    author = "Moore, Gregory W. and Nekrasov, Nikita and Shatashvili, Samson",
    title = "{Integrating over Higgs branches}",
    eprint = "hep-th/9712241",
    archivePrefix = "arXiv",
    reportNumber = "ITEP-TH-62-97, HUTP-97-A089, YCTP-P23-97",
    doi = "10.1007/PL00005525",
    journal = "Commun. Math. Phys.",
    volume = "209",
    pages = "97--121",
    year = "2000"
}

@article{Nekrasov:2003rj,
    author = "Nekrasov, Nikita and Okounkov, Andrei",
    title = "{Seiberg-Witten theory and random partitions}",
    eprint = "hep-th/0306238",
    archivePrefix = "arXiv",
    reportNumber = "ITEP-TH-36-03, PUDM-2003, IHES-P-03-43",
    doi = "10.1007/0-8176-4467-9_15",
    journal = "Prog. Math.",
    volume = "244",
    pages = "525--596",
    year = "2006"
}

@article{Klemm:2002pa,
    author = "Klemm, Albrecht and Marino, Marcos and Theisen, Stefan",
    title = "{Gravitational corrections in supersymmetric gauge theory and matrix models}",
    eprint = "hep-th/0211216",
    archivePrefix = "arXiv",
    reportNumber = "AEI-2002-090, HUTP-02-A060, HU-EP-02-48",
    doi = "10.1088/1126-6708/2003/03/051",
    journal = "JHEP",
    volume = "03",
    pages = "051",
    year = "2003"
}

@article{Shapere:2008zf,
    author = "Shapere, Alfred D. and Tachikawa, Yuji",
    title = "{Central charges of N=2 superconformal field theories in four dimensions}",
    eprint = "0804.1957",
    archivePrefix = "arXiv",
    primaryClass = "hep-th",
    doi = "10.1088/1126-6708/2008/09/109",
    journal = "JHEP",
    volume = "09",
    pages = "109",
    year = "2008"
}

@article{Billo:2013fi,
    author = "Billo, M. and Frau, M. and Gallot, L. and Lerda, A. and Pesando, I.",
    title = "{Deformed N=2 theories, generalized recursion relations and S-duality}",
    eprint = "1302.0686",
    archivePrefix = "arXiv",
    primaryClass = "hep-th",
    reportNumber = "LAPTH-005-13",
    doi = "10.1007/JHEP04(2013)039",
    journal = "JHEP",
    volume = "04",
    pages = "039",
    year = "2013"
}

@article{Fucito:2013fba,
    author = "Fucito, Francesco and Morales, Jose Francisco and Poghossian, Rubik and Ricci Pacifici, Daniel",
    title = "{Exact results in $\mathcal{N} =$ 2 gauge theories}",
    eprint = "1307.6612",
    archivePrefix = "arXiv",
    primaryClass = "hep-th",
    doi = "10.1007/JHEP10(2013)178",
    journal = "JHEP",
    volume = "10",
    pages = "178",
    year = "2013"
}

@inproceedings{Nakajima:2003uh,
    author = "Nakajima, Hiraku and Yoshioka, Kota",
    title = "{Lectures on instanton counting}",
    booktitle = "{CRM Workshop on Algebraic Structures and Moduli Spaces}",
    eprint = "math/0311058",
    archivePrefix = "arXiv",
    month = "11",
    year = "2003"
}

@article{Witten:1995gf,
    author = "Witten, Edward",
    title = "{On $S$-duality in Abelian gauge theory}",
    eprint = "hep-th/9505186",
    archivePrefix = "arXiv",
    reportNumber = "IASSNS-HEP-95-36",
    doi = "10.1007/BF01671570",
    journal = "Selecta Math.",
    volume = "1",
    pages = "383",
    year = "1995"
}

@article{Moore:1997pc,
    author = "Moore, Gregory W. and Witten, Edward",
    title = "{Integration over the u plane in Donaldson theory}",
    eprint = "hep-th/9709193",
    archivePrefix = "arXiv",
    reportNumber = "IASSNS-HEP-97-97, YCTP-P9-97",
    doi = "10.4310/ATMP.1997.v1.n2.a7",
    journal = "Adv. Theor. Math. Phys.",
    volume = "1",
    pages = "298--387",
    year = "1997"
}

@article{Manschot:2019pog,
    author = "Manschot, Jan and Moore, Gregory W. and Zhang, Xinyu",
    title = "{Effective gravitational couplings of four-dimensional $ \mathcal{N} $ = 2 supersymmetric gauge theories}",
    eprint = "1912.04091",
    archivePrefix = "arXiv",
    primaryClass = "hep-th",
    doi = "10.1007/JHEP06(2020)150",
    journal = "JHEP",
    volume = "06",
    pages = "150",
    year = "2020"
}

@article{Russo:2014nka,
    author = "Russo, Jorge G.",
    title = "{$ \mathcal{N} $ = 2 gauge theories and quantum phases}",
    eprint = "1411.2602",
    archivePrefix = "arXiv",
    primaryClass = "hep-th",
    reportNumber = "ICCUB-14-065",
    doi = "10.1007/JHEP12(2014)169",
    journal = "JHEP",
    volume = "12",
    pages = "169",
    year = "2014"
}

@article{Russo:2019ipg,
    author = "Russo, Jorge G.",
    title = "{Properties of the partition function of $ \mathcal{N}=2 $ supersymmetric QCD with massive matter}",
    eprint = "1905.05267",
    archivePrefix = "arXiv",
    primaryClass = "hep-th",
    reportNumber = "ICCUB-19-007",
    doi = "10.1007/JHEP07(2019)125",
    journal = "JHEP",
    volume = "07",
    pages = "125",
    year = "2019"
}

@article{Billo:2017glv,
    author = "Billo, M. and Fucito, F. and Lerda, A. and Morales, J. F. and Stanev, Ya. S. and Wen, Congkao",
    title = "{Two-point correlators in $N =2$ gauge theories}",
    eprint = "1705.02909",
    archivePrefix = "arXiv",
    primaryClass = "hep-th",
    doi = "10.1016/j.nuclphysb.2017.11.003",
    journal = "Nucl. Phys. B",
    volume = "926",
    pages = "427--466",
    year = "2018"
}

@article{Billo:2019job,
    author = "Billo, M. and Fucito, F. and Korchemsky, G. P. and Lerda, A. and Morales, J. -F.",
    title = "{Two-point correlators in non-conformal $ \mathcal{N} $ = 2 gauge theories}",
    eprint = "1901.09693",
    archivePrefix = "arXiv",
    primaryClass = "hep-th",
    reportNumber = "ROM2F/2019/02",
    doi = "10.1007/JHEP05(2019)199",
    journal = "JHEP",
    volume = "05",
    pages = "199",
    year = "2019"
}

@article{doi:10.1143/JPSJ.57.1901,
author = {Hirota ,Ryogo and Ohta ,Yasuhiro and Satsuma ,Junkichi},
title = {Solutios of the Kadomtsev-Petviashvili Equation and the Two-Dimensional Toda Equations},
journal = {Journal of the Physical Society of Japan},
volume = {57},
number = {6},
pages = {1901-1904},
year = {1988},
doi = {10.1143/JPSJ.57.1901},

URL = { 
    
        https://doi.org/10.1143/JPSJ.57.1901
    
    

},
eprint = { 
    
        https://doi.org/10.1143/JPSJ.57.1901
    
    

}
}

@article{Yang_Chen_1998,
doi = {10.1088/0305-4470/31/4/005},
url = {https://doi.org/10.1088/0305-4470/31/4/005},
year = {1998},
month = {jan},
publisher = {},
volume = {31},
number = {4},
pages = {1141},
author = {Yang Chen and Nigel Lawrence},
title = {On the linear statistics of Hermitian random matrices},
journal = {Journal of Physics A: Mathematical and General}
}

@article{BASOR2001214,
title = {Determinants of Hankel Matrices},
journal = {Journal of Functional Analysis},
volume = {179},
number = {1},
pages = {214-234},
year = {2001},
issn = {0022-1236},
doi = {https://doi.org/10.1006/jfan.2000.3672},
url = {https://www.sciencedirect.com/science/article/pii/S0022123600936723},
author = {Estelle L. Basor and Yang Chen and Harold Widom}
}

@article{Basor_2011,
    author = "Basor, E.L. and Chen, Y. and Widom, H.",
    title = "{Hankel Determinants as Fredholm Determinants}",
    eprint = "math/0006070",
    archivePrefix = "arXiv",
    primaryClass = "math.CA",
    journal = "Publications of the Mathematical Sciences Research Institute (MSRI)",
    volume = "40",
    pages = "1-9",
    year = "2001"
}

@ARTICLE{6200861,
  author={Chen, Yang and McKay, Matthew R.},
  journal={IEEE Transactions on Information Theory}, 
  title={Coulumb Fluid, Painlevé Transcendents, and the Information Theory of MIMO Systems}, 
  year={2012},
  volume={58},
  number={7},
  pages={4594-4634},
  doi={10.1109/TIT.2012.2195154}}

@article{Hellerman:2021duh,
    author = "Hellerman, Simeon",
    title = "{On the exponentially small corrections to ${\cal N} = 2$ superconformal correlators at large R-charge}",
    eprint = "2103.09312",
    archivePrefix = "arXiv",
    primaryClass = "hep-th",
    month = "3",
    year = "2021"
}

@article{Grassi:2024bwl,
    author = "Grassi, Alba and Iossa, Cristoforo",
    title = "{Matrix models for extremal and integrated correlators of higher rank}",
    eprint = "2408.07391",
    archivePrefix = "arXiv",
    primaryClass = "hep-th",
    reportNumber = "CERN-TH-2024-134",
    month = "8",
    year = "2024"
}

@article{Brown:2025cbz,
    author = "Brown, Augustus and Galvagno, Francesco and Grassi, Alba and Iossa, Cristoforo and Wen, Congkao",
    title = "{Large charge meets semiclassics in $\mathcal{N}$ = 4 super Yang-Mills}",
    eprint = "2503.02028",
    archivePrefix = "arXiv",
    primaryClass = "hep-th",
    reportNumber = "CERN-TH-2025-016, QMUL-PH-25-03",
    doi = "10.1007/JHEP06(2025)223",
    journal = "JHEP",
    volume = "06",
    pages = "223",
    year = "2025"
}

@article{PhysRevD.7.949,
  title = {Extraction of Energy and Charge from a Black Hole},
  author = {Bekenstein, Jacob D.},
  journal = {Phys. Rev. D},
  volume = {7},
  issue = {4},
  pages = {949--953},
  numpages = {0},
  year = {1973},
  month = {Feb},
  publisher = {American Physical Society},
  doi = {10.1103/PhysRevD.7.949},
  url = {https://link.aps.org/doi/10.1103/PhysRevD.7.949}
}

@article{PhysRevD.7.2333,
  title = {Black Holes and Entropy},
  author = {Bekenstein, Jacob D.},
  journal = {Phys. Rev. D},
  volume = {7},
  issue = {8},
  pages = {2333--2346},
  numpages = {0},
  year = {1973},
  month = {Apr},
  publisher = {American Physical Society},
  doi = {10.1103/PhysRevD.7.2333},
  url = {https://link.aps.org/doi/10.1103/PhysRevD.7.2333}
}

@article{Milne1936,
  title = {Relativity, Gravitation, and World-Structure},
  author = {Milne, E. A.},
  journal = {Astrophysical Journal},
  volume = {83},
  pages = {61},
  year = {1936},
  doi = {10.1086/143701}
}

@article{Hawking1975,
  title = {Particle creation by black holes},
  author = {Hawking, S.  W. },
  journal = {Communications in Mathematical Physics},
  volume = {43},
  number = {3},
  pages = {199--220},
  numpages = {0},
  year = {1975},
  month = {Aug},
  doi = {10.1007/BF02345020},
  url = {https://doi.org/10.1007/BF02345020}
}

@article{PhysRevD.14.2460,
  title = {Breakdown of predictability in gravitational collapse},
  author = {Hawking, S. W.},
  journal = {Phys. Rev. D},
  volume = {14},
  issue = {10},
  pages = {2460--2473},
  year = {1976},
  month = {Nov},
  publisher = {American Physical Society},
  doi = {10.1103/PhysRevD.14.2460},
  url = {https://link.aps.org/doi/10.1103/PhysRevD.14.2460}
}

@article{Hawking1982,
  title = {The unpredictability of quantum gravity},
  author = {Hawking, S.  W. },
  journal = {Communications in Mathematical Physics},
  volume = {87},
  number = {3},
  pages = {395--415},
  year = {1982},
  month = {Dec},
  doi = {10.1007/BF01206031},
  url = {https://doi.org/10.1007/BF01206031}
}

@article{Dima:2024cok,
    author = "Dima, Alexandru and Melis, Marco and Pani, Paolo",
    title = "{Spectroscopy of magnetized black holes and topological stars}",
    eprint = "2406.19327",
    archivePrefix = "arXiv",
    primaryClass = "gr-qc",
    doi = "10.1103/PhysRevD.110.084067",
    journal = "Phys. Rev. D",
    volume = "110",
    number = "8",
    pages = "084067",
    year = "2024"
}

@article{Dima:2025zot,
    author = "Dima, Alexandru and Melis, Marco and Pani, Paolo",
    title = "{Nonradial stability of topological stars}",
    eprint = "2502.04444",
    archivePrefix = "arXiv",
    primaryClass = "gr-qc",
    doi = "10.1103/PhysRevD.111.104001",
    journal = "Phys. Rev. D",
    volume = "111",
    number = "10",
    pages = "104001",
    year = "2025"
}

@article{Melis:2025iaw,
    author = "Melis, Marco and Brito, Richard and Pani, Paolo",
    title = "{Extreme mass ratio inspirals around topological stars}",
    eprint = "2504.16156",
    archivePrefix = "arXiv",
    primaryClass = "gr-qc",
    doi = "10.1103/zng8-9qrn",
    journal = "Phys. Rev. D",
    volume = "111",
    number = "12",
    pages = "124043",
    year = "2025"
}

@article{Heidmann:2025pbb,
    author = "Heidmann, Pierre and Pani, Paolo and Santos, Jorge E.",
    title = "{Asymptotically Flat Rotating Topological Stars}",
    eprint = "2510.05200",
    archivePrefix = "arXiv",
    primaryClass = "hep-th",
    month = "10",
    year = "2025"
}

@article{Cipriani:2024ygw,
    author = "Cipriani, Andrea and Di Benedetto, Carlo and Di Russo, Giorgio and Grillo, Alfredo and Sudano, Giuseppe",
    title = "{Charge (in)stability and superradiance of Topological Stars}",
    eprint = "2405.06566",
    archivePrefix = "arXiv",
    primaryClass = "hep-th",
    doi = "10.1007/JHEP07(2024)143",
    journal = "JHEP",
    volume = "07",
    pages = "143",
    year = "2024"
}

@article{Bianchi:2025uis,
    author = "Bianchi, Massimo and Dibitetto, Giuseppe and Morales, Jose F. and Ruip{\'e}rez, Alejandro",
    title = "{Rotating Topological Stars}",
    eprint = "2504.12235",
    archivePrefix = "arXiv",
    primaryClass = "hep-th",
    month = "4",
    year = "2025"
}

@article{Starobinskil:1974nkd,
    author = "Starobinskil, Alexei A. and Churilov, S. M.",
    title = "{Amplification of electromagnetic and gravitational waves scattered by a rotating ''black hole''}",
    journal = "Sov. Phys. JETP",
    volume = "65",
    number = "1",
    pages = "1--5",
    year = "1974"
}

@article{Bena:2024hoh,
    author = "Bena, Iosif and Di Russo, Giorgio and Morales, Jose Francisco and Ruip\'erez, Alejandro",
    title = "{Non-spinning tops are stable}",
    eprint = "2406.19330",
    archivePrefix = "arXiv",
    primaryClass = "hep-th",
    month = "6",
    year = "2024"
}

@article{Bianchi:2024vmi,
    author = "Bianchi, Massimo and Bini, Donato and Di Russo, Giorgio",
    title = "{Scalar perturbations of topological-star spacetimes}",
    eprint = "2407.10868",
    archivePrefix = "arXiv",
    primaryClass = "gr-qc",
    doi = "10.1103/PhysRevD.110.084077",
    journal = "Phys. Rev. D",
    volume = "110",
    number = "8",
    pages = "084077",
    year = "2024"
}

@article{Bianchi:2024rod,
    author = "Bianchi, Massimo and Bini, Donato and Di Russo, Giorgio",
    title = "{Scalar waves in a topological star spacetime: Self-force and radiative losses}",
    eprint = "2411.19612",
    archivePrefix = "arXiv",
    primaryClass = "gr-qc",
    doi = "10.1103/PhysRevD.111.044017",
    journal = "Phys. Rev. D",
    volume = "111",
    number = "4",
    pages = "044017",
    year = "2025"
}

@article{DiRusso:2025lip,
    author = "Di Russo, Giorgio and Bianchi, Massimo and Bini, Donato",
    title = "{Scalar waves from unbound orbits in a topological star spacetime: PN reconstruction of the field and radiation losses in a self-force approach}",
    eprint = "2502.21040",
    archivePrefix = "arXiv",
    primaryClass = "gr-qc",
    doi = "10.1103/sycf-brn1",
    journal = "Phys. Rev. D",
    volume = "112",
    number = "2",
    pages = "024002",
    year = "2025"
}

@article{Bini:2025qyn,
    author = "Bini, Donato and Di Russo, Giorgio",
    title = "{Characterizing geodesic deviations in a topological star spacetime: Massive, charged, spinning, and stringylike objects}",
    eprint = "2505.13020",
    archivePrefix = "arXiv",
    primaryClass = "gr-qc",
    doi = "10.1103/6xc4-wm2f",
    journal = "Phys. Rev. D",
    volume = "112",
    number = "2",
    pages = "024021",
    year = "2025"
}

@article{Bianchi:2025aei,
    author = "Bianchi, Massimo and Bini, Donato and Di Russo, Giorgio",
    title = "{Scattering angle in a topological star spacetime: A self-force approach}",
    eprint = "2506.04876",
    archivePrefix = "arXiv",
    primaryClass = "gr-qc",
    doi = "10.1103/cz6k-wqpn",
    journal = "Phys. Rev. D",
    volume = "112",
    number = "4",
    pages = "044008",
    year = "2025"
}

@article{Bini:2025ltr,
    author = "Bini, Donato and Di Russo, Giorgio",
    title = "{Topological stars and scalar wave equation: Exact resummation of the renormalized angular momentum in the eikonal limit}",
    eprint = "2506.14442",
    archivePrefix = "arXiv",
    primaryClass = "gr-qc",
    doi = "10.1103/dw5y-4pv8",
    journal = "Phys. Rev. D",
    volume = "112",
    number = "6",
    pages = "064008",
    year = "2025"
}

@article{Bianchi:2022qph,
    author = "Bianchi, Massimo and Di Russo, Giorgio",
    title = "{2-charge circular fuzz-balls and their perturbations}",
    eprint = "2212.07504",
    archivePrefix = "arXiv",
    primaryClass = "hep-th",
    doi = "10.1007/JHEP08(2023)217",
    journal = "JHEP",
    volume = "08",
    pages = "217",
    year = "2023"
}

@article{Bena:2022rna,
    author = "Bena, Iosif and Martinec, Emil J. and Mathur, Samir D. and Warner, Nicholas P.",
    title = "{Fuzzballs and Microstate Geometries: Black-Hole Structure in String Theory}",
    eprint = "2204.13113",
    archivePrefix = "arXiv",
    primaryClass = "hep-th",
    month = "4",
    year = "2022"
}

@article{Sen:1995in,
    author = "Sen, Ashoke",
    title = "{Extremal black holes and elementary string states}",
    eprint = "hep-th/9504147",
    archivePrefix = "arXiv",
    reportNumber = "TIFR-TH-95-19",
    doi = "10.1142/S0217732395002234",
    journal = "Mod. Phys. Lett. A",
    volume = "10",
    pages = "2081--2094",
    year = "1995"
}

@article{Lunin:2001fv,
    author = "Lunin, Oleg and Mathur, Samir D.",
    title = "{Metric of the multiply wound rotating string}",
    eprint = "hep-th/0105136",
    archivePrefix = "arXiv",
    reportNumber = "OHSTPY-HEP-T-01-014",
    doi = "10.1016/S0550-3213(01)00321-2",
    journal = "Nucl. Phys. B",
    volume = "610",
    pages = "49--76",
    year = "2001"
}

@article{Gregory:1993vy,
    author = "Gregory, R. and Laflamme, R.",
    title = "{Black strings and p-branes are unstable}",
    eprint = "hep-th/9301052",
    archivePrefix = "arXiv",
    doi = "10.1103/PhysRevLett.70.2837",
    journal = "Phys. Rev. Lett.",
    volume = "70",
    pages = "2837--2840",
    year = "1993"
}

@article{Jejjala:2005yu,
    author = "Jejjala, Vishnu and Madden, Owen and Ross, Simon F. and Titchener, Georgina",
    title = "{Non-supersymmetric smooth geometries and D1-D5-P bound states}",
    eprint = "hep-th/0504181",
    archivePrefix = "arXiv",
    reportNumber = "DCPT-05-17",
    doi = "10.1103/PhysRevD.71.124030",
    journal = "Phys. Rev. D",
    volume = "71",
    pages = "124030",
    year = "2005"
}

@article{Bianchi:2020yzr,
    author = "Bianchi, M. and Consoli, D. and Grillo, A. and Morales, J. F.",
    title = "{Light rings of five-dimensional geometries}",
    eprint = "2011.04344",
    archivePrefix = "arXiv",
    primaryClass = "hep-th",
    doi = "10.1007/JHEP03(2021)210",
    journal = "JHEP",
    volume = "03",
    pages = "210",
    year = "2021"
}

@article{Bena:2022ldq,
    author = "Bena, Iosif and Martinec, Emil J. and Mathur, Samir D. and Warner, Nicholas P.",
    title = "{Snowmass White Paper: Micro- and Macro-Structure of Black Holes}",
    eprint = "2203.04981",
    archivePrefix = "arXiv",
    primaryClass = "hep-th",
    month = "3",
    year = "2022"
}

@article{Mathur:2024ify,
    author = "Mathur, Samir D. and Mehta, Madhur",
    title = "{The Fuzzball Paradigm}",
    eprint = "2412.09495",
    archivePrefix = "arXiv",
    primaryClass = "hep-th",
    month = "12",
    year = "2024"
}

@article{Mathur:2005zp,
    author = "Mathur, Samir D.",
    editor = "Kiritsis, E.",
    title = "{The Fuzzball proposal for black holes: An Elementary review}",
    eprint = "hep-th/0502050",
    archivePrefix = "arXiv",
    doi = "10.1002/prop.200410203",
    journal = "Fortsch. Phys.",
    volume = "53",
    pages = "793--827",
    year = "2005"
}

@article{Lunin:2001jy,
    author = "Lunin, Oleg and Mathur, Samir D.",
    title = "{AdS / CFT duality and the black hole information paradox}",
    eprint = "hep-th/0109154",
    archivePrefix = "arXiv",
    reportNumber = "OHSTPY-HEP-T-01-019",
    doi = "10.1016/S0550-3213(01)00620-4",
    journal = "Nucl. Phys. B",
    volume = "623",
    pages = "342--394",
    year = "2002"
}

@inproceedings{Bambi:2025wjx,
    author = "Bambi, Cosimo and others",
    title = "{Black hole mimickers: from theory to observation}",
    eprint = "2505.09014",
    archivePrefix = "arXiv",
    primaryClass = "gr-qc",
    month = "5",
    year = "2025"
}

@article{Cardoso:2005gj,
    author = "Cardoso, Vitor and Dias, Oscar J. C. and Hovdebo, Jordan L. and Myers, Robert C.",
    title = "{Instability of non-supersymmetric smooth geometries}",
    eprint = "hep-th/0512277",
    archivePrefix = "arXiv",
    doi = "10.1103/PhysRevD.73.064031",
    journal = "Phys. Rev. D",
    volume = "73",
    pages = "064031",
    year = "2006"
}

@article{Chowdhury:2007jx,
    author = "Chowdhury, Borun D. and Mathur, Samir D.",
    title = "{Radiation from the non-extremal fuzzball}",
    eprint = "0711.4817",
    archivePrefix = "arXiv",
    primaryClass = "hep-th",
    doi = "10.1088/0264-9381/25/13/135005",
    journal = "Class. Quant. Grav.",
    volume = "25",
    pages = "135005",
    year = "2008"
}

@article{Giusto:2004id,
    author = "Giusto, Stefano and Mathur, Samir D. and Saxena, Ashish",
    title = "{Dual geometries for a set of 3-charge microstates}",
    eprint = "hep-th/0405017",
    archivePrefix = "arXiv",
    reportNumber = "OHSTPY-HEP-T-04-005",
    doi = "10.1016/j.nuclphysb.2004.09.001",
    journal = "Nucl. Phys. B",
    volume = "701",
    pages = "357--379",
    year = "2004"
}

@article{Giusto:2004ip,
    author = "Giusto, Stefano and Mathur, Samir D. and Saxena, Ashish",
    title = "{3-charge geometries and their CFT duals}",
    eprint = "hep-th/0406103",
    archivePrefix = "arXiv",
    reportNumber = "OHSTPY-HEP-T-04-006",
    doi = "10.1016/j.nuclphysb.2005.01.009",
    journal = "Nucl. Phys. B",
    volume = "710",
    pages = "425--463",
    year = "2005"
}

@article{Heidmann:2021cms,
    author = "Heidmann, Pierre",
    title = "{Non-BPS floating branes and bubbling geometries}",
    eprint = "2112.03279",
    archivePrefix = "arXiv",
    primaryClass = "hep-th",
    doi = "10.1007/JHEP02(2022)162",
    journal = "JHEP",
    volume = "02",
    pages = "162",
    year = "2022"
}

@article{Heidmann:2022zyd,
    author = "Heidmann, Pierre and Houppe, Anthony",
    title = "{Solitonic excitations in AdS$_{2}$}",
    eprint = "2212.05065",
    archivePrefix = "arXiv",
    primaryClass = "hep-th",
    doi = "10.1007/JHEP07(2023)186",
    journal = "JHEP",
    volume = "07",
    pages = "186",
    year = "2023"
}

@article{Cvetic:1996xz,
    author = "Cvetic, Mirjam and Youm, Donam",
    title = "{General rotating five-dimensional black holes of toroidally compactified heterotic string}",
    eprint = "hep-th/9603100",
    archivePrefix = "arXiv",
    reportNumber = "IASSNS-HEP-96-24, PUPT-1602",
    doi = "10.1016/0550-3213(96)00355-0",
    journal = "Nucl. Phys. B",
    volume = "476",
    pages = "118--132",
    year = "1996"
}

@article{Cvetic:1997uw,
    author = "Cvetic, Mirjam and Larsen, Finn",
    title = "{General rotating black holes in string theory: Grey body factors and event horizons}",
    eprint = "hep-th/9705192",
    archivePrefix = "arXiv",
    reportNumber = "UPR-0752-T",
    doi = "10.1103/PhysRevD.56.4994",
    journal = "Phys. Rev. D",
    volume = "56",
    pages = "4994--5007",
    year = "1997"
}

@article{Giusto:2011fy,
    author = "Giusto, Stefano and Russo, Rodolfo and Turton, David",
    title = "{New D1-D5-P geometries from string amplitudes}",
    eprint = "1108.6331",
    archivePrefix = "arXiv",
    primaryClass = "hep-th",
    reportNumber = "DFPD-11-TH-14, QMUL-PH-11-12",
    doi = "10.1007/JHEP11(2011)062",
    journal = "JHEP",
    volume = "11",
    pages = "062",
    year = "2011"
}

@article{Turton:2012ny,
    author = "Turton, David",
    editor = "Bellucci, Stefano",
    title = "{Black Hole Microstate Geometries from String Amplitudes}",
    eprint = "1203.6378",
    archivePrefix = "arXiv",
    primaryClass = "hep-th",
    doi = "10.1007/978-3-319-00215-6_7",
    journal = "Springer Proc. Phys.",
    volume = "144",
    pages = "255--266",
    year = "2013"
}

@article{Bianchi:2016bgx,
    author = "Bianchi, Massimo and Morales, Jos\`e Francisco and Pieri, Lorenzo",
    title = "{Stringy origin of 4d black hole microstates}",
    eprint = "1603.05169",
    archivePrefix = "arXiv",
    primaryClass = "hep-th",
    reportNumber = "ROM2F-2016-04",
    doi = "10.1007/JHEP06(2016)003",
    journal = "JHEP",
    volume = "06",
    pages = "003",
    year = "2016"
}

@article{Bianchi:2017bxl,
    author = "Bianchi, Massimo and Morales, Jose Francisco and Pieri, Lorenzo and Zinnato, Natale",
    title = "{More on microstate geometries of 4d black holes}",
    eprint = "1701.05520",
    archivePrefix = "arXiv",
    primaryClass = "hep-th",
    doi = "10.1007/JHEP05(2017)147",
    journal = "JHEP",
    volume = "05",
    pages = "147",
    year = "2017"
}

@article{Bena:2016agb,
    author = "Bena, Iosif and Martinec, Emil and Turton, David and Warner, Nicholas P.",
    title = "{Momentum Fractionation on Superstrata}",
    eprint = "1601.05805",
    archivePrefix = "arXiv",
    primaryClass = "hep-th",
    reportNumber = "IPHT-T16-004",
    doi = "10.1007/JHEP05(2016)064",
    journal = "JHEP",
    volume = "05",
    pages = "064",
    year = "2016"
}

@article{Bena:2016ypk,
    author = "Bena, Iosif and Giusto, Stefano and Martinec, Emil J. and Russo, Rodolfo and Shigemori, Masaki and Turton, David and Warner, Nicholas P.",
    title = "{Smooth horizonless geometries deep inside the black-hole regime}",
    eprint = "1607.03908",
    archivePrefix = "arXiv",
    primaryClass = "hep-th",
    reportNumber = "IPHT-T16-067, DFPD-16-TH-11, EFI-16-16, QMUL-PH-16-14, YITP-16-88",
    doi = "10.1103/PhysRevLett.117.201601",
    journal = "Phys. Rev. Lett.",
    volume = "117",
    number = "20",
    pages = "201601",
    year = "2016"
}

@article{Bena:2017xbt,
    author = "Bena, Iosif and Giusto, Stefano and Martinec, Emil J. and Russo, Rodolfo and Shigemori, Masaki and Turton, David and Warner, Nicholas P.",
    title = "{Asymptotically-flat supergravity solutions deep inside the black-hole regime}",
    eprint = "1711.10474",
    archivePrefix = "arXiv",
    primaryClass = "hep-th",
    reportNumber = "IPHT-T17-135, QMUL-PH-17-26, YITP-17-127",
    doi = "10.1007/JHEP02(2018)014",
    journal = "JHEP",
    volume = "02",
    pages = "014",
    year = "2018"
}

@article{Bena:2017upb,
    author = "Bena, Iosif and Turton, David and Walker, Robert and Warner, Nicholas P.",
    title = "{Integrability and Black-Hole Microstate Geometries}",
    eprint = "1709.01107",
    archivePrefix = "arXiv",
    primaryClass = "hep-th",
    reportNumber = "IPHT-T17-134, IPHT-T17/134",
    doi = "10.1007/JHEP11(2017)021",
    journal = "JHEP",
    volume = "11",
    pages = "021",
    year = "2017"
}

@article{Heidmann:2022ehn,
    author = "Heidmann, Pierre and Bah, Ibrahima and Berti, Emanuele",
    title = "{Imaging topological solitons: The microstructure behind the shadow}",
    eprint = "2212.06837",
    archivePrefix = "arXiv",
    primaryClass = "gr-qc",
    doi = "10.1103/PhysRevD.107.084042",
    journal = "Phys. Rev. D",
    volume = "107",
    number = "8",
    pages = "084042",
    year = "2023"
}

@article{Bah:2020pdz,
    author = "Bah, Ibrahima and Heidmann, Pierre",
    title = "{Topological stars, black holes and generalized charged Weyl solutions}",
    eprint = "2012.13407",
    archivePrefix = "arXiv",
    primaryClass = "hep-th",
    doi = "10.1007/JHEP09(2021)147",
    journal = "JHEP",
    volume = "09",
    pages = "147",
    year = "2021"
}

@article{Bah:2021irr,
    author = "Bah, Ibrahima and Dey, Anindya and Heidmann, Pierre",
    title = "{Stability of topological solitons, and black string to bubble transition}",
    eprint = "2112.11474",
    archivePrefix = "arXiv",
    primaryClass = "hep-th",
    doi = "10.1007/JHEP04(2022)168",
    journal = "JHEP",
    volume = "04",
    pages = "168",
    year = "2022"
}

@article{Bonelli:2022ten,
    author = "Bonelli, Giulio and Iossa, Cristoforo and Panea Lichtig, Daniel and Tanzini, Alessandro",
    title = "{Irregular Liouville Correlators and Connection Formulae for Heun Functions}",
    eprint = "2201.04491",
    archivePrefix = "arXiv",
    primaryClass = "hep-th",
    doi = "10.1007/s00220-022-04497-5",
    journal = "Commun. Math. Phys.",
    volume = "397",
    number = "2",
    pages = "635--727",
    year = "2023"
}

@article{Tangherlini:1963bw,
    author = "Tangherlini, F. R.",
    title = "{Schwarzschild field in n dimensions and the dimensionality of space problem}",
    doi = "10.1007/BF02784569",
    journal = "Nuovo Cim.",
    volume = "27",
    pages = "636--651",
    year = "1963"
}

@article{Bianchi:2023rlt,
    author = "Bianchi, Massimo and Di Benedetto, Carlo and Di Russo, Giorgio and Sudano, Giuseppe",
    title = "{Charge instability of JMaRT geometries}",
    eprint = "2305.00865",
    archivePrefix = "arXiv",
    primaryClass = "hep-th",
    doi = "10.1007/JHEP09(2023)078",
    journal = "JHEP",
    volume = "09",
    pages = "078",
    year = "2023"
}

@article{Chandrasekhar:1984siy,
    author = "Chandrasekhar, S.",
    editor = "Bertotti, B. and de Felice, F. and Pascolini, A.",
    title = "{The Mathematical Theory of Black Holes}",
    doi = "10.1007/978-94-009-6469-3_2",
    journal = "Fundam. Theor. Phys.",
    volume = "9",
    pages = "5--26",
    year = "1984"
}

@article{Chakrabarty:2019ujg,
    author = "Chakrabarty, Bidisha and Ghosh, Debodirna and Virmani, Amitabh",
    title = "{Quasinormal modes of supersymmetric microstate geometries from the D1-D5 CFT}",
    eprint = "1908.01461",
    archivePrefix = "arXiv",
    primaryClass = "hep-th",
    doi = "10.1007/JHEP10(2019)072",
    journal = "JHEP",
    volume = "10",
    pages = "072",
    year = "2019"
}

@article{Aminov:2020yma,
    author = "Aminov, Gleb and Grassi, Alba and Hatsuda, Yasuyuki",
    title = "{Black Hole Quasinormal Modes and Seiberg\textendash{}Witten Theory}",
    eprint = "2006.06111",
    archivePrefix = "arXiv",
    primaryClass = "hep-th",
    reportNumber = "RUP-20-18",
    doi = "10.1007/s00023-021-01137-x",
    journal = "Annales Henri Poincare",
    volume = "23",
    number = "6",
    pages = "1951--1977",
    year = "2022"
}

@article{Bianchi:2021xpr,
    author = "Bianchi, Massimo and Consoli, Dario and Grillo, Alfredo and Morales, Jos\`e Francisco",
    title = "{QNMs of branes, BHs and fuzzballs from quantum SW geometries}",
    eprint = "2105.04245",
    archivePrefix = "arXiv",
    primaryClass = "hep-th",
    doi = "10.1016/j.physletb.2021.136837",
    journal = "Phys. Lett. B",
    volume = "824",
    pages = "136837",
    year = "2022"
}

@article{Bonelli:2021uvf,
    author = "Bonelli, Giulio and Iossa, Cristoforo and Lichtig, Daniel Panea and Tanzini, Alessandro",
    title = "{Exact solution of Kerr black hole perturbations via CFT2 and instanton counting: Greybody factor, quasinormal modes, and Love numbers}",
    eprint = "2105.04483",
    archivePrefix = "arXiv",
    primaryClass = "hep-th",
    doi = "10.1103/PhysRevD.105.044047",
    journal = "Phys. Rev. D",
    volume = "105",
    number = "4",
    pages = "044047",
    year = "2022"
}

@article{Consoli:2022eey,
    author = "Consoli, Dario and Fucito, Francesco and Morales, Jose Francisco and Poghossian, Rubik",
    title = "{CFT description of BHs and ECOs: QNMs, superradiance, echoes and tidal responses}",
    eprint = "2206.09437",
    archivePrefix = "arXiv",
    primaryClass = "hep-th",
    doi = "10.1007/JHEP12(2022)115",
    journal = "JHEP",
    volume = "12",
    pages = "115",
    year = "2022"
}

@article{Bianchi:2017sds,
    author = "Bianchi, M. and Consoli, D. and Morales, J. F.",
    title = "{Probing Fuzzballs with Particles, Waves and Strings}",
    eprint = "1711.10287",
    archivePrefix = "arXiv",
    primaryClass = "hep-th",
    doi = "10.1007/JHEP06(2018)157",
    journal = "JHEP",
    volume = "06",
    pages = "157",
    year = "2018"
}

@article{Bena:2020see,
    author = "Bena, Iosif and Mayerson, Daniel R.",
    title = "{Multipole Ratios: A New Window into Black Holes}",
    eprint = "2006.10750",
    archivePrefix = "arXiv",
    primaryClass = "hep-th",
    doi = "10.1103/PhysRevLett.125.221602",
    journal = "Phys. Rev. Lett.",
    volume = "125",
    number = "22",
    pages = "221602",
    year = "2020"
}

@article{Bianchi:2020bxa,
    author = "Bianchi, Massimo and Consoli, Dario and Grillo, Alfredo and Morales, Jos\`e Francisco and Pani, Paolo and Raposo, Guilherme",
    title = "{Distinguishing fuzzballs from black holes through their multipolar structure}",
    eprint = "2007.01743",
    archivePrefix = "arXiv",
    primaryClass = "hep-th",
    doi = "10.1103/PhysRevLett.125.221601",
    journal = "Phys. Rev. Lett.",
    volume = "125",
    number = "22",
    pages = "221601",
    year = "2020"
}

@article{Bianchi:2020miz,
    author = "Bianchi, Massimo and Consoli, Dario and Grillo, Alfredo and Morales, Jose Francisco and Pani, Paolo and Raposo, Guilherme",
    title = "{The multipolar structure of fuzzballs}",
    eprint = "2008.01445",
    archivePrefix = "arXiv",
    primaryClass = "hep-th",
    doi = "10.1007/JHEP01(2021)003",
    journal = "JHEP",
    volume = "01",
    pages = "003",
    year = "2021"
}

@article{Bena:2020uup,
    author = "Bena, Iosif and Mayerson, Daniel R.",
    title = "{Black Holes Lessons from Multipole Ratios}",
    eprint = "2007.09152",
    archivePrefix = "arXiv",
    primaryClass = "hep-th",
    doi = "10.1007/JHEP03(2021)114",
    journal = "JHEP",
    volume = "03",
    pages = "114",
    year = "2021"
}

@article{Mayerson:2020tpn,
    author = "Mayerson, Daniel R.",
    title = "{Fuzzballs and Observations}",
    eprint = "2010.09736",
    archivePrefix = "arXiv",
    primaryClass = "hep-th",
    doi = "10.1007/s10714-020-02769-w",
    journal = "Gen. Rel. Grav.",
    volume = "52",
    number = "12",
    pages = "115",
    year = "2020"
}

@article{Ikeda:2021uvc,
    author = "Ikeda, Taishi and Bianchi, Massimo and Consoli, Dario and Grillo, Alfredo and Morales, Jos\`e Francisco and Pani, Paolo and Raposo, Guilherme",
    title = "{Black-hole microstate spectroscopy: Ringdown, quasinormal modes, and echoes}",
    eprint = "2103.10960",
    archivePrefix = "arXiv",
    primaryClass = "gr-qc",
    doi = "10.1103/PhysRevD.104.066021",
    journal = "Phys. Rev. D",
    volume = "104",
    number = "6",
    pages = "066021",
    year = "2021"
}

@article{Bah:2021jno,
    author = "Bah, Ibrahima and Bena, Iosif and Heidmann, Pierre and Li, Yixuan and Mayerson, Daniel R.",
    title = "{Gravitational footprints of black holes and their microstate geometries}",
    eprint = "2104.10686",
    archivePrefix = "arXiv",
    primaryClass = "hep-th",
    doi = "10.1007/JHEP10(2021)138",
    journal = "JHEP",
    volume = "10",
    pages = "138",
    year = "2021"
}

@article{Bah:2020ogh,
    author = "Bah, Ibrahima and Heidmann, Pierre",
    title = "{Topological Stars and Black Holes}",
    eprint = "2011.08851",
    archivePrefix = "arXiv",
    primaryClass = "hep-th",
    doi = "10.1103/PhysRevLett.126.151101",
    journal = "Phys. Rev. Lett.",
    volume = "126",
    number = "15",
    pages = "151101",
    year = "2021"
}

@article{Ganchev:2021pgs,
    author = "Ganchev, Bogdan and Houppe, Anthony and Warner, Nicholas P.",
    title = "{Q-balls meet fuzzballs: non-BPS microstate geometries}",
    eprint = "2107.09677",
    archivePrefix = "arXiv",
    primaryClass = "hep-th",
    doi = "10.1007/JHEP11(2021)028",
    journal = "JHEP",
    volume = "11",
    pages = "028",
    year = "2021"
}

@article{Ganchev:2023sth,
    author = "Ganchev, Bogdan and Giusto, Stefano and Houppe, Anthony and Russo, Rodolfo and Warner, Nicholas P.",
    title = "{Microstrata}",
    eprint = "2307.13021",
    archivePrefix = "arXiv",
    primaryClass = "hep-th",
    reportNumber = "TTI-MATHPHYS-21",
    doi = "10.1007/JHEP10(2023)163",
    journal = "JHEP",
    volume = "10",
    pages = "163",
    year = "2023"
}

@article{Bah:2021owp,
    author = "Bah, Ibrahima and Heidmann, Pierre",
    title = "{Smooth bubbling geometries without supersymmetry}",
    eprint = "2106.05118",
    archivePrefix = "arXiv",
    primaryClass = "hep-th",
    doi = "10.1007/JHEP09(2021)128",
    journal = "JHEP",
    volume = "09",
    pages = "128",
    year = "2021"
}

@article{Bianchi:2023sfs,
    author = "Bianchi, Massimo and Di Russo, Giorgio and Grillo, Alfredo and Morales, Jose Francisco and Sudano, Giuseppe",
    title = "{On the stability and deformability of top stars}",
    eprint = "2305.15105",
    archivePrefix = "arXiv",
    primaryClass = "gr-qc",
    doi = "10.1007/JHEP12(2023)121",
    journal = "JHEP",
    volume = "12",
    pages = "121",
    year = "2023"
}

@article{Heidmann:2023ojf,
    author = "Heidmann, Pierre and Speeney, Nicholas and Berti, Emanuele and Bah, Ibrahima",
    title = "{Cavity effect in the quasinormal mode spectrum of topological stars}",
    eprint = "2305.14412",
    archivePrefix = "arXiv",
    primaryClass = "gr-qc",
    doi = "10.1103/PhysRevD.108.024021",
    journal = "Phys. Rev. D",
    volume = "108",
    number = "2",
    pages = "024021",
    year = "2023"
}

@article{LIGOScientific:2016aoc,
    author = "Abbott, B. P. and others",
    collaboration = "LIGO Scientific, Virgo",
    title = "{Observation of Gravitational Waves from a Binary Black Hole Merger}",
    eprint = "1602.03837",
    archivePrefix = "arXiv",
    primaryClass = "gr-qc",
    reportNumber = "LIGO-P150914",
    doi = "10.1103/PhysRevLett.116.061102",
    journal = "Phys. Rev. Lett.",
    volume = "116",
    number = "6",
    pages = "061102",
    year = "2016"
}

@article{LIGOScientific:2016lio,
    author = "Abbott, B. P. and others",
    collaboration = "LIGO Scientific, Virgo",
    title = "{Tests of general relativity with GW150914}",
    eprint = "1602.03841",
    archivePrefix = "arXiv",
    primaryClass = "gr-qc",
    reportNumber = "LIGO-P1500213",
    doi = "10.1103/PhysRevLett.116.221101",
    journal = "Phys. Rev. Lett.",
    volume = "116",
    number = "22",
    pages = "221101",
    year = "2016",
    note = "[Erratum: Phys.Rev.Lett. 121, 129902 (2018)]"
}

@article{EventHorizonTelescope:2019dse,
    author = "Akiyama, Kazunori and others",
    collaboration = "Event Horizon Telescope",
    title = "{First M87 Event Horizon Telescope Results. I. The Shadow of the Supermassive Black Hole}",
    eprint = "1906.11238",
    archivePrefix = "arXiv",
    primaryClass = "astro-ph.GA",
    doi = "10.3847/2041-8213/ab0ec7",
    journal = "Astrophys. J. Lett.",
    volume = "875",
    pages = "L1",
    year = "2019"
}

@article{Fioravanti:2020udo,
    author = "Fioravanti, Davide and Rossi, Marco and Shu, Hongfei",
    title = "{$QQ$-system and non-linear integral equations for scattering amplitudes at strong coupling}",
    eprint = "2004.10722",
    archivePrefix = "arXiv",
    primaryClass = "hep-th",
    reportNumber = "NORDITA 2020-035",
    doi = "10.1007/JHEP12(2020)086",
    journal = "JHEP",
    volume = "12",
    pages = "086",
    year = "2020"
}

@article{Rossi:2022mwx,
    author = "Rossi, Marco and Fioravanti, Davide",
    title = "{On the origin of the correspondence between classical and quantum integrable theories}",
    doi = "10.22323/1.414.0439",
    journal = "PoS",
    volume = "ICHEP2022",
    pages = "439",
    month = "11",
    year = "2022"
}

@article{Fioravanti:2021dce,
    author = "Fioravanti, Davide and Gregori, Daniele",
    title = "{A new method for exact results on Quasinormal Modes of Black Holes}",
    eprint = "2112.11434",
    archivePrefix = "arXiv",
    primaryClass = "hep-th",
    month = "12",
    year = "2021"
}

@article{Mathur:2009hf,
    author = "Mathur, Samir D.",
    editor = "Uranga, A. M.",
    title = "{The Information paradox: A Pedagogical introduction}",
    eprint = "0909.1038",
    archivePrefix = "arXiv",
    primaryClass = "hep-th",
    doi = "10.1088/0264-9381/26/22/224001",
    journal = "Class. Quant. Grav.",
    volume = "26",
    pages = "224001",
    year = "2009"
}

@article{Cardoso:2016rao,
    author = "Cardoso, Vitor and Franzin, Edgardo and Pani, Paolo",
    title = "{Is the gravitational-wave ringdown a probe of the event horizon?}",
    eprint = "1602.07309",
    archivePrefix = "arXiv",
    primaryClass = "gr-qc",
    doi = "10.1103/PhysRevLett.116.171101",
    journal = "Phys. Rev. Lett.",
    volume = "116",
    number = "17",
    pages = "171101",
    year = "2016",
    note = "[Erratum: Phys.Rev.Lett. 117, 089902 (2016)]"
}

@article{Cardoso:2017cqb,
    author = "Cardoso, Vitor and Pani, Paolo",
    title = "{Tests for the existence of black holes through gravitational wave echoes}",
    eprint = "1709.01525",
    archivePrefix = "arXiv",
    primaryClass = "gr-qc",
    doi = "10.1038/s41550-017-0225-y",
    journal = "Nature Astron.",
    volume = "1",
    number = "9",
    pages = "586--591",
    year = "2017"
}

@article{Almheiri:2012rt,
    author = "Almheiri, Ahmed and Marolf, Donald and Polchinski, Joseph and Sully, James",
    title = "{Black Holes: Complementarity or Firewalls?}",
    eprint = "1207.3123",
    archivePrefix = "arXiv",
    primaryClass = "hep-th",
    doi = "10.1007/JHEP02(2013)062",
    journal = "JHEP",
    volume = "02",
    pages = "062",
    year = "2013"
}

@article{Schunck:2003kk,
    author = "Schunck, Franz E. and Mielke, Eckehard W.",
    title = "{General relativistic boson stars}",
    eprint = "0801.0307",
    archivePrefix = "arXiv",
    primaryClass = "astro-ph",
    doi = "10.1088/0264-9381/20/20/201",
    journal = "Class. Quant. Grav.",
    volume = "20",
    pages = "R301--R356",
    year = "2003"
}

@article{Mazur:2001fv,
    author = "Mazur, Pawel O. and Mottola, Emil",
    title = "{Gravitational Condensate Stars: An Alternative to Black Holes}",
    eprint = "gr-qc/0109035",
    archivePrefix = "arXiv",
    reportNumber = "LA-UR-01-5067, LA-UR-01-5067",
    doi = "10.3390/universe9020088",
    journal = "Universe",
    volume = "9",
    number = "2",
    pages = "88",
    year = "2023"
}

@article{Solodukhin:2005qy,
    author = "Solodukhin, Sergey N.",
    title = "{Restoring unitarity in BTZ black hole}",
    eprint = "hep-th/0501053",
    archivePrefix = "arXiv",
    reportNumber = "IUB-TH-051",
    doi = "10.1103/PhysRevD.71.064006",
    journal = "Phys. Rev. D",
    volume = "71",
    pages = "064006",
    year = "2005"
}

@article{Balasubramanian:2022lnw,
    author = "Balasubramanian, Vijay and Lawrence, Albion and Magan, Javier M. and Sasieta, Martin",
    title = "{Microscopic Origin of the Entropy of Astrophysical Black Holes}",
    eprint = "2212.08623",
    archivePrefix = "arXiv",
    primaryClass = "hep-th",
    doi = "10.1103/PhysRevLett.132.141501",
    journal = "Phys. Rev. Lett.",
    volume = "132",
    number = "14",
    pages = "141501",
    year = "2024"
}

@article{Balasubramanian:2022gmo,
    author = "Balasubramanian, Vijay and Lawrence, Albion and Magan, Javier M. and Sasieta, Martin",
    title = "{Microscopic Origin of the Entropy of Black Holes in General Relativity}",
    eprint = "2212.02447",
    archivePrefix = "arXiv",
    primaryClass = "hep-th",
    doi = "10.1103/PhysRevX.14.011024",
    journal = "Phys. Rev. X",
    volume = "14",
    number = "1",
    pages = "011024",
    year = "2024"
}

@article{Climent:2024trz,
    author = "Climent, Ana and Emparan, Roberto and Magan, Javier M. and Sasieta, Martin and Vilar L\'opez, Alejandro",
    title = "{Universal Construction of Black Hole Microstates}",
    eprint = "2401.08775",
    archivePrefix = "arXiv",
    primaryClass = "hep-th",
    month = "1",
    year = "2024"
}

@article{Bena:2020yii,
    author = "Bena, Iosif and Eperon, Felicity and Heidmann, Pierre and Warner, Nicholas P.",
    title = "{The Great Escape: Tunneling out of Microstate Geometries}",
    eprint = "2005.11323",
    archivePrefix = "arXiv",
    primaryClass = "hep-th",
    doi = "10.1007/JHEP04(2021)112",
    journal = "JHEP",
    volume = "04",
    pages = "112",
    year = "2021"
}

@article{Cardoso:2019rvt,
    author = "Cardoso, Vitor and Pani, Paolo",
    title = "{Testing the nature of dark compact objects: a status report}",
    eprint = "1904.05363",
    archivePrefix = "arXiv",
    primaryClass = "gr-qc",
    doi = "10.1007/s41114-019-0020-4",
    journal = "Living Rev. Rel.",
    volume = "22",
    number = "1",
    pages = "4",
    year = "2019"
}

@article{DiRusso:2024hmd,
    author = "Di Russo, Giorgio and Fucito, Francesco and Morales, Jose Francisco",
    title = "{Tidal resonances for fuzzballs}",
    eprint = "2402.06621",
    archivePrefix = "arXiv",
    primaryClass = "hep-th",
    doi = "10.1007/JHEP04(2024)149",
    journal = "JHEP",
    volume = "04",
    pages = "149",
    year = "2024"
}

@article{Strominger:1996sh,
    author = "Strominger, Andrew and Vafa, Cumrun",
    title = "{Microscopic origin of the Bekenstein-Hawking entropy}",
    eprint = "hep-th/9601029",
    archivePrefix = "arXiv",
    reportNumber = "HUTP-96-A002, RU-96-01",
    doi = "10.1016/0370-2693(96)00345-0",
    journal = "Phys. Lett. B",
    volume = "379",
    pages = "99--104",
    year = "1996"
}

@article{Bianchi:2018kzy,
    author = "Bianchi, M. and Consoli, Dario and Grillo, Alfredo and Morales, Jose F.",
    title = "{The dark side of fuzzball geometries}",
    eprint = "1811.02397",
    archivePrefix = "arXiv",
    primaryClass = "hep-th",
    doi = "10.1007/JHEP05(2019)126",
    journal = "JHEP",
    volume = "05",
    pages = "126",
    year = "2019"
}

@article{Bah:2022yji,
    author = "Bah, Ibrahima and Heidmann, Pierre and Weck, Peter",
    title = "{Schwarzschild-like topological solitons}",
    eprint = "2203.12625",
    archivePrefix = "arXiv",
    primaryClass = "hep-th",
    doi = "10.1007/JHEP08(2022)269",
    journal = "JHEP",
    volume = "08",
    pages = "269",
    year = "2022"
}

@article{Leaver:1985ax,
    author = "Leaver, E. W.",
    title = "{An Analytic representation for the quasi normal modes of Kerr black holes}",
    doi = "10.1098/rspa.1985.0119",
    journal = "Proc. Roy. Soc. Lond. A",
    volume = "402",
    pages = "285--298",
    year = "1985"
}

@article{Leaver:1990zz,
    author = "Leaver, Edward W.",
    title = "{Quasinormal modes of Reissner-Nordstrom black holes}",
    doi = "10.1103/PhysRevD.41.2986",
    journal = "Phys. Rev. D",
    volume = "41",
    pages = "2986--2997",
    year = "1990"
}

@article{Bianchi:2020des,
    author = "Bianchi, Massimo and Grillo, Alfredo and Morales, Jose Francisco",
    title = "{Chaos at the rim of black hole and fuzzball shadows}",
    eprint = "2002.05574",
    archivePrefix = "arXiv",
    primaryClass = "hep-th",
    doi = "10.1007/JHEP05(2020)078",
    journal = "JHEP",
    volume = "05",
    pages = "078",
    year = "2020"
}

@article{Brito:2015oca,
    author = "Brito, Richard and Cardoso, Vitor and Pani, Paolo",
    title = "{Superradiance}: {New Frontiers in Black Hole
Physics}",
    eprint = "1501.06570",
    archivePrefix = "arXiv",
    primaryClass = "gr-qc",
    doi = "10.1007/978-3-319-19000-6",
    journal = "Lect. Notes Phys.",
    volume = "906",
    pages = "pp.1--237",
    year = "2015"
}

@article{Bianchi:2021yqs,
    author = "Bianchi, Massimo and Di Russo, Giorgio",
    title = "{Turning black holes and D-branes inside out of their photon spheres}",
    eprint = "2110.09579",
    archivePrefix = "arXiv",
    primaryClass = "hep-th",
    doi = "10.1103/PhysRevD.105.126007",
    journal = "Phys. Rev. D",
    volume = "105",
    number = "12",
    pages = "126007",
    year = "2022"
}

@article{Chervonyi:2013eja,
    author = "Chervonyi, Yuri and Lunin, Oleg",
    title = "{(Non)-Integrability of Geodesics in D-brane Backgrounds}",
    eprint = "1311.1521",
    archivePrefix = "arXiv",
    primaryClass = "hep-th",
    doi = "10.1007/JHEP02(2014)061",
    journal = "JHEP",
    volume = "02",
    pages = "061",
    year = "2014"
}

@article{Bena:2015bea,
    author = "Bena, Iosif and Giusto, Stefano and Russo, Rodolfo and Shigemori, Masaki and Warner, Nicholas P.",
    title = "{Habemus Superstratum! A constructive proof of the existence of superstrata}",
    eprint = "1503.01463",
    archivePrefix = "arXiv",
    primaryClass = "hep-th",
    reportNumber = "DFPD-15-TH-06, IPHT-T15-019, QMUL-PH-15-05, YITP-15-12",
    doi = "10.1007/JHEP05(2015)110",
    journal = "JHEP",
    volume = "05",
    pages = "110",
    year = "2015"
}

@article{DiMenza:2014vpa,
    author = "Di Menza, Laurent and Nicolas, Jean-Philippe",
    title = "{Superradiance on the Reissner\textendash{}Nordstr\o{}m metric}",
    eprint = "1411.3988",
    archivePrefix = "arXiv",
    primaryClass = "math-ph",
    doi = "10.1088/0264-9381/32/14/145013",
    journal = "Class. Quant. Grav.",
    volume = "32",
    number = "14",
    pages = "145013",
    year = "2015"
}

@article{Benone:2015bst,
    author = "Benone, Carolina L. and Crispino, Luis C. B.",
    title = "{Superradiance in static black hole spacetimes}",
    eprint = "1511.02634",
    archivePrefix = "arXiv",
    primaryClass = "gr-qc",
    doi = "10.1103/PhysRevD.93.024028",
    journal = "Phys. Rev. D",
    volume = "93",
    number = "2",
    pages = "024028",
    year = "2016"
}

@article{Baake:2016oku,
    author = "Baake, Olaf and Rinne, Oliver",
    title = "{Superradiance of a charged scalar field coupled to the Einstein-Maxwell equations}",
    eprint = "1610.08352",
    archivePrefix = "arXiv",
    primaryClass = "gr-qc",
    doi = "10.1103/PhysRevD.94.124016",
    journal = "Phys. Rev. D",
    volume = "94",
    number = "12",
    pages = "124016",
    year = "2016"
}

@article{Teukolsky:1974yv,
    author = "Teukolsky, S. A. and Press, W. H.",
    title = "{Perturbations of a rotating black hole. III - Interaction of the hole with gravitational and electromagnet ic radiation}",
    doi = "10.1086/153180",
    journal = "Astrophys. J.",
    volume = "193",
    pages = "443--461",
    year = "1974"
}

@article{Danielsson:2017riq,
    author = "Danielsson, U. H. and Dibitetto, G. and Giri, S.",
    title = "{Black holes as bubbles of AdS}",
    eprint = "1705.10172",
    archivePrefix = "arXiv",
    primaryClass = "hep-th",
    reportNumber = "UUITP-15-17",
    doi = "10.1007/JHEP10(2017)171",
    journal = "JHEP",
    volume = "10",
    pages = "171",
    year = "2017"
}

@article{Cardoso:2005vk,
    author = "Cardoso, Vitor and Yoshida, Shijun",
    title = "{Superradiant instabilities of rotating black branes and strings}",
    eprint = "hep-th/0502206",
    archivePrefix = "arXiv",
    doi = "10.1088/1126-6708/2005/07/009",
    journal = "JHEP",
    volume = "07",
    pages = "009",
    year = "2005"
}

@article{Press:1972zz,
    author = "Press, William H. and Teukolsky, Saul A.",
    title = "{Floating Orbits, Superradiant Scattering and the Black-hole Bomb}",
    doi = "10.1038/238211a0",
    journal = "Nature",
    volume = "238",
    pages = "211--212",
    year = "1972"
}

@article{Witten:1981gj,
    author = "Witten, Edward",
    title = "{Instability of the Kaluza-Klein Vacuum}",
    reportNumber = "PRINT-81-0441 (PRINCETON)",
    doi = "10.1016/0550-3213(82)90007-4",
    journal = "Nucl. Phys. B",
    volume = "195",
    pages = "481--492",
    year = "1982"
}

@article{Dibitetto:2020csn,
    author = "Dibitetto, Giuseppe and Petri, Nicol\`o and Schillo, Marjorie",
    title = "{Nothing really matters}",
    eprint = "2002.01764",
    archivePrefix = "arXiv",
    primaryClass = "hep-th",
    doi = "10.1007/JHEP08(2020)040",
    journal = "JHEP",
    volume = "08",
    pages = "040",
    year = "2020"
}

@article{Bomans:2021ara,
    author = "Bomans, Pieter and Cassani, Davide and Dibitetto, Giuseppe and Petri, Nicolo",
    title = "{Bubble instability of mIIA on $\mathrm{AdS}_4\times S^6$}",
    eprint = "2110.08276",
    archivePrefix = "arXiv",
    primaryClass = "hep-th",
    doi = "10.21468/SciPostPhys.12.3.099",
    journal = "SciPost Phys.",
    volume = "12",
    number = "3",
    pages = "099",
    year = "2022"
}

@article{Delgado:2023uqk,
    author = "Delgado, Matilda",
    title = "{The bubble of nothing under T-duality}",
    eprint = "2312.09291",
    archivePrefix = "arXiv",
    primaryClass = "hep-th",
    reportNumber = "IFT-UAM/CSIC-23-146",
    doi = "10.1007/JHEP05(2024)333",
    journal = "JHEP",
    volume = "05",
    pages = "333",
    year = "2024"
}

@article{GarciaEtxebarria:2020xsr,
    author = "Garcia Etxebarria, Inaki and Montero, Miguel and Sousa, Kepa and Valenzuela, Irene",
    title = "{Nothing is certain in string compactifications}",
    eprint = "2005.06494",
    archivePrefix = "arXiv",
    primaryClass = "hep-th",
    doi = "10.1007/JHEP12(2020)032",
    journal = "JHEP",
    volume = "12",
    pages = "032",
    year = "2020"
}

@article{Ooguri:2017njy,
    author = "Ooguri, Hirosi and Spodyneiko, Lev",
    title = "{New Kaluza-Klein instantons and the decay of AdS vacua}",
    eprint = "1703.03105",
    archivePrefix = "arXiv",
    primaryClass = "hep-th",
    reportNumber = "CALT-TH-2017-014, IPMU-17-0041",
    doi = "10.1103/PhysRevD.96.026016",
    journal = "Phys. Rev. D",
    volume = "96",
    number = "2",
    pages = "026016",
    year = "2017"
}

@article{Callan:1985ia,
    author = "Callan, Jr., Curtis G. and Martinec, E. J. and Perry, M. J. and Friedan, D.",
    title = "{Strings in Background Fields}",
    reportNumber = "PRINT-85-0734 (PRINCETON)",
    doi = "10.1016/0550-3213(85)90506-1",
    journal = "Nucl. Phys. B",
    volume = "262",
    pages = "593--609",
    year = "1985"
}

@article{Martinec:2020cml,
    author = "Martinec, Emil J. and Warner, Nicholas P.",
    title = "{The Harder They Fall, the Bigger They Become: Tidal Trapping of Strings by Microstate Geometries}",
    eprint = "2009.07847",
    archivePrefix = "arXiv",
    primaryClass = "hep-th",
    doi = "10.1007/JHEP04(2021)259",
    journal = "JHEP",
    volume = "04",
    pages = "259",
    year = "2021"
}

@article{Ceplak:2021kgl,
    author = "Ceplak, Nejc and Hampton, Shaun and Li, Yixuan",
    title = "{Toroidal tidal effects in microstate geometries}",
    eprint = "2106.03841",
    archivePrefix = "arXiv",
    primaryClass = "hep-th",
    reportNumber = "JHEP03(2022)021",
    doi = "10.1007/JHEP03(2022)021",
    journal = "JHEP",
    volume = "03",
    pages = "021",
    year = "2022"
}

@misc{notebook,
  howpublished = "\url{https://gitlab.com/AleDS/hnn_fuzzball_geodesics}",
}

@article{Cipriani:2025ini,
    author = "Cipriani, Andrea and De Santis, Alessandro and Di Russo, Giorgio and Grillo, Alfredo and Tabarroni, Luca",
    title = "{Hamiltonian neural network approach to fuzzball geodesics}",
    eprint = "2502.20881",
    archivePrefix = "arXiv",
    primaryClass = "hep-th",
    doi = "10.1103/dssv-x49b",
    journal = "Phys. Rev. D",
    volume = "112",
    number = "2",
    pages = "026018",
    year = "2025"
}

@article{Lagaris_1998,
   title={Artificial neural networks for solving ordinary and partial differential equations},
   volume={9},
   ISSN={1045-9227},
   url={http://dx.doi.org/10.1109/72.712178},
   DOI={10.1109/72.712178},
   number={5},
   journal={IEEE Transactions on Neural Networks},
   publisher={Institute of Electrical and Electronics Engineers (IEEE)},
   author={Lagaris, I.E. and Likas, A. and Fotiadis, D.I.},
   year={1998},
   pages={987–1000} }

@article{Cole:2019enn,
    author = "Cole, Alex and Schachner, Andreas and Shiu, Gary",
    title = "{``Searching the Landscape of Flux Vacua with Genetic Algorithms"}",
    eprint = "1907.10072",
    archivePrefix = "arXiv",
    primaryClass = "hep-th",
    reportNumber = "MAD-TH-19-05",
    doi = "10.1007/JHEP11(2019)045",
    journal = "JHEP",
    volume = "11",
    pages = "045",
    year = "2019"
}

@inproceedings{Cole:2021nnt,
    author = "Cole, Alex and Krippendorf, Sven and Schachner, Andreas and Shiu, Gary",
    title = "{``Probing the Structure of String Theory Vacua with Genetic Algorithms and Reinforcement Learning"}",
    booktitle = "{35th Conference on Neural Information Processing Systems}",
    eprint = "2111.11466",
    archivePrefix = "arXiv",
    primaryClass = "hep-th",
    month = "11",
    year = "2021"
}

@article{MacFadden:2024him,
    author = "MacFadden, Nate and Schachner, Andreas and Sheridan, Elijah",
    title = "{``The DNA of Calabi-Yau Hypersurfaces"}",
    eprint = "2405.08871",
    archivePrefix = "arXiv",
    primaryClass = "hep-th",
    reportNumber = "LMU-ASC 06/24",
    month = "5",
    year = "2024"
}

@article{Lanza:2023vee,
    author = "Lanza, Stefano",
    title = "{``Machine learning the breakdown of tame effective theories"}",
    eprint = "2311.03437",
    archivePrefix = "arXiv",
    primaryClass = "hep-th",
    doi = "10.1140/epjc/s10052-024-12988-z",
    journal = "Eur. Phys. J. C",
    volume = "84",
    number = "6",
    pages = "631",
    year = "2024"
}

@article{Lanza:2024mqp,
    author = "Lanza, Stefano",
    title = "{``Neural network learning and Quantum Gravity"}",
    eprint = "2403.03245",
    archivePrefix = "arXiv",
    primaryClass = "hep-th",
    doi = "10.1007/JHEP07(2024)105",
    journal = "JHEP",
    volume = "07",
    pages = "105",
    year = "2024"
}

@article{Deen:2020dlf,
    author = "Deen, Rehan and He, Yang-Hui and Lee, Seung-Joo and Lukas, Andre",
    title = "{``Machine learning string standard models"}",
    eprint = "2003.13339",
    archivePrefix = "arXiv",
    primaryClass = "hep-th",
    reportNumber = "CERN-TH-2020-050, CTPU-PTC-20-06",
    doi = "10.1103/PhysRevD.105.046001",
    journal = "Phys. Rev. D",
    volume = "105",
    number = "4",
    pages = "046001",
    year = "2022"
}

@article{PhysRevD.111.086007,
  title = {Precision string phenomenology},
  author = {Berglund, Per and Butbaia, Giorgi and H\"ubsch, Tristan and Jejjala, Vishnu and Mayorga Pe\~na, Dami\'an and Mishra, Challenger and Tan, Justin},
  journal = {Phys. Rev. D},
  volume = {111},
  issue = {8},
  pages = {086007},
  numpages = {8},
  year = {2025},
  month = {Apr},
  publisher = {American Physical Society},
  doi = {10.1103/PhysRevD.111.086007},
  url = {https://link.aps.org/doi/10.1103/PhysRevD.111.086007}
}

@article{Ashmore:2019wzb,
    author = "Ashmore, Anthony and He, Yang-Hui and Ovrut, Burt A.",
    title = "{``Machine Learning Calabi\textendash{}Yau Metrics"}",
    eprint = "1910.08605",
    archivePrefix = "arXiv",
    primaryClass = "hep-th",
    doi = "10.1002/prop.202000068",
    journal = "Fortsch. Phys.",
    volume = "68",
    number = "9",
    pages = "2000068",
    year = "2020"
}

@article{Jejjala:2020wcc,
    author = "Jejjala, Vishnu and Mayorga Pena, Damian Kaloni and Mishra, Challenger",
    title = "{``Neural network approximations for Calabi-Yau metrics"}",
    eprint = "2012.15821",
    archivePrefix = "arXiv",
    primaryClass = "hep-th",
    doi = "10.1007/JHEP08(2022)105",
    journal = "JHEP",
    volume = "08",
    pages = "105",
    year = "2022"
}

@article{Larfors:2021pbb,
    author = "Larfors, Magdalena and Lukas, Andre and Ruehle, Fabian and Schneider, Robin",
    title = "{``Learning Size and Shape of Calabi-Yau Spaces"}",
    eprint = "2111.01436",
    archivePrefix = "arXiv",
    primaryClass = "hep-th",
    reportNumber = "UUITP-53/21",
    month = "11",
    year = "2021"
}

@article{Berglund:2022gvm,
    author = {Berglund, Per and Butbaia, Giorgi and H\"uubsch, Tristan and Jejjala, Vishnu and Mayorga Pe\~na, Dami\'an and Mishra, Challenger and Tan, Justin},
    title = "{``Machine-learned Calabi\textendash{}Yau metrics and curvature"}",
    eprint = "2211.09801",
    archivePrefix = "arXiv",
    primaryClass = "hep-th",
    doi = "10.4310/ATMP.2023.v27.n4.a3",
    journal = "Adv. Theor. Math. Phys.",
    volume = "27",
    number = "4",
    pages = "1107--1158",
    year = "2023"
}

@article{Berglund:2023ztk,
    author = "Berglund, Per and He, Yang-Hui and Heyes, Elli and Hirst, Edward and Jejjala, Vishnu and Lukas, Andre",
    title = "{``New Calabi\textendash{}Yau manifolds from genetic algorithms"}",
    eprint = "2306.06159",
    archivePrefix = "arXiv",
    primaryClass = "hep-th",
    reportNumber = "QMUL-PH-23-06",
    doi = "10.1016/j.physletb.2024.138504",
    journal = "Phys. Lett. B",
    volume = "850",
    pages = "138504",
    year = "2024"
}

@inproceedings{Anderson:2023viv,
    author = "Anderson, Lara B. and Gray, James and Larfors, Magdalena",
    title = "{``Lectures on Numerical and Machine Learning Methods for Approximating Ricci-flat Calabi-Yau Metrics"}",
    eprint = "2312.17125",
    archivePrefix = "arXiv",
    primaryClass = "hep-th",
    month = "12",
    year = "2023"
}

@article{Hendi:2024yin,
    author = "Hendi, Yacoub and Larfors, Magdalena and Walden, Moritz",
    title = "{Learning group invariant Calabi{\textendash}Yau metrics by fundamental domain projections}",
    eprint = "2407.06914",
    archivePrefix = "arXiv",
    primaryClass = "hep-th",
    reportNumber = "UUITP-21/24",
    doi = "10.1088/2632-2153/adb4bb",
    journal = "Mach. Learn. Sci. Tech.",
    volume = "6",
    number = "1",
    pages = "015050",
    year = "2025"
}

@article{Mirjanic:2024gek,
    author = "Mirjani\'c, Viktor and Mishra, Challenger",
    title = "{``Symbolic Approximations to Ricci-flat Metrics Via Extrinsic Symmetries of Calabi-Yau Hypersurfaces"}",
    eprint = "2412.19778",
    archivePrefix = "arXiv",
    primaryClass = "hep-th",
    month = "12",
    year = "2024"
}

@article{Seong:2023njx,
    author = "Seong, Rak-Kyeong",
    title = "{``Unsupervised machine learning techniques for exploring tropical coamoeba, brane tilings and Seiberg duality"}",
    eprint = "2309.05702",
    archivePrefix = "arXiv",
    primaryClass = "hep-th",
    reportNumber = "UNIST-MTH-23-RS-04",
    doi = "10.1103/PhysRevD.108.106009",
    journal = "Phys. Rev. D",
    volume = "108",
    number = "10",
    pages = "106009",
    year = "2023"
}

@article{Choi:2023rqg,
    author = "Choi, Eugene and Seong, Rak-Kyeong",
    title = "{``Machine learning regularization for the minimum volume formula of toric Calabi-Yau 3-folds"}",
    eprint = "2310.19276",
    archivePrefix = "arXiv",
    primaryClass = "hep-th",
    reportNumber = "UNIST-MTH-23-RS-05",
    doi = "10.1103/PhysRevD.109.046015",
    journal = "Phys. Rev. D",
    volume = "109",
    number = "4",
    pages = "046015",
    year = "2024"
}

@article{Seong:2024wkt,
  title = {Generative AI for brane configurations and coamoeba},
  author = {Seong, Rak-Kyeong},
  journal = {Phys. Rev. D},
  volume = {111},
  issue = {8},
  pages = {086013},
  numpages = {20},
  year = {2025},
  month = {Apr},
  publisher = {American Physical Society},
  doi = {10.1103/PhysRevD.111.086013},
  url = {https://link.aps.org/doi/10.1103/PhysRevD.111.086013}
}

@article{He:2024jao,
title = {Metaheuristic generation of brane tilings},
journal = {Physics Letters B},
volume = {862},
pages = {139365},
year = {2025},
issn = {0370-2693},
doi = {https://doi.org/10.1016/j.physletb.2025.139365},
url = {https://www.sciencedirect.com/science/article/pii/S037026932500125X},
author = {Yang-Hui He and Vishnu Jejjala and Tomás S.R. Silva}
}

@article{Hashemi:2024azx,
    author = "Hashemi, Baran and Corominas, Roderic G. and Giacchetto, Alessandro",
    title = "{``Can Transformers Do Enumerative Geometry?"}",
    eprint = "2408.14915",
    archivePrefix = "arXiv",
    primaryClass = "cs.LG",
    month = "8",
    year = "2024"
}

@article{Cornell:2022enn,
    author = "Cornell, Alan S. and Ncube, Anele and Harmsen, Gerhard",
    title = "{``Using physics-informed neural networks to compute quasinormal modes"}",
    eprint = "2205.08284",
    archivePrefix = "arXiv",
    primaryClass = "physics.comp-ph",
    doi = "10.1103/PhysRevD.106.124047",
    journal = "Phys. Rev. D",
    volume = "106",
    number = "12",
    pages = "124047",
    year = "2022"
}

@article{Jain:2022dsh,
    author = "Jain, Aarjav and Mishra, Challenger and Li\`o, Pietro",
    title = "{``A physics-informed search for metric solutions to Ricci flow, their embeddings, and visualisation"}",
    eprint = "2212.05892",
    archivePrefix = "arXiv",
    primaryClass = "gr-qc",
    month = "11",
    year = "2022"
}

@article{Luna:2022rql,
    author = "Luna, Raimon and Calder\'on Bustillo, Juan and Martinez, Juan Jos\'e Seoane and Torres-Forn\'e, Alejandro and Font, Jos\'e A.",
    title = "{``Solving the Teukolsky equation with physics-informed neural networks"}",
    eprint = "2212.06103",
    archivePrefix = "arXiv",
    primaryClass = "gr-qc",
    doi = "10.1103/PhysRevD.107.064025",
    journal = "Phys. Rev. D",
    volume = "107",
    number = "6",
    pages = "064025",
    year = "2023"
}

@article{Luna:2024spo,
    author = "Luna, Raimon and Doneva, Daniela D. and Font, Jos\'e A. and Lien, Jr-Hua and Yazadjiev, Stoytcho S.",
    title = "{``Quasinormal modes in modified gravity using physics-informed neural networks"}",
    eprint = "2404.11583",
    archivePrefix = "arXiv",
    primaryClass = "gr-qc",
    doi = "10.1103/PhysRevD.109.124064",
    journal = "Phys. Rev. D",
    volume = "109",
    number = "12",
    pages = "124064",
    year = "2024"
}

@article{Geng:2024zzw,
    author = "Geng, Ru and Gao, Yixian and Zu, Jian and Zhang, Hong-Kun",
    title = "{``Graph Attention Hamiltonian Neural Networks: A Lattice System Analysis Model Based on Structural Learning"}",
    eprint = "2412.10821",
    archivePrefix = "arXiv",
    primaryClass = "hep-lat",
    month = "12",
    year = "2024"
}

@ARTICLE{HNN,
       author = {{Greydanus}, Sam and {Dzamba}, Misko and {Yosinski}, Jason},
        title = "{``Hamiltonian Neural Networks"}",
      journal = {arXiv e-prints},
         year = 2019,
        month = jun,
          eid = {arXiv:1906.01563},
        pages = {arXiv:1906.01563},
          doi = {10.48550/arXiv.1906.01563},
archivePrefix = {arXiv},
       eprint = {1906.01563},
 primaryClass = {cs.NE},
       adsurl = {https://ui.adsabs.harvard.edu/abs/2019arXiv190601563G}
}

@article{Harvard,
   title="{``Hamiltonian neural networks for solving equations of motion"}",
   volume="105",
   ISSN={2470-0053},
   url={http://dx.doi.org/10.1103/PhysRevE.105.065305},
   DOI={10.1103/physreve.105.065305},
   number={6},
   journal="Physical Review E",
   publisher={American Physical Society (APS)},
   author={Mattheakis, Marios and Sondak, David and Dogra, Akshunna S. and Protopapas, Pavlos},
   year={2022},
   month=jun }

@misc{solve_ivp,
  title = {$solve\_ivp$ by {SciPy}},
  howpublished = {\url{https://docs.scipy.org/doc/scipy/reference/generated/scipy.integrate.solve_ivp.html}},
}

@misc{keras,
  title="{``Keras"}",
  author={Chollet, Francois and others},
  year={2015},
  howpublished={\url{https://keras.io}},
}

@article{hornik1989multilayer,
  title="{``Multilayer feedforward networks are universal approximators"}",
  author={Hornik, Kurt and Stinchcombe, Maxwell and White, Halbert},
  journal= {Neural networks},
  volume={2},
  number={5},
  pages={359--366},
  year={1989},
  publisher={Elsevier}
}

@book{Goodfellow-et-al-2016,
    title="{``Deep Learning"}",
    author={Ian Goodfellow and Yoshua Bengio and Aaron Courville},
    publisher={MIT Press},
    note={\url{http://www.deeplearningbook.org}},
    year={2016}
}

@misc{tensorflow,
title="{`` {TensorFlow}: Large-Scale Machine Learning on Heterogeneous Systems"}",
url={https://www.tensorflow.org/},
note={Software available from tensorflow.org},
author={
    Martin~Abadi and
    Ashish~Agarwal and
    Paul~Barham and
    Eugene~Brevdo and
    Zhifeng~Chen and
    Craig~Citro and
    Greg~S.~Corrado and
    Andy~Davis and
    Jeffrey~Dean and
    Matthieu~Devin and
    Sanjay~Ghemawat and
    Ian~Goodfellow and
    Andrew~Harp and
    Geoffrey~Irving and
    Michael~Isard and
    Yangqing Jia and
    Rafal~Jozefowicz and
    Lukasz~Kaiser and
    Manjunath~Kudlur and
    Josh~Levenberg and
    Dandelion~Man\'{e} and
    Rajat~Monga and
    Sherry~Moore and
    Derek~Murray and
    Chris~Olah and
    Mike~Schuster and
    Jonathon~Shlens and
    Benoit~Steiner and
    Ilya~Sutskever and
    Kunal~Talwar and
    Paul~Tucker and
    Vincent~Vanhoucke and
    Vijay~Vasudevan and
    Fernanda~Vi\'{e}gas and
    Oriol~Vinyals and
    Pete~Warden and
    Martin~Wattenberg and
    Martin~Wicke and
    Yuan~Yu and
    Xiaoqiang~Zheng},
  year={2015},
}

@misc{Adam,
      title="{``Adam: A Method for Stochastic Optimization"}", 
      author={Diederik P. Kingma and Jimmy Ba},
      year={2017},
      eprint={1412.6980},
      archivePrefix={arXiv},
      primaryClass={cs.LG},
      url={https://arxiv.org/abs/1412.6980}, 
}

@misc{deveney:2021,
      title="{``A deep surrogate approach to efficient Bayesian inversion in PDE and integral equation models"}", 
      author={Teo Deveney and Eike Mueller and Tony Shardlow},
      year={2021},
      eprint={1910.01547},
      archivePrefix={arXiv},
      primaryClass={math.NA},
      url={https://arxiv.org/abs/1910.01547}, 
}

@ARTICLE{raissi:2017,
       author = {{Raissi}, Maziar and {Perdikaris}, Paris and {Karniadakis}, George Em},
        title = "{``Physics Informed Deep Learning (Part I): Data-driven Solutions of Nonlinear Partial Differential Equations"}",
      journal = {arXiv e-prints},
         year = 2017,
        month = nov,
          eid = {arXiv:1711.10561},
        pages = {arXiv:1711.10561},
          doi = {10.48550/arXiv.1711.10561},
archivePrefix = {arXiv},
       eprint = {1711.10561},
 primaryClass = {cs.AI},
       adsurl = {https://ui.adsabs.harvard.edu/abs/2017arXiv171110561R}
}

@article{Louppe:2017ipp,
    author = "Louppe, Gilles and Cho, Kyunghyun and Becot, Cyril and Cranmer, Kyle",
    title = "{``QCD-Aware Recursive Neural Networks for Jet Physics"}",
    eprint = "1702.00748",
    archivePrefix = "arXiv",
    primaryClass = "hep-ph",
    doi = "10.1007/JHEP01(2019)057",
    journal = "JHEP",
    volume = "01",
    pages = "057",
    year = "2019"
}

@ARTICLE{Wu:2022,
       author = {{Wu}, Benjamin and {Hennigh}, Oliver and {Kautz}, Jan and {Choudhry}, Sanjay and {Byeon}, Wonmin},
        title = "{``Physics Informed RNN-DCT Networks for Time-Dependent Partial Differential Equations"}",
      journal = {arXiv e-prints},
         year = 2022,
        month = feb,
          eid = {arXiv:2202.12358},
        pages = {arXiv:2202.12358},
          doi = {10.48550/arXiv.2202.12358},
archivePrefix = {arXiv},
       eprint = {2202.12358},
 primaryClass = {physics.comp-ph},
       adsurl = {https://ui.adsabs.harvard.edu/abs/2022arXiv220212358W}
}

@ARTICLE{Nabian:2020,
       author = {{Nabian}, Mohammad Amin and {Meidani}, Hadi},
        title = "{``Adaptive Physics-Informed Neural Networks for Markov-Chain Monte Carlo"}",
      journal = {arXiv e-prints},
         year = 2020,
        month = aug,
          eid = {arXiv:2008.01604},
        pages = {arXiv:2008.01604},
          doi = {10.48550/arXiv.2008.01604},
archivePrefix = {arXiv},
       eprint = {2008.01604},
 primaryClass = {cs.LG},
       adsurl = {https://ui.adsabs.harvard.edu/abs/2020arXiv200801604N}
}

@ARTICLE{Crammer:2020,
       author = {{Cranmer}, Miles and {Greydanus}, Sam and {Hoyer}, Stephan and {Battaglia}, Peter and {Spergel}, David and {Ho}, Shirley},
        title = "{``Lagrangian Neural Networks"}",
      journal = {arXiv e-prints},
         year = 2020,
        month = mar,
          eid = {arXiv:2003.04630},
        pages = {arXiv:2003.04630},
          doi = {10.48550/arXiv.2003.04630},
archivePrefix = {arXiv},
       eprint = {2003.04630},
 primaryClass = {cs.LG},
       adsurl = {https://ui.adsabs.harvard.edu/abs/2020arXiv200304630C}
}

@ARTICLE{DiHan:2021,
       author = {{Han}, Chen-Di and {Glaz}, Bryan and {Haile}, Mulugeta and {Lai}, Ying-Cheng},
        title = "{``Adaptable Hamiltonian neural networks"}",
      journal = {Physical Review Research},
         year = 2021,
        month = may,
       volume = {3},
       number = {2},
          eid = {023156},
        pages = {023156},
          doi = {10.1103/PhysRevResearch.3.023156},
archivePrefix = {arXiv},
       eprint = {2102.13235},
 primaryClass = {cs.LG},
       adsurl = {https://ui.adsabs.harvard.edu/abs/2021PhRvR...3b3156H}
}

@article{Cipriani:2025ikx,
    author = "Cipriani, Andrea and Di Russo, Giorgio and Fucito, Francesco and Morales, Jos{\'e} Francisco and Poghosyan, Hasmik and Poghossian, Rubik",
    title = "{Resumming post-Minkowskian and post-Newtonian gravitational waveform expansions}",
    eprint = "2501.19257",
    archivePrefix = "arXiv",
    primaryClass = "gr-qc",
    doi = "10.21468/SciPostPhys.19.2.057",
    journal = "SciPost Phys.",
    volume = "19",
    number = "2",
    pages = "057",
    year = "2025"
}

@article{Cipriani:2026xmx,
    author = "Cipriani, Andrea and Nagar, Alessandro and Fucito, Francesco and Morales, Jos{\'e} Francisco",
    title = "{From the confluent Heun equation to a new factorized and resummed gravitational waveform for circularized, nonspinning, compact binaries}",
    eprint = "2602.08833",
    archivePrefix = "arXiv",
    primaryClass = "gr-qc",
    doi = "10.1103/s4tk-z574",
    journal = "Phys. Rev. D",
    volume = "114",
    number = "2",
    pages = "024015",
    year = "2026"
}

@book{Bender:1999box,
    author = "Bender, Carl M. and Orszag, Steven A.",
    title = "{Advanced Mathematical Methods for Scientists and Engineers I}",
    doi = "10.1007/978-1-4757-3069-2",
    publisher = "Springer",
    year = "1999"
}

@article{Pan:2010hz,
    author = "Pan, Yi and Buonanno, Alessandra and Fujita, Ryuichi and Racine, Etienne and Tagoshi, Hideyuki",
    title = "{Post-Newtonian factorized multipolar waveforms for spinning, non-precessing black-hole binaries}",
    eprint = "1006.0431",
    archivePrefix = "arXiv",
    primaryClass = "gr-qc",
    doi = "10.1103/PhysRevD.83.064003",
    journal = "Phys. Rev. D",
    volume = "83",
    pages = "064003",
    year = "2011",
    note = "[Erratum: Phys.Rev.D 87, 109901 (2013)]"
}

@article{Blanchet:2022vsm,
    author = "Blanchet, Luc and Faye, Guillaume and Larrouturou, Francois",
    title = "{The quadrupole moment of compact binaries to the fourth post-Newtonian order: from source to canonical moment}",
    eprint = "2204.11293",
    archivePrefix = "arXiv",
    primaryClass = "gr-qc",
    doi = "10.1088/1361-6382/ac840c",
    journal = "Class. Quant. Grav.",
    volume = "39",
    number = "19",
    pages = "195003",
    year = "2022"
}

@article{Blanchet:2023sbv,
    author = "Blanchet, Luc and Faye, Guillaume and Henry, Quentin and Larrouturou, Francois and Trestini, David",
    title = "{Gravitational-wave flux and quadrupole modes from quasicircular nonspinning compact binaries to the fourth post-Newtonian order}",
    eprint = "2304.11186",
    archivePrefix = "arXiv",
    primaryClass = "gr-qc",
    reportNumber = "DESY-23-044",
    doi = "10.1103/PhysRevD.108.064041",
    journal = "Phys. Rev. D",
    volume = "108",
    number = "6",
    pages = "064041",
    year = "2023"
}

@article{Blanchet:2023bwj,
    author = "Blanchet, Luc and Faye, Guillaume and Henry, Quentin and Larrouturou, Francois and Trestini, David",
    title = "{Gravitational-Wave Phasing of Quasicircular Compact Binary Systems to the Fourth-and-a-Half Post-Newtonian Order}",
    eprint = "2304.11185",
    archivePrefix = "arXiv",
    primaryClass = "gr-qc",
    reportNumber = "DESY-23-043",
    doi = "10.1103/PhysRevLett.131.121402",
    journal = "Phys. Rev. Lett.",
    volume = "131",
    number = "12",
    pages = "121402",
    year = "2023"
}

@article{Nagar:2024oyk,
    author = "Nagar, Alessandro and Chiaramello, Danilo and Gamba, Rossella and Albanesi, Simone and Bernuzzi, Sebastiano and Fantini, Veronica and Panzeri, Mattia and Rettegno, Piero",
    title = "{Effective-one-body waveform model for noncircularized, planar, coalescing black hole binaries. II. High accuracy by improving logarithmic terms in resummations}",
    eprint = "2407.04762",
    archivePrefix = "arXiv",
    primaryClass = "gr-qc",
    doi = "10.1103/PhysRevD.111.064050",
    journal = "Phys. Rev. D",
    volume = "111",
    number = "6",
    pages = "064050",
    year = "2025"
}

@article{Faye:2014fra,
    author = "Faye, Guillaume and Blanchet, Luc and Iyer, Bala R.",
    title = "{Non-linear multipole interactions and gravitational-wave octupole modes for inspiralling compact binaries to third-and-a-half post-Newtonian order}",
    eprint = "1409.3546",
    archivePrefix = "arXiv",
    primaryClass = "gr-qc",
    doi = "10.1088/0264-9381/32/4/045016",
    journal = "Class. Quant. Grav.",
    volume = "32",
    number = "4",
    pages = "045016",
    year = "2015"
}

@book{Maggiore:2018sht,
    author = "Maggiore, Michele",
    title = "{Gravitational Waves. Vol. 2: Astrophysics and Cosmology}",
    isbn = "978-0-19-857089-9",
    publisher = "Oxford University Press",
    month = "3",
    year = "2018"
}

@article{Ivanov:2025ozg,
    author = "Ivanov, Mikhail M. and Li, Yue-Zhou and Parra-Martinez, Julio and Zhou, Zihan",
    title = "{Resummation of Universal Tails in Gravitational Waveforms}",
    eprint = "2504.07862",
    archivePrefix = "arXiv",
    primaryClass = "hep-th",
    reportNumber = "MIT-CTP/5863",
    doi = "10.1103/jzd1-qzkt",
    journal = "Phys. Rev. Lett.",
    volume = "135",
    number = "14",
    pages = "141401",
    year = "2025"
}

@article{LISA:2022kgy,
    author = "Arun, K. G. and others",
    collaboration = "LISA",
    title = "{New horizons for fundamental physics with LISA}",
    eprint = "2205.01597",
    archivePrefix = "arXiv",
    primaryClass = "gr-qc",
    doi = "10.1007/s41114-022-00036-9",
    journal = "Living Rev. Rel.",
    volume = "25",
    number = "1",
    pages = "4",
    year = "2022"
}

@article{Gair:2012nm,
    author = "Gair, Jonathan R. and Vallisneri, Michele and Larson, Shane L. and Baker, John G.",
    title = "{Testing General Relativity with Low-Frequency, Space-Based Gravitational-Wave Detectors}",
    eprint = "1212.5575",
    archivePrefix = "arXiv",
    primaryClass = "gr-qc",
    doi = "10.12942/lrr-2013-7",
    journal = "Living Rev. Rel.",
    volume = "16",
    pages = "7",
    year = "2013"
}

@article{Yunes:2013dva,
    author = "Yunes, Nicol{\'a}s and Siemens, Xavier",
    title = "{Gravitational-Wave Tests of General Relativity with Ground-Based Detectors and Pulsar Timing-Arrays}",
    eprint = "1304.3473",
    archivePrefix = "arXiv",
    primaryClass = "gr-qc",
    doi = "10.12942/lrr-2013-9",
    journal = "Living Rev. Rel.",
    volume = "16",
    pages = "9",
    year = "2013"
}

@article{RevModPhys.52.299,
  title = {Multipole expansions of gravitational radiation},
  author = {Thorne, Kip S.},
  journal = {Rev. Mod. Phys.},
  volume = {52},
  issue = {2},
  pages = {299--339},
  numpages = {0},
  year = {1980},
  month = {Apr},
  publisher = {American Physical Society},
  doi = {10.1103/RevModPhys.52.299},
  url = {https://link.aps.org/doi/10.1103/RevModPhys.52.299}
}

@inproceedings{Damour:1982wm,
    author = "Damour, Thibault",
    title = "{GRAVITATIONAL RADIATION AND THE MOTION OF COMPACT BODIES}",
    booktitle = "{Les Houches Summer School on Gravitational Radiation}",
    reportNumber = "PRINT-82-0836 (MEUDON)",
    month = "8",
    year = "1982"
}

@article{Blanchet:2013haa,
    author = "Blanchet, Luc",
    title = "{Post-Newtonian Theory for Gravitational Waves}",
    eprint = "1310.1528",
    archivePrefix = "arXiv",
    primaryClass = "gr-qc",
    doi = "10.12942/lrr-2014-2",
    journal = "Living Rev. Rel.",
    volume = "17",
    pages = "2",
    year = "2014"
}

@ARTICLE{1973ApJ...185..635T,
       author = {{Teukolsky}, Saul A.},
        title = "{Perturbations of a Rotating Black Hole. I. Fundamental Equations for Gravitational, Electromagnetic, and Neutrino-Field Perturbations}",
      journal = {Astrophysical Journal},
         year = 1973,
        month = oct,
       volume = {185},
        pages = {635-648},
          doi = {10.1086/152444},
       adsurl = {https://ui.adsabs.harvard.edu/abs/1973ApJ...185..635T}
}

@article{PhysRevD.47.1497,
  title = {Gravitational radiation from a particle in circular orbit around a black hole. I. Analytical results for the nonrotating case},
  author = {Poisson, Eric},
  journal = {Phys. Rev. D},
  volume = {47},
  issue = {4},
  pages = {1497--1510},
  numpages = {0},
  year = {1993},
  month = {Feb},
  publisher = {American Physical Society},
  doi = {10.1103/PhysRevD.47.1497},
  url = {https://link.aps.org/doi/10.1103/PhysRevD.47.1497}
}

@article{PhysRevD.49.4016,
  title = {Gravitational waves from a point particle in circular orbit around a black hole: Logarithmic terms in the post-Newtonian expansion},
  author = {Tagoshi, Hideyuki and Nakamura, Takashi},
  journal = {Phys. Rev. D},
  volume = {49},
  issue = {8},
  pages = {4016--4022},
  numpages = {0},
  year = {1994},
  month = {Apr},
  publisher = {American Physical Society},
  doi = {10.1103/PhysRevD.49.4016},
  url = {https://link.aps.org/doi/10.1103/PhysRevD.49.4016}
}

@article{Poisson:1994yf,
    author = "Poisson, Eric and Sasaki, Misao",
    title = "{Gravitational radiation from a particle in circular orbit around a black hole. 5: Black hole absorption and tail corrections}",
    eprint = "gr-qc/9412027",
    archivePrefix = "arXiv",
    doi = "10.1103/PhysRevD.51.5753",
    journal = "Phys. Rev. D",
    volume = "51",
    pages = "5753--5767",
    year = "1995"
}

@article{Shibata:1994jx,
    author = "Shibata, Masaru and Sasaki, Misao and Tagoshi, Hideyuki and Tanaka, Takahiro",
    title = "{Gravitational waves from a particle orbiting around a rotating black hole: PostNewtonian expansion}",
    eprint = "gr-qc/9409054",
    archivePrefix = "arXiv",
    reportNumber = "OU-TAP-9, KUNS-1268",
    doi = "10.1103/PhysRevD.51.1646",
    journal = "Phys. Rev. D",
    volume = "51",
    pages = "1646--1663",
    year = "1995"
}

@article{Tagoshi:1994sm,
    author = "Tagoshi, Hideyuki and Sasaki, Misao",
    title = "{PostNewtonian expansion of gravitational waves from a particle in circular orbit around a Schwarzschild black hole}",
    eprint = "gr-qc/9405062",
    archivePrefix = "arXiv",
    reportNumber = "KUNS-1262",
    doi = "10.1143/PTP.92.745",
    journal = "Prog. Theor. Phys.",
    volume = "92",
    pages = "745--772",
    year = "1994"
}

@article{Tanaka:1996lfd,
    author = "Tanaka, Takahiro and Tagoshi, Hideyuki and Sasaki, Misao",
    title = "{Gravitational waves by a particle in circular orbits around a Schwarzschild black hole: 5.5 postNewtonian formula}",
    eprint = "gr-qc/9701050",
    archivePrefix = "arXiv",
    reportNumber = "OU-TAP-44, NAOJ-TH-AP-1996-3",
    doi = "10.1143/PTP.96.1087",
    journal = "Prog. Theor. Phys.",
    volume = "96",
    pages = "1087--1101",
    year = "1996"
}

@article{Mano:1996vt,
    author = "Mano, Shuhei and Suzuki, Hisao and Takasugi, Eiichi",
    title = "{Analytic solutions of the Teukolsky equation and their low frequency expansions}",
    eprint = "gr-qc/9603020",
    archivePrefix = "arXiv",
    reportNumber = "OU-HET-238",
    doi = "10.1143/PTP.95.1079",
    journal = "Prog. Theor. Phys.",
    volume = "95",
    pages = "1079--1096",
    year = "1996"
}

@article{Mino:1997bx,
    author = "Mino, Yasushi and Sasaki, Misao and Shibata, Masaru and Tagoshi, Hideyuki and Tanaka, Takahiro",
    title = "{Black hole perturbation: Chapter 1}",
    eprint = "gr-qc/9712057",
    archivePrefix = "arXiv",
    doi = "10.1143/PTPS.128.1",
    journal = "Prog. Theor. Phys. Suppl.",
    volume = "128",
    pages = "1--121",
    year = "1997"
}

@article{Suzuki:1998vy,
    author = "Suzuki, Hisao and Takasugi, Eiichi and Umetsu, Hiroshi",
    title = "{Perturbations of Kerr-de Sitter black hole and Heun's equations}",
    eprint = "gr-qc/9805064",
    archivePrefix = "arXiv",
    reportNumber = "EPHOU-98-005, OU-HET-296",
    doi = "10.1143/PTP.100.491",
    journal = "Prog. Theor. Phys.",
    volume = "100",
    pages = "491--505",
    year = "1998"
}

@article{Fujita:2010xj,
    author = "Fujita, Ryuichi and Iyer, Bala R.",
    title = "{Spherical harmonic modes of 5.5 post-Newtonian gravitational wave polarisations and associated factorised resummed waveforms for a particle in circular orbit around a Schwarzschild black hole}",
    eprint = "1005.2266",
    archivePrefix = "arXiv",
    primaryClass = "gr-qc",
    doi = "10.1103/PhysRevD.82.044051",
    journal = "Phys. Rev. D",
    volume = "82",
    pages = "044051",
    year = "2010"
}

@article{Fujita:2011zk,
    author = "Fujita, Ryuichi",
    title = "{Gravitational radiation for extreme mass ratio inspirals to the 14th post-Newtonian order}",
    eprint = "1104.5615",
    archivePrefix = "arXiv",
    primaryClass = "gr-qc",
    doi = "10.1143/PTP.127.583",
    journal = "Prog. Theor. Phys.",
    volume = "127",
    pages = "583--590",
    year = "2012"
}

@article{Fujita:2012cm,
    author = "Fujita, Ryuichi",
    title = "{Gravitational Waves from a Particle in Circular Orbits around a Schwarzschild Black Hole to the 22nd Post-Newtonian Order}",
    eprint = "1211.5535",
    archivePrefix = "arXiv",
    primaryClass = "gr-qc",
    doi = "10.1143/PTP.128.971",
    journal = "Prog. Theor. Phys.",
    volume = "128",
    pages = "971--992",
    year = "2012"
}

@article{PhysRevD.47.1511,
  title = {Gravitational radiation from a particle in circular orbit around a black hole. II. Numerical results for the nonrotating case},
  author = {Cutler, Curt and Finn, Lee Samuel and Poisson, Eric and Sussman, Gerald Jay},
  journal = {Phys. Rev. D},
  volume = {47},
  issue = {4},
  pages = {1511--1518},
  numpages = {0},
  year = {1993},
  month = {Feb},
  publisher = {American Physical Society},
  doi = {10.1103/PhysRevD.47.1511},
  url = {https://link.aps.org/doi/10.1103/PhysRevD.47.1511}
}

@article{Damour:2008gu,
    author = "Damour, Thibault and Iyer, Bala R. and Nagar, Alessandro",
    title = "{Improved resummation of post-Newtonian multipolar waveforms from circularized compact binaries}",
    eprint = "0811.2069",
    archivePrefix = "arXiv",
    primaryClass = "gr-qc",
    doi = "10.1103/PhysRevD.79.064004",
    journal = "Phys. Rev. D",
    volume = "79",
    pages = "064004",
    year = "2009"
}

@article{Damour:2009kr,
    author = "Damour, Thibault and Nagar, Alessandro",
    title = "{An Improved analytical description of inspiralling and coalescing black-hole binaries}",
    eprint = "0902.0136",
    archivePrefix = "arXiv",
    primaryClass = "gr-qc",
    doi = "10.1103/PhysRevD.79.081503",
    journal = "Phys. Rev. D",
    volume = "79",
    pages = "081503",
    year = "2009"
}

@article{Buonanno:2009zt,
    author = "Buonanno, Alessandra and Iyer, Bala and Ochsner, Evan and Pan, Yi and Sathyaprakash, B. S.",
    title = "{Comparison of post-Newtonian templates for compact binary inspiral signals in gravitational-wave detectors}",
    eprint = "0907.0700",
    archivePrefix = "arXiv",
    primaryClass = "gr-qc",
    doi = "10.1103/PhysRevD.80.084043",
    journal = "Phys. Rev. D",
    volume = "80",
    pages = "084043",
    year = "2009"
}

@article{Damour:2012ky,
    author = "Damour, Thibault and Nagar, Alessandro and Bernuzzi, Sebastiano",
    title = "{Improved effective-one-body description of coalescing nonspinning black-hole binaries and its numerical-relativity completion}",
    eprint = "1212.4357",
    archivePrefix = "arXiv",
    primaryClass = "gr-qc",
    doi = "10.1103/PhysRevD.87.084035",
    journal = "Phys. Rev. D",
    volume = "87",
    number = "8",
    pages = "084035",
    year = "2013"
}

@article{Bautista:2023sdf,
    author = "Bautista, Yilber Fabian and Bonelli, Giulio and Iossa, Cristoforo and Tanzini, Alessandro and Zhou, Zihan",
    title = "{Black hole perturbation theory meets CFT2: Kerr-Compton amplitudes from Nekrasov-Shatashvili functions}",
    eprint = "2312.05965",
    archivePrefix = "arXiv",
    primaryClass = "hep-th",
    doi = "10.1103/PhysRevD.109.084071",
    journal = "Phys. Rev. D",
    volume = "109",
    number = "8",
    pages = "084071",
    year = "2024"
}

@article{Aminov:2023jve,
    author = "Aminov, Gleb and Arnaudo, Paolo and Bonelli, Giulio and Grassi, Alba and Tanzini, Alessandro",
    title = "{Black hole perturbation theory and multiple polylogarithms}",
    eprint = "2307.10141",
    archivePrefix = "arXiv",
    primaryClass = "hep-th",
    reportNumber = "CERN-TH-2023-110",
    doi = "10.1007/JHEP11(2023)059",
    journal = "JHEP",
    volume = "11",
    pages = "059",
    year = "2023"
}

@article{Fucito:2023afe,
    author = "Fucito, Francesco and Morales, Jose Francisco",
    title = "{Post Newtonian emission of gravitational waves from binary systems: a gauge theory perspective}",
    eprint = "2311.14637",
    archivePrefix = "arXiv",
    primaryClass = "gr-qc",
    doi = "10.1007/JHEP03(2024)106",
    journal = "JHEP",
    volume = "03",
    pages = "106",
    year = "2024"
}

@article{Bini:2023fiz,
    author = "Bini, Donato and Damour, Thibault and Geralico, Andrea",
    title = "{Comparing one-loop gravitational bremsstrahlung amplitudes to the multipolar-post-Minkowskian waveform}",
    eprint = "2309.14925",
    archivePrefix = "arXiv",
    primaryClass = "gr-qc",
    doi = "10.1103/PhysRevD.108.124052",
    journal = "Phys. Rev. D",
    volume = "108",
    number = "12",
    pages = "124052",
    year = "2023"
}

@article{Bini:2024ijq,
    author = "Bini, Donato and Damour, Thibault and Geralico, Andrea",
    title = "{Gravitational bremsstrahlung waveform at the fourth post-Minkowskian order and the second post-Newtonian level}",
    eprint = "2407.02076",
    archivePrefix = "arXiv",
    primaryClass = "gr-qc",
    doi = "10.1103/PhysRevD.110.064035",
    journal = "Phys. Rev. D",
    volume = "110",
    number = "6",
    pages = "064035",
    year = "2024"
}

@article{Bini:2024rsy,
    author = "Bini, Donato and Damour, Thibault and De Angelis, Stefano and Geralico, Andrea and Herderschee, Aidan and Roiban, Radu and Teng, Fei",
    title = "{Gravitational waveforms: A tale of two formalisms}",
    eprint = "2402.06604",
    archivePrefix = "arXiv",
    primaryClass = "hep-th",
    doi = "10.1103/PhysRevD.109.125008",
    journal = "Phys. Rev. D",
    volume = "109",
    number = "12",
    pages = "125008",
    year = "2024"
}

@article{Jakobsen:2021smu,
    author = "Jakobsen, Gustav Uhre and Mogull, Gustav and Plefka, Jan and Steinhoff, Jan",
    title = "{Classical Gravitational Bremsstrahlung from a Worldline Quantum Field Theory}",
    eprint = "2101.12688",
    archivePrefix = "arXiv",
    primaryClass = "gr-qc",
    reportNumber = "HU-EP-21/03-RTG",
    doi = "10.1103/PhysRevLett.126.201103",
    journal = "Phys. Rev. Lett.",
    volume = "126",
    number = "20",
    pages = "201103",
    year = "2021"
}

@article{Mougiakakos:2021ckm,
    author = "Mougiakakos, Stavros and Riva, Massimiliano Maria and Vernizzi, Filippo",
    title = "{Gravitational Bremsstrahlung in the post-Minkowskian effective field theory}",
    eprint = "2102.08339",
    archivePrefix = "arXiv",
    primaryClass = "gr-qc",
    doi = "10.1103/PhysRevD.104.024041",
    journal = "Phys. Rev. D",
    volume = "104",
    number = "2",
    pages = "024041",
    year = "2021"
}

@article{Cristofoli:2021vyo,
    author = "Cristofoli, Andrea and Gonzo, Riccardo and Kosower, David A. and O'Connell, Donal",
    title = "{Waveforms from amplitudes}",
    eprint = "2107.10193",
    archivePrefix = "arXiv",
    primaryClass = "hep-th",
    doi = "10.1103/PhysRevD.106.056007",
    journal = "Phys. Rev. D",
    volume = "106",
    number = "5",
    pages = "056007",
    year = "2022"
}

@article{DiVecchia:2023frv,
    author = "Di Vecchia, Paolo and Heissenberg, Carlo and Russo, Rodolfo and Veneziano, Gabriele",
    title = "{The gravitational eikonal: From particle, string and brane collisions to black-hole encounters}",
    eprint = "2306.16488",
    archivePrefix = "arXiv",
    primaryClass = "hep-th",
    reportNumber = "CERN-TH-2023-108, NORDITA 2023-026, QMUL-PH-23-09, UUITP-14/23",
    doi = "10.1016/j.physrep.2024.06.002",
    journal = "Phys. Rept.",
    volume = "1083",
    pages = "1--169",
    year = "2024"
}

@article{Brandhuber:2023hhy,
    author = "Brandhuber, Andreas and Brown, Graham R. and Chen, Gang and De Angelis, Stefano and Gowdy, Joshua and Travaglini, Gabriele",
    title = "{One-loop gravitational bremsstrahlung and waveforms from a heavy-mass effective field theory}",
    eprint = "2303.06111",
    archivePrefix = "arXiv",
    primaryClass = "hep-th",
    reportNumber = "QMUL-22-28,SAGEX-22-32-E",
    doi = "10.1007/JHEP06(2023)048",
    journal = "JHEP",
    volume = "06",
    pages = "048",
    year = "2023"
}

@article{Herderschee:2023fxh,
    author = "Herderschee, Aidan and Roiban, Radu and Teng, Fei",
    title = "{The sub-leading scattering waveform from amplitudes}",
    eprint = "2303.06112",
    archivePrefix = "arXiv",
    primaryClass = "hep-th",
    reportNumber = "LCTP-23-04",
    doi = "10.1007/JHEP06(2023)004",
    journal = "JHEP",
    volume = "06",
    pages = "004",
    year = "2023"
}

@article{Elkhidir:2023dco,
    author = "Elkhidir, Asaad and O'Connell, Donal and Sergola, Matteo and Vazquez-Holm, Ingrid A.",
    title = "{Radiation and reaction at one loop}",
    eprint = "2303.06211",
    archivePrefix = "arXiv",
    primaryClass = "hep-th",
    doi = "10.1007/JHEP07(2024)272",
    journal = "JHEP",
    volume = "07",
    pages = "272",
    year = "2024"
}

@article{Georgoudis:2023lgf,
    author = "Georgoudis, Alessandro and Heissenberg, Carlo and Vazquez-Holm, Ingrid",
    title = "{Inelastic exponentiation and classical gravitational scattering at one loop}",
    eprint = "2303.07006",
    archivePrefix = "arXiv",
    primaryClass = "hep-th",
    reportNumber = "NORDITA 2023-010, UUITP-03/23",
    doi = "10.1007/JHEP06(2023)126",
    journal = "JHEP",
    volume = "2023",
    number = "06",
    pages = "126",
    year = "2023"
}
 
\end{document}